\documentclass[12pt,twoside]{mitthesis}
\usepackage{lgrind}
\usepackage{cmap}
\usepackage[T1]{fontenc}
\def\all{all}
\ifx\files\all  \else %\typeout{Including only \files.} \includeonly{\files} \fi

\newcommand{\figures}{Figures/}

\usepackage{lineno}
\usepackage[breaklinks,colorlinks=true, allcolors=blue]{hyperref}
\usepackage{graphicx}
\usepackage{eqnarray}
\usepackage{amssymb}
\usepackage{amsmath}
\usepackage{float}
\usepackage{multirow}
\usepackage{makecell}
\usepackage{tikz}
\usepackage{subfig}
\usepackage{comment}
\usepackage{cite}
\usepackage{adjustbox}
\usepackage{upgreek}
\usepackage{makecell}
\usepackage{silence}
\usepackage{float}
\floatstyle{plaintop}
\restylefloat{table}

\newcommand{\uB}{MicroBooNE}

\newcommand{\CCIpOpi}{CC1p0$\pi$}

\newcommand{\CosThetaMu}{$\cos(\theta_{\mu}$)}

\newcommand{\eep}{\mbox{($e,e'p$)}} 
\newcommand{\ee}{\mbox{$(e,e')$}}

\newcommand{\pmperp}{$P_T$}

\newcommand{\NOVA}{NO$\nu$A}
\newcommand{\Minerva}{MINER$\nu$A}
\newcommand{\eGenie}{eGENIE}
\newcommand{\susa}{SuSAv2}
\newcommand{\Signal}{CC1p0$\pi$\,\,}

\newcommand{\version}{v08_00_00_52}
\newcommand{\BreakdownFigSize}{0.32}

\begin{document}

\title{Lepton-Nucleus Scattering Measurements for\\Neutrino Interactions and Oscillations}

\author{Afroditi Papadopoulou\\\vspace{0.4cm}B.S., National Kapodistrian University of Athens (2016)}

\department{Department of Physics}

\degree{Doctor of Philosophy}

\degreemonth{May}
\degreeyear{2022}
\thesisdate{May 12, 2022}

\supervisor{Or Hen}{Associate Professor}

\chairman{Deepto Chakrabarty}{Associate Department Head of Physics, MIT}

\maketitle

\cleardoublepage

\setcounter{savepage}{\thepage}
\begin{abstractpage}
Currently running and forthcoming precision neutrino oscillation experiments aim to unambiguously determine the neutrino mass ordering, the charge-parity violating phase in the lepton sector and the possible existence of physics Beyond the Standard Model. 
To have an understanding of all the effects necessary for the success of these experiments, lepton-nucleus interactions must be modeled in unprecedented detail. 
With this thesis, expertise in both neutrino and electron cross-section modeling and analysis was leveraged in order to make fundamental and critical improvements to our understanding of these interactions. 
The outlined work takes a significant step towards this high-precision measurement era with three complementary approaches.
Cross sections are reported using neutrino data sets from the MicroBooNE liquid argon time projection chamber detector at Fermi National Laboratory, as well as electron scattering data from the CLAS detector at Thomas Jefferson National Laboratory.
Furthermore, the modeling development of the commonly used GENIE event generator is presented.

\begin{figure}[H]
\centering  
\includegraphics[width=0.94\linewidth]{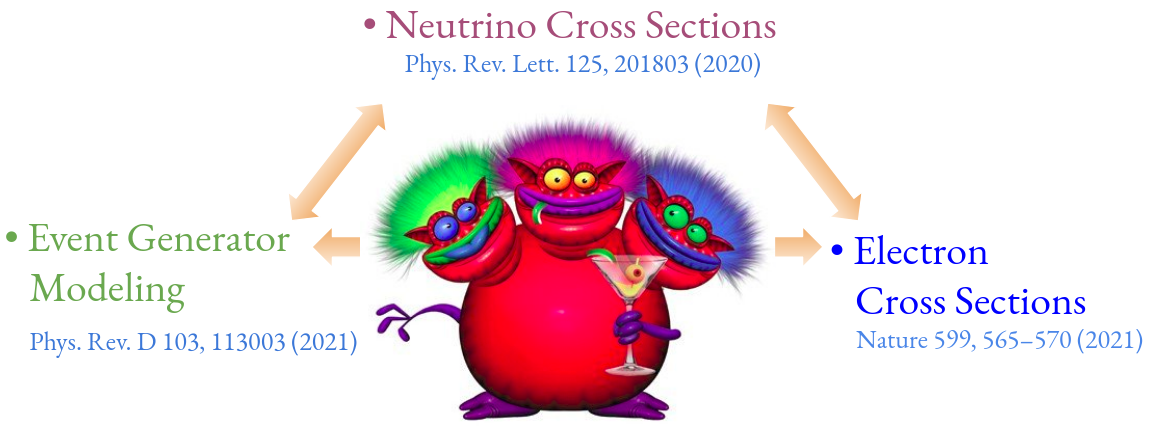}
%\caption{.}
\label{abstractmonster}
\end{figure}
\end{abstractpage}

\cleardoublepage

\section*{Acknowledgments}

%Steven Gardiner 
%Kirsty Duffy
%Michael Kirby
%Andrew Mastbaum
%Steven Dolan
%Stephen Dytman
%Gerallyn Zeller
%Florian
%Minerba
%Rey
%Carlos Ayerbe Gayoso
%Cathy Modica

%%%%%%%%%%%%%%%%%

First and foremost, my eternal gratitude goes to my advisor, Prof. Or Hen. 
Or has provided me with all the resources necessary to proceed in my academic career. 
I am also grateful to Prof. Lawrence Weinstein. 
Their guidance has been crucial in my formation as a physicist.

I’m grateful to all the members of our group for their invaluable support and to all the MIT graduate students in my year.
Many thanks to all the members of the MicroBooNE, ``Electrons-For-Neutrinos'', CLAS, GENIE, and GlueX collaborations for all their input and guidance.
I’m also grateful to my entire family and all my friends for their continuous support and encouragement.
\pagestyle{plain}
  % -*- Mode:TeX -*-
%% This file simply contains the commands that actually generate the table of
%% contents and lists of figures and tables.  You can omit any or all of
%% these files by simply taking out the appropriate command.  For more
%% information on these files, see appendix C.3.3 of the LaTeX manual. 
\tableofcontents
%\newpage
%\listoffigures
%\newpage
%\listoftables

\chapter{Introduction}\label{intro}

%%%%%%%%%%%%%%%%%%%%%%%%%%%%%%%%%%%%%%%%%%%%%%%%%%%%%%%%%%%%%%%%%%

\section{Neutrinos In The Standard Model}\label{NuPhys}

The Standard Model (SM) of particle physics was developed to describe the particles that
are considered to be fundamental and their interactions~\cite{PhysRevD.98.030001}. 
Though an extremely accurate theory, we are already aware of particles exhibiting behaviors not predicted within the scope of the SM. 
These peculiar particles are called neutrinos, which travel enormous distances before they interact with matter, exclusively via weak interactions. 
The recent realization of neutrino oscillations was a historic discovery~\cite{PhysRevLett.81.1562,SNO:2002tuh}. 
The SM had been incredibly successful, yet it requires neutrinos to be massless. 
The new observations clearly showed that neutrinos oscillate between different identities. 
This behavior is driven by their non-zero masses and is indicative of new physics Beyond the Standard Model.

\begin{figure}[htb!]
\centering  
\includegraphics[width=0.6\linewidth]
{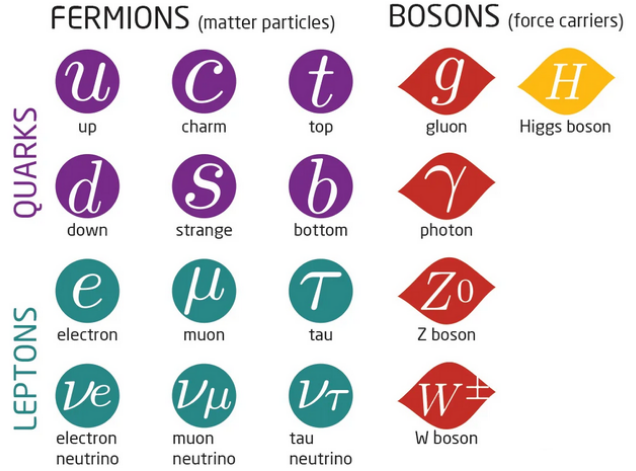}
\caption{The Standard Model of particle physics illustrating the three generations of fermions, the gauge bosons, and the scalar Higgs boson. Figure adapted from ScienceAlert~\cite{sciencealert}.}
\label{sm}
\end{figure}

Though very challenging to detect them, the study of neutrinos is a promising venue towards the generalisation of the SM. 
The latter, initially developed more than 40 years ago, introduced the notion of fundamental particles. 
Those particles are either the building blocks of matter referred to as fermions with semi-integer spins, or mediators of the interactions referred to as bosons with integer spins. 
The collection of the fundamental particles and the carriers is shown in figure~\ref{sm}.

\begin{figure}[htb!]
\centering  
\includegraphics[width=0.4\linewidth]
{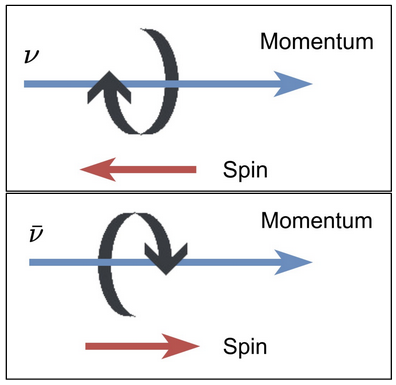}
\caption{Graphic illustrating the left-handed (right-handed) nature of neutrinos (anti-neutrinos) via the orientation of the momentum-spin vectors. Figure adapted from~\cite{VanDePontseele:2020tqz}.}
\label{LHN}
\end{figure}

Prior to the SM, the electroweak theory incorporated neutrinos ($\nu_{e}$ , $\nu_{\mu}$, $\nu_{\tau}$) as left-handed partners of the three families of charged leptons
(e, $\mu$, $\tau$) and the corresponding anti-neutrinos as right-handed partners, which is illustrated in figure~\ref{LHN}. 
The experimental verification that neutrinos (anti-neutrinos) are left-handed (right-handed) was established with the Goldhaber experiment~\cite{PhysRev.109.1015}. 

In the context of the SM, neutrinos are assumed to be massless and their individual lepton number is conserved. Under the assumption that neutrinos are massless, only the left-handed component of the Dirac spinor interacts via the weak force and right-handed components are completely absent. 
Prior to the discovery of neutrino oscillations, there was no experimental evidence of violation of the individual lepton number. 
A typical example of such searches is the muon- and electron-number violating decay $\mu^{\pm} \rightarrow e^{\pm} + \gamma$, where only upper limits have been placed on the branching ratio~\cite{PhysRevD.65.112002}.
Yet, the experimental discovery of neutrino oscillations established that the conservation of individual lepton number is not universal. 
That raises a wealth of questions related to the possible right-handed nature and the non-zero mass of neutrinos. 

If neutrinos are massive, helicity is not exactly conserved. A massive neutrino can have its spin and momentum anti-aligned, and that would correspond to a left-handed neutrino. 
However, in the reference frame of an observer travelling faster than the neutrino itself, the neutrino spin and momentum can be aligned, and that would correspond to a right-handed particle. 
On the other hand, if neutrinos are massless, they must travel at the speed of light.
Since there exists no reference frame travelling faster than that, the neutrino helicity cannot change.

At the same time, the neutrino masses are significantly smaller compared to other particles. 
The latest measurements by KATRIN indicate that electrons are $\mathcal{O}(10^{6})$ more massive than neutrinos~\cite{NatureKATRIN}.
Massless particles require no extra terms to be added in the SM Lagrangian. 
On the other hand, the tiny neutrino masses constitute a riddle that requires fine tuning of the additional Lagrangian terms. 
Instead, many theorists argue in favor of a fundamental reason why the neutrino masses are so small.

In order to accommodate a non-zero neutrino mass, there are two potential approaches. 
The first one includes new particles, namely Dirac neutrinos, and the second introduces a new particle type, namely Majorana neutrinos~\cite{majorana}. 

A Dirac neutrino can acquire mass via its coupling to the Higgs field. 
Those particles that interact with the Higgs field change helicity and, thus, left-handed particles become right-handed and vice versa. 
Experimental results to-date indicate that interacting neutrinos are left-handed~\cite{PhysRev.109.1015}.
A potential extension of the SM includes such right-handed neutrinos that obtain mass via the Higgs field. 
However, these neutrinos do not have electroweak charge, thus they interact only via the mixing with the left-handed counterpart.
If indeed neutrinos are Dirac particles and obtain their mass due to the coupling with the Higgs field, their masses should be comparable to the other fermions predicted by the SM.
However, the neutrino interaction coupling can be tuned to retrieve masses that are comparable with the experimental observations. 

Another possibility would be the introduction of Majorana particles.
With this approach, there is no distinction between neutrinos and anti-neutrinos, with the latter being plausible since neutrinos are electrically neutral. 
Massive neutrinos can be accommodated in this extension of the SM.
Earlier the argument that an observer traveling at the speed of light might observe a flip of the neutrino helicity was used. 
However, if neutrinos are their own anti-particles, there is no helicity change within this massive neutrino hypothesis.
The simplest SM extension with Majorana-like particles is the type 1 seesaw mechanism~\cite{AKHMEDOV2000215}. 
In this framework, a left-handed neutrino interacts with the Higgs boson and a really heavy right-handed neutrino is briefly produced.
The latter further interacts with the Higgs field to produce a light left-handed Majorana neutrino.

%%%%%%%%%%%%%%%%%%%%%%%%%%%%%%%%%%%%%%%%%%%%%%%%%%%%%%%%%%%%%%%%%%%%%%%%%%%%%%%%

\section{Neutrino Oscillations}\label{NuOsc}

Neutrinos are produced in specific flavor eigenstates, namely $\nu_{e}$ , $\nu_{\mu}$ or $\nu_{\tau}$. 
The neutrino mass eigenstates can be expressed as a superposition of flavor eigenstates and vice versa.
The flavor eigenstates evolve as a function of the distance that neutrinos travel, the neutrino energy and the neutrino mass differences squared. 
Such an evolution is referred to as ``neutrino oscillation'' and is a natural consequence of the fact that a neutrino flavor state is composed of multiple mass eigestates. 
For simplicity, the derivation of the formalism for the two-neutrino oscillation case with masses $m_{i}$ for i $\in$ $\{$1,2$\}$ is outlined here. The neutrino wavefunction is treated as a quantum-mechanical plane wave $\psi$ that evolves in time and space as:

\begin{equation}
\psi(L) = \psi(0) \cdot e^{i \cdot p \cdot x}, 
\label{wave}
\end{equation}

with $p \cdot x = Et - \vec{p} \cdot \vec{x}$ corresponding to the Lorentz-invariant phase and with ($E$,$\vec{p}$\,) being the energy and the three-momentum, respectively.
For a particle of mass $m$, its momentum $p$ can be obtained in the highly relativistic limit where $E \gg m$ via a Taylor expansion as: 

\begin{equation}
p = \sqrt{E^{2} - m^{2}} \approx E - m^{2} / 2E.
\label{TaylorMom}
\end{equation}

If a neutrino travels some distance $L$ in vacuum, a phase shift is introduced in its wavefunction given by equation~\ref{wave}, and equation~\ref{TaylorMom} becomes:

\begin{equation}
    \nu_{i}(L) = \nu_{i}(0) \cdot e^{-i \cdot m_{i}^{2} \cdot L/2E}
    \label{evolution}
\end{equation}

where all the constants have been absorbed in the global phase $\nu_{i}(0)$. 
It is important to stress the fact that these expressions are accurate for plane waves propagating in vacuum.
Additional phase shifts have to be introduced when neutrinos travel through a high density material.

In the two-neutrino case, the flavor eigenstates $\nu_{e}$ and $\nu_{\mu}$ can be expressed as a linear superposition of the mass eigenstates $\nu_{1}$ and $\nu_{2}$:

\begin{equation}
\begin{split}
    \nu_{e} = cos\theta \cdot \nu_{1} + sin\theta \cdot \nu_{2},\\
    \nu_{\mu} = -sin\theta \cdot \nu_{1} + cos\theta \cdot \nu_{2},    
\end{split}    
\end{equation}

where $\theta$ is the mixing angle between the two states. 
Consider now a pure beam of electron neutrinos produced at the source, effectively at a distance of $L$ = 0. 
The wavefunction evolves following the formalism in equation~\ref{evolution},

\begin{equation}
    \nu_{e}(L) = cos\theta \cdot e^{-i \cdot m_{1}^{2} \cdot L/2E} \cdot \nu_{1}(0) + sin\theta \cdot e^{-i \cdot m_{2}^{2} \cdot L/2E} \cdot \nu_{2}(0). 
    \label{EvolutionLMass}
\end{equation}

Given that in our detectors the products of weak interactions are reconstructed based on the neutrino flavor, equation~\ref{EvolutionLMass} needs to be rewritten using the flavor basis,

\begin{equation}
\begin{split}
    \nu_{e}(L) = [cos^{2}\theta \cdot e^{-i \cdot m_{1}^{2} \cdot L/2E} + sin^{2}\theta \cdot e^{-i \cdot m_{2}^{2} \cdot L/2E} ] \cdot \nu_{e}(0)\\
    - sin\theta \cdot cos\theta \cdot [e^{i \cdot m_{1}^{2} \cdot L/2E} - e^{i \cdot m_{2}^{2} \cdot L/2E}]  \cdot \nu_{\mu}(0)
\end{split}    
\end{equation}

The probability of detecting a neutrino of a given flavor is obtained via the square of the amplitude,

\begin{equation}
    P_{\nu_{e} \rightarrow \nu_{e}} = \langle \nu_{e}(L) | \nu_{e}(L) \rangle = 1 - sin^{2}(2\theta) sin^{2}(\frac{\Delta m^{2} L }{4E}),  
\end{equation}

\begin{equation}
    P_{\nu_{e} \rightarrow \nu_{\mu}}  = \langle \nu_{e}(L) | \nu_{\mu}(L) \rangle = sin^{2}(2\theta) sin^{2}(\frac{\Delta m^{2} L }{4E}),  
    \label{OscProbE}
\end{equation}

where $\Delta m^{2} = m_{2}^{2} - m_{1}^{2}$ is the neutrino mass difference squared. 
Provided that $\theta \neq$ 0 or $\pi$/2 and $\Delta m^{2} \neq$ 0 for oscillations to take place, the neutrino beam evolves as a function of $L$/$E$.
The amplitude of this oscillation is given by $sin^{2}(2\theta)$. 
The wavelength, expressed in commonly used units, is obtained by

\begin{equation}
    1.27 \frac{\Delta m^{2}\,[eV^{2}] L\,[km]}{E\,[GeV]}.   
\label{MaxMix}    
\end{equation}

The experimental observation of neutrino oscillations serves as proof of evidence that the initial state indeed undergoes a phase shift. 
The explanation for such a shift is that at least some neutrinos have non-zero mass and that the transformation between the mass and the flavor eigenstates involves a non-zero mixing angle.
Equation~\ref{OscProbE} gives the neutrino oscillation probability for the simplified case of only two types of neutrinos.  
The full three-flavor probability for $\nu_\mu\rightarrow \nu_e$ oscillation in vacuum is given by

\begin{eqnarray}
P_{\nu_\mu\rightarrow \nu_e}(E,L)&\approx&
A\sin^2\frac{\Delta m^2_{13}L}{4E} \\
&&-B\cos\left(\frac{\Delta m^2_{13}L}{4E} +\delta_{CP}\right)
\sin\frac{\Delta m^2_{13}L}{4E}, \nonumber
\end{eqnarray}

where $\Delta m^2_{13} = m_{1}^2 - m_{3}^2$ is the neutrino mass difference squared that determines the oscillation wavelength as a function of $L$/$E$ and $\delta_{CP}$ is the charge-parity (CP) symmetry violating phase~\cite{Freund:2001pn,cervera2000,cervera2001}.  
The coefficients $A$ and $B$ depend primarily on the neutrino oscillation mixing angles,

\begin{eqnarray}
A=\sin^2\theta_{23}\sin^22\theta_{13},
B=-\frac{\sin2\theta_{12}\sin2\theta_{23}}{2\sin\theta_{13}}\sin\frac{\Delta
m^2_{21}L}{4E}\sin^22\theta_{13}. 
\end{eqnarray}

The different flavor neutrinos (labelled $\nu_e,\nu_\mu$ and $\nu_\tau$) are linear combinations of the different mass neutrinos labelled $1,2,3$.

The outlined neutrino mixing can be generalized to N neutrino eigenstates. 
The probability of finding an $\alpha$ flavor eigenstate neutrino in a flavor eigenstate $\beta$, after traveling some distance $L$, is given by:

\begin{equation}
\begin{split}
    P_{\alpha \rightarrow \beta} = \delta_{\alpha\beta} - 4 \sum_{j > i} Re[U^{*}_{\alpha i} U_{\beta i} U_{\alpha i} U^{*}_{\beta j}] sin^{2}\left(\left[\frac{1.27\,GeV}{eV^{2}\,km}\right]\frac{\Delta m_{ji}^{2} L}{E}\right)\\
    + 2 \sum_{j > i} Im[U^{*}_{\alpha i} U_{\beta i} U_{\alpha i} U^{*}_{\beta j}] sin\left(\left[\frac{2.54\,GeV}{eV^{2}\,km}\right]\frac{\Delta m_{ji}^{2} L}{E}\right)
    \label{OscProb}
\end{split}    
\end{equation}

where U is the N$\times$N unitary neutrino mixing matrix and $\Delta m_{ji}^{2} = m^{2}_{j} - m^{2}_{i}$~\cite{Fund}.
The corresponding antineutrino oscillation probability can be obtained by replacing $U\rightarrow U^{\dagger}$.

Within the widely accepted neutrino model, there exist three active neutrinos~\cite{2006257}, resulting into two
squared mass splittings, $\Delta m_{21}^{2}$ and $\Delta m_{32}^{2}$. 
The corresponding 3 $\times$ 3 mixing matrix, referred to as the Pontecorvo-Maki-Nakagawa-Sakata (PNMS) matrix, relates the mass eigenstates ($\nu_{1}$, $\nu_{2}$, $\nu_{3}$) to the flavor eigenstates ($\nu_{e}$, $\nu_{\mu}$, $\nu_{\tau}$):

\begin{equation}
\begin{pmatrix}
\nu_{e}\\
\nu_{\mu}\\
\nu_{\tau}\\
\end{pmatrix}
=
\begin{pmatrix}
U_{e1} & U_{e2} & U_{e3}\\
U_{\mu1} & U_{\mu2} & U_{\mu3}\\
U_{\tau1} & U_{\tau2} & U_{\tau3}\\
\end{pmatrix}
\cdot
\begin{pmatrix}
\nu_{1}\\
\nu_{2}\\
\nu_{3}\\
\end{pmatrix}
\end{equation}

The mixing of the mass-flavor eigenstates is parametrised with three mixing angles ($\theta_{12}$, $\theta_{23}$, $\theta_{13}$) and the $\delta_{CP}$ violating phase.
The best measured values of the these angular parameters to date are obtained from global
fits as documented in the latest Review of Particle Physics edition~\cite{PhysRevD.98.030001} and listed in table~\ref{OscParTable}.

\begin{center}
\begin{table}[htb!]
\centering
\begin{tabular}{ c c c }
\hline
\hline
& Normal Ordering & Inverted Ordering\\
\hline
\hline
$sin^{2}\theta_{12}$ & $0.320^{+0.020}_{-0.016}$ & $0.320^{+0.020}_{-0.016}$ \\
$sin^{2}\theta_{23}$ & $0.547^{+0.020}_{-0.030}$ & $0.551^{+0.018}_{-0.030}$ \\
$sin^{2}\theta_{13}$ & $(2.160^{+0.083}_{-0.069})\cdot 10^{-2}$ & $(2.220^{+0.074}_{-0.076})\cdot 10^{-2}$ \\  
$\Delta m_{21}^{2}$ & $(7.55^{+0.02}_{-0.16})\cdot 10^{-5} \,eV^{2}$ & $(7.55^{+0.02}_{-0.16})\cdot 10^{-5} \,eV^{2}$ \\
$\Delta m_{32}^{2}$ & $(2.42^{+0.03}_{-0.03})\cdot 10^{-3} \,eV^{2}$ & $(-2.50^{+0.04}_{-0.03})\cdot 10^{-3} \,eV^{2}$ \\  
\hline
\hline
\end{tabular}
\caption{Summary table of the 3$\nu$ oscillation parameters. The 1$\sigma$ intervals for both the case normal are inverse ordering are shown. Table adapted from~\cite{PhysRevD.98.030001}.}
\label{OscParTable}
\end{table} 
\end{center}

The $\delta_{CP}$ phase and the neutrino mass hierarchy ($m_{2}$ < $m_{3}$ or the other way around) are still not determined.
A potential way to resolve these open questions is via the Mikheyev-Smirnov-Wolfenstein (MSW) effect~\cite{MSW}.
In the presence of matter, the neutrino wavefunction propagation is modified.
When a neutrino travels through a dense material, the energy-momentum relationship is affected via coherent interactions with matter particles.
Given that regular matter on Earth contains electrons, but not muons or taus, charged current interactions with the medium only affect the electron neutrino propagation. 
This interaction flavor dependence results in measurable changes in the contribution of the different flavors. 
Thus, the MSW effect is exploited to perform long-baseline neutrino experiments.
The Deep Underground Neutrino Experiment (DUNE) is a forthcoming multi-billion-dollar international experiment aiming to resolve the aforementioned open questions.
To successfully achieve that goal, DUNE will utilize an intense muon beam, with a near detector located at Fermi National Laboratory IL and a far detector $\approx$ 1300\,km away from the neutrino source at Sanford Underground Research Facility SD. 

Before DUNE starts taking data, a number of other neutrino experiments will have already provided constraints for $\delta_{CP}$ and the mass hierarchy~\cite{10.3389/fspas.2018.00036}.
For these results to be obtained, different neutrino oscillation experiments utilize a number of neutrino sources covering a wide range of energies.

Reactor neutrino experiments exploit the large electron anti-neutrino fluxes produced in nuclear reactors by $\beta$ decays of heavy nuclei (nuclear fissions of $^{235}U$, $^{238}U$, $^{239}Pu$, $^{241}Pu$). 
The typical energy scale of reactor $\overline{\nu}_{e}$’s is a few MeVs.

Atmospheric neutrino experiments take advantage of neutrinos produced by cosmic rays.
The latter interact with the upper layers of the atmosphere producing a large flux of pions and kaons. 
These decay in the atmosphere into muons and muon neutrinos. 
These muons might decay into electrons, electron anti-neutrinos, and muon neutrinos before they reach the Earth. 
Atmospheric neutrino experiments aim to detect these muon neutrinos. 
The energy of these neutrinos spans a really wide range up to $\mathcal{O}$(100)\,GeV.
In some cases, like in IceCube~\cite{PhysRevD.91.072004}, even neutrinos with PeV energies can be detected.
Such neutrinos cover distances between $\mathcal{O}$(10)\,km - for neutrinos produced in the upper layers of the atmosphere directly above the detector - to $\mathcal{O}(10^{4})$\,km - for netrinos that are produced on the other side of the Earth and travel through the core.
Atmospheric experiments provide the current best limit of $\Delta m^{2}_{32} = 2.56^{+0.13}_{-0.11}\times 10^{-3} eV^{2}$~\cite{PhysRevD.98.030001}.

Solar neutrino experiments detect the electron neutrinos produced in the Sun core due to nuclear fusion processes. 
Solar neutrino experiments are designed to detect the $\nu_{e}$'s produced via these processes.
The currently best limit corresponds to an extremely small value of $\Delta m^{2}_{21} = 7.37^{+0.59}_{-0.44} \times 10^{-5} eV^{2}$~\cite{PhysRevD.98.030001}, much smaller than the $\Delta m^{2}_{32}$ splitting mentioned above.

Accelerator-based experiments use muon neutrino beams produced via the decay of primarily pions produced when a proton beam hits a heavy target. 
Such experiments are further classified into appearance and disappearance experiments.
The former search for electron neutrinos oscillated from the initial muon neutrino beam.
The latter look for the reduction of muon neutrino interactions due to oscillations. 

Different experiments are designed to be sensitive to different values of $\Delta m^{2}$ by choosing the appropriate $L$/$E$ ratio. 
Building on equation~\ref{MaxMix}, the value of $\Delta m^{2}$ for which

\begin{equation}
\frac{\Delta m^{2} L}{2 E} \simeq 1
\label{baseline}
\end{equation}

corresponds to the $\Delta m^{2}$ sensitivity of a given experiment. 

Neutrino oscillation experiments are further classified depending on the average value of this $L$/$E$ ratio. 
Using equation~\ref{baseline}, short baseline experiments with $L/E \lesssim$ 1 km/GeV are sensitive to $\Delta m^{2} \gtrsim  1\,eV^{2}$. 
Long baseline experiments where $L/E \lesssim 10^{3}$ km/GeV are sensitive to 
$\Delta m^{2} \gtrsim 10^{-3}\,eV^{2}$.

%%%%%%%%%%%%%%%%%%%%%%%%%%%%%%%%%%%%%%%%%%%%%%%%%%%%%%%%%%%%%%%%%%%%%%%%%%%

\section{Long-Baseline Accelerator-Based Neutrino\texorpdfstring{\newline}{ }Oscillation Experiments}\label{longbase}

\begin{figure}[htb!]
\centering  
\includegraphics[width=0.7\linewidth]
{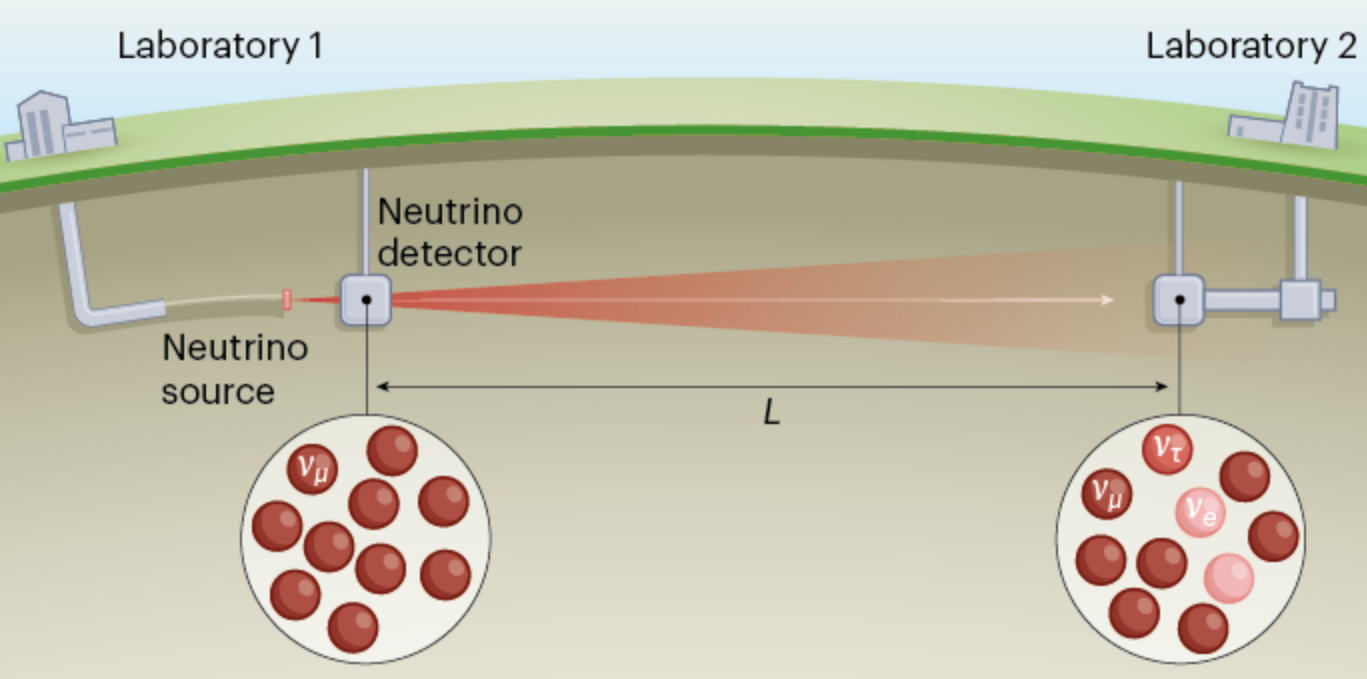}
\caption{The experimental setup of an accelerator-based long-baseline neutrino oscillation experiment.}
\label{LongBase}
\end{figure}

Long-baseline accelerator-based experiments consist of a near detector (ND) positioned close to the neutrino source, and a far detector (FD), which is positioned close to the oscillation maximum,, as shown in figure~\ref{LongBase}.
The number of interactions of flavor $\alpha$ in the near detector ($N_{ND}^{\alpha}$) is obtained as

\begin{equation}
N^{ND}_{\alpha}(E_{reco}) = \int \Phi_{\alpha}^{ND}(E_{true}) \sigma_{\alpha}(E_{true}) \epsilon_{\alpha}(E_{true}) f^{ND}(E_{true}, E_{reco}) dE_{true}
    \label{NDRate}
\end{equation}

where $\Phi_{\alpha}^{ND}(E_{true})$ is the neutrino flux of flavor $\alpha$ close to the source, $\sigma_{\alpha}(E_{true})$ is the cross section for a given flavor $\alpha$, $\epsilon_{\alpha}(E_{true})$ is the reconstruction efficiency for flavor $\alpha$, and $f^{ND}(E_{true}, E_{reco})$ is the detector response function, describing how $E_{true}$ is mapped to $E_{reco}$ in the near detector.
For $\nu_{\alpha} \rightarrow \nu_{\beta}$ oscillations, the number of events in the far detector can be obtained as

\begin{equation}
N^{FD}_{\alpha\rightarrow\beta}(E_{reco}) = \int \Phi_{\alpha}^{FD}(E_{true}) P_{\alpha\rightarrow\beta}(E_{true}) \sigma_{\beta}(E_{true}) \epsilon_{\beta}(E_{true}) f^{FD}(E_{true}, E_{reco}) dE_{true}
    \label{FDRate}
\end{equation}

where $N^{FD}_{\alpha\rightarrow\beta}(E_{reco})$ is the number of $\beta$ flavor interactions, $\Phi_{\alpha}^{FD}(E_{true})$ is the neutrino flux of flavor $\alpha$, $P_{\alpha\rightarrow\beta}(E_{true})$ is the oscillation probability for $\alpha\rightarrow\beta$, $\sigma_{\beta}(E_{true})$ is the cross section for flavor $\beta$, $\epsilon_{\beta}(E_{true})$ is the reconstruction efficiency for flavor $\beta$, and $f^{FD}(E_{true}, E_{reco})$ is the detector response function, describing how $E_{true}$ is reconstructed in $E_{reco}$.

Accelerator-produced neutrino beams predominantly contain muon neutrinos~\cite{AguilarArevalo:2008yp}. 
Therefore, long-baseline accelerator-based neutrino oscillation experiments focus on muon neutrino disappearance and electron neutrino appearance studies.
The former are sensitive to the oscillation parameters $\theta_{23}$ and $\Delta m^{2}_{23}$, while the latter have sensitivity to $\theta_{13}$ and $\delta_{CP}$~\cite{Abi:2020qib}.
At a distance $L$ from the neutrino production point, some muon neutrinos will oscillate to electron neutrinos, resulting in fluxes of approximately

\begin{eqnarray}
\Phi^{FD}_e(E_{true})&\propto& \;\;\;\;\;\;\;\; P_{\nu_\mu\rightarrow \nu_e}(E_{true},L)\; \Phi^{ND}_\mu(E_{true}), \\ \nonumber
\Phi^{FD}_\mu(E_{true})&\propto& \left[1-P_{\nu_\mu\rightarrow \nu_e}(E_{true},L)\right]\Phi^{ND}_\mu(E_{true}),
\end{eqnarray}

where the proportionality constant depends on the experiment geometry, that can be affected by the different experimental acceptances at the ND and FD locations, and $P_{\nu_\mu\rightarrow \nu_e}$ is the electron neutrino appearance probability.
$\nu_\mu\rightarrow \nu_e$ oscillations are thus observed by measuring the neutrino fluxes $\Phi^{FD}_{e}(E_{true})$ and $\Phi^{FD}_{\mu}(E_{true})$ as a function of energy and distance.  
The three-flavor oscillation equations are similar but include additional terms.  
Charge-parity (CP) symmetry violation in the leptonic sector would add a phase $\delta_{CP}$ to the three-flavor oscillation with an opposite sign for neutrinos and anti-neutrinos~\cite{fukugita86,T2KNature20}.

\begin{figure} [htb!]
\begin{center}
\includegraphics[width=\linewidth]{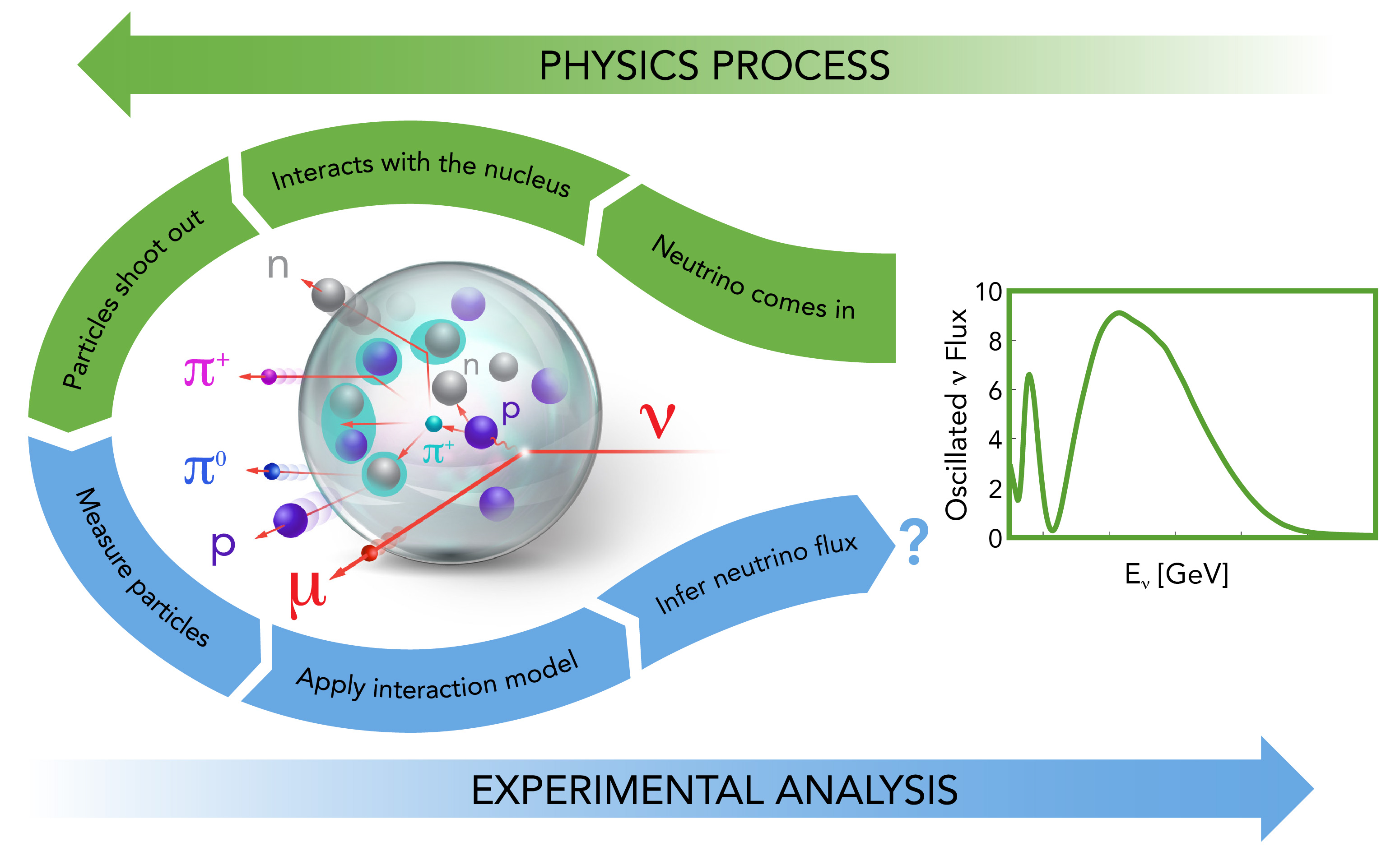}
\end{center}
\caption{\label{fig:Fig1}Neutrino energy spectra reconstruction depends on our
ability to model the interaction of neutrinos with atomic nuclei and
the propagation of particles through the atomic nucleus. This flow
chart shows the process, starting with an oscillated far-detector
 incident-energy spectrum (green), differentiating the physical
neutrino interactions (green arrows) from the
experimental analysis (blue arrows), and ending up with an
inferred incident-energy spectrum that hopefully matches the actual one. 
}
\end{figure}

Therefore, the precision to which oscillation parameters can be determined experimentally depends on our ability to extract $\Phi^{FD}_{\alpha}(E)$ from $N^{FD}_{\alpha}(E_{reco})$, as can be seen in equation~\ref{FDRate} and is graphically illustrated in figure~\ref{fig:Fig1}.

While experimental effects are generally understood and can be minimized using improved detectors, nuclear effects are irreducible and must be accounted for using theoretical models, typically implemented in neutrino event generators.
Thus, the experimental sensitivity is largely determined by the accuracy of the theoretical models used to calculate the interaction cross sections and the accuracy of the energy reconstruction.  
The available models have many free parameters that are poorly constrained and are ``tuned'' by each neutrino experiment.  
Current oscillation experiments report significant systematic uncertainties due to these interaction models~\cite{T2K,Alvarez-Ruso:2017oui,T2KNature20,NOvA:2018gge}. 
Simulations further show that energy reconstruction errors can lead to significant biases in extracting $\delta_{CP}$ at DUNE~\cite{Ankowski:2015kya}.  
There is a robust theoretical effort to improve these models~\cite{Rocco2020,PhysRevD.101.033003,PhysRevLett.116.192501}.

Since there are no mono-energetic high-energy neutrino beams, these models cannot be tested for individual neutrino energies.
Instead, experiments tune models to their near-detector data, where the unoscillated flux $\Phi^{ND}_{\alpha}(E_{true})$ is calculated from hadronic reaction rates~\cite{PhysRevD.94.092005,NovaNearDetector,T2KNearDetector}.
While highly informative, such integrated constraints are insufficient to ensure that the models are correct for each value of $E_{true}$.
Thus, for precision measurements using a broad-energy neutrino beam, the degree to which the near-detector data alone can constrain models is unclear for a number of reasons.
In some cases, the near and far detectors use different target nuclei, which demands the cross-section extrapolation between the two targets.
At the same time, the fluxes at the two detectors are not identical due to their distance from the source, and the fact that neutrinos oscillate.
The near detector measures neutrino interactions originating from pion decays in a long pipe with a length of $\mathcal{O}$(50\,m), while the far detector measures neutrinos from a much smaller solid angle.
Thus, the particle acceptance is different between the two detectors.
Furthermore, the selection efficiencies $\epsilon_{\alpha}(E_{true})$ in equations~\ref{NDRate} and~\ref{FDRate}, both for the signal and the background events at the two detectors, are model dependent and different.

Yet another challenge for neutrino oscillation experiments is the accuracy of the true neutrino energy reconstruction that enters the oscillation probability formula shown in equation~\ref{OscProb}.
The neutrino energy is reconstructed using the measured energy deposition of the final state particles. However, energy losses due to the energy of neutral particles, particles below thresholds, and inactive detector regions have to be taken into account.
Inevitably, assumptions about these effects have to be made based on the underlying modeling choices and on the detector capabilities.
This procedure introduces systematic uncertainties that might limit the precise reconstruction of the true neutrino energy.
Thus, the success of forthcoming neutrino experiments like DUNE relies on the accurate identification and reconstruction of all particles produced.
Hence, tracking detectors with low detection thresholds are key elements.

Furthermore, neutrinos traveling through matter experience a potential due to the coherent elastic scattering with electrons and nucleons.
Coherent scattering takes place when a neutrino wavefunction interacts with matter as a whole.
The implication of this behavior is that neutrinos and antineutrinos are affected in different ways, due to the lack of positrons in regular matter. 
This effect can mimic a CP violating picture with $P(\nu_{\alpha}\rightarrow\nu_{\beta}) \neq P(\bar{\nu}_{\alpha}\rightarrow\bar{\nu}_{\beta})$, though it contains no fundamental information related to the matter-antimatter asymmetry.
Therefore, accounting for matter effects in oscillation experiments that aim to extract the oscillation parameters is crucial.

All these open questions need to be addressed in order to ensure the success of forthcoming high-precision neutrino oscillation experiments.
Hence, this thesis progresses in that direction by improving the understanding of lepton-nucleus interactions described in sections~\ref{NuNucleusInt} and~\ref{LepXS}.

%%%%%%%%%%%%%%%%%%%%%%%%%%%%%%%%%%%%%%%%%%%%%%%%%%%%%%%%%%%%%%%%%%%%%%%%%

\section{Neutrino-Nucleus Interactions}\label{NuNucleusInt}

Neutrino-nucleus interactions are extremely complicated processes that become even more complex on heavier nuclei, such as on argon, the target nucleus used in liquid argon time projection chambers.
These interactions are approximated by deploying a two-step approach.

\begin{figure}[htb!]
\centering  
\includegraphics[width=\linewidth]
{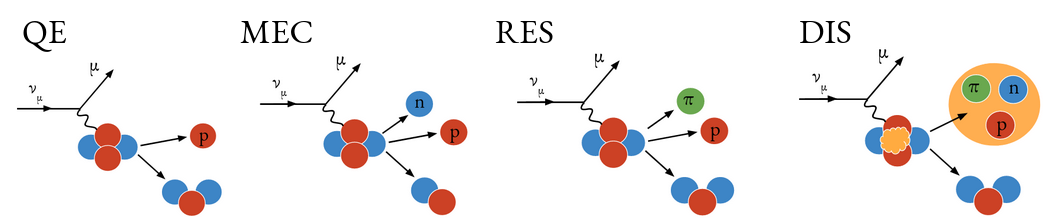}
\caption{The four main interaction processes for neutrino-nucleus scattering events.}
\label{IntModes}
\end{figure}

The first step involves the primary neutrino interaction with a single nucleon or a pair of nucleons with the nucleons being treated as quasi-free objects. 
There are four main types of neutrino scattering processes that play a major role in the energy spectrum of neutrino experiments like MicroBooNE and DUNE, namely quasielastic (QE), meson exchange currents (MEC), resonant production (RES), and deep inelastic scattering (DIS). 
These four interaction types and the relevant outgoing particles are shown in figure~\ref{IntModes}. 

\begin{figure}[htb!]
\centering  
\includegraphics[width=0.7\linewidth]
{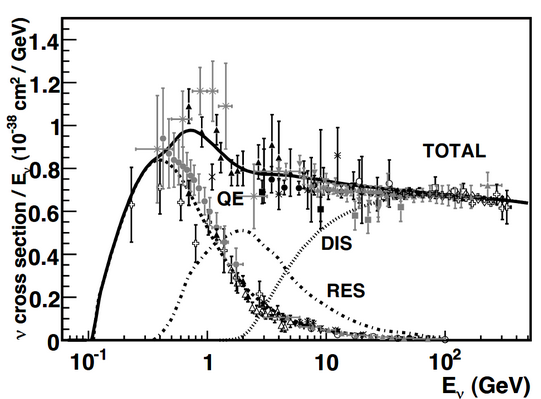}
\caption{Neutrino cross-section evolution as a function of the neutrino energy illustrating the energy range where each one of the four main processes dominates. Figure adapted from~\cite{RevModPhys.84.1307}.}
\label{IntXSec}
\end{figure}

Each one of those processes dominates in different energy ranges, as illustrated in figure~\ref{IntXSec}.
At neutrino energies below $\approx$ 1\,GeV, QE interactions are the ones that dominate. 
With this process, leptons scatter off, and liberate, a single nucleon from the target nucleus.
In the region around $\mathcal{O}$(1 GeV), a set of nuclear forces include the exchange of virtual mesons between two (or more) nucleons, which is referred to as 2 particle-2 hole effects (2p2h).
Such an effect has major contributions from MEC events that lead to the emission of two nucleons from the primary neutrino interaction point. 
For interaction energies greater than the $\Delta$ baryon mass of 1232\,MeV, RES interactions become energetically allowed.
Such processes are the dominant ones in the energy range between 1-4\,GeV.
With such an interaction type, the struck nucleon is brought into an excited state that is called a baryon resonance.
The resonance deexcites and that leads to the emission of a single nucleon and a single pion in the final state.
At even higher energies starting at $\approx$ 4\,GeV, the incoming neutrino scatters off a quark in the nucleus.
Such an energy transfer results in the production of multiple hadrons in the final state.
Interference terms across the different interaction processes need to be added coherently at the quantum-mechanical amplitude level.
However, commonly used neutrino event generators, such as GENIE~\cite{Genie2010}, do not include such interference terms between the different reaction modes.
In other words, the total cross section is obtained by adding the individual cross sections $\sigma_i(E)$ incoherently.

\begin{figure}[htb!]
\centering  
\includegraphics[width=0.7\linewidth]
{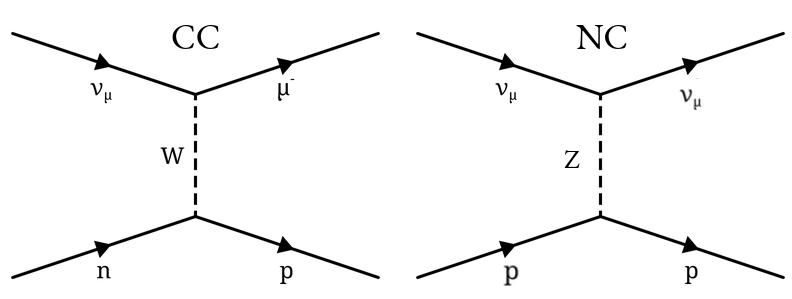}
\caption{Feyman diagrams illustrating charged current (CC) and neutral current (NC) processes.}
\label{currents}
\end{figure}

For neutrino scattering events, interaction processes can take place in the form of either charged current (CC) or neutral current (NC) interactions. 
The former includes the exchange of charged W bosons, and the latter of neutral Z bosons respectively.
The corresponding Feyman diagrams are shown in figure~\ref{currents}.

Beyond the possible interactions and processes mentioned above, neutrino-nucleon interactions become even more sophisticated due to the fact that the interaction takes place in a dense nuclear medium, such as an argon nucleus.  
These complex nuclear effects can be further classified into initial and final state interactions.

\begin{figure}[htb!]
\centering  
\includegraphics[width=0.6\linewidth]
{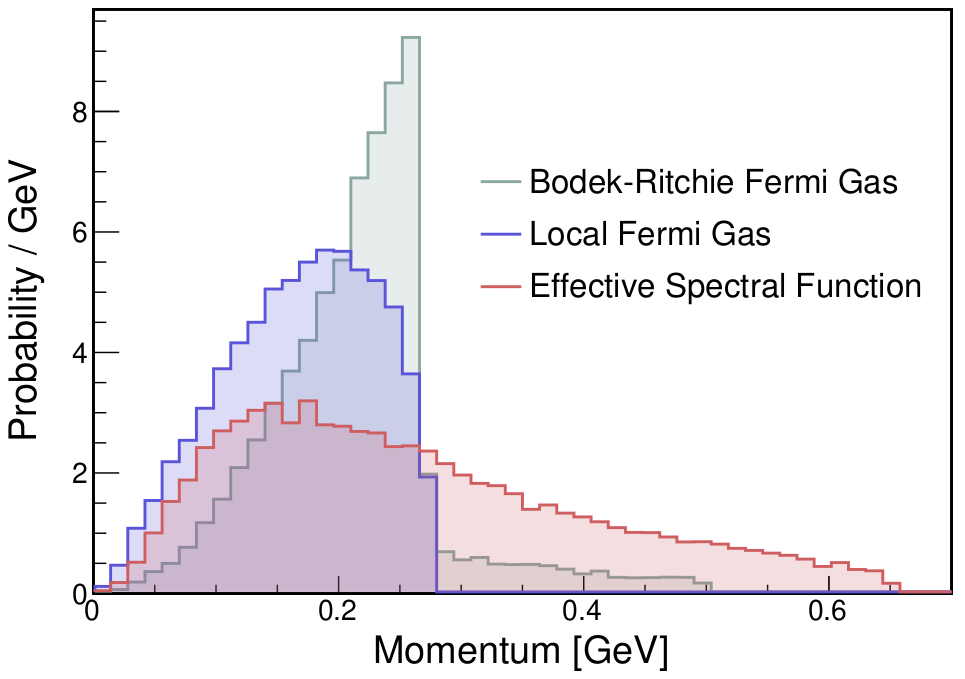}
\caption{Nucleon momentum distribution options available in commonly used neutrino event generators. Figure adapted from~\cite{MarcoThesis}.}
\label{momdistr}
\end{figure}

Initial state interactions are associated with the nucleon-nucleon correlations in the target nucleus.
The nucleons inside the target nucleus are not at rest.
Their momentum distribution can be approximated in a number of ways, typical examples of which would be a Bodek-Richie Fermi Gas~\cite{PhysRevD.23.1070}, a Local Fermi Gas~\cite{Carrasco:1989vq}, or an Effective Spectral Function~\cite{PhysRevC.74.054316}, as illustrated in figure~\ref{momdistr}.
The precise behavior of the Fermi motion is unknown, and it results in the smearing of the reconstructed value of the true neutrino energy.

\begin{figure}[htb!]
\centering  
\includegraphics[width=0.6\linewidth]
{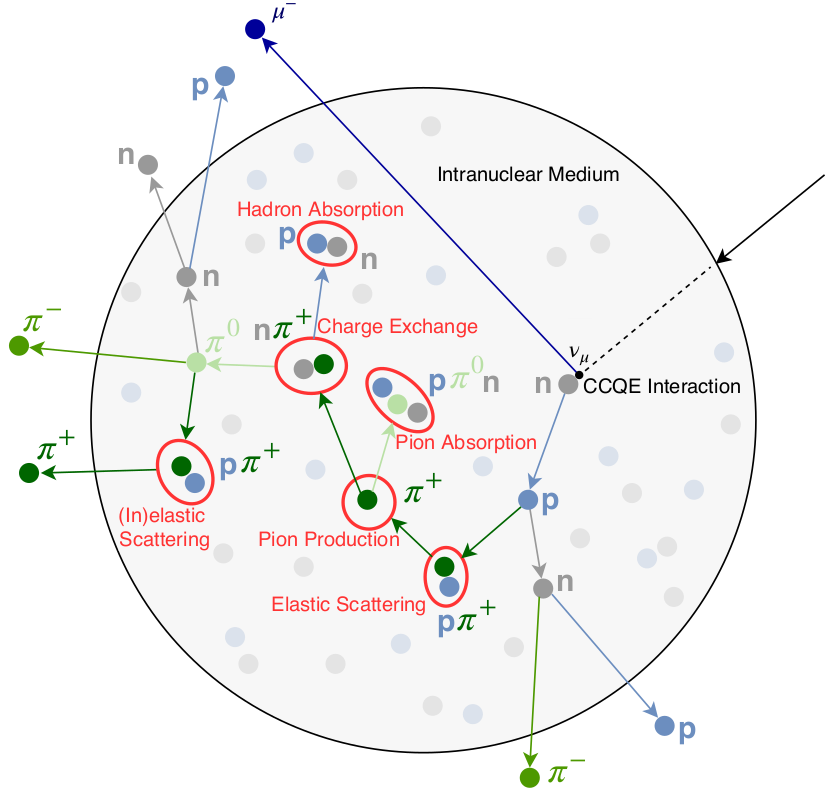}
\caption{Schematic illustration the wealth of possible final state interactions that the final state hadrons can undergo. Figure adapted from~\cite{LarsThesis}.}
\label{fsi}
\end{figure}

In the second part of the factorization process, after the primary neutrino-nucleus interaction, the outgoing nucleons can undergo a wealth of final state interactions while transversing the nuclear medium, before they exit the nucleus.
Such re-interactions might result in processes such as the emission of further hadrons, the absorption of the initially emitted hadrons, charge exchange processes, and/or acceleration/deceleration of the primary hadrons, as illustrated in figure~\ref{fsi}.

All the effects described above have a direct impact on the ability to accurately reconstruct the energy of the incoming neutrino using the properties of the final state particles.

%%%%%%%%%%%%%%%%%%%%%%%%%%%%%%%%%%%%%%%%%%%%%%%%%%%%%%%%%%%

\section{Connections To Electron Scattering}\label{LepXS}

Neutrinos and electrons interact with atomic nuclei by exchanging intermediate vector bosons, a massive $W^\pm$ or $Z$ for the neutrino and a massless photon for the electron.  
Electrons interact via a vector current $j^\mu_{EM}=\bar u\gamma^\mu u$ and neutrinos interact via vector and axial-vector $j_{CC}^\mu=\bar u\gamma^\mu(1 - \gamma^5)u\frac{-ig_W}{2\sqrt{2}}$ currents.

Fundamental considerations give an inclusive $(e,e')$ electron-nucleon elastic scattering cross section that depends on only two structure functions~\cite{Katori_2017},

\begin{eqnarray}
\frac{d^2\sigma^{e}}{dxdQ^2}=\frac{4\pi\alpha^2}{Q^4}\left[\frac{1-y}{x}F^{e}_2(x,Q^2)
  + y^2F^{e}_1(x,Q^2)\right] \,.
\label{eq:sigmaeAPrime}
\end{eqnarray}

\noindent Here $F^{e}_1$ and $F^{e}_2$ are the standard electromagnetic vector structure functions,  $Q^2={\bf q}^2-\nu^2$ is the squared momentum transfer and {\bf q} and $\nu$ are the three-momentum and energy transfers, $x=Q^2/(2m\nu)$ is the Bjorken scaling variable, $m$ is the nucleon mass, $y=\nu/E_e$ is the electron fractional energy loss, and $\alpha$ is the fine structure constant.  
This formula shows the simplest case where $Q^2 \gg m^2$.

The corresponding inclusive charged-current (CC) $(\nu,l^\pm)$ neutrino-nucleon elastic cross section has a similar form, where $l^\pm$ is the outgoing charged lepton~\cite{Amaro_2020}.  
The vector part of the current is subject to the same fundamental considerations as above, but the axial-vector part of the current does not conserve parity.  
This leads to a third, axial, structure function,

\begin{eqnarray}
\begin{aligned}
\frac{d^2\sigma^{\nu}}{dxdQ^2}=&\frac{G_F^2}{2\pi}\left[\frac{1-y}{x}F^{\nu}_2(x,Q^2)
  + y^2F^{\nu}_1(x,Q^2) \right. \\
  &\left. - y(1-y/2)F^{\nu}_3(x,Q^2)\right] \,.
\label{eq:sigmaNuAPrime}
\end{aligned}
\end{eqnarray}

\noindent Here $F^{\nu}_1$ and $F^{\nu}_2$ are the parity-conserving neutrino-nucleus vector structure functions, $F^{\nu}_3$ is the axial structure function, and $G_F$ is the Fermi constant.  
The vector form factors, $F^{\nu}_1$ and $F^{\nu}_2$, have both vector-vector and axial-axial contributions.

\begin{figure}[htb!]
\centering  
\includegraphics[width=0.6\linewidth]
{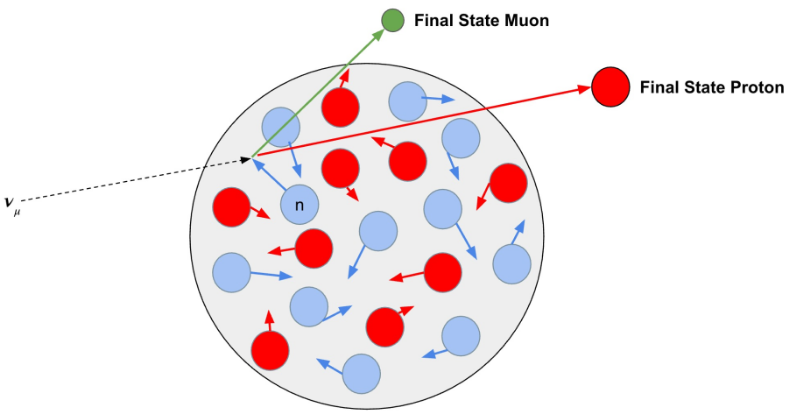}
\caption{Quasielastic lepton-nucleus scattering where the outgoing nucleon does not reinteract as it leaves the nucleus.}
\label{qenofsi}
\end{figure}

These simple equations are very similar for electron-nucleus scattering.
In the limit of electron-nucleon elastic scattering ($x=1$), the two structure functions reduce to the Dirac and Pauli form factors, which are linear combinations of the electric $G_E(Q^2)$ and magnetic $G_M(Q^2)$ form factors.   
Neutrino-nucleon elastic scattering has an additional axial form factor.  
In the simplest case where a lepton scatters quasielastically from a nucleon in the nucleus and the nucleon does not reinteract as it leaves the nucleus shown in figure~\ref{qenofsi}, the lepton-nucleus cross section is the integral over all initial state nucleons,

\begin{eqnarray}
\begin{aligned}
\frac{d\sigma}{dE d\Omega}=&\int_{\mathbf{p_i}}\int_{E_b}
d^3\mathbf{p_i} dE_b K
S(\mathbf{p_i},E_b)\frac{d\sigma^{free}}{d\Omega} \\
&\quad\delta^3(\mathbf{q}-\mathbf{p_f}-\mathbf{p_r})\delta(\omega
- E_b - T_f - T_r),
\end{aligned}
\end{eqnarray}

where $\mathbf{p_i}$ and $\mathbf{p_f}=\mathbf{q}+\mathbf{p_i}$ are the initial and final momenta of the struck nucleon in the absense of any reinteraction, $\mathbf{p_r}=-\mathbf{p_i}$  is the momentum of the recoil $A-1$ nucleus, $E_b$ is the nucleon binding energy, $S(\mathbf{p_i},E_b)$ is the probability of finding a nucleon in the nucleus with momentum $\mathbf{p_i}$ and binding energy $E_b$, $T_f$ and $T_r$ are the kinetic energies of the final state nucleon and $A-1$ system, $d\sigma^{free}/d\Omega$ is the lepton-bound nucleon elastic cross section, and $K$ is a known kinematic factor.

This simple picture is complicated by nucleon reinteractions which change the overlap integral between the initial and final states, and thus the cross section.
The latter further changes the momentum and angle of the outgoing nucleon.
Thus, to calculate even the simplest type of lepton-nucleus interaction, the momentum and binding energy distribution of all nucleons in the nucleus need to be known, as well as how the outgoing nucleon wave function is distorted by the nucleon-nucleus potential, and how the outgoing nucleon kinematics is changed by final state interactions.  

Electron-nucleus scattering is much easier to understand than neutrino-nucleus scattering for three reasons: (a) electron beams have a single, well-known, energy;
(b) electron experiments typically have low statistical uncertainty because electron
  beams have high flux, and electron-nucleus cross sections are far
  higher than their neutrino counterparts; and
(c) electron cross sections are purely vector.
The strength of the interaction is very different: $4\pi\alpha^2/Q^4$ for electrons versus $G_F^2/(2\pi)$ for neutrinos, where the factor of $1/Q^4$ in the electron cross section is due to the exchanged boson mass (i.e. massless photon) in the propagator.  
When compensating for the factor of $1/Q^4$, the shapes of the electron- and neutrino-nucleus cross sections are very similar~\cite{PhysRevD.103.113003}.
Nuclear medium effects such as nucleon motion, binding energy, two-body currents, and final state interactions are similar or identical.
Therefore electron-nucleus scattering can be used to constrain models of neutrino-nucleus scattering.  
Any model which fails to accurately describe electron-nucleus (vector-vector) scattering data cannot be used with confidence to simulate neutrino-nucleus (vector-vector + axial-axial + vector-axial) interactions.
Thus, models of the neutrino-nucleus cross section must be able to describe the more limited electron-nucleus cross section.

%%%%%%%%%%%%%%%%%%%%%%%%%%%%%%%%%%%%%%%%%%%%%%%%%%%%%%%%%%%

\section{Thesis Scope}\label{scope}

\begin{figure}[htb!]
\centering  
\includegraphics[width=0.94\linewidth]{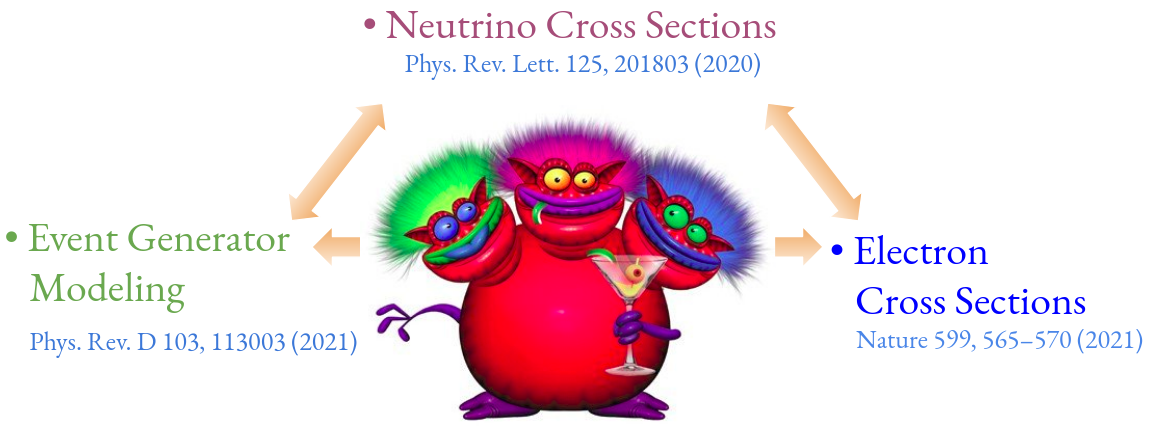}
\caption{Graphic illustration of the three complementary approaches used in this thesis to improve our understanding of lepton-nucleus interactions.}
\label{intromonster}
\end{figure}

As detailed in section~\ref{longbase}, the success of future precision neutrino oscillation experiments depends on an unprecedented understanding of neutrino-nucleus interactions.
Insufficient knowledge of either the energy reconstruction or the cross section will limit the experimental precision.
In this thesis, the expertise in both neutrino and electron cross-section modeling and analysis is leveraged in order to alleviate this insufficient knowledge using three complementary approaches, illustrated in figure~\ref{intromonster}.

Namely, neutrino scattering data sets from the MicroBooNE detector at Fermi National Laboratory were analysed (chapter~\ref{ubexp}).
The first measurements of exclusive cross sections with a single proton and no pions detected in the final state were reported.
These results were used to identify regions where modeling improvements are required and specific nuclear effects can be studied in detail (chapter~\ref{xsec}).
The exact same event topology was investigated using electron scattering data sets.
For the connection across neutrinos and electrons to be established, significant modeling improvements took place to ensure a consistent modeling across the two particle species in the commonly used GENIE event generator (chapter~\ref{genie}).
Building on those improvements, the ``Electrons-For-Neutrinos'' analysis reported on the first use of wide phase-space electron scattering data sets from the CLAS detector at Thomas Jefferson National Laboratory (chapter~\ref{e2a}).
The analysis identified significant shortcomings in our lepton-nucleus interaction understanding by reporting cross sections as a function of energy reconstruction methods and testing the validity of models commonly used in neutrino oscillation analyses (chapter~\ref{e4v}).

\chapter{The MicroBooNE Experiment At Fermi National Laboratory}\label{ubexp}

%%%%%%%%%%%%%%%%%%%%%%%%%%%%%%%%%%%%%%%%%%%%%%%%%%%%%%%%%%%%%%%%

\section{Fermilab Neutrino Beamlines}\label{beam}

Fermi National Laboratory (Fermilab) takes advantage of artificially produced neutrino beams in order to study neutrino oscillations. 
The advantage of that choice is the better control over the energy spectrum range and the neutrino flavor content. 
Thus, the spectrum can be tuned so that, given the distance of the detector from the source, the experiment is located at an oscillation maximum or minimum, depending on the objective of the intended measurement~\cite{KOPP2007101}.
On top of that, a narrower energy spectrum can be obtained via the off-axis technique~\cite{OffAxis}.

The neutrino beam production includes a well-defined procedure that is outlined below for the Fermilab beamlines.
The Fermilab linac is used to accelerate a proton beam up to a kinetic energy of 400 MeV.
Then the Booster synchrotron is used to further accelerate the protons up to a kinetic energy of 8 GeV.

\begin{figure}[htb!]
\centering  
\includegraphics[width=0.7\linewidth]
{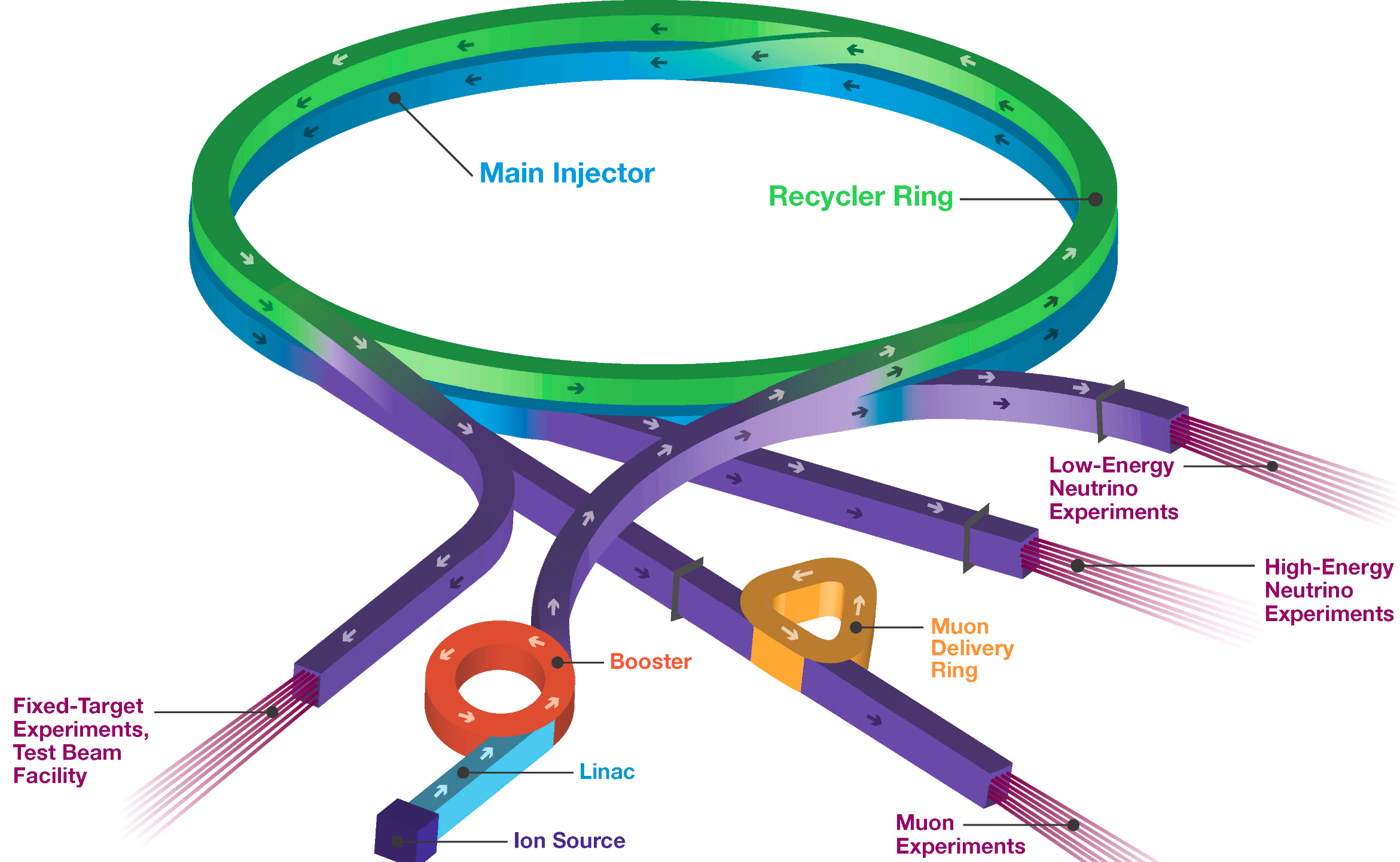}
\caption{The Fermilab accelerator complex showing the accelerator components and the different beams. Figure adapted from~\cite{fnalaccelerator}.}
\label{fermiaccdiv}
\end{figure}

For the Booster Neutrino Beam (BNB), a kicker extracts the protons in a single turn and those protons are redirected to the relevant target hall. 
For all the other experiments, the protons are redirected to the Main Injector.
The latter is the highest energy US-based accelerator facility and the source of neutrinos from the Main Injector (NuMI), the source of muons for the muon campus experiments, and the forthcoming high intensity DUNE neutrino beam.
Figure~\ref{fermiaccdiv} illustrates the series of the relevant accelerator components and the different beams. 

\begin{figure}[htb!]
\centering  
\includegraphics[width=0.7\linewidth]
{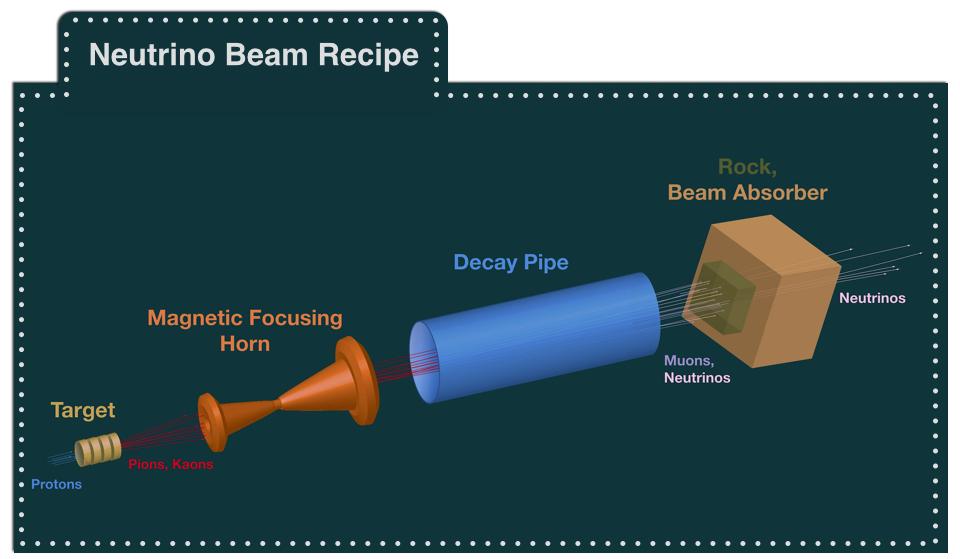}
\caption{The ingredients for a neutrino beam include the accelerated protons, the target, the magnetic horn, the decay pipe, and the absorbers. Figure adapted from~\cite{fnalaccelerator}.}
\label{MakeNu}
\end{figure}

The BNB takes advantage of the 8 GeV protons from the Booster synchrotron.
These protons collide with a beryllium target which is located within a pulsed electromagnet called horn.
Such collisions lead to the production of mesons, mainly $\pi^{\pm}$ , $K^{\pm}$, and $K^{0}$.
The channel that dominates is p + Be $\rightarrow$ $\pi^{+}$ + X, with X corresponding to all the other hadrons produced out of the interaction.
When operating in neutrino mode, the horn focuses the positively charged particles.
The focused particle beam then enters the decay pipe via a concrete-made collimator with a length of $\approx$ 2\,m.
In the decay pipe, some of the particles decay, a process that results in the production of neutrinos primarily via the channel $\pi^{+}$ $\rightarrow$ $\mu^{+}$ + $\nu_{\mu}$.
The massive particles are stopped using a beam absorber made of steel and concrete.
On the other hand, neutrinos transverse the absorber and, at the end of this process, a neutrino-dominated beam is obtained.
A graphic illustration of the process is shown in figure~\ref{MakeNu}.

\begin{figure}[htb!]
\centering  
\includegraphics[width=0.7\linewidth]
{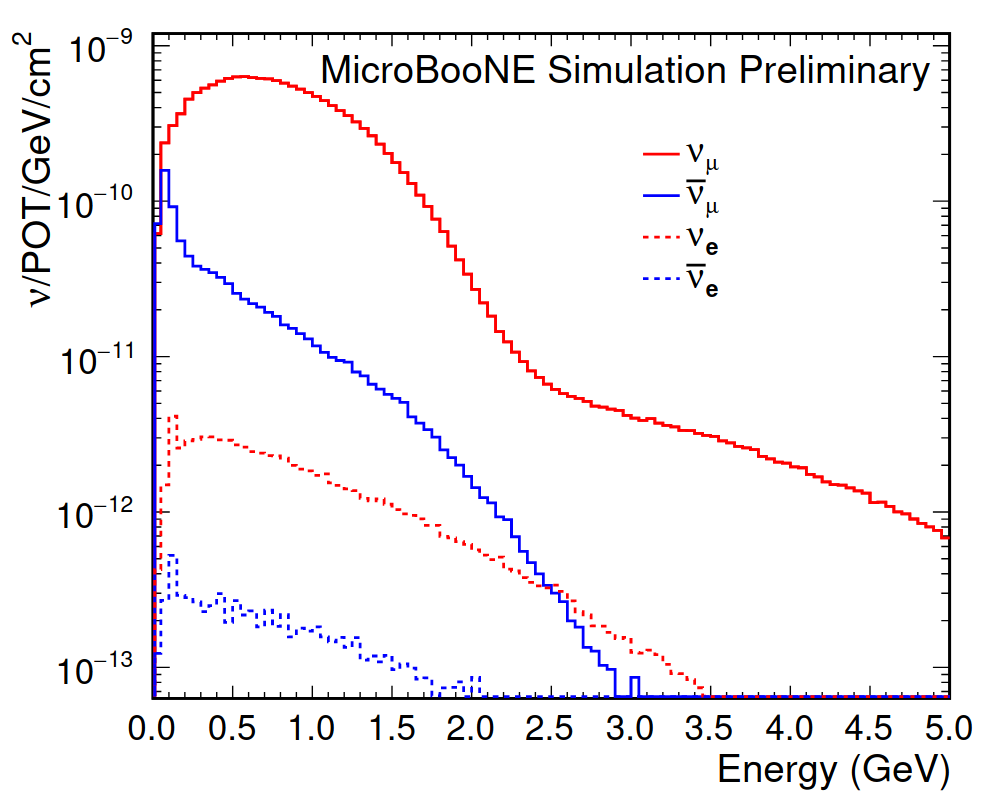}
\caption{The BNB neutrino flux prediction through the MicroBooNE detector for $\nu_{\mu}$,  $\bar{\nu}_{\mu}$, $\nu_{e}$, and  $\bar{\nu}_{e}$. 
A TPC volume with dimensions 2.56\,m $\times$ 2.33\,m $\times$ 10.37\,m is assumed. Figure adapted from~\cite{PhysRevD.79.072002}.}
\label{BNB}
\end{figure}

Apart from the desired $\nu_{\mu}$ beam, neutrinos can also be produced via the decay of anti-muons coming out of the proton-target collision and the channel $\mu^{+} \rightarrow e^{+} + \nu_{e} + \bar{\nu}_{\mu}$~\cite{PhysRevD.79.072002}.
Such interactions are the main sources of intrinsic $\nu_{e}$ contamination to the main $\nu_{\mu}$ beam.
Similar hadron decays result in further contamination of the $\nu_{\mu}$ beam.
The main source of the $\bar{\nu}_{\mu}$ contamination comes from $\pi^{-}$'s which are not separated from the main beam by the horn.
Figure~\ref{BNB} illustrates the BNB energy spectrum while operated in neutrino mode.

The Fermilab-based NuMI neutrino beamline is used by the MINOS, Miner$\nu$a, ArgoNeuT, NO$\nu$A, and MINOS+ experiments~\cite{ADAMSON2016279,Evans,doi:10.7566/JPSCP.12.010006,PhysRevD.99.012002,PhysRevLett.123.151803,PhysRevLett.122.091803,PhysRevD.80.073001}.
On top of that, the MiniBooNE and MicroBooNE experiments use the beamline for off-axis studies.
In this case, neutrinos are produced via the collision of a 120\,GeV beam on a graphite target.
Due to the large angle with respect to the NuMI beam dump, MicroBooNE collects a significant number of low-energy neutrinos originating from kaons decaying at rest.
Using the off-axis NuMI beamline, MicroBooNE records events with a narrower energy spectrum and with a higher electron-neutrino contribution.

%%%%%%%%%%%%%%%%%%%%%%%%%%%%%%%%%%%%%%%%%%%%%%%%%

\section{The MicroBooNE Detector}\label{reco}

The Micro Booster Neutrino Experiment (MicroBooNE) at Fermilab was proposed to succeed MiniBooNE in order to resolve the Low Energy Excess (LEE)~\cite{PhysRevLett.121.221801}.
It is located $\approx$ 20 m away from the MiniBooNE detector, thus establishing almost the same $L$/$E$ ratio.
MicroBooNE uses the same beamline as MiniBooNE, therefore the two experiments share similar flux uncertainties.
At the core of the MicroBooNE detector, there exists a Liquid Argon Time Projection Chamber (LArTPC). 
This detector technology offers high spacial and momentum resolution, allowing an unprecedented energy reconstruction accuracy and a precise particle identification.

\begin{figure}[htb!]
\centering  
\includegraphics[width=0.7\linewidth]
{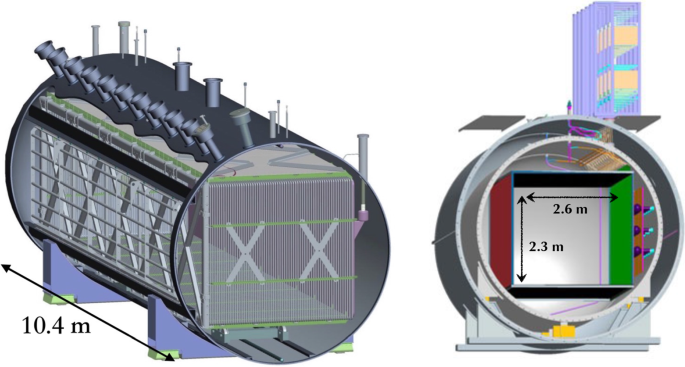}
\caption{Schematic illustration of the MicroBooNE detector and its dimensions. Figure adapted from~\cite{Adams:2018lzd}.}
\label{uBDet}
\end{figure}

Some neutrinos from the two beamlines enter the MicroBooNE LArTPC and interact with the argon nuclei.
Out of these interactions, charged and neutral particles are produced.
When the charged particles transverse the liquid argon, the argon atoms get excited and they further ionize the medium, a process that results in the emission of ionization electrons.
In the presence of a strong electric field of 273\,V/cm, the ionization electrons drift towards the anode plane, where three wire planes are located. Apart from the ionization electrons, scintillation light is also produced and is collected by 32 photomultiplier tubes (PMTs), which are located behind the anode plane.
Both the TPC and the light collection system are embedded withing a cylindrical 170 ton liquid argon cryostat. 
Figure~\ref{uBDet} shows the size of the detector.

The readout electronics are embedded withing the liquid argon in order to significantly reduce the electronic noise.
The analogue-to-digital conversion (ADC) and PMT electronics are located outside the cryostat. 

\begin{figure}[htb!]
\centering  
\includegraphics[width=0.7\linewidth]
{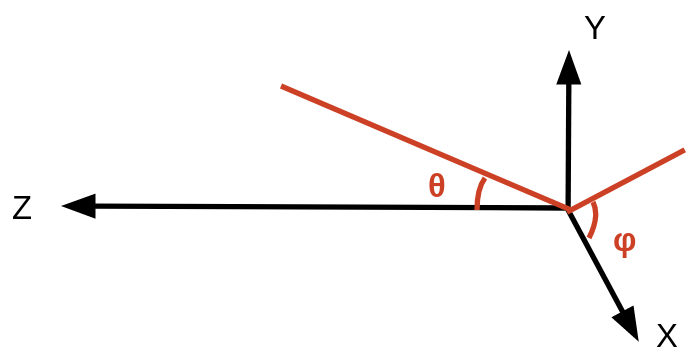}
\caption{The MicroBooNE coordinate system.}
\label{uBCoord}
\end{figure}

Given the baseline of $\approx$ 470\,m, neutrinos in the two beamlines need to be characterized by two angles in the MicroBooNE coordinate system. 
As shown in figure~\ref{uBCoord}, $\theta$ is the angle with respect to the z-direction along the beam direction.
Furthermore, $\phi$ defines the orientation with respect to the XY-plane, orthogonal to the beam direction.

\begin{figure}[htb!]
\centering  
\includegraphics[width=0.8\linewidth]
{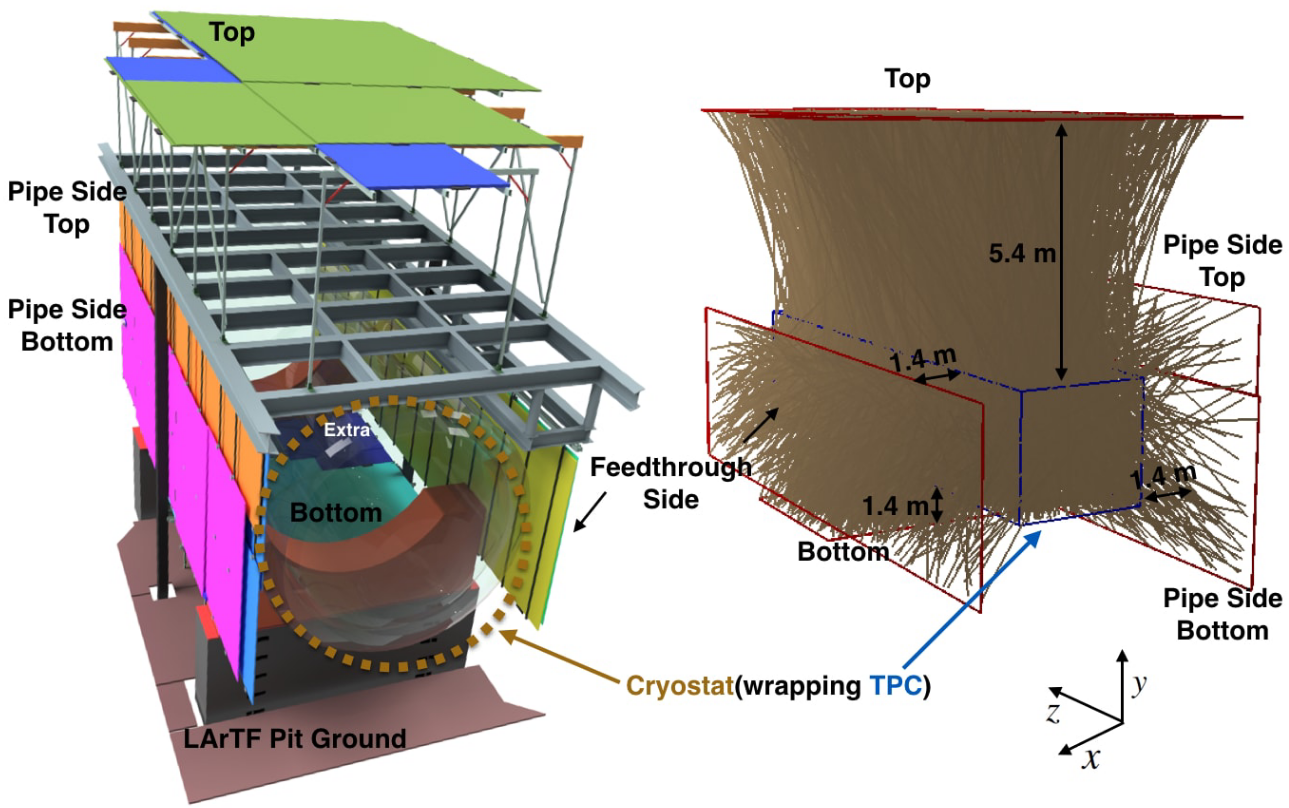}
\caption{The design of CRT planes as part of the MicroBooNE detector. Simulation of cosmic rays crossing
the CRT, the brown lines represent possible cosmic ray trajectories. There are four CRT planes: top plane,
bottom plane, pipe side plane and feedthrough side plane. The beam direction is along the z axis. Figure adapted from~\cite{uBCRT}.}
\label{uBCRTFig}
\end{figure}

A Cosmic Ray Tagger (CRT) system shown in figure~\ref{uBCRTFig} was installed in 2017~\cite{uBCRT} to improve on the rejection of cosmics, which constitute the greatest source of backgrounds on MicroBooNE.   
This detector sub-system consists of 73 scintillating modules made of interleaved layers of scintillating plastic strips situated on the top, bottom, and two sides parallel to the neutrino beam~\cite{uBCRT}.
Based on simulation predictions from CORSIKA~\cite{corsika} and GEANT~\cite{Geant4}, an estimated coverage of 85\% is obtained. 
The CRT installation aimed to improve on the identification and rejection of the dominant cosmic background. 
Cosmic muons transversing the CRT result in the production of scintillation light that can be reconstructed as hits on the CRT channels.
Such hits allow the identification of the cosmic-induced muon tracks with a time precision of $\approx$ 100\,ns.
The latter offers a complementary way to resolve the x-direction ambiguity in the TPC reconstruction.

%\begin{figure}[htb!]
%\centering  
%\includegraphics[width=0.49\linewidth]{\figures uBCRTTop.png}
%\includegraphics[width=0.49\linewidth]{\figures uBCRTSide.png}
%\caption{The top and bottom CRT planes are composed of modules arranged in two layers with
%24 and 9 modules, respectively. The two side planes are composed of modules arranged in two layers. The pipe side and feedthrough side planes consist of 30 and 13 modules, respectively. Figure adapted from~\cite{uBCRT}.}
%\label{ubcrttopside}
%\end{figure}

%%%%%%%%%%%%%%%%%%%%%%%%%%%%%%%%%%%%%%%%%%%%%%%%%%%%

\section{Liquid Argon Time Projection Chambers}\label{lartpctech}

\begin{figure}[htb!]
\centering  
\includegraphics[width=0.7\linewidth]
{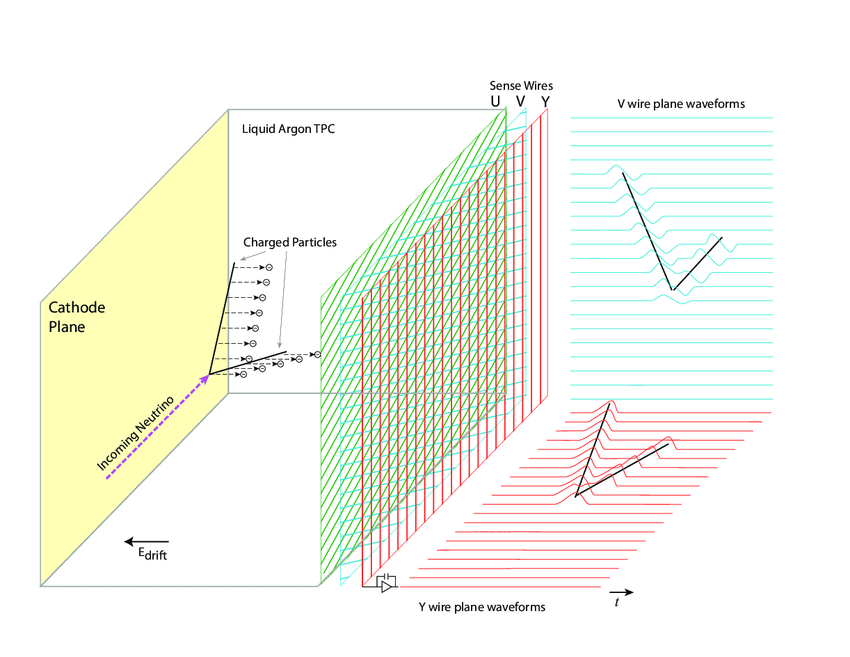}
\caption{Working principle of a LArTPC detector. Figure adapted from~\cite{CAVANNA20181}.}
\label{lartpc}
\end{figure}

The TPC technology was introduced in the 1970s by David Nygren~\cite{TPC}.
Carlo Rubbia designed a LArTPC in 1977 using the same TPC principles as Nygren, but with liquid argon instead of gas~\cite{LArTPCConcept}.
Figure~\ref{lartpc} shows the working principle of a LArTPC.
The cuboid volume of the LArTPC is filled with ultra pure liquid argon.
The presence of a high-voltage cathode on one side of the detector and a grounded anode on the other side create a homogeneous electric field.

When a neutrino interaction takes place in the TPC, charged particles are produced in the final state.
Along their propagation path in LAr, such particles excite and ionize the argon nuclei, a process that results in the emission of ionization electrons.
In the presence of a strong electric field, these ionization electrons drift towards the anode plane.
To ensure an electron drift time of $\mathcal{O}$(ms) before recombination effects take place, a very small $O_{2}$ contamination of 10 parts per trillion has to be guaranteed. 

\begin{figure}[htb!]
\centering  
\includegraphics[width=0.35\linewidth]
{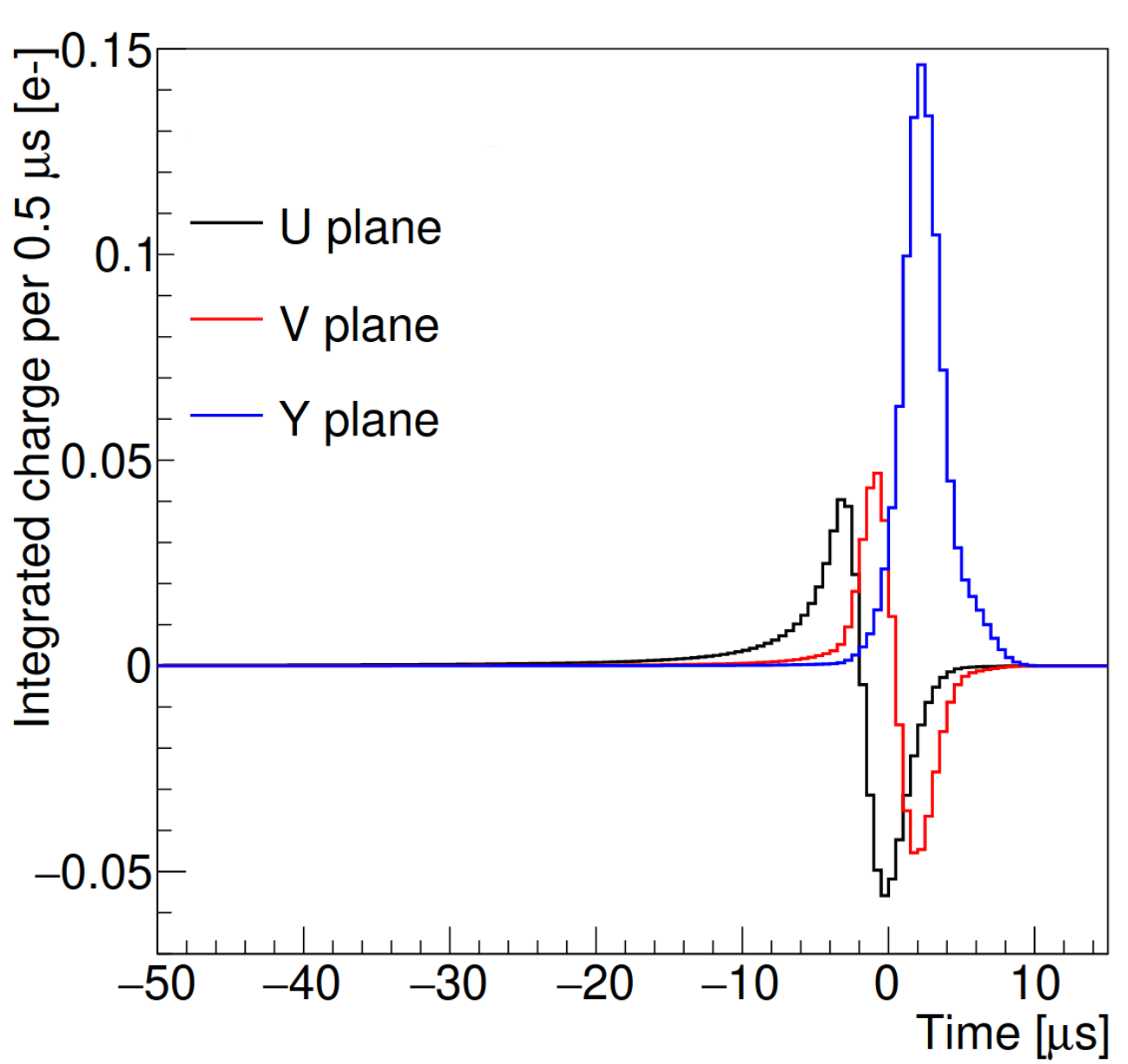}
\includegraphics[width=0.61\linewidth]
{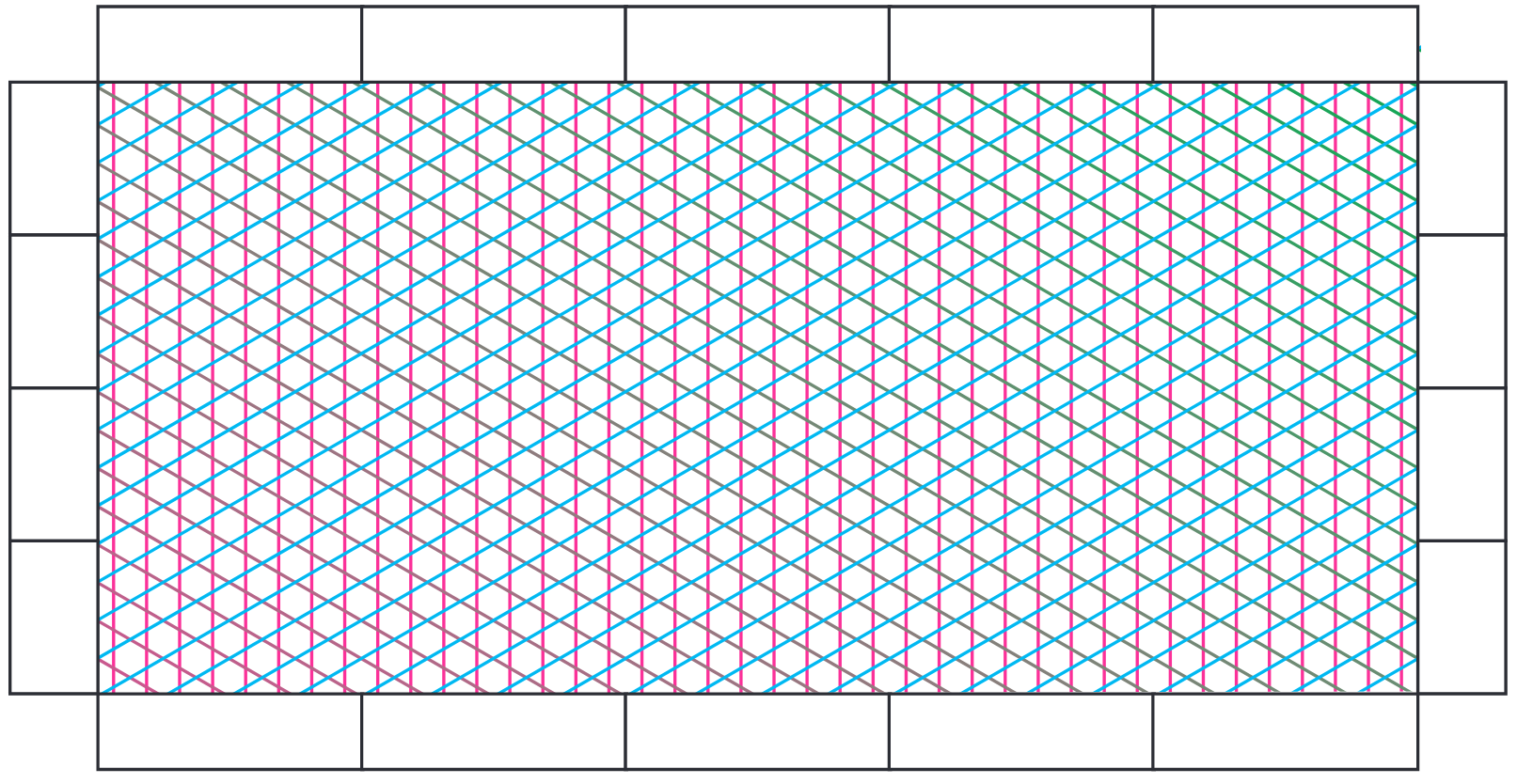}
\caption{(Left) bipolar (U and V induction planes) and unipolar (Y collection plane) signal induction on the three MicroBooNE planes. 
Figure adapted from~\cite{Acciarri_2017}.
(Right) schematic view of the wire planes. The vertical collection Y wires are shown in pink, the induction U wires, angled at +60$^{o}$ are shown in blue and the induction V wires, angled at -60$^{o}$, are shown in green. Figure adapted from~\cite{read}.}
\label{indsignal}
\end{figure}

On the anode plane, there exist three wire planes with a 3\,mm spacing where the clouds of the arriving ionization electrons create signals.
The first two planes correspond to induction planes and are oriented $\pm60^{o}$ with respect to the vertical axis.
In order to obtain 3D views of the particle trajectories, at least two planes with different orientations are required.
The third collection plane removes the ambiguities due to dead wires.
On top of that, the calorimetric and tracking abilities are improved.
Bias voltages are applied on the wire planes so that the two induction planes satisfy the transparency condition outlined in~\cite{grid,Acciarri_2017}.
The condition requires that the drifting electrons pass the two induction planes and are fully captured on the third collection plane.
The drifting charge induces a bipolar signal on the two induction planes and is collected on the third one, where a unipolar signal is produced, as shown in figure~\ref{indsignal} (left). 
The signal area is proportional to the ionization.

The MicroBooNE field cage has a height of 2.3\,m, a width of 2.6\,m and a length along the beam direction of 10.4\,m.
The liquid argon is kept at a pressure of 1.2\,atm, a boiling temperature of 89\,K, and a resulting density of 1.38\,g/cm$^{3}$.
The TPC is merged into the liquid argon.
The active TPC volume is 86\,tons.
The cathode is kept at -70\,kV and the anode is grounded, which results into a homogeneous electric field of 273\,V/cm. 
That translates into a drift velocity of 1.14\,m/ms.
Thus, the MicroBooNE readout window is 2.3\,ms. 
The induction planes U and V are biased at -110\,V and the collection V plane is biased at 230\,V.
The induction planes include 2400 wires and the collection plane consists of 3456 wires. 
The distance between both the different planes and the plane wires is 3\,mm. Figure~\ref{indsignal} (right) shows the schematic view of the wire planes.

Within the liquid argon, the wire signals are fed into a front end ASIC.
Intermediate amplifiers further amplify the signal and pass it to the feed-through.
Then, outside the cryogenic environment, the signal is digitized by readout modules with a frequency of 16\,Hz.
The next step is to downsample the signal to lower frequencies of 2\,Hz~\cite{Chen:2012ysa}.

There exist three 1.6\,ms signal readout windows for each event.
These frames are further truncated to the range between -0.4\,ms to 2.7\,ms.
The hardware-defined trigger 0 time is obtained from the accelerator division.
Given the 2.3\,ms MicroBooNE drift time, the extra buffer of 0.4\,ms makes sure that there is enough time to isolate the neutrino interaction from the cosmic rays that arrive close to the neutrino beam trigger time.

\begin{figure}[htb!]
\centering  
\includegraphics[width=0.7\linewidth]
{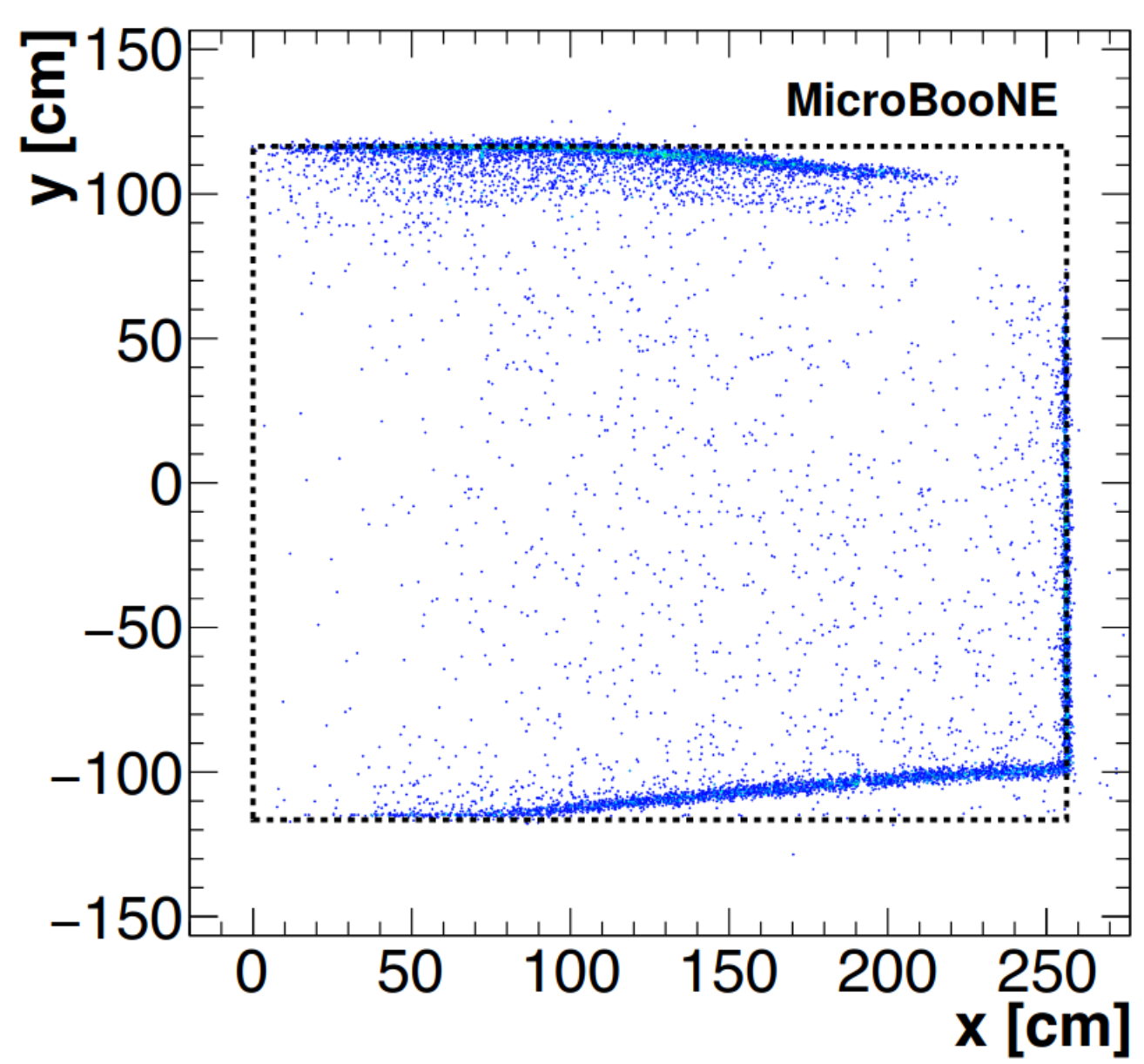}
\caption{Entry/exit points of cosmic muon tracks with a signal from a muon counter located outside of the cryostat.  
In the absence of space charge effects, the points should be located along the TPC boundaries indicated by the dashed lines. Figure adapted from~\cite{SCEPaper}.}
\label{ubsceeffects}
\end{figure}

The ion drift velocity in liquid argon is $\approx$ 5\,mm/s, orders of magnitude smaller than the electron one.
Thus, the argon ions result in the build-up of positive charge in the LArTPC for minutes.
On top of that, the continuous interaction of cosmic rays in the TPC at $\approx$ 5\,kHz also results in a continuous build-up of positively charged argon ions.
The existence of this positive charge leads to a distortion of the homogeneous electric field within the TPC, as shown in figure~\ref{ubsceeffects}.
This distortion is referred to as ``space charge effect'' (SCE)~\cite{SCEPaper}, which leads to a displacement of the reconstructed signal ionization source by up to $\mathcal{O}$(10\,cm).

%%%%%%%%%%%%%%%%%%%%%%%%%%%%%%%%%%%%%%%%%%%%%%%%%%%%%%%%%%%%%%%%%

\section{Scintillation Light}\label{scintlight}

\begin{figure}[htb!]
\centering  
\includegraphics[width=\linewidth]
{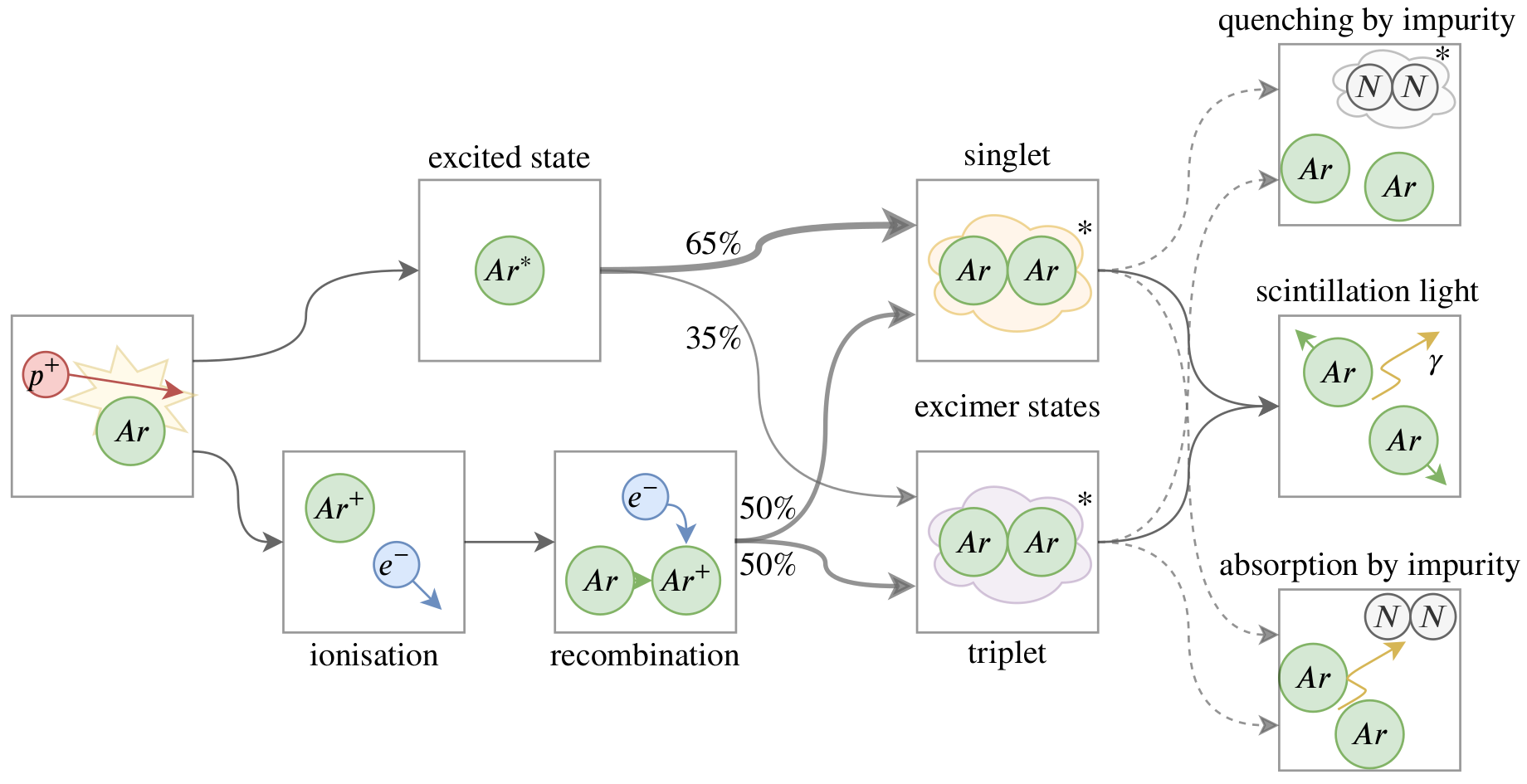}
\caption{The production of scintillation light in liquid argon. A charged particle can either excite or ionise the argon. Figure adapted from~\cite{VanDePontseele:2020tqz}.}
\label{ubscintlight}
\end{figure}

The scintillation light production and propagation is almost instantaneous and of the order of $\mathcal{O}$(ns). 
Thus, collecting the emitted scintillation light is crucial to identify the time that the event took place and to determine the x-position along the drift direction in the TPC. 
When charged particles transverse the liquid argon, scintillation light is produced that results in the deposition of $\approx$ 10$^{4} \, \gamma$ / MeV.
This light can be produced via two mechanisms, shown in figure~\ref{ubscintlight}.

The first mechanism is the self-trapped exciton luminescence.
In that case, charged particles transverse the liquid argon and they leave some of the argon atoms in an excited state called excitons.
These states are molecules with another argon atom with a short lifetime and are called dimers or excimers.
Roughly 65\% of them are in a singlet state $^{1}\Sigma_{u}$ and the remaining ones are in a triplet state $^{3}\Sigma_{u}$.

With the second mechanism, the charged particles ionize the argon atoms and that results in the creation of free electrons.
These electrons recombine with the positive argon ions, a process that also creates excited dimers.
With this mechanism, the probability of creating either singlets or triplets is equal.

The singlet states result in the emission of fast scintillation light with a decay time of $\approx$ 6\,ns.
The triplet states result in a slow component and in a decay time of $\approx$ 1.5\,$\mu$s.
Both the fast and the slow components have a peak wavelength at 128\,nm in the Vacuum Ultra-Violet (VUV) region.
Both states have an energy minimum which is equivalent to a distance between the atoms of $\approx$ 2.8\,\r{A}~\cite{ArMinEn}.
The liquid argon inter-atom separation is $\approx$ 4\,\r{A}, which is greater than the one that corresponds to the excimer energy minimum.
Thus, liquid argon is transparent to its own light and, therefore, the light can be detected over long distances.

The second mechanism is related to the recombination luminescence and relies on the free electron and argon ion local density.
Thus, recombination effects are stronger for particles with higher energy deposition per unit length.
This implies a particle-type dependence of the energy deposition in the detector to the light-equivalent production.
That also means that the scintillation light yield depends on the choice for the electric field.
A strong electric field will reduce the recombination effects and, therefore, the recombination-induced luminescence.
The amount of scintillation light can be further reduced due to impurities that result into quenching and absorption effects.
With that in mind, the liquid argon purity in MicroBooNE is monitored to ensure an absorption length greater than the 2.5\,m-long TPC width.

\begin{figure}[htb!]
\centering  
\includegraphics[width=4cm,height=5cm]{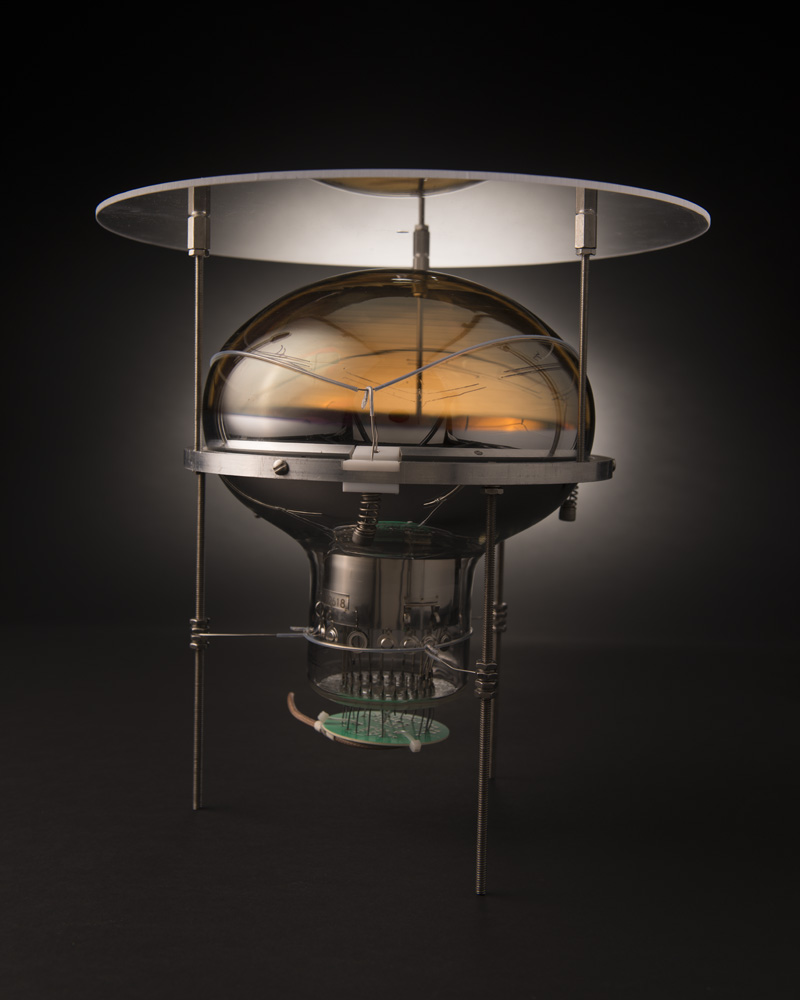}
\includegraphics[width=0.49\linewidth]{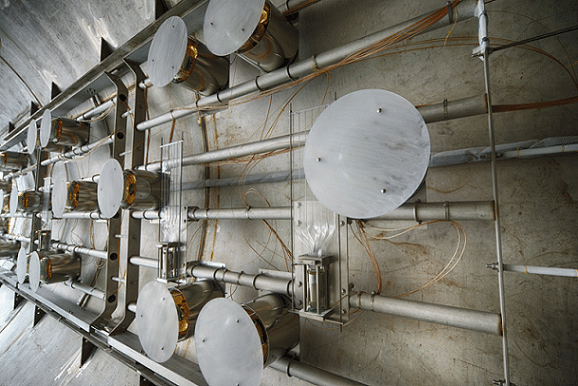}
\caption{The MicroBooNE light collection system with the 32 PMTs. Figure adapted from~\cite{ubpmtpic}.}
\label{ubpmt}
\end{figure}

The MicroBooNE light detection system consists of 32 8-inch Hamamatsu cryogenic PMTs located behind the anode plane, shown in figure~\ref{ubpmt}.
A Tetra-Phenyl Butadiene (TPB) coated acrylic plate is located in front of the PMT.
The TPB converts the 128\,nm light argon scintillation light to 425\,nm visible light. 
The tube is surrounded by mu-metal to shield the electronics against any magnetic fields.
The PMTs operate at a voltage of 1300\,V and have a gain of 10$^{7}$~\cite{gain}.
The PMT signal is sent to a splitter board, where it gets pre-amplified and digitized.
The signals have a 60\,ns rise time, much smaller than the spill windows for the BNB (1.6\,$\mu$s) and NuMI (10\,$\mu$s) beams.

%%%%%%%%%%%%%%%%%%%%%%%%%%%%%%%%%%%%%%%%%%%%%%%%%%%%%%%%%%%%%%%%%%%%%%%%%%%%%%%%%%%

\section{Hardware And Software Triggers}\label{trigger}

\begin{figure}[htb!]
\centering  
\includegraphics[width=0.9\linewidth]{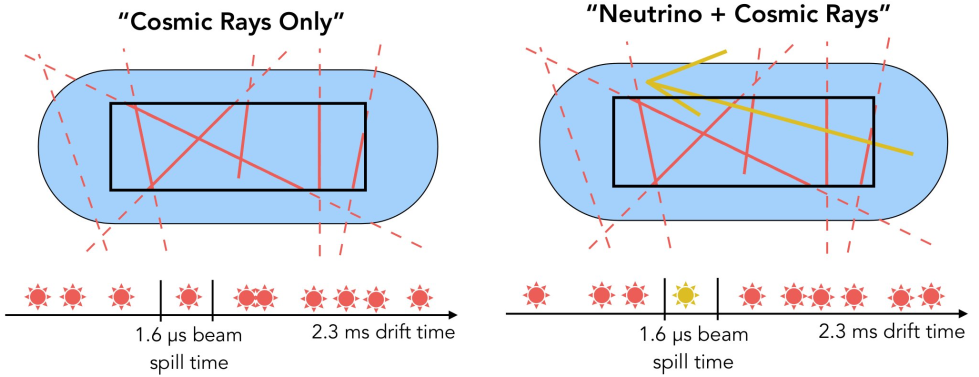}
\caption{(Left) cosmic-induced event that was stored because of the coincidence of a
1.6\,$\mu$s accelerator BNB signal and light detected by the PMTs. 
(Right) neutrino-induced event where the light was coming from a neutrino interaction. Figure adapted from~\cite{DelTuttoThesis}.}
\label{ubtrigger}
\end{figure}

Even when accelerator-induced neutrino beams transverse the MicroBooNE detector, it is highly unlikely that any neutrino will interact due to the extremely small neutrino-nucleus cross sections.
Each BNB bunch includes $\approx$ 1.2 $\times$ $10^{12}$ protons and the TPC surface area is $\approx$ 60 $\times$ $10^{3}$ cm$^{2}$.
That yields a neutrino interaction every $\approx$ 500 bunches~\cite{AguilarArevalo:2008yp}.
In order to significantly reduce the amount of data that is stored on tape, a minimal amount of optical activity is required.
Regardless, background events with light activity within the beam spill arrival time are still the ones that dominate.
Figure~\ref{ubtrigger} shows the differences between two such events~\cite{DelTuttoThesis}.
The first one (left) corresponds to pure cosmic activity that arrives in coincidence with the beam spill.
The second one (right) includes a neutrino interaction of interest, although there is still a significant cosmic contamination.

Cosmic-induced background events are accounted for by recording events when the neutrino beam is off.
That leads to the existence of two trigger types.
The first one, referred to as ``Beam On'', corresponds to data samples recorded when the beam is on and the light system yielded a signal.
The second one, referred to as ``Beam Off'' or ``External BNB'' (ExtBNB), uses the same configuration as in the Beam On case, but those data sets are recorded when the beam is off.

The MicroBooNE readout system is triggered with the arrival of a signal~\cite{read}.
The signal might originate from the BNB/NuMI accelerator clocks.
Alternatively, there is a function generator in the trigger rack producing pulses at a fixed frequency.
The latter is used to record Beam Off events.
The trigger board sends a signal to all the readout crates to start recording data.
On top of that, the trigger keeps track of the trigger type and the time that the signal was received.
The accelerator-based signals are produced in couples.
The former (early signal) vetos Beam Off triggers just before the beam triggers.
That process aims to avoid any trigger overlap that might result in the reduction of the exposure to the beam. 
The latter signal is the one used to trigger the readout.
The MicroBooNE TPC readout is completely unbiased, thus all the time ticks are stored and the readout is not zero-suppressed.
The PMT readout is biased though, with data being stored during specific time intervals determined by the ``discriminators''.

There exist two PMT data-taking configurations that differ by their duration and by the suppression level.
The duration is frequently expressed in tick units, with each tick corresponding to 15.625 ns.
For the BNB triggering, the beam discriminator starts the data recording simultaneously for all PMTs.
That happens by replicating the trigger signal and by redirecting it to all the PMT boards.
The duration of this beam window is 23.4 $\mu$s or 1500 optical ticks.
On average, the neutrino arrival time is $\approx$ 4 $\mu$s after the opening of the beam window.
The cosmic discriminators span a range outside the 23.4 $\mu$s window.
The ultimate goal of the cosmic discriminators is to suppress the amount of data that is recorded over a long time interval, once the signal for a triggered event has arrived.
The duration of the cosmic window is 4.8 ms and spans the range of [-1.6,+3.2] ms.
Only waveforms with more than 130 ADC counts are stored.
That number corresponds to $\approx$ 7 photo-electrons (PE).
The accelerator signals are meticulously timed so that neutrinos originating from the 1.6\,$\mu$s BNB spill arrive in MicroBooNE during this time window of 23.4\,$\mu$s.
When the beamline is fully operational, that takes place $\approx$ 5 times/s.
The majority of the triggered events based on the hardware-driven signal do not include neutrino interactions.
Thus, a software trigger is further applied to determine whether an event is a neutrino candidate or not.
This trigger uses PMT optical waveforms and searches for light activity within the 1.6\,$\mu$s BNB beam-spill.
That action takes place after the TPC data has been sent to the DAQ crates.

%%%%%%%%%%%%%%%%%%%%%%%%%%%%%%%%%%%%%%%%%%%%%%%%%%%%%%%%%%%%%%%%%%%%%%%%%%%%%%%%%%%

\section{Optical Event Reconstruction}\label{opreco}

\begin{figure}[htb!]
\centering  
\includegraphics[width=0.9\linewidth]{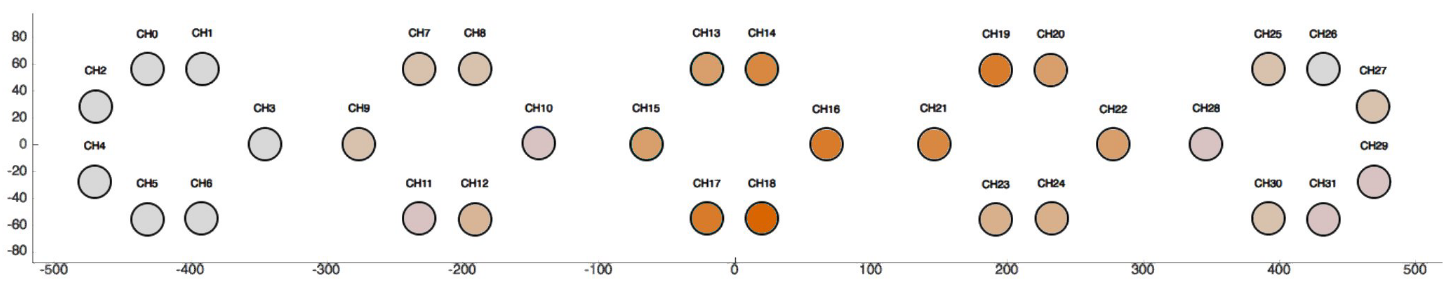}
\caption{Optically reconstructed flash object recorded by the MicroBooNE PMT light collection system. The dark orange regions represent a higher PE yield. Figure adapted from~\cite{read}.}
\label{ubpmtorient}
\end{figure}

\begin{figure}[htb!]
\centering  
\includegraphics[width=0.5\linewidth]{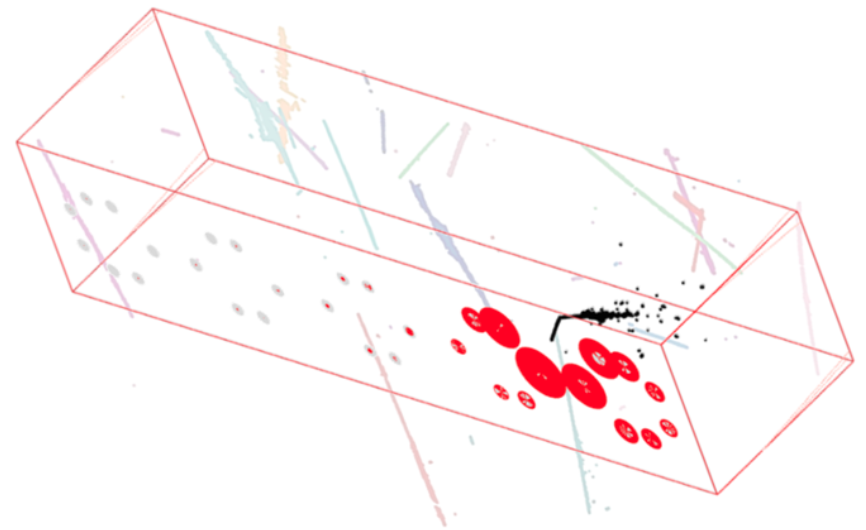}
\caption{Neutrino-induced tracks (black) are matched to the corresponding light signals collected by PMTs (red circles) and are clearly separated from the cosmic-induced ones (dimmed color tracks). Figure adapted from~\cite{WCcosmicpic}.}
\label{ubflashmatch}
\end{figure}

In order to perform a MicroBooNE data analysis, the low-level PMT signals have to be converted to high-level data products.
That is achieved by using the raw PMT waveforms as an input quantity and, based on these, flashes are reconstructed.
Such flashes are indicative of optical activity in the TPC recorded by multiple PMTs simultaneously.
This light activity is collected by the 32 8-inch PMTs oriented as shown in figure~\ref{ubpmtorient}.
The neutrino-induced light activity results in a higher PE yield compared to the cosmic-induced one, as shown in figure~\ref{ubflashmatch}.

The PMTs have a low ($\approx$ 20 ADC/PE) and a high ($\approx$ 2 ADC/PE) gain readout~\cite{read}.
The optical reconstruction merges the two streams into a corrected waveform which corrects saturated high-gain pulses based on the information obtained from the low-gain pulses.
Depending  on the discriminator type, a different baseline estimation is used.  
The cosmic discriminator uses a constant baseline, called pedestal.
The beam discriminator uses a time dependent baseline estimate.
This is a more accurate estimation and addresses potential overestimations or underestimations in the signal baseline.
Once the pedestal has been identified, the waveform ADC counts are used.
Those pulses above threshold are identified and propagated to the next stage for the ``hit'' reconstruction.
These pulses result in the creation of data products based on this optical hit reconstruction.
Each optical hit refers to optical PMT activity, namely the number of produced PEs and the event time. 
The optical hits are clustered and the PMT PE production is summed in order to reconstruct flashes~\cite{WireCellEventSelection}.
%Each time range is divided in intervals of 30\,ns.
%Potential coincidental optical hits across multiple PMTs (at least three) are searched for.
%In case of coincidental hits, an 8\,$\mu$s integration window is used to collect the interaction-related late light.
%If multiple flashes are identified within a time window, the one with the greatest number of PEs is stored.
%We are particularly interested in flashes that are reconstructed during the 1.6\,$\mu$s long beam spill arrival window.
%A constant background subtraction of 2 PEs per PMT takes place across all reconstructed flashes.
%This subtraction accounts for the 0.25\,MHz dark noise integrated over the 8\,$\mu$s time window (8\,$\mu$s $\times$ 0.25\,MHz = 2 PEs).

%\begin{figure}[htb!]
%\centering  
%\includegraphics[width=0.8\linewidth]{\figures uBFlash.png}
%\caption{Example with two reconstructed flashes using PMT waveforms from the beam discriminator. The black curves correspond to the deconvolved PE spectra for each one of the 32 PMTs. The red lines show the flash start times and the red bands show the 8\,$\mu$s flash windows.}
%\label{ubflash}
%\end{figure}

%%%%%%%%%%%%%%%%%%%%%%%%%%%%%%%%%%%%%%%%%%%%%%%%%%%%%%%%%%%%%%%%%%%%%%%%%%%%%%%%%%

\section{TPC Signal Processing}\label{tpcsignal}

While drifting towards the anode plane, the ionization electrons repel each other.
That results in a diffused signal arriving on the anode plane and the level of the diffusion depends on the position that the interaction took place.
Those drifting electrons result in the induction of current on the neighboring wires.
Therefore, waveforms produced on a specific wire might have an effect on those produced on a neighboring wire.
The current induced on the wires is amplified and shaped by the ASICs located within the liquid argon.

\begin{figure}[htb!]
\centering  
\includegraphics[width=0.9\linewidth]{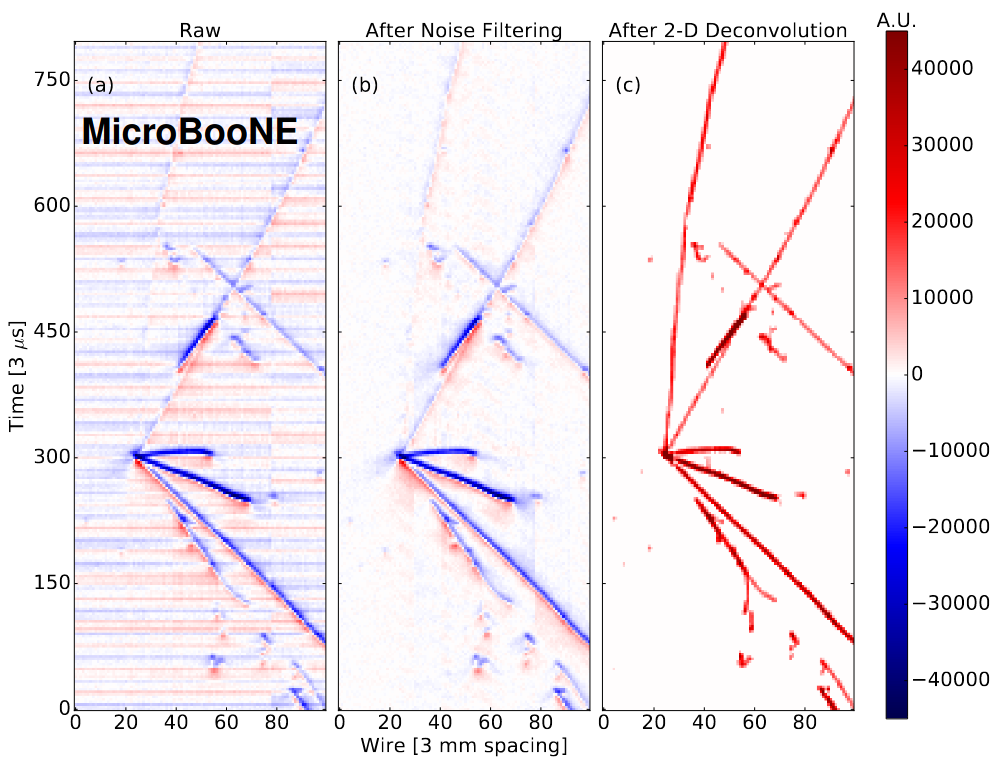}
\caption{Candidate neutrino event display from MicroBooNE data on one of the induction planes. (a) The raw waveform image. (b) The image after noise-filtering. (c) The image after 2D deconvolution. The image quality near the neutrino interaction vertex significantly improves after the 2D deconvolution and the latter leads to improvements in the pattern recognition. Figure adapted from~\cite{Adams:2018gbi}.}
\label{ubtpcsignalprocessing}
\end{figure}

The objective of the noise filtering and signal processing on MicroBooNE is to convert those raw digitized waveforms into the number of ionization electrons passing through a specific wire plane at a given time~\cite{Adams_2018,Adams:2018gbi}.
For that to be achieved, the first step is the application of noise filters to remove the external noise and the electronics-induced one.
The major sources of external noise originate from the TPC drift high-voltage power supply and the low-voltage regulators for the front-end ASICs~\cite{Acciarri_2017}.
Then, the application of a deconvolution of the digitized TPC wire signals follows.
That takes place in two dimensions, with the first one being over time and the second one being the effect across multiple wires.  
A region of interest (ROI) is identified based on the deconvoluted charge distribution.
The ionization charge is obtained with a linear baseline subtraction within the start/end bins of the ROI window.
At the final step, the processed signals accounting for the number of electrons on a given wire at a certain time are used as input to the MicroBooNE event reconstruction.
The signals are calibrated before the conversion to deposited energy is performed.
Figure~\ref{ubtpcsignalprocessing} illustrates the necessity of signal processing before any reconstruction is enabled.

%%%%%%%%%%%%%%%%%%%%%%%%%%%%%%%%%%%%%%%%%%%%%%%%%%%%%%%%%%%%%%%%%%%%%

\section{Pandora Reconstruction Framework}\label{pandora}

The processed signals from the previous stage are used as the input for higher level reconstruction of objects such as vertices, tracks, and showers.
In order to reconstruct objects in the TPC, the Pandora reconstruction framework is deployed~\cite{Acciarri:2017hat}.
Pandora uses pattern-recognition algorithms, along with the use of multiple algorithms completing specific tasks for a given topology.
The waveforms obtained with 4.8\,ms window are used as an input to the reconstruction framework.

\begin{figure}[htb!]
\centering  
\includegraphics[width=0.6\linewidth]{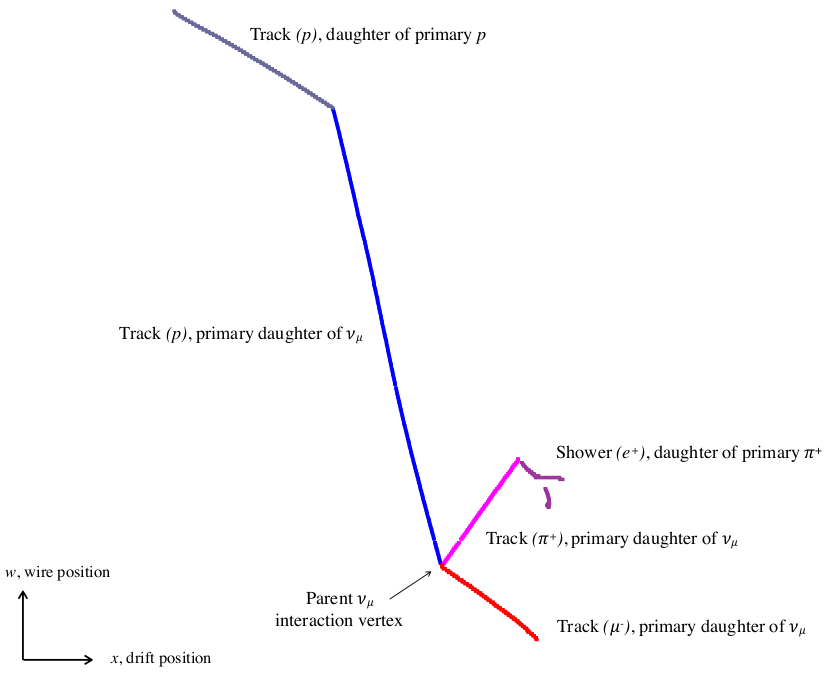}
\caption{Illustration of the hierarchical structure of particles reconstructed for a simulated charged current $\nu_{\mu}$ event in MicroBooNE. The interaction includes a muon, proton and charged pion in the visible final state. Figure adapted from~\cite{Acciarri:2017hat}.}
\label{ubparentdaughter}
\end{figure}

The first step includes fitting the processed waveforms with Gaussian distributions to each peak.
This fitting process results into the creation of a 2D hit.
Then, PandoraCosmic is a track-focused selection that aims to tag the cosmic muons.
The selection results in the creation of a cosmic-free hit collection.
PandoraNu identifies neutrino interaction vertices and uses them to reconstruct tracks and showers originating from the vertex.
A parent neutrino particle is defined and the reconstructed objects are added as daughter particles.
Creating a ``slice'' is the first step that Pandora performs.
A slice is defined as a collection of reconstructed particles originating from the same interaction. 
For the creation of the slices, the PandoraCosmic algorithm is first ran over all the hits in order to identify the cosmic-induced muon tracks and the associated $\delta$-rays and Michel electrons under a cosmic hypothesis.
The obvious cosmic activity of through-going muon tracks is identified based on the geometric information.
The remaining hits are used by the PandoraNu algorithm and objects are reconstructed under the neutrino hypothesis, creating a slice.
Each slice is reconstructed using both hypotheses.
For the interactions to be reconstructed in a three-dimensional space, Pandora requires information from at least two wire planes~\cite{Acciarri:2017hat}.
The 2D hits are clustered on each wire plane and for each slice.
A collection of 3D candidate vertices is produced by identifying locations that project to the same points of the 2D clusters.
All the candidate vertices are propagated into a Support Vector Machine (SVM) selection and the candidate with the highest score is isolated.
The cluster matching algorithms are ran around this candidate vertex on each plane and are compared to improve the matching of the reconstructed objects~\cite{Acciarri:2017hat}.
With this process, a collection of reconstructed Particle Flow Particles (PFParticles) is constructed.
Such a PFParticle is created by combining 2D cluster objects on the three wire planes.
Each one of those PFParticles is associated with a vertex location and has a collection of 3D points, which contain the charge details from the relevant 2D hits.
These 3D points are referred to as SpacePoints.
The reconstructed PFParticles in the neutrino slice are organized with a hierarchical structure based on parent-daughter assocciations, as shown in figure~\ref{ubparentdaughter}.

%\begin{figure}[htb!]
%\centering  
%\includegraphics[width=\linewidth]{\figures uBPandora.png}
%\caption{Illustration of the Pandora output data products, as clustered by the LArSoft Event Data Model. The dashed lines illustrate the hierarchy evolution via the PFParticle interface. The solid arrows illustrate the associations between the PFParticles and other reconstructed products. Figure adapted from~\cite{Acciarri:2017hat}.}
%\label{ubpandora}
%\end{figure}

%\begin{figure}[htb!]
%\centering  
%\includegraphics[width=0.95\linewidth]{\figures uBPandoraReco.png}
%\caption{Collection of topologies considered by the 3D track reconstruction. (a) An unambiguous grouping of clusters. (b) An ambiguous clustering where the most appropriate grouping of 2D clusters can be identified. (c) Ambiguous clustering, which can be resolved by splitting the U cluster at the indicated position. (d) Ambiguous clustering, which can be resolved by merging the two V cluster fragments.}
%\label{ubpandorareco}
%\end{figure}

A candidate neutrino PFParticle is assigned at the very top of the hierarchy by the PandoraNu algorithm.
That candidate neutrino will be having at least one daughter PFParticle.
These daughter particles are assigned a score that classifies them as either-track-like or shower-like objects.
A Support Vector Machine uses the collection of hits to determine the nature of reconstructed object.
Track-like objects have a score close to 1 and shower-like objects score closer to 0.
Based on that score, a shower- or a track-like data product is constructed for each particle.

\begin{figure}[htb!]
\centering  
\includegraphics[width=0.7\linewidth]{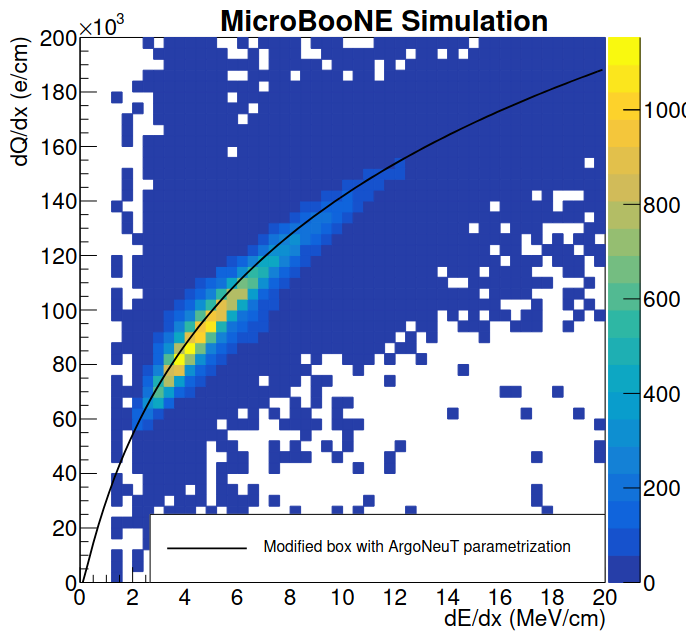}
\caption{Illustration of the measured dQ/dx vs dE/dx distribution with the modified recombination model in the MC simulation with the ArgoNeuT parametrization.
Figure adapted from~\cite{Acciarri_2013}.}
\label{ubdqdxvsdedx}
\end{figure}

In the case of a track-like classification, Pandora uses a linear fit, described in detail in~\cite{Acciarri:2017hat}, and returns the direction and the position of each point across the particle trajectory in 3D. 
For each one of the points, the charge deposition dQ/dx and the residual range - the distance from the end of the track - are stored.
This approach allows accurate measurements of dx that might include deflections and displacements due to space charge effects.
For track-like objects, dQ/dx is converted to dE/dx using the inverse Modified Box model~\cite{Acciarri_2013}, as shown in figure~\ref{ubdqdxvsdedx}.
The advantage of this model is that the non-linear dependence of the local density of ions is taken into consideration.

For shower-like objects, Pandora creates a 3D cone along the hit collection with a fixed 3D orientation, solid angle and length.
The shower energy is obtained and calibrated using the same techniques as the ones used on the $\pi^{0}$ reconstruction~\cite{caratelli_neutralpi}.
Furthermore, showers are fitted with a Kalman filter~\cite{kalman_filter}.
With this fit, hits that are longitudinally or transversely displaced from the main shower cone are removed.
This fitting process returns a track-like object.
Therefore, the calorimetric tools mentioned in the previous paragraph become available for shower-like objects too.

%%%%%%%%%%%%%%%%%%%%%%%%%%%%%%%%%%%%%%%%%%%%%%%%%%%%%%%%%%%%%%%%%%%%%

\section{Cosmic Overlay Simulation}\label{eventsim}

\begin{figure}[htb!]
\centering  
\includegraphics[width=\linewidth]{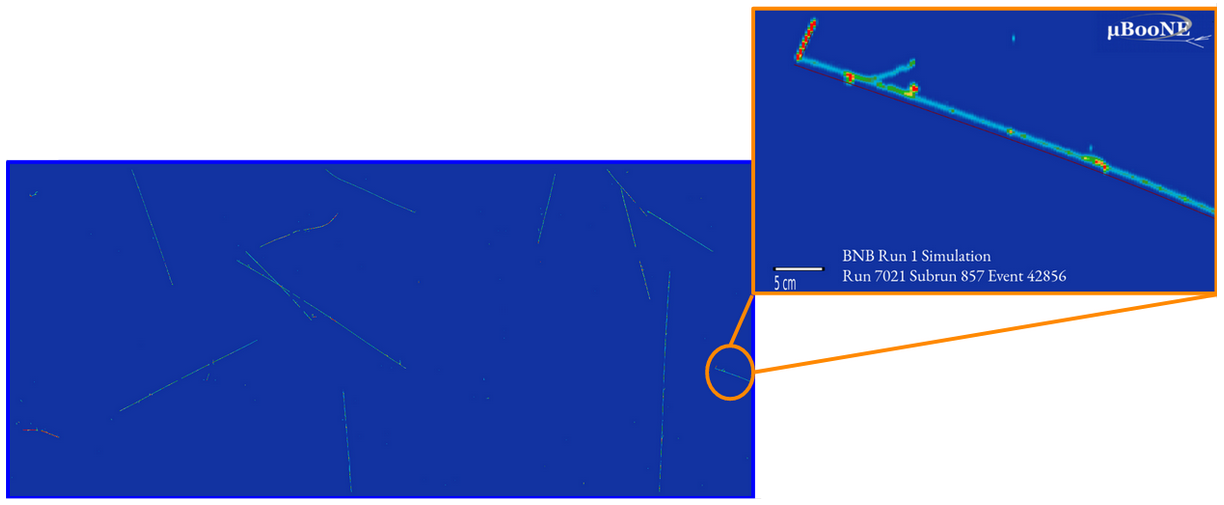}
\caption{MicroBooNE event display of an event in a cosmic overlay sample. A simulated neutrino event (orange box) is overlaid on top of cosmic events (blue box).}
\label{uboverlay}
\end{figure}

For the purposes of the oscillation and cross section analyses on MicroBooNE, neutrino-induced interactions in our detector need to be induced.
However, MicroBooNE is a surface detector dominated by the cosmic activity.
In order to simulate this cosmic contamination as accurately as possible, real cosmic events collected with the unbiased trigger are used.
This trigger stores events outside the beam-related trigger windows and does not demand the existence of any optical activity in a specific part of the detector.
Such events are overlaid on top of GENIE simulated neutrino interactions~\cite{geniev3highlights}. 
These resulting samples are referred to as ``cosmic overlays'' and an example event display is shown in figure~\ref{uboverlay}.

For the simulated part of the cosmic overlays, the reconstructed information is ``backtracked'' to the underlying truth-level information.
That is achieved by associating the hits on each plane to the GEANT particles that resulted in the production of these hits.
Thus, it is feasible to relate reconstructed PFParrticles to the underlying simulated interaction products.
It is further possible to identify the amount of charge originating from the cosmic part of the overlay samples.
Reconstructed track- and shower-like objects are matched to a simulated object when they have more hits in common than any of the other simulated particles or cosmic tracks in a given event.

Motivated by the analyzer's involvement in the development and validation of these samples, the analyses presented in sections~\ref{CCQE} and~\ref{CC1p} were the first ones to adopt the aforementioned overlay technique.
Due to the success in accurately describing the cosmic background, this technique is currently used as the default simulation option across the MicroBooNE collaboration.

%%%%%%%%%%%%%%%%%%%%%%%%%%%%%%%%%%%%%%%%%%%%%%%%%%%%%%%%%%%%%%%%%%%%%

\chapter{MicroBooNE Quasielastic-like Cross-Section Results\texorpdfstring{\newline}{ }\normalsize{[Phys. Rev. Lett. 125, 201803 (2020)]}}\label{xsec}

%%%%%%%%%%%%%%%%%%%%%%%%%%%%%%%%%%%%%%%%%%%%%%%%%%%%%%%%%%%%%%%%

\section{First Measurement of Differential Charged Current Quasielastic-like Scattering Cross Sections}\label{CCQE}

\subsection{Quasielastic-like Neutrino Data Analysis}\label{CCQEDataAna}

As outlined in section~\ref{NuNucleusInt}, understanding the interaction of neutrinos with argon nuclei is of particular importance as a growing number of neutrino oscillation experiments employ liquid argon time projector chamber (LArTPC) neutrino detectors.
Experimentally, the energy of interacting neutrinos is determined from the measured momenta of particles that are emitted following the neutrino interaction in the detector.
Many accelerator-based oscillation studies focus on measurements of charged-current (CC) neutrino-nucleon quasielastic (QE) scattering interactions~\cite{Anderson:2012jds,Nakajima:2010fp,Aguilar-Arevalo:2013dva,Abe:2014iza,Carneiro:2019jds,Abratenko:2019jqo,Fiorentini:2013ezn,Betancourt:2017uso,Walton:2014esl,Abe:2018pwo}, where the neutrino removes a single intact nucleon from the nucleus without producing any additional particles.
This choice is guided by the fact that CCQE reactions can be reasonably well approximated as two-body interactions, and their experimental signature of a correlated muon-proton pair is relatively straightforward to measure.
Therefore, precise measurements of CCQE processes are expected to allow precise reconstruction of neutrino energies with discovery-level accuracy~\cite{Mosel:2013fxa}.

A working definition for identifying CCQE interactions in experimental measurements requires the identification of a neutrino interaction vertex with an outgoing lepton, exactly one outgoing proton, and no additional particles.
These events are referred to herein as ``CCQE-like''.
This definition can include contributions from non-CCQE interactions that lead to the production of additional particles that are absent from the final state due to nuclear effects, such as pion absorption, or have momenta that are below the experimental detection threshold.
Pre-existing data on neutrino CCQE-like interactions came from experiments using various energies and target nuclei~\cite{Formaggio:2013kya}. 
These primarily included measurements of CCQE-like muon neutrino ($\nu_\mu$) cross sections for interactions where a muon and no pions were detected, with~\cite{Fiorentini:2013ezn,Betancourt:2017uso,Walton:2014esl,Abe:2018pwo} and without~\cite{Anderson:2012jds,Nakajima:2010fp,Aguilar-Arevalo:2013dva,Abe:2014iza,Carneiro:2019jds,Abratenko:2019jqo} requiring the additional detection of a proton in the final state.
While most relevant for LArTPC based oscillation experiments, no measurements of CCQE-like cross sections on argon with the detection of a proton in the final state existed until 2021.

This analysis presents the first measurement of exclusive CCQE-like neutrino-argon interaction cross sections, measured using the MicroBooNE LArTPC~\cite{PhysRevLett.125.201803}.
Our data serve as the first study of exclusive CCQE-like differential cross sections on argon as well as a benchmark for theoretical models of $\nu_{\mu}$-Ar interactions, which are key for performing a precise extraction of oscillation parameters by future LArTPC oscillation experiments.
We focused on a specific subset of CCQE-like interactions, denoted here as CC1p0$\pi$, where the contribution of CCQE interactions is enhanced~\cite{Adams:2018lzd}.
These include charged-current $\nu_{\mu}$-Ar scattering events with a detected muon and exactly one proton, with momenta greater than 100\,MeV/$c$ and 300\,MeV/$c$, respectively. 
The measured muon-proton pairs were required to be co-planar with small missing transverse momentum and minimal residual activity near the interaction vertex that was not associated with the measured muon or proton.
For these CC1p0$\pi$ events, the flux-integrated $\nu_{\mu}$-Ar total and differential cross sections in muon and proton  momentum and angle were extracted.
The relevant cross sections were further reported as a function of the calorimetric measured energy and the reconstructed momentum transfer.

The measurement used data from the \uB\  LArTPC detector~\cite{Acciarri:2016smi}, which is the first of a series of LArTPCs to be used for precision oscillation measurements~\cite{Antonello:2015lea,Tortorici:2018yns,Abi:2020wmh,Abi:2020evt,Abi:2020oxb,Abi:2020loh}.
As described in chapter~\ref{ubexp}, the \uB\ detector has an active mass of 85 tons and is located along the Booster Neutrino Beam (BNB) at Fermilab, 463\,m downstream from the target.
The BNB energy spectrum extends to 2 GeV and peaks around 0.7 GeV~\cite{AguilarArevalo:2008yp}.
A neutrino is detected by its interaction with an argon nucleus in the LArTPC.
The secondary charged particles produced in the interaction travel through the liquid argon, leaving a trail of ionization electrons that drift horizontally  and  transverse  to  the  neutrino beam direction in an electric field of 273 V/cm, to a system of three anode wire planes located 2.5\,m from the cathode plane detailed in section~\ref{lartpctech}.
The Pandora tracking package~\cite{Acciarri:2017hat} described in section~\ref{pandora} is used to form individual particle tracks from the measured ionization signals.
Particle momenta are determined from the measured track length for protons and multiple Coulomb scattering pattern for muons~\cite{Abratenko:2017nki}.
	
The analysis presented here was performed on data collected from the BNB beam, with an exposure of $4.59 \times 10^{19}$ protons on target (POT). 
At nominal running conditions, one neutrino interaction is expected in $\approx$ 500 BNB beam spills. 
The trigger, based on the scintillation light detected by the 32 photomultiplier tubes (PMTs), increases the fraction of recorded spills with a neutrino interaction to $\approx 10 \%$. 
Application of additional software selection further rejects background events, mostly from cosmic muons, to provide a sample that contains a neutrino interaction in $\approx$ 15\% of the selected spills~\cite{Kaleko:2013eda,Adams:2018gbi}.  
A CCQE-like event selection, further cosmic rejection and neutrino-induced background rejection cuts, described in detail in~\cite{Adams:2018lzd}, are applied. 
Muon-proton pair candidates are identified by requiring two tracks with a common vertex and an energy deposition profile consistent with a proton and a muon~\cite{Adams:2016smi}. 
Further cuts on the track pair opening angle ($|\Delta \theta_{\mu,p} - 90^\circ| < 55^\circ$) and the muon and proton track lengths ($l_\mu > l_p$) reduce the cosmic background rate to less than $1 \%$~\cite{Adams:2018lzd}.
	
The selected CC1p0$\pi$ event definition includes events with any number of 
protons with momenta below 300\,MeV/$c$, neutrons at any momenta, and charged pions with momentum lower than 70\,MeV/$c$.
The minimal proton momentum requirement of 300\,MeV/$c$  is guided by its stopping range in LAr and corresponds to five wire pitches in the TPC, to ensure  an efficient particle identification.
	
To avoid contributions from cosmic tracks, our CC1p0$\pi$ selection considered only pairs of tracks with a fully-contained proton candidate, and a fully or partially contained muon candidate in the fiducial volume of the MicroBooNE detector. 
The fiducial volume is defined by 3 < $x$ < 253 cm, -110  < $y$ < 110  cm, and 5 < $z$ <  1031\,cm.
The  $x$ axis points  along  the  negative  drift  direction  with  $0$ cm  placed  at  the  anode  plane,  $y$ points vertically upward with $0$ cm at the center of the detector, and $z$ points  along  the  direction  of  the  beam,  with  $0$ cm  at  the  upstream edge of the detector.
Tracks are fully contained if both the start point and end point are within this volume, and partially contained if only the start point is within this volume.	

\begin{figure}[htb!]
\centering  
\includegraphics[width=0.7\linewidth]
	{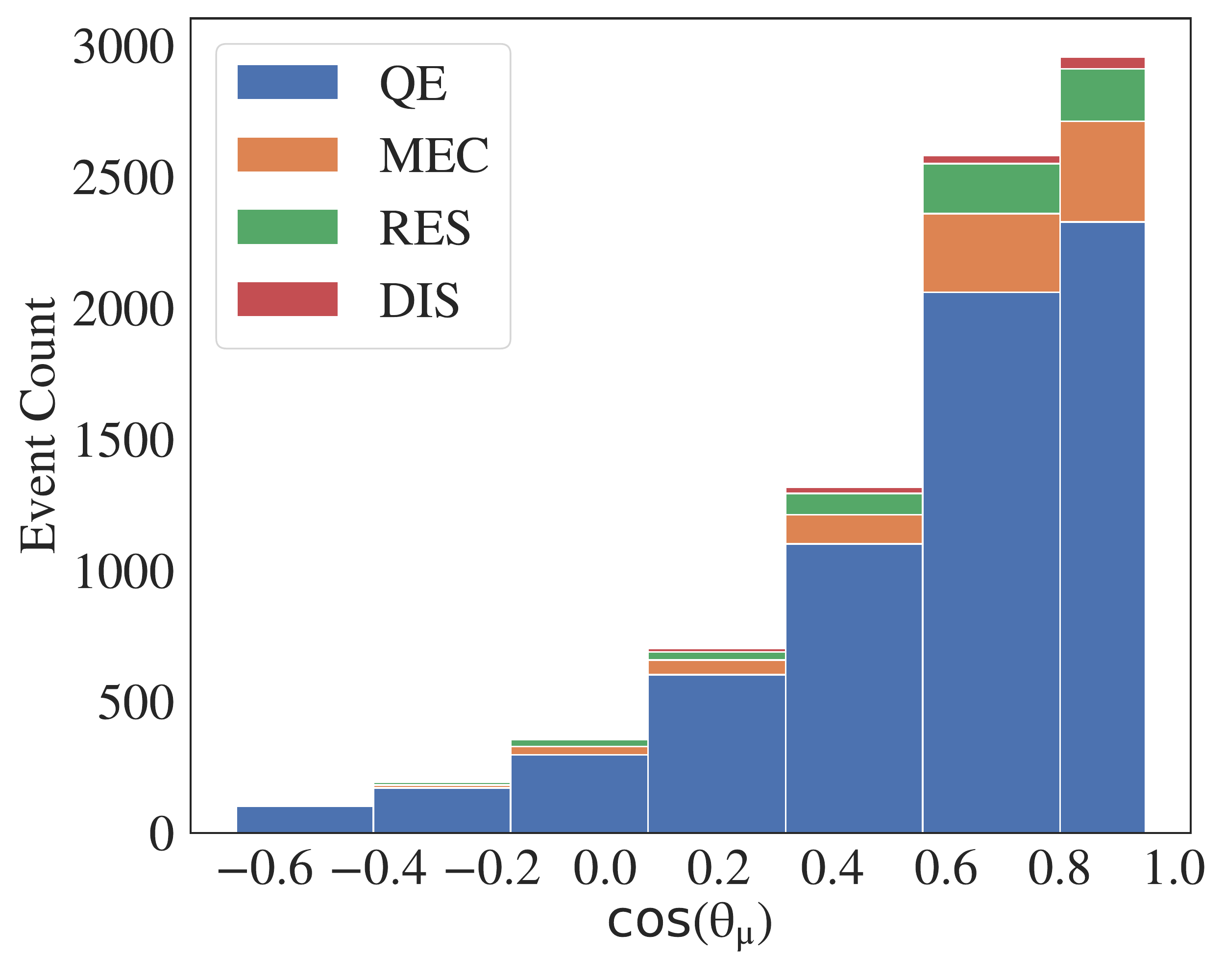}
	 \caption{Interaction breakdown of the cos$\theta_{\mu}$ plot illustrating the dominance of CCQE interactions after the application of our selection cuts.}
	 \label{CosThetaMuBreakDown}
\end{figure}

We limited our analysis to a phase space region where the detector response  to our signal is well understood and its effective detection efficiency is higher than 2.5\%.
This corresponds to $0.1 < p_{\mu} < 1.5$  GeV/$c$, $0.3 < p_{p} < 1.0$ GeV/$c$, $-0.75 < \cos\theta_{\mu} < 0.95$, and $\cos\theta_{p} >  0.15$.
Additional kinematical selections were used to enhance the contribution of CCQE interactions in our \CCIpOpi\ sample. 
These include requiring that the measured muon-proton pairs be coplanar ($|\Delta \phi_{\mu,p} - 180^\circ| < 35^\circ$) relative to the beam axis, have small missing transverse momentum relative to the beam direction ($p_T = |\vec{p}^{\,\mu}_T + \vec{p}^{\,p}_T| < 350$ MeV/$c$), and have a small energy deposition around the interaction vertex that is not associated with the muon or proton tracks.
This event selection results in a CCQE dominated sample, where table \ref{tab:BreakDown} shows the fractional contribution for each interaction channel.
Figure~\ref{CosThetaMuBreakDown} shows the relevant interaction breakdown for the entire sample of selected events as a function of cos$\theta_{\mu}$. The same nominal MC sample was also used to compute the purity.

After the application of the event selection requirement on the data sample, we retained 410 \CCIpOpi\ candidate events.
It is estimated that our \CCIpOpi\ CCQE-like event selection purity equals $\approx$ 84\%~\cite{Adams:2018lzd}, with $\approx$ 81\% of the measured events originating from an underlying CCQE interaction as defined by the GENIE event generator.
The efficiency for detecting \CCIpOpi\ events, out of all generated \CCIpOpi\ with an interaction vertex within our fiducial volume, was estimated using our Monte Carlo (MC) simulation and equals $\approx$ 20\%~\cite{Adams:2018lzd}.
We note that this efficiency includes acceptance effects, as the typical LArTPC efficiency for reconstructing a contained high-momentum proton or muon track is grater than $\approx$ 90\%~\cite{Acciarri:2017hat}.

\begin{center}
\begin{table}[htb!]
\centering
\begin{tabular}{ c c }
  \hline
  \hline
  \makecell{Interaction\\Mode} & \makecell{Fractional\\Contribution (\%)} \\
  \hline
  \hline
QE   & 81.1   \\
MEC  & 10.9  \\
RES  & 6.6   \\
DIS  & 1.4   \\
  \hline
  \hline
 \end{tabular}
 \caption{Interaction breakdown after the application of our selection cuts.}
 \label{tab:BreakDown}
\end{table} 
\end{center}

\begin{figure}[htb!]
\centering  
\includegraphics[width=0.7\linewidth]
{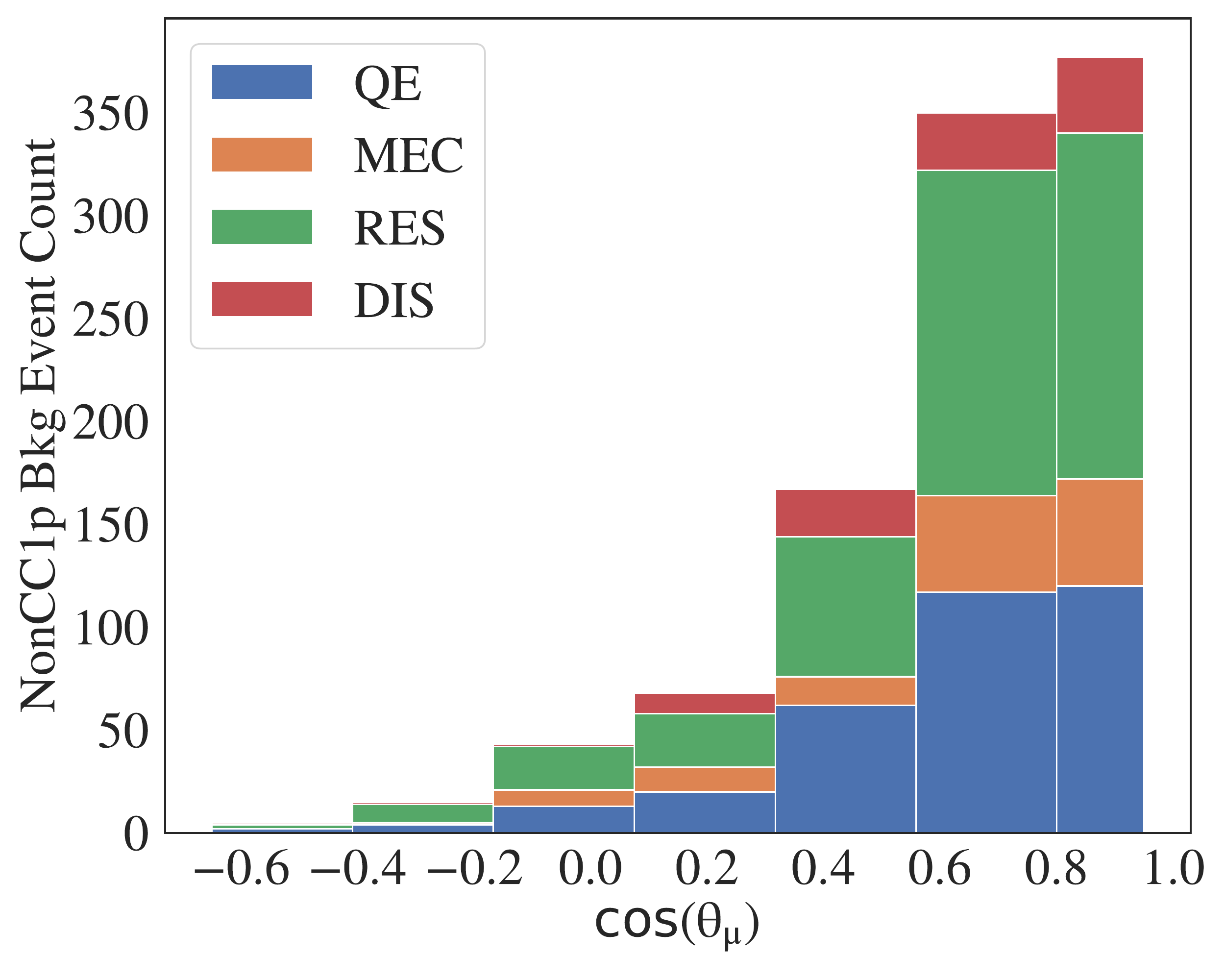}
\caption{Interaction breakdown of the cos$\theta_{\mu}$ plot illustrating the dominance of RES interactions after the application of our selection cuts for the non-CC1p0$\pi$ background part of the MC sample.}
\label{NonCC1pCosThetaMuBreakDown}
\end{figure}

Single differential cross sections are reported in measured proton and muon kinematics. 
The differential cross section is given by:
\begin{equation}
\label{eq:Xsec}
	\frac{\mathrm{d}\sigma}
	{\mathrm{d}X_n }
	=
	\frac{N^\textrm{on}_n - N^\textrm{off}_n - B_{n}}{ \epsilon_{n} \cdot \Phi_\nu \cdot N_{\textrm{target}} \cdot \Delta^p_{n}},
\end{equation}

where $X$ stands for the kinematical variable that the cross section is differential in and $n$ marks the cross-section bin.
In each bin $n$, $N_n^\textrm{on}$ is the number of measured events when the beam is on, $N_n^\textrm{off}$ is the number of measured events when the beam is off and cosmic-induced background events are collected, $B_n$ is the non-\Signal beam-related background estimated from MC, $N_{\textrm{target}}$ is the number of scattering nuclei, $\Phi_\nu$ is the integrated incoming neutrino flux, $\Delta^\mu_{n}$ and $\Delta^p_{n}$ are the differential bin widths, and $\epsilon_{n}$ is the effective particle detection efficiency.
We note that the high \CosThetaMu\ bin has large beam-related background corresponding $B_n$ in equation~\ref{eq:Xsec}, which is estimated using the GENIE v2.12.2 based MC simulation and is presented in figure~\ref{NonCC1pCosThetaMuBreakDown}.

As the detection efficiency is a multidimensional function of the interaction vertex and the particle momentum and direction, the data were binned in three-dimensional momentum, in-plane, and out-of-place angle bins with the effective detection efficiency calculated for each such bin separately and integrated over the interaction vertex in the detector.
The efficiency was extracted based on simulation and is defined as the ratio of the number of reconstructed CC1p0$\pi$ events to the number of true generated CC1p0$\pi$ events with a vertex inside our fiducial volume in bin n.  
This procedure accounts for bin migration effects such that cross-sections are obtained as a function of true kinematical variables, as opposed to experimentally reconstructed ones. 

The proton and muon efficiencies were extracted independently of each other, such that, when the cross-section is differential in muon kinematics, the proton kinematics is integrated over and vise versa.
This is done due to the limited data and simulation statistics and is justified since the proton and muon efficiencies are largely independent in the region of interest. 
The effect of residual correlations is accounted for in the systematic uncertainties.
We further note that the missing transverse momentum requirement increases the sensitivity of our efficiency corrections to the meson exchange current (MEC) and final state interaction (FSI) models used in our simulations. We accounted for the model sensitivity in our systematic studies detailed below.

The neutrino flux was predicted using the flux simulation of the MiniBooNE collaboration that used the same beamline~\cite{Aguilar-Arevalo:2013dva}. 
We accounted for the small distance between MiniBooNE and \uB.
Neutrino  cross  section  modeling uncertainties were estimated using the GENIE framework of event reweighting~\cite{Andreopoulos:2009rq,Andreopoulos:2015wxa} with its standard reweighting parameters.  
For both cross section and flux systematics, we use a multisim technique \cite{Roe:2007hw},  which  consists  of  generating many MC replicas, each one called a ``universe''\kern-.2em, where model parameters are varied  within their uncertainties. 
Each universe represents a different reweighting. 
The simultaneous reweighting of all model parameters allows the correct treatment of their correlations.	

A different model is followed for detector model systematic uncertainties, that are dominated by individual detector parameters.
Unisim samples~\cite{Roe:2007hw} were generated, where one detector parameter was varied each time by $1\sigma$.  
We then examined the impact of each parameter variation on the extracted cross sections, by obtaining the differences with respect to the central value on a bin-by-bin basis.
We note that the detection efficiency used for the cross section extraction is re-evaluated for each variation separately, including bin migration corrections. 
This procedure therefore accounted for the systematic uncertainty in these corrections due to both the cross-section and detector response modeling.
We then defined the total detector $1\sigma$ systematic uncertainty by summing in quadrature the effect of each individual variation.

%One exception to this process is the systematic uncertainty due to induced charge effects mentioned above that include the data-driven correction and are thus estimated separately.
%The extracted cross sections are expected to be independent of the azimuthal angle $\phi$. 
%However, the simple model used to simulate the effect of induced charge on neighboring TPC wires leads to a low reconstruction efficiency of tracks perpendicular to the wire planes ($\phi \approx 0$ and $\phi \approx \pm\pi$) that created an artificial $\phi$ dependence to the cross section. 
%We correct for this effect using an iterative procedure. 
%We first reweight events with a muon track falling in the $\phi \approx 0$ bin and $|\sin\theta|>0.3$ to the weighted average of the cross sections in all other bins of $\phi_{\mu}$ where  $|\sin\theta|>0.3$. 
%Due to the coplanarity requirement, this reweighting affects the distribution of $\phi_{p} \approx \pm \pi$. 
%We repeat the process starting from a proton track with $\phi_{p} \approx 0$ until the cross section change is less than 0.01$\%$, typically after 5 iterations.

\begin{figure}[htb!]
\centering  
\includegraphics[width=0.48\linewidth]{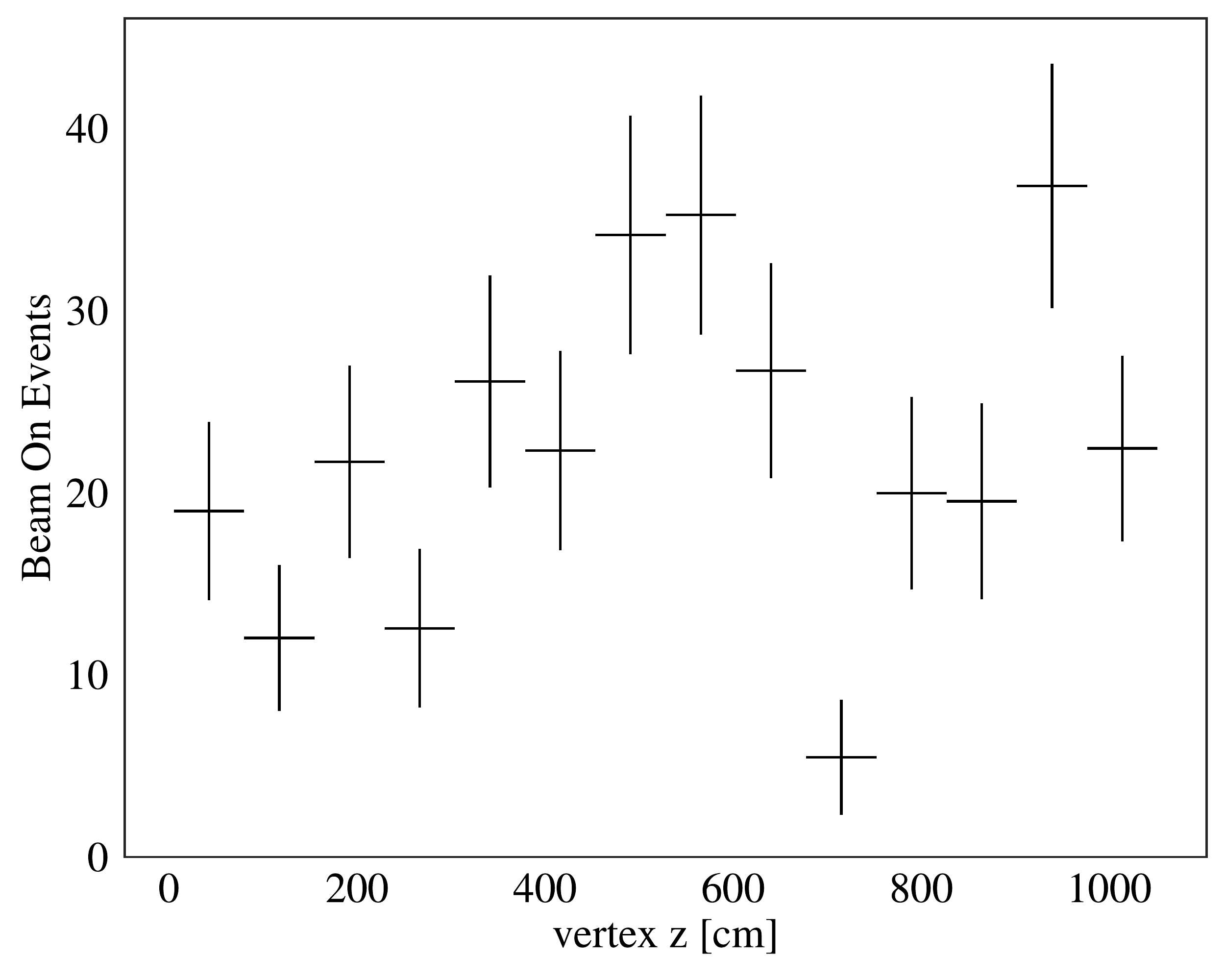}
\includegraphics[width=0.48\linewidth]{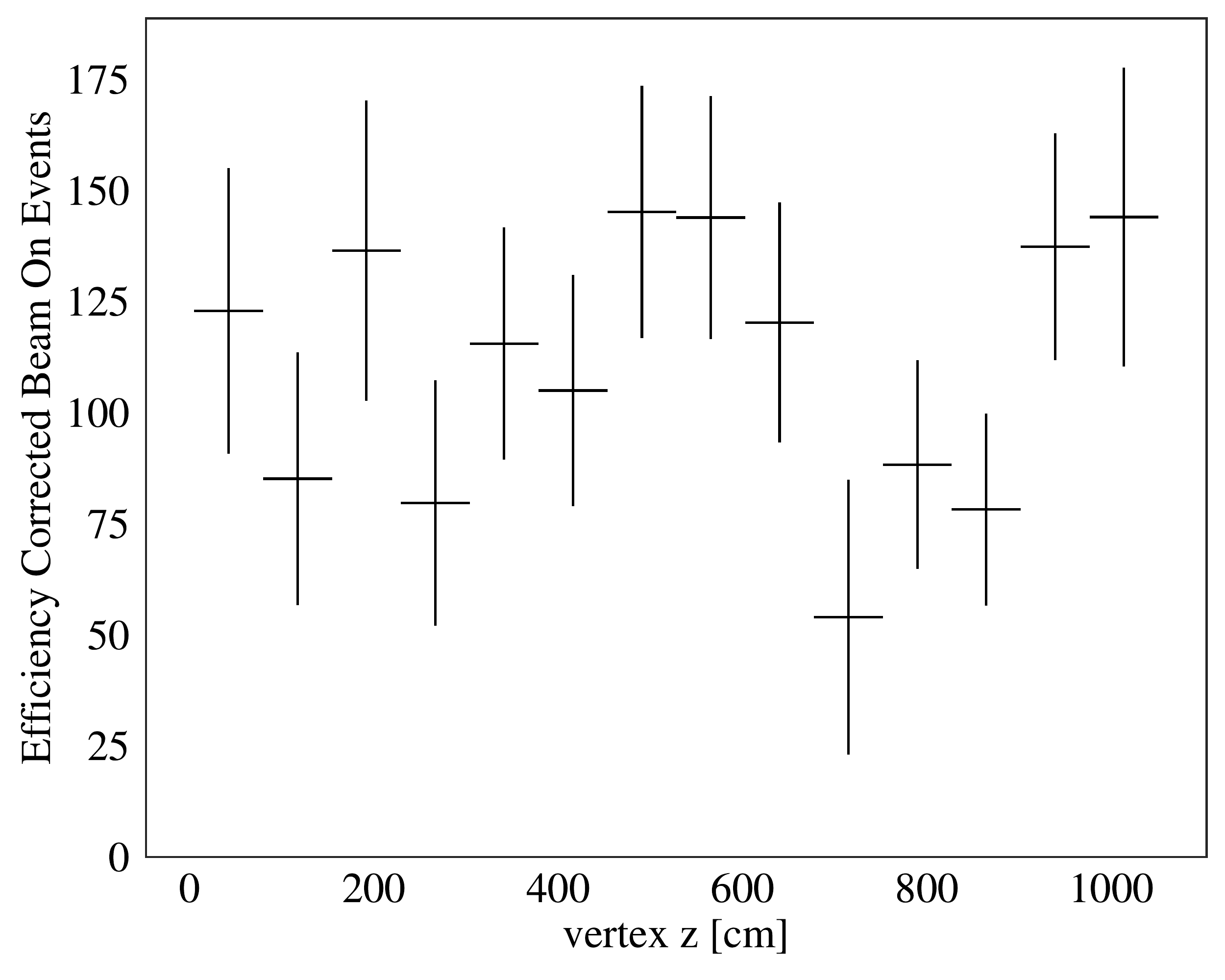} 
\caption{Vertex $z$ distribution for the measured events, after the beam related MC background has been subtracted, before (left) and after (right) detection efficiency corrections. No small-$z$ enhancement is observed and, with efficiency corrections, the measured distribution is consistent with that of a uniform neutrino interaction vertex.}
\label{fig:VertexZ}
\end{figure}

A dedicated MC simulation was used to estimate possible background from events in which a neutrino interacts outside the \uB\ cryostat but produce particles that enter the TPC and pass the event selection cuts~\cite{Abratenko:2019jqo}. 
No such events were found in that study, which is also supported by our observation that the $z$-vertex distributions for the measured events follows a uniform distribution, as can be seen in figure~\ref{fig:VertexZ}.
The measured $z$-vertex distribution, after the beam related MC background has been subtracted, does not show an excess at low-$z$, which indicates that background events from interactions upstream of the detector are not accidentally entering our selection, which would show up as a small-$z$ enhancement in our vertex distribution. 
The deficit at $z$ = 700\,cm is due to dead wires in our detector and its effect has been incorporated in our simulation. 

The MC simulation used to estimate the backgrounds and effective efficiency contains real cosmic data overlayed onto a neutrino interaction simulation that uses GENIE~\cite{Andreopoulos:2009rq,Andreopoulos:2015wxa} to simulate both the signal events and the beam backgrounds~\cite{Adams:2018lzd}. 
For the simulated portion, the particle propagation is based on GEANT4~\cite{Geant4}, while the simulation of the MicroBooNE detector is performed in the LArSoft framework~\cite{Pordes:2016ycs,Snider:2017wjd}. 
The beam-related background subtracted from the candidate \CCIpOpi\ events in the data sample is simulated.

%%%%%%%%%%%%%%%%%%%%%%%%%%%%%%%%%%%%%%%%%%%%%%%%%%%%%%%%%%%%%%%%%%%

\subsection{Quasielastic-like Cross-Section Results}\label{CCQEResults}

\begin{figure}[htb!]
\centering  
\includegraphics[width=0.7\linewidth]{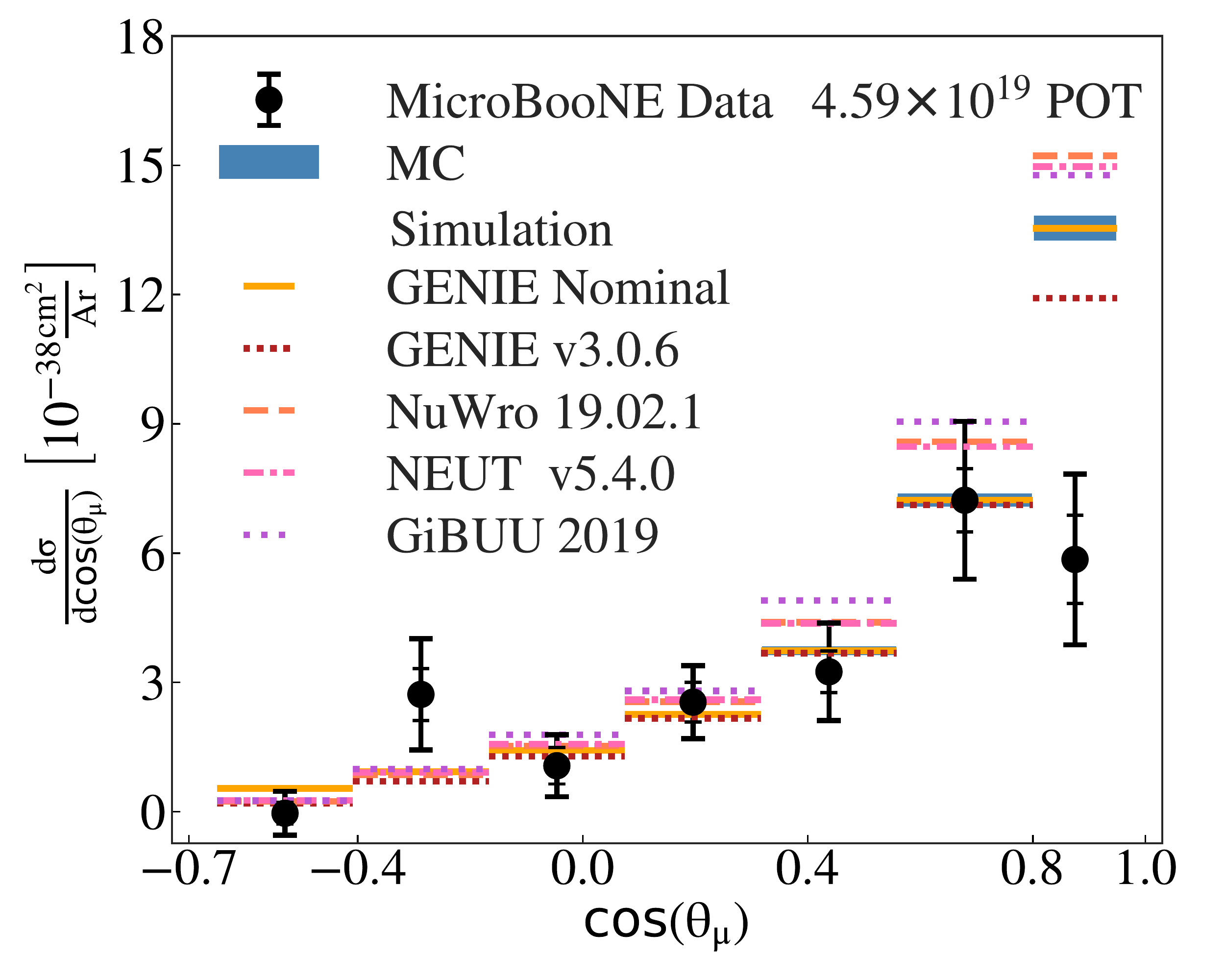}\\
\caption{The flux integrated single differential \CCIpOpi\ cross sections as a function of the cosine of the measured muon scattering angle.
Inner and outer error bars show the statistical and total (statistical and systematic) uncertainty at the 1$\sigma$, or 68\%, confidence level. Colored lines show the results of theoretical absolute cross section calculations using different event generators (without passing through a detector simulation). 
The blue band shows the extracted cross section obtained from analyzing MC events propagated through our full detector simulation. The width of the band denotes the simulation statistical uncertainty.}
\label{fig:Xsec_1D}
\end{figure}

Figure~\ref{fig:Xsec_1D} shows the flux integrated single differential \CCIpOpi\ cross section as a function of the cosine of the measured muon scattering angle.
The data were compared to several theoretical calculations and to our GENIE-based MC prediction.
This prediction is the result of analyzing a sample of MC events produced using our ``nominal'' GENIE model and propagated through the full detector simulation in the same way as data.
This model (GENIE v2.12.2)~\cite{Andreopoulos:2009rq,Andreopoulos:2015wxa} treats the nucleus as a the Bodek-Ritchie Fermi Gas (RFG), used the Llewellyn-Smith CCQE scattering prescription~\cite{LlewellynSmith:1971uhs}, the empirical MEC model~\cite{Katori:2013eoa}, the Rein-Sehgal resonance (RES) model, the coherent (COH) scattering model~\cite{Rein:1980wg}, and a data driven FSI model denoted as ``hA''~\cite{Mashnik:2005ay}.
 
In addition, theoretical predictions by several other event generators are shown at the cross-section level without any detector effects~\cite{Stowell_2017}.
These include GENIE v2.12.2 and v3.0.6~\cite{Andreopoulos:2009rq,Andreopoulos:2015wxa}, NuWro 19.02.1~\cite{GolanNuWro:2008yp}, and NEUT v5.4.0~\cite{Hayato:2008yp}.
The agreement between the ``nominal'' GENIE calculation (v2.12.2) and the MC prediction constitutes a closure test for our analysis. 
The other generators all improve on GENIE v2.12.2 by using updated nuclear interaction models, among which is the use of a Local Fermi Gas model (LFG)~\cite{Carrasco:1989vq} and Random Phase Approximation (RPA) correction~\cite{RPA}.
GENIE v3.0.6 and NEUT also include Coulomb corrections for the outgoing muon~\cite{Engel:1997fy}.
The theoretical models implemented in these event generators include free parameters that are typically fit to data, with different generators using different data sets.
We also consider the GiBUU 2019~\cite{Mosel:2008yp} event generator which fundamentally differs from the others due to its use of a transport equation approach.
A brief discussion of the underlying model configuration used in the different event generator predictions included in this analysis is shown below.

\begin{itemize}

\item \underline{GENIE Nominal}: Uses the aforementioned GENIE v2.12.2 modeling.	

\item \underline{NuWro 19.02.1}~\cite{GolanNuWro:2008yp}: Using the LFG ground state model~\cite{Carrasco:1989vq}, the Llewellyn-Smith CCQE scattering prescription~\cite{LlewellynSmith:1971uhs}, the Transverse Enhancement model for two--body currents~\cite{Bodek:2017hat}, the Adler-Rarita-Schwinger formalism to calculate the $\Delta$ resonance explicitly~\cite{Graczyk:2008yp}, the BS COH~\cite{Berger:2008xs} scattering model and an intranuclear cascade model for FSI.

\item \underline{NEUT v5.4.0}~\cite{Hayato:2008yp}: Using the LFG ground state model~\cite{Carrasco:1989vq}, the Nieves CCQE scattering prescription~\cite{Nieves:2012yz}, the Nieves MEC model~\cite{Schwehr:2016pvn}, the BS RES~\cite{Nowak:2009se,Kuzmin:2003ji,Berger:2007rq,Graczyk:2007bc} and Rein-Sehgal COH~\cite{Rein:1980wg} scattering models, and FSI with Oset medium correction for pions~\cite{Andreopoulos:2009rq,Andreopoulos:2015wxa}.

\item \underline{GiBUU 2019}: Using somewhat similar models, but unlike other generators, those are implemented in a coherent way, by solving the Boltzmann-Uehling-Uhlenbeck transport equation~\cite{Mosel:2008yp}. The models include: Local Fermi Gas model~\cite{Carrasco:1989vq}, standard CCQE expression~\cite{Leitner:2006ww}, empirical MEC model and a dedicated spin dependent resonance amplitude calculation following the MAID analysis~\cite{Mosel:2019vhx}. The DIS model is as in PYTHIA~\cite{Sjostrand:2006za} and the FSI treatment is different as the hadrons propagate through the residual nucleus in a nuclear potential which is consistent with the initial state. 
\end{itemize}

As can be seen in figure~\ref{fig:Xsec_1D}, all models are in overall good agreement with our data, except for the highest \CosThetaMu\ bin with \CosThetaMu$>$ 0.8, where the measured cross section is significantly lower than the theoretical predictions.
This discrepancy cannot be explained by the systematic uncertainties and is therefore indicative of an issue with the theoretical models. 
Specifically, high \CosThetaMu\ correspond to low momentum transfer events which were previously observed to not be well reproduced by theory in inclusive reactions~\cite{Abratenko:2019jqo,Carneiro:2019jds} and is now also seen in exclusive reactions.

As the differential cross sections in proton kinematics and muon momentum include contributions from all muon scattering angles, their agreement with the theoretical calculation is affected by this disagreement in the forward muon scattering angle.
Therefore, for the results presented below, we repeated the cross-section extraction exercise twice, where the first time we included all the events that satisfy our selection criteria, and the second time we excluded those events with \CosThetaMu$>$ 0.8.
The corresponding integrated measured CC1p0$\pi$ cross sections are summarized in table~\ref{IntegratedXSec}. 
The same table also lists the $\chi^2$  for the agreement of the different models with the data for differential cross sections for the full available phase-space and for \CosThetaMu\ $ < 0.8$.
The values reported in the table are the simple sum of those $\chi^2$ values obtained for each distribution separately. 
Systematic uncertainties and correlations were accounted for using covariance matrices.

\begin{table}[htb!]
\caption{ Integrated cross section values and $\chi^2$ values for the agreement between the measured  cross sections and various event generators.
Results are listed for the full measured phase space and for a limited one of $\cos(\theta_{\mu}) < 0.8$.}
\centering
\resizebox{0.6\textwidth}{!}{%
\centering
\begin{tabular}{|c|c|c|c|}
\hline
\multicolumn{2}{|c|}{\multirow{2}{*}{}} & \multicolumn{2}{c|}{Integrated Cross Section $[10^{-38} $cm$^2]$}  \\ 
\multicolumn{2}{|c|}{\multirow{2}{*}{}} & \multicolumn{2}{c|}{(Differential Cross Section  $\chi^2$/d.o.f)}  \\ \cline{3-4} 
\multicolumn{2}{|c|}{}                  & $-0.75<\cos(\theta_\mu)<0.95$   & $-0.75<\cos(\theta_\mu)<0.8$                   \\ \hline
\multicolumn{2}{|c|}{Data CC$1p0\pi$}       & 4.93 $\pm$ 1.55           & 4.05 $\pm$ 1.40                                                                                                                      \\ \hline 
\multirow{4}{*}{\rotatebox{90}{Generators\;\;\;\;\;}}  & GENIE Nominal           & 6.18~(63.2/28)                      & 4.04~(30.1/27)                      \\ 
                                             & GENIE v3.0.6            & 5.45~(34.6/28)                      & 3.66~(21.4/27)                        \\
                                             & NuWro 19.02.1           & 6.67~(76.7/28)                      & 4.39~(29.9/27)                    \\ 
                                             & NEUT v5.4.0             & 6.64~(78.5/28)                      & 4.39~(32.2/27)                    \\ 
                                             & GiBUU 2019              & 7.00~(82.2./28)                      & 4.78~(40.0/27)                     \\ \hline
\end{tabular}}
\label{IntegratedXSec}
\end{table}

As can be seen in table~\ref{IntegratedXSec}, GENIE v3.0.6 is the only model that reaches a $\chi^2$/d.o.f. close to unity for the full phase-space.
It is also the closest model to the data at the highest \CosThetaMu\ bin.
For all other models, the $\chi^2$/d.o.f. in the \CosThetaMu\ $ < 0.8$ sample is reduced by a factor of $\sim 2$ as compared to the full phase-space sample. 
GENIE v3.0.6 shows a smaller reduction in this case, and GiBUU 2019 obtains a consistently higher $\chi^2$/d.o.f. for both the full and limited phase-space samples.

\begin{figure}[htb!]
\centering  
\includegraphics[width=0.32\linewidth]{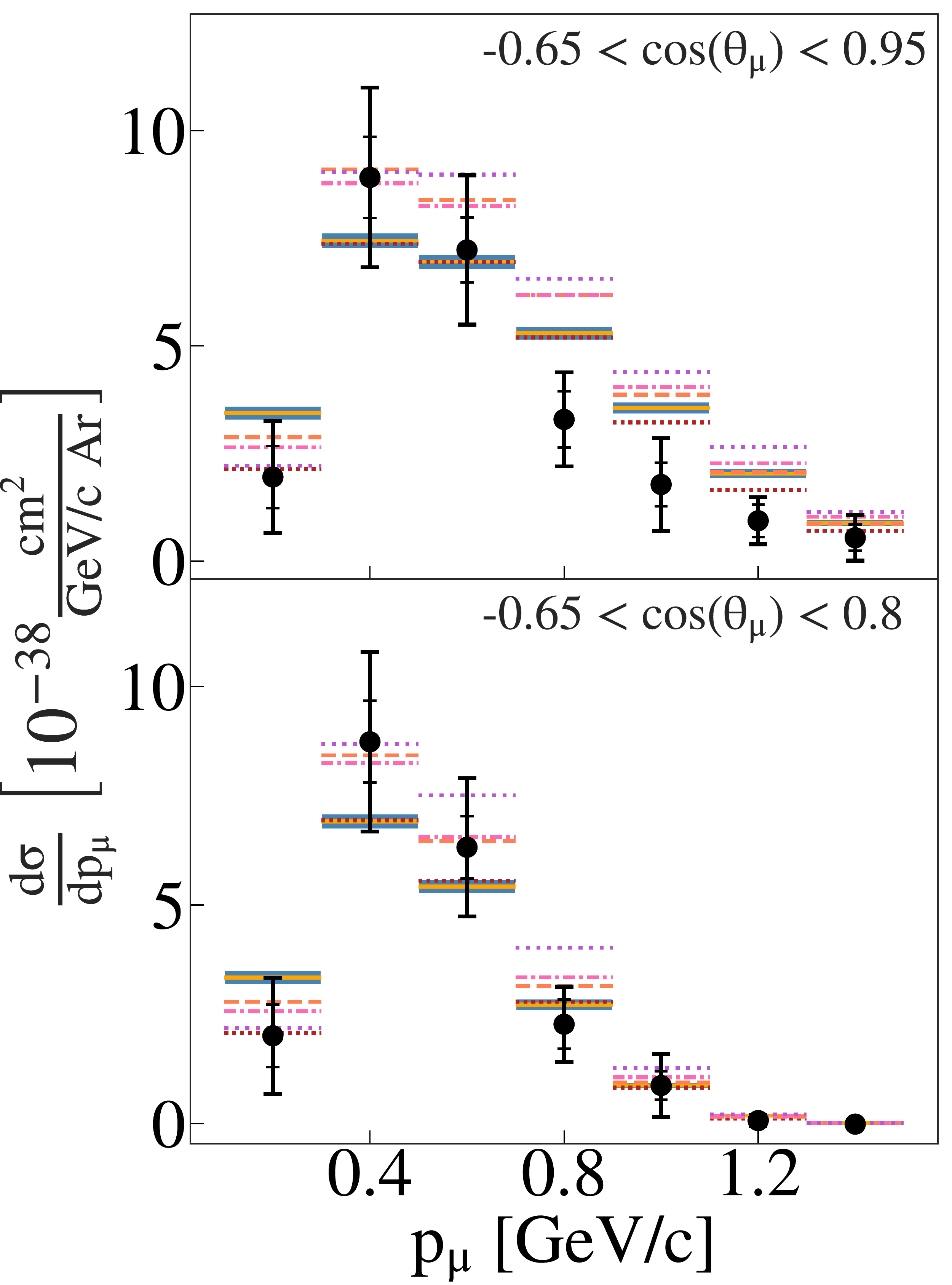}
\includegraphics[width=0.32\linewidth]{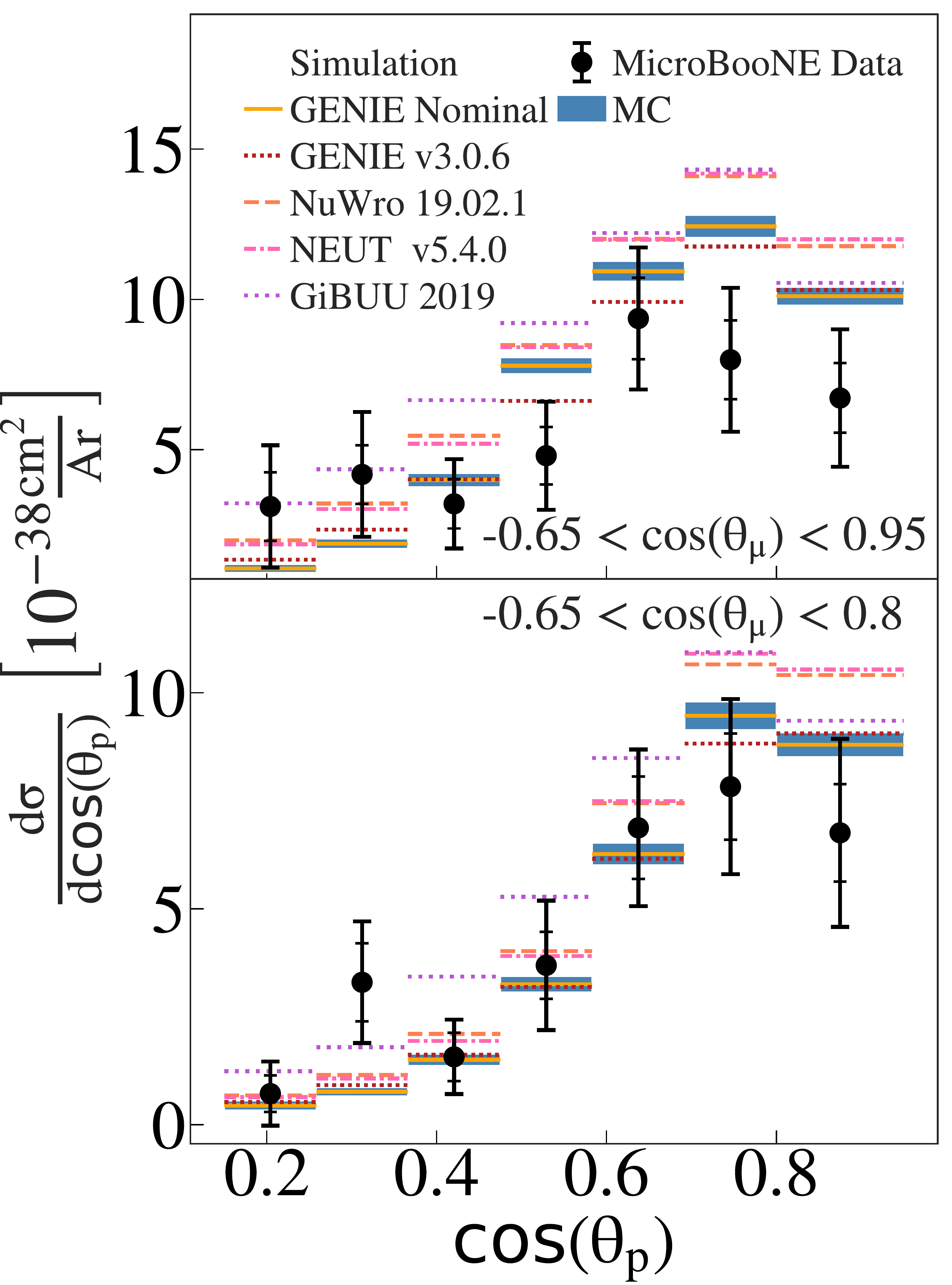}		
\includegraphics[width=0.32\linewidth]{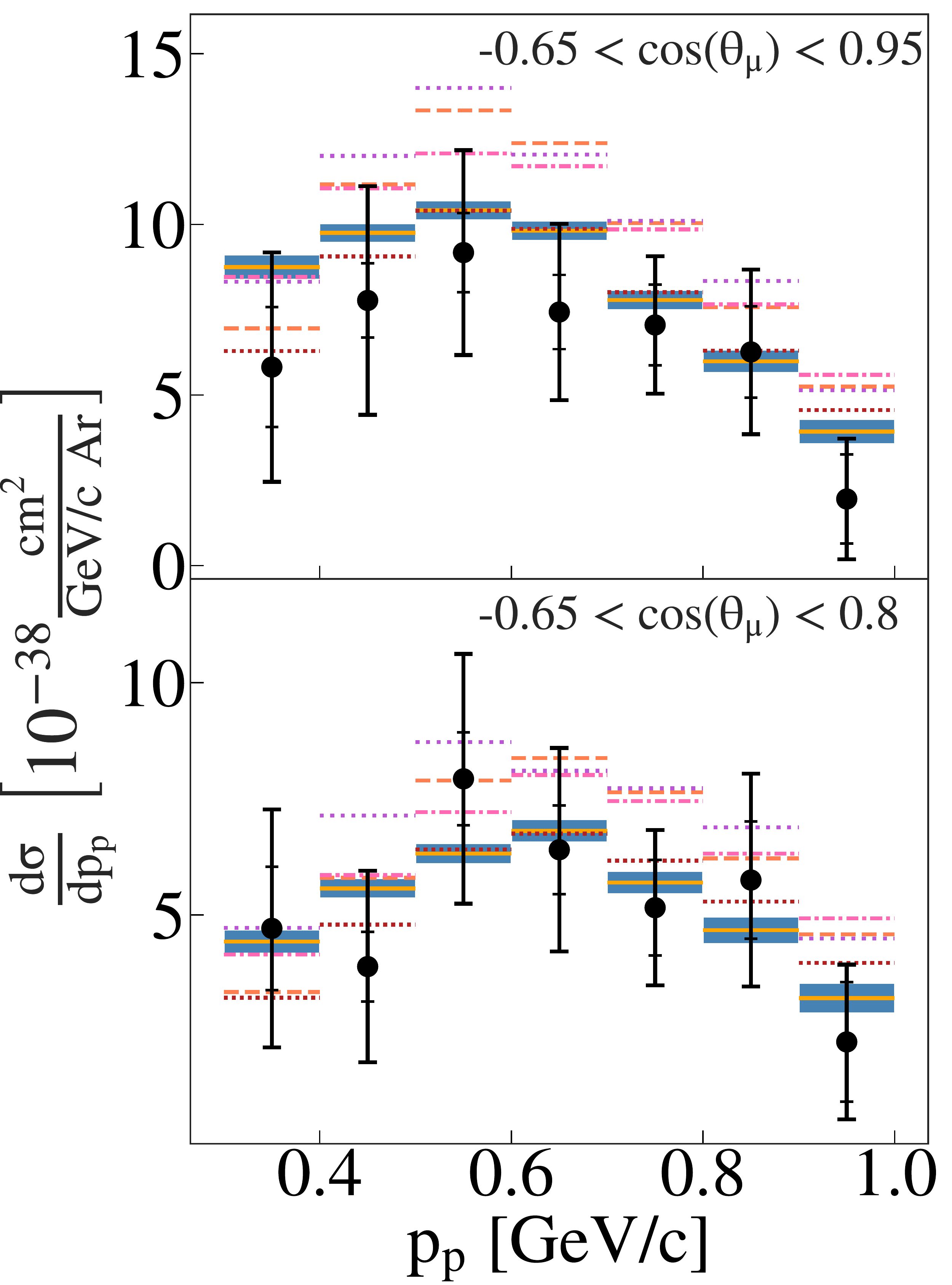}
\caption{As figure~\ref{fig:Xsec_1D},  but for the differential cross sections as a function of measured muon momentum (left) and measured proton scattering angle (middle) 
and momentum (right). 
Cross sections are shown for the full measured phase-space (top) and for events with cos$(\theta_\mu) < 0.8$ (bottom). 
Inner and outer error bars show the statistical and total (statistical and systematic) uncertainty at the 1$\sigma$, or 68\%, confidence level. 
Colored lines show the results of theoretical absolute cross section calculations using different event generators (without passing through a detector simulation). 
The blue band shows the extracted cross section obtained from analyzing MC events passed through our full detector simulation.}
\label{fig:Xsec_1D_without_last_bin}
\end{figure}

Figure~\ref{fig:Xsec_1D_without_last_bin} shows this comparison between the relevant cross sections in the full available phase-space (top) and in the case where events with \CosThetaMu\ $ > 0.8$ are excluded (bottom). 
Removing this part of the phase-space significantly improves the agreement between data and theory.
The improved agreement with the data observed for GENIE v3.0.6, especially for the full phase-space sample, is intriguing. 
Specifically, GENIE v3.0.6 and NEUT v5.4.0 are quite similar, using the same nuclear, QE, and MEC models, which are the most significant processes in our energy range.
They do differ in the Coulomb corrections that only GENIE v3.0.6 and NEUT have, their free parameter tuning process, and the implementation of RPA correction, that are known to be important at low momentum transfer~\cite{RPA}.
Our data indicates that these seemingly small differences can have a highly significant impact, as seen in table~\ref{IntegratedXSec}.	

\begin{figure}[htb!]
\centering  
\includegraphics[width=0.39\linewidth]{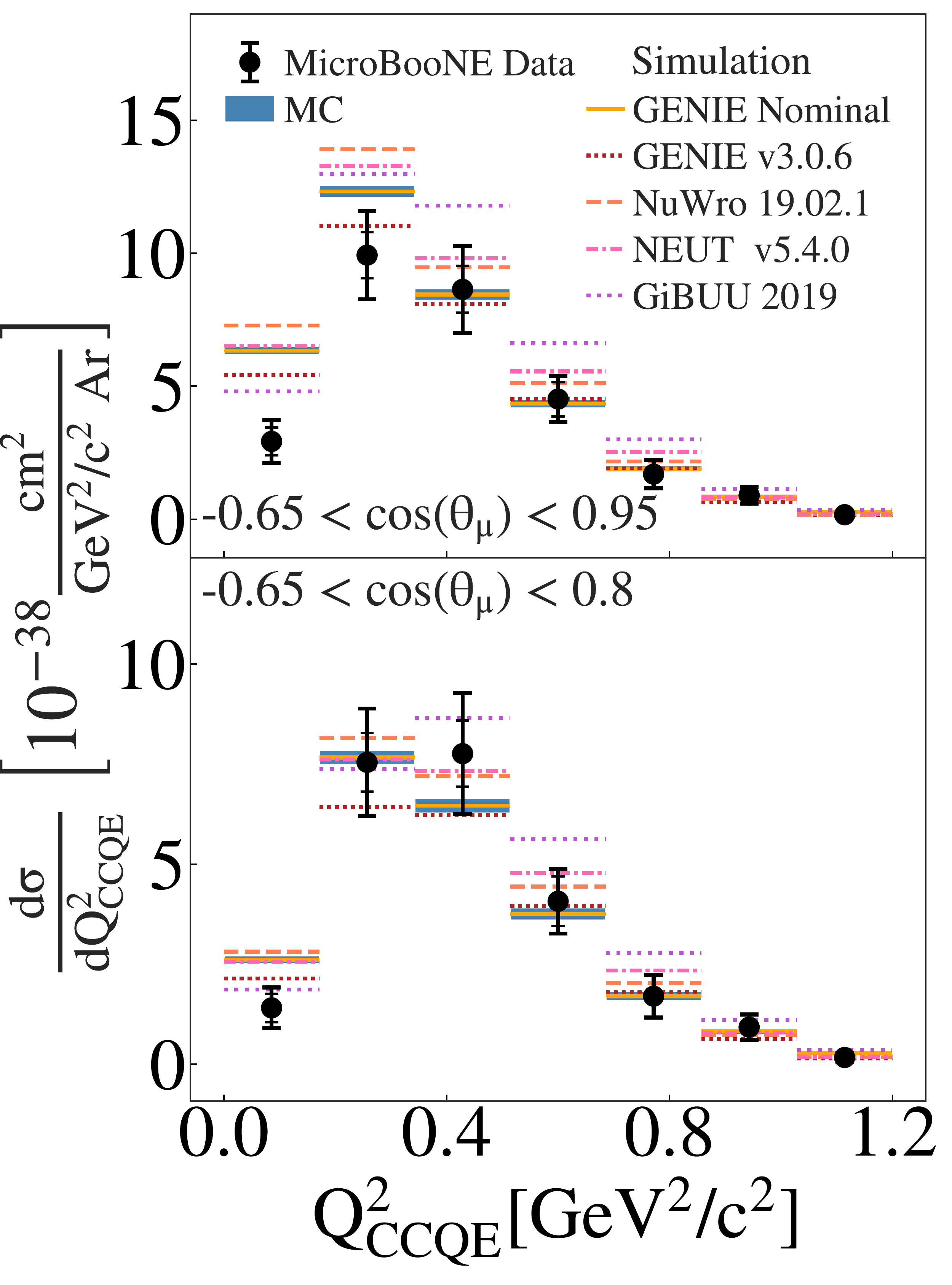}
\includegraphics[width=0.39\linewidth]{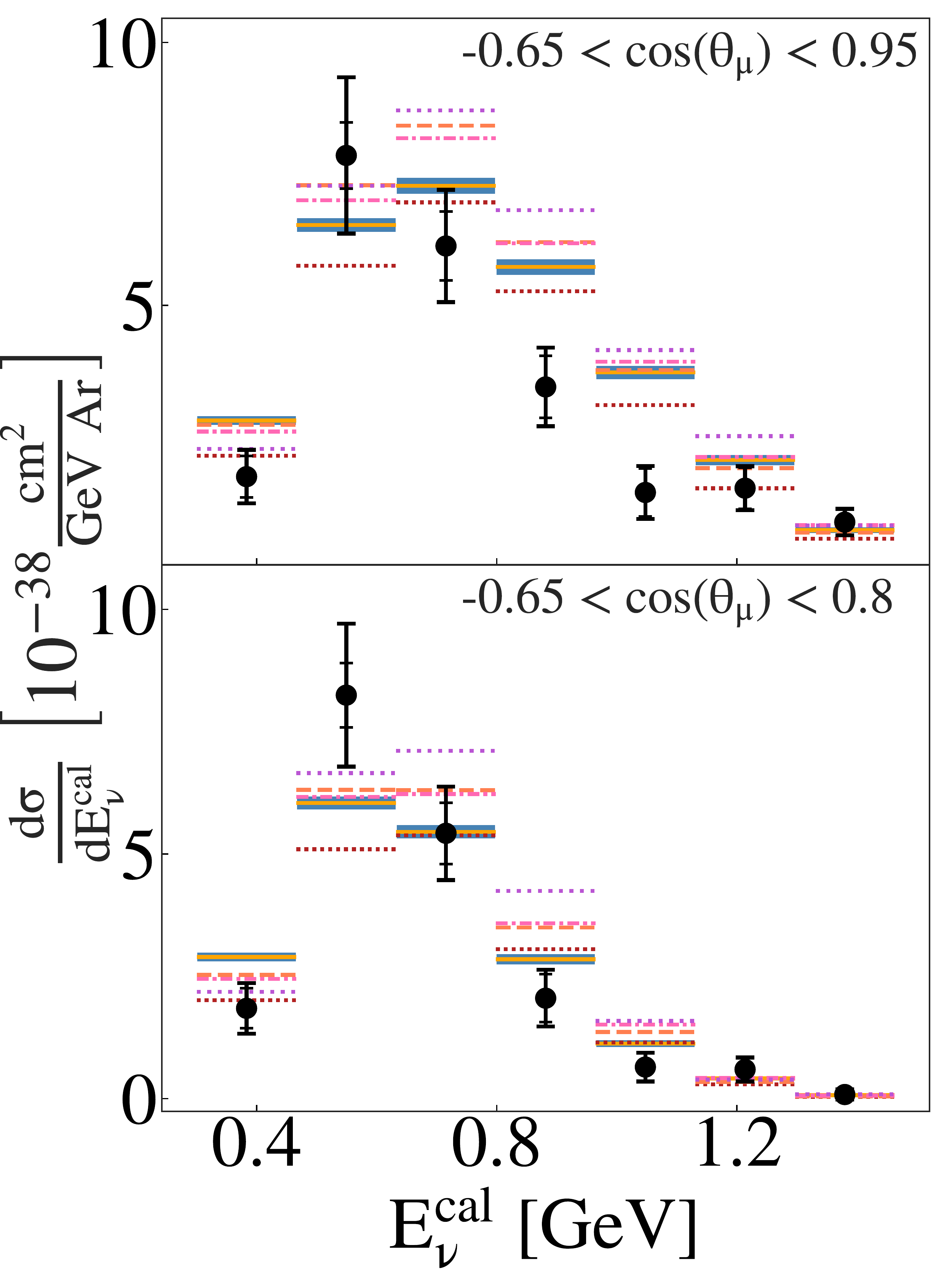} 		
\caption{The flux integrated single differential \CCIpOpi\ cross sections as a function of $Q^{2}_{CCQE}  = (E^{cal}_{\nu} - E_{\mu})^{2} - (\vec{p}_{\nu} - \vec{p}_{\mu})^{2}$ and $E_{\nu}^{cal}  = E_{\mu} + T_{p} + BE$, 
where $BE =  40$ MeV and $\vec{p}_{\nu} = (0, 0, E^{cal}_{\nu})$. 
%Inner and outer error bars show the statistical and total (statistical and systematic) uncertainty at the 1$\sigma$, or 68\%, confidence level. 
Colored lines show the results of theoretical absolute cross section calculations using different event generators (without passing through a detector simulation). 
The blue band shows the extracted cross section obtained from analyzing MC events passed through our full detector simulation.}
\label{fig:Xsec_1D_Ev_Q2}
\end{figure}

Lastly, figure~\ref{fig:Xsec_1D_Ev_Q2} shows the flux-integrated single differential cross sections as a function of calorimetric measured energy and reconstructed momentum transfer, with and without events with \CosThetaMu\ $> 0.8$. 
The former is defined as $E_{\nu}^{cal} = E_{\mu} + T_{p} + BE$, and the latter as $Q^{2}_{CCQE} = (\vec{p}_{\nu} - \vec{p}_{\mu})^{2} - (E^{cal}_{\nu} - E_{\mu})^{2}$, where E$_{\mu}$ is the muon energy, T$_{p}$ is the proton kinetic energy, BE = 40 MeV is the effective nucleon binding energy for argon, and $\vec{p}_{\nu} = (0, 0, E^{cal}_{\nu})$ is the reconstructed interacting neutrino momentum. 
$E_{\nu}^{cal}$ is often used as a proxy for the true neutrino energy.
Overall, good agreement is observed between data and calculations for these complex variables, even for the full event sample without the \CosThetaMu\ $ < 0.8$  requirement.

%The statistical uncertainty of our measurement is 15.9\%.
The systematic uncertainty of our measurement summed up to 26.2\% and included contributions from the neutrino flux prediction and POT estimation (18.7\%), detector response modeling (18.4\%), imperfect proton and muon efficiency decoupling (5.7\%), and neutrino interaction cross section modeling (7.1\%).

%%%%%%%%%%%%%%%%%%%%%%%%%%%%%%%%%%%%%%%%%%%%%%%%%%%%%%%%%%%%%%%%%%%%%%%%

\subsection{Quasielastic-like Cross-Section Analysis Conclusions}\label{QEConcl}

In summary, the first measurement of $\nu_{\mu}$ CCQE-like differential cross sections on argon was reported for event topologies with a single muon and a single proton detected in the final state using data sets from the MicroBooNE LArTPC. 
The data are in good agreement with simulation predictions, except at small muon scattering angles that correspond to low-momentum-transfer reactions.
This measurement confirmed and constrained calculations essential for the extraction of oscillation parameters and highlights kinematic regimes where improvement of theoretical models is required. 
The benchmarking of exclusive \CCIpOpi\,cross sections on argon presented here suggests that measurements of \CCIpOpi\,interactions are a suitable choice for use in precision neutrino oscillation analyses, especially after theoretical models are reconciled with the small scattering angle data.

%%%%%%%%%%%%%%%%%%%%%%%%%%%%%%%%%%%%%%%%%%%%%%%%%%%%%%%%%%%%%%%%%%%%%%%%

\section{First Multidimensional Measurement Of Kinematic Imbalance Cross Sections On Argon}\label{CC1p}

%%%%%%%%%%%%%%%%%%%%%%%%%%%%%%%%%%%%%%%%%%%%%%%%%%%%%%%%%%%%%%%%%%%%%%%%

\subsection{Kinematic Imbalance Neutrino Data Analysis}\label{CC1pDataAna}

\begin{figure}[htb!]
\centering  
\includegraphics[width=0.7\linewidth]{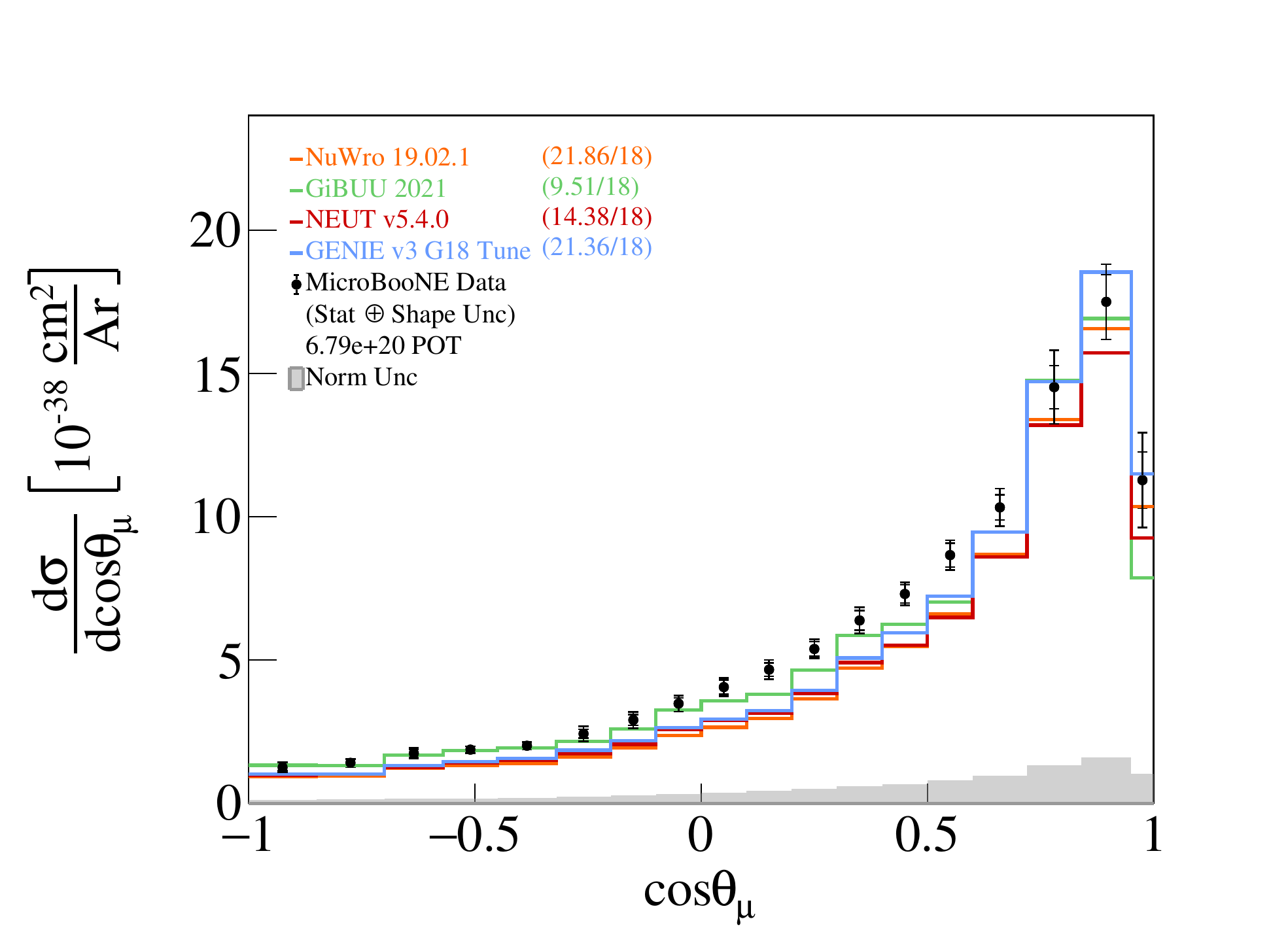}
\caption{Muon angular distribution after the implementation of the analysis framework improvements. No data-MC disagreement is observed in the forward direction.}
\label{mcc9muoncostheta}
\end{figure}

Over the course of two years (2019-2021), the MicroBooNE collaboration made significant improvements to the pre-existing analysis framework.
These improvements provided high statistics neutrino-argon data sets, improved signal processing~\cite{Adams:2018gbi}, reduced detector systematics~\cite{WireMod}, a theory-driven interaction modeling~\cite{geniev3highlights}, and the creation of the first MicroBooNE tune~\cite{GENIEKnobs}.
Figure~\ref{mcc9muoncostheta} illustrates the improved data-MC agreement after the implementation of these changes as a function of cos$\theta_{\mu}$, where the disagreement in the forward direction is longer observed.
The improved picture at cos$\theta_{\mu} \approx$ 1 is primarily driven by the improved modeling of the MC beam related backgrounds. 

Motivated by these improvements, in this analysis the first study of kinematic imbalance variables on argon, which are sensitive to nuclear effects, is reported.
These variables are studied using \Signal events within a neutrino slice, as defined by the Pandora reconstruction framework and detailed in section~\ref{pandora}. 
Exactly one muon with 0.1 < $p_{\mu}$ < 1.2\,GeV/c, exactly one proton with 0.3 < $p_{p}$ < 1\,GeV/c, no charged pions above the 0.07\,GeV/c threshold, and no other mesons of any momenta are required in the final state originating from charged-current $\nu_{\mu}$-Ar scattering events.	
The existence of any number of neutrons, electrons or photons is allowed.

\begin{figure}[htb!]
\centering  
\includegraphics[width=10cm]{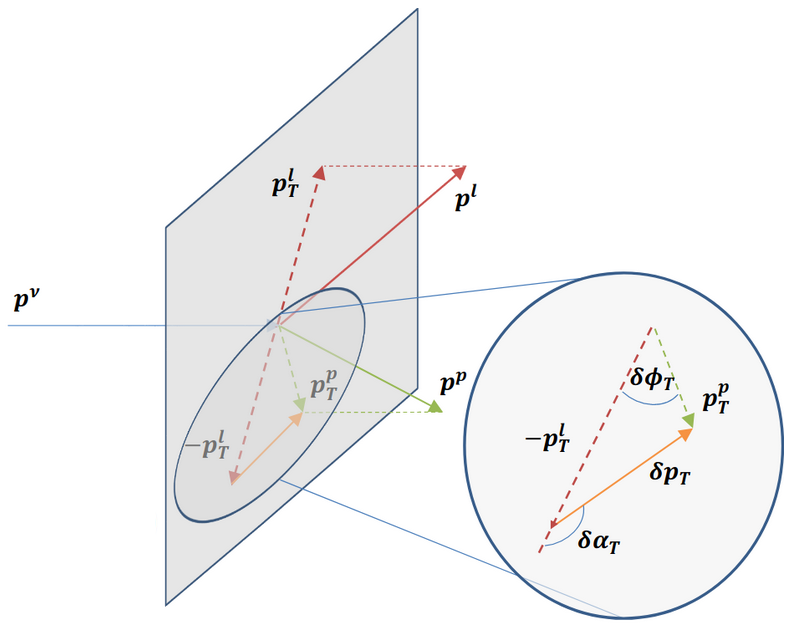}		
\caption{Schematic illustration of the single transverse variables $\delta p_{T}$, $\delta\alpha_{T}$ and $\delta\phi_{T}$. Figure adapted from~\cite{DolanTKITrento}.}
\label{TransVar}
\end{figure}

Such kinematic imbalance variables of interest include the transverse variables (TVs), namely $\delta p_{T}$, $\delta \alpha_{T}$ and $\delta \phi_{T}$~\cite{PhysRevLett.121.022504,PhysRevD.103.112009}. 
These are built specifically to characterize and minimize the  degeneracy  between  the  nuclear  effects  most pertinent  to  long-baseline  oscillation  experiments.
In particular, the TVs facilitate the possible identification of the  Fermi motion of the initial state nucleon, the final state re-interactions of the nucleons in the nucleus and the multi-nucleon interactions (2p2h).
As shown in figure~\ref{TransVar}, the TVs are defined by projecting the lepton and proton momentum  on  the  plane  perpendicular  to  the  neutrino  direction.   

In  the  absence  of  any  nuclear  effects,  the  proton and muon momenta are equal and opposite in this plane and therefore the measured difference between their projections is a direct probe of nuclear effects in quasi-elastic (QE) events.
$\delta\vec{p}_{T}$ can be fully characterized in terms of the vector magnitude ($\delta p_{T}$) and the two angles ($\delta \alpha_{T}$ and $\delta \phi_{T}$):
        
\begin{equation}
\label{deltapt}
\begin{array}{c}
\delta p_{T} = |\vec{p}_{T}\,^{\ell} + \vec{p}_{T}\,^{p}|\\
\end{array}
\end{equation}        
	
\vspace{-1cm}	
	
\begin{equation}
\label{deltaalphat}
\begin{array}{c}
\delta \alpha_{T} = arccos(\frac{- \vec{p}_{T}\,^{\ell} \cdot \delta \vec{p}_{T}}{p_{T}\,^{\ell} \cdot \delta p_{T}})\\
\end{array}
\end{equation} 

\vspace{-1cm}
	
\begin{equation}
\label{deltaphit}
\begin{array}{c}
\delta \phi_{T} = arccos(\frac{- \vec{p}_{T}\,^{\ell} \cdot \vec{p}_{T}\,^{p}}{p_{T}\,^{\ell} \cdot p_{T}\,^{p}})\\
\end{array}
\end{equation}        	
        
where $\vec{p}_{T}\,^{\ell}$ and $\vec{p}_{T}\,^{p}$ are, respectively, the projections of the momentum  of  the  outgoing  lepton  and  proton  on  the transverse plane.  
Different nuclear effects alter the distributions of the TVs in different and predictable ways.
Measurements of the TVs therefore have a unique sensitivity to identify nuclear effects.  
This  allows  cross sections  extracted  using  these observables to act as a powerful tool to tune and distinguish nuclear models.  
Furthermore, in case of disagreement, the TV distributions provide useful hints on the possible causes of the discrepancies.  

Another kinematic variable of interest corresponds to the total momentum of the struck nucleon.
The formalism introduced in~\cite{PhysRevLett.121.022504} is adopted.
This formalism provides an approximation for the longitudinal component of the struck nucleon momentum shown in equation~\ref{DPLFinalmain}, which is derived in appendix~\ref{totstruckmom},

\begin{equation}
\begin{split}
\delta p_{L} = \frac{1}{2}R - \frac{m_{A-1}^{2} + \delta p_{T}^{2}}{2R}.
\end{split}
\label{DPLFinalmain}
\end{equation} 	
 	
For simplicity, we defined
 	
\begin{equation}
\begin{split}
R \equiv m_{A} + p^{\mu}_{L} + p^{p}_{L} - E^{\mu} - E^{p}
\end{split}
\label{Longmain}
\end{equation}
 	
Combining information from both the longitudinal and the transverse components gave us access to an approximation for the total struck nucleon momentum

\begin{equation}
\begin{split}
p_{n,proxy} = \sqrt{\delta p_{L}^{2} + \delta p_{T}^{2}}
\end{split}
\label{totmain}
\end{equation}	
 	
\begin{figure}[htb!]
\centering  
\includegraphics[width=0.5\linewidth]{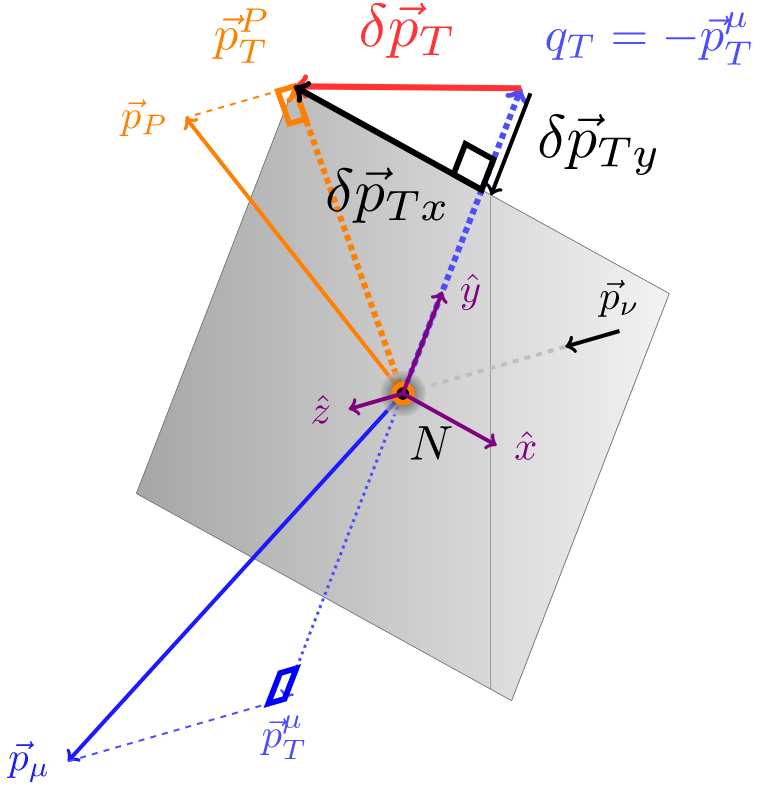}
\caption{Schematic illustration of $\delta p_{Tx}$ and $\delta p_{Ty}$. Figure adapted from~\cite{PhysRevD.101.092001}.}
\label{Graphicxy}
\end{figure}

The muon-proton momentum imbalances introduced in~\cite{PhysRevD.101.092001} parallel and transverse to $\delta\vec{p}_{T}$, as shown in figure~\ref{Graphicxy}, are explored,
 	
\begin{equation}
\begin{split}
\delta p_{Tx} = (\hat{p}_{\nu} \times \hat{p}_{T}^{\mu}) \cdot \delta\vec{p}_{T}\\
\delta p_{Ty} = - \hat{p}_{T}^{\mu} \cdot \delta\vec{p}_{T},
\end{split}
\label{imbalanceVect}
\end{equation}	
 	
and, in terms of the magnitudes,
 	
\begin{equation}
\begin{split}
\delta p_{Tx} = \delta p_{T} \cdot sin\delta\alpha_{T}\\
\delta p_{Ty} = \delta p_{T} \cdot cos\delta\alpha_{T}.
\end{split}
\label{imbalance}
\end{equation}	

\begin{figure}[htb!]
\centering  
\includegraphics[width=0.49\linewidth]{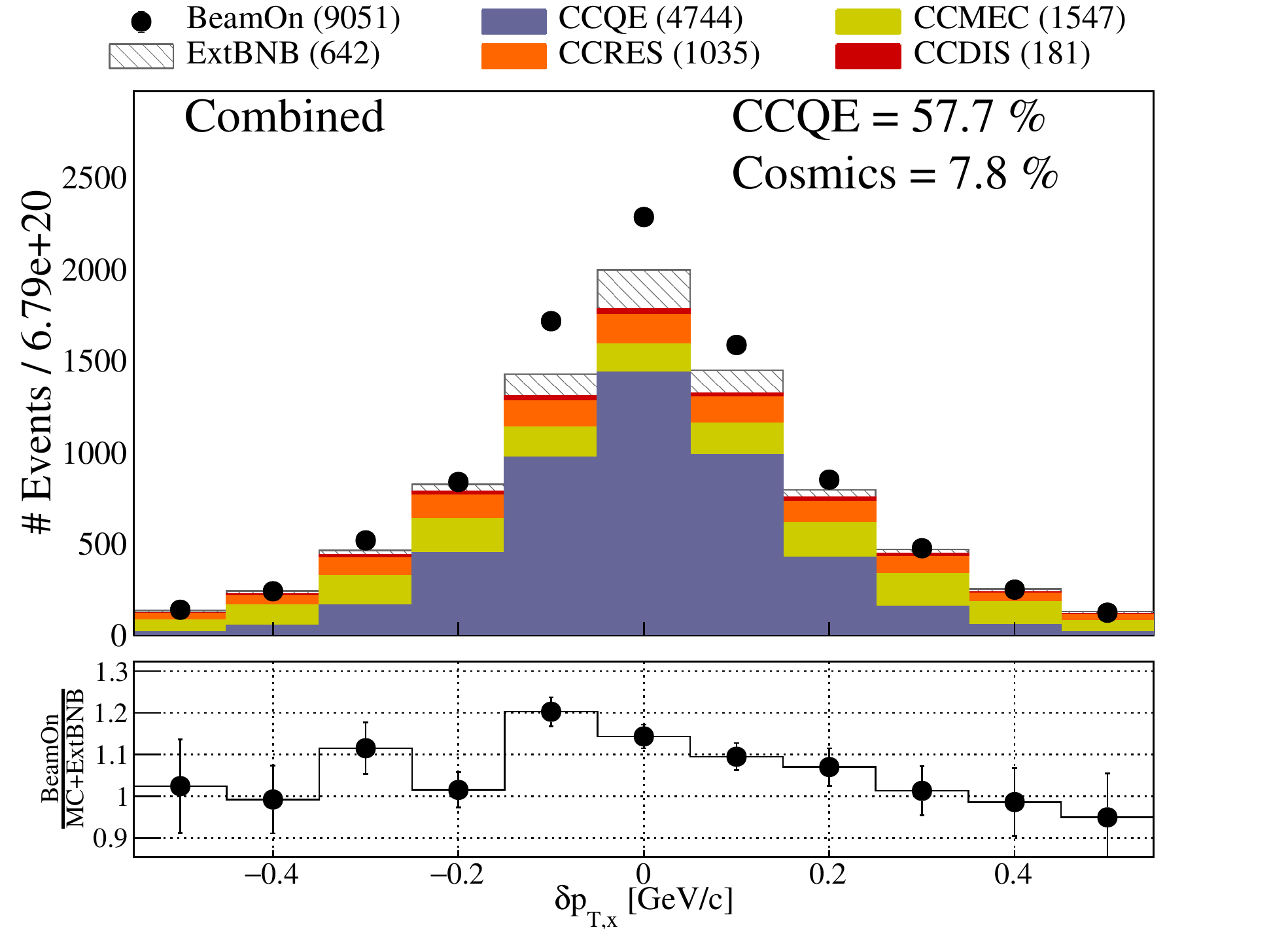}
\includegraphics[width=0.49\linewidth]{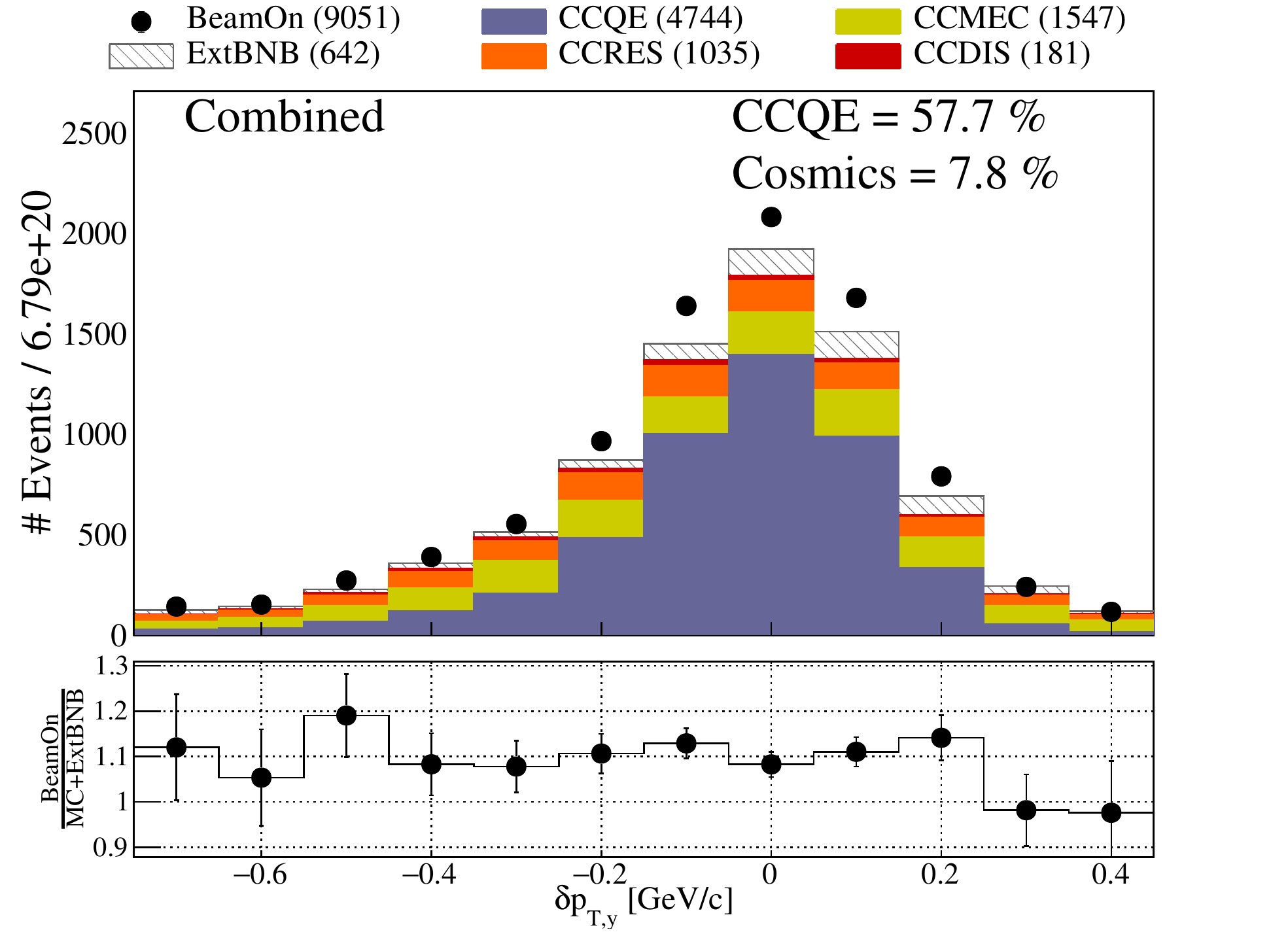}
\caption{Interaction breakdown of the \Signal events as a function of $\delta p_{Tx}$ (left) and $\delta p_{Ty}$ (right). The data correspond to the ``combined'' MicroBooNE runs 1-3.}
\label{InteBreakDeltaPtxy}
\end{figure}

The measured $\delta p_{Tx}$ event distribution shown in figure~\ref{InteBreakDeltaPtxy} (left) using the ``combined'' MicroBooNE runs 1-3 exhibit a QE peak near 0. 
If the interaction had occurred on a free nucleon, then a delta function would be expected at 0 because the muon and proton final states must balance. 
The width of the QE peak mostly results from the Fermi motion.	
If no significant deviation is assumed in the non-QE distributions originating from MEC and RES/DIS events, then data-MC discrepancies could imply an overestimation of the argon Fermi momentum, and/or a difference in the total fraction of the FSI contribution.

Unlike the $\delta p_{Tx}$ distribution, a non-QE tail is observed towards the negative $\delta p_{Ty}$ values shown in figure~\ref{InteBreakDeltaPtxy} (right). 
Inelastic events such as 2p2h, resonance, and DIS are inefficient at transferring the lepton momentum to the final state nucleons, since multiple initial states particles are often involved. 
Therefore, the protons tagged in the non-QE events will in general have less momenta then the muons and the $\delta p_{Ty}$ distribution is shifted to the left.

To avoid multiple cosmic contributions and tracks from trajectories that exit the detector but their end-points are incorrectly reconstructed around its edges, a fiducial volume of

\begin{equation}
\label{FV}
\begin{array}{c}
10 < x < 246 , -105 < y < 105 , 10 < z < 1026 \textrm{ cm}
\end{array}
\end{equation}

is defined.		 
Candidate muon and proton tracks that were fully contained in this region were considered and their momenta were obtained based on their range~\cite{MuInAr,osti139791}.

The log-likelihood ratio particle identification (LLR PID) score method~\cite{log} is used to obtain our muon and proton candidates.
%Using truth-level information from the MC samples, figure~\ref{LLRScore} shows that the muon LLR PID score tends to be greater than the proton one.
The candidate track with the greater LLR PID score was assigned the label of the candidate muon, while the one with the smaller 3-plane loglikelihood was our candidate proton.

%\begin{figure}[htb!]
%\centering  
%\includegraphics[width=0.7\linewidth]{\figures PID_BreakDown_RecoLLRPIDPlot_NoCuts_Combined_v08_00_00_52}
%\caption{Particle identification using the LLR PID score method.}
%\label{LLRScore}
%\end{figure}

To minimize the contribution of misreconstructed tracks, we took advantage of the fact that we had two muon momentum reconstruction methods available for contained tracks, namely the momentum from range~\cite{osti139791} and the one from Multiple Coulomb Scattering (MCS)~\cite{Abratenko:2017nki}. 
A quality cut was applied on the contained muons by requiring the range and MCS momenta to be in agreement within 25\%.
	
In order to avoid flipped tracks, it was further required that the distance between the track start points and the vertex is smaller than the corresponding distance between the track end points and 
the vertex.
It was also required that the distance between the start points of the two candidate tracks is smaller than the one between the two end points. 
%We verified that the matching between reconstructed and true variables is performing as expected.
%That is illustrated in with the diagonal form shown in figure~\ref{TwoDeltaPT} where we used \Signal MC events as a function of $\delta p_{T}$.	
 %\begin{figure}[htb!]
%\centering  
%\includegraphics[width=0.7\linewidth]{\figures PtCanvas_Combined.pdf}		
%\caption{Reco-to-truth matching using \Signal events for $\delta$p$_{T}$.}
%\label{TwoDeltaPT}
%\end{figure}

The maximal possible signal contribution was ensured, while the majority of the cosmic contamination and the beam related MC backgrounds were rejected, by requiring that the proton LLP PID score is less than 0.05.
%That is achieved by maximizing the product of the purity and the efficiency, where the %purity is defined as
	
%\begin{equation}
%\label{purity}
%Purity = \frac{MC^{\textrm{\Signal}}_{Reco}}{MC_{Reco} + Dirt_{Reco} + ExtBNB_{Reco}},
%\end{equation}	
	
%and the efficiency as
	
%\begin{equation}
%\label{efficiency}
%Efficiency = \frac{MC^{\textrm{\Signal}}_{Reco}}{MC^{\textrm{\Signal}}_{True}}.
%\end{equation}	
	
%We studied the effect of cutting on different values of the candidate proton LLR PID score, which has a strong discrimination power rejecting non-\Signal and cosmic events. 
%Maximizing the purity $\times$ efficiency product yielded an optimal cut at proton LLP PID score = 0.05, which is indicated by the dashed line in figure~\ref{llrcont}.	
	
%\begin{figure}[htb!]
%\centering  
%\includegraphics[width=0.7\linewidth]{\figures %THStack_BreakDown_RecoProtonLLRPIDPlot_Combined_\version_NoCuts.pdf}		
%\caption{Candidate proton LLR PID score before the application of any selection cuts. The dashed line at 0.05 indicates the place where we place our cut to reject the MC backgrounds and the cosmics dominant in the region .}
%\label{llrcont}
%\end{figure}	

\begin{figure}[htb!]
\centering  
\includegraphics[width=0.49\linewidth]{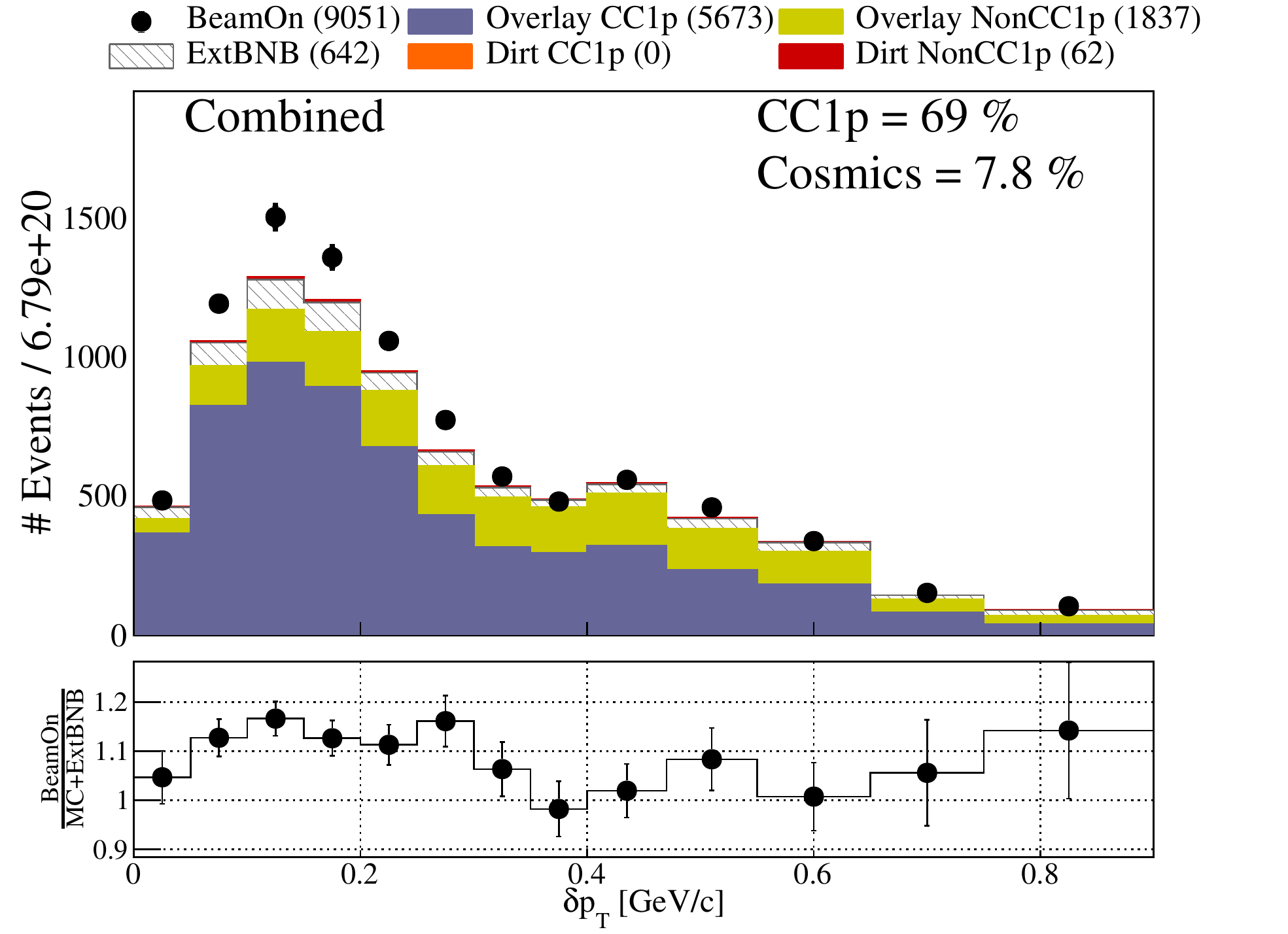}
\includegraphics[width=0.49\linewidth]{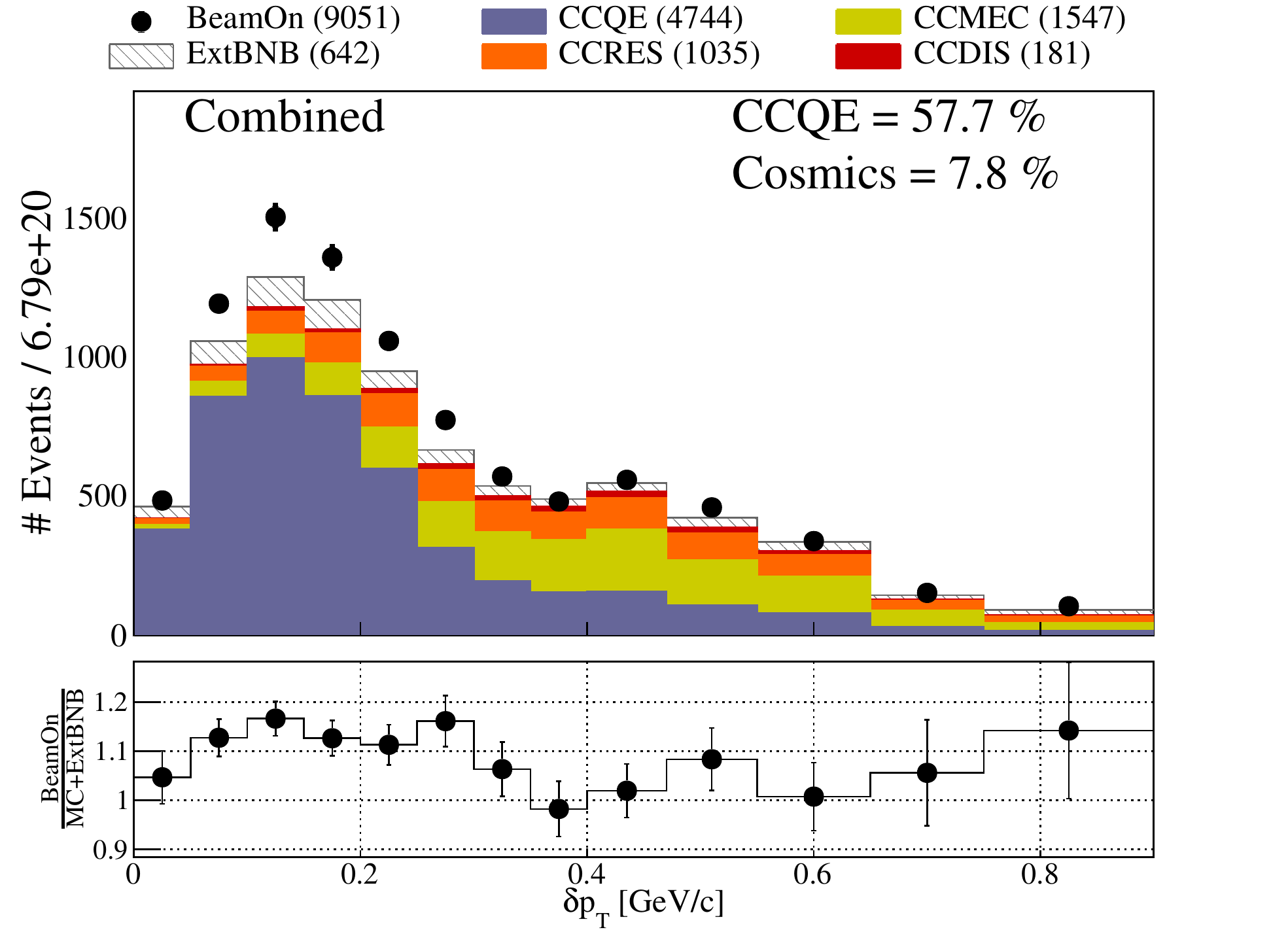}
\caption{Topological (left) and interaction (right) breakdown after the application of the event selection for $\delta p_{T}$.}
\label{ubbreakdown}
\end{figure}

The application of our event selection resulted in 9051 candidate events in our data sample.
Using the MC, it was estimated that our event selection yielded a purity of $\approx$ 70\% and an efficiency of $\approx$ 10\%.
There was also some contribution from the remaining cosmic contamination ($\approx$ 8\%).
After the application of the event selection, topological and interaction breakdowns for the kinematic variables of interest, such as the ones shown in figure~\ref{ubbreakdown} for $\delta p_{T}$, were obtained.

%%%%%%%%%%%%%%%%%%%%%%%%%%%%%%%%%%%%%%%%%%%%%%%%%%%%%%%%%%%%%%%%%%%%%%%%

\subsection{Cross-Section Extraction Technique}\label{CC1pxsec}

The  unfolded  cross-section results  reported  in  this  analysis  took advantage of  the Wiener-SVD  unfolding~\cite{Tang_2017}. 
This method  combines  the  use  of  the  singular  value  decomposition  (SVD)  unfolding  and  a Wiener filter.  
SVD unfolding~\cite{H_cker_1996}, such as the Tikhonov regularisation~\cite{Schmitt_2012}, unfolds a distribution by minimising a $\chi^{2}$ function comparing a prediction to data.  
To avoid the large variance introduced, a penalty term is added to regularise the curvature (second derivative) of the results.  
The strength of such a term is determined by finding an appropriate trade-off between the bias and the variance between the data and the MC.
More details on the technique are included in appendix~\ref{svdapp}.
In order to report the cross-section results, two key ingredients are required, namely the response and covariance matrices.

\begin{figure}[ht]
\centering 
\includegraphics[width=0.7\linewidth]{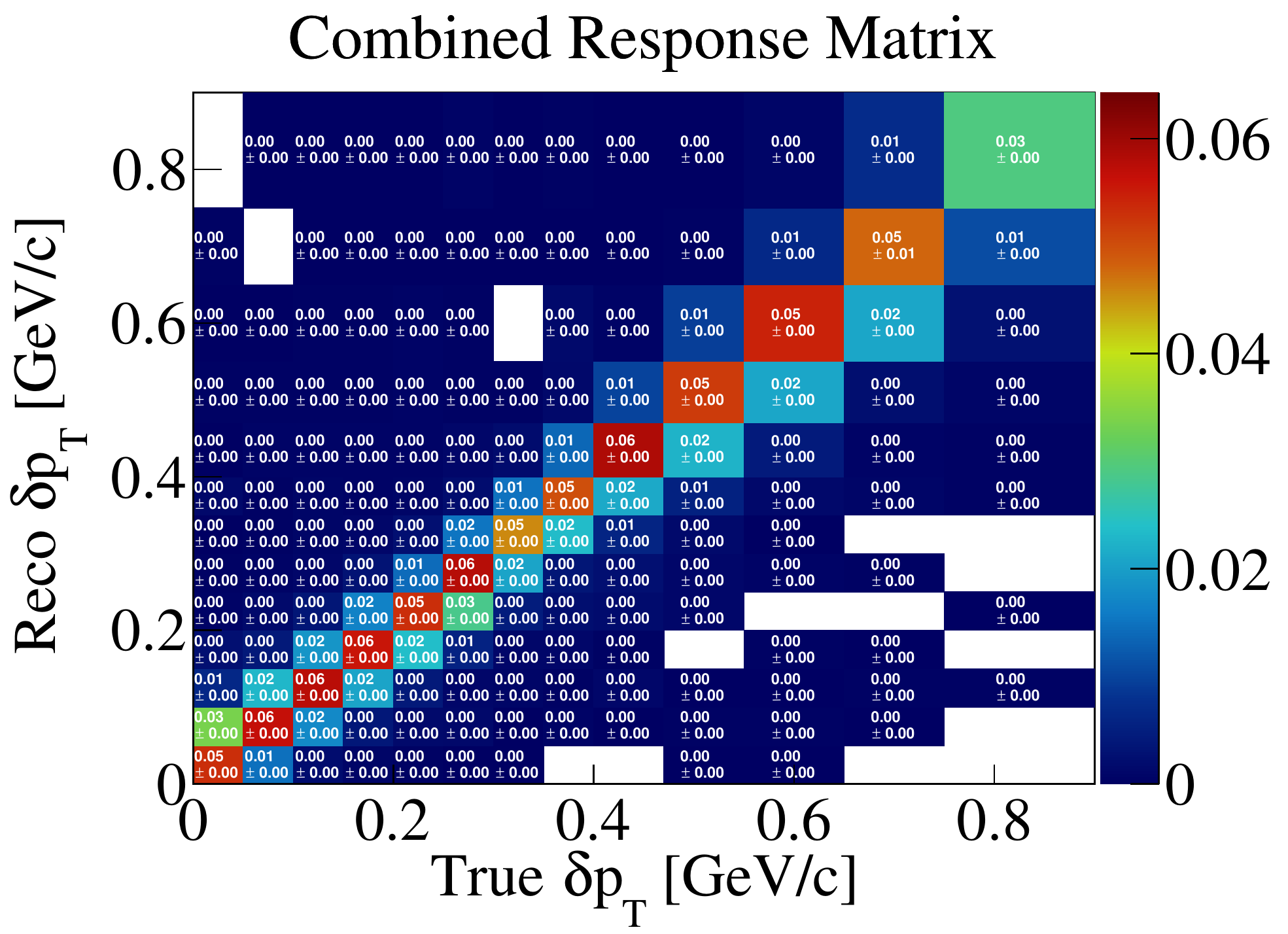}		
\caption{Response matrices of $\delta p_{T}$ using the selected \Signal MC events.}
\label{DeltaPT_response}
\end{figure}

The construction of the response matrices uses the selected MC \Signal events to construct a two-dimensional (2D) object, where each entry in true bin i and reconstructed bin j ($N^{true\,i,reco\,j}$) is divided by the true number of events generated in bin i ($S^{true\,i}$).
These response matrices serve as ``2D local efficiencies'', as defined in equation~\ref{respmatr} and can be seen in figure~\ref{DeltaPT_response} for $\delta p_{T}$.
	
\begin{equation}
M_{ij} = \frac{N^{true\,i,reco\,j}}{S^{true\,i}}
\label{respmatr}
\end{equation}
    
The method  also uses  a  covariance  matrix  constructed from  the  MC  flux normalised  event  rate  as  an input.   
The total covariance matrix incorporates information related to the systematic and statistical uncertainties.  
The flux-normalized MC event rates in reconstructed space were obtained as
  
\begin{equation}
\tilde{\sigma}^{reco\,i}=\frac{N^{reco\,i}}{\Phi_{\nu}^{CV} \times N_{targets}},   
\label{rate}
\end{equation}  
    
where $N^{reco\,i} = M_{ij} \times S^{true\,j} + B^{reco\,i}$ is the total number of reconstructed events in bin i, $M_{ij}$ is the response matrix corresponding to reco bin i and true bin j as defined in equation~\ref{respmatr}, $S^{true\,j}$ is the true signal without any detector or reconstruction effects in bin j, and $B^{reco\,i}$ is the total number of reconstructed beam-related MC background events in bin i. 
Substituting $N^{reco\,i}$ into equation~\ref{rate} yields
    
\begin{equation}
\tilde{\sigma}^{reco\,i}=\frac{M_{ij}^{univ} \times S^{true\,j\,CV} + B^{reco\,i\,univ}}{\Phi_{\nu}^{CV} \times N_{targets}}.   
\label{exprate}
\end{equation}  
    
For  each  systematic  variation,  each  term  in  equation~\ref{exprate} labelled  with ``univ” is  reweighted/modified and each term labelled with “CV” is fixed to the central value. 
The integrated flux, $\Phi_{\nu}^{CV}$, remains fixed for each variation.
    
In the case of the cross section variations, the calculation of the response matrix in each universe is slightly modified via the normalization to the true signal in a given universe $S^{true\,j\,univ}$,
    
\begin{equation}
M_{ij}^{univ}=\frac{N^{true\,j,reco\,i\,univ}}{S^{true\,j\,univ}}   
\label{fluxRespM}
\end{equation} 
    
This treatment of the systematic uncertainties addresses both the signal and the beam related background uncertainties. 

Using these flux-normalized event rates $\tilde{\sigma}$, a covariance matrix, $E_{ij}$, can be calculated using the central value and $N_{univ}$ multisims with the covariance formalism, 
	
\begin{equation}
\label{Cov}
E_{ij} = \frac{1}{N_{univ}} \sum_{s=0}^{N_{univ}} (\tilde{\sigma}^{univ}_{i} - \tilde{\sigma}^{CV}_{i}) (\tilde{\sigma}^{univ}_{j} - \tilde{\sigma}^{CV}_{j})
\end{equation}
     
where $N_{univ}$ is number of alternative universes, $\tilde{\sigma}_{i}^{univ}$ corresponds to the variation and $\tilde{\sigma}_{i}^{CV}$ to the central-value prediction.

The unfolding model uncertainty is accessed by comparing the data spectra unfolded with G18 to the data spectra unfolded with two alternative G18 configurations.
The former one did not include the effect of the MicroBooNE tune and the latter included an additional weight of 2 on the MEC events.
The spread between the three configurations on a bin-by-bin basis normalized to $\sqrt(2)$ is assigned as an additional uncertainty~\cite{BaBarStat}.

The statistical uncertainty of our measurement is 1.5\%.
The total uncertainty sums to 13\% and includes contributions from the neutrino flux prediction (7.3\%), unfolding model uncertainty (7.3\%), neutrino interaction cross section modeling (5\%), detector response modeling (4.9\%), POT estimation (2.3\%), number-of-scattering-targets (1.15\%), reinteractions (1\%), and out-of-cryostat interaction modeling (0.2\%).

\begin{figure}[htb!]
\centering 
\includegraphics[width=0.7\linewidth]{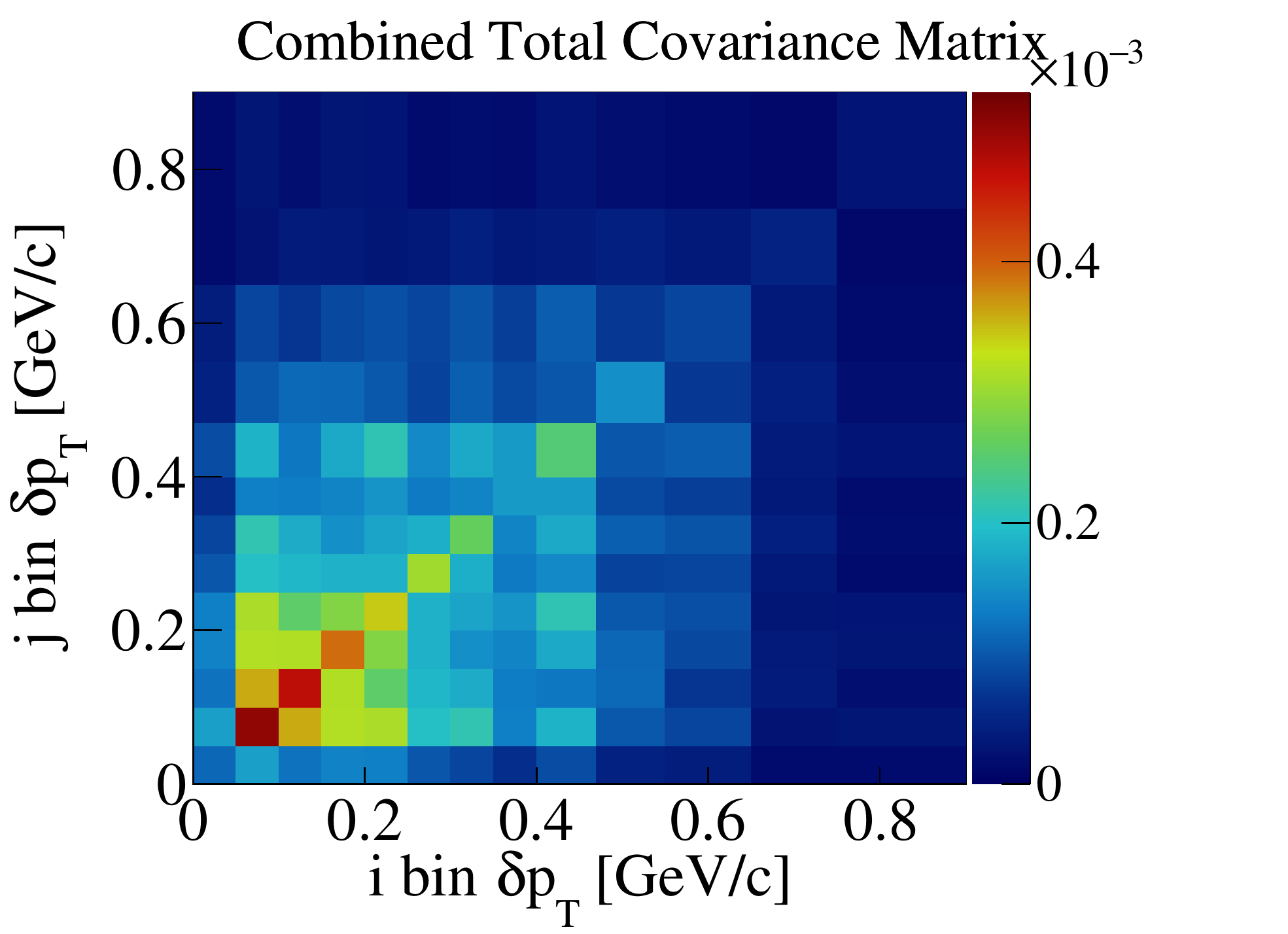}
\caption{Total covariance matrix for $\delta p_{T}$.} 
\label{CovDeltaPT}
\end{figure}

Figure~\ref{CovDeltaPT} shows the total covariance matrix due to the aforementioned sources of uncertainty for $\delta p_{T}$.

\begin{figure}[htb!]
\centering 
\includegraphics[width=0.7\linewidth]{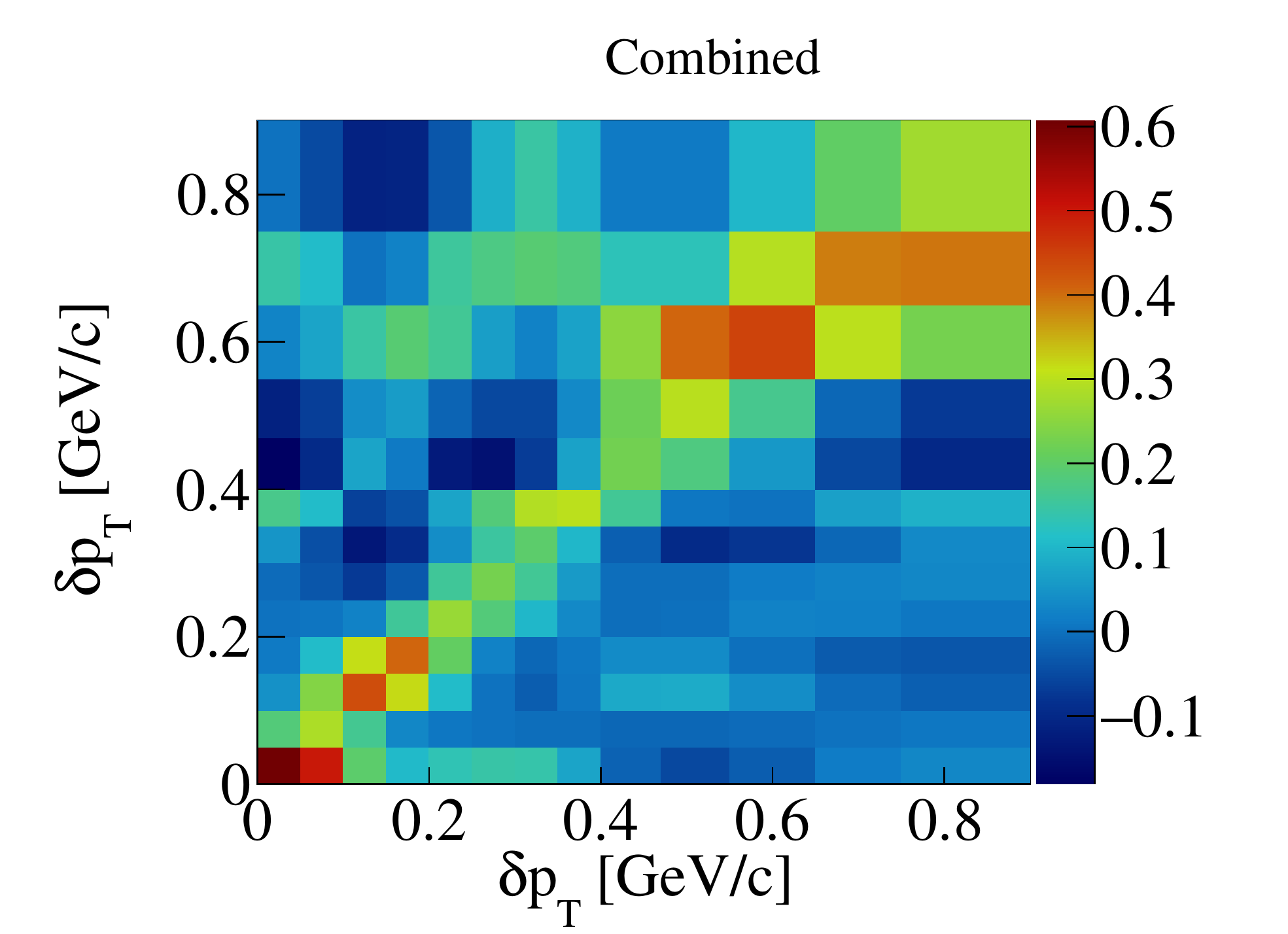}
\caption{Additional smearing matrix $A_{c}$ for $\delta p_{T}$.} 
\label{AcDeltaPT}
\end{figure}

The Wiener SVD unfolding machinery returns  an  unfolded  data  cross  section  along  with  an  unfolded covariance matrix and an additional smearing matrix, $A_{c}$. 
The corresponding $A_{c}$ matrix for $\delta p_{T}$ is shown in figure~\ref{AcDeltaPT}. 
The smearing matrix $A_{c}$ contains information about the regularisation of the measurement and is applied to the true model cross section predictions when compared to the data. 
Therefore, the result of our measurement lives in a ``regularized'' phase-space, which is not identical to the true phase-space. 

%%%%%%%%%%%%%%%%%%%%%%%%%%%%%%%%%%%%%%%%%%%%%%%%%%%%%%%%%%%%%%%%%%%%%%%%

\subsection{Event Generator Modeling And Configurations}\label{model}

The extracted cross sections were compared to GENIE v3.0.6 G18\_10a\_02\_11a (G18) and the theory-driven GiBUU 2021 (GiBUU) event generator.
Additional comparisons to the corresponding events generators when FSI are turned off were also included (G18 No FSI and GiBUU No FSI).
G18 uses the Local Fermi Gas model~\cite{Carrasco:1989vq}, the Nieves CCQE scattering prescription~\cite{Nieves:2012yz} which includes Coulomb corrections for the outgoing muon~\cite{Engel:1997fy} and Random Phase Approximation correction~\cite{RPA}, the Nieves MEC model~\cite{Schwehr:2016pvn}, the KLN-BS RES~\cite{Nowak:2009se,Kuzmin:2003ji,Berger:2007rq,Graczyk:2007bc} and Berger-Sehgal COH~\cite{Berger:2008xs} scattering models, the hA2018 FSI model~\cite{Ashery:1981tq}, and the T2K tune weights~\cite{GENIEKnobs}.

GiBUU uses somewhat similar models, but, unlike GENIE, those are implemented in a coherent way, by solving the Boltzmann-Uehling-Uhlenbeck transport equation~\cite{Mosel:2008yp}. 
The modeling includes the Local Fermi Gas model~\cite{Carrasco:1989vq}, a standard CCQE expression~\cite{Leitner:2006ww}, an empirical MEC model and a dedicated spin dependent resonances amplitude calculation following the MAID analysis~\cite{Mosel:2019vhx}. 
The DIS model is as in PYTHIA~\cite{Sjostrand:2006za} and the FSI treatment is different as the hadrons propagate through the residual nucleus in a nuclear potential which is consistent with the initial state.

Apart from the nominal G18 prediction, we further included a comparison to the recently added theory driven GENIE v3.0.6 G21\_11b\_00\_000 configuration (G21 hN).
The latter uses the SuSAv2 model for QE and MEC interactions~\cite{PhysRevD.101.033003}, and the hN2018 FSI model~\cite{hN2018}.
The modeling options for RES, DIS, and COH interactions are the same as for G18.	
We investigated the effect of the FSI modeling choice by comparing the G21hN results to the ones obtained with G21 hA, where the hA2018 FSI model was used instead, and to G21 G4 with the recently coupled Geant4 FSI framework~\cite{Wright:2015xia}.

Lastly, our results present the comparison between the nominal G18 LFG model and predictions using the same G18 modeling configuration but different nuclear model options available in the GENIE event generator, namely the Bodek-Ritchie Fermi Gas (G18 RFG)~\cite{PhysRevD.23.1070} and an Effective Spectral Function (G18 EffSF)~\cite{PhysRevC.74.054316}.
Furthermore, the prediction without Random Phase Approximation (RPA) effects was used for comparison (G18 No RPA)~\cite{RPA}.

%%%%%%%%%%%%%%%%%%%%%%%%%%%%%%%%%%%%%%%%%%%%%%%%%%%%%%%%%%%%%%%%%%%%%%%%%

\subsection{Kinematic Imbalance Differential Cross-Section Results}\label{CC1p1D}

\begin{figure*}[htb!]
\centering 
\includegraphics[width=0.49\linewidth]{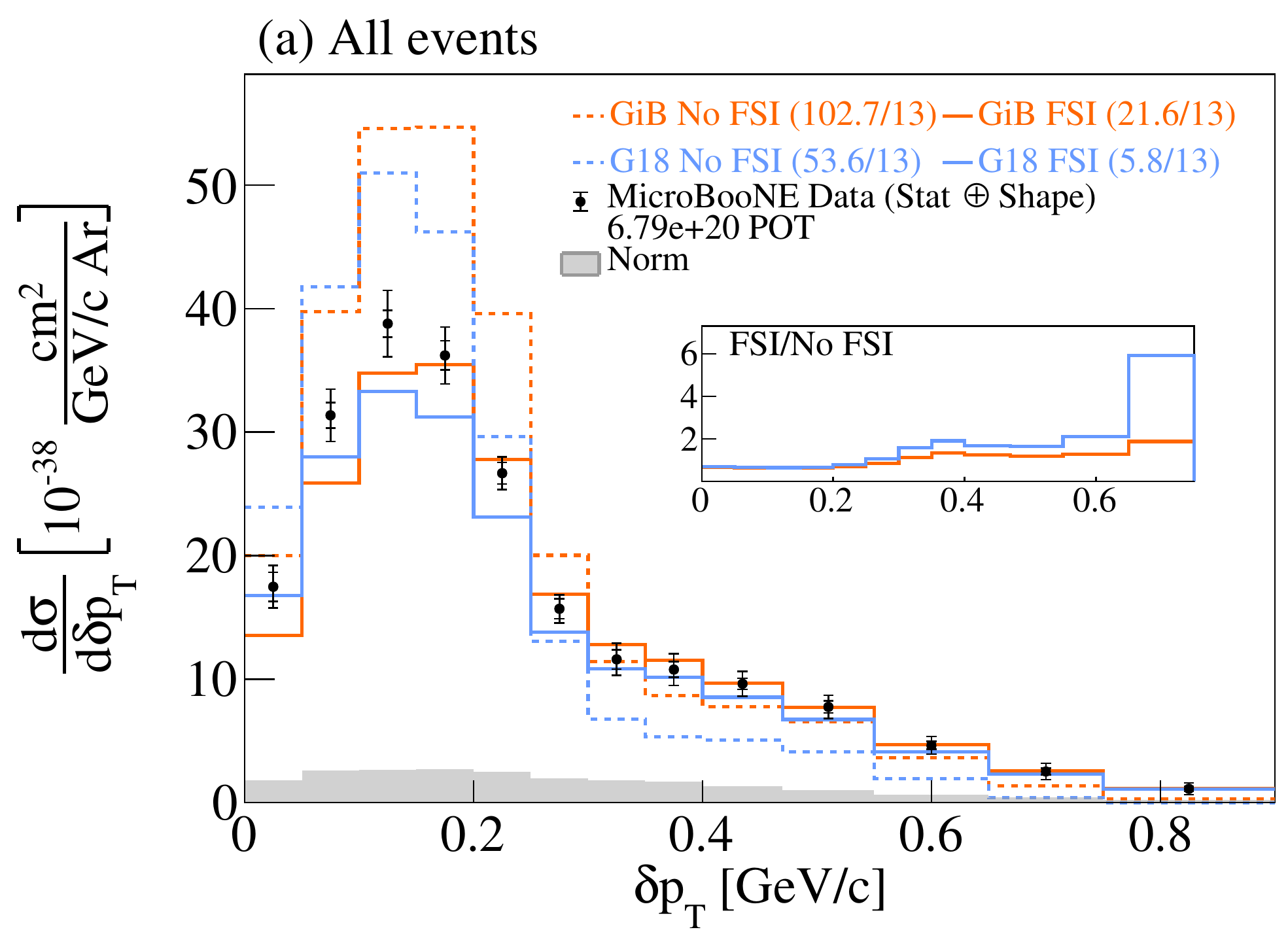}\\
\includegraphics[width=0.49\linewidth]{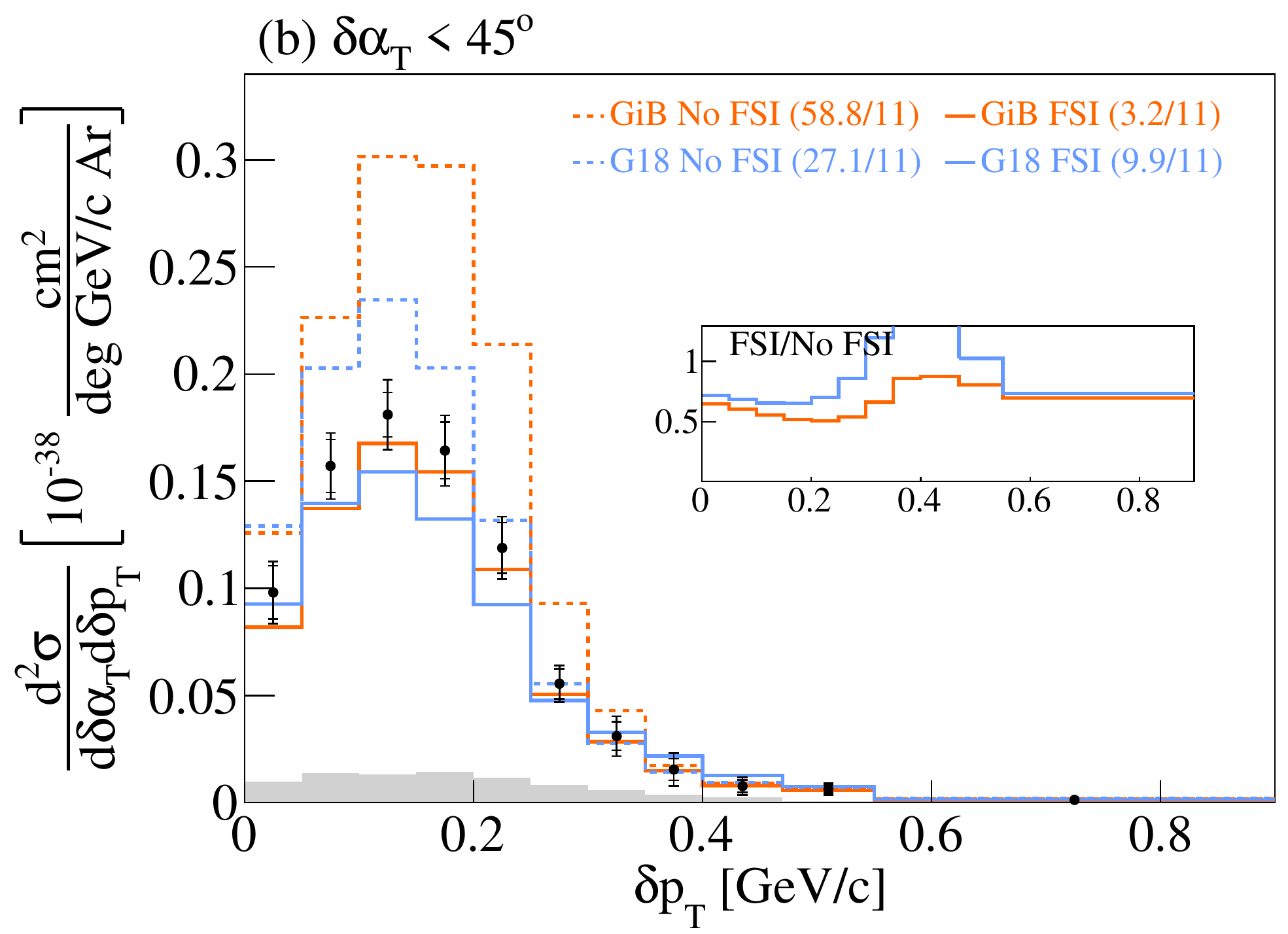}
\includegraphics[width=0.49\linewidth]{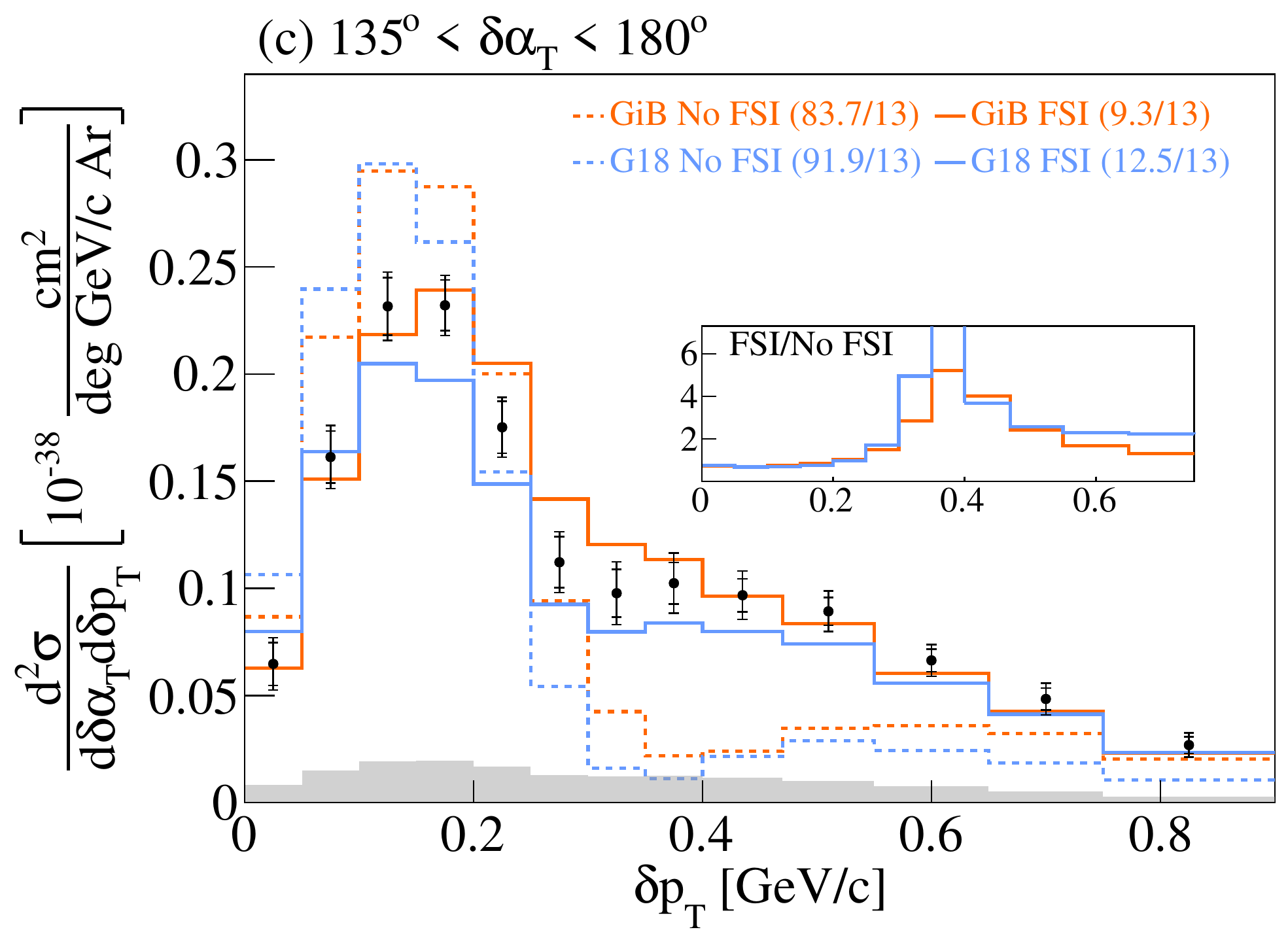}
\caption{The flux-integrated single- (top) and double- in $\delta\alpha_{T}$ bins (bottom) differential \CCIpOpi\ cross sections as a function of the transverse missing momentum $\delta p_{T}$. 
Inner and outer error bars show the statistical and total (statistical and shape systematic) uncertainty at the 1$\sigma$, or 68\%, confidence level. 
The gray band shows the normalization systematic uncertainty.
Colored lines show the results of theoretical absolute cross section calculations with and without FSI based on the GENIE and GiBUU event generators.}
\label{DeltaPTInDeltaAlphaT}
\end{figure*}

The single- and double- in $\delta\alpha_{T}$ bins differential unfolded cross sections as a function of $\delta p_{T}$ are presented in figure~\ref{DeltaPTInDeltaAlphaT}.
The single-differential results as a function of $\delta p_{T}$ using all the events that satisfy our selection are shown in the top panel.
The peak height of both generator predictions is $\approx$ 30\% higher when FSI effects are turned off.
Yet, all distributions illustrate a transverse missing momentum tail that extends beyond the Fermi momentum whether FSI effects are activated or not.
The ratio between the generator predictions with and without FSI is shown in the insert and illustrates significant shape variations across the range of interest.
The double-differential result using events with $\delta\alpha_{T} < 45^{o}$ shown in the bottom left panel of figure~\ref{DeltaPTInDeltaAlphaT} is dominated by events that primarily occupy the region up to the Fermi momentum and do not exhibit a high momentum tail.
The corresponding ratio insert illustrates a fairly uniform behavior indicative of transparency effects ranging between 50-70\%.
The double-differential results using events with $135^{o} < \delta\alpha_{T} < 180^{o}$ is shown in the bottom right panel of figure~\ref{DeltaPTInDeltaAlphaT} and illustrate the high transverse missing momentum up to 1\,GeV/c. 
The case without FSI effects is strongly disfavored and the ratio insert illustrates strong shape variations.
Therefore, the high $\delta\alpha_{T}$ region is an appealing candidate for neutrino experiments to benchmark and tune the FSI modeling in event generators. 

\begin{figure*}[htb!]
\centering 
\includegraphics[width=0.49\linewidth]{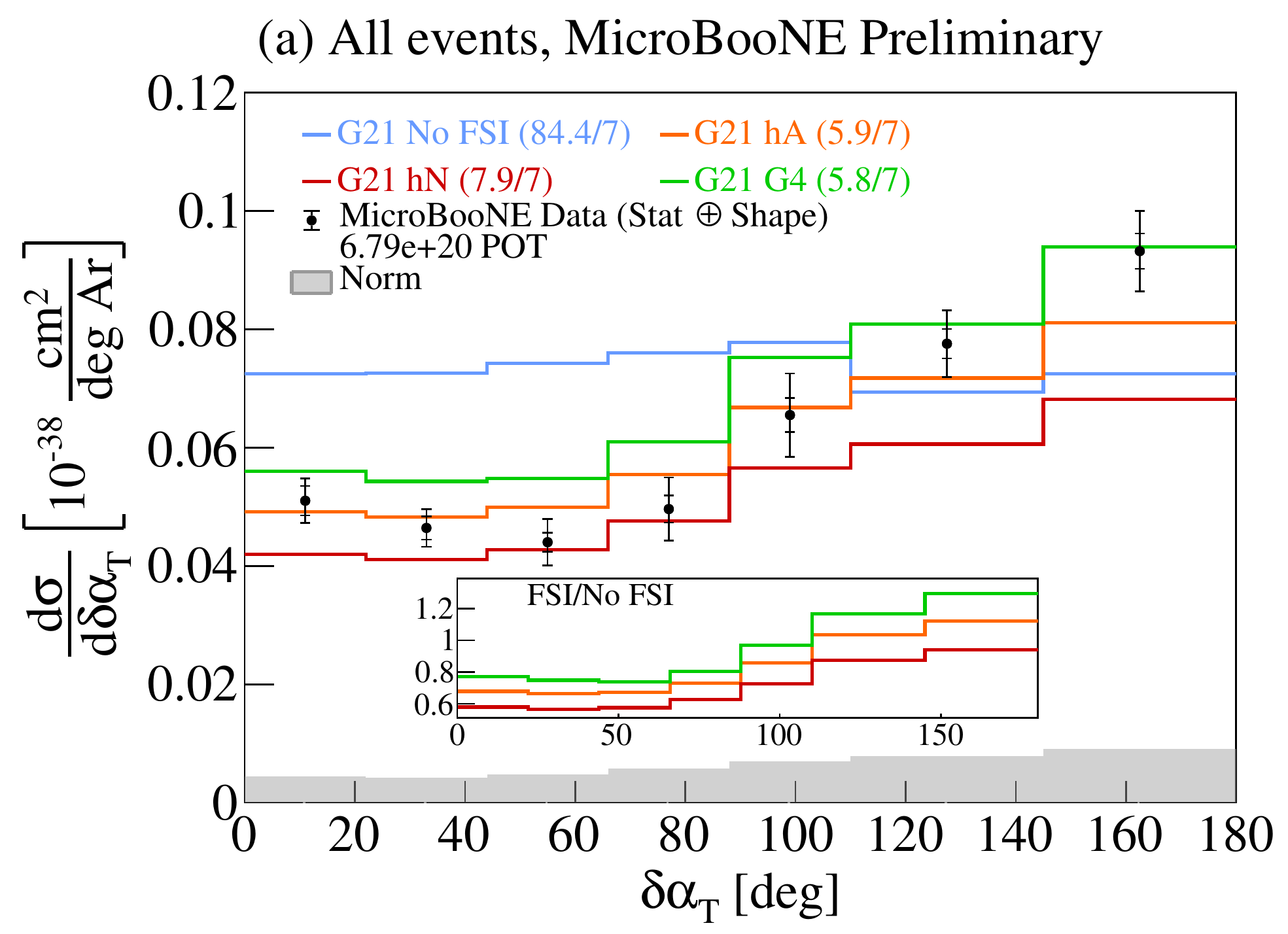}\\
\includegraphics[width=0.49\linewidth]{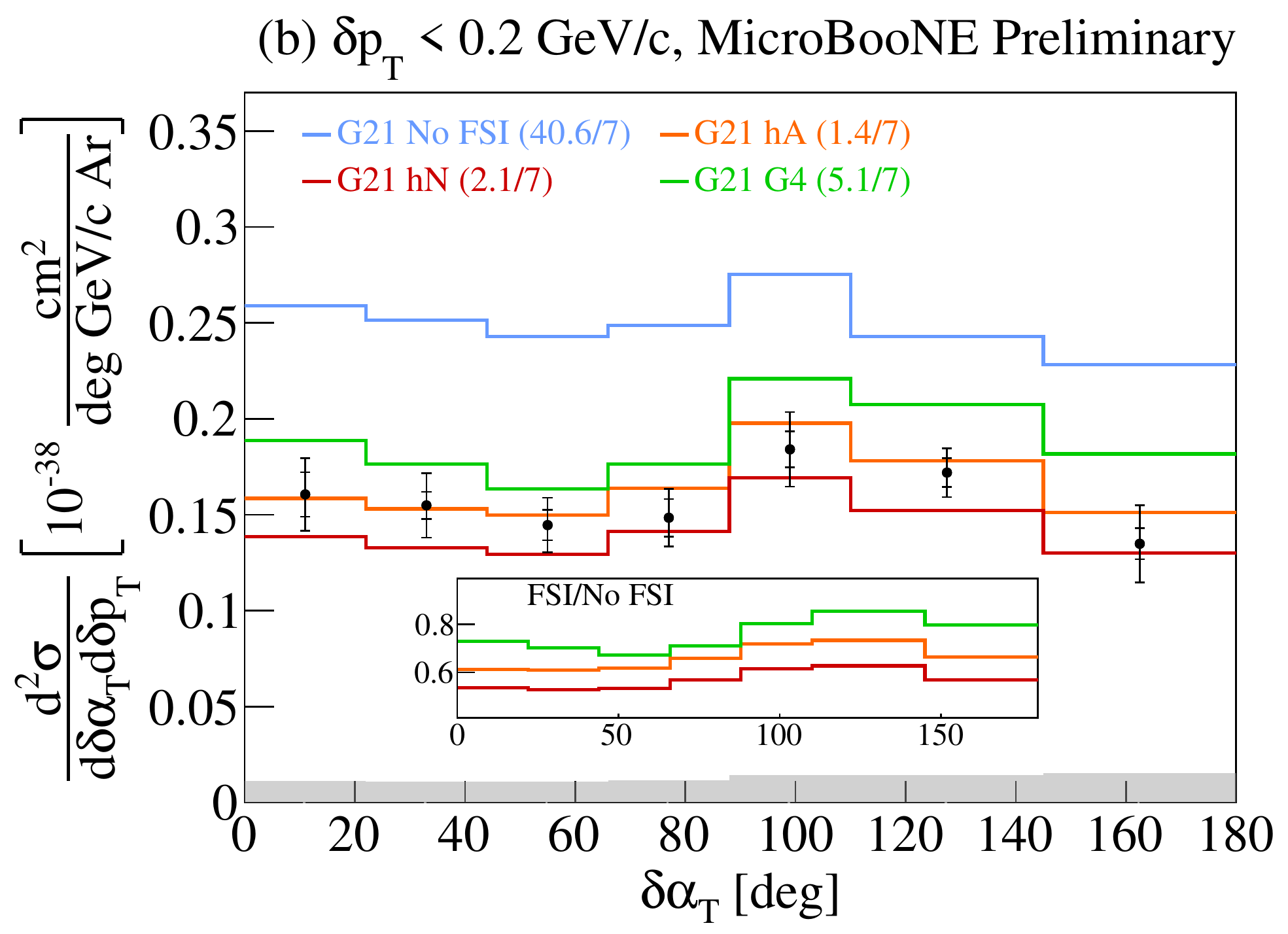}
\includegraphics[width=0.49\linewidth]{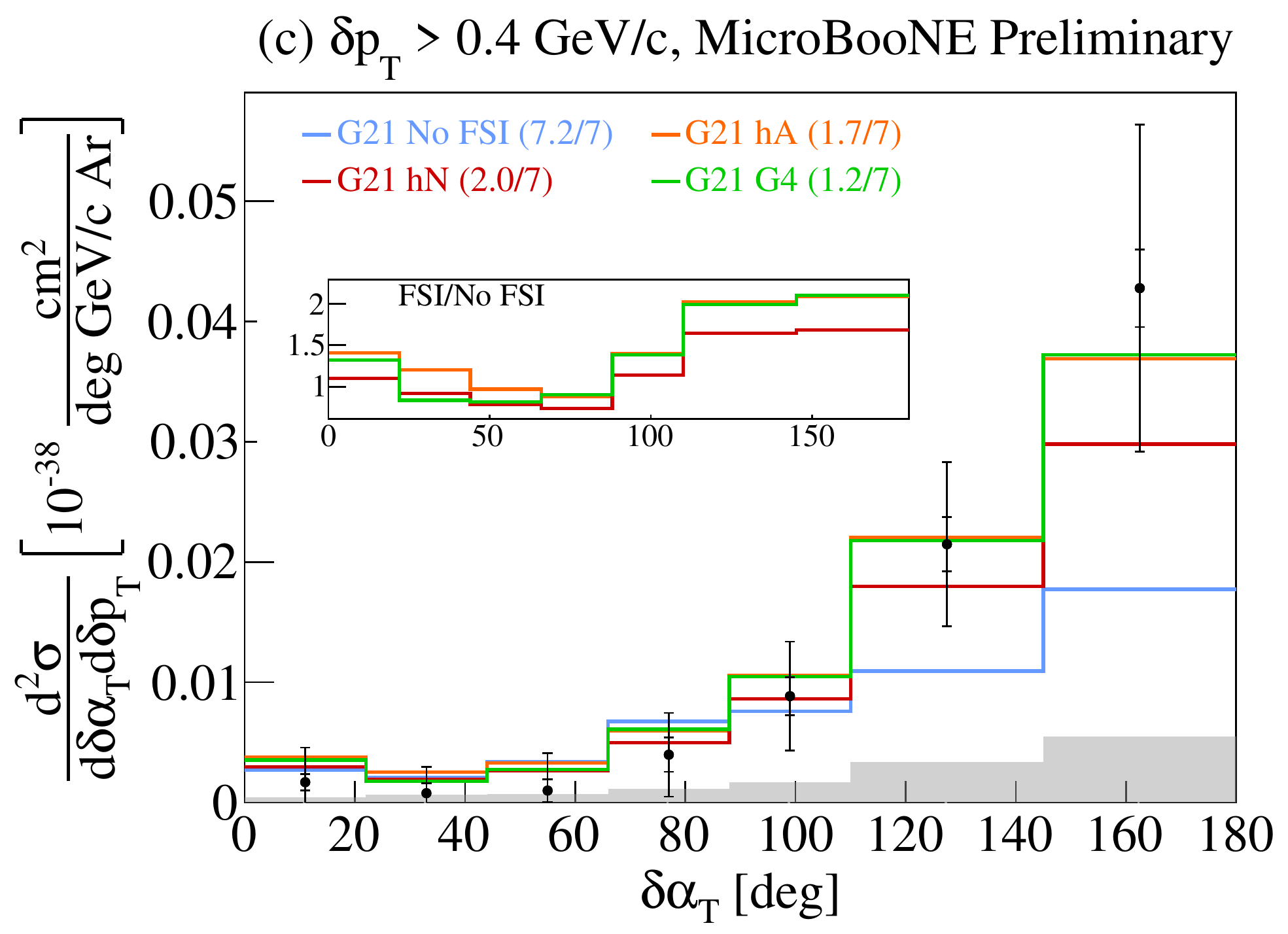}
\caption{The flux-integrated single- (top) and double- in $\delta p_{T}$ bins (bottom) differential \CCIpOpi\ cross sections as a function of the angle $\delta\alpha_{T}$. 
Inner and outer error bars show the statistical and total (statistical and shape systematic) uncertainty at the 1$\sigma$, or 68\%, confidence level. 
The gray band shows the normalization systematic uncertainty.
Colored lines show the results of theoretical absolute cross section calculations with a number of FSI modeling options based on the GENIE event generator.}
\label{DeltaAlphaTInDeltaPT}
\end{figure*}

The single-differential results as a function of $\delta\alpha_{T}$ using all the events that satisfy our selection are shown in top panel of figure~\ref{DeltaAlphaTInDeltaPT}.
The result without FSI illustrates a uniform behavior across the whole distribution and is disfavored.
The addition of FSI effects leads to a $\approx$ 30\% asymmetry around $\delta\alpha_{T} = 90^{o}$ due to the fact that the proton in our selection undergoes FSI.
The three FSI models used here for comparison result in a comparable performance, also shown in terms of the ratio plot of the different FSI options to the prediction without FSI.
The double-differential result using events with $\delta p_{T} <$ 0.2\,GeV/c shown in the bottom left panel of figure~\ref{DeltaAlphaTInDeltaPT} illustrates a uniform distribution indicative of the suppressed FSI impact in that part of the phase-space.
The double-differential result using events with $\delta p_{T} >$ 0.4\,GeV/c is shown in the bottom right panel of figure~\ref{DeltaAlphaTInDeltaPT} and illustrates the presence of strong FSI effects.
The case without FSI effects is disfavored and the asymmetry around $90^{o}$ is significantly enhanced.
Therefore, the high $\delta\alpha_{T}$ region is an appealing candidate for neutrino experiments to benchmark and tune the FSI modeling in event generators.

\begin{figure*}[htb!]
\centering 
\includegraphics[width=0.49\linewidth]{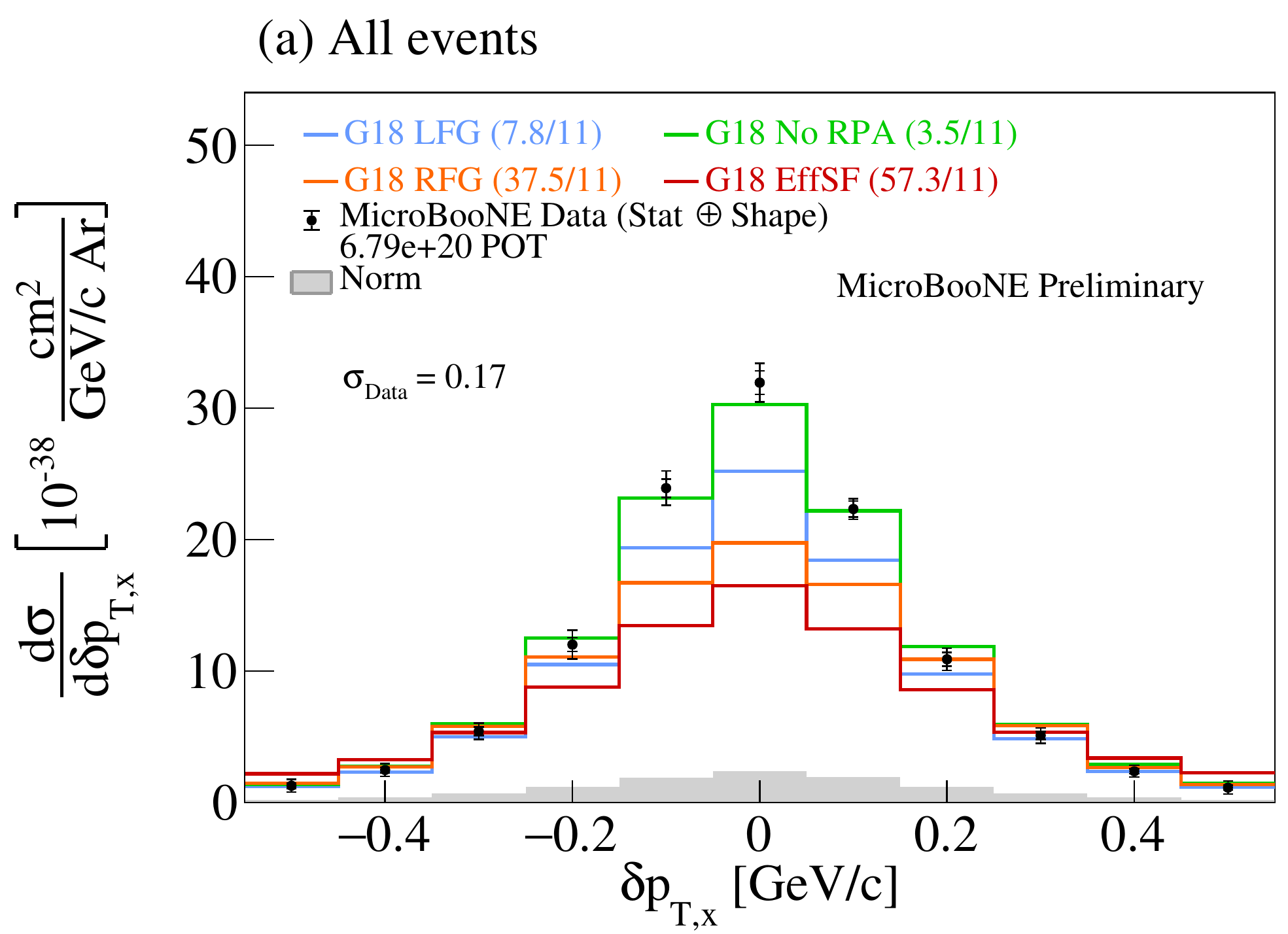}\\
\includegraphics[width=0.49\linewidth]{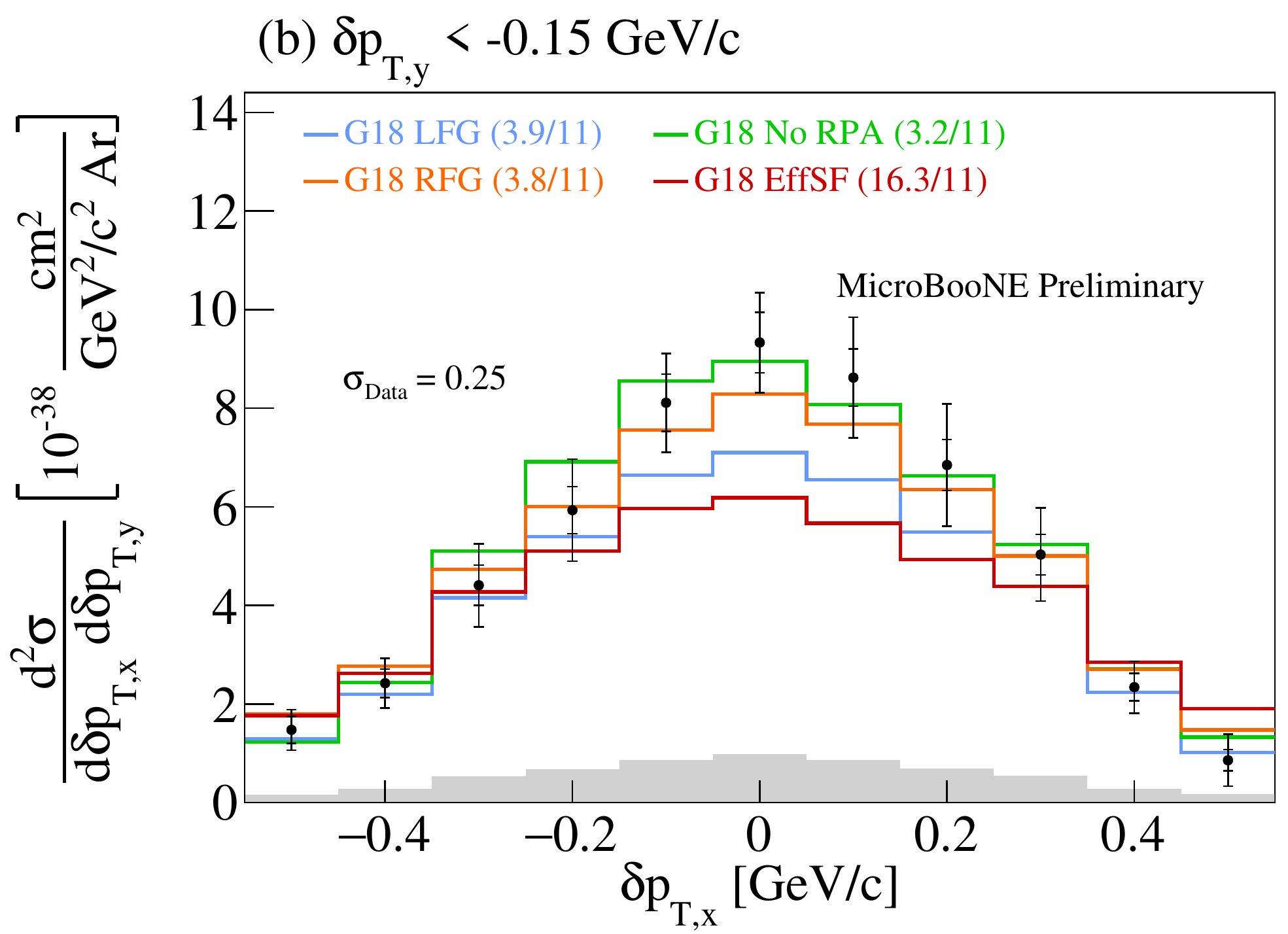}
\includegraphics[width=0.49\linewidth]{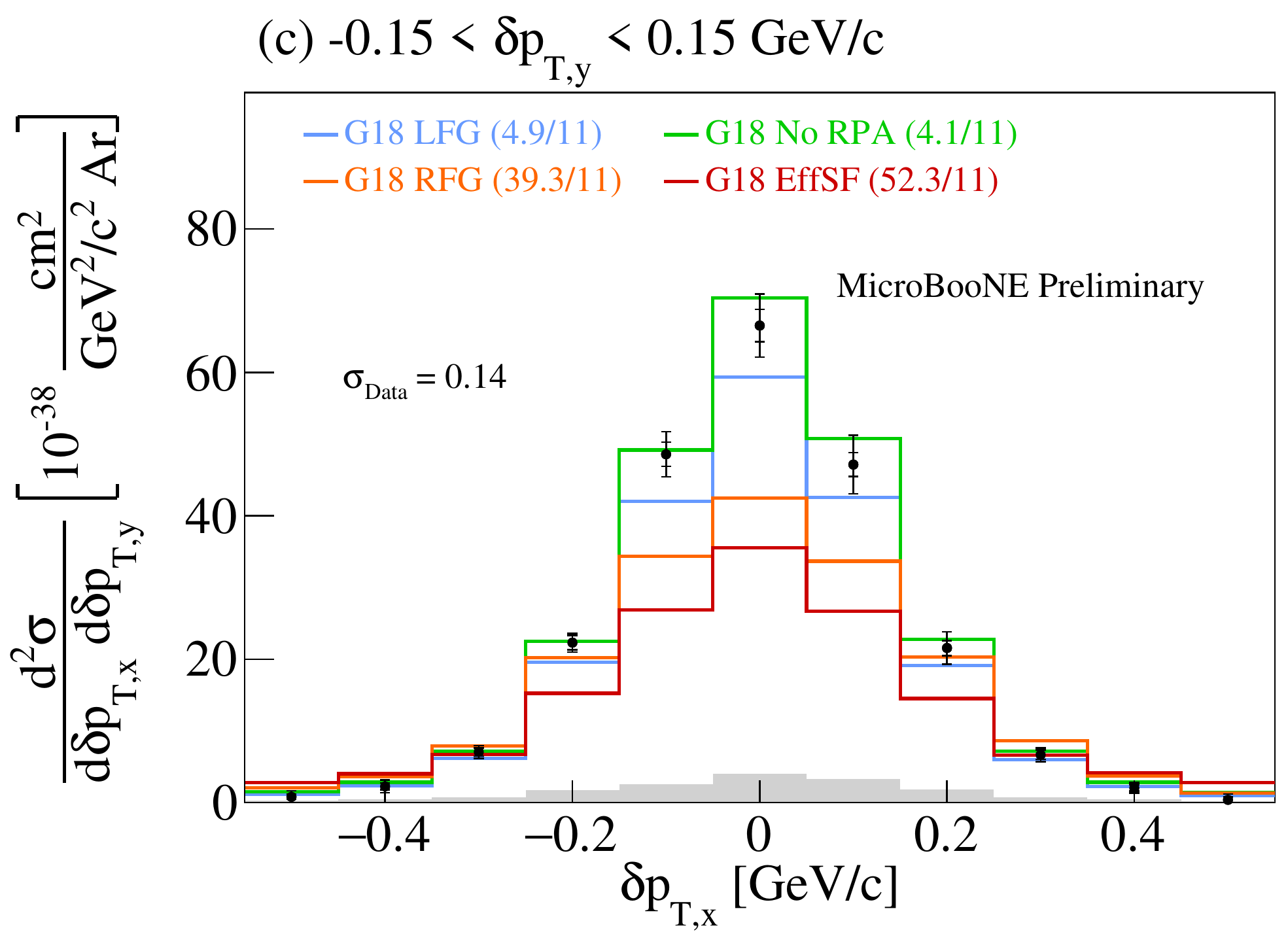}
\caption{The flux-integrated single- (top) and double- in $\delta p_{T,y}$ bins (bottom) differential \CCIpOpi\ cross sections as a function of the angle $\delta p_{T,x}$. 
Inner and outer error bars show the statistical and total (statistical and shape systematic) uncertainty at the 1$\sigma$, or 68\%, confidence level. 
The gray band shows the normalization systematic uncertainty.
Colored lines show the results of theoretical absolute cross section calculations with a number of event generators.}
\label{DeltaPtxInDeltaPty}
\end{figure*}

Lastly, figure~\ref{DeltaPtxInDeltaPty} shows the single- (top) and double- in $\delta p_{T,y}$ bins (bottom) differential unfolded cross sections as a function of $\delta p_{T,x}$.
The event distributions of $\delta p_{T,x}$ and $\delta p_{T,y}$ have already been presented in figure~\ref{ubbreakdown}.
The single differential result (top panel) illustrates a fairly broad symmetric distribution centered around 0.
The double-differential result for events where $\delta p_{T,y} <$ -0.15\,GeV/c (bottom left panel) illustrates an even broader distribution where all predictions yield comparable results.
Unlike the asymmetric part of the $\delta p_{T,y}$ tail, the double-differential result for events with -0.15 $< \delta p_{T,y} <$ 0.15\,GeV/c (bottom right panel) shows a much narrower peak which strongly depends on the choice of the underlying model and the addition or not of nuclear effects, such as RPA ones.
The G18 LFG and G18 No RPA predictions are favored in that part of the phase-space.

\begin{figure}[H]
\centering 
\includegraphics[width=\BreakdownFigSize\linewidth]{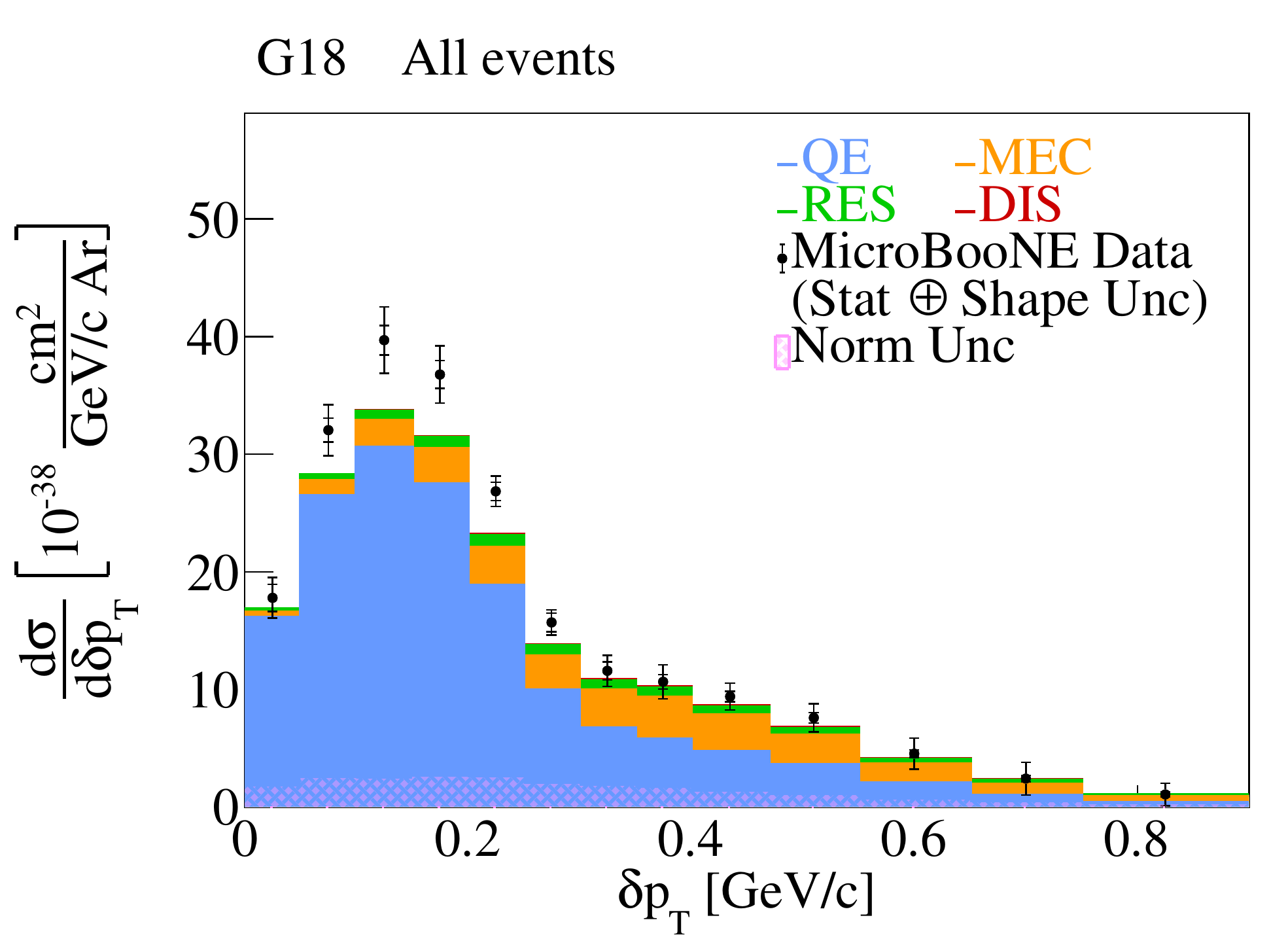}
\includegraphics[width=\BreakdownFigSize\linewidth]{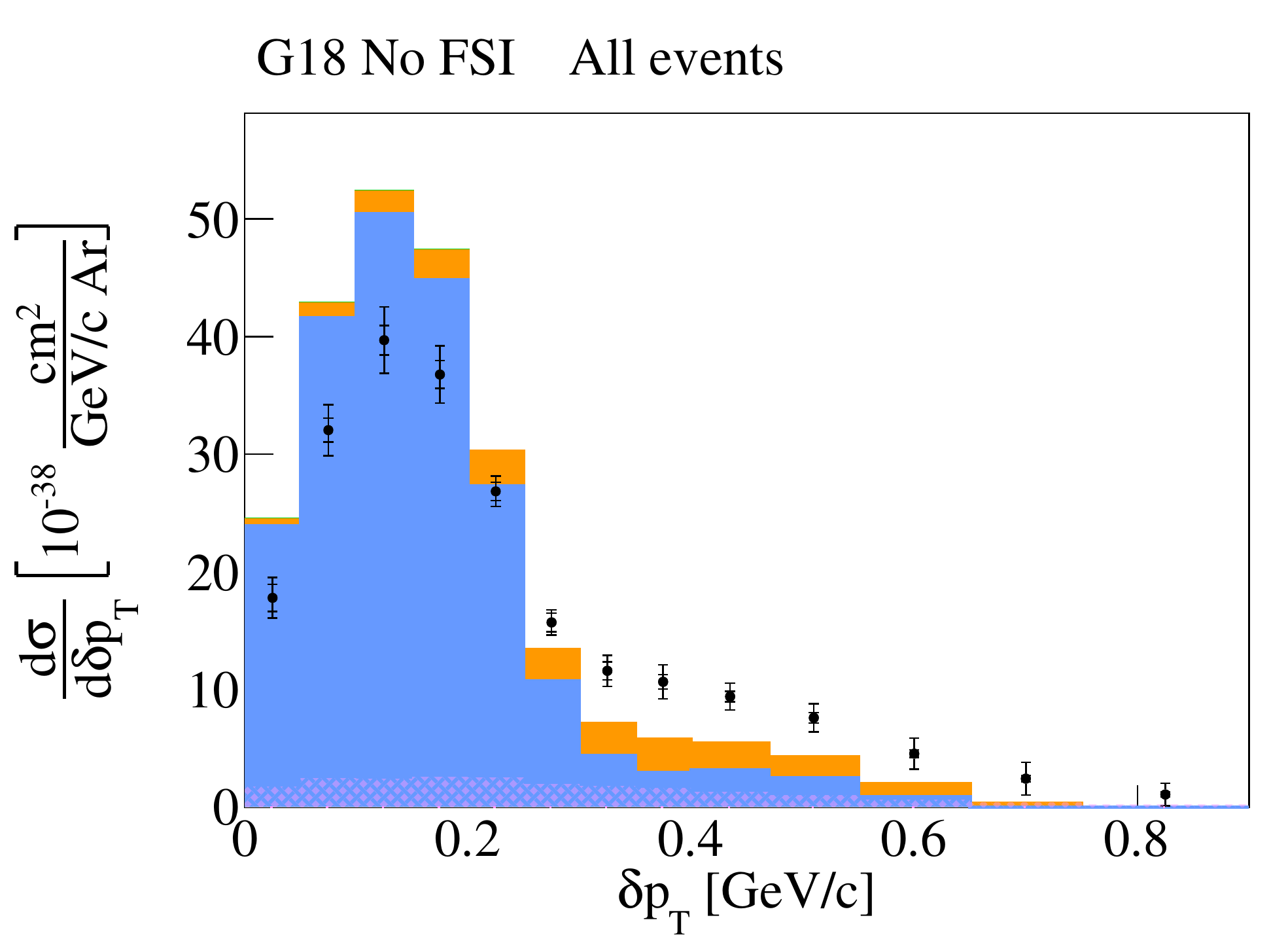}\\
\includegraphics[width=\BreakdownFigSize\linewidth]{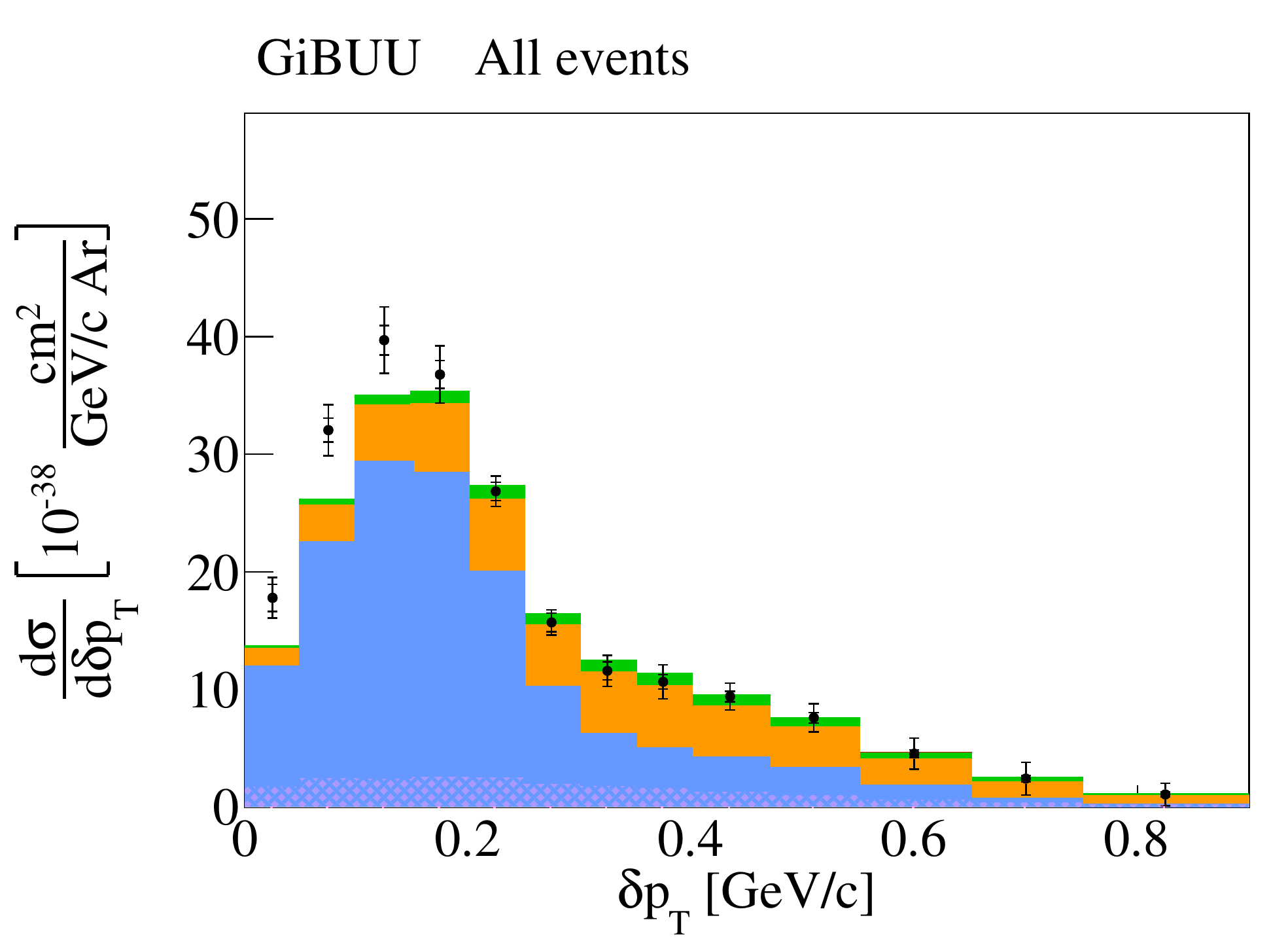}
\includegraphics[width=\BreakdownFigSize\linewidth]{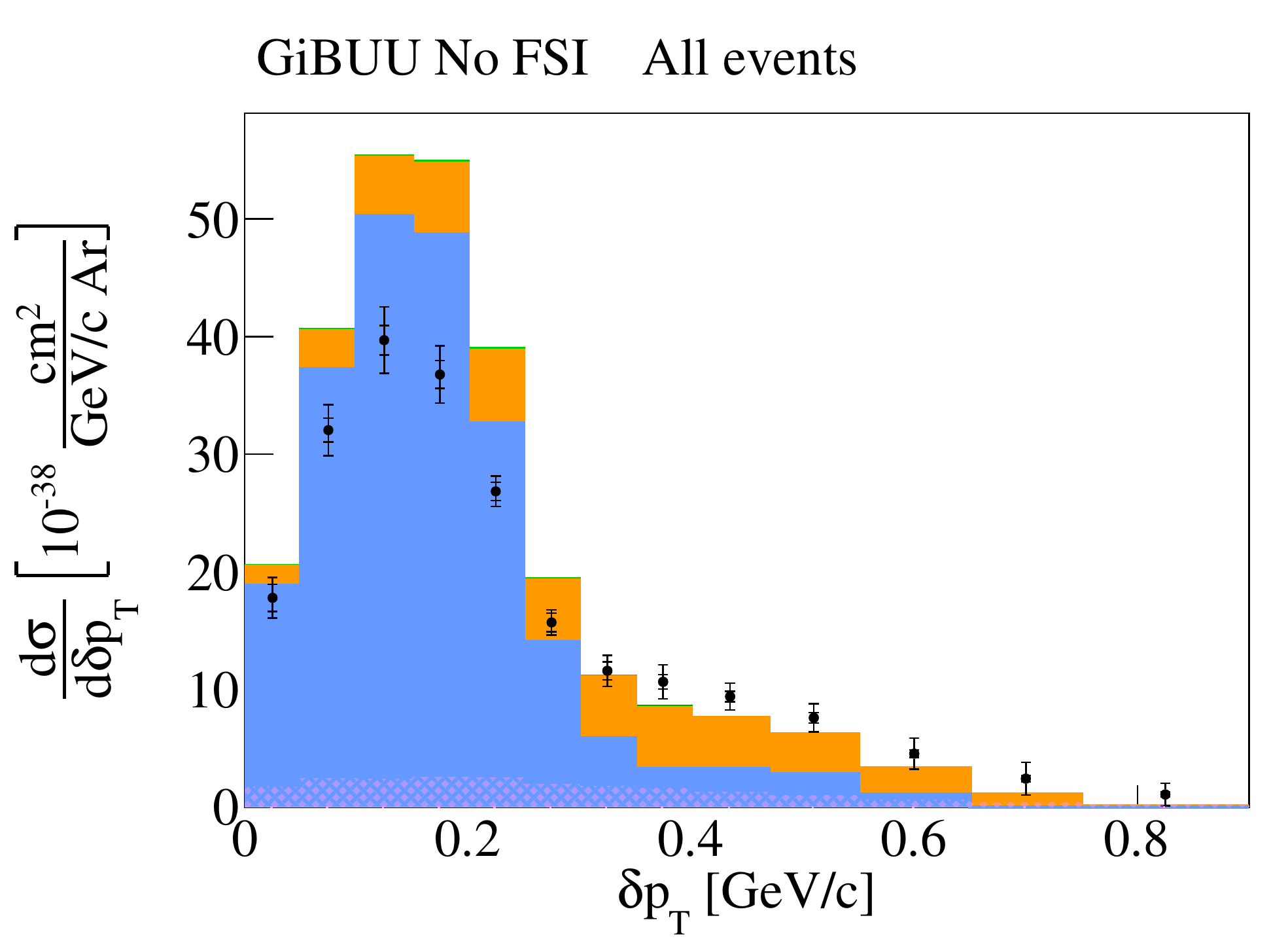}

\caption{Cross-section interaction breakdown for all the selected events. 
The breakdown is shown for (top left) the G18 configuration with FSI effects, (top right) the G18 configuration without FSI effects, (bottom left) GiB with FSI effects, and (bottom right) GiB without FSI effects.
}

\label{DeltaPTBreakdown_All}
\end{figure}

\begin{figure}[H]
\centering

\includegraphics[width=\BreakdownFigSize\linewidth]{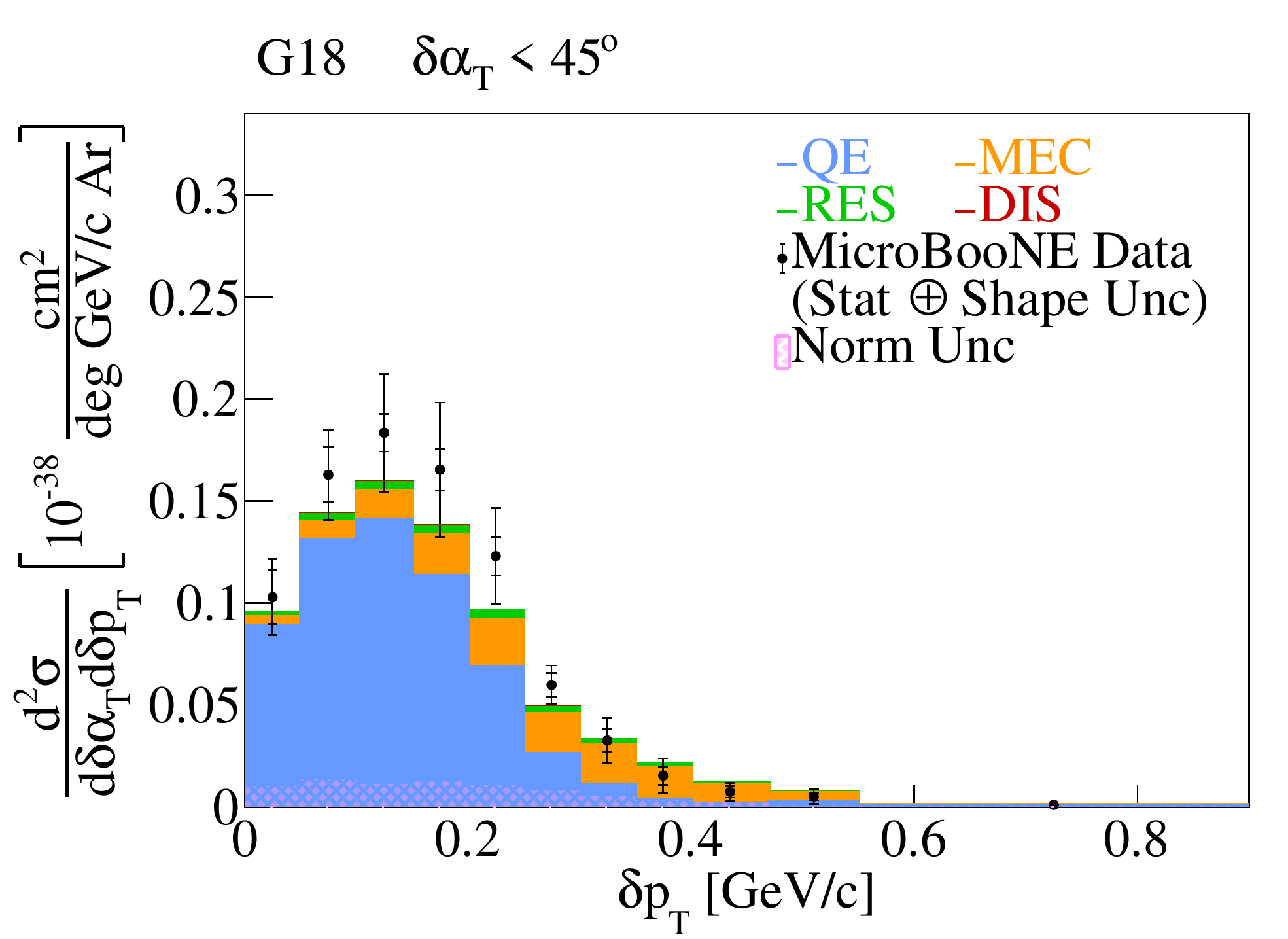}
\includegraphics[width=\BreakdownFigSize\linewidth]{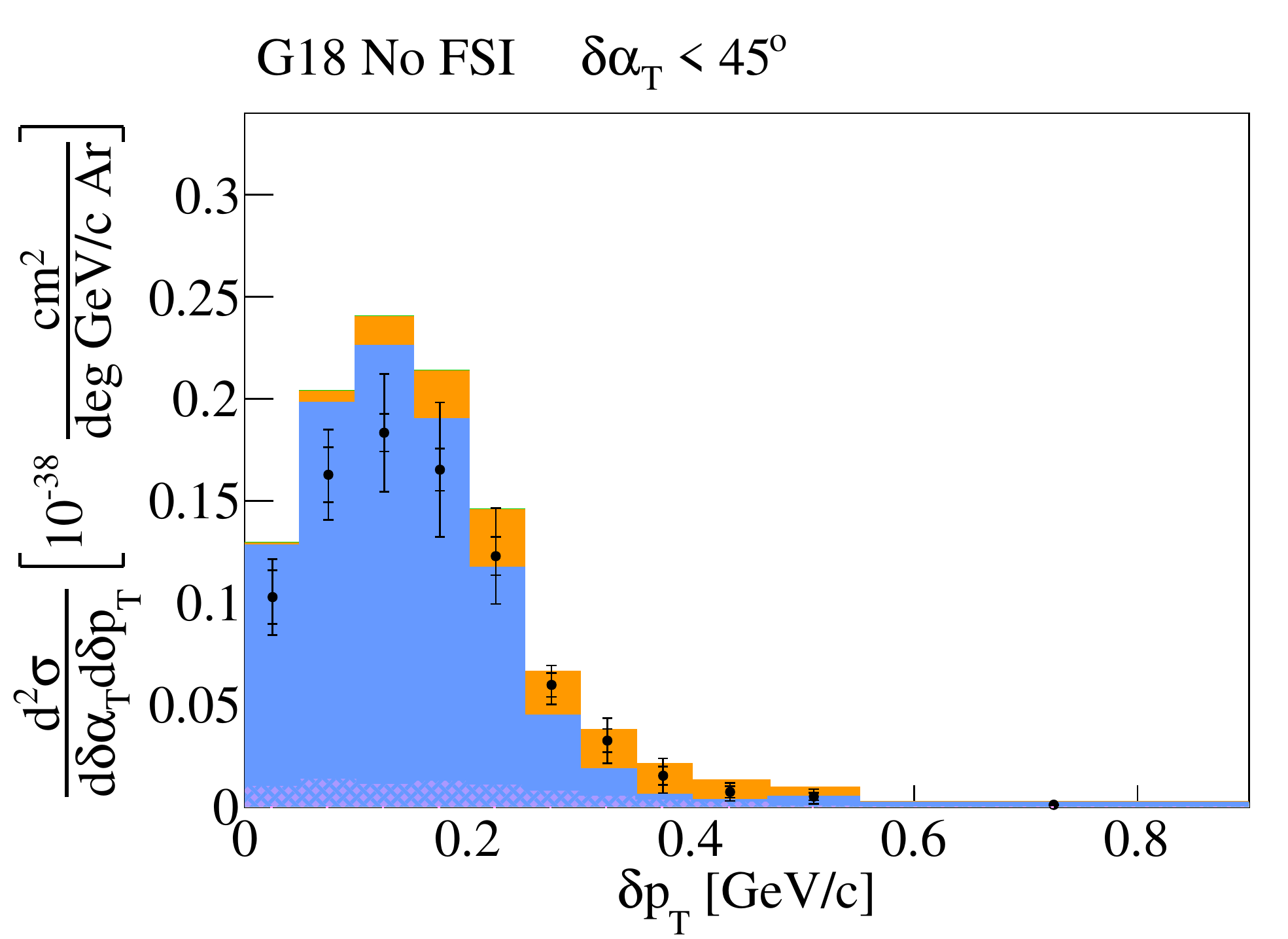}\\
\includegraphics[width=\BreakdownFigSize\linewidth]{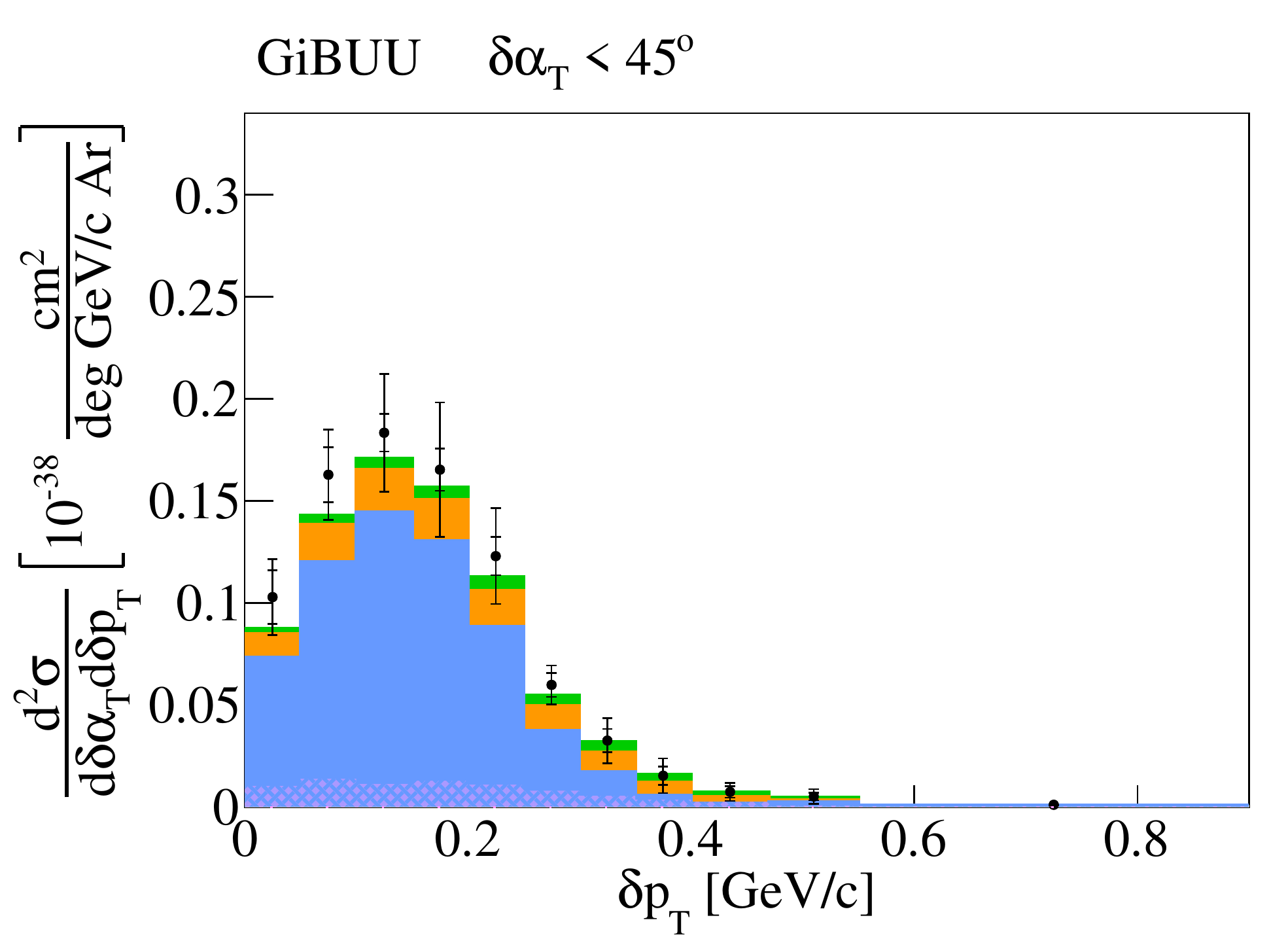}
\includegraphics[width=\BreakdownFigSize\linewidth]{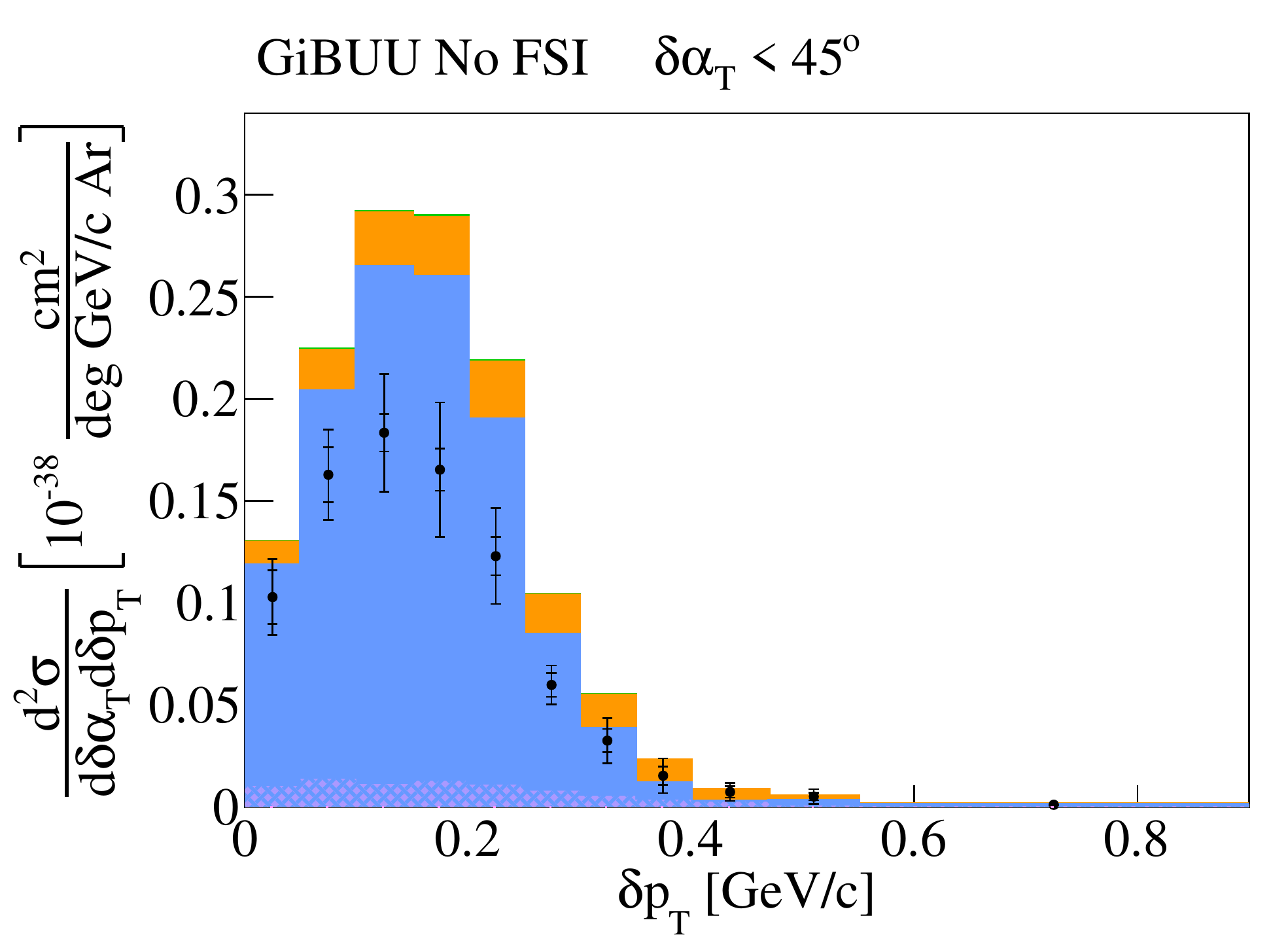}

\caption{Cross-section interaction breakdown for events with $\delta\alpha_{T} < 45^{o}$. 
The breakdown is shown for (top left) the G18 configuration with FSI effects, (top right) the G18 configuration without FSI effects, (bottom left) GiB with FSI effects, and (bottom right) GiB without FSI effects.
}

\label{DeltaPTBreakdown_Slice1}
\end{figure}

\begin{figure}[H]
\centering

\includegraphics[width=\BreakdownFigSize\linewidth]{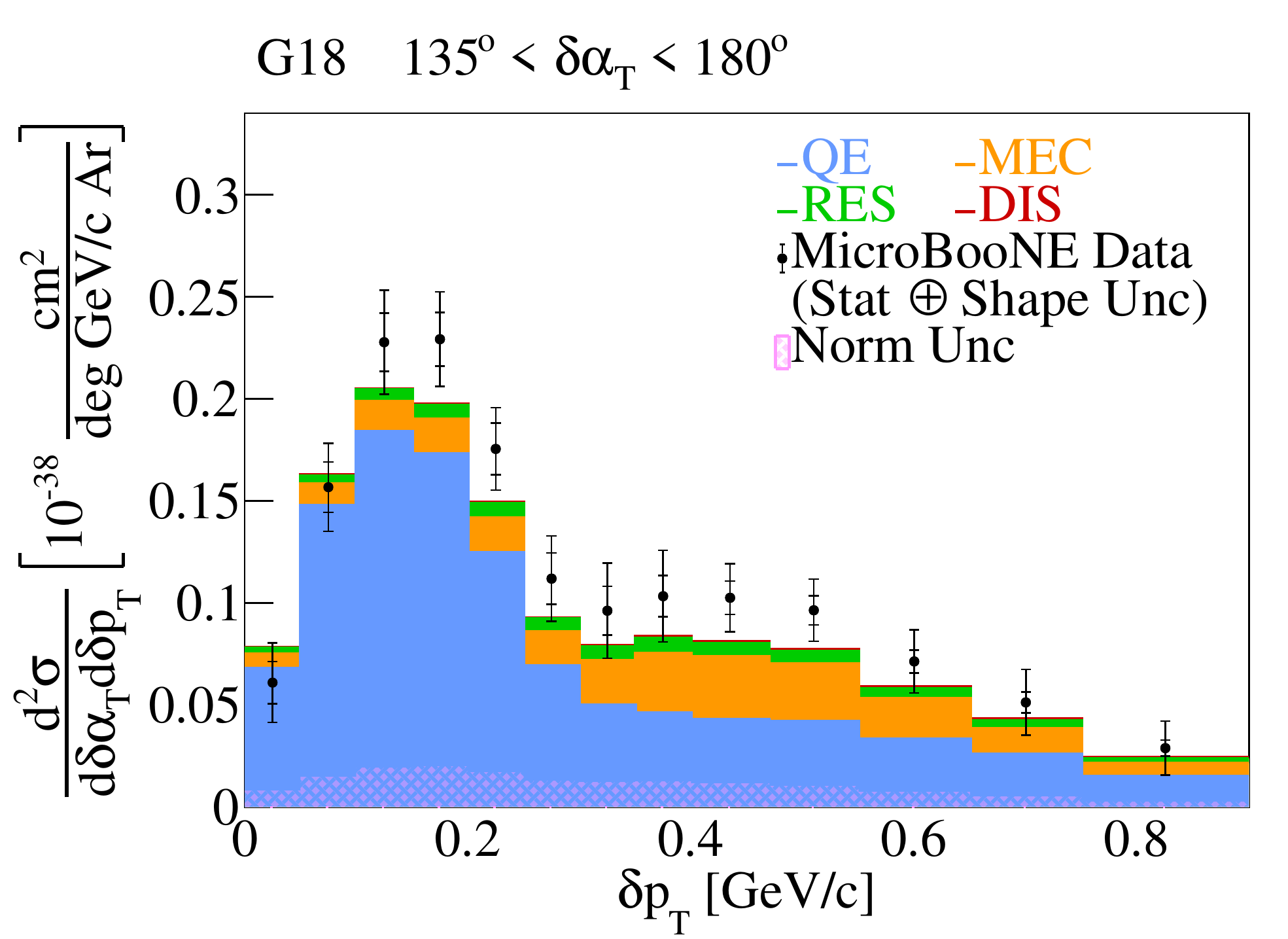}
\includegraphics[width=\BreakdownFigSize\linewidth]{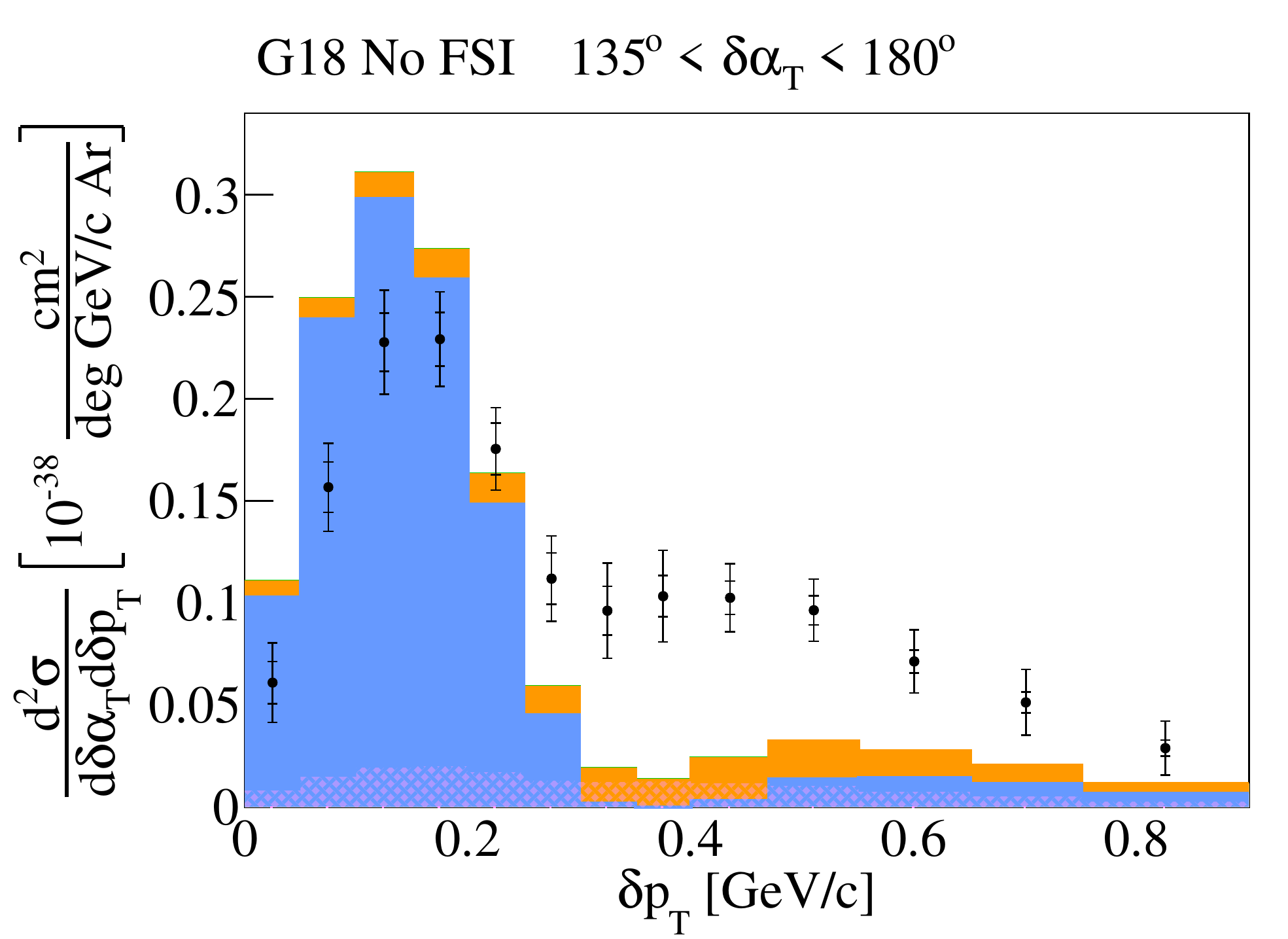}\\
\includegraphics[width=\BreakdownFigSize\linewidth]{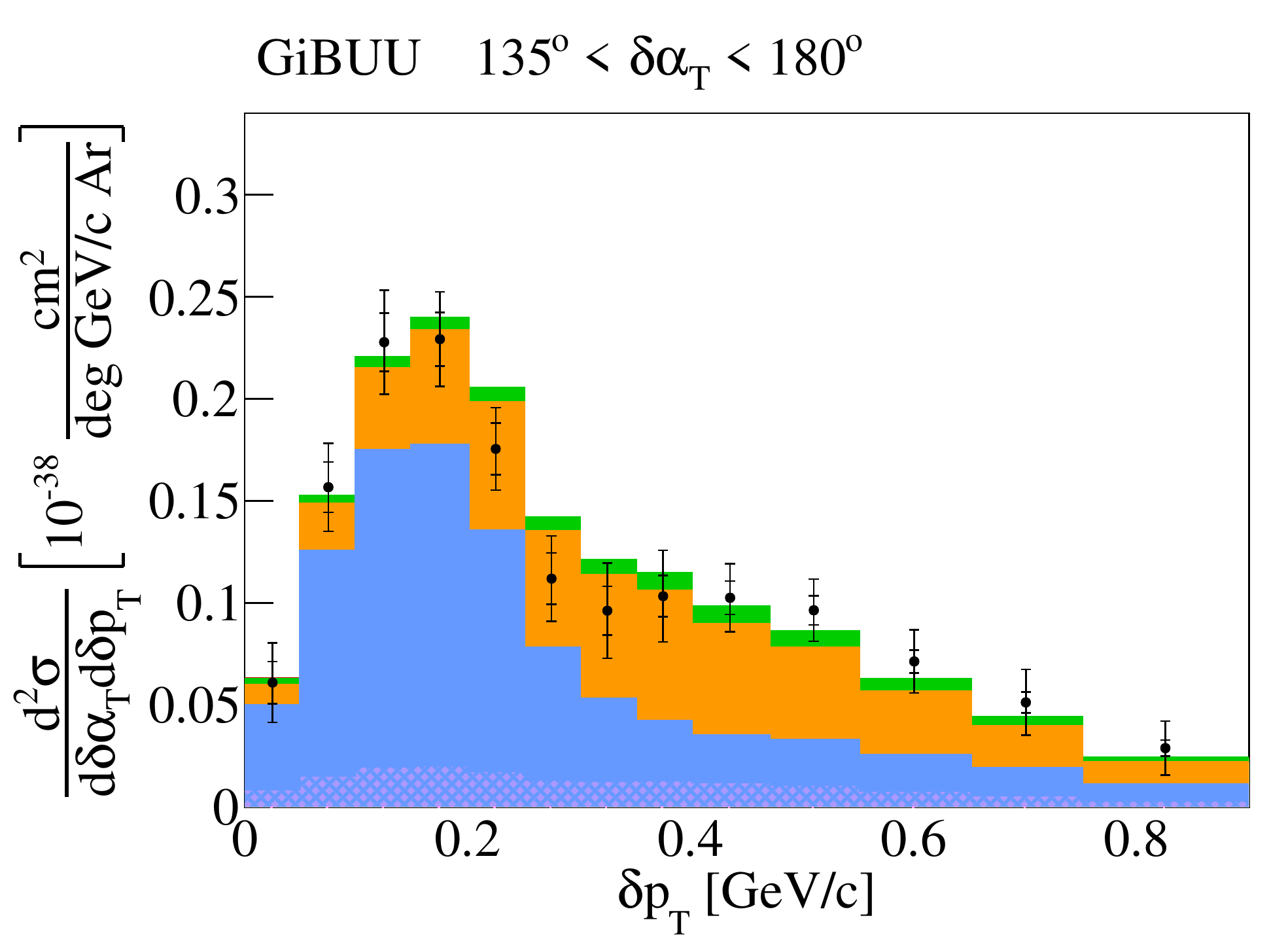}
\includegraphics[width=\BreakdownFigSize\linewidth]{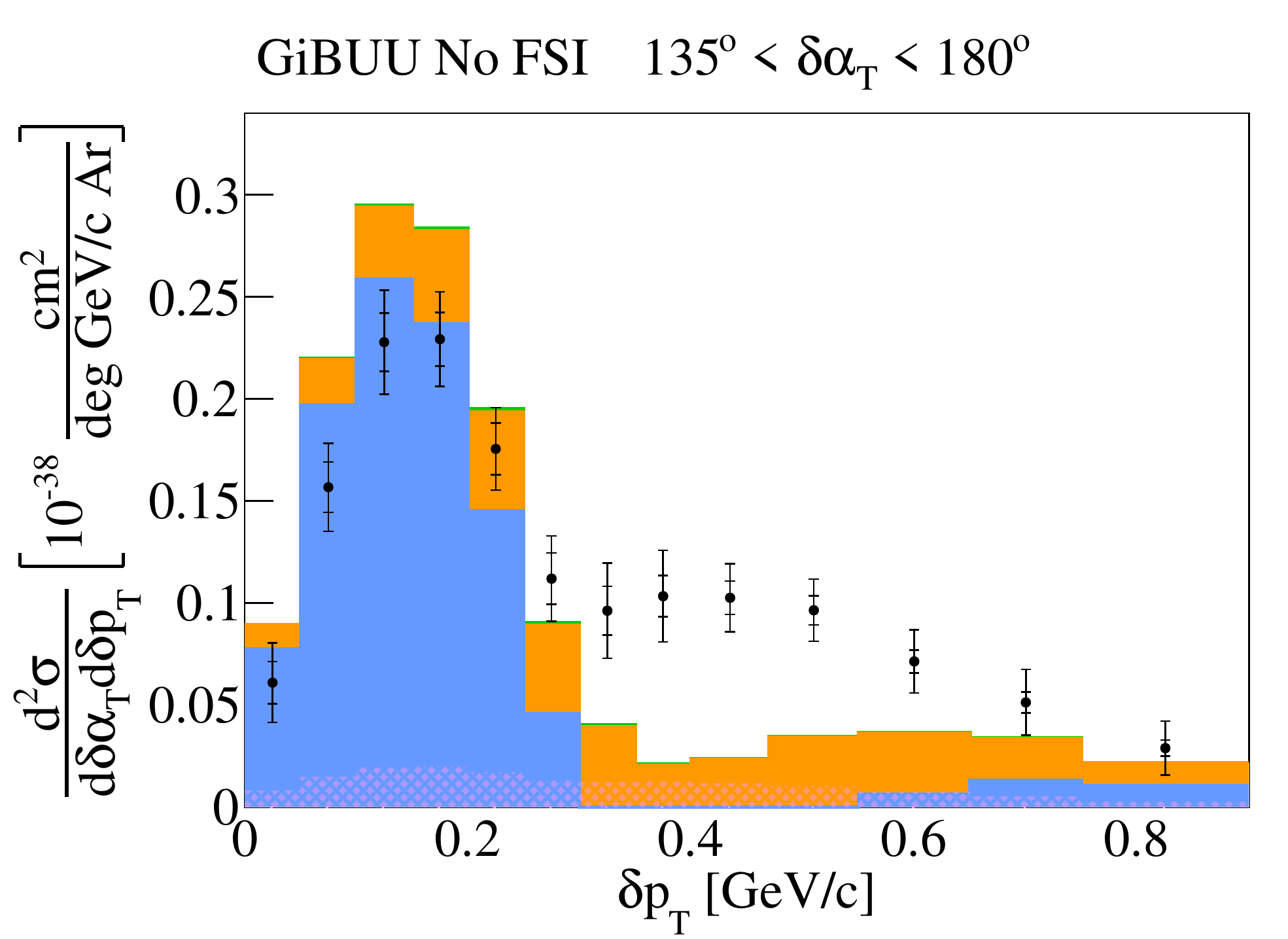}

\caption{Cross-section interaction breakdown for events with $135^{o} < \delta\alpha_{T} < 180^{o}$. 
The breakdown is shown for (top left) the G18 configuration with FSI effects, (top right) the G18 configuration without FSI effects, (bottom left) GiB with FSI effects, and (bottom right) GiB without FSI effects.
}

\label{DeltaPTBreakdown_Slice3}
\end{figure}

\begin{figure}[H]
\centering 
\includegraphics[width=\BreakdownFigSize\linewidth]{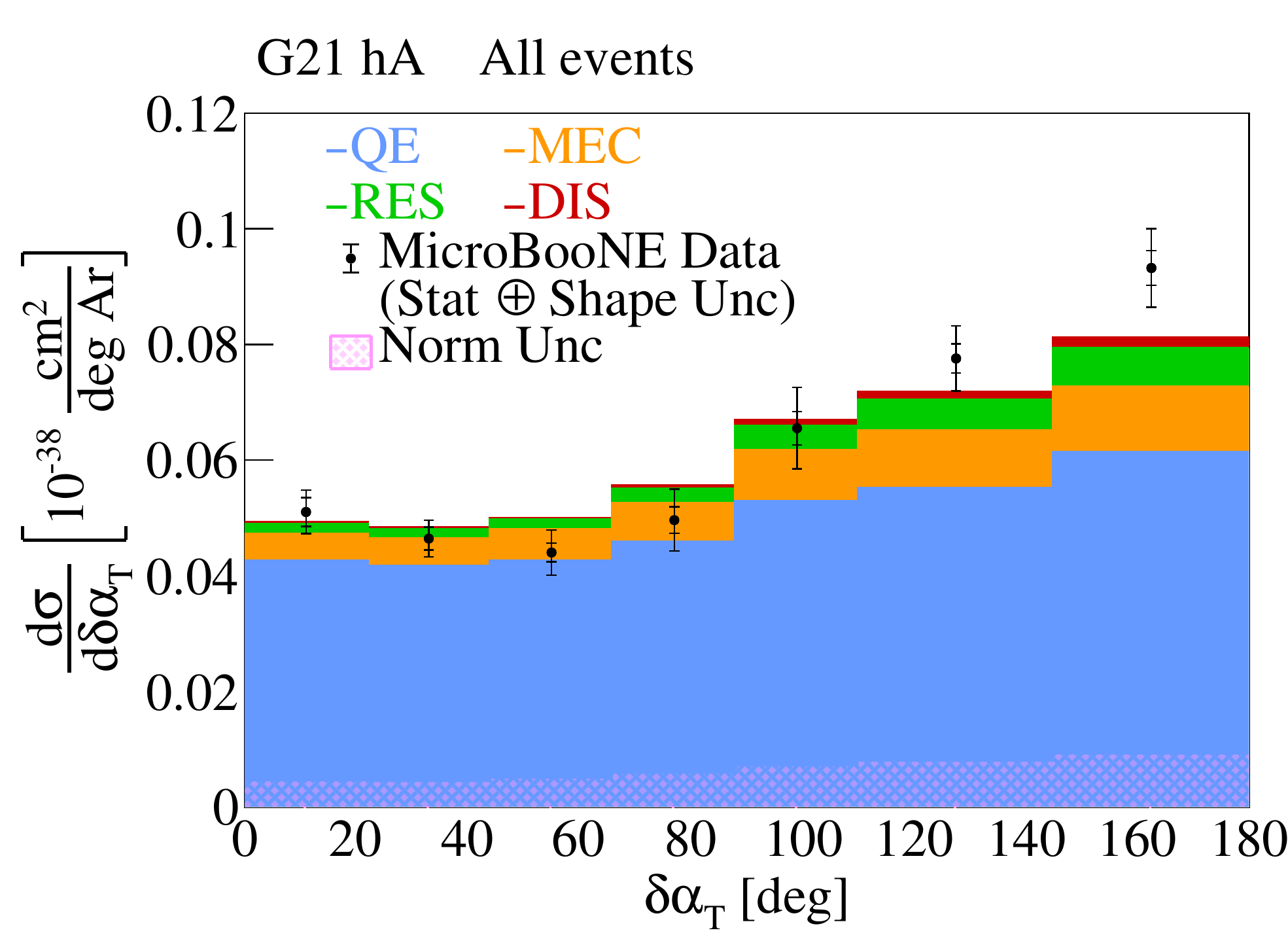}
\includegraphics[width=\BreakdownFigSize\linewidth]{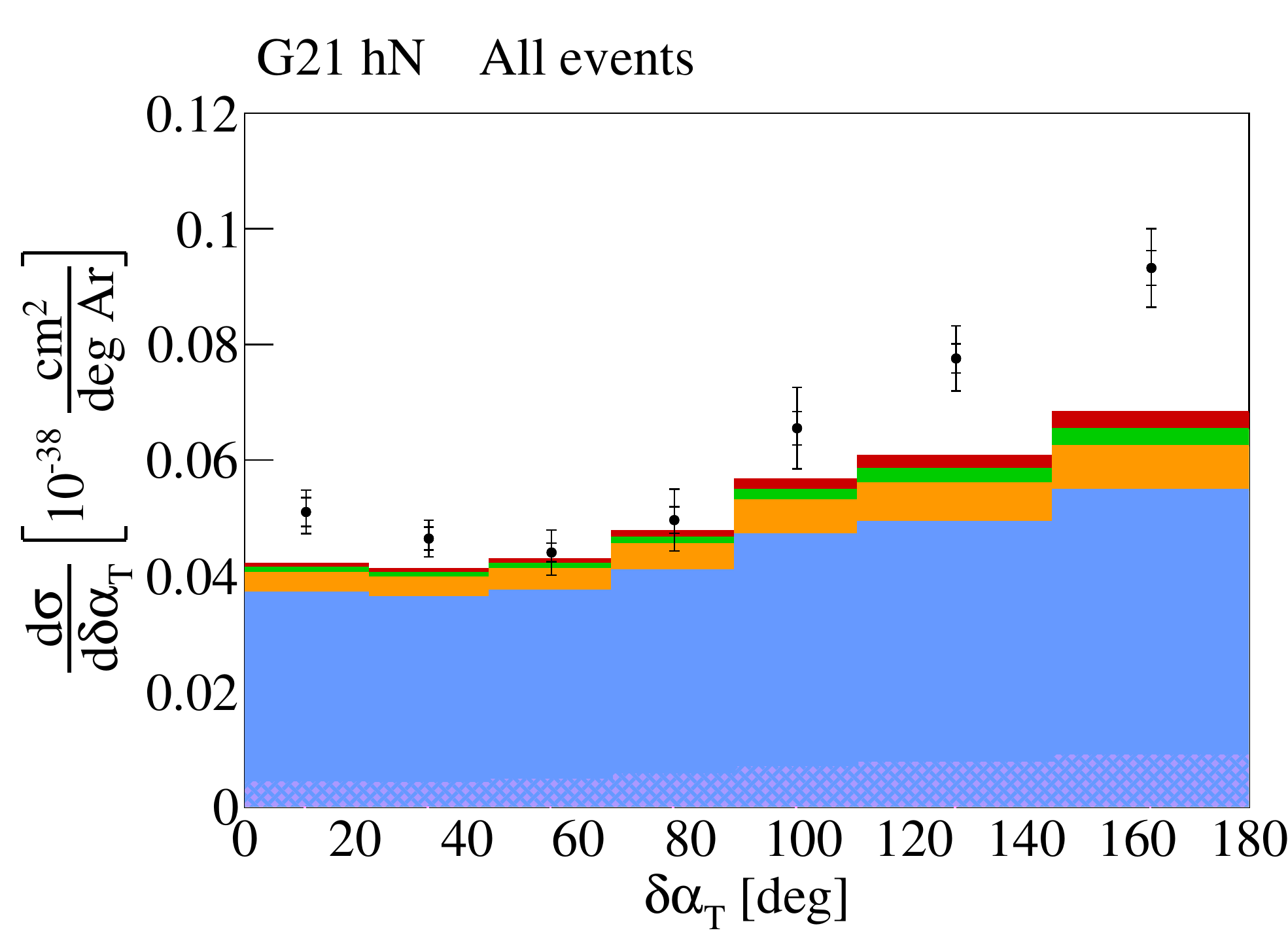}\\
\includegraphics[width=\BreakdownFigSize\linewidth]{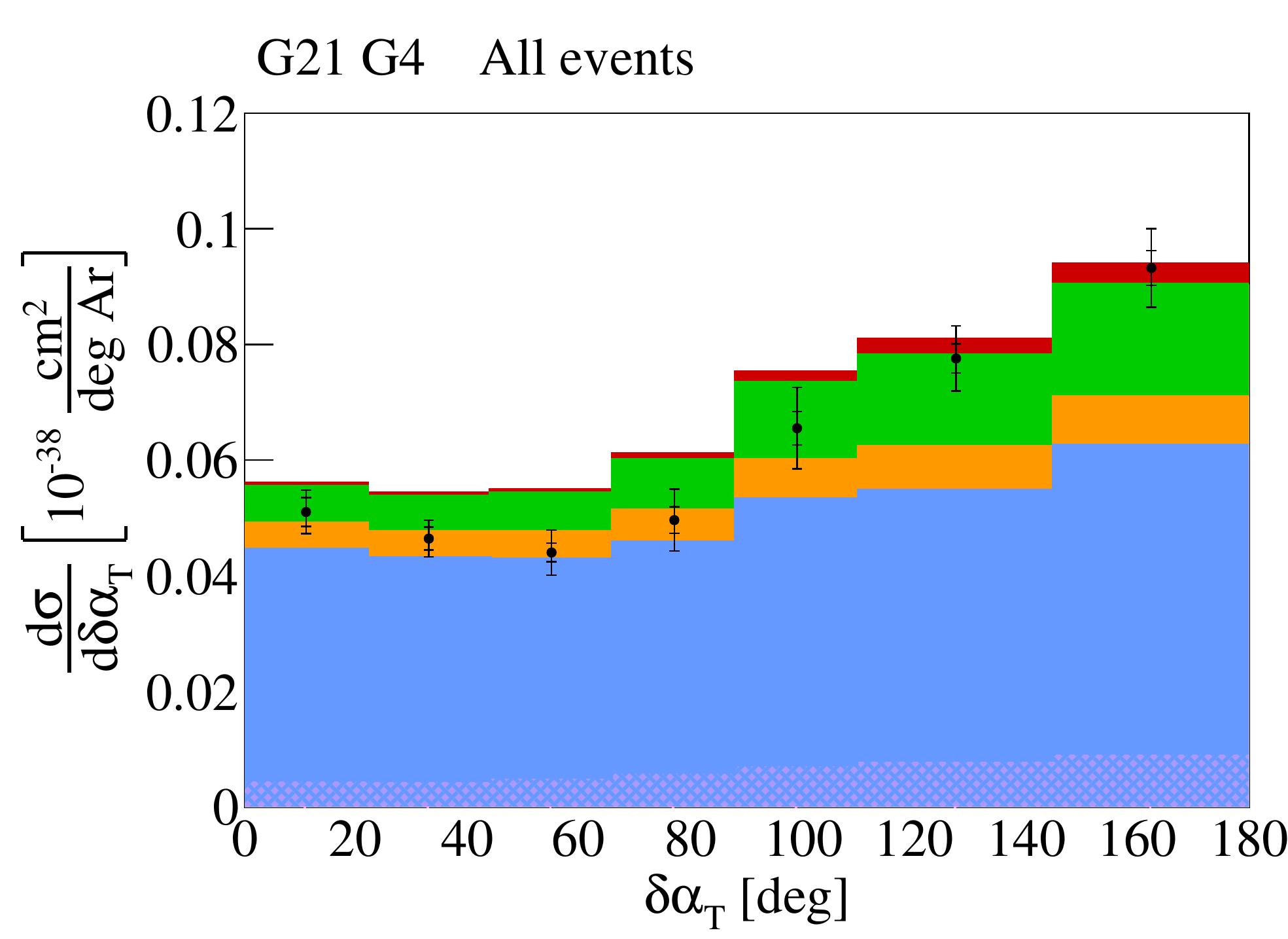}
\includegraphics[width=\BreakdownFigSize\linewidth]{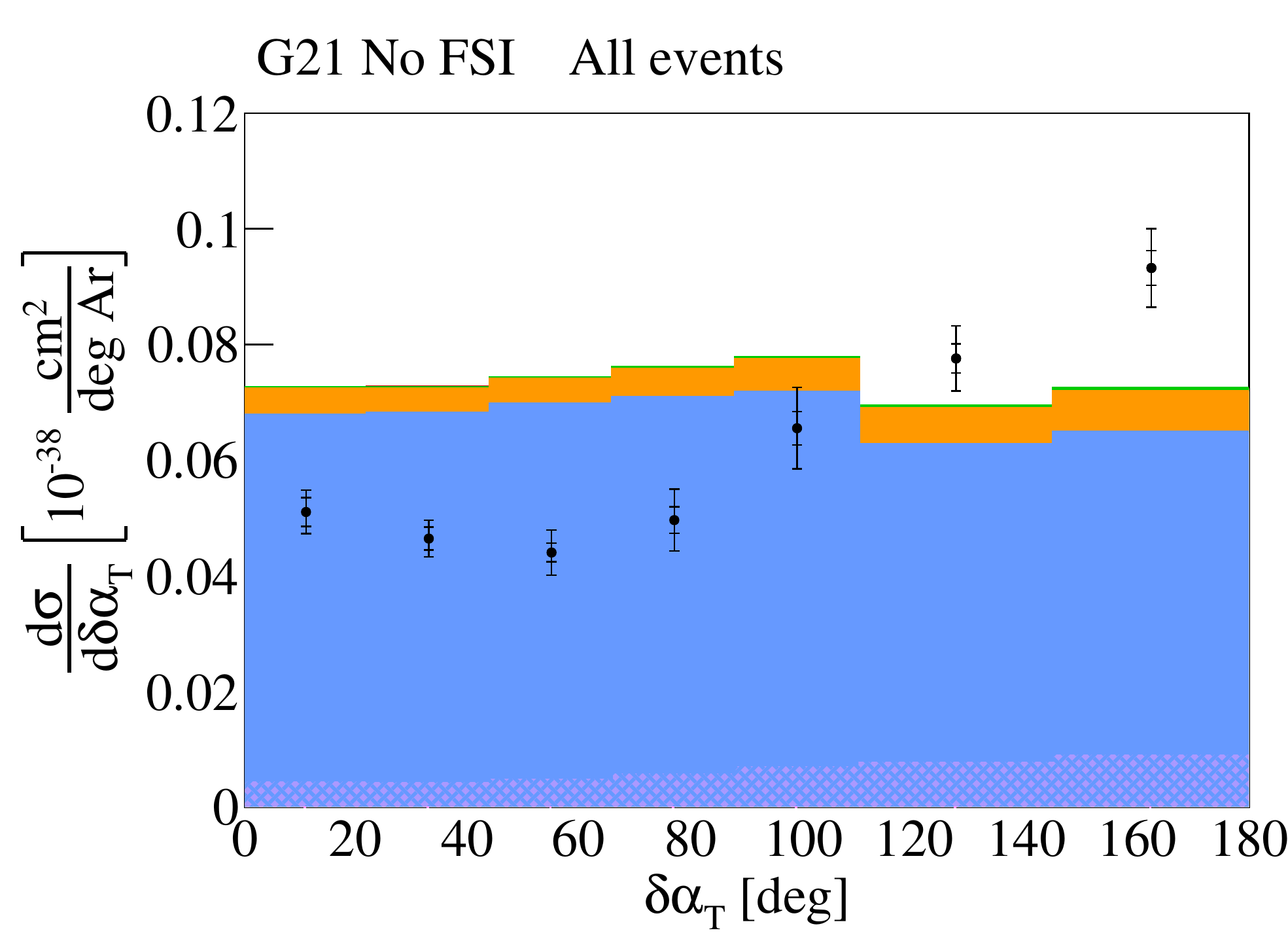}

\caption{Cross-section interaction breakdown for all the selected events. 
The breakdown is shown for (top left) the G21 hA configuration with the hA2018 FSI model, (top right) the G21 hN configuration with the hN FSI model, (bottom left) the G21 G4 configuration with the G4 FSI model, and (bottom right) the G21 No FSI configuration without FSI effects.}

\label{DeltaAlphaTBreakdown_All}
\end{figure}

\begin{figure}[H]
\centering
\includegraphics[width=\BreakdownFigSize\linewidth]{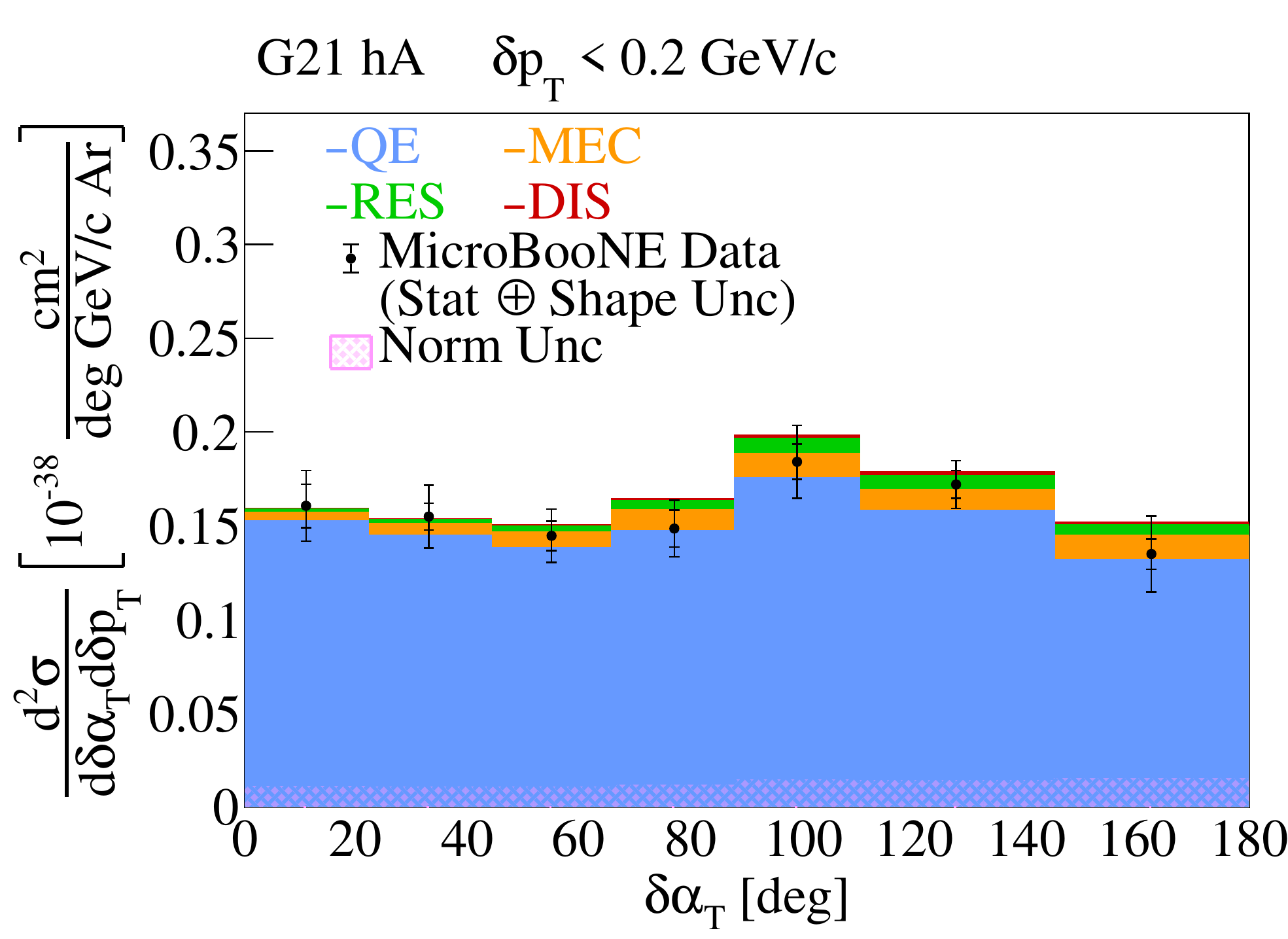}
\includegraphics[width=\BreakdownFigSize\linewidth]{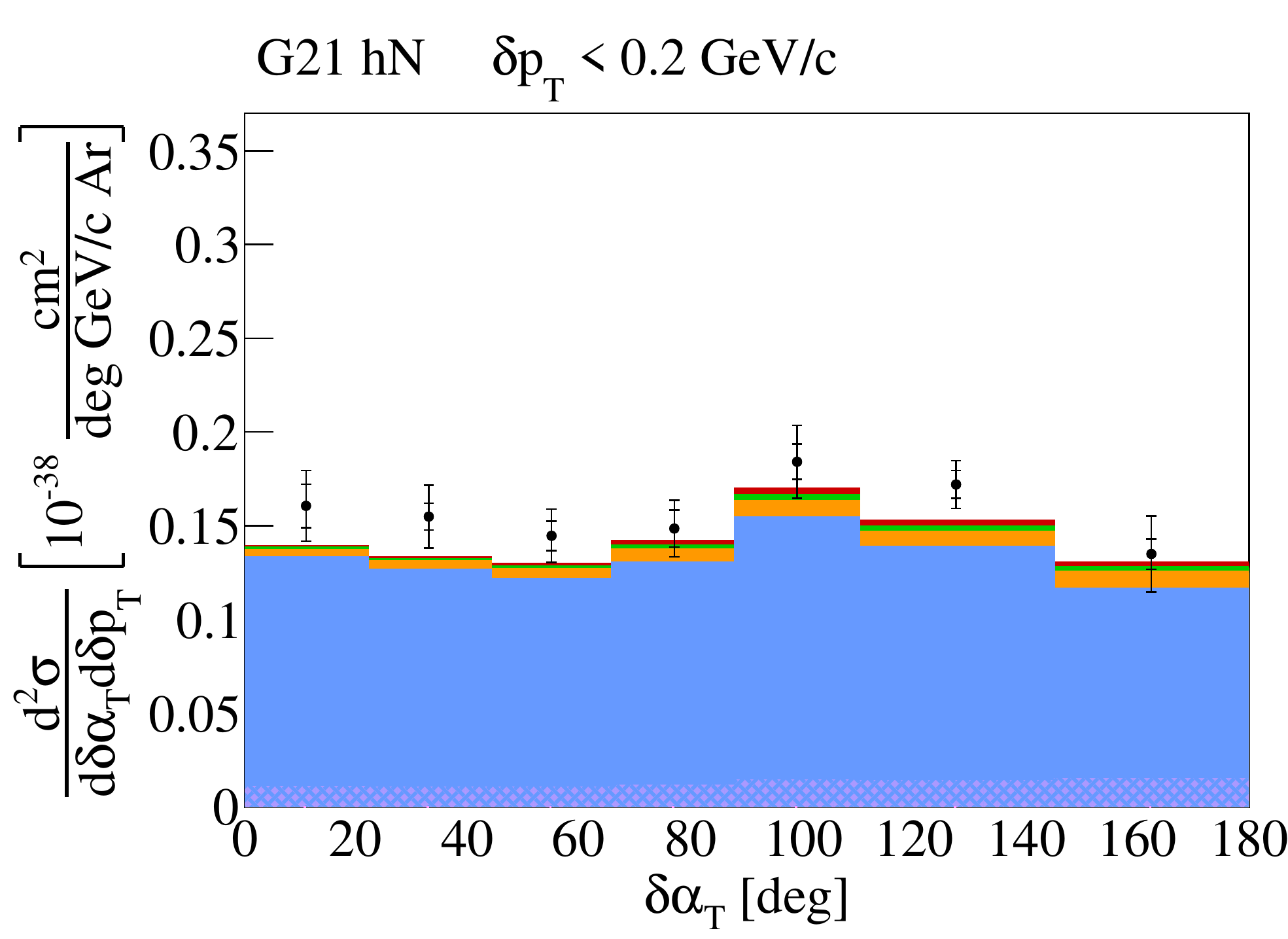}\\
\includegraphics[width=\BreakdownFigSize\linewidth]{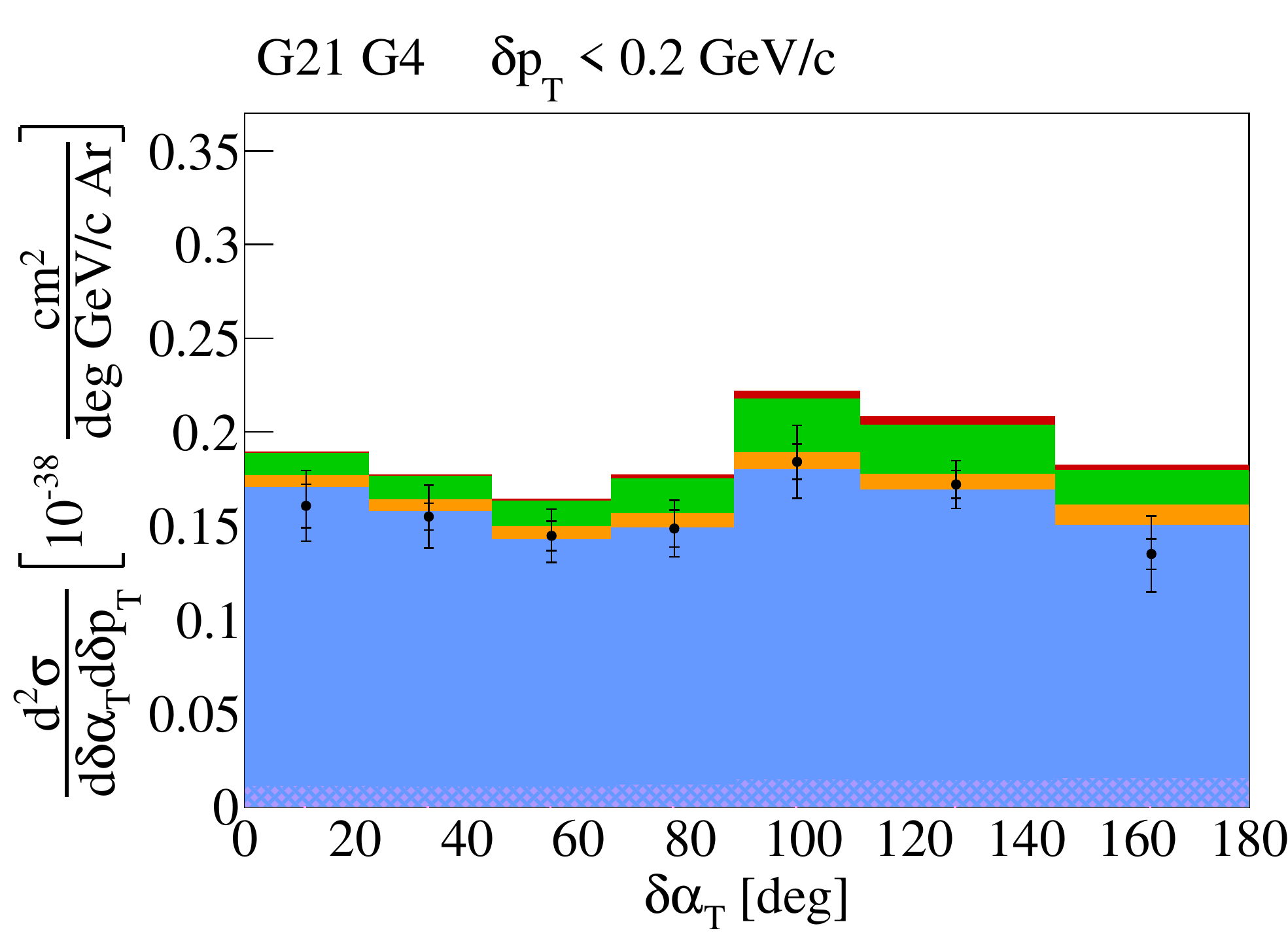}
\includegraphics[width=\BreakdownFigSize\linewidth]{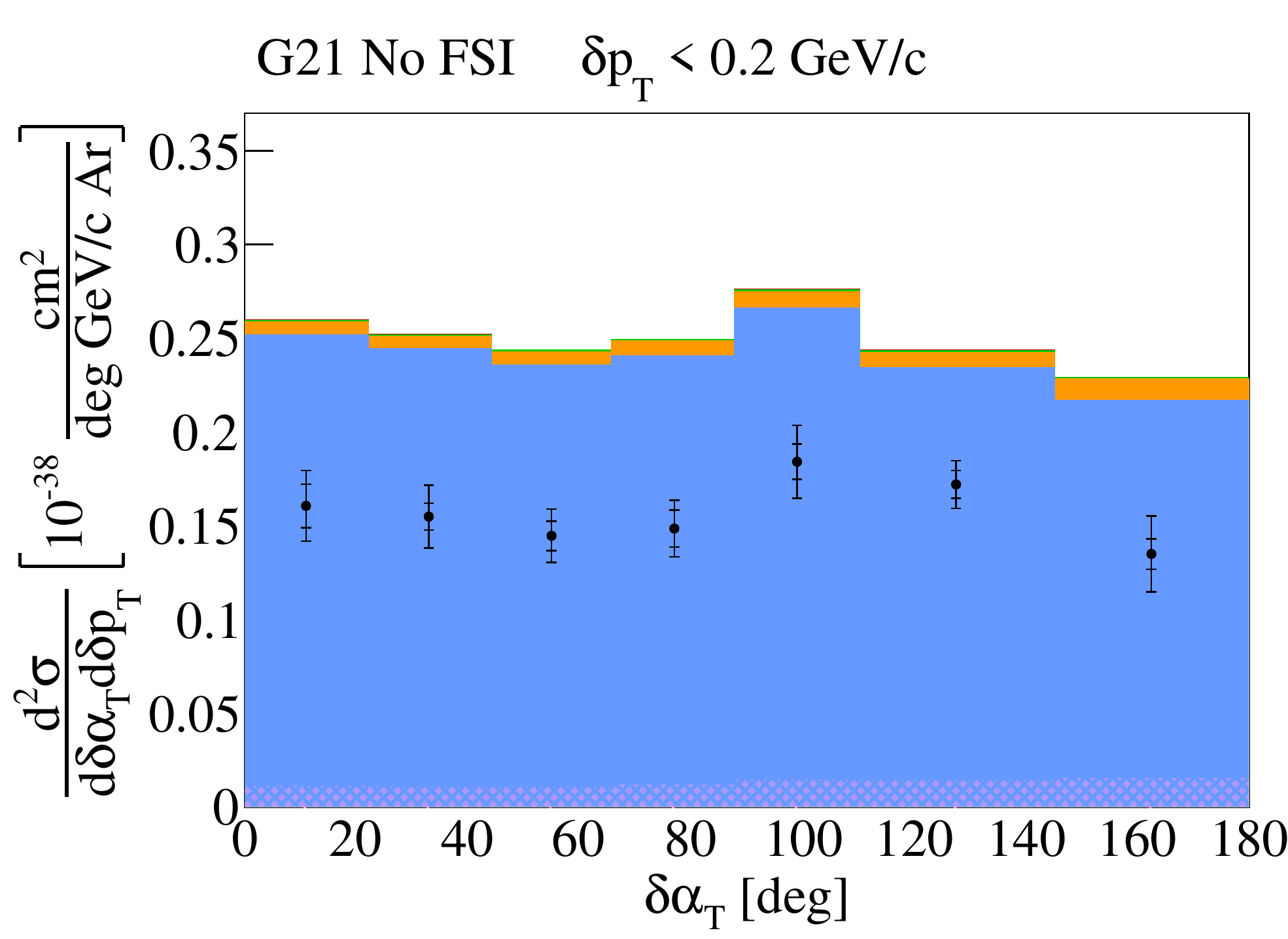}

\caption{Cross-section interaction breakdown for events with $\delta p_{T} <$ 0.2\,GeV/c. 
The breakdown is shown for (top left) the G21 hA configuration with the hA2018 FSI model, (top right) the G21 hN configuration with the hN FSI model, (bottom left) the G21 G4 configuration with the G4 FSI model, and (bottom right) the G21 No FSI configuration without FSI effects.
}

\label{DeltaAlphaTBreakdown_Slice1}
\end{figure}

\begin{figure}[H]
\centering
\includegraphics[width=\BreakdownFigSize\linewidth]{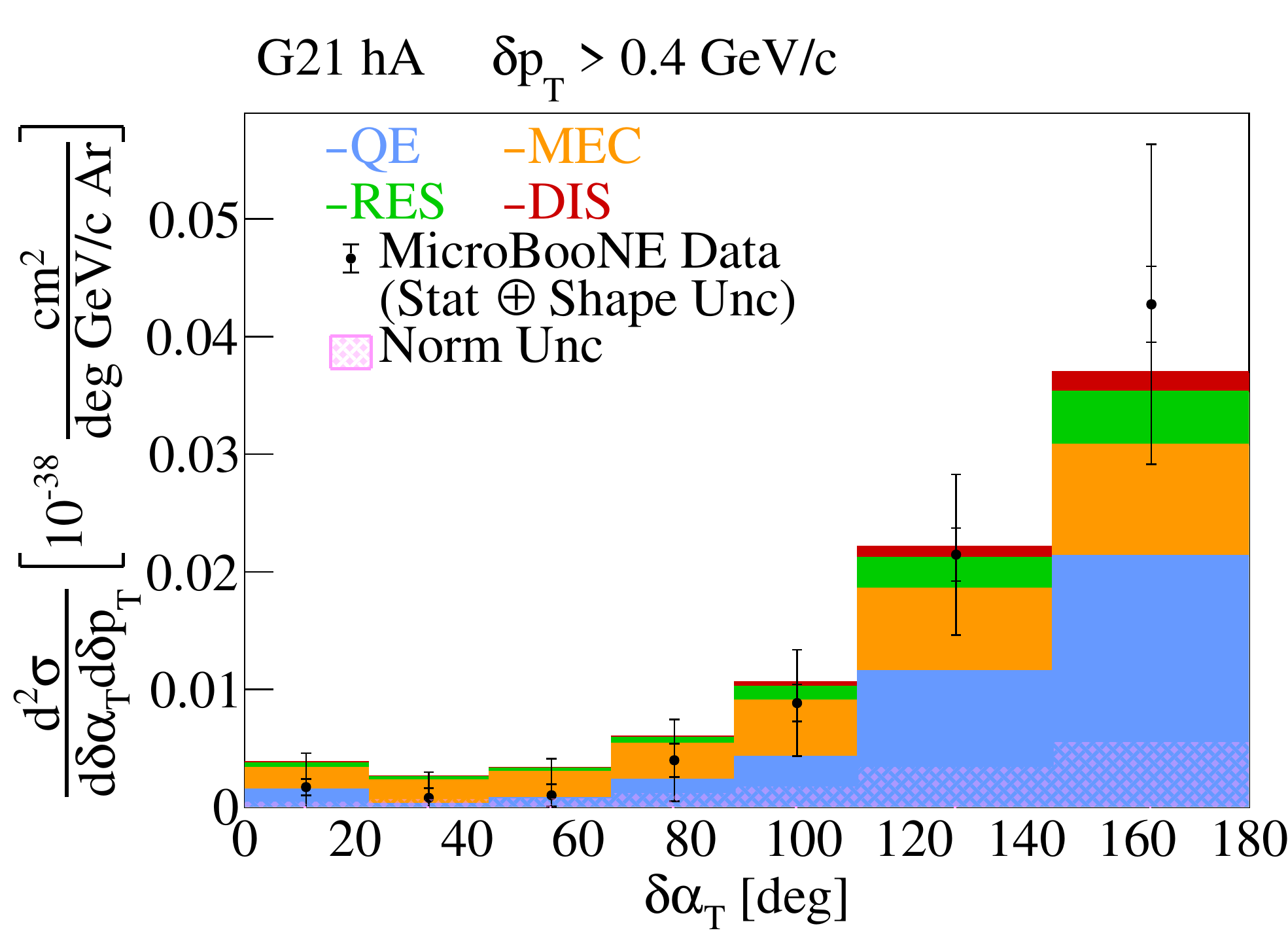}
\includegraphics[width=\BreakdownFigSize\linewidth]{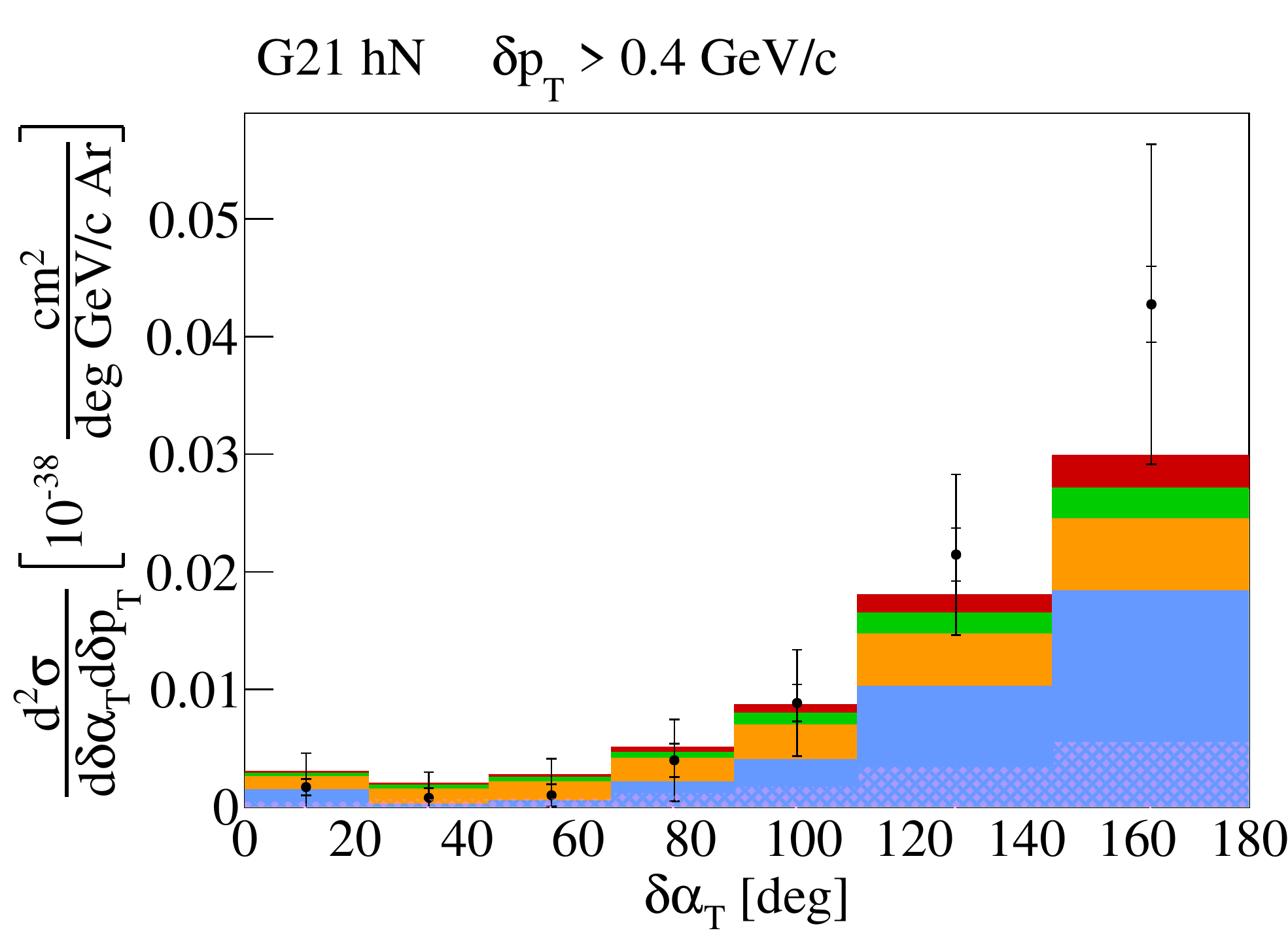}\\
\includegraphics[width=\BreakdownFigSize\linewidth]{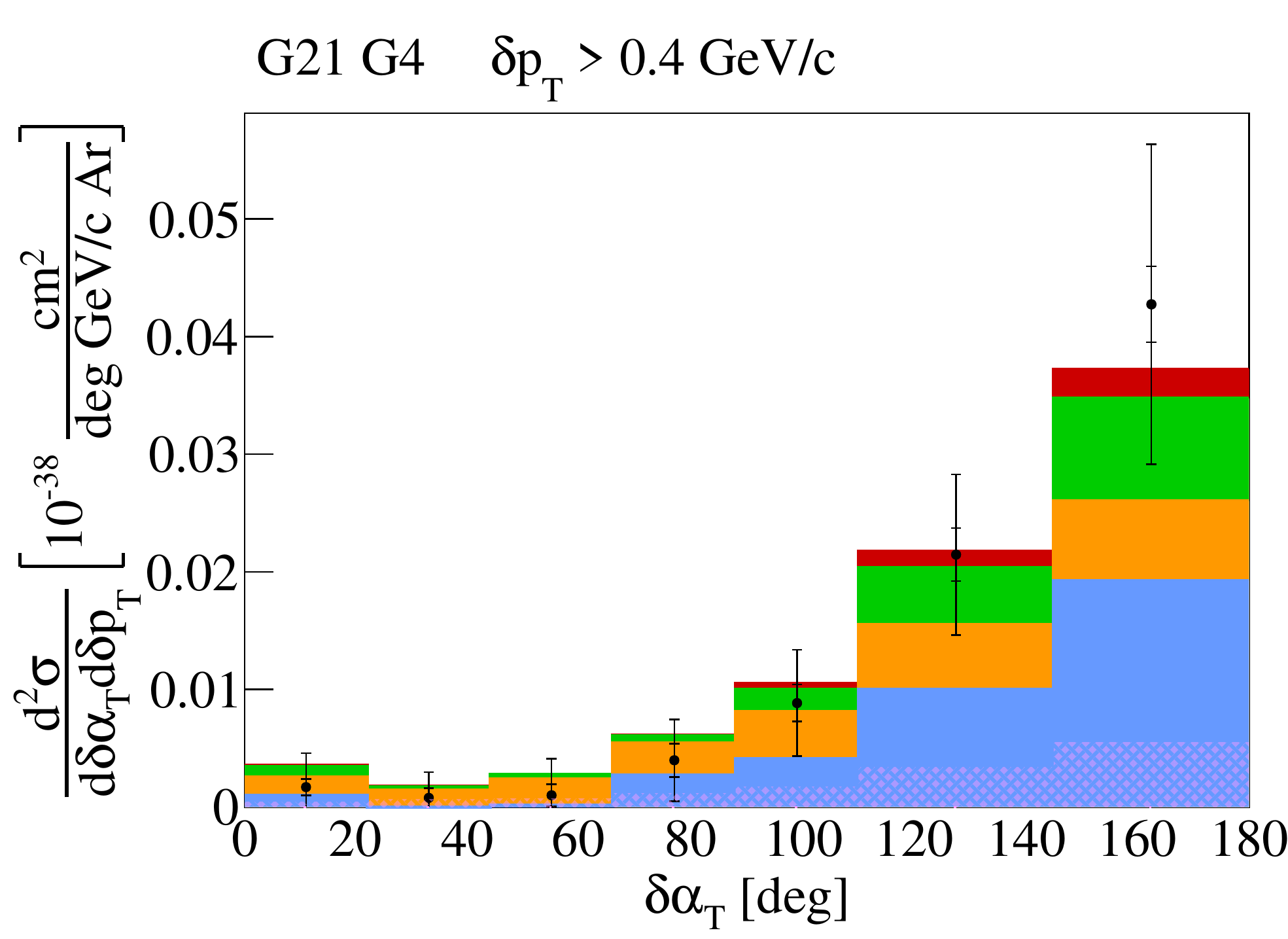}
\includegraphics[width=\BreakdownFigSize\linewidth]{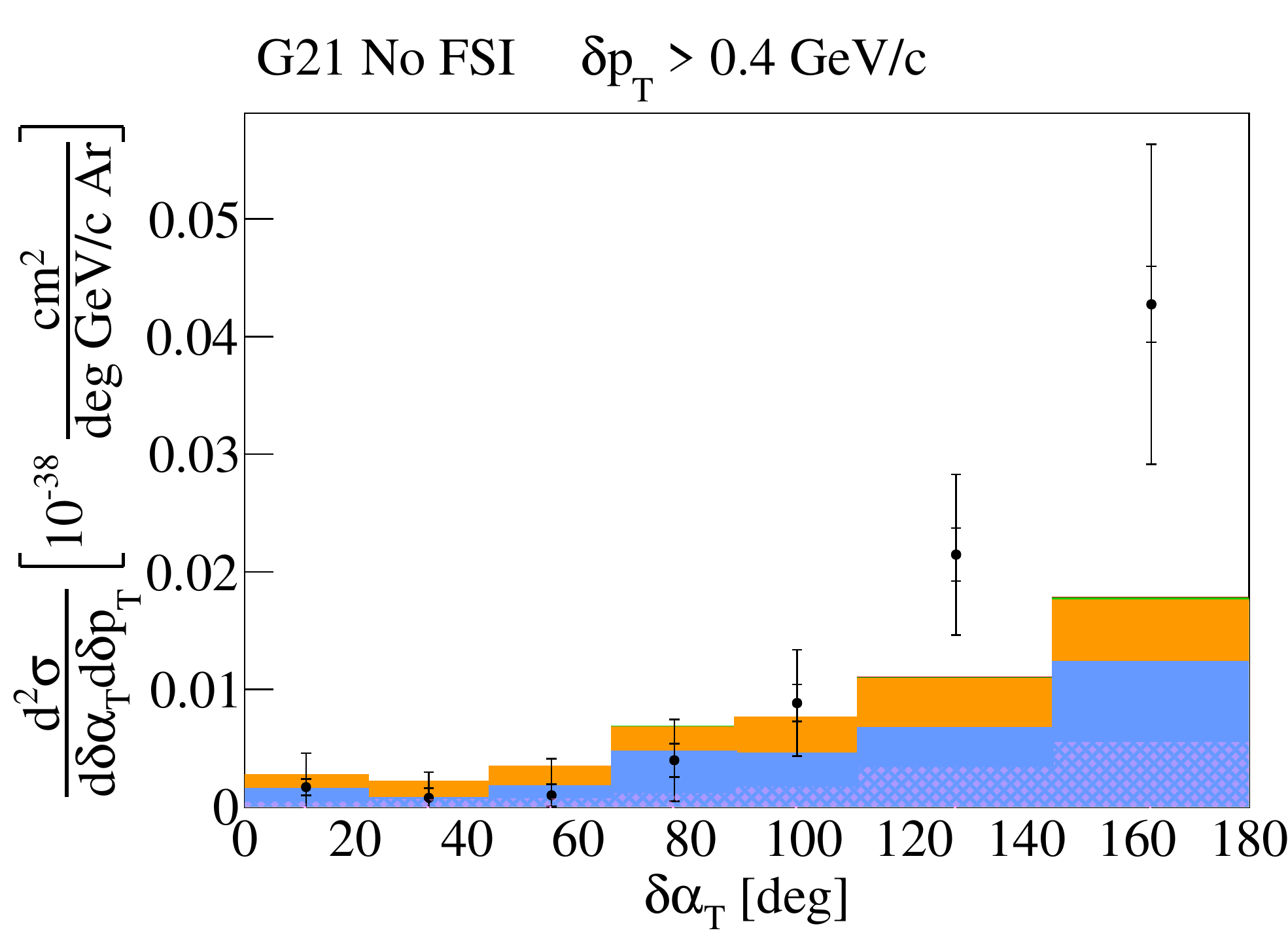}

\caption{Cross-section interaction breakdown for events with $\delta p_{T} >$ 0.4\,GeV/c. 
The breakdown is shown for (top left) the G21 hA configuration with the hA2018 FSI model, (top right) the G21 hN configuration with the hN FSI model, (bottom left) the G21 G4 configuration with the G4 FSI model, and (bottom right) the G21 No FSI configuration without FSI effects.
}

\label{DeltaAlphaTBreakdown_Slice3}
\end{figure}

\begin{figure}[H]
\centering 
\includegraphics[width=\BreakdownFigSize\linewidth]{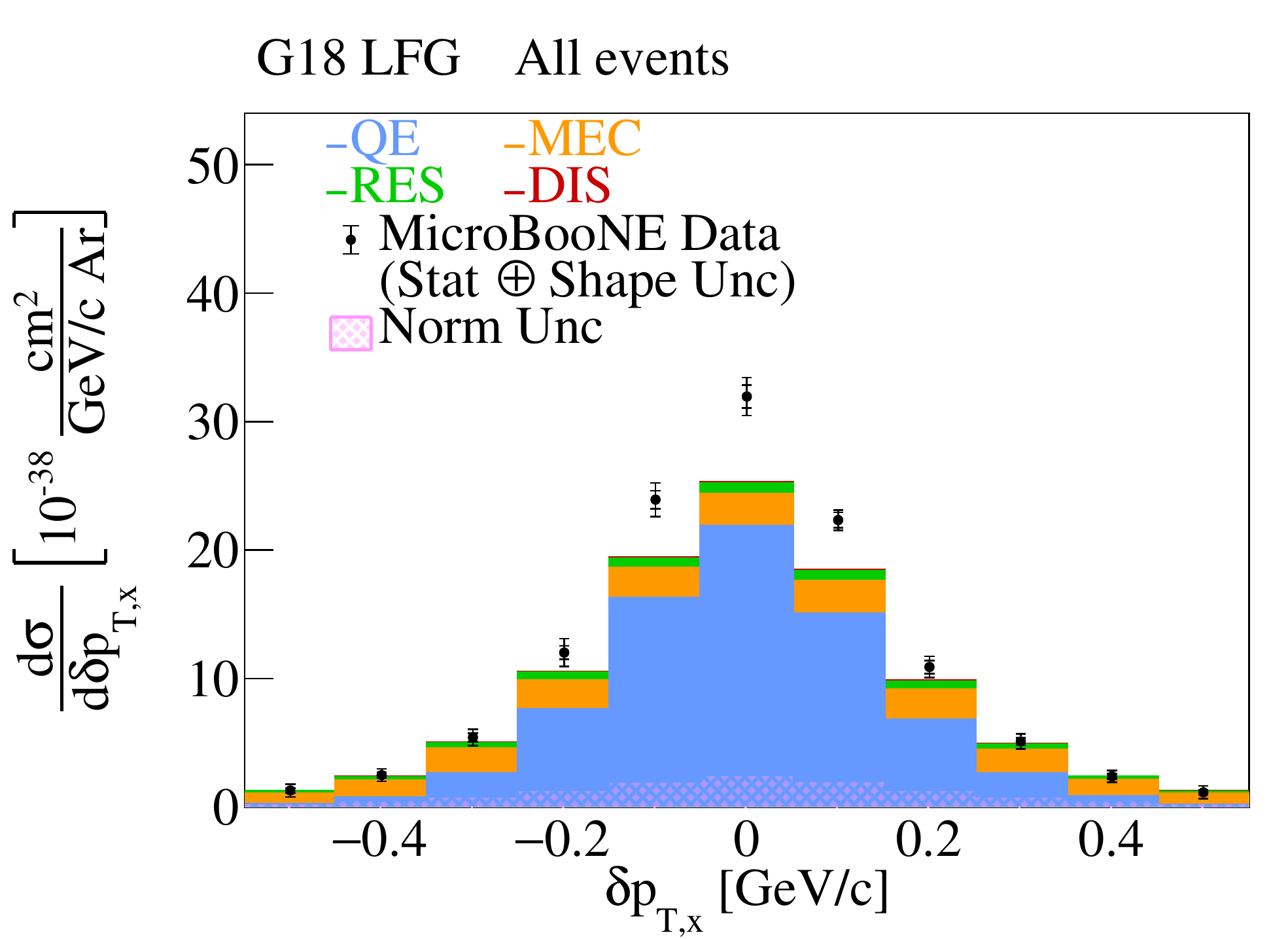}
\includegraphics[width=\BreakdownFigSize\linewidth]{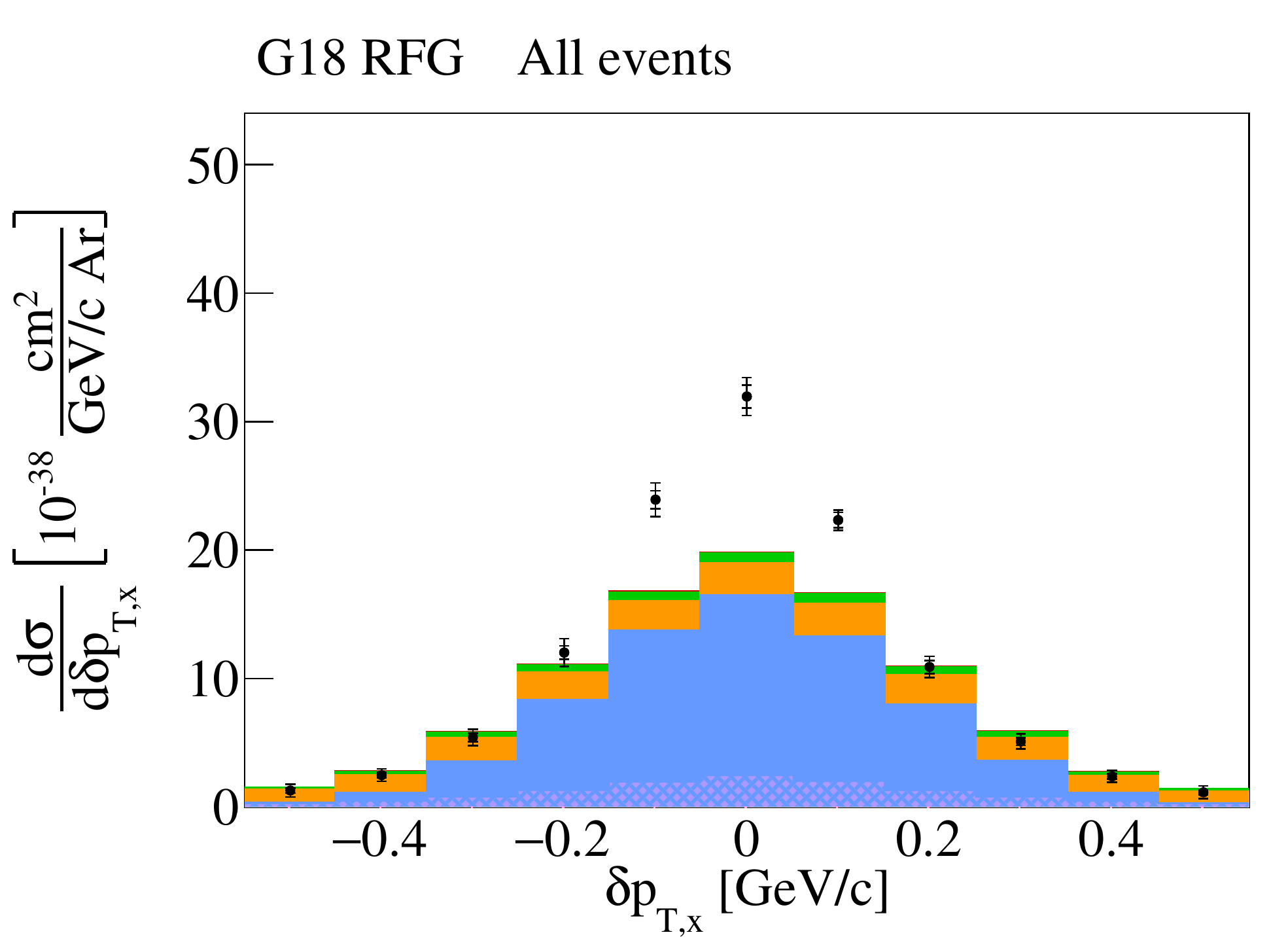}\\
\includegraphics[width=\BreakdownFigSize\linewidth]{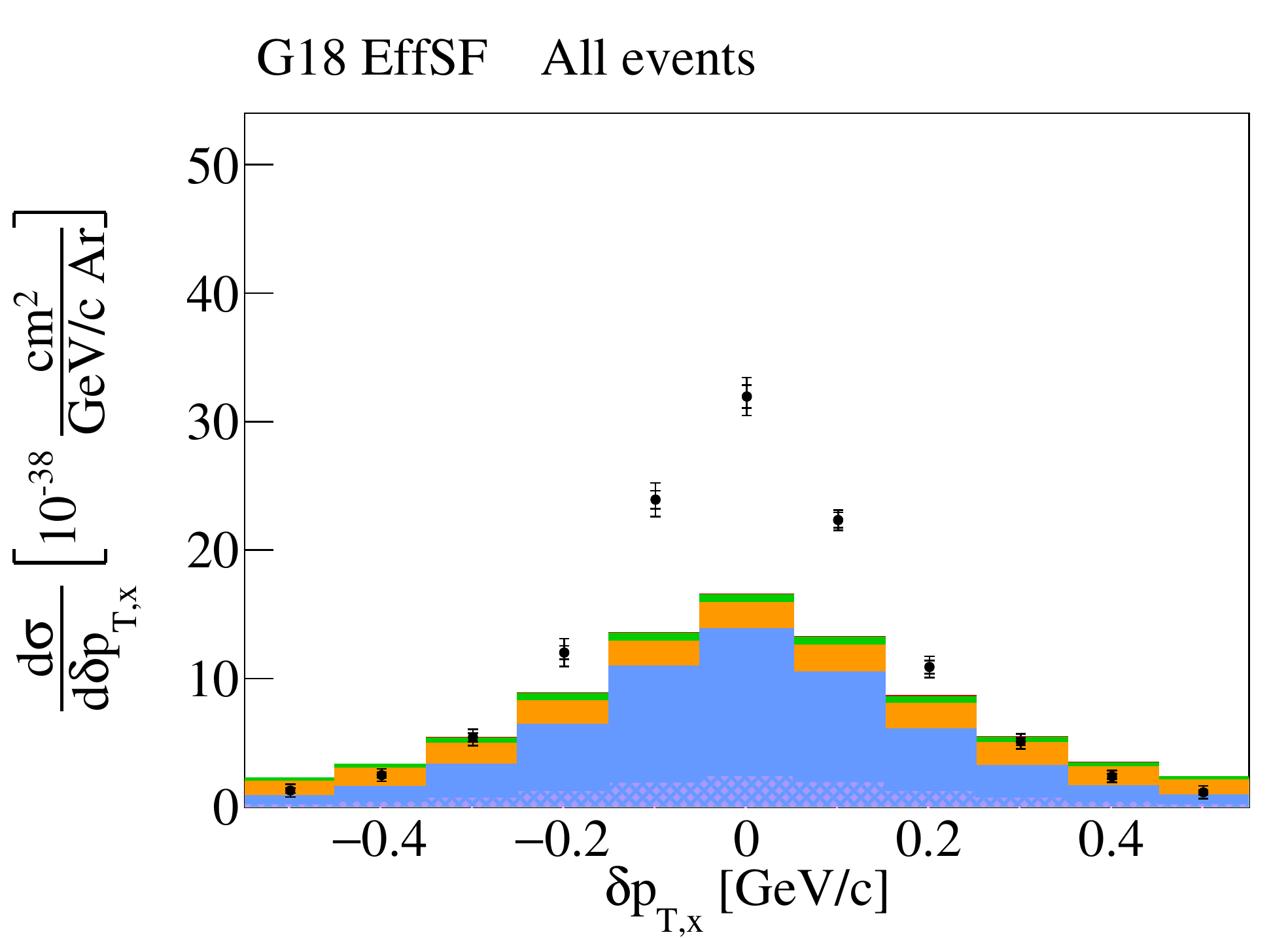}
\includegraphics[width=\BreakdownFigSize\linewidth]{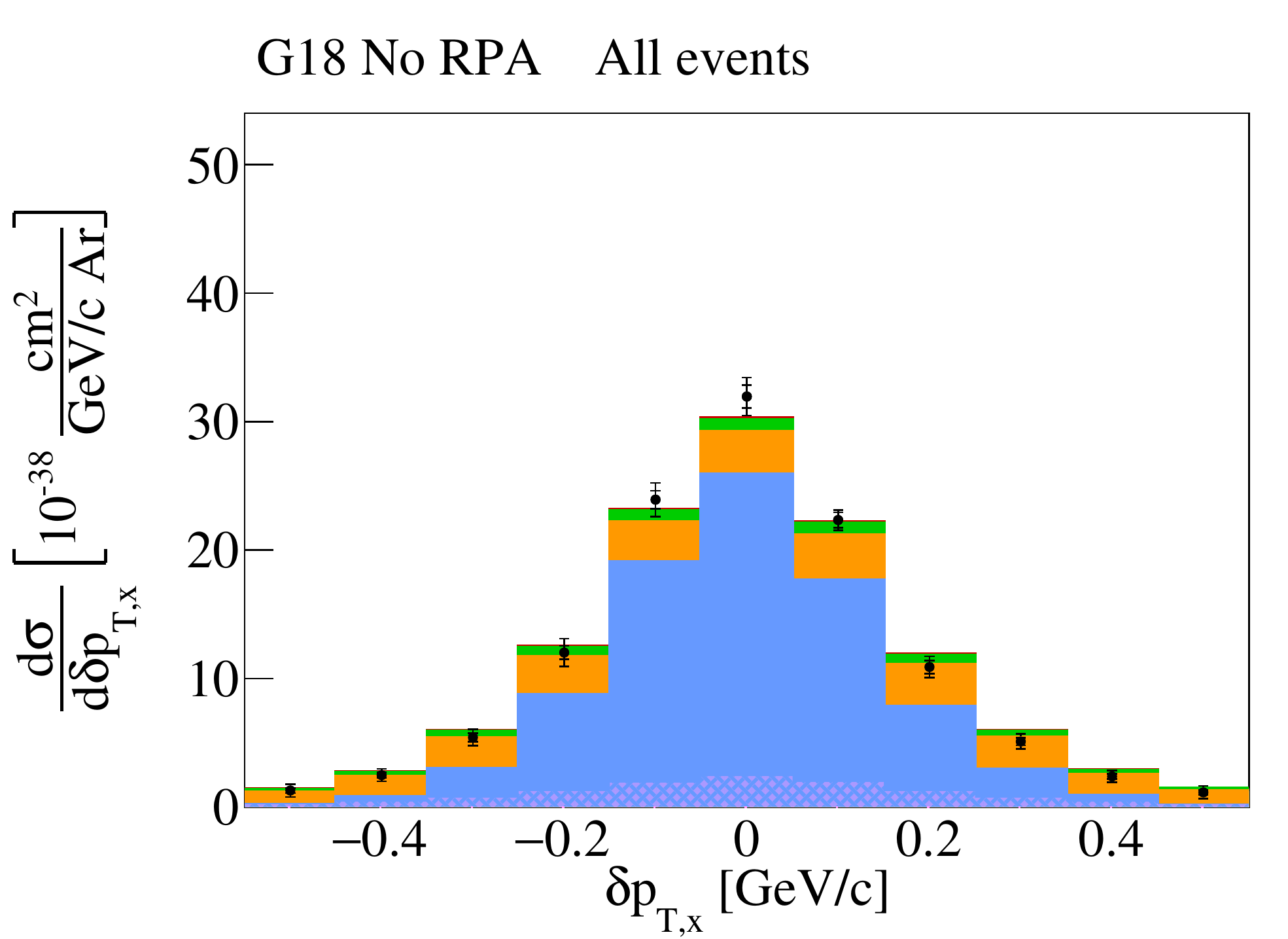}

\caption{Cross-section interaction breakdown for all the selected events. 
The breakdown is shown for (top left) the G18 LFG configuration, (top right) the G18 RFG configuration, (bottom left) the G18 EffSF configuration, and (bottom right) the G18 No RPA configuration.
}

\label{DeltaPtxBreakdown_All}
\end{figure}

\begin{figure}[H]
\centering 
\includegraphics[width=\BreakdownFigSize\linewidth]{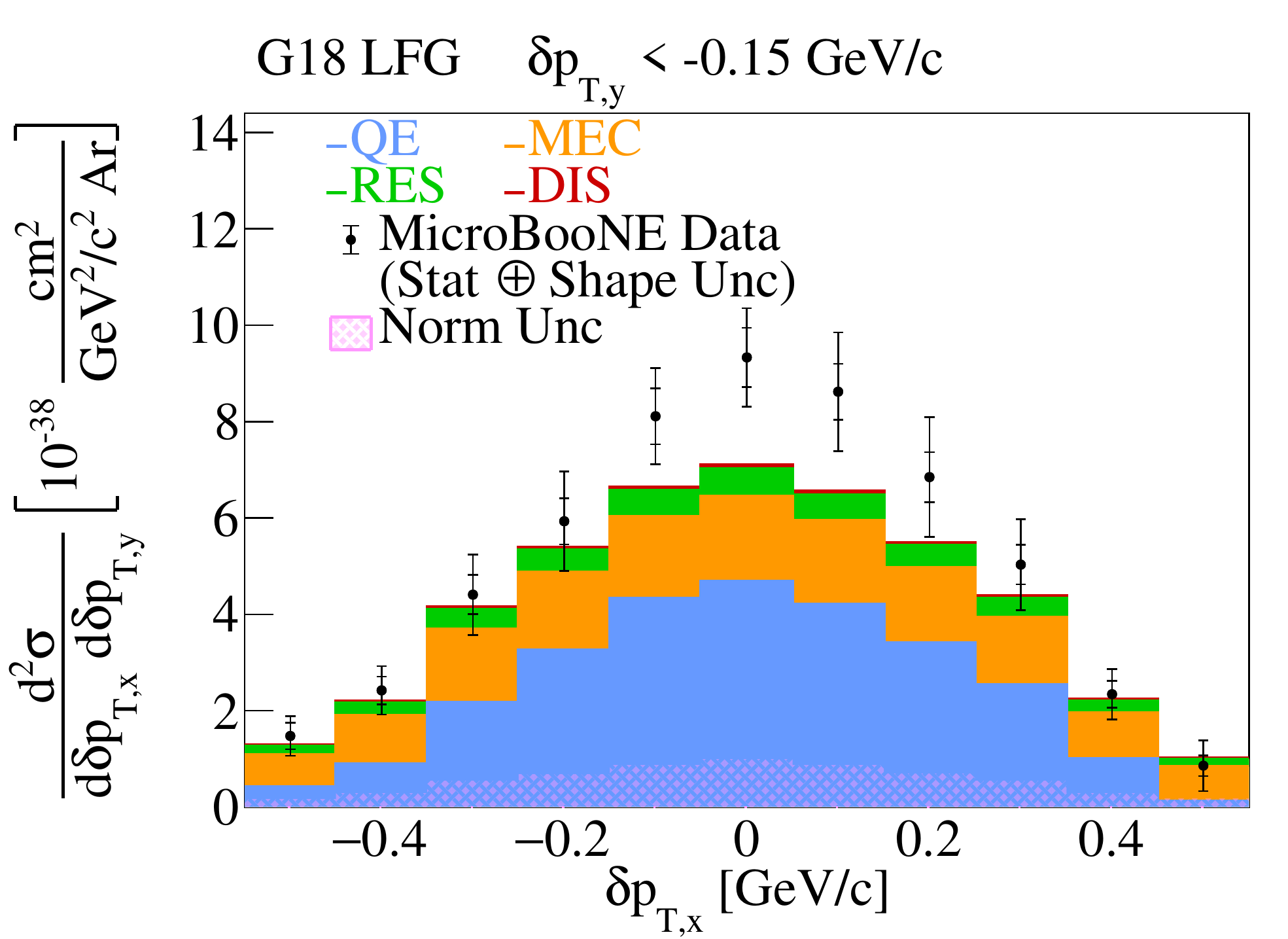}
\includegraphics[width=\BreakdownFigSize\linewidth]{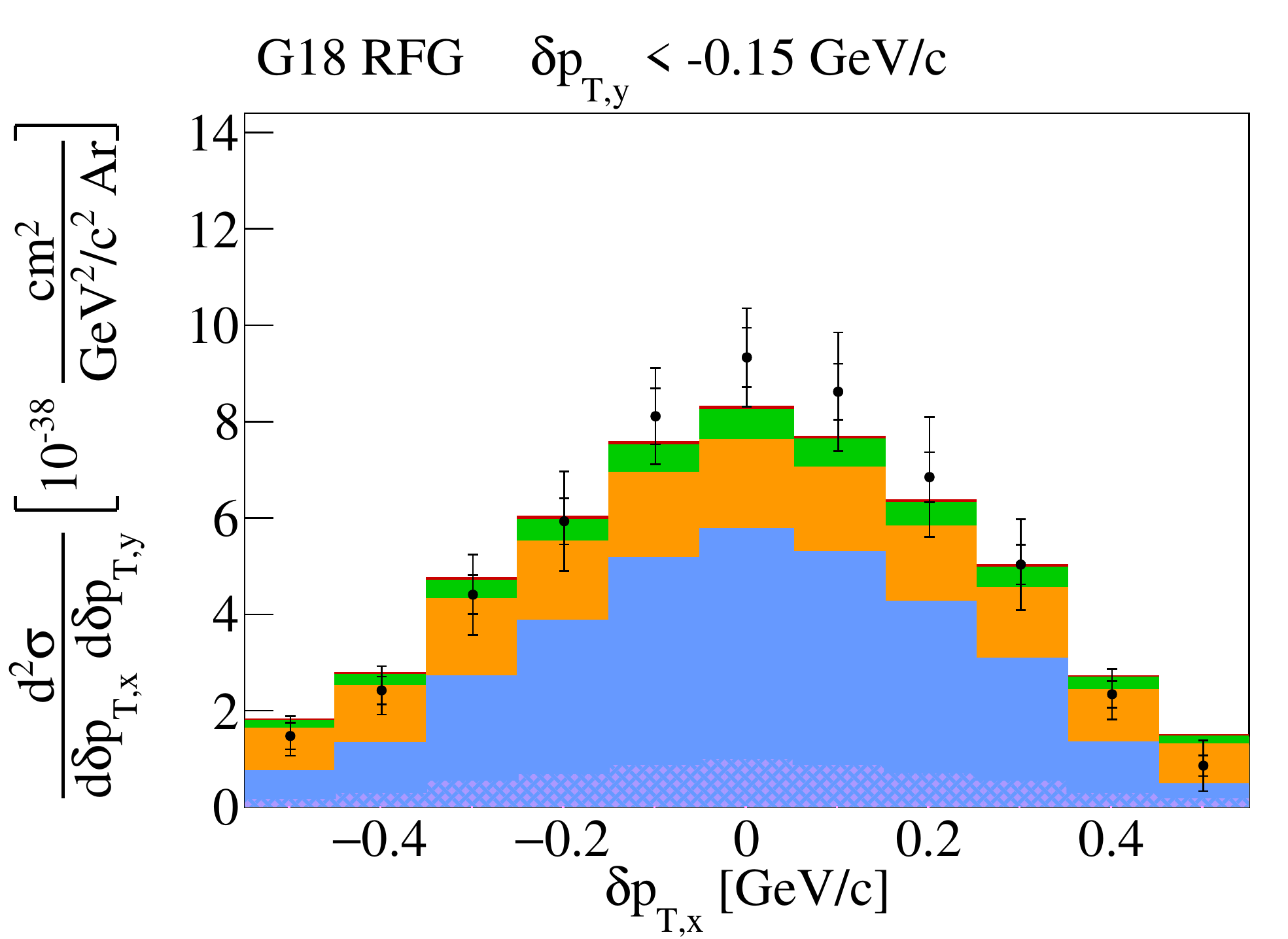}\\
\includegraphics[width=\BreakdownFigSize\linewidth]{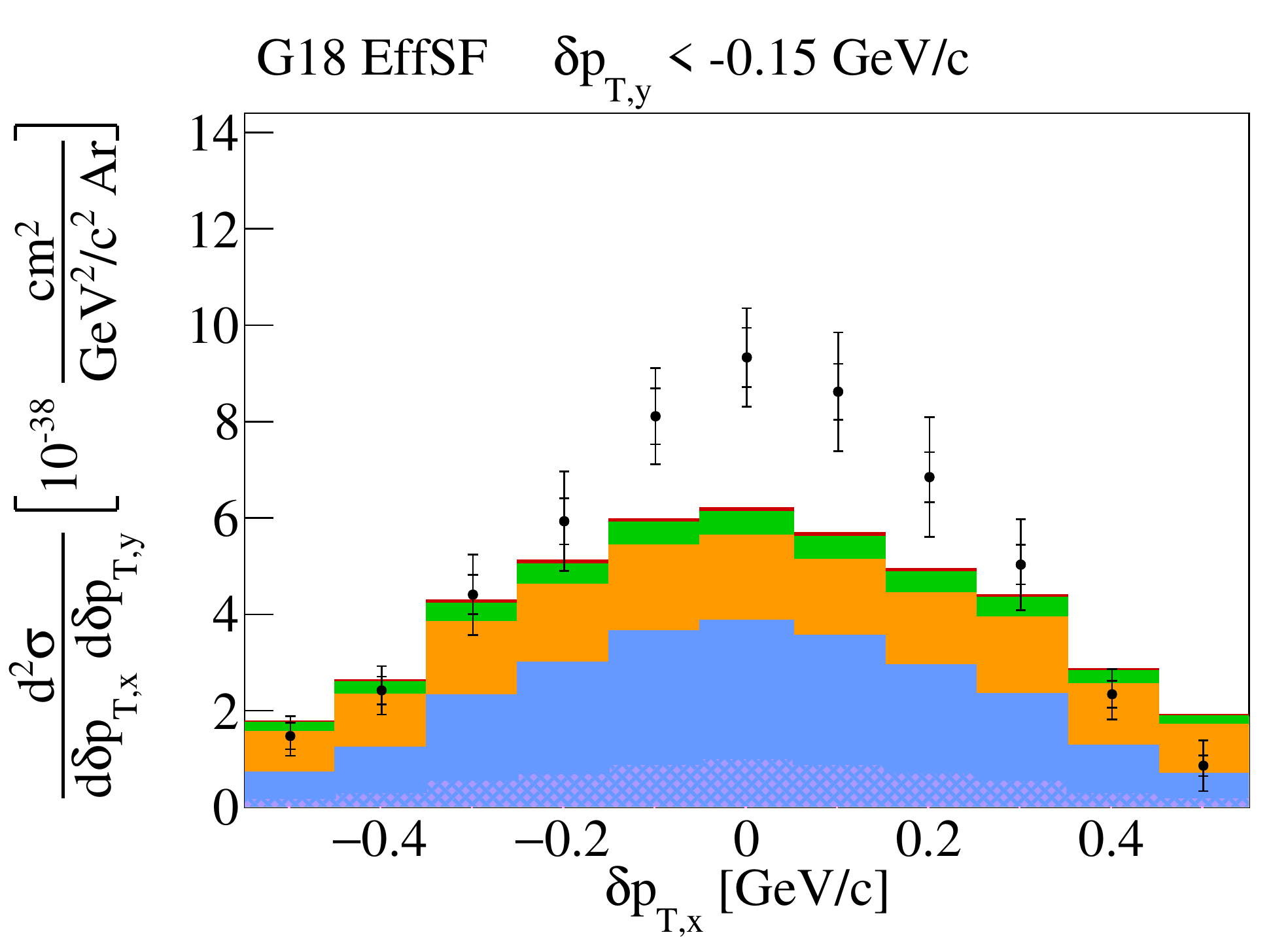}
\includegraphics[width=\BreakdownFigSize\linewidth]{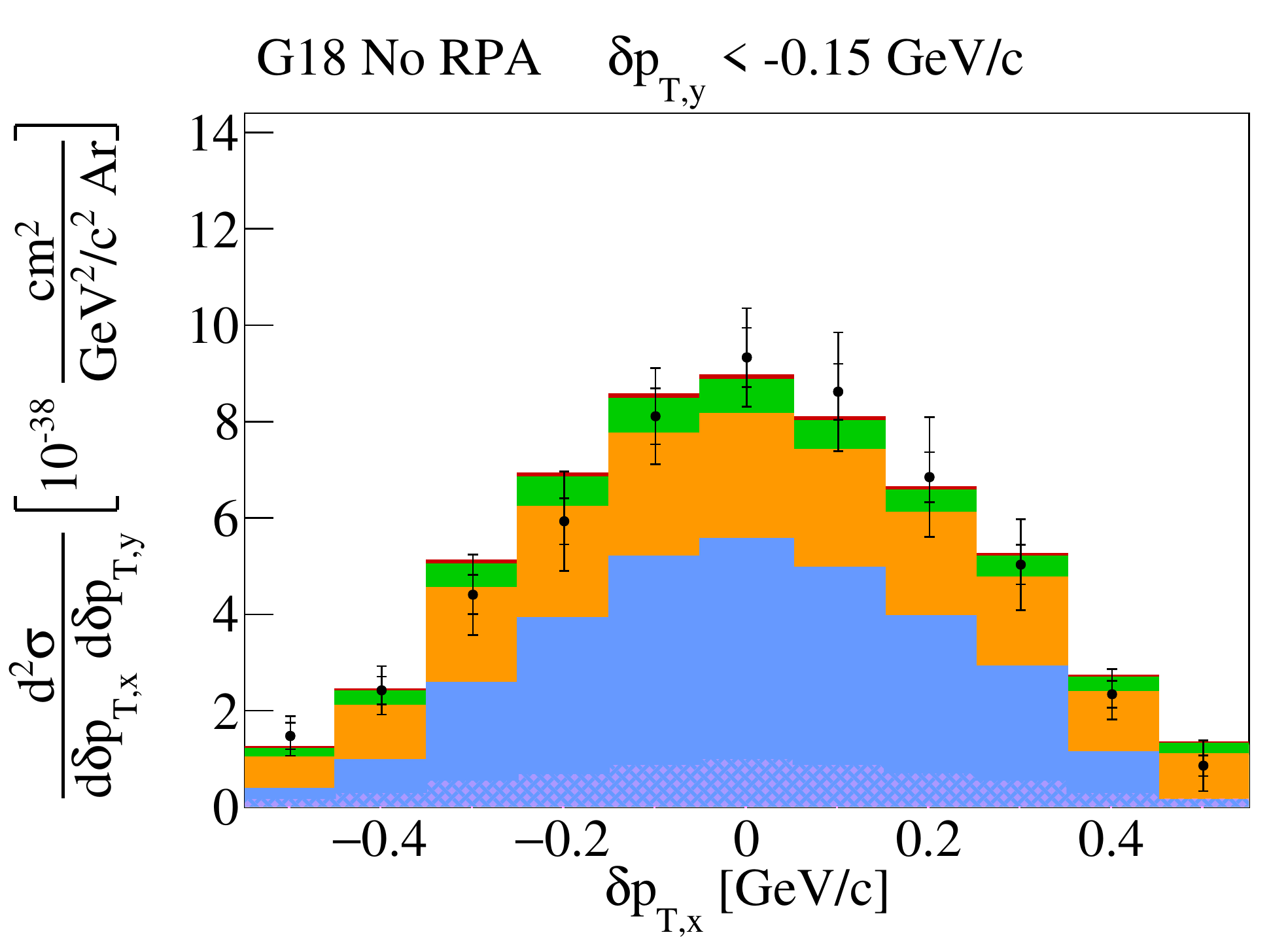}

\caption{Cross-section interaction breakdown for events with $\delta p_{T,y} <$ -0.15\,GeV/c. 
The breakdown is shown for (top left) the G18 LFG configuration, (top right) the G18 RFG configuration, (bottom left) the G18 EffSF configuration, and (bottom right) the G18 No RPA configuration.
}

\label{DeltaPtxBreakdown_Slice1}
\end{figure}

\begin{figure}[H]
\centering
\includegraphics[width=\BreakdownFigSize\linewidth]{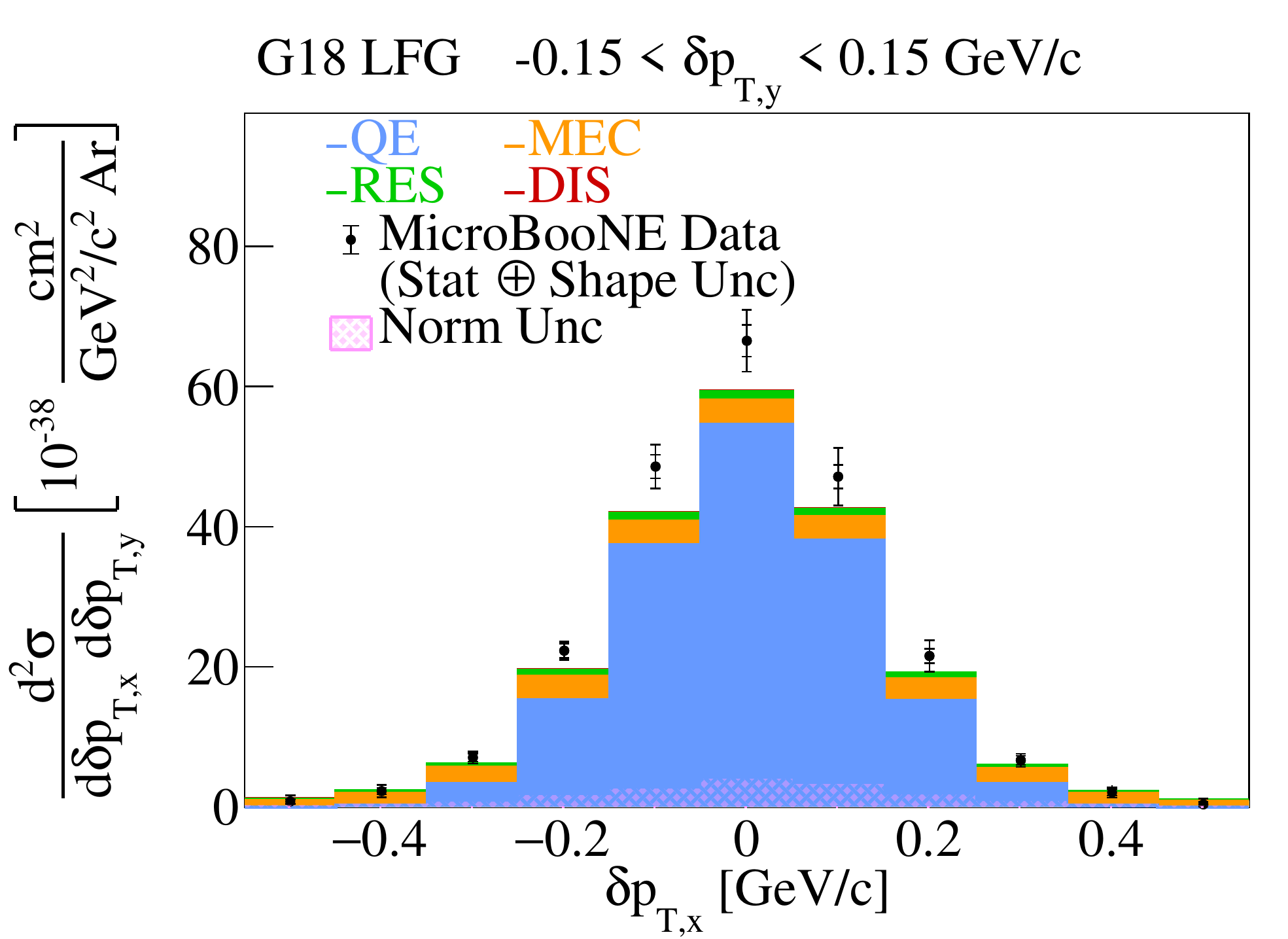}
\includegraphics[width=\BreakdownFigSize\linewidth]{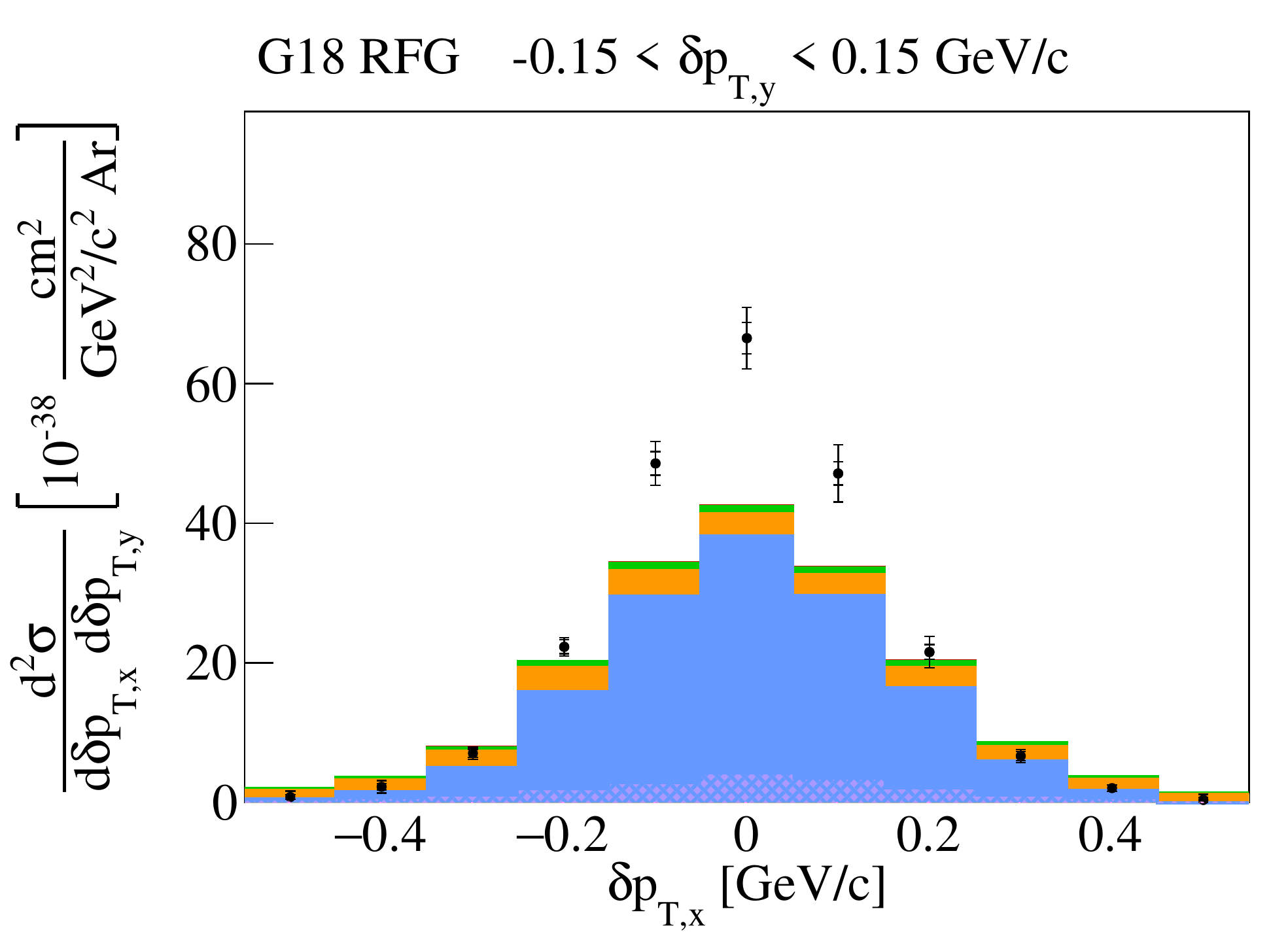}\\
\includegraphics[width=\BreakdownFigSize\linewidth]{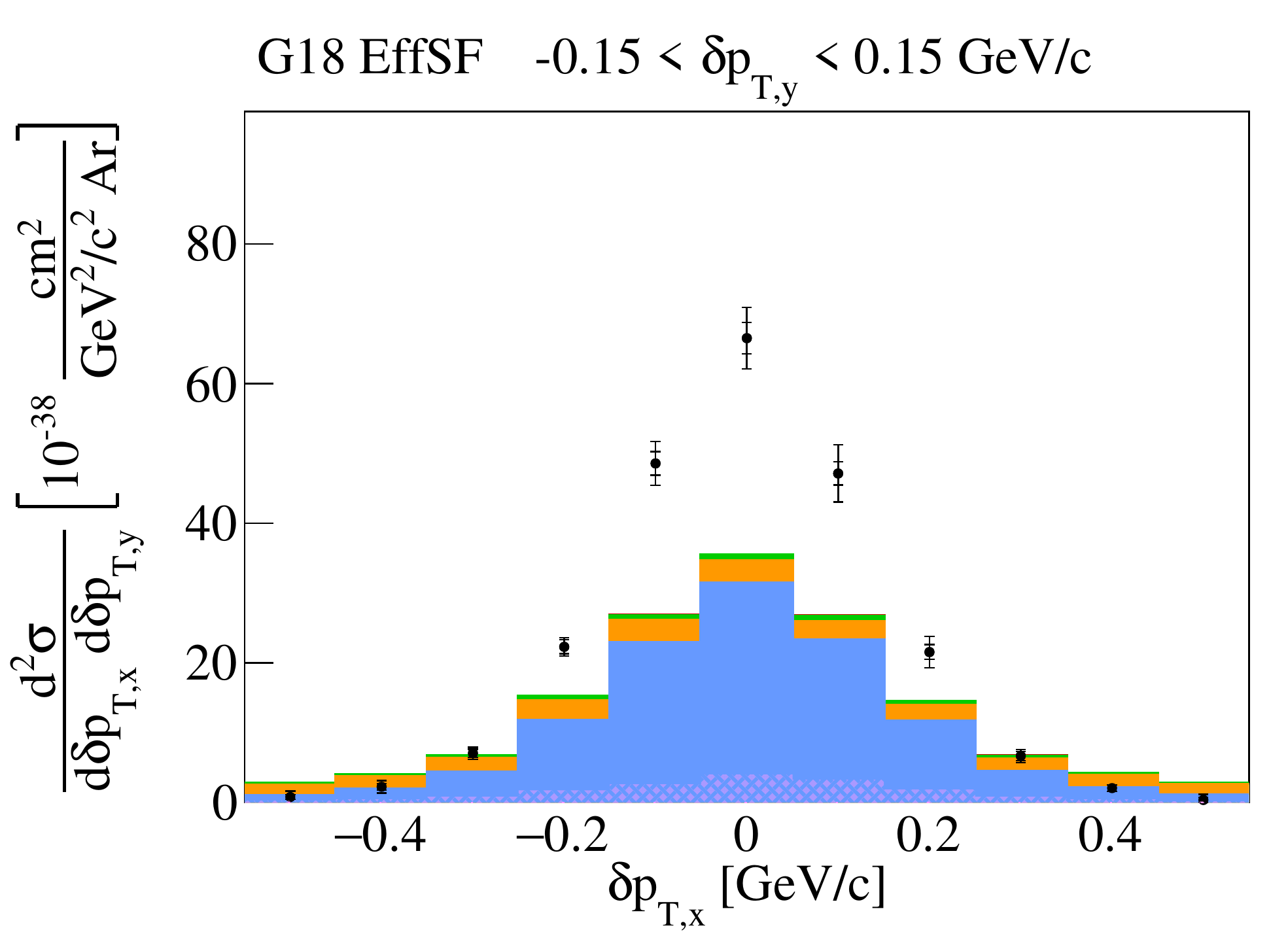}
\includegraphics[width=\BreakdownFigSize\linewidth]{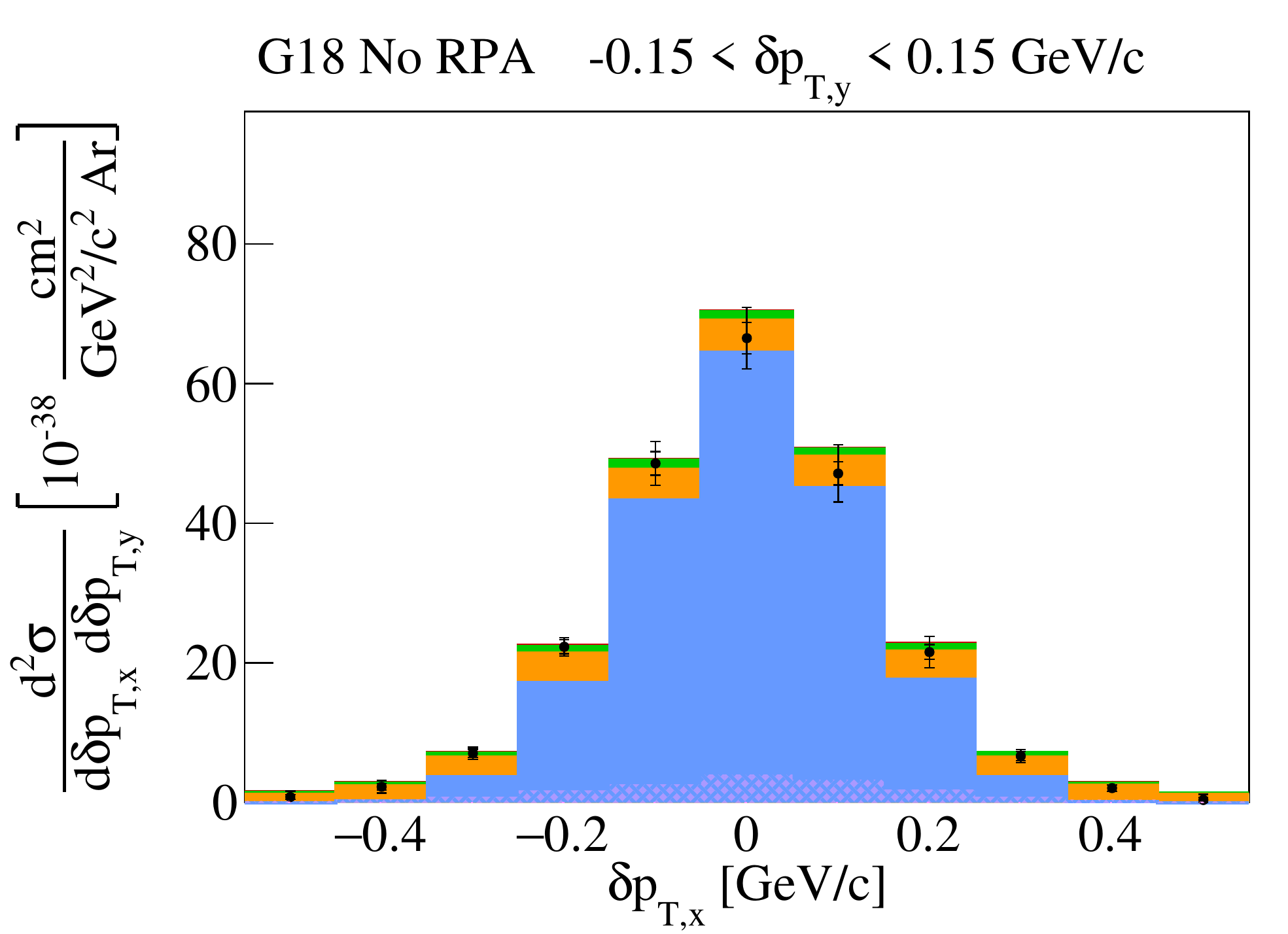}

\caption{Cross-section interaction breakdown for events with -0.15 $< \delta p_{T,y} <$ 0.15\,GeV/c. 
The breakdown is shown for (top left) the G18 LFG configuration, (top right) the G18 RFG configuration, (bottom left) the G18 EffSF configuration, and (bottom right) the G18 No RPA configuration.
}

\label{DeltaPtxBreakdown_Slice2}
\end{figure}

The $\chi^{2}$ per degree of freedom (d.o.f.) data comparison for each prediction shown on the results in figures~\ref{DeltaPTInDeltaAlphaT},~\ref{DeltaAlphaTInDeltaPT}, and~\ref{DeltaPtxInDeltaPty} takes into account the total covariance matrix including the off-diagonal elements.
Figures~\ref{DeltaPTBreakdown_All} - \ref{DeltaPtxBreakdown_Slice2} show in the interaction breakdown of the aforementioned results.

%%%%%%%%%%%%%%%%%%%%%%%%%%%%%%%%%%%%%%%%%%%%%%%%%%%%%%%%%%%%%%%%%%%%%%%%

\subsection{Kinematic Imbalance Cross-Section Analysis Conclusions}\label{KineImbConcl}

The first measurement of $\nu_\mu$ \CCIpOpi\ single and double differential cross sections on argon as a function of kinematic imbalance variables for event topologies with a single muon and a single proton detected in the final state was reported.
The unfolded data results were compared to a number of event generators, available model configurations and FSI modeling options.
This measurement identified regions of the phase-space which are ideal to provide constraints for nuclear and final state interaction effects in generator predictions essential for the extraction of oscillation  parameters.

%%%%%%%%%%%%%%%%%%%%%%%%%%%%%%%%%%%%%%%%%%%%%%%%%%%%%%%%%%%%%%%%%%%%%%%%

\section{Prospects With Future Neutrino Experiments}\label{NuProsp}

The MicroBooNE experiment is a crucial step in the understanding of the underlying neutrino-argon interactions that will be used to drastically reduce the uncertainties of forthcoming high precision neutrino oscillation measurements. 
Furthermore, MicroBooNE is the first LArTPC in a neutrino beam with automated event reconstruction and selection.

\begin{figure}[htb!]
\centering  
\includegraphics[width=0.7\linewidth]{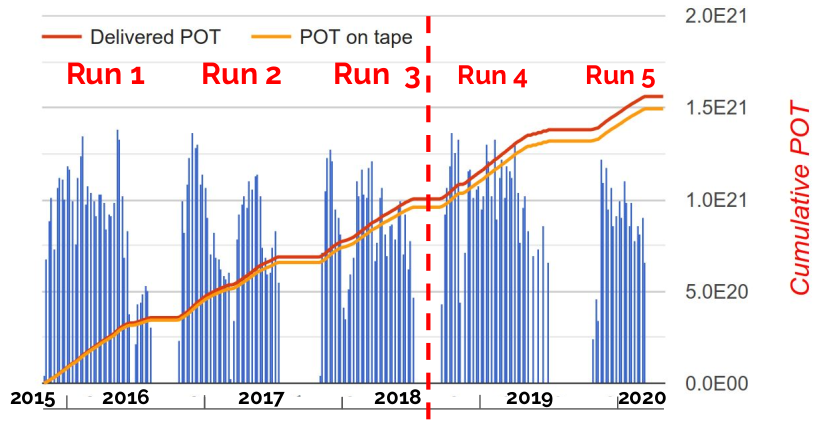}
\caption{MicroBooNE total Protons on Target (POT) collected with the Booster Neutrino Beam (BNB) during the five run periods. In this thesis, the first three run periods were used.}
\label{ubpot}
\end{figure}

This work paved the path towards precision $\nu_{\mu}$ CC cross section measurements with a single proton and no pions in the final state.
However, only three out of the five available run periods are used in the results presented in this thesis, as shown in figure~\ref{ubpot}.
Within the next year, runs 4 and 5 will also become available.
Therefore, the statistical uncertainties will be further reduced, and the path to further multidimensional analyses will be further explored. 

Beyond MicroBooNE, experiments of the Short-Baseline Neutrino (SBN) Program at Fermilab, namely ICARUS and SBND, have already turned (or will very soon turn) on.
The progress of SBN, which shares the same technology and beam as MicroBooNE, will largely benefit from the LArTPC expertise developed by the MicroBooNE collaboration.
With these experiments, the largest neutrino-argon scattering data sets will be collected and will be used to test the performance of our theory predictions in multiple variables simultaneously.

Despite the abundance of neutrino-argon cross sections that will be reported at the SBN program, the relevant energy range is significantly lower than the DUNE energy spectrum.
However, there is a wealth of cross-section opportunities that will become available once the ArgonCube demonstrator, a 2$\times$2 grid engineering prototype for the LArTPC DUNE  Near Detector (ND) module design, is commissioned in Fall 2022.
ArgonCube will provide sufficiently high statistics to measure cross sections and to evaluate the corresponding systematic uncertainties using both multi-particle channels and multidimensional analyses and will pave the path towards the final design of the DUNE ND.

All these neutrino experimental efforts will be complemented by the continuous benchmarking of the neutrino event generators predictions against external data sets, such as against electron scattering data sets.
A major step in this direction is made in chapter~\ref{genie} with the unification of the event generation process and of the modeling across the two particle species.
Furthermore, chapters~\ref{e2a} and~\ref{e4v} detail the analysis of electron scattering data sets from the CLAS detector at Thomas Jefferson Laboratory following neutrino data analysis methods and the testing against the performance of commonly used GENIE event generator.
\chapter{Inclusive Electron Scattering And The GENIE Event Generator\texorpdfstring{\newline}{ }\normalsize{[Phys. Rev. D 103, 113003 (2021)]}}\label{genie}

\section{Electron-Nucleus Modeling Development}\label{modeling} 

As already discussed in section~\ref{NuOsc}, the extraction of neutrino mixing parameters from neutrino oscillation experiments~\cite{DUNE,T2K,nova19} relies on comparing the energy-dependent neutrino event distribution for a particular neutrino flavor near the neutrino production point with the corresponding event distribution at a significant distance away. 
In practice, the yield at each neutrino energy is extracted from the measured neutrino-nucleus interactions in a detector, as reconstructed from the measured particles ejected in the neutrino-nucleus interaction.  
This requires detailed knowledge of the neutrino-nucleus interactions.

\begin{figure} [htb!]
\centering
\includegraphics[width=\textwidth]{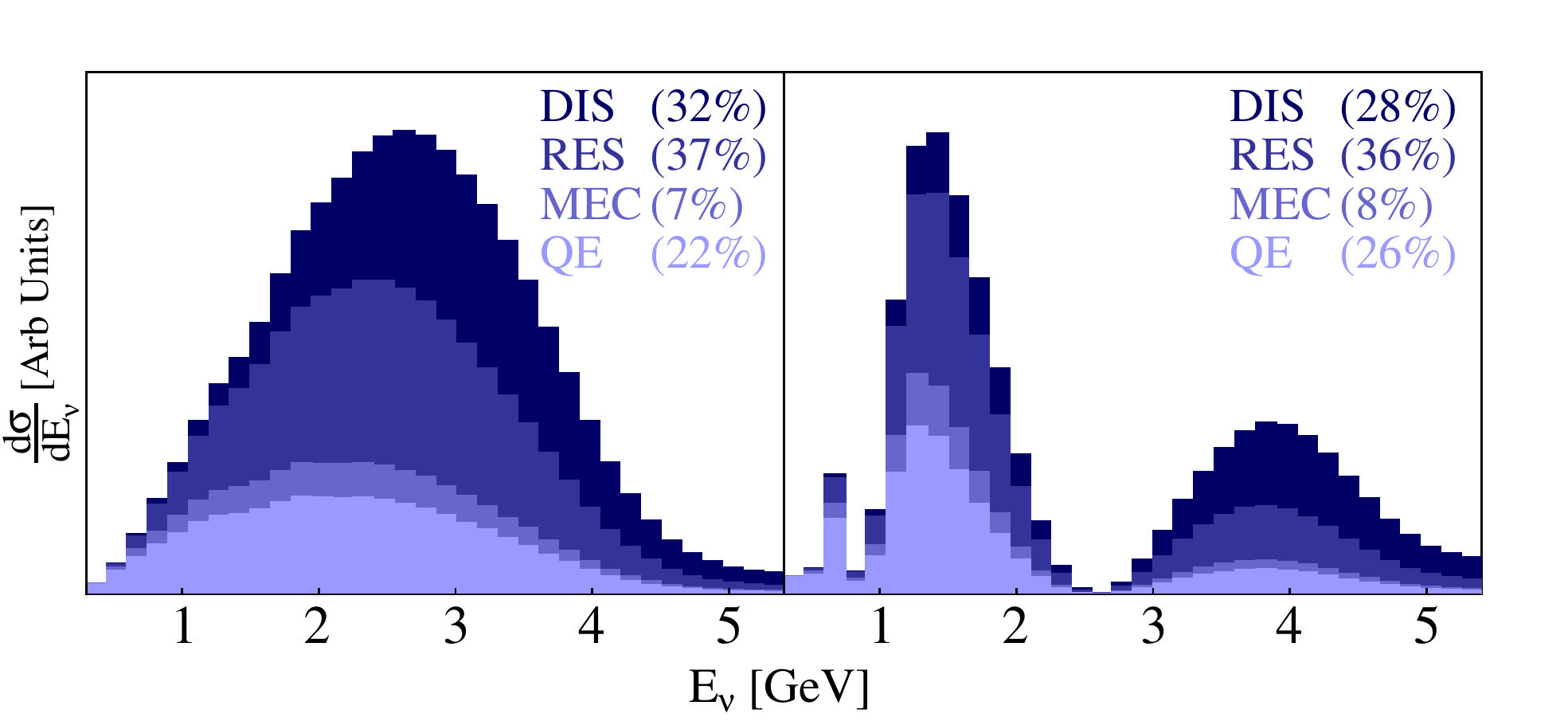}
\caption{\label{Fluxes}Charged-current cross sections as a function
  of neutrino energy obtained using GENIE for muon neutrino scattering
  using the DUNE near detector (left) and far detector (right) oscillated
  fluxes.  The shaded bands show the
  fractional contribution for each interaction mechanism,
  quasielastic scattering (QE), meson-exchange currents (MEC),
  resonance excitation (RES), and deep inelastic scattering (DIS).
  See text for details of the interaction mechanisms. The numbers in
  parentheses indicate the percentage of the cross section due to each
  interaction mechanism.}
\end{figure}

Unfortunately, measuring the neutrino-nucleus interaction is difficult due to the wide energy spread of accelerator-produced neutrino beams~\cite{DUNEFlux}, as can be seen in figure~\ref{Fluxes}, and the tiny neutrino-nucleus cross section.  
A relatively small body of data has been published~\cite{Alvarez-Ruso:2017oui}, which suffers from poor statistics and is flux-averaged over a wide range of neutrino energies. 
This data is then supplemented with theoretical models and implemented into event generator codes such as GENIE~\cite{Genie2010}, which is extensively used across US-based neutrino experiments, to simulate the neutrino-nucleus interactions across a wide range of energies and targets.
GENIE simulations are then used to aid in extraction of the incident neutrino flux as a function of energy from the neutrino-nucleus scattering events measured in neutrino detectors.
However, the theoretical models need to describe many different interaction
processes for medium to heavy nuclei (typically C, O, or Ar) where 
nuclear effects complicate the interactions.  
As a result, the uncertainties in the extraction of oscillation parameters are often dominated by the lack of knowledge of the neutrino-nucleus interactions~\cite{T2K,nova19}.

Figure~\ref{Fluxes} shows such a wide energy spectrum for the DUNE near detector flux-averaged cross sections (left) and the far detector oscillated flux-averaged cross sections (right) using one model configuration in GENIE.
All four neutrino-nucleus reaction mechanisms contribute significantly and all four need to be well understood.  
This is especially true because different reaction mechanisms contribute differently in the different oscillation peaks.  
Understanding one reaction mechanism better than the others could have significant implications for oscillation analyses.

\begin{figure}[htb!]
\centering
\includegraphics[width=0.7\textwidth]{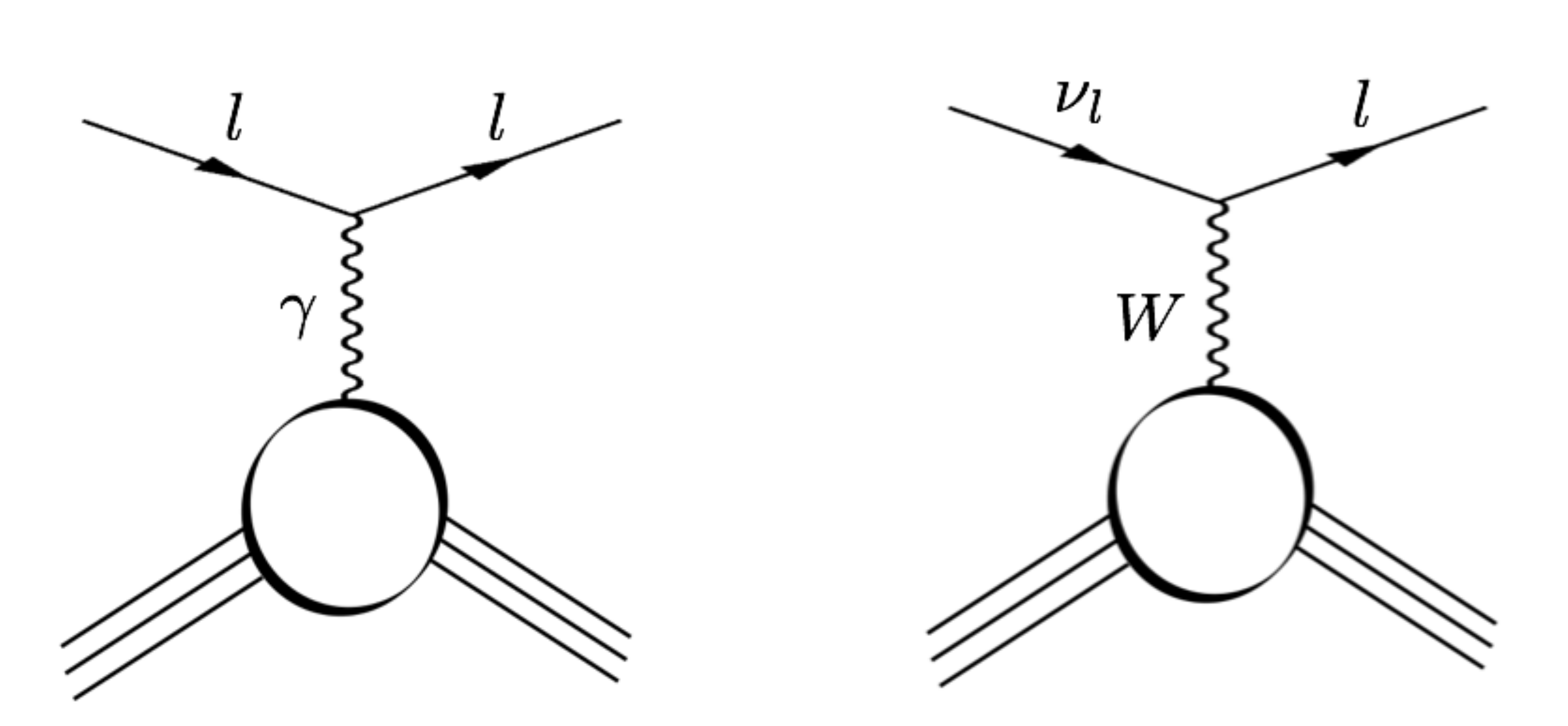}
\caption{(Left) electron-nucleus inclusive scattering via one-photon
  exchange and (right) charged current neutrino-nucleus inclusive
  scattering via $W$ exchange with a final state charged lepton.}
\label{fig:eAnuA}
\end{figure}

To improve our understanding of neutrino-nucleus interactions, we can take advantage of the fact that neutrinos and electrons are both leptons.
Thus, they interact with atomic nuclei in similar ways via the same reaction mechanisms, as illustrated in figure~\ref{fig:eAnuA} and detailed in sections~\ref{NuNucleusInt} and~\ref{LepXS}.  

\begin{figure}[htb!]
\centering
\includegraphics[height=3cm]{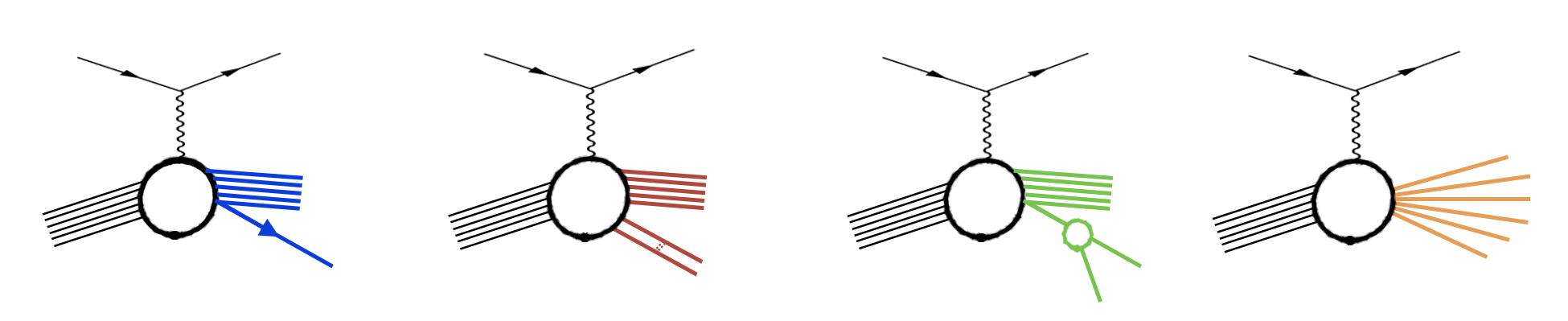}
\caption{Reaction mechanisms for lepton-nucleus scattering (a)
  quasielastic scattering (QE) where one nucleon is knocked out of the
  nucleus, (b) 2p2h where two nucleons are knocked out of the
  nucleus, (c) RES resonance production where a nucleon is excited to
  a resonance which decays to a nucleon plus meson(s), and (d) DIS where
 the lepton interacts with a quark in the nucleon.}
\label{fig:ReacMech}
\end{figure}

The most common lepton-nucleus interaction mechanisms are shown in figure~\ref{fig:ReacMech} include: (a) quasielastic (QE) scattering from individual moving nucleons in the nucleus; (b) two-nucleon knockout, due to interactions with a meson being exchanged between two nucleons referred to two-particle two-hole excitations, 2p2h or its major component, meson exchange currents (MEC);  (c) interactions which leave the struck nucleon in an excited resonance state (RES); and (d) nonresonant interactions with a quark within the nucleon (DIS).

All these interaction processes have to be described in detailed in event generators like GENIE.
GENIE started as an event generator that could exclusively handle neutrino interactions.  
In recognition of the importance of electron scattering, the latter was added as a new option in close conjunction with the neutrino scattering section.  
As much as possible, the neutrino cross section references vector and axial contributions separately and uses the same modeling for vector interactions as the electron section.  
Some models were developed separately for electrons and others were developed for both applications in tandem.

An earlier electron version of GENIE (v2.12.10) had already been tested by comparing with inclusive $(e,e')$ data~\cite{Ankowski:2020qbe}.  
Although the QE peak was well-described for a variety of energies and nuclei, the RES region was poorly described.  
However, the establishment of full compatibility between the electron and neutrino versions was then still in its early stages.

With this work, we significantly improved both neutrino and electron versions of GENIE to address these and other issues.  
We fixed significant errors in the previous version, including an error in the Mott cross section in the electron QE Rosenbluth interaction, a missing Lorentz boost in the MEC interaction affecting both electron and neutrino interactions, and incorrect electron couplings used in the RES interactions.  
We worked to better integrate the electron and neutrino codes for QE and MEC models.
We also added more up-to-date models such as SuSAv2~\cite{Amaro:2019zos}.  
These changes have been incorporated in the latest GENIE version (v3.2.0).  
We refer to the electron-scattering component of the widely-used GENIE~\cite{Genie2010} event generator as eGENIE.

\begin{figure}[htb!]
\centering
\includegraphics[width=0.7\textwidth]{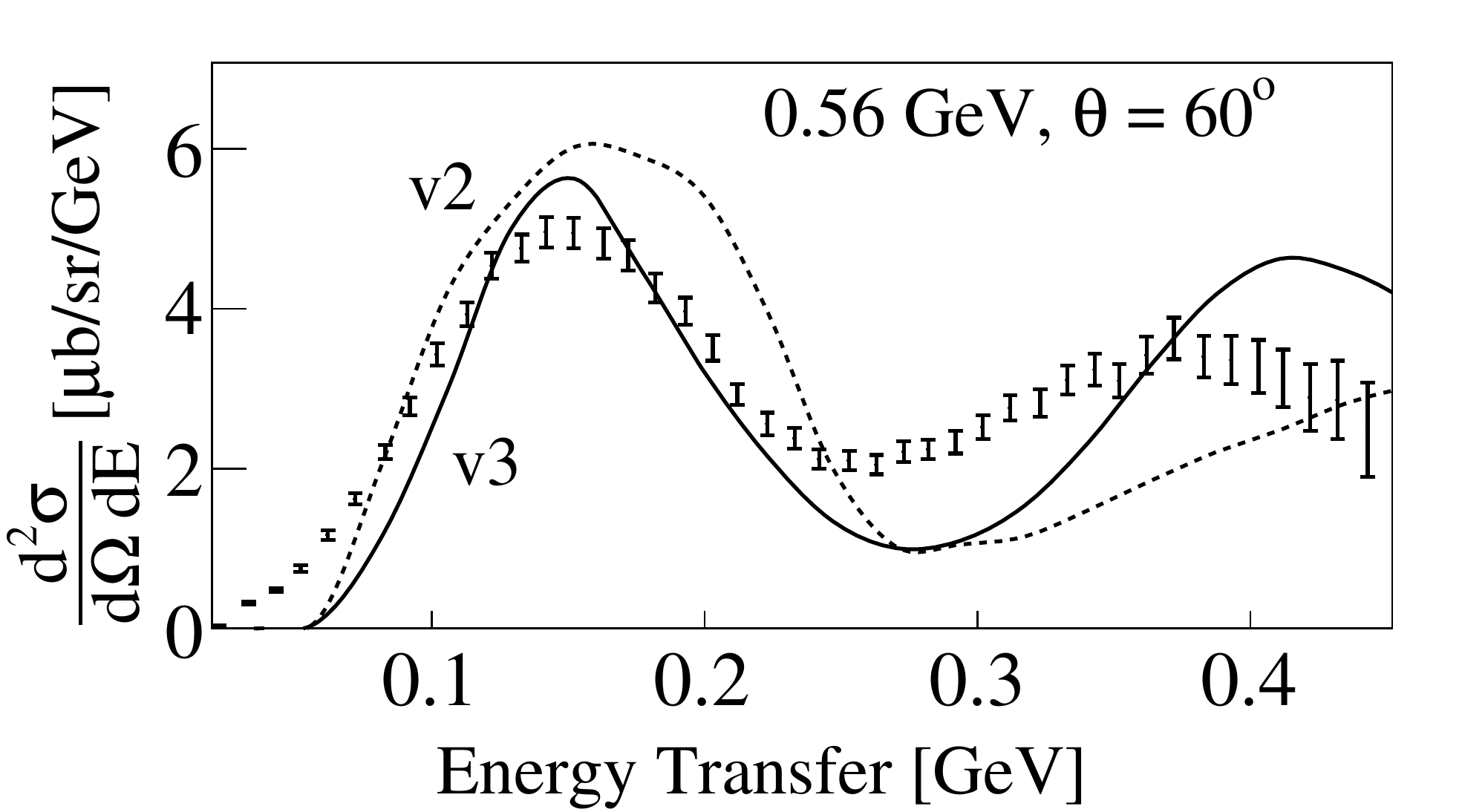}
\caption{Comparison between GENIE v2 and v3 descriptions of inclusive
  C$(e,e')$ scattering cross sections  at $E_0=0.56$ GeV, $\theta_e = 60^\circ$ and $Q^2_{QE}\approx
  0.24$ GeV$^2$.  Black points show the data, solid 
black line 
  shows the GENIE v3 results and dashed black line shows the GENIE
  v2 results.}
\label{fig:Improvements}
\end{figure}

The GENIE improvements can be seen in figure~\ref{fig:Improvements}~\cite{Barreau:1983ht}.  
The QE peak (at $\omega\approx$ 0.15\,GeV/c) predicted by the older GENIE v2 is too large and is slightly shifted to higher energy transfer than the data, while the first simulated resonance peak is at a larger energy transfer than the one observed in the data.  
The QE peak predicted by the updated GENIE v3 has about the correct integral and is at the correct energy transfer (but is slightly too narrow) and the first resonance peak is located at $m_\Delta - m \approx 300$ MeV beyond the QE peak, as expected.
Details of the calculations and of the discrepancies between GENIE v3 and the data are discussed in detail below.

In this analysis, we specifically focused on testing our knowledge of the electron-nucleus cross section by benchmarking eGENIE against existing inclusive electron scattering data for different target nuclei, beam energies and scattering angles.  
The goals are very similar to the ones in reference~\cite{Ankowski:2020qbe}, but we test a much more modern version of eGENIE and we also compare different models within eGENIE.
In addition, if eGENIE describes electron-nucleus scattering well, then it would be an improvement on the former empirical fit~\cite{Bosted:2012qc} and would be valuable for helping simulate a variety of electron experiments.

For fixed incident beam energy and scattered electron angle, the dominant process changes from QE at low energy transfer ($\omega\approx Q^2/2m$) through MEC to RES and to DIS at high energy transfer. 
Therefore, examining the agreement of eGENIE with data as a function of energy transfer can provide valuable insight into the specific shortcomings of the eGENIE models and their implementations.  
This separation according to the underlying physics interactions gives valuable insights which are not presently possible with neutrino cross sections, because only broad-energy beams are available.

The GENIE simulation framework offers several models of the nuclear ground state, multiple models for each of the electron- or neutrino-nucleus scattering mechanisms accompanied by various tunable model parameters, and a number of models for hadronic final state interactions (FSI), i.e., intranuclear rescattering of the outgoing hadrons~\cite{Dytman:2021ohr,Genie2010,Andreopoulos:2015wxa}.  
We describe the different models relevant for this work and the electron-specific effects that we accounted for during the eGENIE development below~\cite{PhysRevD.103.113003}.

Since our goal is to use electron scattering data to validate neutrino interaction modeling in GENIE, the GENIE code for electron and neutrino interactions are unified in many places.
The neutrino interacts with a nucleus via the weak interaction and massive $W$ or $Z$ exchange, whereas the electron interacts mostly electromagnetically via massless photon exchange, as shown in figure~\ref{fig:eAnuA}. 
This causes the cross sections to differ by an overall factor of

\begin{equation}
  \frac{8\pi^2\alpha^2}{G_F^2}\frac{1}{Q^4}
  \label{eq:factor}
\end{equation}

when equations~\ref{eq:sigmaeAPrime} and~\ref{eq:sigmaNuAPrime} are compared.
In the code, both interactions use the same nuclear ground state and many of the nuclear reaction effects, such as FSI, are very similar or identical.
Except for mass effects and form factors, the electron nucleus cross section can be obtained by setting the axial part of the interaction to zero.  
We also accounted for isoscalar and isovector terms appropriately.

Many of the models reported in this work, except for SuSAv2, use the GENIE implementation of the Local Fermi gas (LFG) model to describe the nuclear ground state. 
In the simplest Fermi gas model, nucleons occupy all momentum states up to the global Fermi momentum $k_F$ with equal probability.  
In the LFG model, the Fermi momentum at a given radial position depends on the local nuclear density obtained from measurements of nuclear charge densities. 
To account for this radial dependence, GENIE selects an initial momentum for the struck nucleon by first sampling an interaction location $r$ inside the nucleus according to the nuclear density. 
The nucleon momentum is then drawn from a Fermi distribution using the local Fermi momentum $k_F(r)$.

Another commonly used nuclear model is the Relativistic Fermi Gas (RFG). 
Here a global momentum distribution is used for the entire nucleus, independent of the interaction location in the nucleus. 
However, a high-momentum tail of nucleons with momenta above the Fermi-momentum is included. 
This tail is meant to approximately account for the effects of two-nucleon short-range correlations~\cite{hen15b, Hen:2016kwk} and follows a $1/k^4$ distribution, where $k$ is the nucleon momentum.

We consider two distinct sets of eGENIE configurations: 

\begin{itemize}
\item {\bf G2018}, which uses the LFG nuclear model, the Rosenbluth cross section for QE scattering, and the empirical MEC model~\cite{Katori:2013eoa}. 
This model set is formally marked as the \texttt{G18\_10a\_02\_11a} configuration of GENIE v3.

\item {\bf GSuSAv2}~\cite{PhysRevD.101.033003}, which follows the universal SuSAv2 super-scaling approach to lepton scattering.
This new model set is included in the latest GENIE v3.2.0 release as the \texttt{GEM21\_11b\_00\_000} configuration for electron scattering and \texttt{G21\_11b\_00\_000} for neutrino interactions. 
\end{itemize}

In both model sets, RES is modeled using the Berger-Sehgal model~\cite{Berger:2007rq} and DIS reactions are modeled using Bodek and Yang\cite{Bodek2003}.  
The models are described in more detail below.

In QE interactions, a lepton scatters on a single nucleon, removing it from the spectator $A-1$ nucleus unless final-state interactions lead to reabsorption.
The electron QE interaction in the G2018 configuration of GENIE uses the Rosenbluth cross section with the vector structure function parametrization of reference~\cite{Bradford:2006yz}.  
We corrected the implementation of this model for eGENIE and modified
the cross section to account for the identified issues. 
This electron QE cross section differs in important ways from the Valencia CCQE model \cite{ValenciaModel} used in the G2018 configuration for neutrinos.
Most notably, the Rosenbluth treatment lacks medium polarization corrections.

A new QE model in GENIE, based on the SuSAv2 approach \cite{Megias:2016lke,PhysRevD.101.033003,Megias:2016fjk}, uses superscaling to write the inclusive cross section in terms of a universal function independent of momentum transfer and nucleus. 
For EM scattering, the scaling function may be expressed in the form

\begin{eqnarray}
f(\psi')=k_F\frac{\frac{d^2\sigma}{d\Omega_e
    d\nu}}{\sigma_{Mott}(v_LG_L^{ee'}+V_TG_T^{ee'})} \,,
\end{eqnarray}

where $\psi'$ is a dimensionless scaling variable, $k_F$ is the nuclear Fermi momentum, the denominator is the single-nucleon elastic cross section, $v_L$ and $v_T$ are known functions of kinematic variables, and $G_L^{ee'}(q,\omega)$ and $G_T^{ee'}(q,\omega)$ are the longitudinal and transverse nucleon structure functions linearly related  to $F_1^e$ and $F_2^e$~\cite{Caballero:2006wi}. 
For eGENIE, we extended the original neutrino implementation \cite{PhysRevD.101.033003} to the electron case.

The original SuSAv2 QE cross section calculations used a Relativistic Mean Field (RMF) model of the nuclear ground state~\cite{Gonzalez-Jimenez:2019qhq,Gonzalez-Jimenez:2019ejf}. 
This approach includes the effects of the real part of the nucleon-nucleus potential on the outgoing nucleons which creates a ``distorted'' nucleon momentum distribution.

Although GENIE lacks the option to use an RMF nuclear model directly, we achieve approximate consistency with the RMF-based results by using a two-step strategy for QE event generation. 
First, an energy and scattering angle for the outgoing lepton are sampled according to the inclusive double-differential cross section. 
This cross section is computed by interpolating precomputed values of the nuclear responses $G_L^{ee'}(q,\omega)$ and $G_L^{ee'}(q,\omega)$ which are tabulated on a two-dimensional grid in $(q, \omega)$ space. 
The responses were obtained using the original RMF-based SuSAv2 calculation.
Second, the outgoing nucleon kinematics are determined by choosing its initial momentum from an LFG distribution. 
The default nucleon binding energy used in GENIE for the LFG model is replaced for SuSAv2 with an effective value tuned to most closely duplicate the RMF distribution. 
The outgoing nucleon kinematics are not needed for the comparisons to inclusive $(e,e')$ data shown in this work.

MEC describes an interaction that results in the ejection of two nucleons from the nucleus, thus is often referred to as 2p2h.  
It typically proceeds via lepton interaction with a pion being exchanged between two nucleons or by interaction with a nucleon in an Short Range Correlated (SRC) pair.  
MEC is far less understood than other reaction mechanisms because, unlike the others, it involves scattering from two nucleons simultaneously. 
GENIE has several models for MEC.

The G2018 configuration of eGENIE uses the empirical MEC model~\cite{Katori:2013eoa} that is useable for both electron- and neutrino-nucleus scattering. 
It assumes that the MEC peak for inclusive scattering has a Gaussian distribution in $W$ and is located between the QE and first RES peaks. 
Although both versions of the model use the same effective form factors, the amplitude of the MEC peak was tuned separately to electron and neutrino scattering data. 
This model was developed in the context of empirically fitting GENIE to MiniBooNE inclusive neutrino scattering data and is still used for neutral-current interactions~\cite{Katori:2013eoa}.
For charged-current neutrino interactions, $\nu$GENIE G2018 uses the very different Valencia 2p2h model~\cite{ValenciaModel,GENIEValenciaMEC} instead of the empirical model.

For the description of the 2p2h MEC contributions, the SuSAv2 model uses the fully relativistic calculations from~\cite{Simo_2017}.
This treatment allows for a proper separation of neutron-proton and proton-proton pairs in the final state via the analysis of the direct-exchange interference terms~\cite{RUIZSIMO2016124}. 
This approach is capable of reproducing the nuclear dynamics and superscaling properties observed in inclusive electron-nucleus scattering reactions~\cite{PhysRevLett.82.3212,PhysRevC.60.065502,PhysRevC.99.042501}.
The latter serves as a robust test for nuclear models.
It further provides an accurate description of existing neutrino data~\cite{PhysRevD.94.013012,PhysRevD.94.093004,PhysRevC.99.042501,Megias_2018,PhysRevD.99.113002}. 
As in the case for the SuSAv2 QE model, we extended the original GENIE implementation of SuSAv2 MEC for neutrinos to the electron case for eGENIE~\cite{PhysRevD.101.033003,Megias:2014qva,Amaro:2017eah,Megias:2016lke}.

The SuSAv2 MEC approach is the only fully relativistic model that can be extended without approximations to the full-energy range of interest for neutrino scattering events.
Therefore, it is a very promising modeling choice for present and future neutrino experiments for one of the least understood interaction channels. 

RES production in GENIE is simulated using the Berger-Sehgal model~\cite{Berger:2007rq}, in which the lepton interacts with a single moving nucleon and excites it to one of 16 resonances.  
The cross sections are calculated based on the Feynman-Kislinger-Ravndal (FKR) model~\cite{fkr1971}, without any interferences between them.
Form factors are derived separately for vector and axial probes~\cite{Rein:1980wg} but have not been updated to include recent electron scattering results.

The GENIE treatment of DIS used in this work is based on that of Bodek and Yang~\cite{Bodek2003}. 
Hadronization is modeled using an approach which transitions gradually as a function of the hadronic invariant mass $W$ between the AGKY model~\cite{Yang2009} and the PYTHIA 6 model~\cite{PYTHIA6}. 
At low $W$ values, the Bodek-Yang differential cross section is scaled by tunable parameters that depend on the multiplicity of hadrons in the final-state~\cite{Andreopoulos:2015wxa}. 

Integration of the RES and DIS contributions is complicated by the need for a model of nonresonant meson production. 
There is no definite separation of RES and DIS contributions.  
GENIE makes a sharp cutoff at $W=1.93$ GeV in the latest tune and uses a suppression factor to enable usage of the Bodek-Yang cross section at low $W$ in place of a true nonresonant model.
These features were recently retuned by the GENIE collaboration using measurements of charged-current $\nu_\mu$ and $\bar{\nu}_\mu$ scattering on deuterium~\cite{geniecollaboration2021neutrinonucleon}. 
The W cutoff and suppression factors apply to both electron- and neutrino-nucleus models.

Final state interactions of outgoing hadrons with the residual nuclei are calculated in eGENIE using the INTRANUKE~\cite{Dytman:2021ohr,dytman2011fsi} package and one of two options. 
The first, hA, an empirical data-driven method, uses the cross-section of pions and nucleons with {\it nuclei} as a function of energy up to 1.2 GeV and the CEM03~\cite{mashnik2006fsi} calculation for higher energies.  
The second, hN, is a full intra-nuclear cascade calculation of the interactions of pions, kaons, photons, and nucleons with nuclei.  
In the hN model, each outgoing particle can interact successively with any or all the nucleons it encounters on its path leaving the nucleus, and any particles created in those interactions can also subsequently reinteract.  
The ability of the two models to describe hadron-nucleus data is very similar.
The eGENIE G2018 configuration uses the hA FSI model, while GSuSAv2 uses hN. 
However, the choice of FSI model has no effect on the inclusive cross sections considered in the present work.

%%%%%%%%%%%%%%%%%%%%%%%%%%%%%%%%%%%%%%%%%%%%%%%%%%%%%%%%%%%%

\section{Inclusive Electron Scattering Data Comparisons}\label{comparison}

To test eGENIE, we compared inclusive electron scattering data to theoretical predictions made using two different program configurations which differ in their choice of QE and MEC models.
Namely, G2018 adopts the Rosenbluth model for QE interactions and the empirical Dytman model for MEC events, while GSuSAv2 uses SuSAv2 for both QE and MEC interactions.

\begin{figure}[htb!]
\centering
\includegraphics[width=\textwidth]{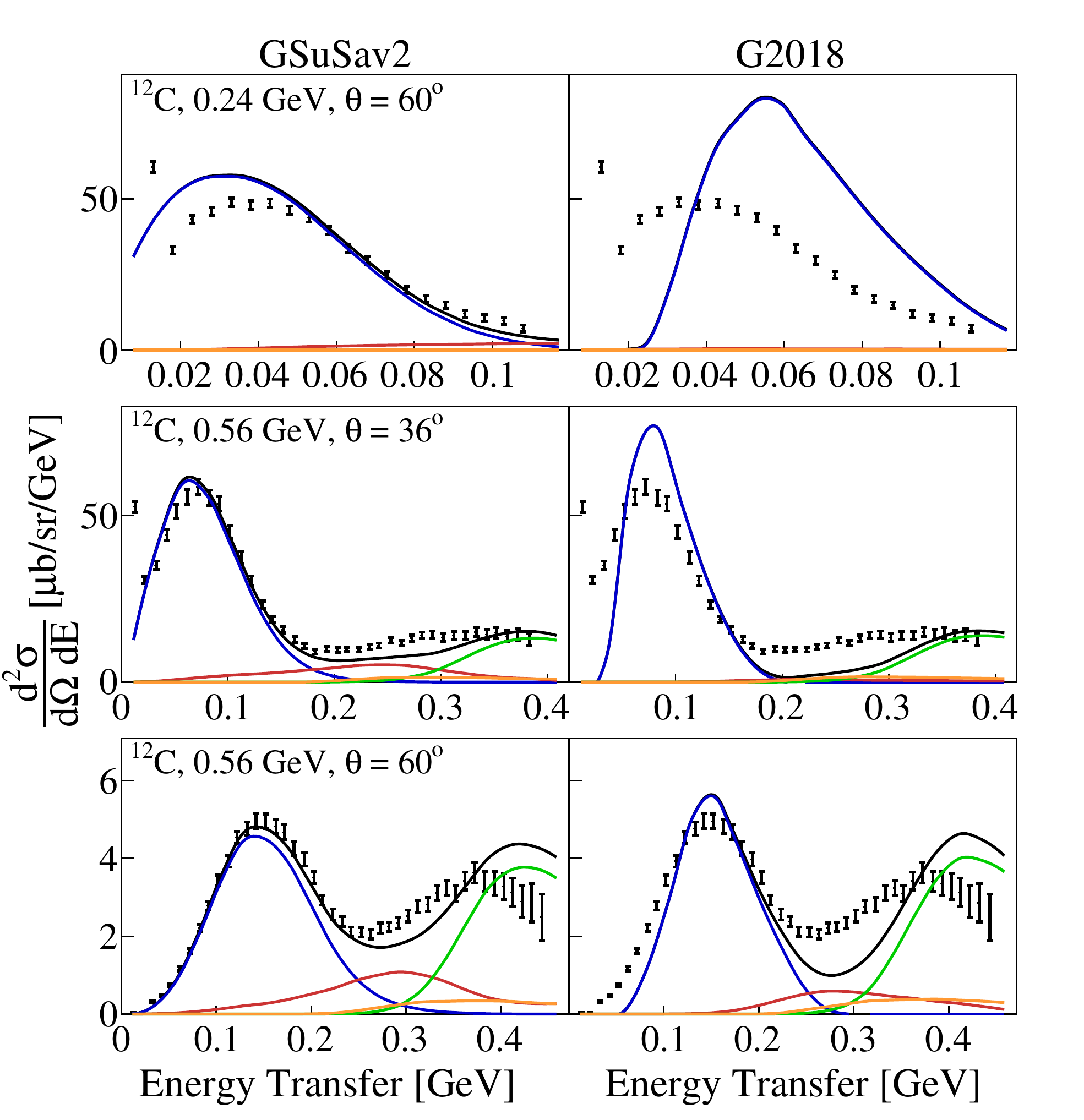}
\caption{Comparison of inclusive C$(e,e')$ scattering cross sections for data
  and for GENIE.  (left) data vs GSuSAv2 and (right) data vs G2018.
  (top) $E_0=0.24$ GeV, $\theta_e = 60^\circ$ and $Q^2_{QE}\approx
  0.05$ GeV$^2$ (middle) $E_0=0.56$ GeV, $\theta_e = 36^\circ$ and $Q^2_{QE}\approx
  0.11$ GeV$^2$, and (bottom) $E_0=0.56$ GeV, $\theta_e = 60^\circ$ and $Q^2_{QE}\approx
  0.24$ GeV$^2$.  Black points show the data, solid 
black lines
  show the total GENIE prediction, colored lines show the contribution
of the different reaction mechanisms: (blue) QE, (red) MEC, (green)
RES and (orange) DIS.}
\label{fig:C1}
\end{figure}

\begin{figure}[htb!]
\centering
\includegraphics[width=\textwidth]{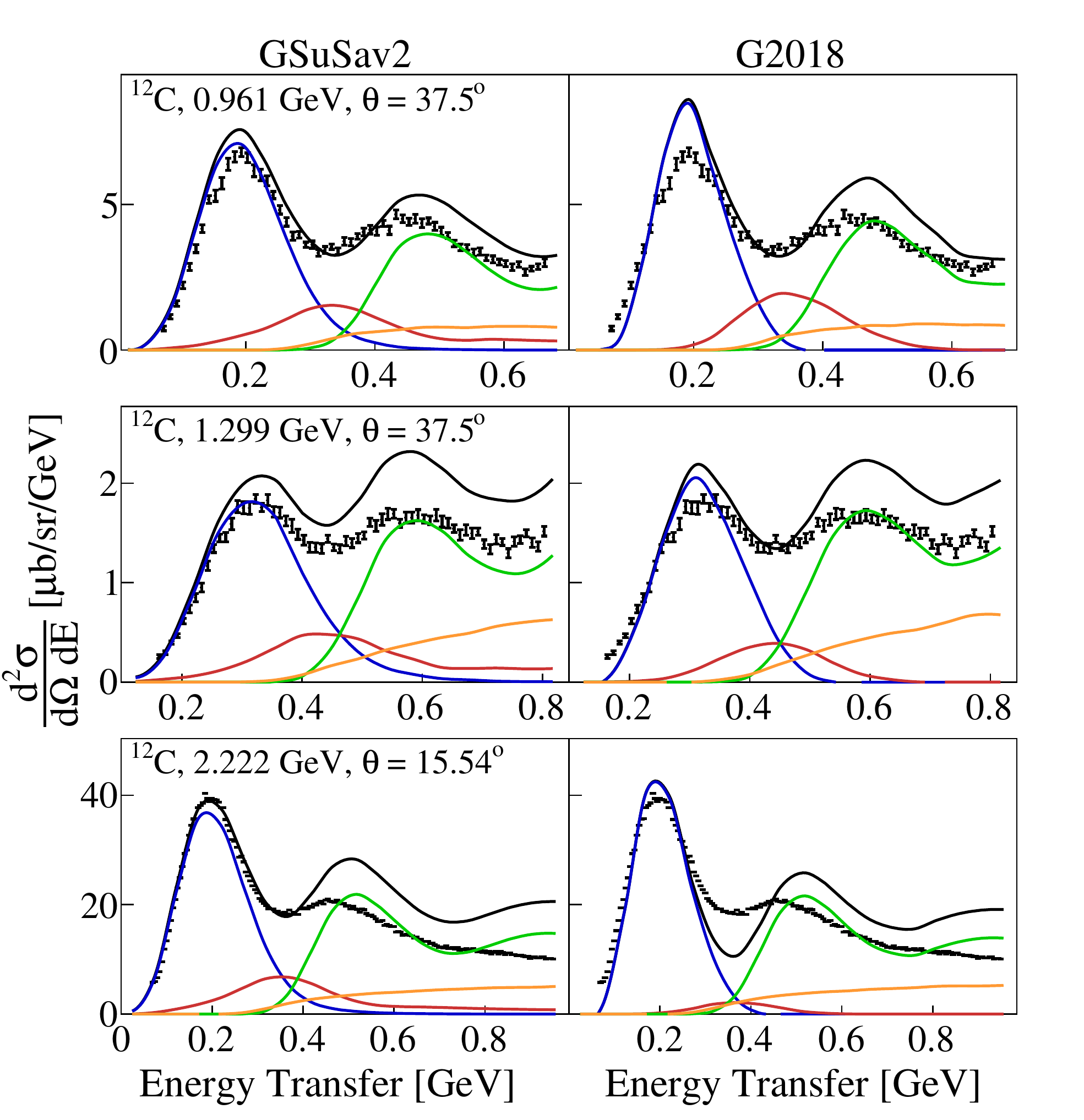}
\caption{Comparison of inclusive C$(e,e')$ scattering cross sections
  for data and for GENIE.  (left) data vs GSuSAv2 and (right) data vs
  G2018.  (top) $E_0=0.96$ GeV, $\theta_e = 37.5^\circ$ and
  $Q^2_{QE}\approx 0.32$ GeV$^2$, (middle)
  $E_0=1.30$ GeV, $\theta_e = 37.5^\circ$ and $Q^2_{QE}\approx 0.54$
  GeV$^2$, and (bottom) $E_0=2.22$ GeV,
  $\theta_e = 15.5^\circ$ and $Q^2_{QE}\approx 0.33$
  GeV$^2$.  Black points show the data, solid black
  lines show the total GENIE prediction, colored lines show the
  contribution of the different reaction mechanisms: (blue) QE, (red) MEC, (green)
RES and (orange) DIS.}
\label{fig:C2}
\end{figure}

\begin{figure}[htb!]
\centering
\includegraphics[width=\textwidth]{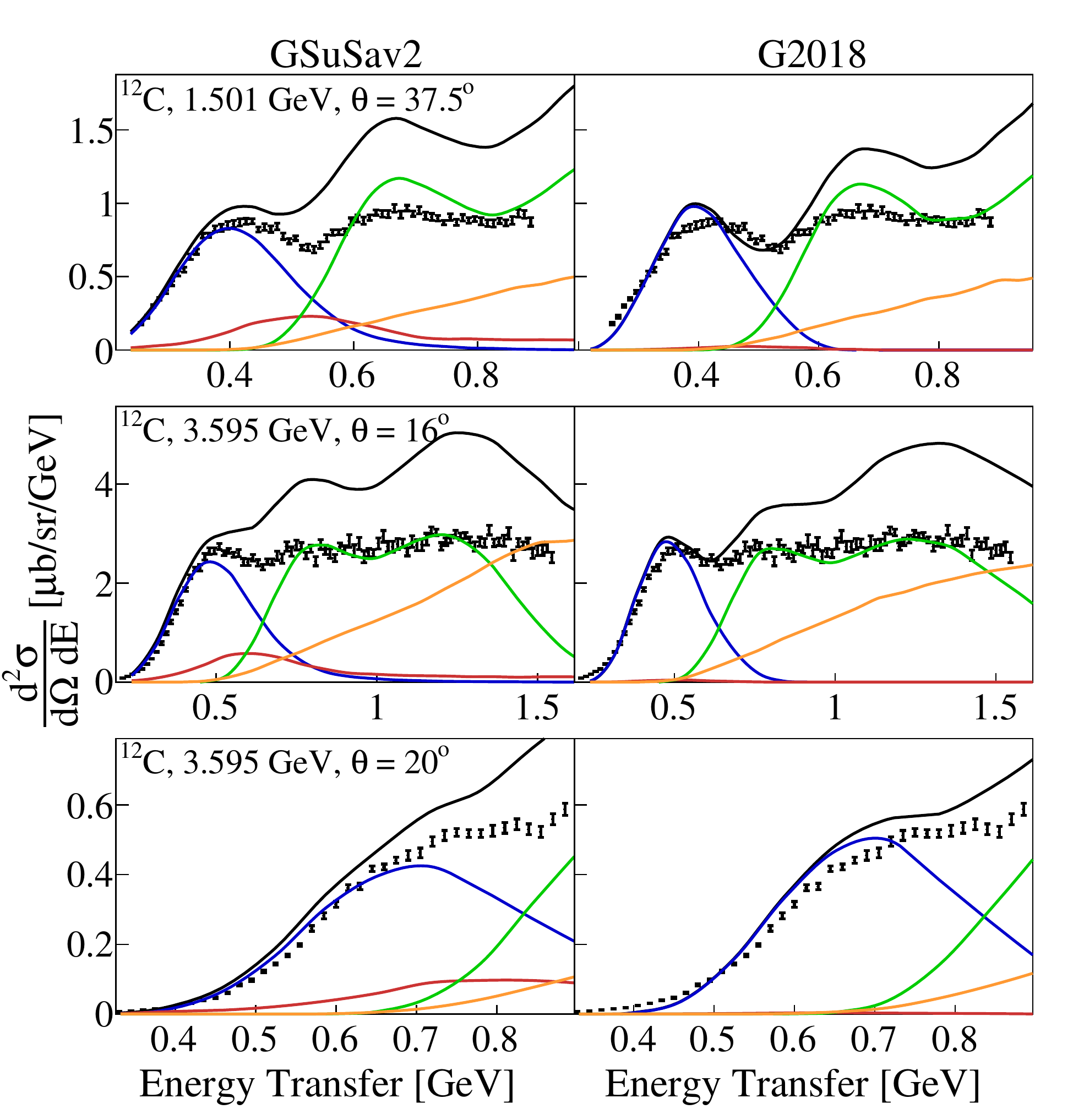}
\caption{Comparison of inclusive C$(e,e')$ scattering cross sections
  for data and for GENIE.  (left) data vs GSuSAv2 and (right) data vs
  G2018.  (top) $E_0=1.501$ GeV, $\theta_e = 37.5^\circ$ and
  $Q^2_{QE}\approx 0.92$ GeV$^2$, (middle)
  $E_0=3.595$ GeV, $\theta_e = 16^\circ$ and $Q^2_{QE}\approx 1.04$
  GeV$^2$, and (bottom) $E_0=3.595$ GeV,
  $\theta_e = 20^\circ$ and $Q^2_{QE}\approx 1.3$
  GeV$^2$.  Black points show the data, solid black
  lines show the total GENIE prediction, colored lines show the
  contribution of the different reaction mechanisms: (blue) QE, (red) MEC, (green)
RES and (orange) DIS.}
\label{fig:C3}
\end{figure}

Figures~\ref{fig:C1}, \ref{fig:C2} and \ref{fig:C3} show the inclusive C$(e,e')$ cross sections for a wide range of beam energies and scattering angles compared to the G2018 and GSuSAv2 models~\cite{Barreau:1983ht,PhysRevLett.62.1350,Dai:2018xhi,Day:1993md,PhysRevC.99.054608,nicolescu2009,PhysRevLett.85.1186,malace2006,PhysRevD.12.1884}.

\begin{figure}[htb!]
\centering
\includegraphics[width=\textwidth]{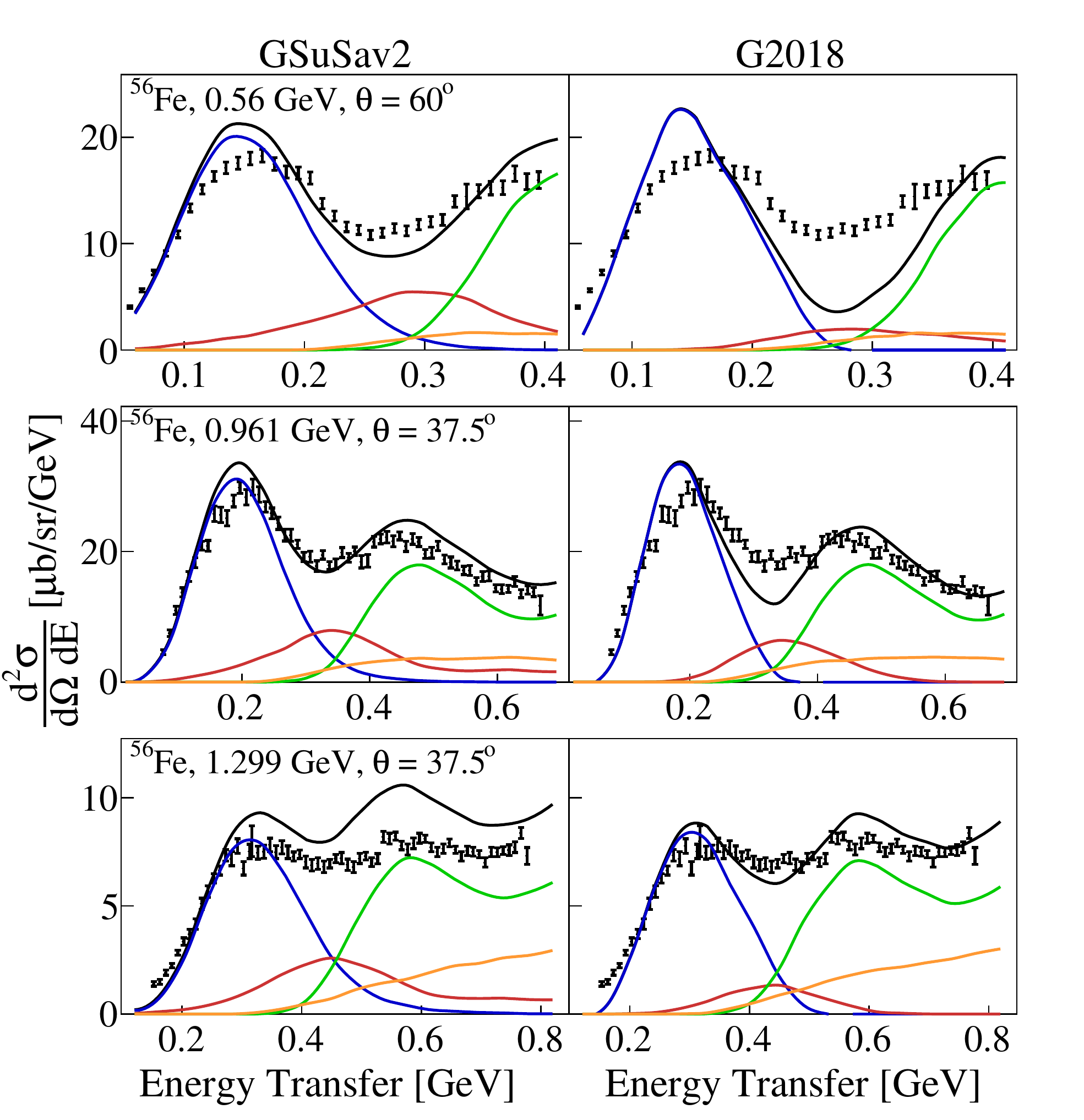}
\caption{Comparison of inclusive Fe$(e,e')$ scattering cross sections
  for data and for GENIE.  (left) data vs GSuSAv2 and (right) data vs
  G2018.  (top) Fe$(e,e')$, $E_0=0.56$ GeV, $\theta_e = 60^\circ$ and
  $Q^2_{QE}\approx 0.24$ GeV$^2$, (middle)
  Fe$(e,e')$, $E_0=0.96$ GeV, $\theta_e = 37.5^\circ$ and
  $Q^2_{QE}\approx 0.32$ GeV$^2$, (bottom)
  Fe$(e,e')$, $E_0=1.30$ GeV, $\theta_e = 37.5^\circ$ and
  $Q^2_{QE}\approx 0.54$ GeV$^2$.  Black
  points show the data, solid black lines show the total GENIE
  prediction, colored lines show the contribution of the different
  reaction mechanisms: (blue) QE, (red) MEC, (green)
RES and (orange) DIS.}
\label{fig:Fe}
\end{figure}

\begin{figure}[htb!]
\centering
\includegraphics[width=\textwidth]{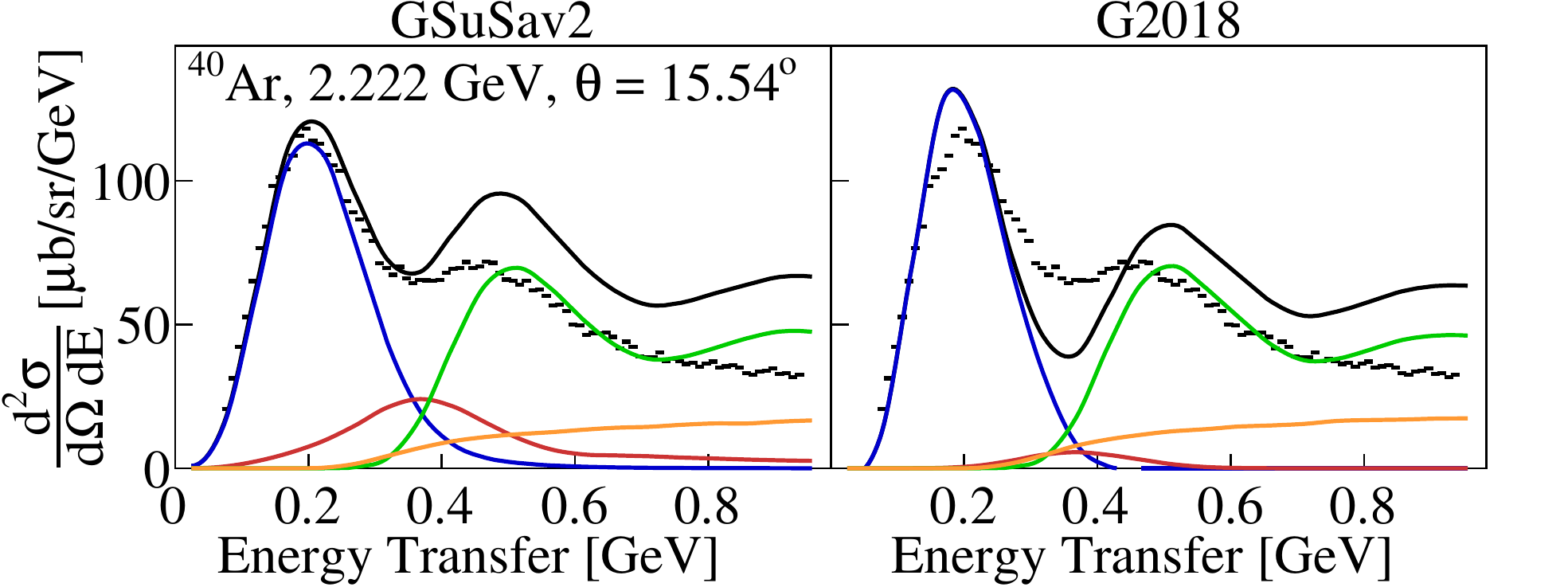}
\caption{Comparison of inclusive Ar$(e,e')$ scattering cross sections for 
data
  and for GENIE at $E_0=2.22$ GeV, $\theta_e = 15.5^\circ$ and $Q^2_{QE}\approx
  0.33$ GeV$^2$.  (left) data vs GSuSAv2 and (right) data vs G2018.
   Black points show the data, solid black lines
  show the total GENIE prediction, colored lines show the contribution
of the different reaction mechanisms:  (blue) QE, (red) MEC, (green)
RES and (orange) DIS.}
\label{fig:Ar}
\end{figure}

The QE peak is the one at lowest energy transfer ($\nu\approx Q^2/2m$) in each plot.  
The next peak at about 300\,MeV larger energy transfer corresponds to the $\Delta(1232)$ excitation and the ``dip-region'' is between the two peaks.  
The $\Delta$ peak in the data is separated from the QE peak by less than the 300\,MeV $\Delta$-nucleon mass difference, indicating that it is shifted in the nuclear medium.  
This shift is more visible at lower momentum transfer where the $\Delta$ peak is more prominent.

GSuSAv2 clearly describes the QE and dip-regions much better than G2018, especially at the three lowest momentum transfers, as shown in  figure~\ref{fig:C1}.  
G2018 has particular difficulty describing the data for $E_0=0.24$ GeV and $\theta_e=60^\circ$, where $Q^2=0.05$ GeV$^2$ at the QE peak.  
G2018 also predicts too small a width for the QE peak and too small a MEC contribution for $E_0=0.56$ GeV and $\theta_e=60^\circ$.
GSuSAv2 describes both features far better.

At intermediate momentum transfers shown in figure~\ref{fig:C2}, GSuSAv2 describes the data somewhat better than G2018, although it overpredicts the dip-region cross section at $E_0=1.299$ GeV and $\theta_e=37.5^\circ$.
The MEC contribution for G2018 appears to be much too small for $E_0=2.222$ GeV and $\theta_e = 15.54^\circ$ ($Q^2_{QE}=0.33$ GeV$^2$).
Both model sets significantly disagree with the data in the resonance region, where they use the same RES and DIS models. 
The 0.961 GeV, 37.5$^\circ$ and the 2.222 GeV, $15.54^\circ$ data are taken at almost identical $Q^2_{QE}$.  
The lower beam-energy data is more transverse, since it is at larger scattering angle. 
The GSuSAv2 MEC contribution is similar for both data sets, but the G2018 MEC contribution is far smaller for the higher beam-energy data.  
The GSuSAv2 MEC contribution describes the dip-region better in the higher beam-energy data set.
The RES model appears to agree with the data  slightly better for the lower beam-energy, more transverse, data set.

At the highest momentum transfers ($Q^2\approx 1$ GeV$^2$) shown in figure~\ref{fig:C3}, the disagreement at the larger energy transfers is far greater.  
The empirical G2018 MEC model contributions are negligible, in marked contrast to the GSuSAv2 MEC contributions.  
The RES and DIS contributions are very significant at high $Q^2$ and in general the GENIE model is larger than the data in the region dominated by RES interactions~\cite{Ankowski:2020qbe}.  
In addition, GENIE does not include the nuclear medium dependent $\Delta$-peak shift, so that the predicted location of the $\Delta$-peak is at larger energy transfer than that of the data.

Figure~\ref{fig:Fe} shows the inclusive Fe$(e,e')$ cross sections for several beam energies and scattering angles compared to the G2018 and GSuSAv2 models. 
The GSuSAv2 model describes the QE region better for all three data sets.  
As described previously, the GSuSAv2 MEC model is independently calculated.  
The empirical G2018 MEC model was fit using GENIE v2 QE and RES models. 
The fit will have to be redone once the QE and RES models stabilize.  
The GSuSAv2 MEC contributions are significantly larger than the empirical G2018 MEC contributions and match the dip-region data far better at $Q^2_{QE}=0.24$ and $0.32$ GeV$^2$.  
However, it overpredicts the dip-region cross section at $Q^2_{QE}=0.54$ GeV$^2$.  
The RES and DIS models describe the Fe data better than the C data at large energy transfers.

Figure~\ref{fig:Ar} shows the inclusive Ar$(e,e')$ cross sections for $E_0=2.222$ GeV and $\theta_e=15.54^\circ$~\cite{PhysRevC.99.054608} compared to the G2018 and GSuSAv2 models.  
The GSuSAv2 model reproduces the data very well in the QE-peak region and the G2018 reproduces the data moderately well.  
The GSuSAv2 MEC model describes the dip-region much better than the G2018 model.  
Again, there is significant disagreement with the RES and DIS models at larger energy transfers.

\begin{figure}[htb!]
\centering
\includegraphics[width=\textwidth]{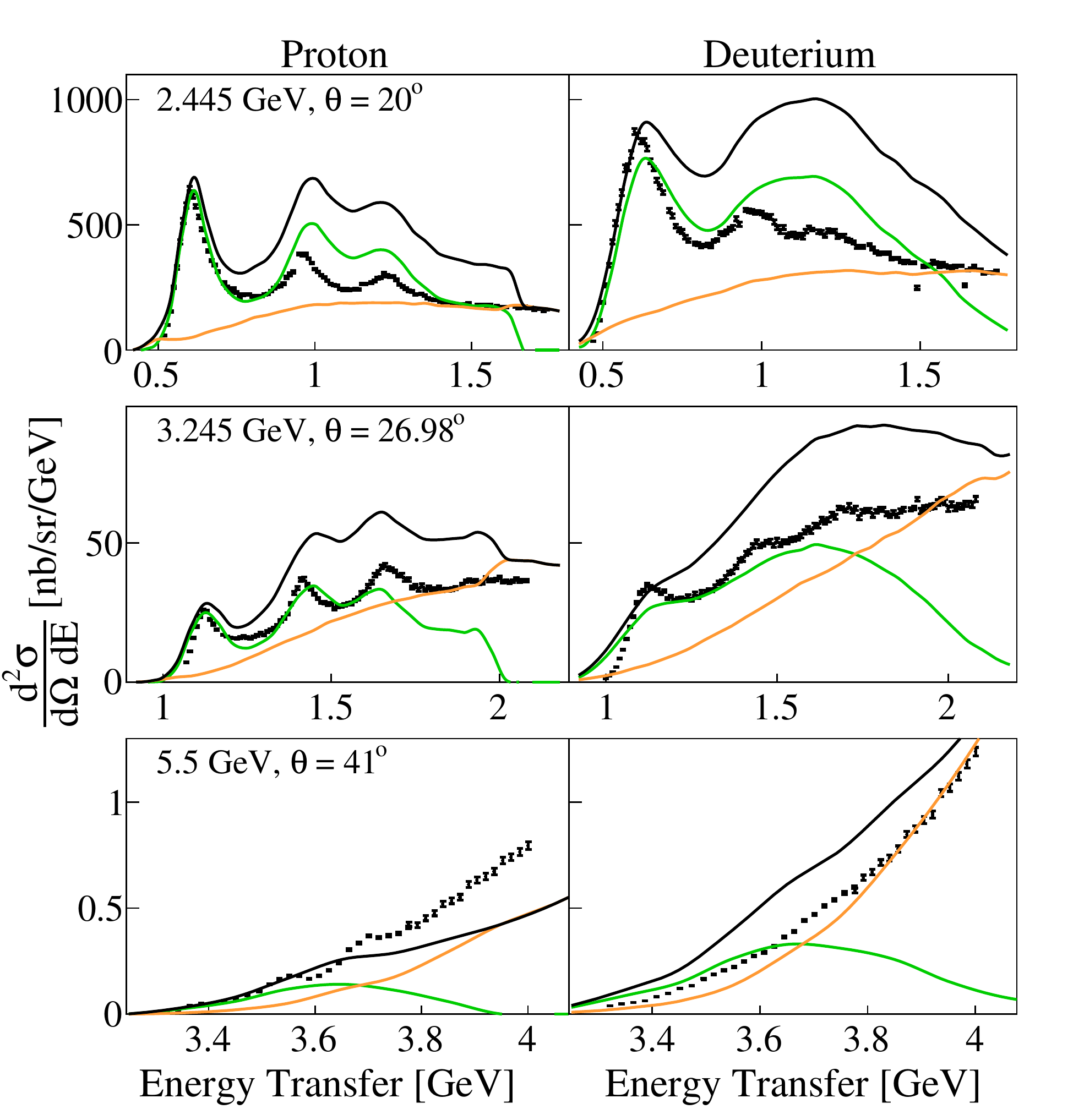}
\caption{Comparison of inclusive proton (left) and deuterium (right)
  $(e,e')$ scattering cross sections for data and for GENIE using
  G2018. (top) $E_0=2.445$ GeV and $\theta_e = 20^\circ$, (middle)
  $E_0=3.245$ GeV and $\theta_e = 26.98^\circ$, and (bottom) $E_0=5.5$
  GeV and
  $\theta_e =
  41^\circ$.
  Black points show the data, solid black lines show the total GENIE
  prediction, colored lines show the contribution of the different
  reaction mechanisms: (green)
RES and (orange) DIS.  The first peak at lowest energy transfer is the
$\Delta(1232)$ resonance.}
\label{fig:Hydrogen}
\end{figure}

\begin{figure}[htb!]
\centering
\includegraphics[width=\textwidth]{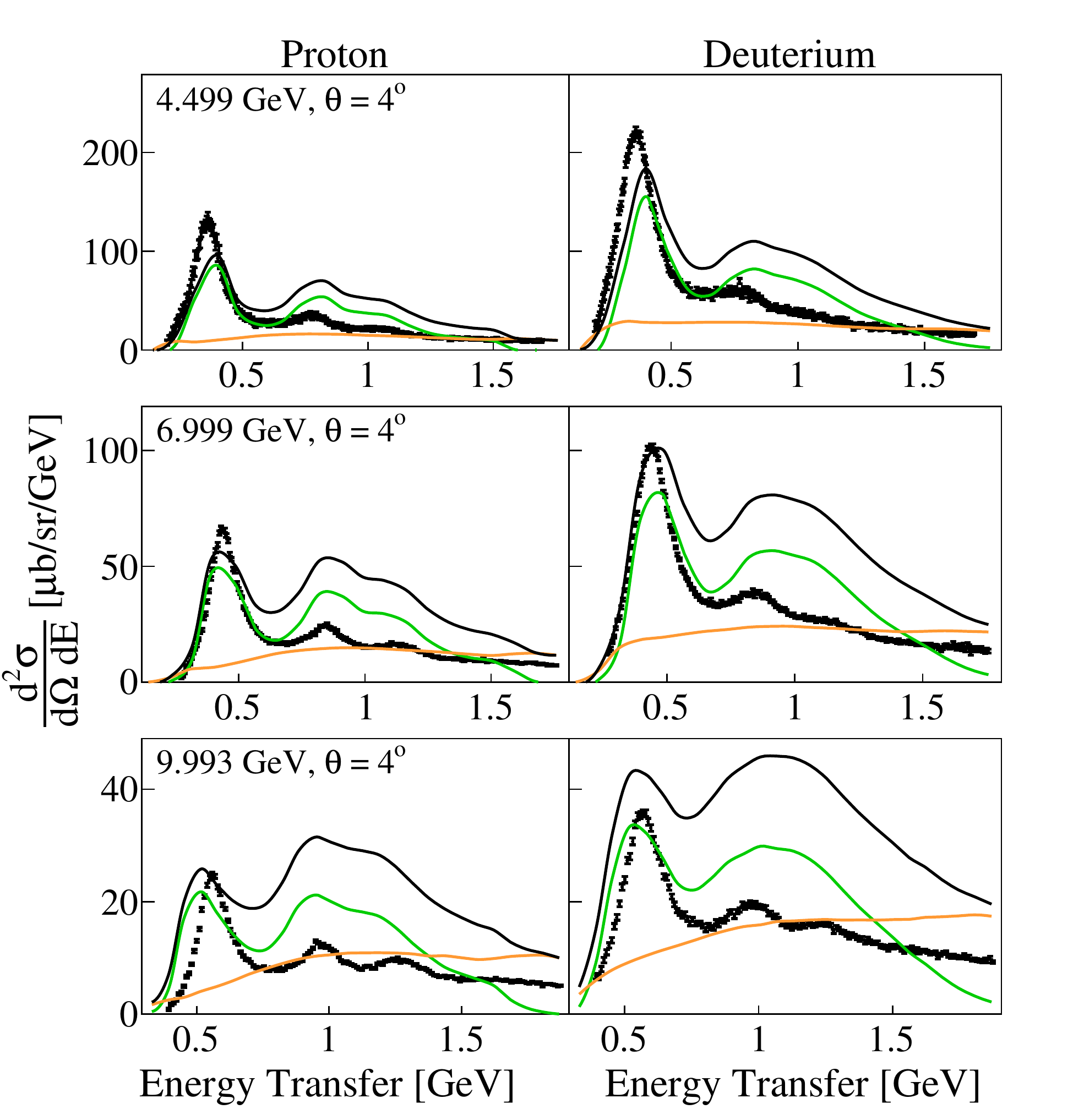}
\caption{Comparison of inclusive proton (left) and deuterium (right)
  $(e,e')$ scattering cross sections for data and for GENIE using
  G2018. (top) $E_0=4.499$ GeV and $\theta_e = 4^\circ$, (middle)
  $E_0=6.699$ GeV and $\theta_e = 4^\circ$, and (bottom) $E_0=9.993$
  GeV and
  $\theta_e = 4^\circ$.
  Black points show the data, solid black lines show the total GENIE
  prediction, colored lines show the contribution of the different
  reaction mechanisms: (green)
RES and (orange) DIS.  The first peak at lowest energy transfer is the
$\Delta(1232)$ resonance.}
\label{fig:HydrogenSLAC}
\end{figure}

The quality of the agreement between data and GENIE depends more on the beam energy and angle than on the target mass from C to Fe.
There is a possible momentum-transfer dependent shift in the location of the SuSAv2 QE peak in Fe due to the extrapolation via scaling from C to Fe.
The GSuSAv2 QE model generally describes the data as well as or better than the G2018 model.

The GSuSAv2 MEC model appears to be significantly superior to the empirical MEC model, especially at $Q^2 < 0.5$ GeV$^2$ or at smaller scattering angles.  
The empirical MEC contribution is often much smaller than needed to explain the dip-region cross section.
However, as an empirical model, it can be tuned to better describe the data.

eGENIE dramatically overpredicts the large-energy transfer data at higher momentum transfers ($Q^2>0.5$ GeV$^2$), indicating issues with the RES (Berger-Sehgal) and DIS (Bodek and Yang) models used.  

This discrepancy at larger momentum and energy transfers is due to the elementary electron-nucleon cross section in the RES and DIS regions, rather than to the nuclear models, since eGENIE also significantly overpredicts the proton and deuteron cross sections.
That is the case especially above the $\Delta$ peak, as shown in figures~\ref{fig:Hydrogen} and \ref{fig:HydrogenSLAC}. 
The discrepancy becomes even more pronounced due to the double counting of processes common across the two interaction channels.
This shows that tuning the RES and DIS models to neutrino data \cite{geniecollaboration2021neutrinonucleon} is not sufficient to constrain the vector part of the cross section.

%%%%%%%%%%%%%%%%%%%%%%%%%%%%%%%%%%%%%%%%%%%%%%%%%%%%%%%%%%%%%

\section{Implications For Neutrinos}

\begin{figure} [htb!]
\centering
\includegraphics[width=0.49\textwidth]{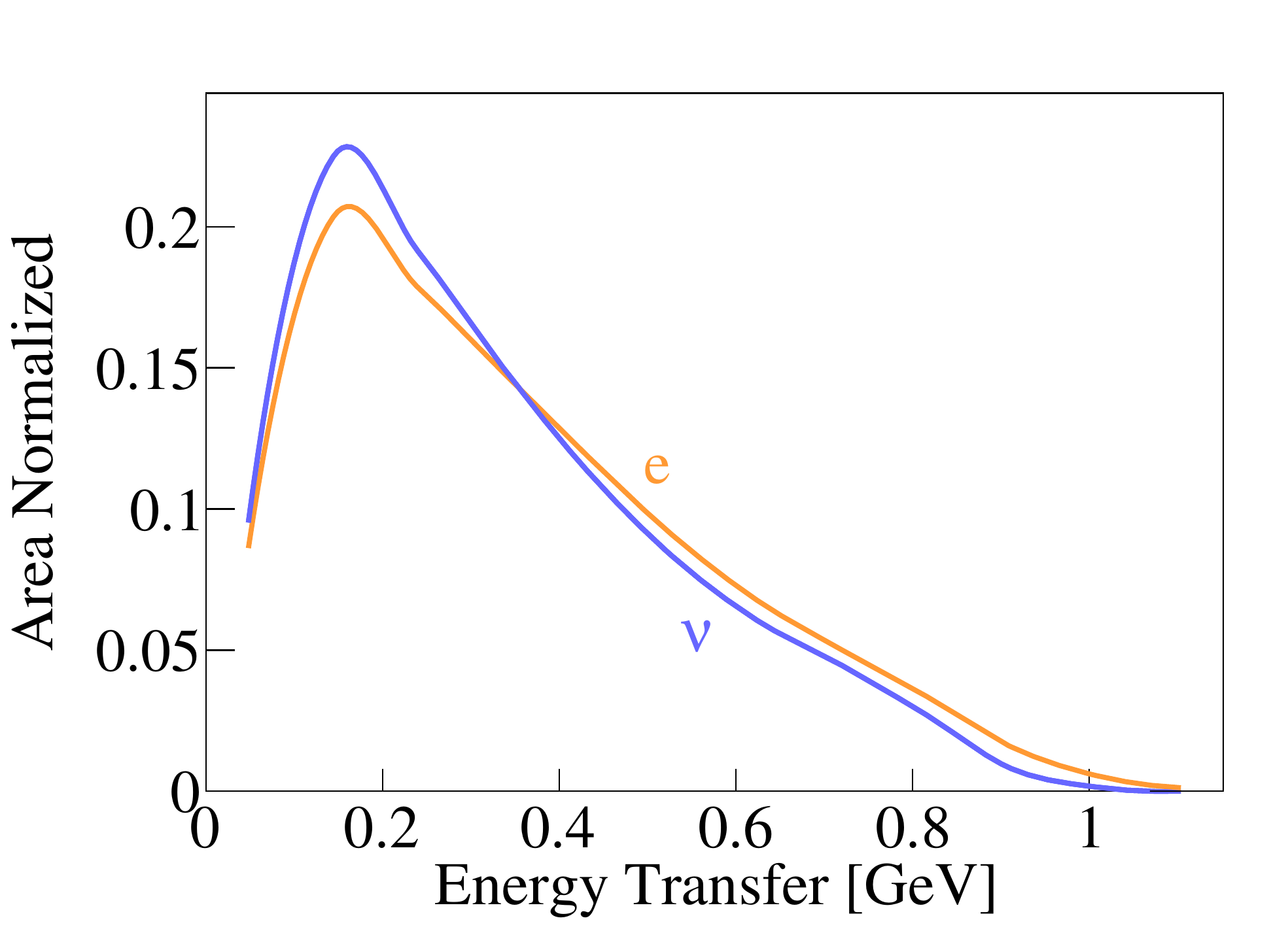}
\includegraphics[width=0.49\textwidth]{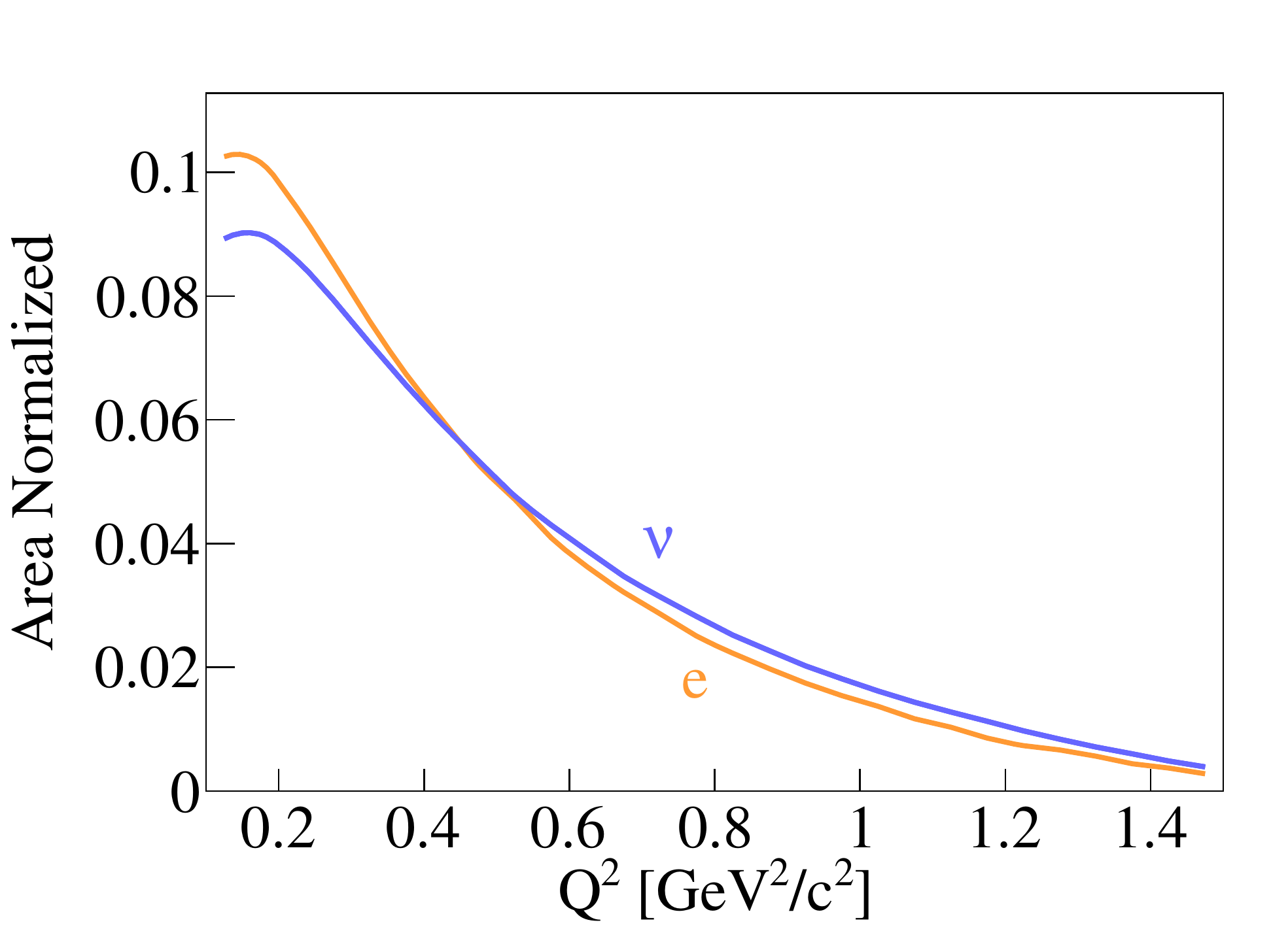}
\caption{\label{e_nu_similarities} Comparison of semi-exclusive 1.16
  GeV lepton-carbon scattering for $Q^2\ge 0.1$ GeV$^2$.  The number
  of generated events is plotted versus energy transfer (left) and
  4-momuntm transfer squared (right) for events with exactly one
  proton with $P_{p} \geq$ 300 MeV/c, no charged pions with
  $P_{\pi} \geq$ 70 MeV/c and no neutral pions or photons of any
  momentum for eGENIE electrons (orange) and GENIE CC $\nu_{\mu}$
  (blue). The electron events have been weighted by $Q^{4}$.  Both
  curves are area normalized.}
\end{figure}

Electron-scattering data can be a very effective tool for testing neutrino event generators due to the similarity between the interactions.   
Figure~\ref{e_nu_similarities} shows the remarkably similar cross-section shapes for electron-nucleus and neutrino-nucleus scattering for semi-exclusive 1.16 GeV lepton-carbon scattering  with exactly one proton with $Q^2\geq 0.1$~GeV$^2$ and $P_{p}  \geq$ 300 MeV/c, no charged pions with $P_{\pi} \geq$ 70 MeV/c and no neutral pions or photons of any momenta.
This corresponds approximately to  the JLab CLAS detector thresholds.  
When comparing electron and neutrino distributions, the electron events are each weighted by $Q^4$ to reflect the difference in the electron and neutrino elementary interactions.

Exploiting these similarities within the same code is invaluable for minimizing the systematic uncertainties of future high-precision neutrino-oscillation experiments. 
Oscillation analysis uncertainties exceeding 1\% for signal and 5\% for backgrounds may substantially degrade the experimental sensitivity to CP violation and mass hierarchy~\cite{DUNE}. 
Such uncertainties already include the relevant neutrino-nucleus interaction uncertainties.
These uncertainties are driven by the choices of the nuclear models and cross-section configurations available in event generators like GENIE.

\begin{figure} [htb!]
\centering
\includegraphics[width=\textwidth]{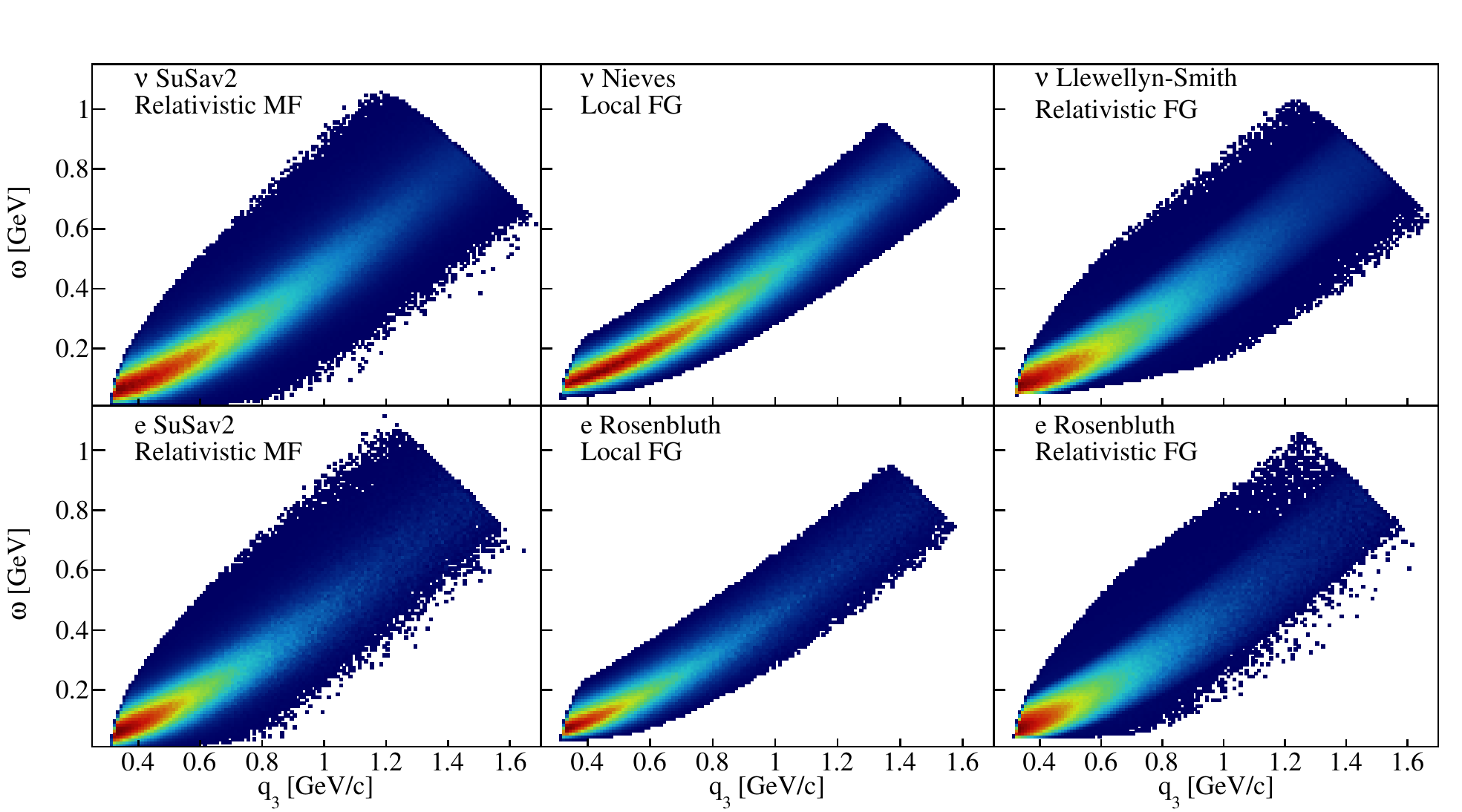}
\caption{\label{QE_Nuclear_Models} Number of simulated events for QE
  scattering on $^{12}$C at 1.161 GeV with $Q^{2} \geq$ 0.1 shown as a
  function of the energy transfer $\omega$ and the momentum transfer
  $q_{3}=\vert\vec q\thinspace\vert$ for all the available nuclear models in GENIE for neutrinos
  (top) and for electrons (bottom). (left) the GSuSAv2
  model which uses a Relativistic Mean Field momentum distribution,
  (middle) the Nieves or Rosenbluth cross section with the Local Fermi
  Gas momentum distribution, and (right) the Llewellyn-Smith or
  Rosenbluth cross section with the Relativistic Fermi Gas momentum
  distribution. The electron events have been
  weighted by $Q^{4}$.}
\end{figure}

Figure~\ref{QE_Nuclear_Models} shows that there is a larger difference among QE scattering models than there is between QE electron and neutrino scattering using the same nuclear model.  
All six panels show a ``ridge'', a maximum in the cross section as a function of energy transfer and momentum transfer.  
The length of the ridge, namely the decrease in intensity as the energy and momentum transfers increase, reflects the momentum transfer dependence of the nucleon form factors used in the cross section model. 
The width of the distribution perpendicular to the ridge reflects the width of the nuclear momentum distribution.

\begin{figure} [htb!]
\centering
\includegraphics[width=0.7\textwidth]{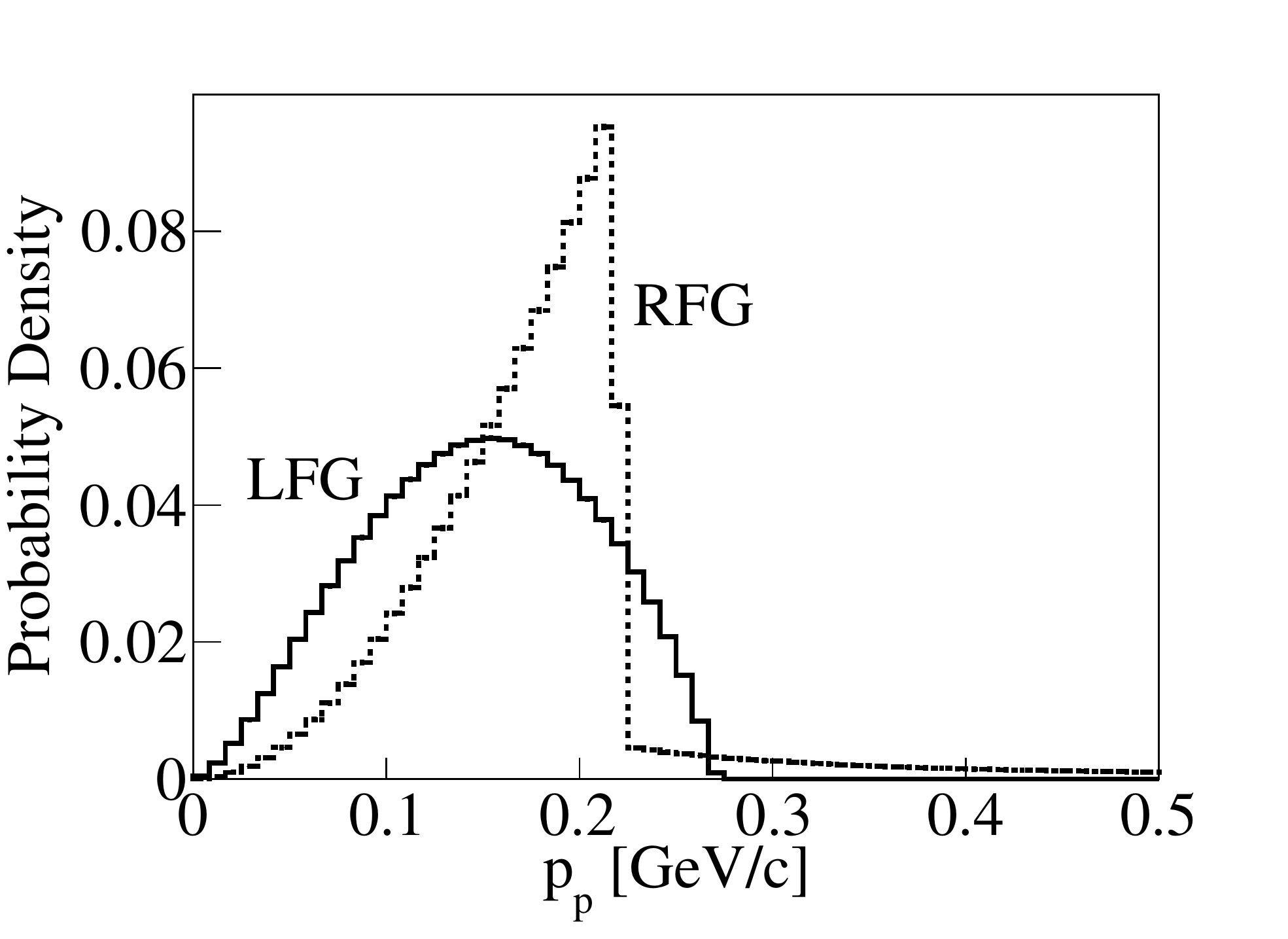}
\caption{\label{fig:LfgVsRfg} Initial momentum distribution of
  protons in simulated QE C$(e,e'p)$ events at $E=1.161$ GeV for the
  Local Fermi Gas (solid histogram) and Relativistic Fermi Gas (dotted
  histogram) models.  The two curves
  are normalized to have the same area.}
\end{figure}

The momentum distribution of the LFG model cuts off at about 260 MeV/c for C, whereas the RMF and the RFG models have ``tails'' that extend to much larger momenta, as shown in figure~\ref{fig:LfgVsRfg}.  
The Nieves cross section decreases more slowly with momentum transfer than the others.
For GSuSAv2, the electron cross section appears to decrease slightly faster with momentum transfer than the neutrino cross section, possibly reflecting differences in the axial and vector nucleon form factors. 

Our ability to use the GENIE code to transfer knowledge gained from electron scattering depends critically on the implementation of its components.
Because of its modular design, all reaction models in GENIE use the same nuclear model, for instance RFG or LFG.
Although the electron scattering capability was added after the initial code release, many of the reaction models used electron scattering data to construct  the vector components of neutrino interactions.  
This was true for the RES~\cite{Rein:1980wg,Berger:2007rq} and the DIS~\cite{Bodek2003} interactions.
The difference between vector neutrino- and electron-scattering is an overall factor illustrated in equation~\ref{eq:factor} and an appropriate change in the form factors.

\begin{figure} [htb!]
\centering
\includegraphics[width=\textwidth]{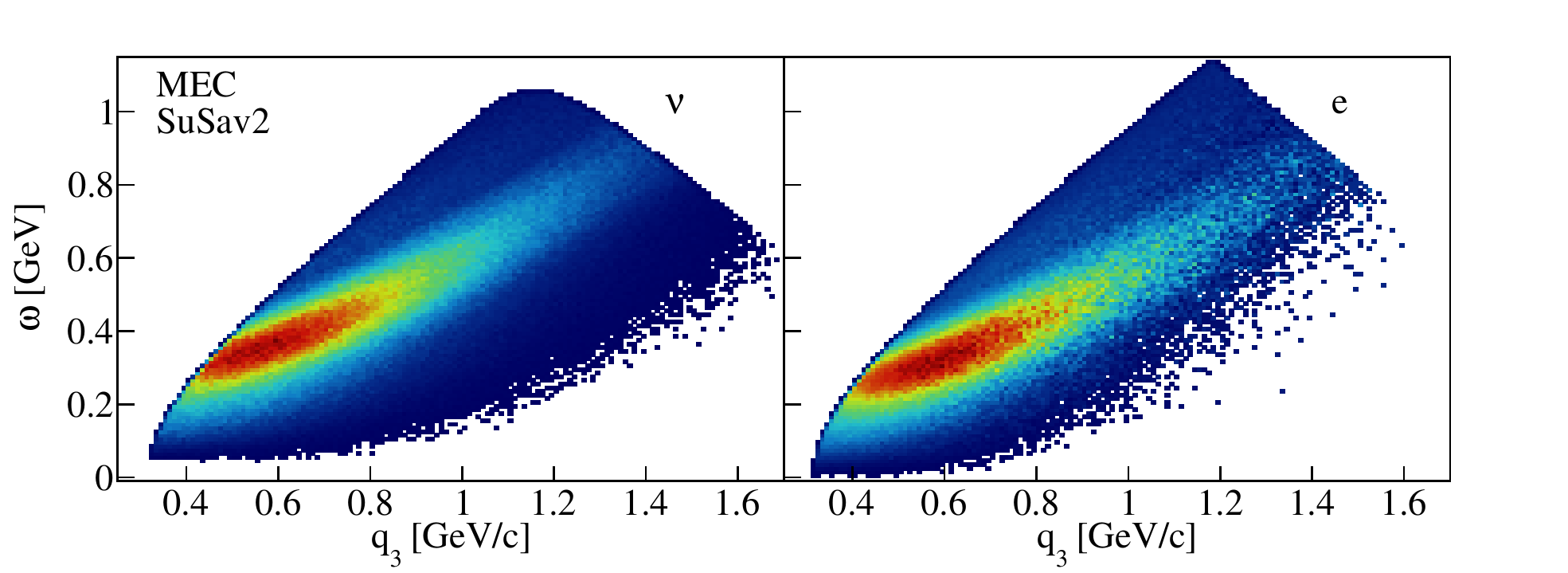}
\caption{\label{MEC_Q0_Q3} Number of simulated events as a function of
  the energy transfer $\omega$ and of the momentum transfer
  $q_{3}=\vert\vec q\thinspace\vert$ for
  neutrinos (left) and for electrons (right) using GSuSav2 for MEC
  interactions. The electron events have been scaled by $Q^{4}$ and
  all the samples have been generated with $Q^{2} \geq$ 0.1.}
\end{figure}

Similarly, figure~\ref{MEC_Q0_Q3} shows that the distribution of MEC events is very similar for electrons and for neutrinos within the same model.  
Thus, measurements of electron scattering will be able to significantly constrain models of neutrino scattering.

Both QE and MEC  models use the same vector form factors for neutrino and for electron scattering.  
QE models can use nucleon form factors from electron scattering~\cite{Bradford:2006yz}, but MEC models must calculate the form factors. 

The GENIE DIS cross section comes from the Bodek-Yang model~\cite{Bodek2003} for the full cross section which extends to $\pi$N threshold.  
The cross section is scaled in the RES region so that it agrees with neutrino-deuterium data~\cite{geniecollaboration2021neutrinonucleon}.
Since a single factor is used to fit the model to the neutrino data, the high-quality electron-proton and electron-deuterium data will be poorly described.  
While the total neutrino cross section and some of the hadronic content of the final state are loosely constrained by the neutrino-deuterium data, the vector component of the models is poorly constrained.

The QE models describe the data reasonably well in the low energy-transfer region.  
Similarly, the largest energy-transfer portions of figures~\ref{fig:C3} and \ref{fig:Hydrogen} show a reasonable agreement between GENIE and data.  
However, at intermediate energy transfer, the RES modeling disagrees with the data for both nuclear and nucleon targets, also observed in~\cite{Ankowski:2020qbe}.
This is due to the use of RES form factors that are not up-to-date and the way the nonresonant contribution was modeled.  

Improvements are in progress but are not simple and therefore not available at this time.  
A possible short-term fix would be to include the electron-proton and electron-deuterium inclusive electron-scattering models of Bosted and Christy~\cite{Christy:2007ve,Bosted:2007xd}.  
Alternatively, the vector resonant form factors could be updated using electroproduction data from JLab and elsewhere. 
A fit to that data is available~\cite{maid} and partially implemented in GENIE, but it does not include nonresonant scattering.  
A more comprehensive solution would be to use the recent DCC model~\cite{Kamano:2016bgm,Nakamura:2013zaa} to simultaneously describe both resonant and nonresonant scattering of both electrons and neutrinos.

The comparisons presented in this work are focused on inclusive electron cross-section measurements.
Yet, forthcoming neutrino oscillation experiments like DUNE will use 4$\pi$-coverage liquid argon time projection chamber tracking detectors in order to investigate exclusive interaction channels.
Until 2021, no prior tests of the event generator performance against such exclusive channels existed with electron scattering events.
Chapters~\ref{e2a} and~\ref{e4v} present the work that alleviated this shortage using wide phase-space electron scattering data sets from Hall B and the e2a experiment at JLab.
\chapter{The CLAS Electrons-For-Neutrinos Experiment At Thomas Jefferson National Laboratory}\label{e2a}

\section{Continuous Electron Beam Accelerator Facility}\label{cebaf}

To test the event generator performance against wide phase-space exclusive interaction channels with electron-nucleus scattering events, data sets from Hall B and the e2a experiment at Thomas Jefferson Laboratory (JLab) in Newport News VA were used.

\begin{figure} [htb!]
\begin{center}
\includegraphics[width=0.7\linewidth]{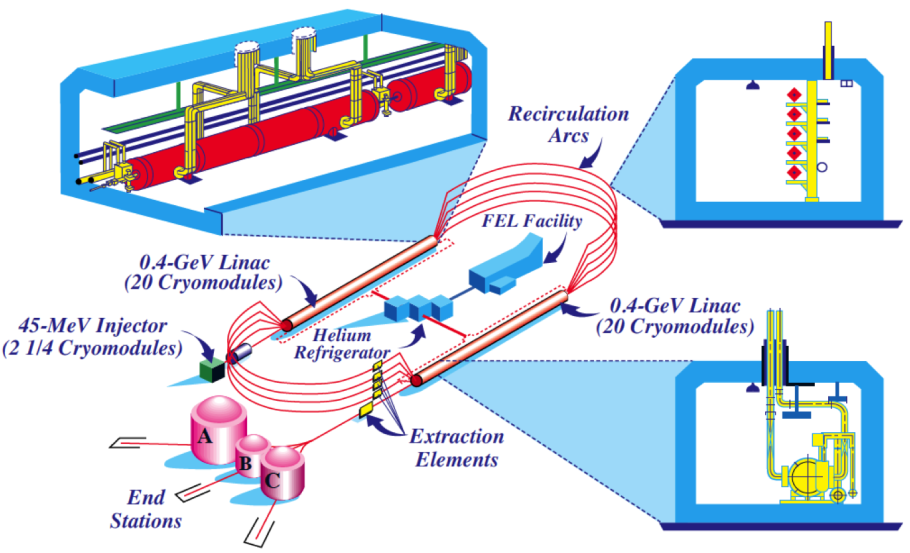}
\end{center}
\caption{Schematic view of the accelerator facility and the experimental halls at Jefferson Lab. Figure adapted from~\cite{cebafreview}.}
\label{CebafFig}
\end{figure}

The Continuous Electron Beam Accelerator Facility (CEBAF) electron accelerator~\cite{cebafref} provided electron beams up to 6 GeV to three experimental halls, referred to as halls A, B, and C, until 2015. 
CEBAF has since been upgraded to 12\,GeV and a new hall (Hall D) was constructed in the context of the 12\,GeV upgrade.
A laser shone on a gallium arsenide cathode to emit photo-electrons.
The electrons were then accelerated by cavities operating at 1.497\,GHz.
Each of the halls received electron bunches every 2\,ns.
Those bunches can have different electron densities so that each hall could be supplied with a different beam current.
A schematic of the facility is shown in figure~\ref{CebafFig}.

The electrons are accelerated through the beamline. 
The latter consists of two parallel linacs that are connected with two arcs with curvature radii of 80\,m.
Each one of the linacs could increase the electron beam energy by 550\,MeV.
The beam was circulated up to 5 times, for a maximum energy of 6\,GeV.
The beam energy spread was $\approx$ 10$^{-4}$.

%%%%%%%%%%%%%%%%%%%%%%%%%%%%%%%%%%%%%%%%%%%%%%%%%%%%%%%%%%%%%%%

\section{The CEBAF Large Acceptance Spectrometer}\label{clasapp}

\begin{figure} [htb!]
\begin{center}
\includegraphics[width=\linewidth]{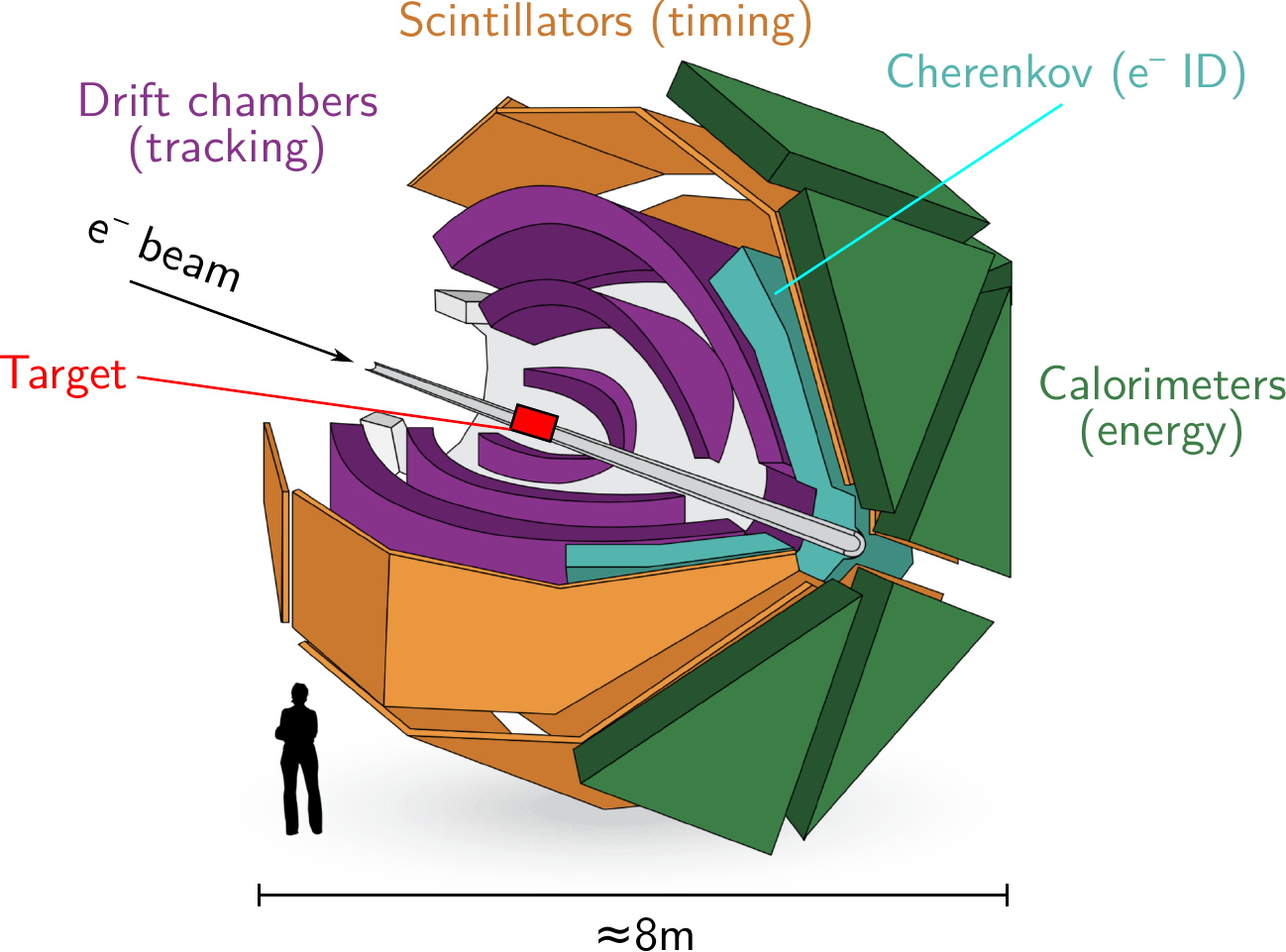}
\end{center}
 \caption{\label{Clas6Det}  Drawing of the CLAS detector showing the sector structure and the different detectors. The beam enters from the upper left side. The target is located at the center of the detector.}
\end{figure}

The e2a data sets were collected at Hall B using the CEBAF Large Acceptance Spectrometer (CLAS) spectrometer shown in figure~\ref{Clas6Det}.
The spectrometer had $\approx$ 50\% angular coverage and ran at luminosities up to 10$^{34}$\,cm$^{-2}$\,sec$^{-1}$.

CLAS used a toroidal magnet for momentum reconstruction.  
The magnet divided CLAS into six almost identical sectors.  
Each sector had three regions of drift chambers (DC) for charged particle trajectory measurements, threshold Cherenkov counters (CC) for electron identification, scintillator counters (SC) for timing and for charged hadron identification, and electromagnetic calorimeters (EC) for electron identification and for photon and neutron detection.
The $\theta$-angle coverage for the DC is $8^{o}-140^{o}$, for the SC $9^{o}-140^{o}$, and for the EC $8^{o}-45^{o}$.

%\begin{figure} [htb!]
%\begin{center}
%\includegraphics[width=\linewidth]{\figures clas_cs_eep.pdf}
%\end{center}
% \caption{\label{Schem} Schematic view of the CLAS detector components in 2D.}
%\end{figure}

%%%%%%%%%%%%%%%%%%%%%%%%%%%%%%%%%%%%%%%%%%%%%%%%%%%%%%%%%%%%%%%%%%%%%%%%%

\section{Toroidal Magnet}\label{magnet}

The torus magnet consisted of six sectors and was made of iron-free super conducting coils oriented around the beam axis~\cite{osti_282366}.
The torus had a 2\,T maximum magnetic field. The coils had a length of 5\,m, and a width of 2.5\,m.
The magnetic field was 5 times smaller at large angles.
The central region of the spectrometer had no magnetic field allowing for polarized target operation.

\begin{figure} [htb!]
\begin{center}
\includegraphics[width=0.4\linewidth]{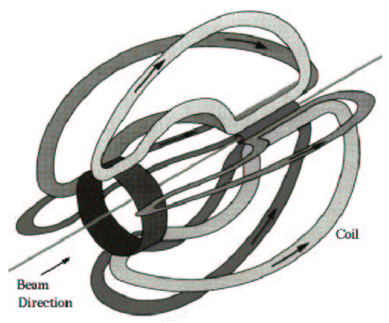}
\end{center}
 \caption{\label{Clasmagnet}The CLAS superconducting toroidal magnet. Figure adapted from~\cite{osti_282366}.}
\end{figure}

%%%%%%%%%%%%%%%%%%%%%%%%%%%%%%%%%%%%%%%%%%%%%%%%%%%%%%%%%%%%%%%%%%%%%%%%%

\section{Drift Chambers}\label{DriftChambers}

\begin{figure} [htb!]
\begin{center}
\includegraphics[width=0.35\linewidth]{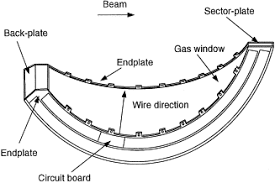}
\includegraphics[width=0.35\linewidth]{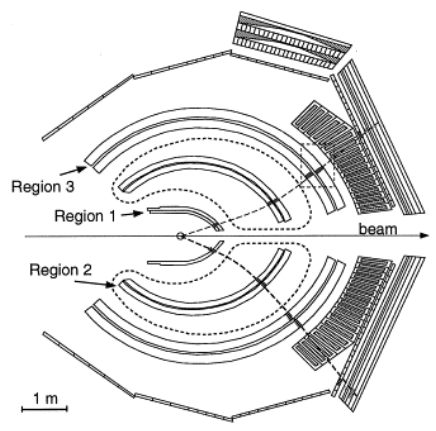}
\end{center}
 \caption{\label{clasdriftchambers}(Left) Illustration of the region 3 drift chamber structure for one of the CLAS sectors. (Right) Schematic representation of the thee drift chamber regions. Figures adapted from~\cite{DriftChambersCLAS}.}
\end{figure}

The Drift Chambers (DC) are used for tracking and momentum measurements of the charged
particles produced out of the interactions. 
The DC have the structure shown in figure~\ref{clasdriftchambers} (left) and more details can be found in~\cite{DriftChambersCLAS}.

Each one of the modules is filled with a gas mixture of 90\% Argon and 10\% CO$_{2}$.
The charged particles that transverse the DC ionize the gas mixture.
The produced electrons drift towards the anode wires and the corresponding ions drift towards the cathode wires.
The relevant electrical signal is used to determine the distance from the charged particle trajectory to the wires.

As the ionization electrons approach the wires, the increased electric field results in the multiplication of the electron-atom collisions.
The latter results in the production of more electrons by a factor of $\approx 10^{4}$ and results into an ``avalanche''.
The CO$_{2}$ molecule serves as a quenching gas and prevents the creation of secondary avalanches.

The DC consist of three radial regions as shown in figure~\ref{clasdriftchambers} (right).
Regions 1 and 3 are located in a low magnetic field region.
Each one of these regions is made of two superlayers, one axial and one stereo (at a stereo angle of $6^{o}$), to allow for complete coordinate reconstruction.
The layers consist of six layers of 192 hexagonal drift cells with the 20-micron sense wire at the center and six shared field wires creating the electric field, with the exception of the innermost layer in region 1, which is made of four layers.
The DC structure provides a position resolution $\approx$ 400 $\mu$m, $\approx$ 5 mrad for the angular variables and $\approx$ 1\% (0.5\%) for the hadron (electron) momenta.

%%%%%%%%%%%%%%%%%%%%%%%%%%%%%%%%%%%%%%%%%%%%%%%%%%%%%%%%%%%%%%%%%%%%%%%

\section{Electromagnetic Calorimeter}\label{ecal}

\begin{figure} [htb!]
\begin{center}
\includegraphics[width=0.6\linewidth]{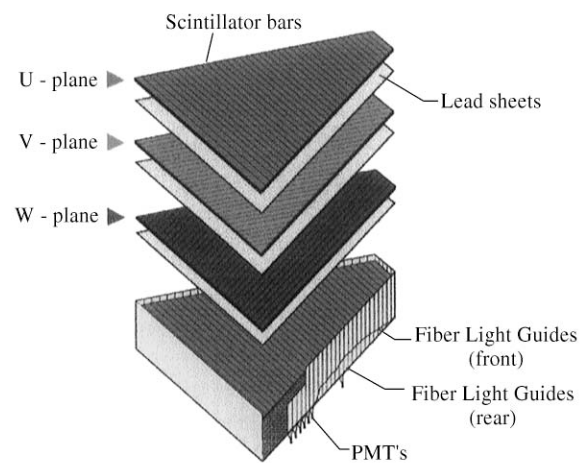}
\end{center}
 \caption{\label{ECalPng}Schematic view of a CLAS electromagnetic calorimeter module. Figures adapted from~\cite{Amarian:2001kk}.}
\end{figure}

The electromagnetic calorimeter (EC)~\cite{Amarian:2001kk} was used to trigger on and to identify the electons.
It further had a high momentum neutron detection efficiency of $\approx$ 50\% and a high photon detection efficiency for energies above 300\,MeV.
It was also used to measure neutral particles with an efficiency of $\approx$ 60\%.
It consists of 6 modules corresponding to each one of the detector sectors.
Each one of the modules has a triangular shape with a projected vertex at the CLAS target location $\approx$ 5\,m.
Each element has a thickness corresponding to 16 radiation lengths and is made of 39 layers of alternating 10\,mm thick scintillator strips and 2.2\,mm of lead sheets.
All the layers have triangular shapes.
In each layer, the scintillator strips are parallel to a given side of the triangle.
Each subsequent layer is rotated by 120$^{o}$ with respect to the previous layer.
That results in the formation of three views, namely u,v, and w, as can be seen in figure~\ref{ECalPng}. 

Each view consists of 13 layers.
The scintillator material thickness for each EC module is 39\,cm and the lead one is 8.4\,cm.
This proportionality between the scintillator and the lead layers leads to one third of the shower energy being deposited in the scintillator part.
The time resolution for electrons and neutrons is  200 and 600\,ps, respectively.
The 13 layers are combined into an inner (5 layers) and an outer (8 layer) stack in order to provide longitudinal sampling of the showers and a hadron identification.

The EC energy resolution is given by the expression

\begin{equation}
\frac{\Delta E}{E} = 0.003 + \frac{0.093}{\sqrt{E\,[GeV]}}    
\end{equation}

%%%%%%%%%%%%%%%%%%%%%%%%%%%%%%%%%%%%%%%%%%%%%%%%%%%%%%%%%%%%%%%%%%%%%%%%%%%%%%%%%

\section{Cherenkov Counters}\label{drift}

\begin{figure} [htb!]
\begin{center}
\includegraphics[width=0.6\linewidth]{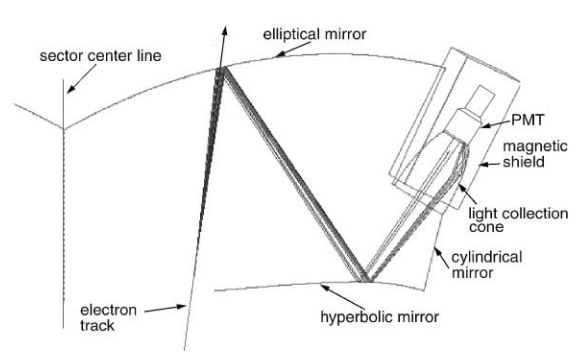}
\end{center}
 \caption{\label{CherPng}Optical arrangement of one of the optical modules
of the CLAS Cherenkov detector, showing the optical and light
collection components. Figures adapted from~\cite{Adams:2001kk}.}
\end{figure}

There are six Cherenkov Counters (CC)~\cite{Adams:2001kk}, which are used to trigger on the electrons and to separate between pions and electrons.
A schematic of the CC is shown in figure~\ref{CherPng}.

Each CC covers the polar angular range of $\theta$ = 8 - 45$^{o}$ and consists of eighteen regions.
Each region is made of two modules called segments.
The CC is filled with perfluorobutane ($C_{4}F_{10}$) gas with an index of refraction of n = 1.00153.
That yields a 2.5\,GeV energy threshold for pions and a 10\,MeV threshold for electrons.
When the particle velocity (v) is greater than the speed of light (c) in the detector medium (c/n), Cherenkov light is emitted.
The produced light is then directed to the light collections cone using elliptical and hyperbolic mirrors in order to get focused to the PMT.

%%%%%%%%%%%%%%%%%%%%%%%%%%%%%%%%%%%%%%%%%%%%%%%%%%%%%%%%%%%%%%%%%%%%%%

\section{Time-Of-Flight Detector}\label{tof}

\begin{figure} [htb!]
\begin{center}
\includegraphics[width=0.6\linewidth]{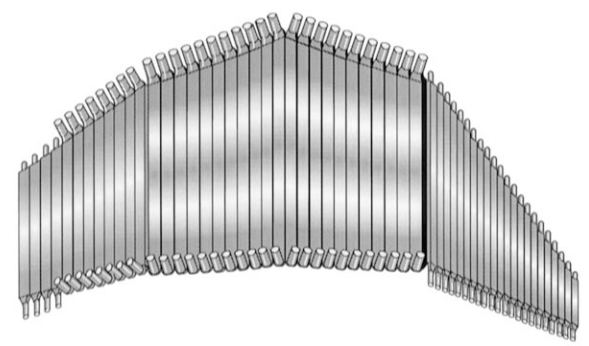}
\end{center}
 \caption{\label{TofPng}Schematic view of the TOF counters in one sector illustrating the grouping into four panels. Figure adapted from~\cite{Smith:2001kk}.}
\end{figure}

The time-of-flight detectors (TOF)~\cite{Smith:2001kk} were made of scintillator paddles and are shown in figure~\ref{TofPng}.
Each scintillator was 5\,cm thick and 10-15\,cm wide.
They ranged in length between 25-450\,cm.
Using the time-of-flight (t) from the TOF subsystem and the distance (d) based on tracking information from the DC, the velocity of the charged particles would be obtained (v = d/t).
The particle mass would be calculated by combining the information of the particle momentum from the DC, as shown in equation~\ref{mass}.

\begin{equation}
m = \frac{p \sqrt{1 - \beta^{2}}}{\beta}    
\label{mass}
\end{equation}

There exist 57 TOF paddles in each sector of the detectororiented perpendicular to the beamline.
Each one covers $\approx$ 2$^{o}$.
The overall coverage in $\theta$ starts at 8$^{o}$ and goes up to 142$^{o}$, organized in four panels.
There is a PMT attached at the end of each paddle.
The time resolution for electrons in the TOF is 163\,ps.

%%%%%%%%%%%%%%%%%%%%%%%%%%%%%%%%%%%%%%%%%%%%%%%%%%%%%%%%%%%%%%%%%%%%%%%%%%%%%%%%%

\section{E2a Targets}\label{e2adetails}

The Hall B e2a experiment ran between April 15-May 27 1999.
During the data taking period, data were collected using the energies and targets presented in table~\ref{TableLum}.

\begin{center}
\begin{table}[htb!]
 \centering
\begin{tabular}{ c c c c c}
  \hline
  \hline
  Target & Energy & Areal Density [g/cm$^{2}$] &  Integrated Charge [mC]\\ 
  \hline
  \hline
  $^{3}$He  & 1.159 & 0.2770 & 0.08\\
  $^{3}$He  & 2.257 & 0.2770 & 0.08\\
  $^{3}$He  & 4.453 & 0.2770 & 0.11\\  
  $^{4}$He  & 2.257 & 0.5375 & 1.17 \\
  $^{4}$He  & 4.453 & 0.5375 & 0.98 \\  
  $^{12}$C  & 1.159 & 0.1786 & 0.08 \\
  $^{12}$C  & 2.257 & 0.1786 & 1.79 \\
  $^{12}$C  & 4.453 & 0.1786 & 2.84 \\  
  $^{56}$Fe & 2.257 & 0.1181 & 0.22\\
  $^{56}$Fe & 4.453 & 0.1181 & 0.31\\  
\hline
  \hline
 \end{tabular}
 \caption{Target areal densities and integrated charges for the $e4\nu$ data sets.}
 \label{TableLum}
\end{table} 
\end{center}

The 2.257 and 4.453\,GeV data sets used a torus current of 2250\,A, while the 1.159\,GeV ones were obtained with a 750\,A torus current.
The beam current ranged between 3-18\,nA.
The solid targets ($^{12}$C and $^{56}$Fe) were 0.9 $\times$ 0.9\,cm$^{2}$ square plates with a thickness of 1\,mm.
The liquid targets ($^{3}$He and $^{4}$He) were stored in cylindrical-shaped vessel with a diameter of 2.8\,cm.
More details on the used data sets can be found in the already published analyses~\cite{Osipenko:2010sb,Egiyan:2005hs,Protopopescu:2004vx,Stavinsky:2004ky,Niyazov:2003zr,Egiyan:2003vg}.

%%%%%%%%%%%%%%%%%%%%%%%%%%%%%%%%%%%%%%%%%%%%%%%%%%%%%%%%%%%%%%%%%%%%%%%%%%

\chapter{Electrons-For-Neutrinos Results\texorpdfstring{\newline}{ }\normalsize{[Nature 599, 565–570 (2021)]}}\label{e4v}

\section{Electron Data Mining Analysis}\label{e4vDataAna}

As discussed in section~\ref{LepXS}, the incomplete lepton-nucleus knowledge can be leveraged using the fact that electrons and neutrinos interact similarly with nuclei.
While previous work has compared these interaction models with inclusive electron scattering~\cite{Ankowski:2020qbe,PhysRevD.103.113003}, this analysis is the first comparison of electron scattering data with these interaction models where events with one or more detected hadrons are used~\cite{e4vNature21}.
The data presented here can therefore test and constrain neutrino-nucleus interaction models to be used in analysis of neutrino oscillation measurements.

For the purposes of the ``Electrons-For-Neutrinos'' (e4$\nu$) analysis, electron scattering data sets from the CEBAF Large Acceptance Spectrometer (CLAS)~\cite{Mecking:2003zu} at the Thomas Jefferson National Accelerator Facility (JLab) were used.
The components of the detector are described in detail in chapter~\ref{e2a}.

\begin{figure} [htb!]
\begin{center}
\includegraphics[width=0.49\linewidth]{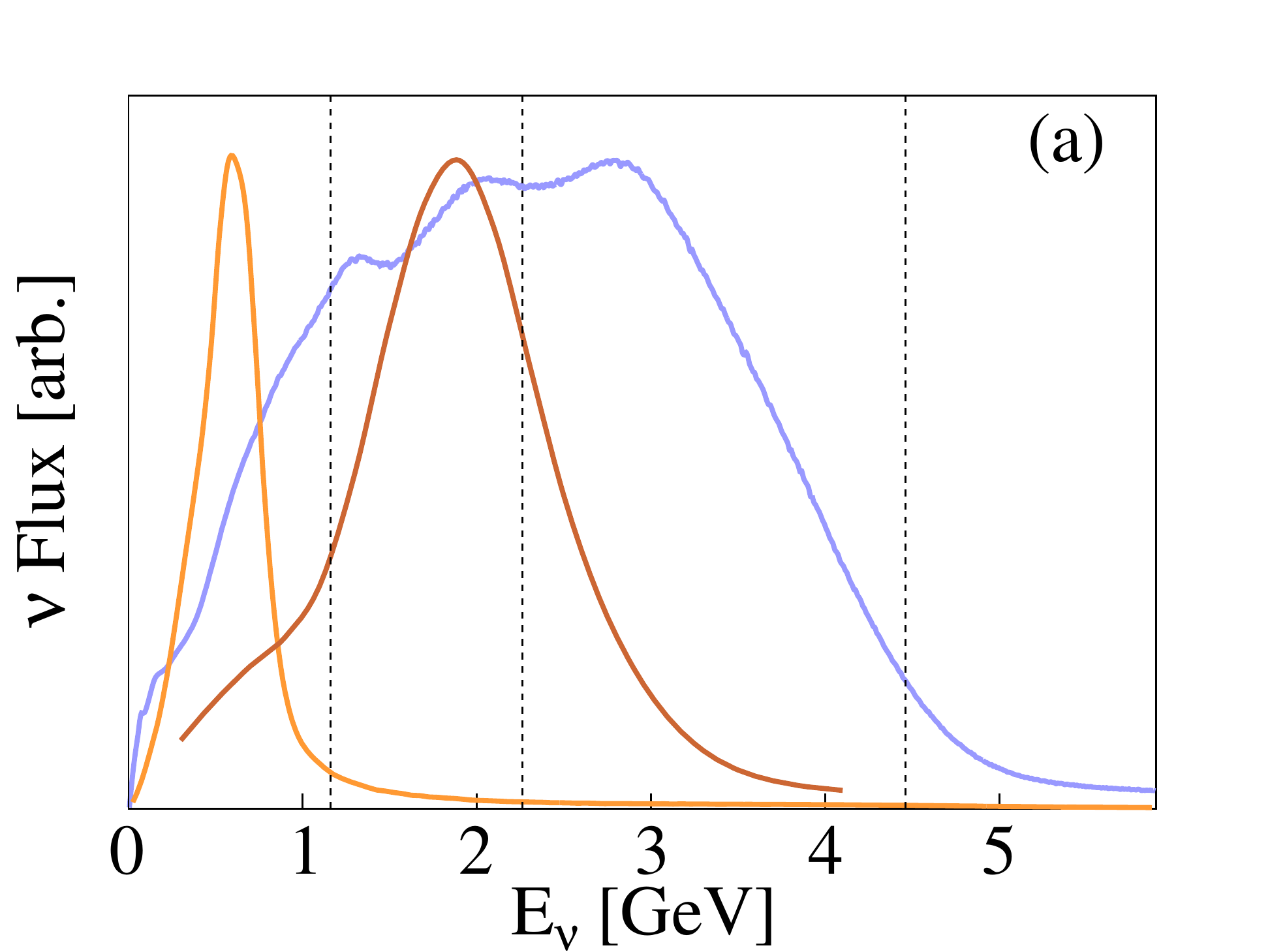}
\includegraphics[width=0.49\linewidth]{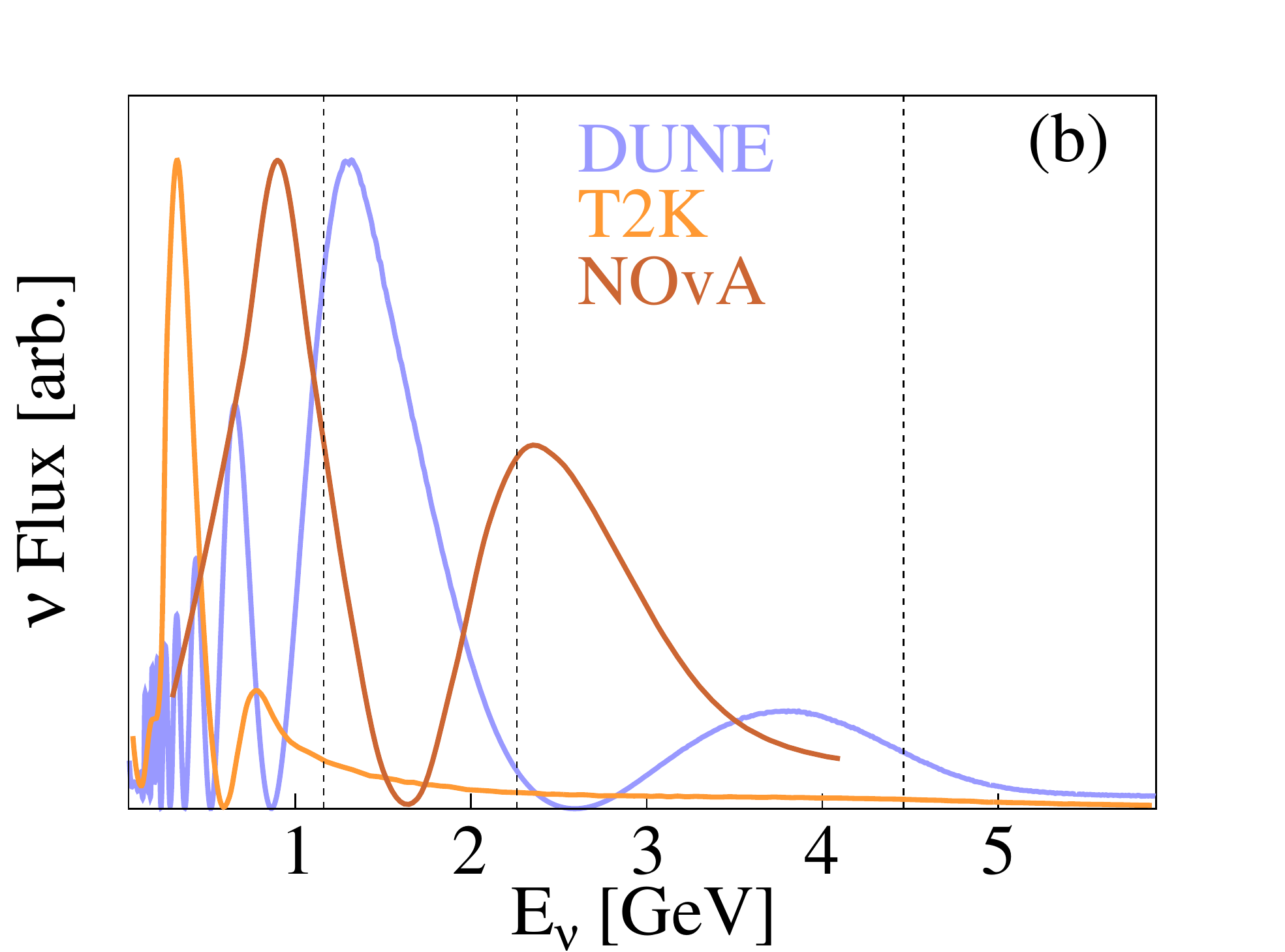}
\end{center}
\caption{\label{fig:Ebeams}The expected energy distribution of different
  $\nu_\mu$ beams, (left) before oscillation at the near detector and
  (right) after oscillation at the far detector.  The vertical lines show the
  three electron beam energies of this measurement.  The NO$\nu$A 
  far-detector beam flux is calculated using the near detector
  flux and the neutrino oscillation parameters from~\cite{PhysRevD.98.030001}.}
\end{figure}

The e2a data sets detailed in section~\ref{e2adetails} were used, which were measured in 1999 and reported in many published analyses~\cite{Osipenko:2010sb,Egiyan:2005hs,Protopopescu:2004vx,Stavinsky:2004ky,Niyazov:2003zr,Egiyan:2003vg}.
These include electron scattering events on $^4$He, $^{12}$C, and $^{56}$Fe nuclei at beam energies of 1.159, 2.257 and 4.453\,GeV. 
The beam energy equaled the injector energy plus the pass number times the linac energy.  
The three-pass beam energy was measured using the Hall A arc measurement and the four pass energy was measured using the Hall C arc measurement.  
These gave a central linac energy of 1.0979~GeV and the Hall B one-, two-, and four-pass beam energies of $1.159, 2.257$, and $4.453$~GeV, respectively.  
An uncertainty of $2\times10^{-3}$ was assigned to these energies, based on the difference between the Hall A and Hall C measurements.
The incident energies used in this analysis span the range of typical accelerator-based neutrino beam energies~\cite{t2kcollaboration2021improved,Abi:2020qib}, as can be seen in figure~\ref{fig:Ebeams}.
The carbon data are relevant for scintillator-based experiments such as \Minerva{} and \NOVA{}~\cite{ALIAGA2014130} and similar to the oxygen in water-based Cherenkov detectors such as Super-Kamiokande (SK)~\cite{T2K,T2KNature20} and HK~\cite{Abe:2018uyc}.  
The iron is similar to the argon in the liquid argon time projection chambers of MicroBooNE \cite{Acciarri:2016smi}, the Fermilab short-baseline oscillation program~\cite{Antonello:2015lea} and DUNE~\cite{Acciarri:2015uup}.  
Many nuclear interaction processes are mass dependent, so it is important to measure a range of target nuclei.

Electrons with energies $E_e \ge 0.4, 0.55$ and $1.1$~GeV for $E_{beam} = 1.159, 2.257$, and $4.453$~GeV respectively, and angles $15^\circ \le \theta_e\le 45^\circ$ were detected.
Protons with momenta $p_p\ge 300$~MeV/$c$ and angles $\theta_p\ge 10^\circ$, charged pions with momenta $p_{\pi}\ge 150$~MeV/$c$ and angles $\theta_{\pi+}\ge 10^\circ$ and $\theta_{\pi-}\ge 22^\circ$, and photons with energy $E_\gamma\ge 300$~MeV and $8\le\theta_\gamma\le 45^\circ$ were detected.  
Separate fiducial cuts were applied for electrons, negatively-charged pions, positively-charged particles, and photons, to select momentum-dependent regions of CLAS where the detection efficiency was constant and close to one.  
These hadron detection thresholds are similar to those of neutrino detectors~\cite{Betancourt:2017uso}, however neutrino detectors have full angular coverage and lower lepton energy thresholds.

Apart from the aforementioned momentum and angular cuts, additional angular outlines are applied to account for the detector acceptance.
The minimum electron angle as a function of electron momentum $p$ for each beam energy was determined as

\begin{equation}
\theta_e^{1.1} \ge 17^\circ + \frac{7^\circ}{p \thinspace\hbox{[GeV]}}
\label{eq:thetaemin_low}
\end{equation}

\vspace{-0.5cm}

\begin{equation}
\theta_{e}^{2.2} \ge 16^\circ + \frac{10.5^\circ}{p \thinspace\hbox{[GeV]}} 
\label{eq:thetaemin_mid}
\end{equation}

\vspace{-0.5cm}

\begin{equation}
\theta_{e}^{4.4} \ge 13.5^\circ + \frac{15^\circ}{p \thinspace\hbox{[GeV]}} 
\label{eq:thetaemin_high}
\end{equation}

and the minimum $\pi^-$ angle as 

\vspace{-0.5cm}

\begin{eqnarray}
\theta_{\pi^-}^{1.1}\ge 17^\circ + \frac{4^\circ}{p \thinspace\hbox{[GeV]}} 
\label{eq:thetaemin_piminus_lowE}
\end{eqnarray}

and

\vspace{-0.5cm}

\begin{eqnarray}
\theta_{\pi^-}^{2.2,4.4}\ge 25^\circ + \frac{7^\circ}{p \thinspace\hbox{[GeV]}} 
\label{eq:thetaemin_piminus_highE}
\end{eqnarray}

for $p_{\pi^-}<0.35$ GeV/c and

\vspace{-0.5cm}

\begin{eqnarray} 
\theta_{\pi^-}^{2.2,4.4}\ge 16^\circ + \frac{10^\circ}{p \thinspace\hbox{[GeV]}} 
\label{eq:thetaemin_piminus_highP}
\end{eqnarray}

for $p_{\pi^-}\ge 0.35$ GeV/c.  
The minimum $\pi^+$ and proton angle was $\theta > 12^\circ$ for all data sets and momenta.

The momentum and charge of the outgoing charged particles were obtained from their measured positions in the drift chambers and the curvature  of their trajectories in the magnetic field.  
Electrons were identified by requiring that the track originated in the target, produced a time-correlated signal in the Cherenkov counter, and deposited enough energy in the electromagnetic calorimeter.  
The charged pions and protons were separated by requiring that the track originated in the target and that the measured time of flight agreed with that calculated from the particle's momentum and assumed mass.
Photons were identified by requiring a signal in the electromagnetic calorimeter which implied a velocity greater than $\approx$ 0.96\,c~\cite{Hen:2014nza}.

Elastic electron scattering from hydrogen was used to correct the electron momentum as a function of angle for uncertainties in the CLAS magnetic field and in the tracking chamber locations.  
These corrections also significantly narrowed  the elastic peak width.  
Typical correction factors were less than 1\%.  
The momentum correction factors at lower scattered electron energies were tested using the H$(e,e'\pi^{+})X$ and $^3$He$(e,e'pp)X$ reactions and found that they gave the correct missing mass for the undetected neutron~\cite{e4vNature21}.

%\begin{figure} [htb!]
%\begin{center}
%\begin{tikzpicture}
%\draw (0, 0) node[inner sep=0] {\includegraphics[width=0.49\linewidth,height=6cm]{\figures MissingMass3He_epp.png}};
%\draw (3., 2.5) node {(a)};
%\end{tikzpicture}
%\begin{tikzpicture}
%\draw (0, 0) node[inner sep=0] {\includegraphics[width=0.49\linewidth,height=6cm]{\figures MissingMassH_epi.png}};
%\draw (3., 2.5) node {(b)};
%\end{tikzpicture}
%\end{center}
% \caption{\label{Efig:EnergyCal}  (left) The 2.257 GeV
%   $^3$He$(e,e'pp)X$ missing mass for (solid histogram) data and
%   (dashed histogram) simulation, and (right) the
%   H$(e,e'\pi^+)X$ missing mass for (black) data and (red) fit to data. 
%   Figure adapted from~\cite{e4vNature21}.}
%\end{figure}

Low momentum protons were corrected for  energy losses traversing the target and detector material.  
The CLAS GEANT Monte Carlo (MC) was used to simulate the proton energy loss in CLAS as a function of  proton momentum.  
The maximum correction was about 20 MeV/c for a proton momentum of 300\,MeV/c.  
The correction was negligible for protons with momenta greater than 600\,MeV/c.

The results from the e2a data sets were compared to predictions from the GENIE~\cite{Genie2010} simulation, which is used by most neutrino experiments in the USA and has an electron-scattering version (\eGenie) that was recently overhauled to be consistent with the neutrino counterpart ($\nu$GENIE), as detailed in~\cite{PhysRevD.103.113003}.  
GENIE includes quasi-elastic lepton scattering (QE), meson exchange currents (MEC), resonance production (RES) and deep inelastic scattering (DIS), as well as rescattering via final state interactions (FSI) of the outgoing hadrons.  
The two GENIE configurations already presented in section~\ref{modeling} were compared.
These include the significantly improved G2018 setup which reproduces measured neutrino~\cite{Adams:2019iqc} and electron inclusive cross sections, and newly implemented \susa{} that uses modern, theoretically-inspired, recently-implemented QE and MEC models~\cite{PhysRevD.101.033003}.
The resulting simulated events were then analyzed using the same code as the data  and the two were compared.

\begin{figure} [htb!]
\begin{center}
\includegraphics[width=0.7\linewidth]{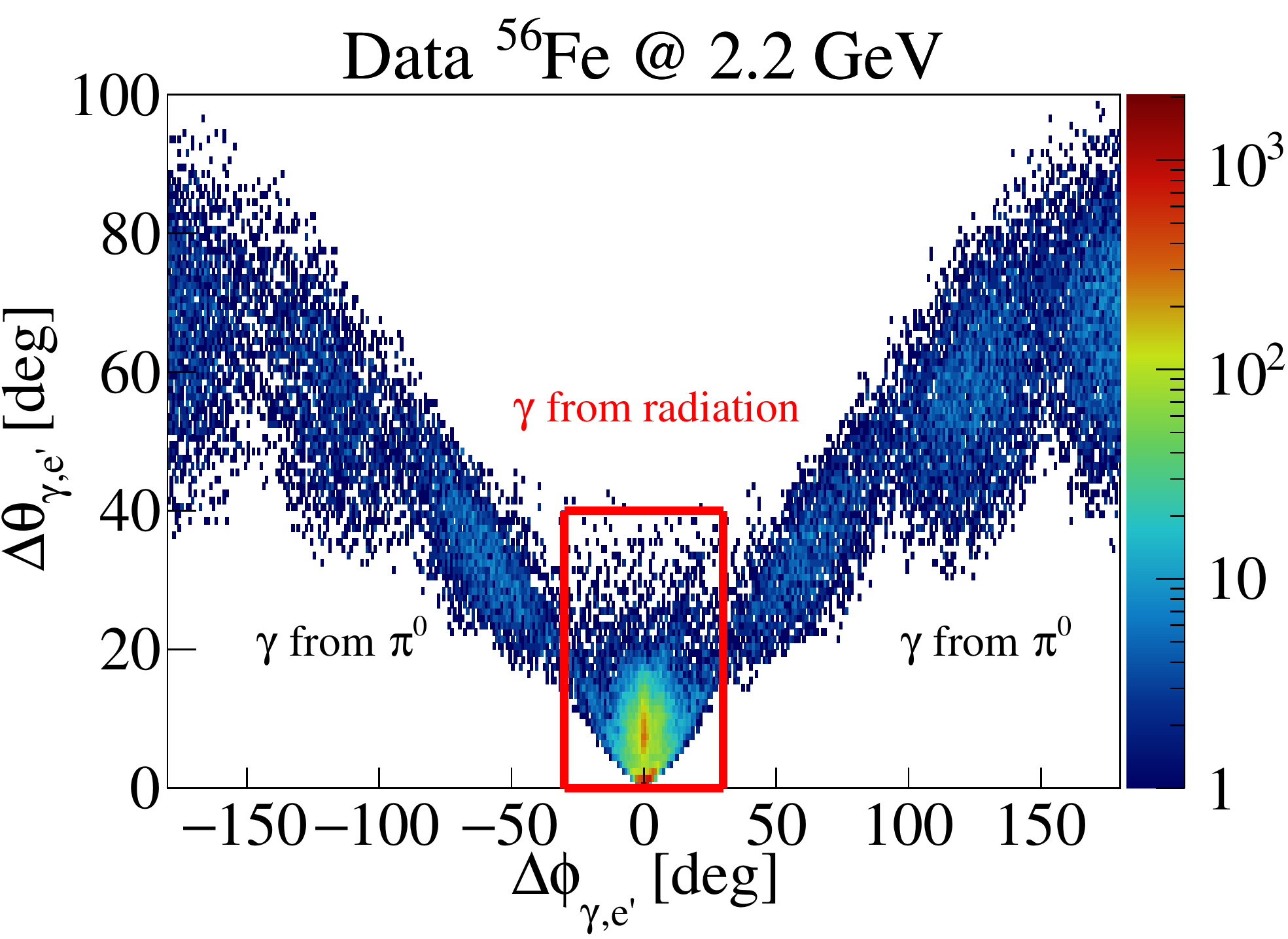}
\end{center}
\caption{\label{batman}$\Delta\theta_{\gamma,e'}$ as a function of $\Delta\phi_{\gamma,e'}$. The red box indicated the region with radiated photons which was removed in our analysis.}
\end{figure}

Electrons, unlike neutrinos, radiate bremsstrahlung photons in the electric field of the nucleus.
Events where the photons from scattered-electron radiation were detected in CLAS were vetoed.
It was assumed that the photons came from either radiation by the outgoing electron approximately parallel to its motion or from $\pi^0$ decay.  
The radiated photons were identified by requiring that they be detected within $\Delta\phi_{\gamma,e'}\le 30^\circ$ and $\Delta\theta_{\gamma,e'}\le 40^\circ$ of the scattered electron and removed them from the data set. 
The events that were removed are indicated by the red box in figure~\ref{batman}.

\begin{figure}[htb!]
\centering
\includegraphics[width=0.7\textwidth]{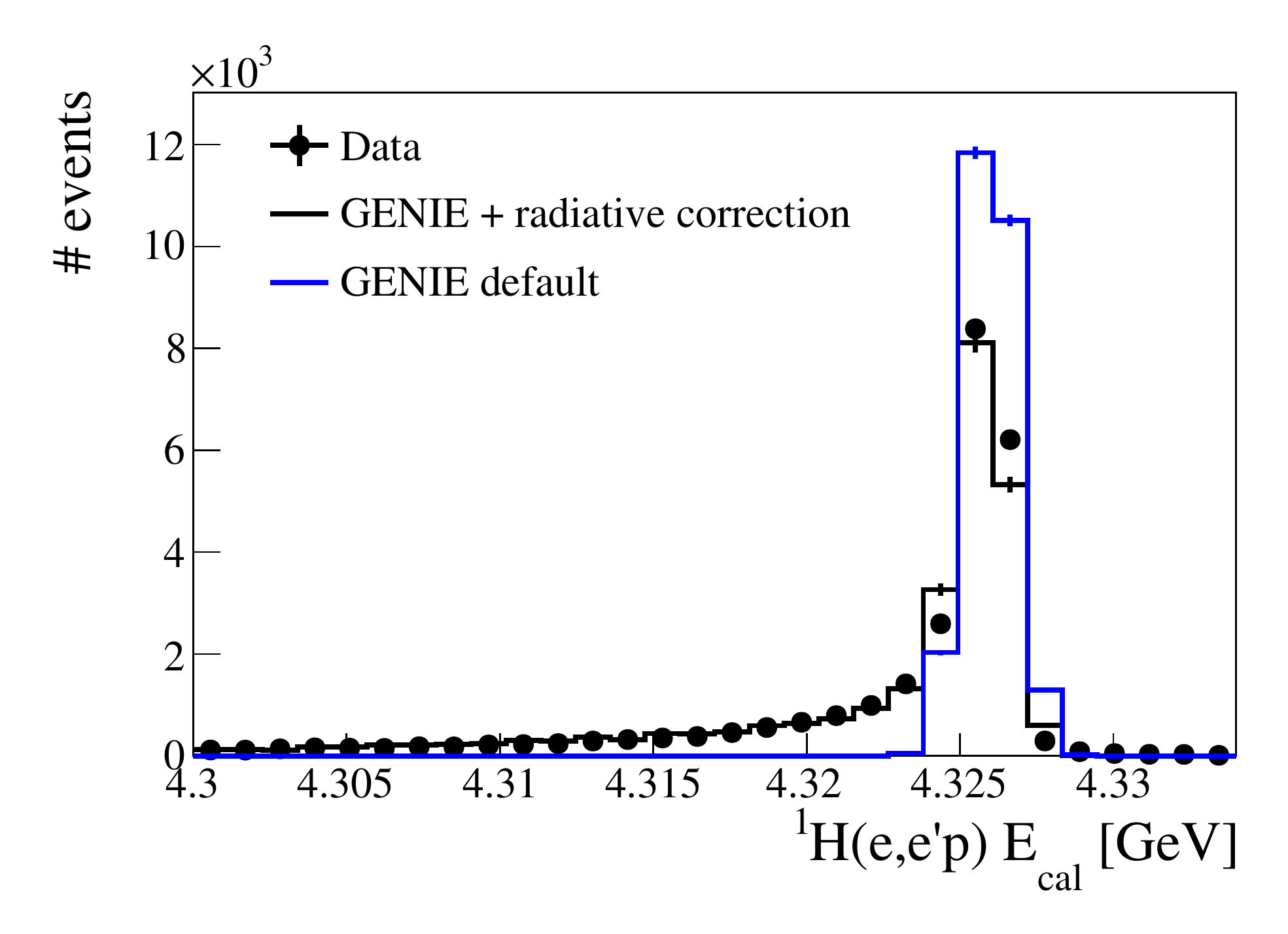}
\caption{  Number of
events vs $E_{cal}= E_{e'} +T_p$ the scattered electron energy plus
proton kinetic energy
  for 4.32 GeV H$(e,e'p)$. Black points are data, the blue
  histogram shows the unradiated GENIE prediction and the black histogram
  shows the GENIE prediction with electron radiation.  The GENIE
  calculations have been scaled to have the same integral as the data.}
\label{fig:radTail}
\end{figure}

The incoming and outgoing electrons can each radiate a real photon, which changes the kinematics of the interaction or the detected particles, and there can be vertex or propagator corrections that change the cross section.
When comparing electron scattering data to models, either the data or
the model needs to be corrected for radiative effects.  
Published electron scattering cross sections are  typically corrected for radiative effects, but this correction is complicated and somewhat model-dependent.
A framework for electron radiative corrections in GENIE was implemented for the first time to allow comparisons to nonradiatively corrected data.  
The framework allows electron radiation, which can change the kinematics of the event by changing either the incident or scattered electron energy through radiation of a real photon.  
We modeled external radiation in the same way as the JLab SIMC event generator~\cite{SimcRadiation,Mo:1968cg}. 
The implementation takes advantage of the peaking approximation that greatly simplifies the calculation of the angular distribution of the emitted photon radiation by making the assumption that radiation \textit{along} the direction of a given particle can be interpreted as radiation \textit{due} to that particle~\cite{PhysRevC.64.054610}.
Future versions of eGENIE will incorporate cross section changes due to vertex and propagator corrections.
The radiative correction procedure was validated by comparing a simulated sample to electron scattering from protons at JLab. 
Figure~\ref{fig:radTail} shows the data compared to the GENIE simulation with and without radiative corrections~\cite{Cruz-Torres:2019bqw}.  
The radiatively corrected calculation is clearly much closer to the data. 
This correction can be used for comparisons with nonradiatively corrected data.  

The primary focus of this analysis was events with one electron and zero pions or photons from $\pi^0$ decay above threshold, which are referred to as (e,e')$_{0\pi}$.  
That choice was made to maximize the contribution of QE events where the incident lepton scattered from a single nucleon in the nucleus, as is done in many neutrino oscillation analyses~\cite{PhysRevD.98.030001,Katori:2016yel}.
Furthermore, events with one detected electron, one proton, and zero pions, denoted here as (e,e'p)$_{1p0\pi}$ were examined.
These events were expected to be dominated by well-understood QE events.  

\begin{figure} [htb!]
\begin{center}
\includegraphics[width=0.6\linewidth]{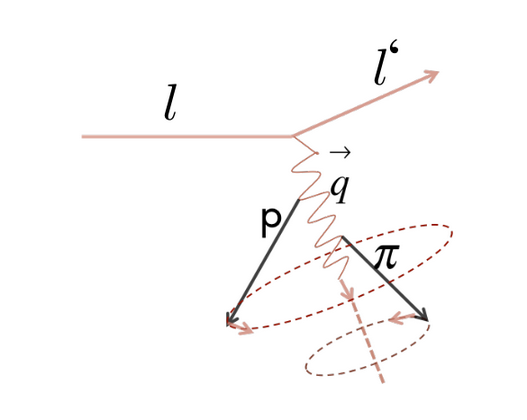}
\end{center}
\caption{\label{datadrivencorr}Schematic illustration of the data driven background correction using detected (e,e'p$\pi$) events.}
\end{figure}

Because the CLAS geometrical coverage is incomplete ($\approx$ 50\%), undetected pions and photons needed to be subtracted to obtain true (e,e')$_{0\pi}$ and (e,e'p)$_{1p0\pi}$ event samples.
The undetected pion, photon, and proton (if applicable) contribution was quantified from the events with detected pions, photons or protons.  
The pion-production cross section was assumed to be independent of $\phi_{q\pi}$.
The latter is the angle between the electron-scattering plane (the plane containing the incident and scattered electrons and the virtual photon) and the hadron plane (the plane containing the virtual photon and pion). 
The data-driven correction for these undetected hadrons is outlined below for the (e,e'p)$_{1p0\pi}$ case.
For each detected $(e,e'p\pi)$ event, the proton-pion pair was rotated around the momentum transfer direction $\vec{q}$ randomly multiple times, as can be seen in figure~\ref{datadrivencorr}.
For each rotation, it was determined whether the particle was within the fiducial region of the detector.  
The particle acceptance would then be $A_\pi=N_{det}/N_{rot}$, where $N_{rot}$ is the number of rotations and $N_{det}$ is the number of times the pion would have been detected.  
The corresponding number of undetected $(e,e'p\pi)$ events for that detected $(e,e'p\pi)$ event is $(N_{rot}-N_{det})/N_{det}$.  
That was used as a weight to subtract for the undetected pion events. 
For example, if one specific $(e,e'p\pi)$ event would have been detected 250 times out of 1000 rotations, then it was inferred that for each detected event, there were three more that were not detected.  
The appropriate variables were calculated for that event and subtracted it from the corresponding distributions with a weight of three.  
That was done separately for $\pi^+,\pi^-$ and photons in order to obtain a true (e,e'p)$_{1p0\pi}$ sample.
The same process is repeated for the (e,e')$_{0\pi}$ channel, where the corrections included undetected protons.

Higher multiplicity events were also accounted for, such as for events with two detected $\pi^\pm$ or photons.
When these events were rotated, each rotated event could have been detected as a $2\pi$ event, a $1\pi$ event, or a $0\pi$ event.  
If it appeared as a $0\pi$ event, its contribution was subtracted from the various $0\pi$ spectra as described above.  
If it appeared as a $1\pi$ event, it was included in the set of $1\pi$ events with the appropriate negative weight.
It was then treated as a regular $1\pi$ event, which was then rotated and added to the $0\pi$ data set.  
Some of the detected $1\pi$ events were actually $2\pi$ events with an undetected pion. 
When the effect of these events was accounted for, there were fewer true $1\pi$ events left.
This reduced the contamination of the $1\pi$ events in the $0\pi$ channel.

\begin{figure} [htb!]
\begin{center}
\includegraphics[width=0.49\linewidth]{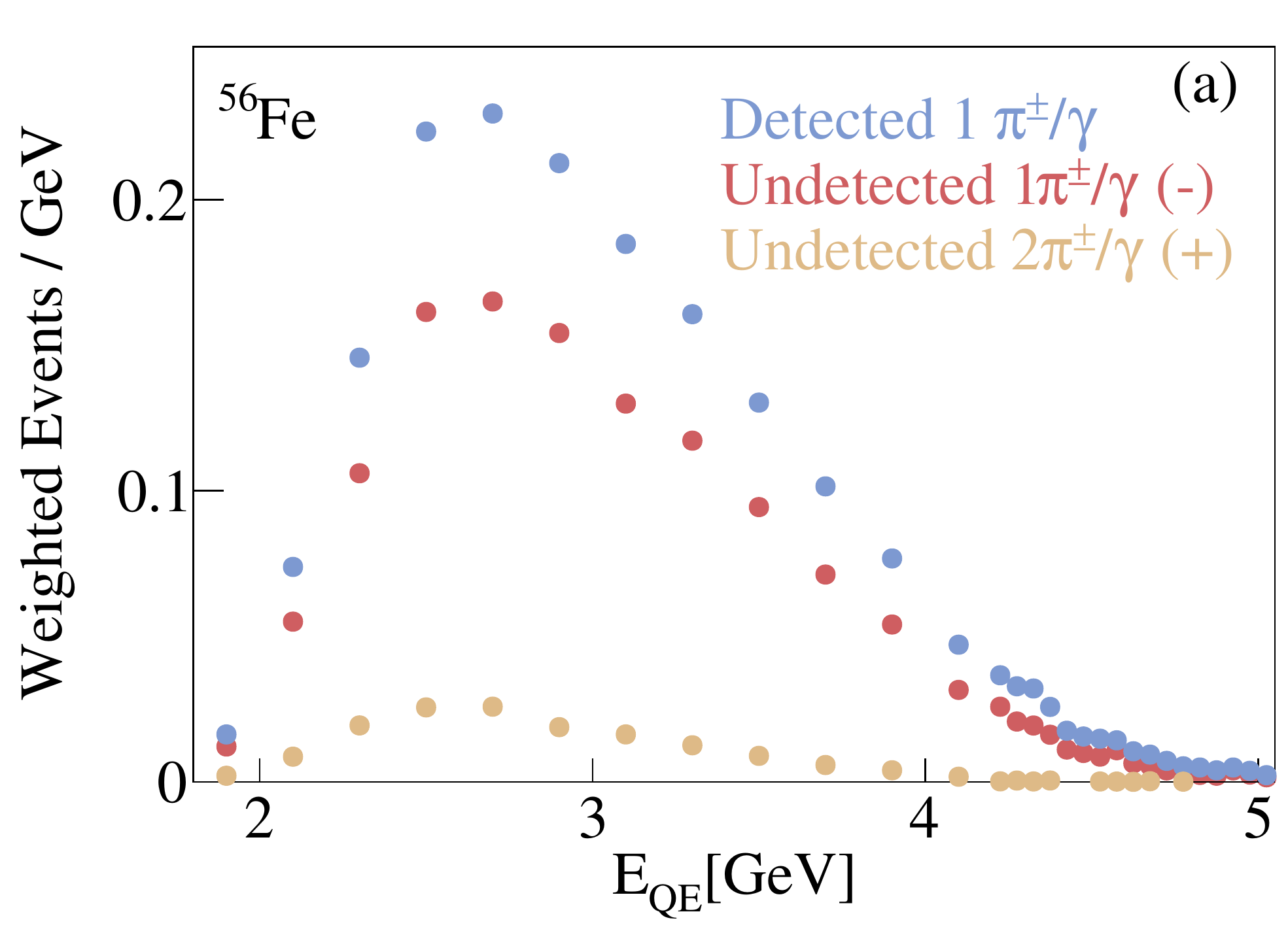}
\includegraphics[width=0.49\linewidth]{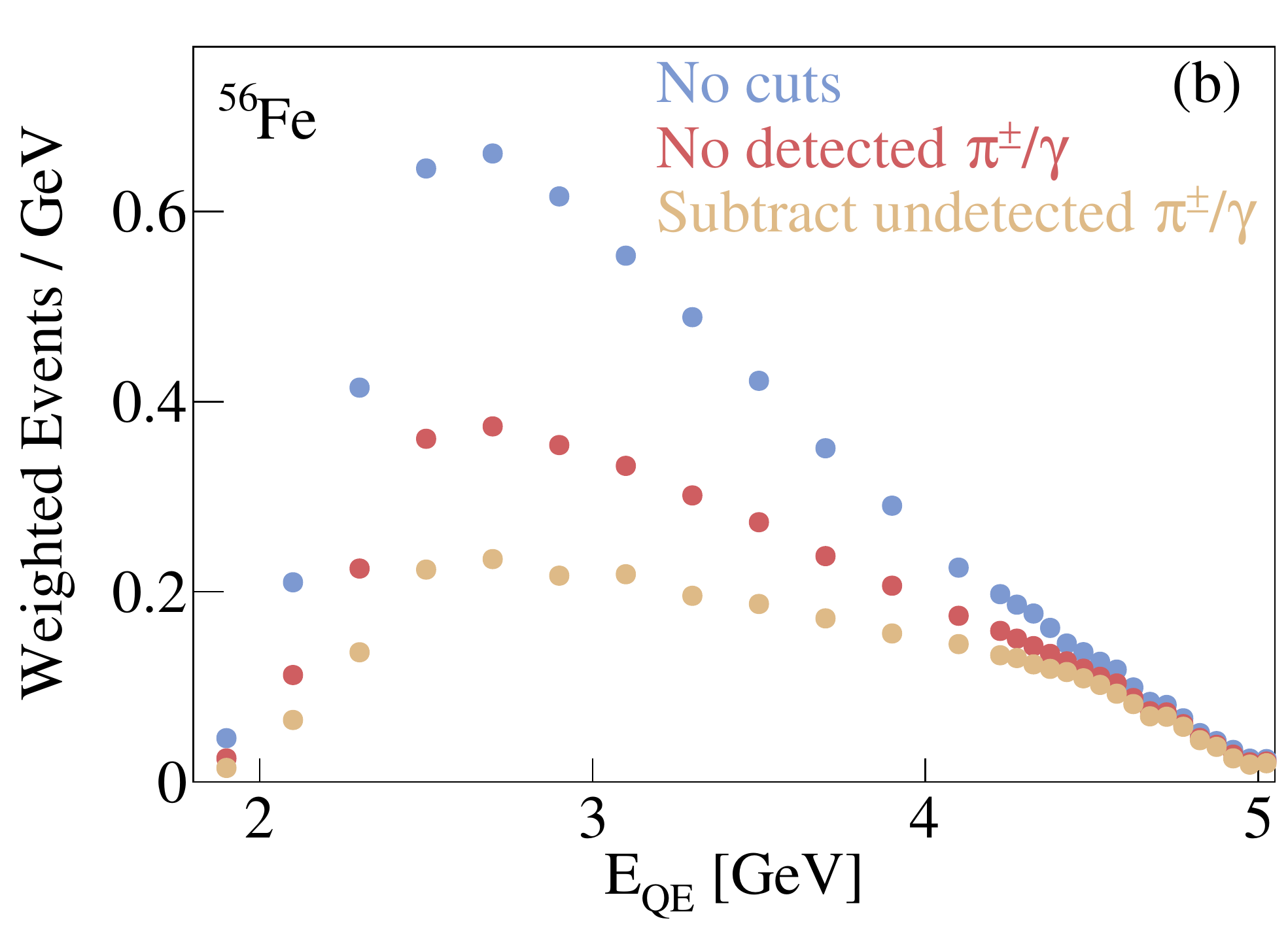}
 \end{center}
 \caption{\label{Efig:subtract} The effect of undetected pion
   subtraction.  The number of weighted events as a function of
   reconstructed energy $E_{QE}$ for 4.453~GeV Fe\ee{} events for
   (left) events with a detected $\pi^\pm$ or photon (blue), events
   with one (red) or two (light brown) undetected $\pi^\pm$ or photons
   and (right) all $(e,e'X)$ events with detected or undetected
   $\pi^\pm$ or photon (blue), \ee{} events with no detected $\pi^\pm$
   or photon (red), and \ee{} events after subtraction for undetected
   $\pi^\pm$ or photon (light brown).}
\end{figure}

In practice, the process was initiated with the highest multiplicity events.
Then, their contributions to each of the detected lower multiplicity channels were subtracted.  
The process was repeated recursively by rotating the higher multiplicity events.
In this way, their contributions to the lower multiplicity channels were determined and subtracted, and then each of the lower multiplicity channels in turn were considered.
Event multiplicities up to three pions and photons (total) for the \ee{}$_{0\pi}$ channel and up to three protons, pions and photons (total) for the  \eep$_{1p0\pi}$ channel, where the subtraction converged, were considered.  
The effects of the subtraction and its convergence can be seen in figure~\ref{Efig:subtract} for $E_{QE}$.
The number of events with an undetected $\pi^\pm$ or photon is about equal to the number of events with a detected $\pi^\pm$ or photon, consistent with the $\approx$ 50\% CLAS geometrical acceptance.  
The effect of including two $\pi^\pm$ or photon events is much less than that of the one $\pi^\pm$ or photon events and the effect of including three $\pi^\pm$ or photon events is negligible.

\begin{figure} [htb!]
\begin{center}
\includegraphics[width=0.6\linewidth]{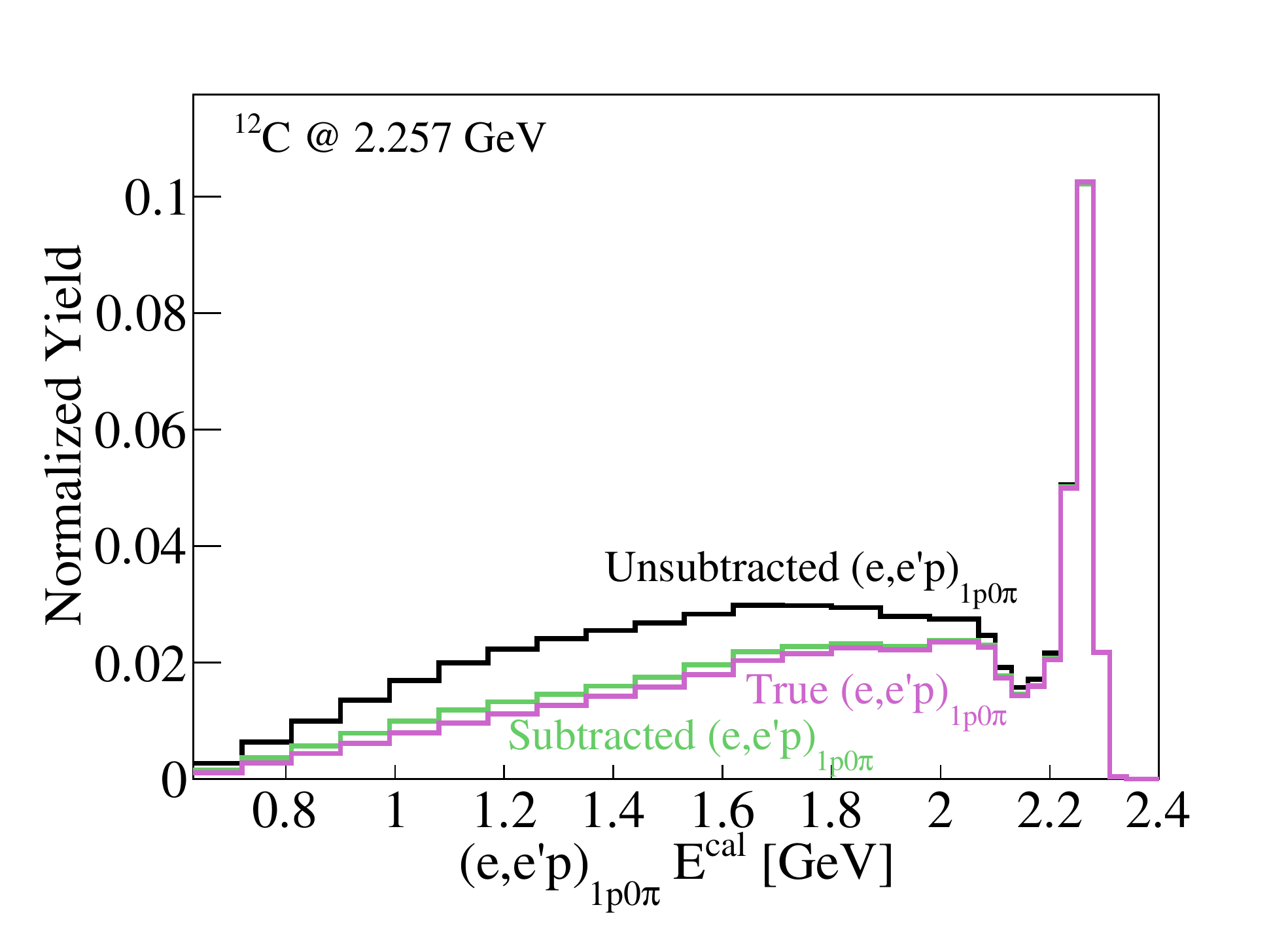}
\end{center}
\caption{\label{Efig:closure}Illustration of the successful closure test of the data driven correction for undetected particles as a function of $E^{Cal}$ using the (e,e'p)$_{1p0\pi}$ channel on $^{12}$C at $E_{beam}$ = 2.257\,GeV. The contribution of the unsubtracted (e,e'p)$_{1p0\pi}$ spectrum (black) is reduced to the subtracted (e,e'p)$_{1p0\pi}$ spectrum (magenta), which is in reasonable agreement with the true (e,e'p)$_{1p0\pi}$ spectrum (green).}
\end{figure}

\begin{figure} [htb!]
\begin{center}
\includegraphics[width=0.6\linewidth]{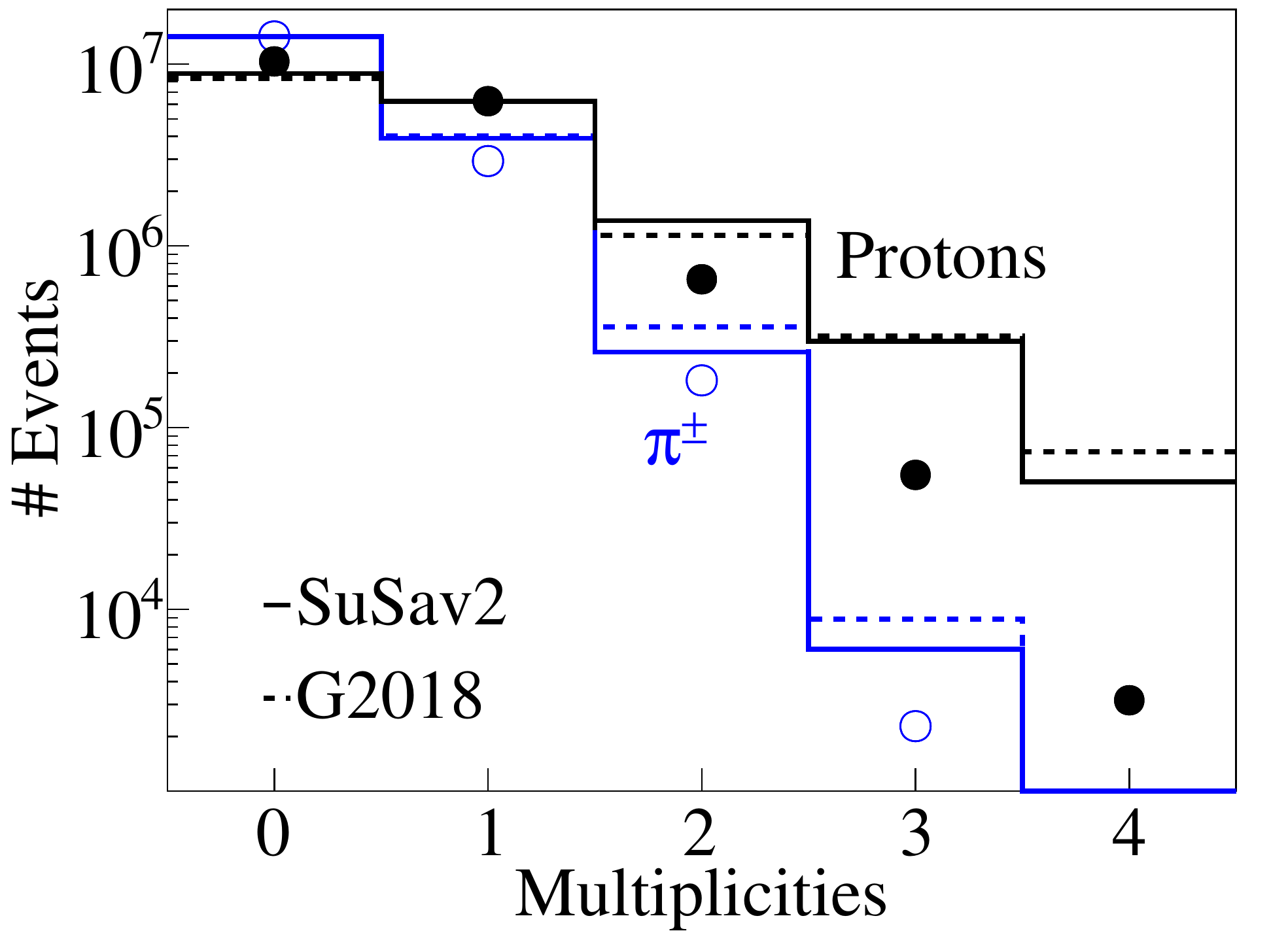}
\end{center}
\caption{\label{Efig:mult} The proton (black) and charged pion
  (blue) multiplicities for data (points), SuSav2 (solid histogram)
  and G2018 (dashed histogram) for 2.257~GeV carbon. Error bars
  show the 68\% ($1\sigma$) confidence limits for the statistical and
  point-to-point systematic uncertainties added in quadrature. Error
  bars are not shown when they are smaller than the size of the data
  point. Normalization uncertainties of 3\% not shown.}
\end{figure}

The subtraction method was tested by applying it to eGENIE events.  
The resulting subtracted spectra agreed reasonably with the true $1p0\pi$ spectra as can be seen in figure~\ref{Efig:closure}.
The method diverges for total hadron multiplicities greater than four due to the proton and pion multiplicity differences shown in figure~\ref{Efig:mult}.  
It is clear that \eGenie{} dramatically overpredicts the number of events with large  proton and pion multiplicities.

\begin{figure} [htb!]
\begin{center}
\includegraphics[width=\linewidth]{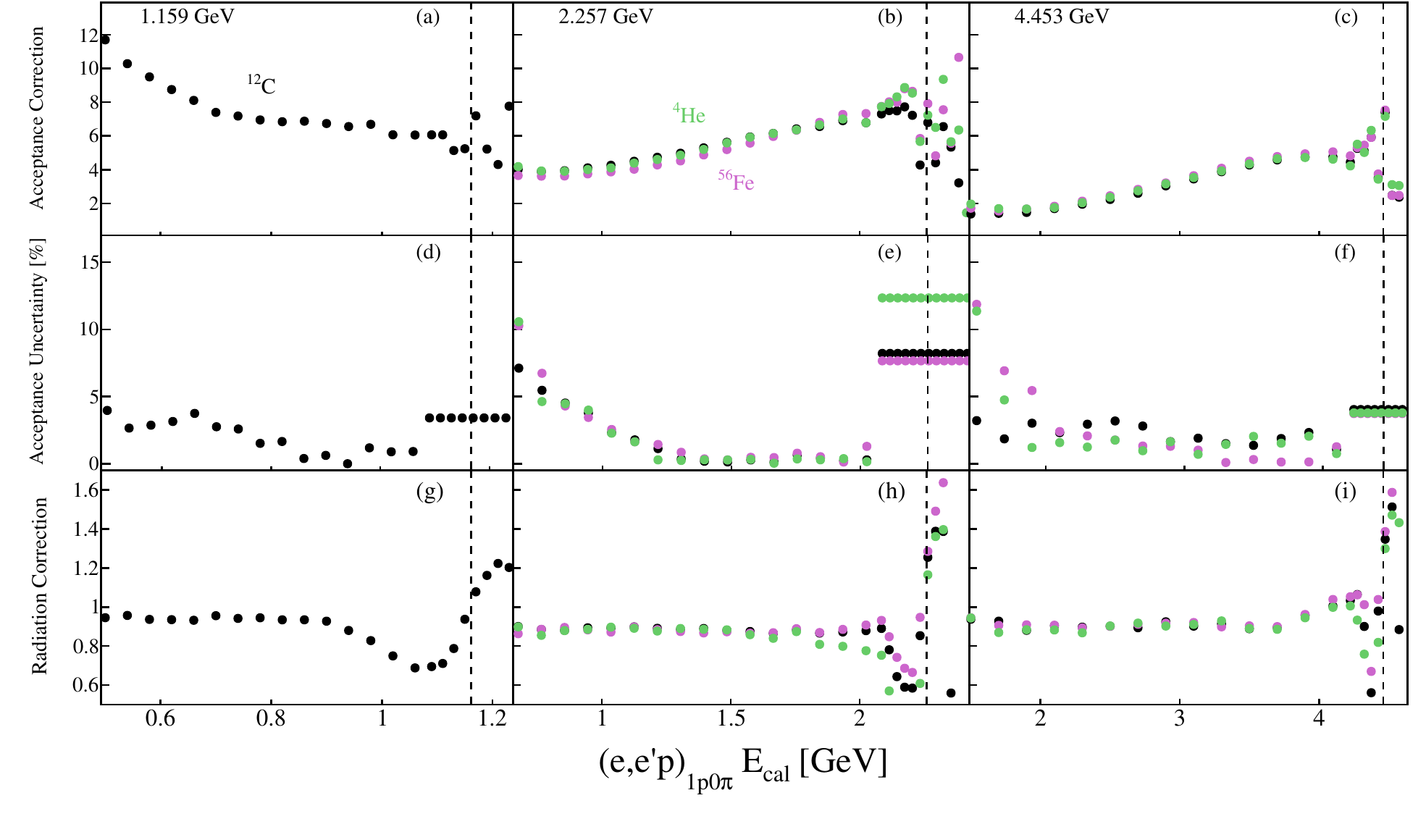}
\end{center}
\caption{\label{Efig:AccCorr} (Top row) Acceptance correction factors,
  (middle row) acceptance correction factor uncertainties, and (bottom
  row) electron radiation correction factors plotted vs E$_{cal}$ for
  the three incident beam energies.  
  Results for carbon are shown in
  black, helium in green and iron in magenta.  The left column (a,d,g)
shows the 1.1 GeV results, the middle column (b,e,h) shows the 2.2 GeV
results and the right column (c,f,i) shows the 4.4 GeV results.}
\end{figure}

\begin{figure} [htb!]
\begin{center}
\includegraphics[width=0.49\linewidth]{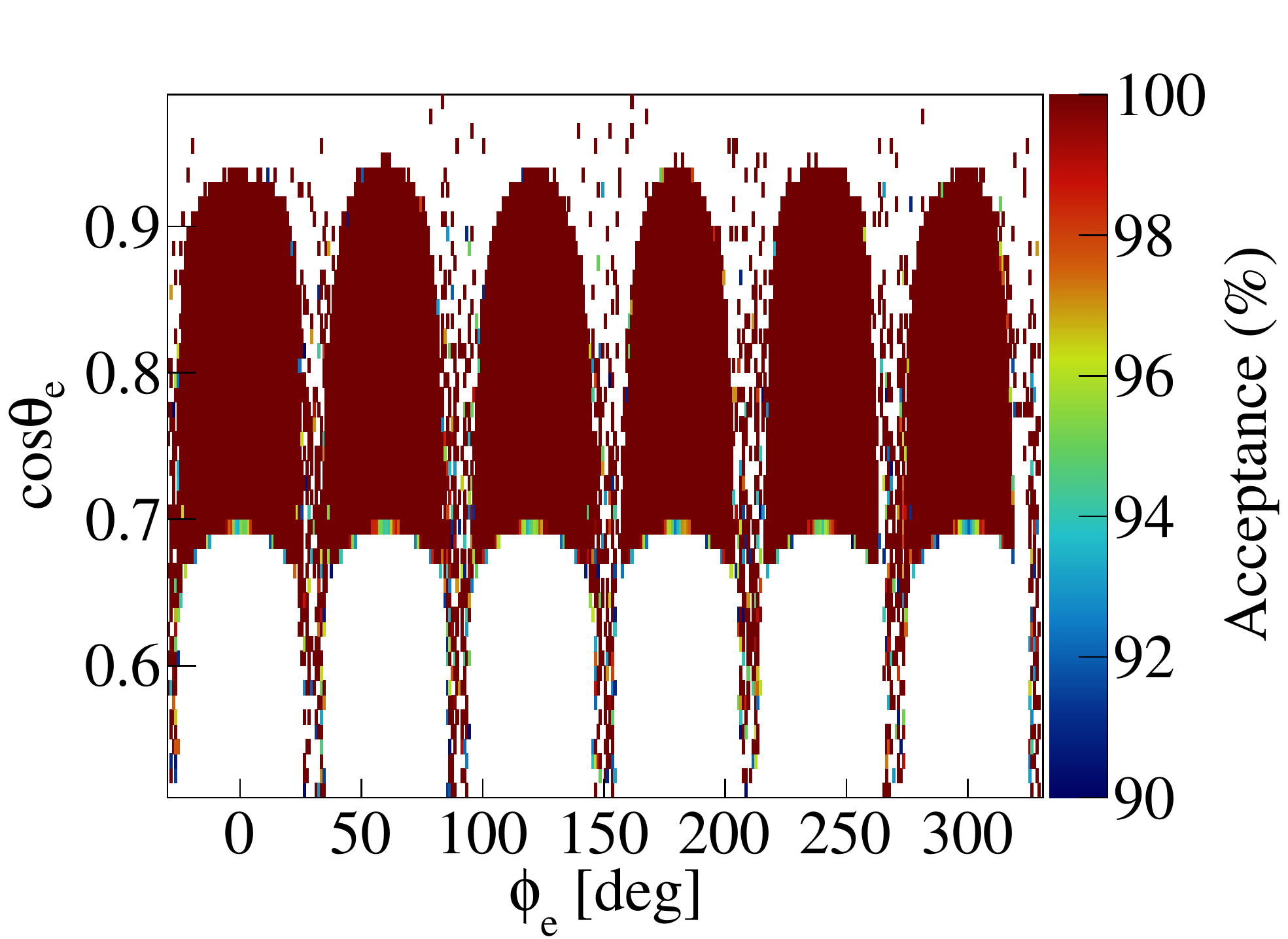}
\includegraphics[width=0.49\linewidth]{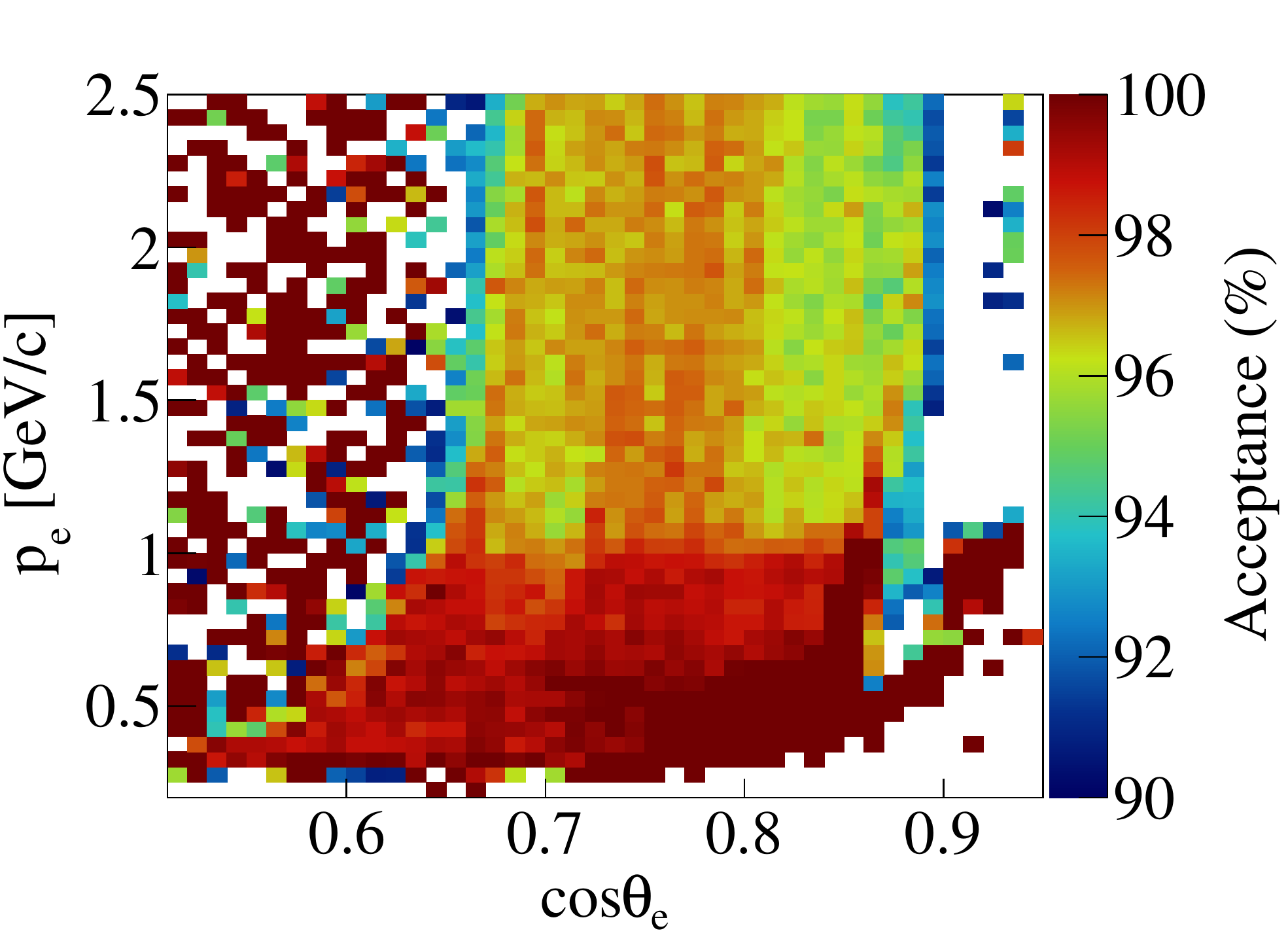}
\end{center}
\caption{\label{Efig:AccMaps}Electron acceptance maps for $^{12}$C at $E_{beam}$ = 1.159\,GeV as a function of (left) cos$\theta_{e}$ vs $\phi_{e}$ and (right) cos$\theta_{e}$ vs momentum $p_{e}$.}
\end{figure}

The CLAS acceptance maps were used to determine the probability that each particle produced by eGENIE was detected as a function of the momentum, the angular orientation, and the particle species. Figure~\ref{Efig:AccMaps} shows the electron acceptance map for $^{12}$C at $E_{beam}$ = 1.159\,GeV as a function of (left) cos$\theta_{e}$ vs $\phi_{e}$ and (right) cos$\theta_{e}$ vs momentum $p_{e}$ illustrating an acceptance greater than 90\% across the majority of the detector fiducial volume.

The particle momenta were smeared with an effective CLAS resolution.
Namely, electrons and proton momentum resolutions of 0.5\% and 1\%, respectively, for the 2.257 and 4.453 GeV data and 1.5\% and 3\% for the 1.159 GeV data, which was taken with a lower torus magnetic field, were used.

The cross section as a function of variables of interest for particles above the minimum angles shown in equations~\ref{eq:thetaemin_low}-\ref{eq:thetaemin_piminus_highP} was determined in several steps.  
All events were first weighted by a factor of $Q^4$ to account for the major difference in electron- and neutrino-nucleus scattering. 
The number of weighted events was then determined and corrected (if appropriate) for events with undetected pions, photons, and extra protons.
The background-subtracted event distribution was divided by the number of target nuclei per area and the number of incident beam electrons to get the normalized yield.
The delivered integrated beam charge was measured using the CLAS Faraday Cup~\cite{Mecking:2003zu}.
Electron radiation effects were corrected for by multiplying the resulting spectra by the ratio of \eGenie{} without electron radiation divided by \eGenie{} with electron radiation, as shown in panels g-i of figure~\ref{Efig:AccCorr}.  
This includes a multiplicative factor to account for the effects of internal radiation.
Electron and proton acceptance and other detector effects were corrected for using \eGenie.
The acceptance correction factor is the ratio of the number of true signal events without detector effects to the number of true signal events with detector effects.  
The detector effects included momentum resolution, fiducial cuts, and acceptance map effects.  
The fiducial cuts determine the useful areas of the detector as a function of particle momenta and angles, and the acceptance maps describe the efficiency of the detector as  a function of particle momenta and angles.  
This factor corrects the effective electron and proton solid angles to almost $4\pi$.  
It excludes all electrons, pions and protons below their minimum angles defined in equation~\ref{eq:thetaemin_low}-\ref{eq:thetaemin_piminus_highP}.
The acceptance correction factor was obtained using both G2018 and SuSav2 shown in panels a-c of figure~\ref{Efig:AccCorr} as the bin-by-bin average of the two configurations.
The G2018 results were shifted so that the energy reconstruction peaks lined up at the correct beam energy.
Finally, the bin width division was taken into account.

\begin{figure} [htb!]
\begin{center}
\includegraphics[width=0.7\linewidth]{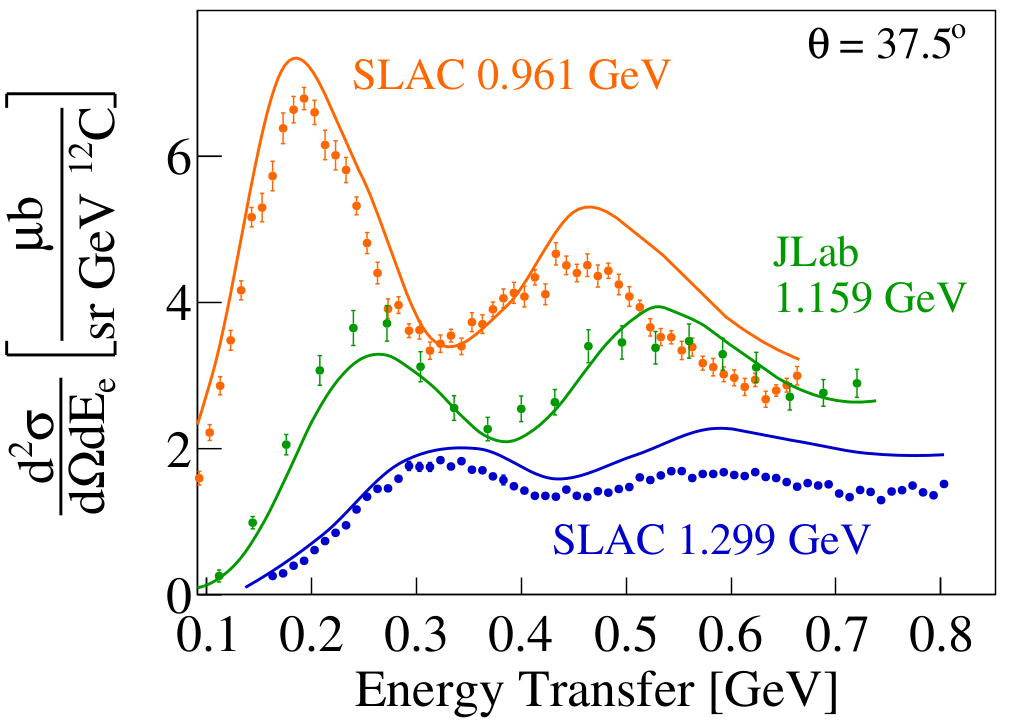}
\end{center}
\caption{\label{Efig:incl}
  Comparison between the inclusive C$(e,e')$ cross sections measured
  at 37.5$^\circ$ for data (points) and SuSav2 (lines) for the 0.961
  and 1.299 GeV SLAC data and our 1.159 GeV
  CLAS data.}
\end{figure}

In order to perform a sanity check of our cross-section extraction procedure, the inclusive 1.159 37.5$^\circ$ GeV C$(e,e')$ cross section was determined as follows:

\begin{eqnarray}
\frac{d\sigma}{d\Omega d\omega} = \frac{N_e}{\Delta\Omega N_i N_t}
\label{inclusive_xsec}
\end{eqnarray}

where $N_e$ is the number of detected electrons in Sector 1 within $36^\circ \le \theta_e\le 39^\circ$ and a $12^\circ$ range in $\phi_e$, $\Delta\Omega = \sin\theta_e d\theta_e d\phi_e = 6.68$ msr, $N_i$ is the number of incident electrons, and $N_t= 0.179$ g/cm$^2 = 8.95\times10^{-9}$ nuclei/$\mu$b.   
The extracted cross section was further compared to measurements from SLAC~\cite{PhysRevLett.62.1350}, as can be seen in  figure~\ref{Efig:incl}.
The measured JLab cross section is in reasonable agreement with the GENIE predictions and also consistent with the SLAC measurements at lower and higher energies~\cite{PhysRevLett.62.1350}.

Several major sources of systematic uncertainties were considered, including the angular dependence of the pion-production cross section for the undetected-pion subtraction, the effects of fiducial cuts on undetected particle subtraction, photon identification cuts, the sector-to-sector variation of the data to eGENIE ratio, the model-dependence of the acceptance correction, and uncertainties in the normalization measurement.
Table~\ref{tab:syst} shows the summary of the total systematic uncertainties used in the e4$\nu$ analysis.

\begin{table}[htb!]
\centering
\begin{tabular}{|c|c|}
\hline
Source   & Uncertainty (\%) \\ 
\hline
\makecell{Detector acceptance\\Identification cuts\\Number of rotations\\$\phi_{q\pi}$ cross-section dependence} & \makecell{2, 2.1, 4.7\\(@1.1,2.2,4.4\,GeV)} \\
\hline
Sector dependence & 6 \\
\hline
Acceptance correction & 2-15\\
\hline
Overall normalization & 3\\
\hline
Electron inefficiency & 2\\
\hline
\end{tabular}
\caption{Summary of the total systematic uncertainties used in the e4$\nu$ analysis.}
\label{tab:syst}	
\end{table}
  
When events containing pions were rotated around the momentum transfer vector, the cross section was assumed to not depend on $\phi_{q\pi}$.  
The $\phi_{q\pi}$ independence of the pion-production cross section was tested by weighting the subtraction using the measured $\phi_{q\pi}$-dependent $H(e,e'p\pi)$ cross sections of reference~\cite{Markov20}.  
This changed the subtracted spectra by about 1\% and was included as a systematic uncertainty.

The subtraction of events with undetected pions depends on the CLAS acceptance for such particles.  
The final spectrum should be independent of the CLAS pion acceptance.  
The effect of varying the CLAS acceptance on the undetected particle subtraction was estimated by comparing the results using the nominal fiducial cuts and using fiducial cuts with the $\phi$ acceptance in each CLAS sector reduced by $6^\circ$ or about 10-20\%.
This changed the resulting subtracted spectra by about 1\% at 1.159 and 2.257~GeV and by 4\% at 4.453~GeV.
This difference was included as a point-to-point systematic uncertainty. 
  
The photon identification cuts were also varied.  
Photons were also identified as neutral particle hits in the calorimeter with a velocity greater than $2\sigma$ ($3\sigma$ at 1.159~GeV) below the mean of the photon velocity peak at $v=c$.  
This limit was varied by $\pm0.25\sigma$.  
This gave an uncertainty in the resulting subtracted spectra of 0.1\%, 0.5\% and 2\%  at 1.159, 2.257 and 4.453~GeV, respectively. 

CLAS had six almost identical sectors. 
The primary difference among the sectors is the distribution of dead detector channels.  
These dead channels were accounted for in our fiducial cuts and in our acceptance maps, where the effect of the dead detectors on the particle detection efficiency was measured and applied that efficiency to the particles generated in the eGENIE simulation. 
If our fiducial cuts and acceptance maps completely accounted for the effect of the dead and inefficient detector channels, then the ratio of data to eGENIE should be the same for all six sectors. 
Sectors with anomalous data to eGENIE ratios were discarded.
More precisely, sectors 3/5 were discarded at $E_{beam}$ = 1.159\,GeV, as well as sectors 3/4/5 at $E_{beam}$ = 2.257\,GeV.
All the sectors were used at $E_{beam}$ = 4.453\,GeV.
The variance of the ratios for the remaining sectors was used as a measure of the uncertainty in the measured normalized yields.  
This gave a point-to-point systematic uncertainty of 6\%.

The acceptance correction factor uncertainty was obtained using both G2018 and SuSav2 and their bin-by-bin difference divided by $\sqrt{12}$.
The uncertainty was averaged over the entire peak to avoid large uncertainties due to small misalignments, as shown in panels d-f of figure~\ref{Efig:AccCorr}.

The overall normalization was determined using inclusive 4.4 GeV H$(e,e')$ measurements.  
The measured and simulated H$(e,e')$ cross sections agreed to within an uncertainty of 3\%, which is used as a normalization uncertainty~\cite{osipenkoThesis}.

The statistical uncertainty and the point-to-point systematic uncertainties were added in quadrature and displayed on the data points.
The total point-to-point systematic uncertainties ranged between 7-25\%, with the largest uncertainties for the smallest cross sections.

%%%%%%%%%%%%%%%%%%%%%%%%%%%%%%%%%%%%%%%%%%%%%%%%%%%%%%%%%

\section{Incident Energy Reconstruction Results}\label{e4vEreco}

There are two general approaches for reconstructing the incident neutrino energy based on the particle detection capabilities of the neutrino detector.

Water Cherenkov detectors only measure charged leptons and pions.
If the neutrino scattered quasi-elastically from a stationary nucleon in the nucleus, its energy can be reconstructed from the measured lepton as:

\begin{equation}
E_{QE} = \frac{2M_N\epsilon + 2M_N E_l - m_l^2}{2(M_N-E_l+k_l\cos\theta_l)},
\label{eq:eqe}
\end{equation}

where $\epsilon \approx 20$~MeV is the average nucleon separation energy, $M_N$ is the nucleon mass, and $(m_l, E_l, k_l, \theta_l)$ are the scattered lepton mass, energy, momentum, and angle.  

\begin{figure} [htb!]
\begin{center}
\includegraphics[width=0.7\linewidth]{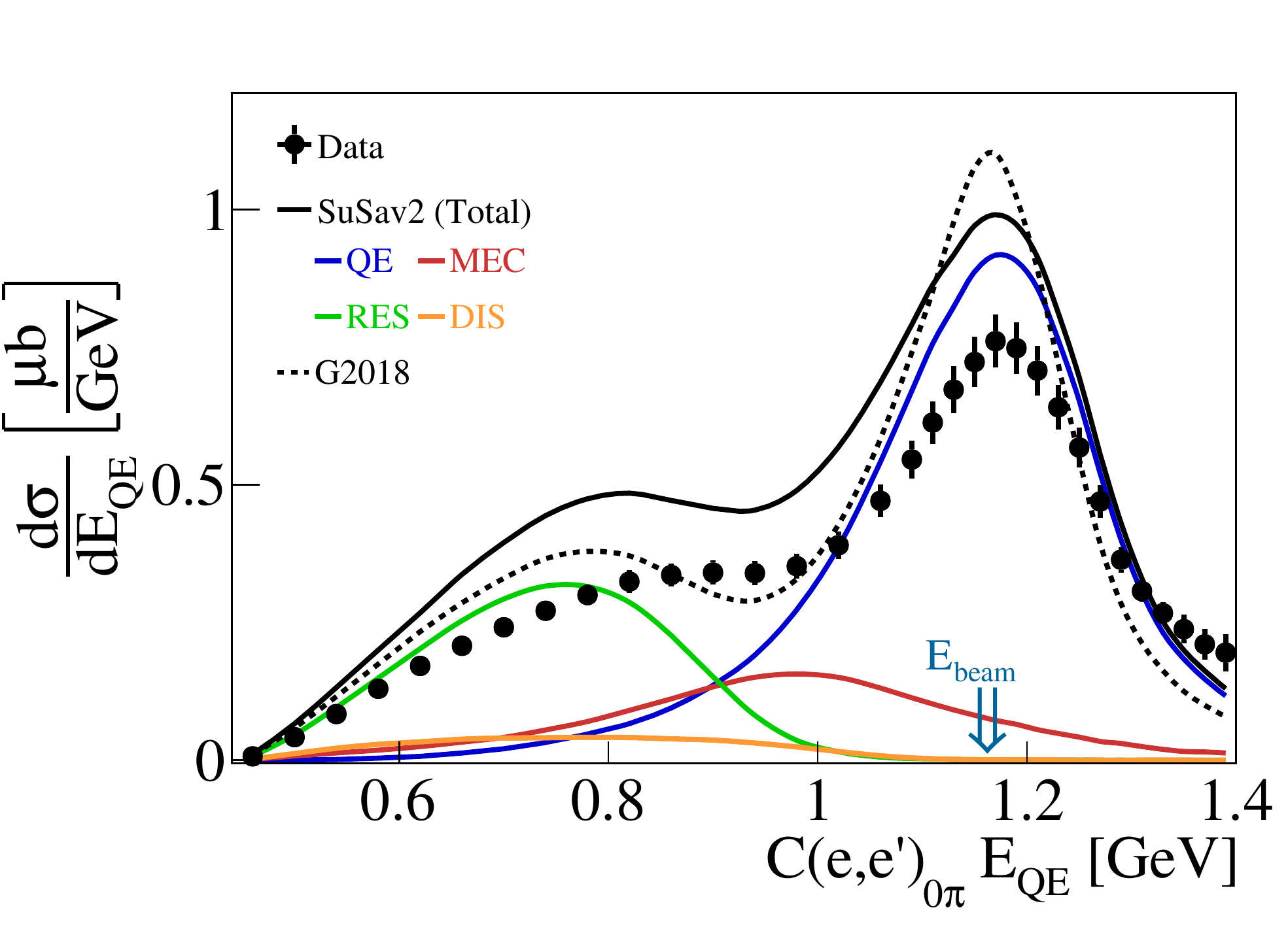}\hspace{-0.1cm}
\end{center}
\caption{\label{fig:Erec}The 1.159~GeV C\ee$_{0\pi}$ cross section
    plotted as a function of the reconstructed energy
    $E_{QE}$ for data (black points), GENIE SuSAv2 (solid black curve)
    and GENIE G2018 (dotted black curve). The
    colored lines show the contributions of different processes to the
    GENIE SuSAv2 cross section: QE (blue), MEC (red), RES (green) and DIS
    (orange).    Error bars show the 68\% (1$\sigma$) confidence
    limits for the statistical and point-to-point
    systematic uncertainties added in quadrature.  Error
    bars are not shown when they are smaller than the size of the data
    point. Normalization uncertainty of 3\% not shown.
    }
\end{figure}

Figure~\ref{fig:Erec} shows the $E_{QE}$ distribution for 1.159~GeV C\ee{}$_{0\pi}$ events, which are most relevant for T2K and HK.  
A broad peak is observed centered at the real beam energy with a large tail extending to lower energies.  
The peak is doppler-broadened by the  motion of the nucleons in the nucleus.  
The tail is caused by non-QE reactions that pass the $\ee_{0\pi}$ selection.
The tail is cut off at the lowest energies by the CLAS minimum detected electron energy of 0.4~GeV. 
The \susa{} \eGenie{} peak has the correct width, but is somewhat larger than the data.  
It overestimates  the tail by about 25\%.  
The G2018 eGENIE peak also exceeds the data, but is too narrow, with a Gaussian width of $\sigma=76$~MeV, compared to 89~MeV for the data.
This is due to inexact modeling of the nuclear ground state momentum distribution.  
The tail dips below the data at around 0.9~GeV, and is larger than the data at lower reconstructed energies.  
Neither model describes the data quantitatively well.

\begin{comment}
Figure~\ref{fig:EqeSlices} shows the same mismodelling in terms of the $E_{QE}$ resolution in specific regions of the phase space as a function of the outgoing lepton energy and scattering angle.

\begin{figure} [htb!]
\begin{center}
    \includegraphics[width=\linewidth]{\figures 12C_1_161_T2KEQEResoCanvas_EePrimeAndCosThetaSlices_2D.pdf}
\end{center}
\caption{\label{fig:EqeSlices}The $A\eep_{0\pi}$ cross section plotted as a
    function of the reconstructed quasielastic energy $E_{QE}$ in energy and angular slices simultaneously for
    data (black points), SuSAv2 (black solid curve) and G2018 (black
    dotted curve).}
\end{figure}
\end{comment}

\begin{figure} [htb!]
\begin{center}
    \includegraphics[width=\linewidth]{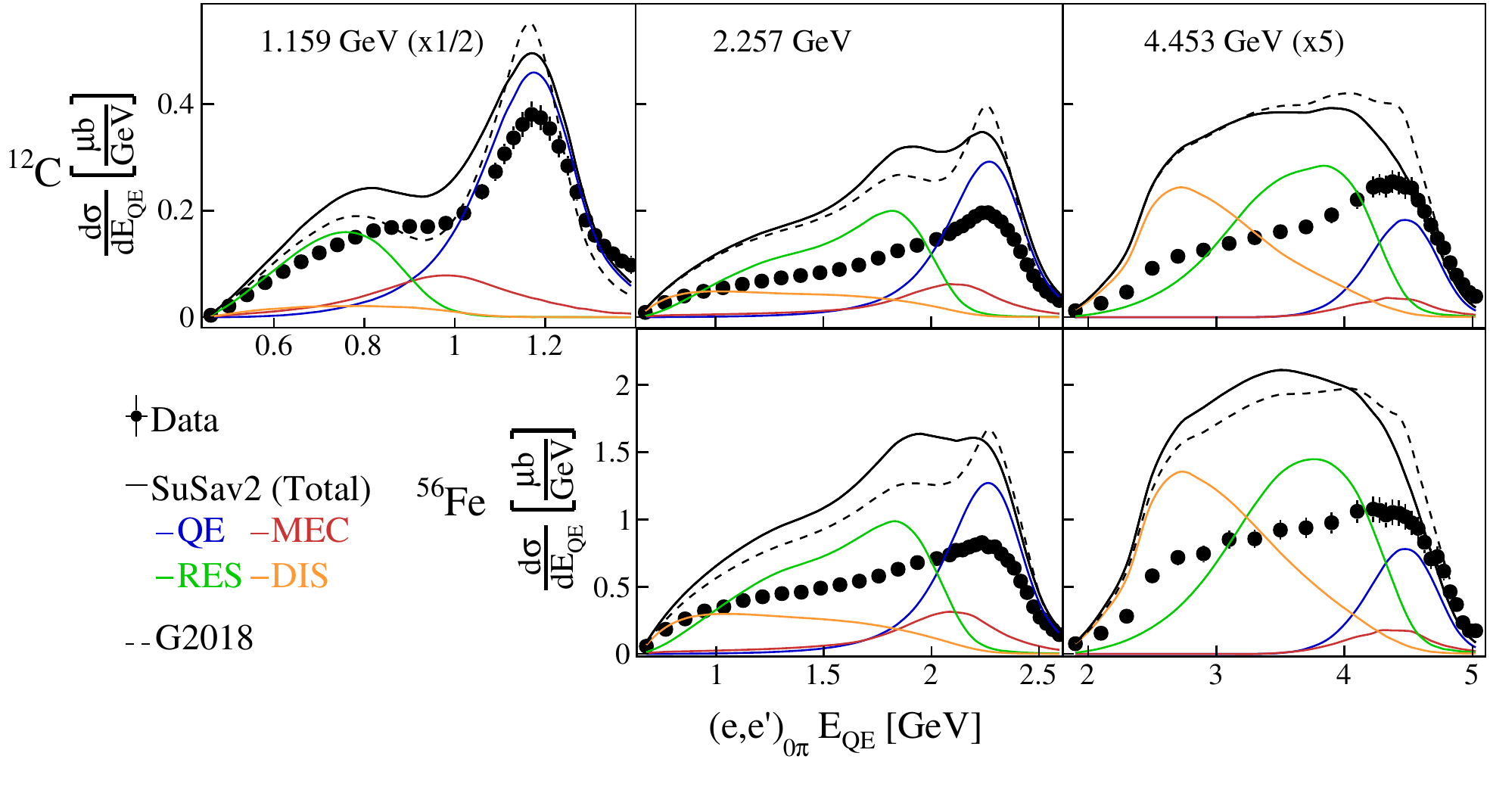}
\end{center}
\caption{\label{fig:Eqe}The $A\eep_{0\pi}$ cross section plotted as a
    function of the reconstructed quasielastic energy $E_{QE}$ for
    data (black points), SuSAv2 (black solid curve) and G2018 (black
    dotted curve).  Different panels show results for different beam
    energy and target nucleus combinations: (top row) Carbon target
    at (left to right) 1.159, 2.257 and 4.453~GeV, and (bottom) Iron
    target at (left) 2.257 and (right) 4.453~GeV incident beam.  The
    1.159 GeV yields have been scaled by 1/2 and the 4.453 GeV yields have
    been scaled by 5 to have the same vertical scale. Colored lines
    show the contributions of different processes to the SuSAv2 GENIE
    simulation: QE (blue), MEC (red), RES (green) and DIS (orange).
    Error bars show the 68\% (1$\sigma$) confidence limits for the
    statistical and point-to-point systematic uncertainties added in
    quadrature.  Error bars are not shown when they are smaller than
    the size of the data point. Normalization uncertainties of 3\%
    not shown.}
\end{figure}

Figure~\ref{fig:Eqe} shows the cross section as a function of  $E_{QE}$ for 1.159, 2.257 and 4.453~GeV C\ee{}$_{0\pi}$ events and 2.257 and 4.453~GeV Fe\ee{}$_{0\pi}$ events.
It is clear that the mismodeling already observed in the 1.159~GeV C\ee$_{0\pi}$ sample becomes even more pronounced for higher energies and heavier nuclei.

Tracking detectors measure all charged particles above their detection thresholds.  
The ``calorimetric'' incident neutrino energy is then the sum of all the detected particle energies:

\begin{equation}
E_{cal} = \sum E_i + \epsilon,
\label{eq:ecal}
\end{equation}

where $E_i$ are the detected nucleon kinetic energies and the lepton and meson total energies and $\epsilon$ is the average total removal energy for the detected particles.
This quantity $\epsilon$, used in reconstructing the incident energies in equations~\ref{eq:eqe} and~\ref{eq:ecal}, was determined from the data.  
It is defined using the difference in the binding energies for knocking a proton out of nucleus $A$ as $\epsilon = \vert M_A - M_{A-1} - m_p\vert + \Delta\epsilon$.  
The removal energy correction $\Delta\epsilon$ was adjusted so that the peaks in the $E_{cal}$ spectrum for low transverse missing momentum events reconstructed to the correct beam energy.  
It was found that $\Delta\epsilon=5$ and 11\,MeV for $^{12}$C and $^{56}$Fe, respectively, which are consistent with average excitation energies from single-nucleon knockout from nuclei.

\begin{figure} [htb!]
\begin{center}
    \includegraphics[width=\linewidth]{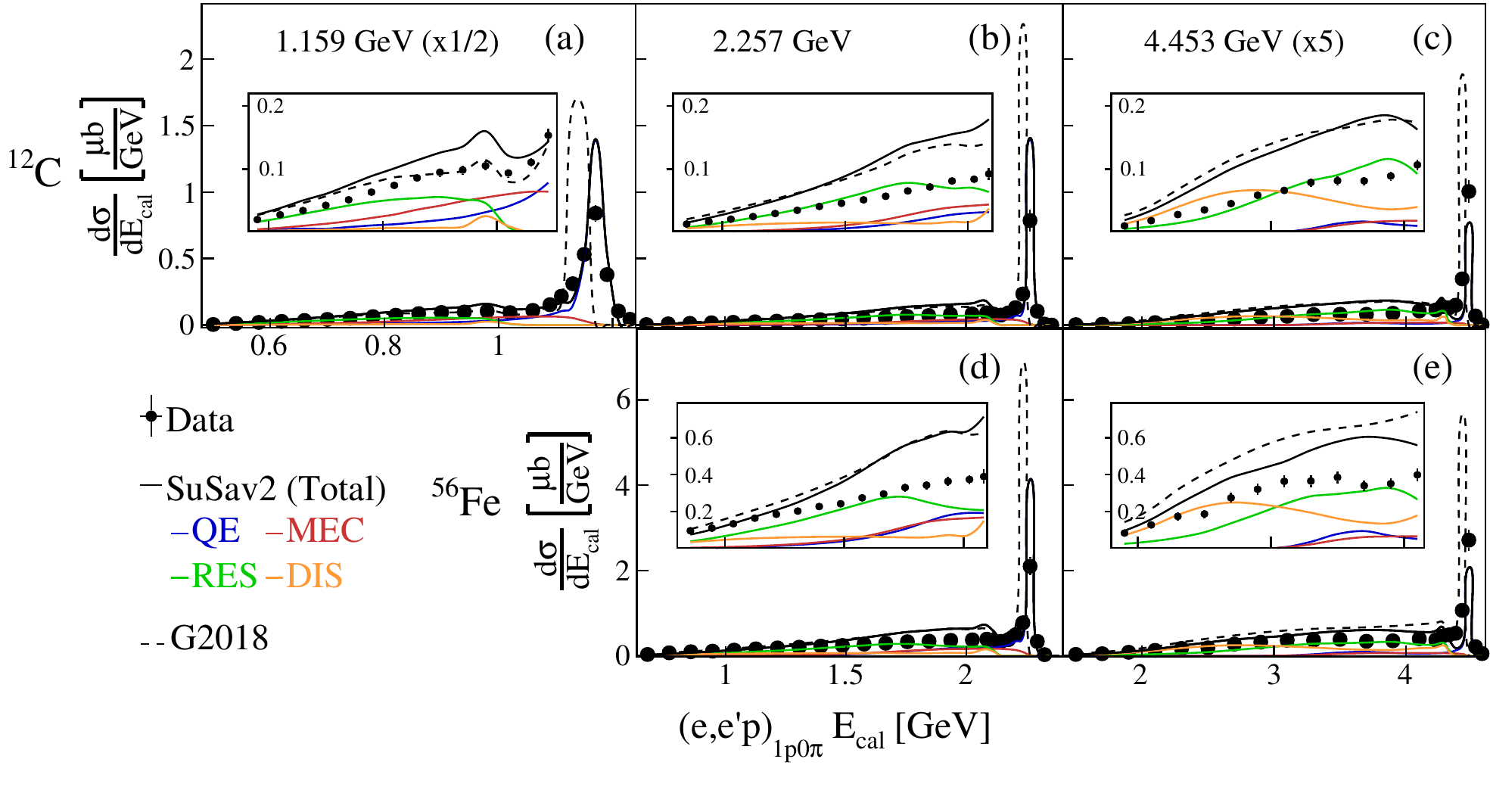}
\end{center}
\caption{\label{fig:Ecal}The $A\eep_{1p0\pi}$ cross section plotted as a
    function of the reconstructed calorimetric energy $E_{cal}$ for
    data (black points), SuSAv2 (black solid curve) and G2018 (black
    dotted curve).  Different panels show results for different beam
    energy and target nucleus combinations: (top row) Carbon target
    at (left to right) 1.159, 2.257 and 4.453~GeV, and (bottom) Iron
    target at (left) 2.257 and (right) 4.453~GeV incident beam.  The
    1.159 GeV yields have been scaled by 1/2 and the 4.453 GeV yields have
    been scaled by 5 to have the same vertical scale.  The insets show
    the cross sections with the same horizontal scale and an
    expanded vertical scale. Colored lines
    show the contributions of different processes to the SuSAv2 GENIE
    simulation: QE (blue), MEC (red), RES (green) and DIS (orange).
    Error bars show the 68\% (1$\sigma$) confidence limits for the
    statistical and point-to-point systematic uncertainties added in
    quadrature.  Error bars are not shown when they are smaller than
    the size of the data point. Normalization uncertainties of 3\%
    not shown.
    }
\end{figure}

Figure~\ref{fig:Ecal} shows the cross section as a function of  $E_{cal}$ for 1.159, 2.257 and 4.453~GeV C\eep{}$_{1p0\pi}$ events and 2.257 and 4.453~GeV Fe\eep{}$_{1p0\pi}$ events.
All spectra show a sharp peak at the real beam energy, followed by a large tail at lower energies.  
For carbon, only 30-40\% of the events reconstruct to within 5\% of the real beam energy, as illustrated in table~\ref{tab:erec}.

	\begin{table}[htb!]
	\begin{center}
	\begin{tabular}{|c|c|c|c|c|c|c|c|}
	\hline
	\multicolumn{2}{|c|}{}     & \multicolumn{2}{c|}{1.159~GeV}                                 & \multicolumn{2}{c|}{2.257~GeV}  & \multicolumn{2}{c|}{4.453~GeV}                                   \\ \cline{3-8}
	\multicolumn{2}{|l|}{}     & \multicolumn{1}{l|}{Peak} & \multicolumn{1}{l|}{Peak} & \multicolumn{1}{l|}{Peak}   & \multicolumn{1}{l|}{Peak} & \multicolumn{1}{l|}{Peak} & \multicolumn{1}{l|}{Peak} \\ 
	\multicolumn{2}{|l|}{}     & \multicolumn{1}{l|}{Fraction} & \multicolumn{1}{l|}{Sum [$\mu$b]} & \multicolumn{1}{l|}{Fraction}   & \multicolumn{1}{l|}{Sum [$\mu$b]} & \multicolumn{1}{l|}{Fraction} & \multicolumn{1}{l|}{Sum [$\mu$b]} \\ \hline
	%\multirow{3}{*}{$^{4}$He}  & Data   & -  & -    & 41  & 0.48 & 38 &  0.15  \\ 
	%			               & SuSAv2 & -  & -    & 45  & 1.31 & 22 &  0.14  \\ 
     %                          & G2018  & -  & -    & 39  & 0.93 & 24 &  0.16  \\ \hline
	\multirow{3}{*}{$^{12}$C}  & Data   & 39 & 4.13 & 31  & 1.26 & 32 &  0.34 \\  
				               & SuSAv2 & 44 & 5.33 & 27  & 1.76 & 12 &  0.20 \\ 
				               & G2018  & 51 & 6.53 & 37  & 2.44 & 23 &  0.43 \\ \hline
	\multirow{3}{*}{$^{56}$Fe} & Data   & -  & -    & 20  & 3.73 & 23 &  1.01 \\  
				               & SuSAv2 & -  & -    & 21  & 5.28 & 10 &  0.58 \\ 
				               & G2018  & -  & -    & 30  & 8.22 & 19 &  1.48 \\ \hline
	\end{tabular}
	\end{center}
 \caption{$(e,e'p)_{1p0\pi}$ events
   reconstructed to the correct beam energy.  Peak Fraction refers to the
   fraction of events reconstructed to the correct beam energy and
   Peak Sum refers to the integrated weighted cross section (as
   shown in Fig.~\ref{fig:Ecal}) reconstructed to the correct beam energy.  The
   peak integration  windows are $1.1 \le E_{cal} \le 1.22$ GeV, $2.19\le
   E_{cal} \le 2.34$ GeV, and $4.35\le E_{cal}\le 4.60$ GeV,
   respectively, for the three incident beam energies.
   \susa{} is not intended
   to model nuclei lighter than $^{12}$C.}
 \label{tab:erec}	
	\end{table}

For iron this fraction is only 20-25\%, highlighting the crucial need to well model the low-energy tail of these distributions.  
\eGenie{} overpredicts the fraction of events in the peak at 1.159 GeV and significantly underpredicts it at 4.453 GeV.
\eGenie{} using \susa{} dramatically overpredicts the peak cross section at 1.159 and 2.257 GeV, and significantly underestimates the peak cross section at 4.453 GeV, as shown in table~\ref{tab:erec}. 
eGENIE using the older G2018 models overestimates the peak cross section at all three incident energies.  
It also reconstructs the peak position (i.e. the incident energy) to be 10, 25 and 36~MeV too low for $^4$He, C and Fe, respectively, at all three beam
energies.  
This is due to an error in the G2018 QE modeling. 
This beam-energy dependence of the data-GENIE discrepancy could have significant implications for the neutrino flux reconstruction.

At 1.159 GeV, eGENIE using \susa{} slightly overpredicts the low energy tail and \eGenie{} using G2018 is reasonably close.  
Both models dramatically overpredict the low energy tail at the higher beam energies shown in the insets of figure~\ref{fig:Ecal}.  
The tail seems to be dominated by RES and DIS at 4.453\,GeV that did not result in the production of other charged particles above detection threshold.  
This overprediction has already been observed in inclusive electron scattering from the proton and deuteron, and thus appears to be due to the electron-nucleon interaction, rather than to the nuclear modeling~\cite{PhysRevD.103.113003}.

\begin{figure} [htb!]
\begin{center}
\subfloat{\includegraphics[width=0.4\linewidth]{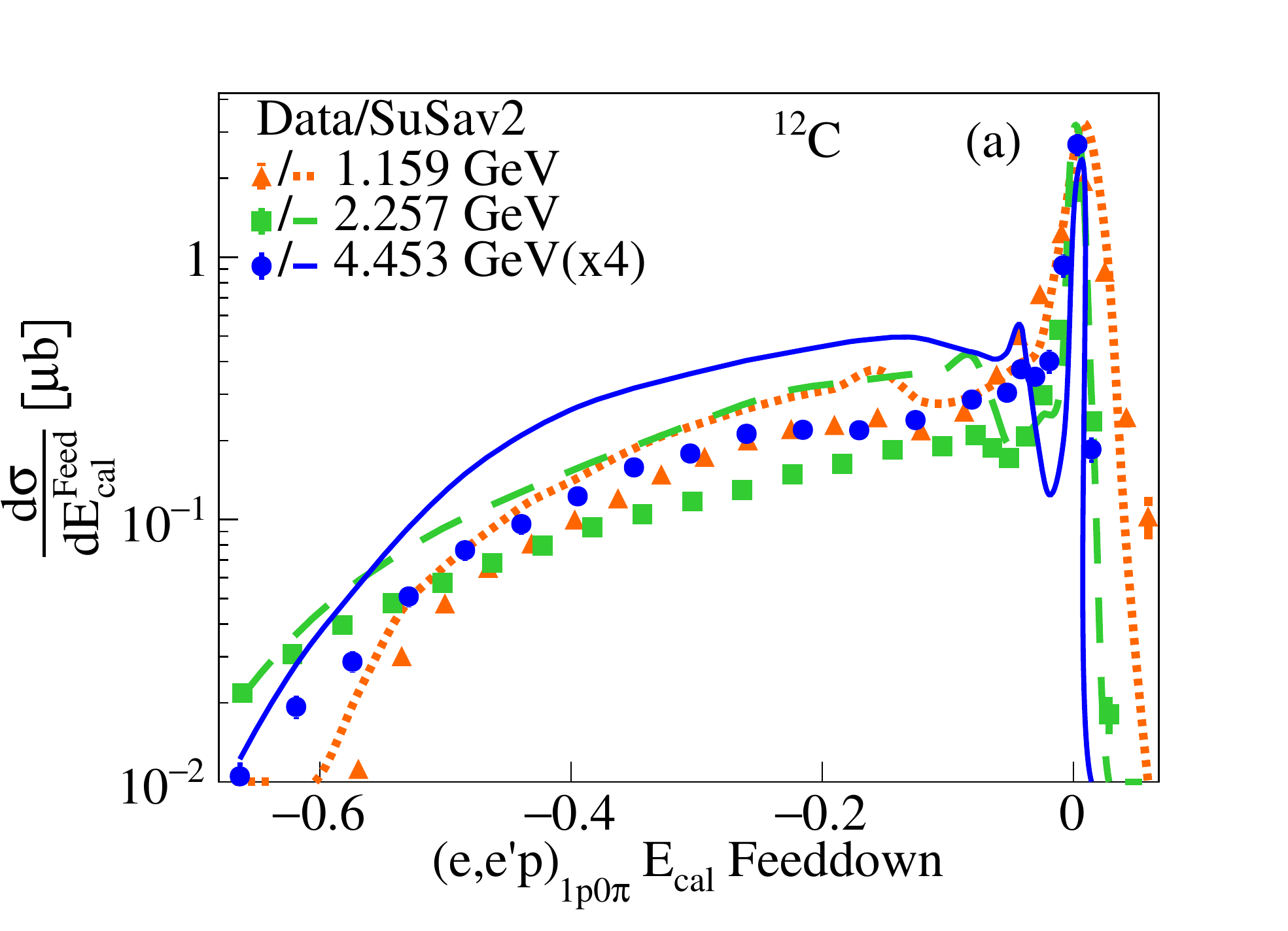}}
\subfloat{\includegraphics[width=0.4\linewidth]{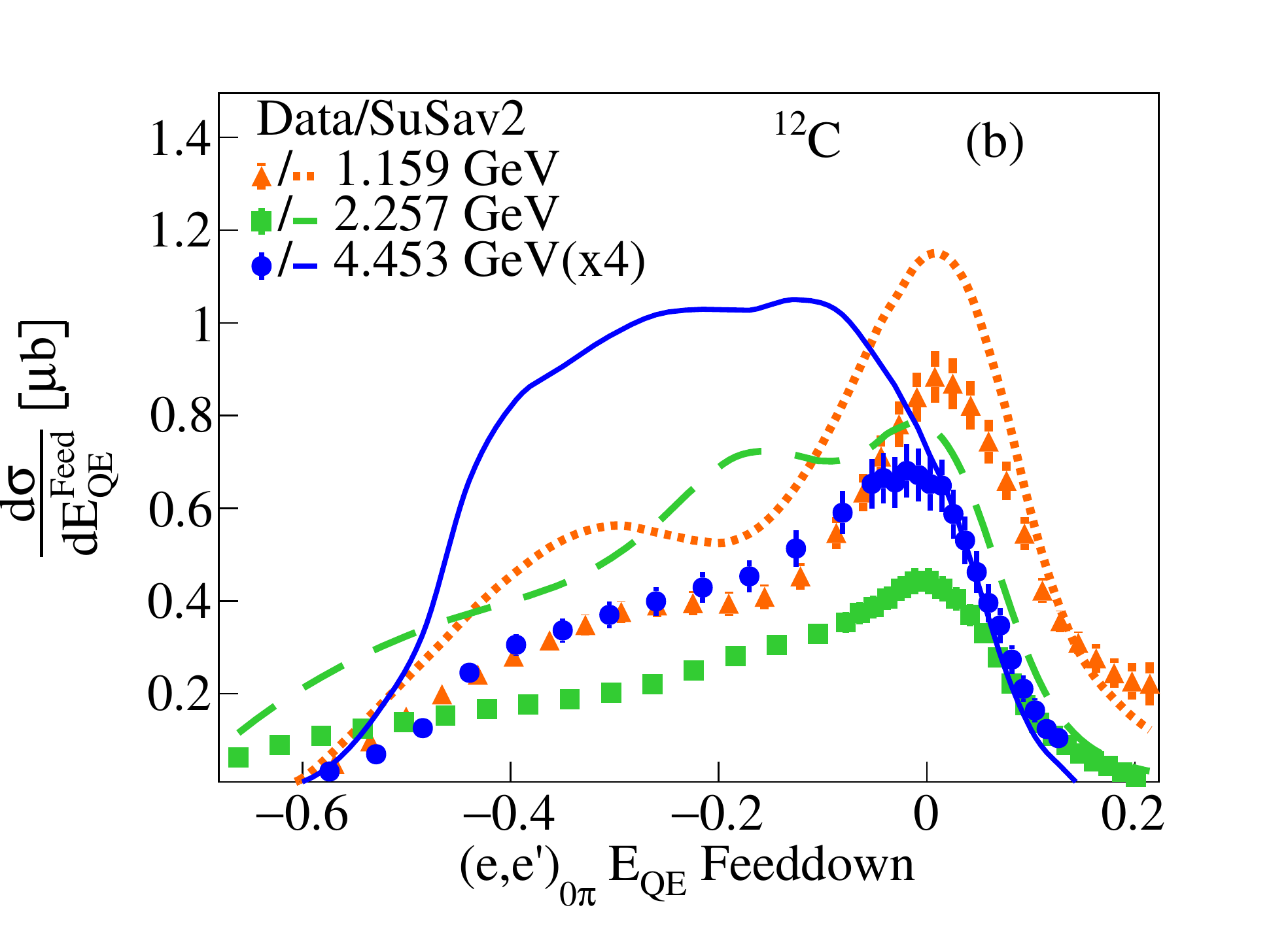}}\\
\end{center}
\caption{\label{Efig:Ereco2_C} 
Energy feed-down cross-sections 
$(E_{rec}-E_{true})/E_{true}$ 
for data
  (points) and SuSav2 (lines) for 1.159~GeV (red triangles and dotted
  lines), 2.257~GeV (green squares and dashed lines) and 4.453~GeV
  (blue dots and solid lines) on carbon for (a) $E^{cal}$, and (b) $E^{QE}$.
}
\end{figure}

\begin{figure} [htb!]
\begin{center}
\subfloat{\includegraphics[width=0.4\linewidth]{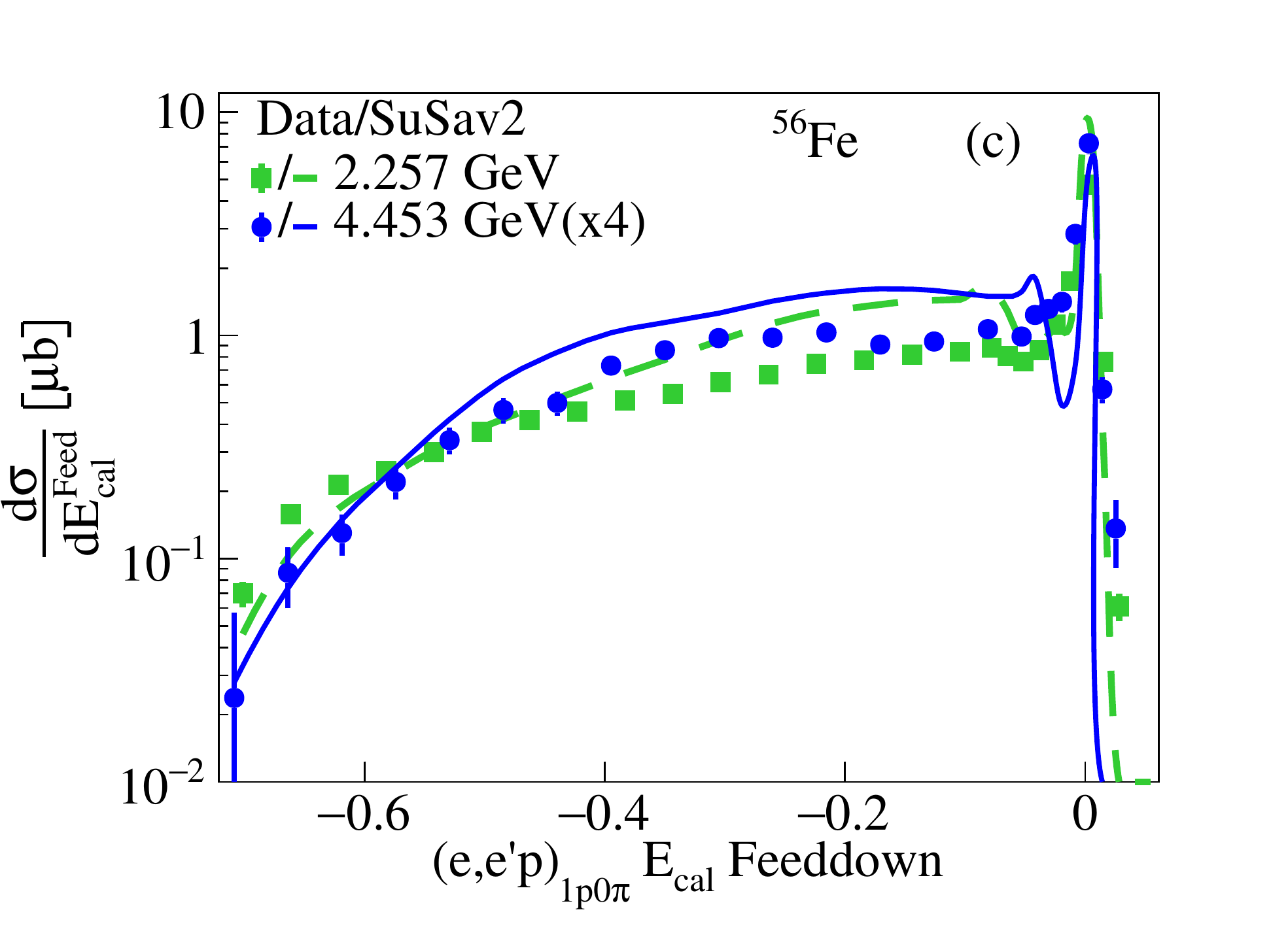}}
\subfloat{\includegraphics[width=0.4\linewidth]{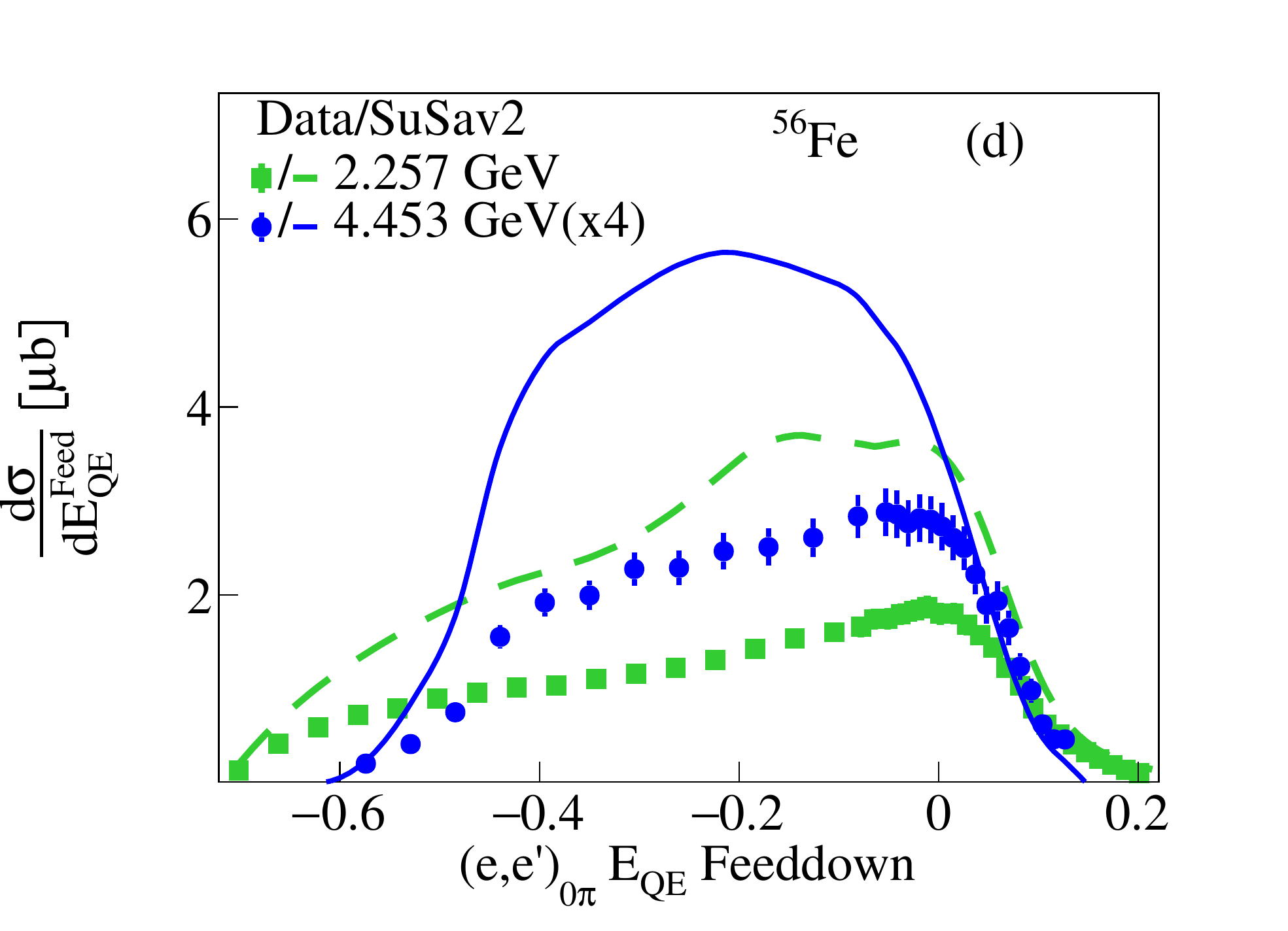}}\\
\end{center}
\caption{\label{Efig:Ereco2_Fe} 
Energy feed-down cross-sections 
$(E_{rec}-E_{true})/E_{true}$ 
for data
  (points) and SuSav2 (lines) for 1.159~GeV (red triangles and dotted
  lines), 2.257~GeV (green squares and dashed lines) and 4.453~GeV
  (blue dots and solid lines) on iron for (c) $E^{cal}$, and (d) Fe
  $E^{QE}$.
}
\end{figure}

\susa{} describes the peak cross section - the part of the cross section that reconstructs to the correct beam energy - equally well for C and for Fe, while G2018 over estimates the peak cross section more for Fe than for C. 
Both models predict a greater peak fraction relative to the data for Fe than for C, particularly at 2.2 GeV, as shown in table~\ref{tab:erec}.	
While the \ee$_{0\pi}$ QE reconstruction of equation~\ref{eq:eqe} gives a much broader peak at the true beam energy than the calorimetric energy $E_{cal}$ due to the effects of nucleon motion, as shown in figures~\ref{Efig:Ereco2_C} and~\ref{Efig:Ereco2_Fe}, it has the same tail of lower energy events for the same \eep$_{1p0\pi}$ data set.

%%%%%%%%%%%%%%%%%%%%%%%%%%%%%%%%%%%%%%%%%%%%%%%%%%%

\section{Kinematic Imbalance Results}\label{e4vSTV}

Neutrino experiments use the ``transverse variables'' (TVs) outlined in section~\ref{CC1pDataAna} to enhance their sensitivity to different aspects of the reaction mechanism.
These TVs are independent of the neutrino energy and use the momentum of the detected particles transverse to the incident lepton~\cite{PhysRevC.94.015503,PhysRevD.98.032003,PhysRevLett.121.022504} as shown in equation~\ref{STV}, where $\vec P_{T}^{\,e'}$ and $\vec P_T^p$ are the three-momenta of the detected lepton and proton perpendicular to the direction of the incident lepton, respectively.
The $\vec{P}_{T}$ vector is intended to characterize the nuclear ground state, $\delta\alpha_T$ the FSI and $\Delta\phi_T$ is intended to probe regions where MEC events dominate~\cite{PhysRevC.94.015503,PhysRevD.98.032003,PhysRevLett.121.022504}.

\begin{eqnarray}
\vec P_{T} &=&\vec P_{T}^{\,e'}+\vec P_T^p  \\
  \delta \alpha_T&=& \arccos(-\frac{\vec P_{T}^{\,e'}\cdot\vec P_{T}
  }{P_{T}^{e'} P_{T}})\\
  \delta \phi_T&=& \arccos(-\frac{\vec P_{T}^{e'}\cdot\vec P_T^p
  }{P_{T}^{e'} P_T^p})
  \label{STV}
\end{eqnarray}

Purely QE events without final state interactions, where the lepton scattered from a bound moving proton, will have small \pmperp, consistent with the  motion of the struck nucleon.  
Events with small \pmperp\,\,should thus reconstruct to the correct incident energy.
Non-QE events, where neutral or sub-detection-threshold charged particles were produced, will have larger \pmperp\,\,and will not reconstruct to the correct incident energy.  
\pmperp\,\,is thus an ideal observable for tuning reaction models to ensure they correctly account for non-QE processes.

\begin{figure} [htb!]
\begin{center}
	    \includegraphics[width=0.49\linewidth]{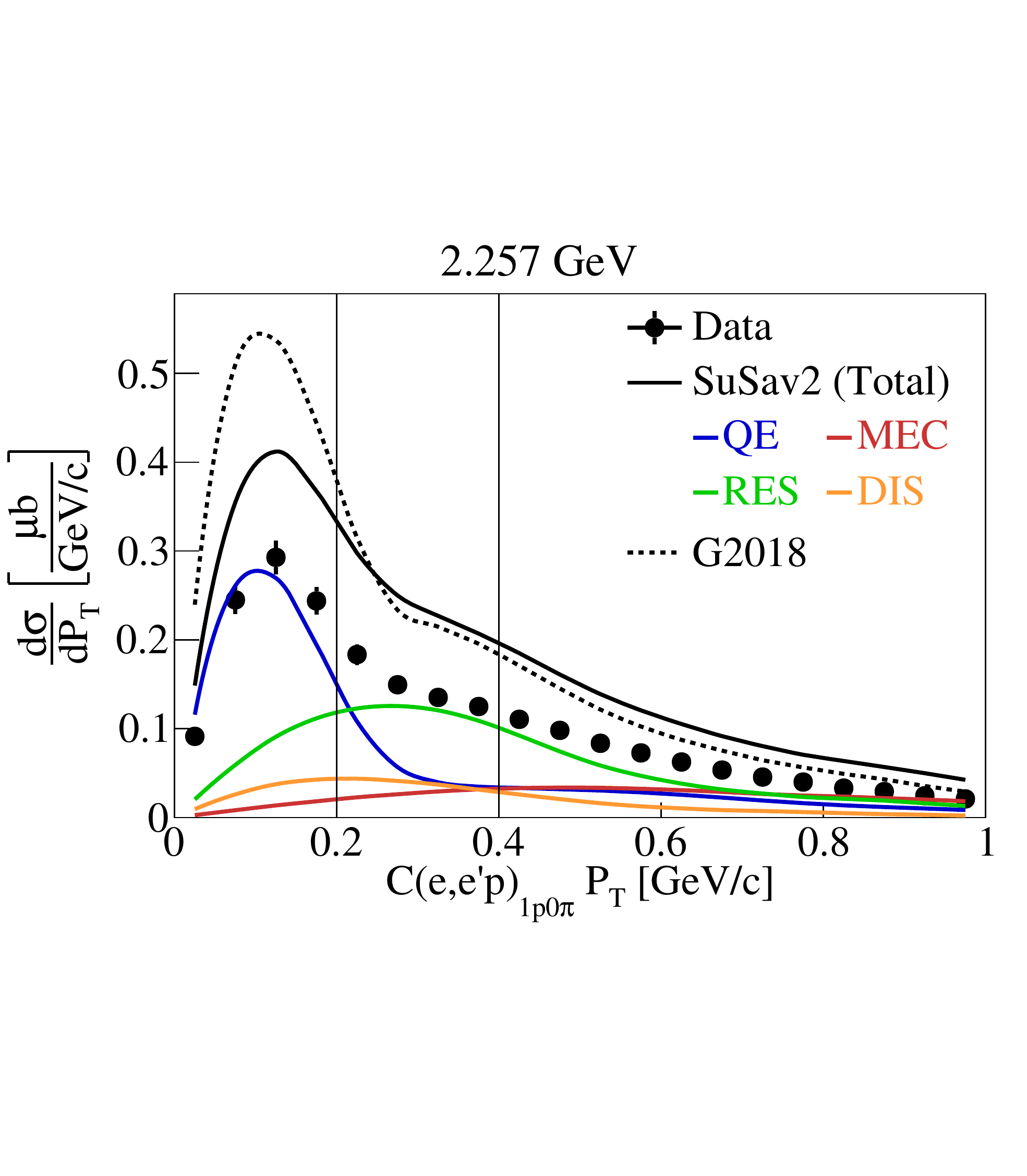}
	    \includegraphics[width=0.49\linewidth]{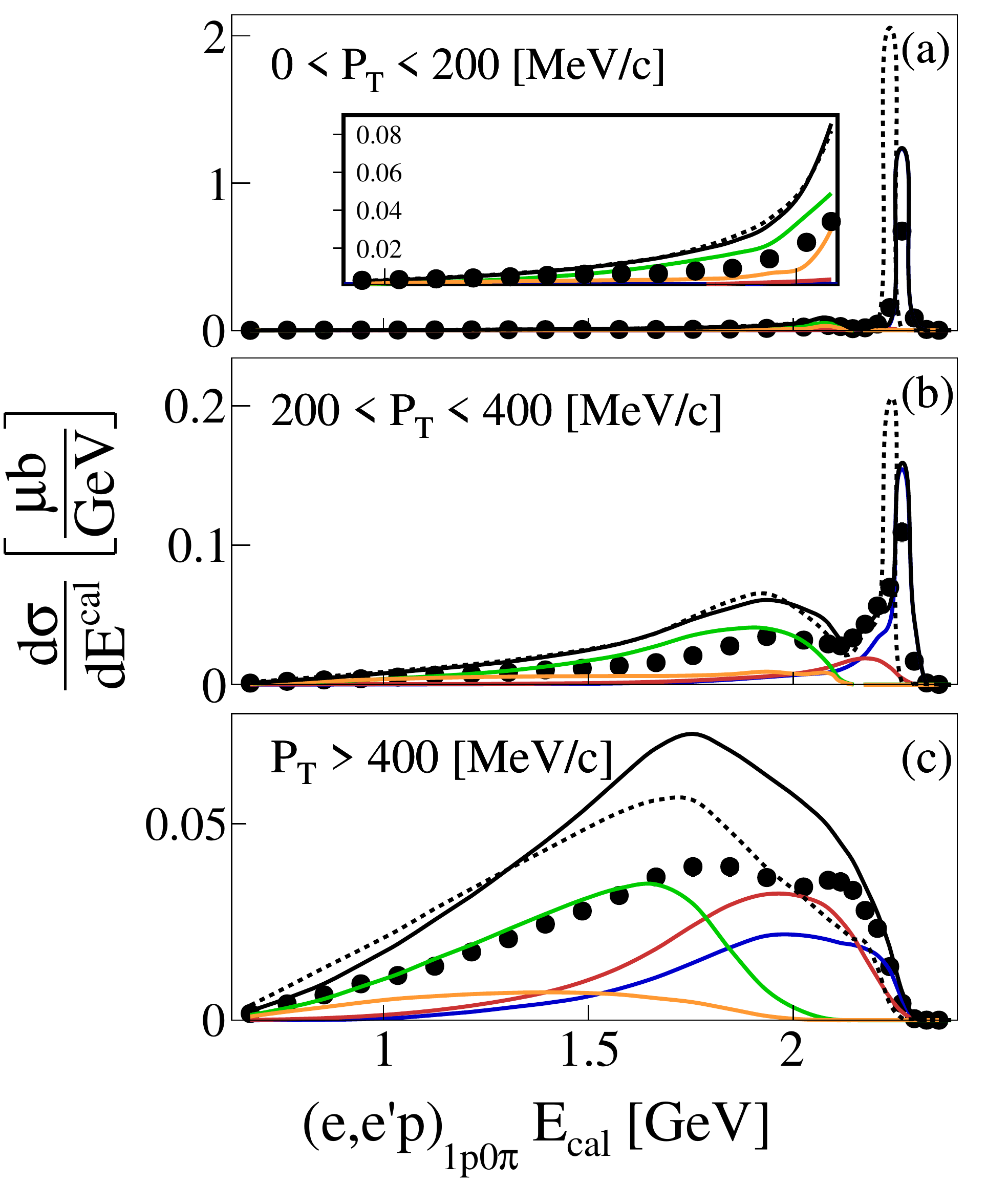}
\end{center}
\caption{\label{fig:pperp}(Left) the 2.257~GeV C$(e,e'p)_{1p0\pi}$ cross section plotted versus missing transverse momentum,
\pmperp, for data (black points), SuSav2 (black solid line) and G2018 (black dashed line).
The vertical lines at 200~MeV/$c$ and at 400~MeV/$c$ separate the three bins in \pmperp.  
Colored lines show the contributions of different processes to the SuSAv2 GENIE simulation: QE (blue), MEC (red), RES (green) and DIS (orange).  (Right) The cross section plotted versus the
calorimetric energy $E_{cal}$ for different bins in \pmperp:
(top) \pmperp $< 200$~MeV/$c$,
(middle) 200~MeV/$c$ 
$\le$ \pmperp $\le$ 400~MeV/$c$, 
and (bottom) \pmperp > 400~MeV/$c$.
    }
\end{figure}

\begin{figure} [htb!]
\begin{center}
\includegraphics[width=\linewidth]{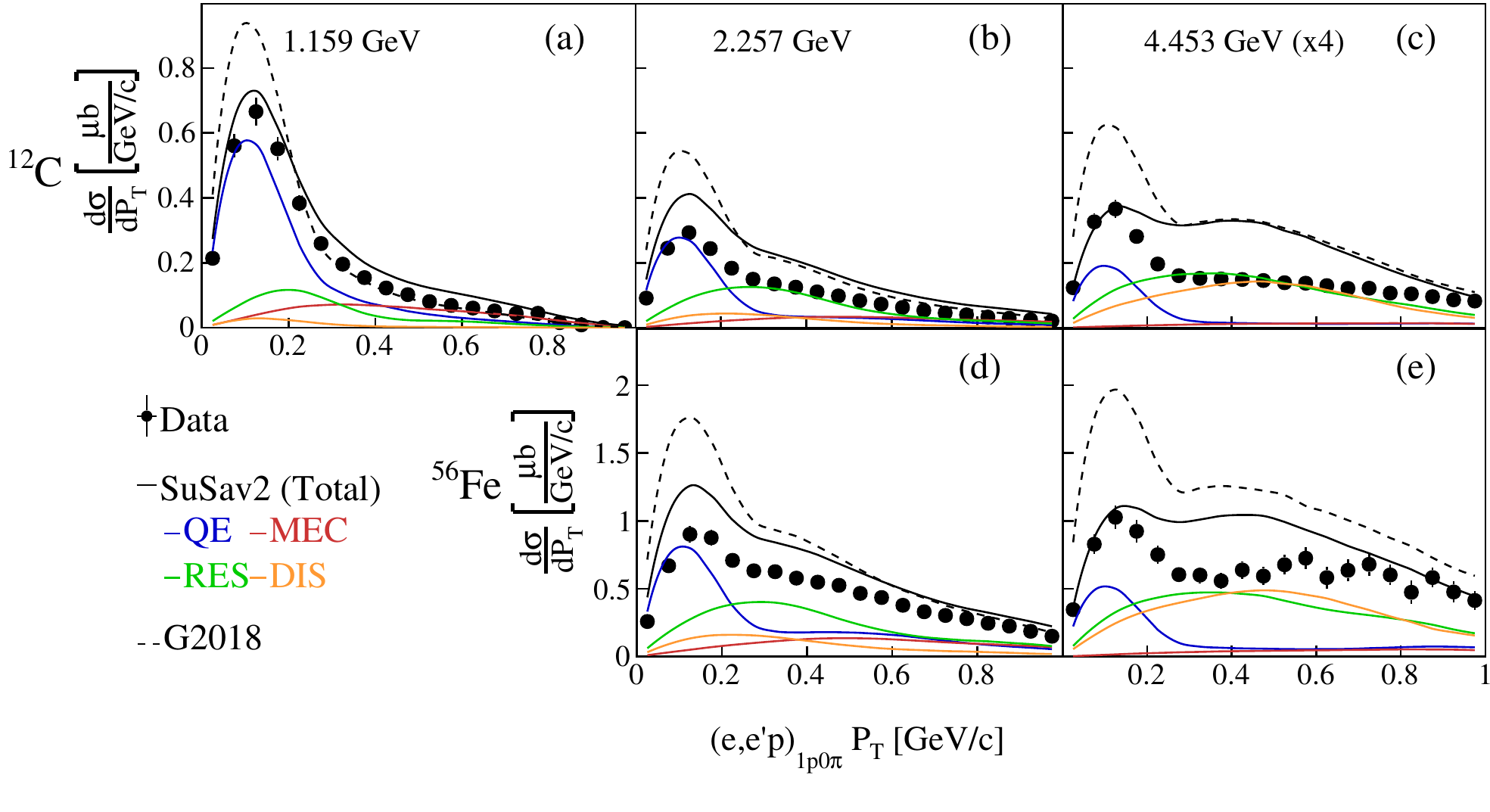}
\end{center}
\caption{\label{fig:pperp2} The cross section plotted vs transverse
  missing momentum $P_{T}$ for data (black points), SuSAv2 (black
  solid curve) and G2018 (black dotted curve). Different panels show
  results for different beam energy and target nucleus combinations:
  (top row) Carbon target at (left to right) 1.159, 2.257 and 4.453
  GeV, and (bottom) Iron target at (left) 2.257 and (right) 4.453 GeV. The 
  4.453 GeV yields have been scaled by four to have the same vertical
  scale. Colored lines show the
  contributions of different processes to the SuSAv2 GENIE simulation:
  QE (blue), MEC (red), RES (green) and DIS (orange).
}
\end{figure}

The \pmperp\,\,distribution for 2.257~GeV C\eep$_{1p0\pi}$ is shown in figure~\ref{fig:pperp} and the other targets and energies are shown in figure~\ref{fig:pperp2}.  
Both data and eGENIE peak at relatively low momenta, as expected, and both have a large tail extending out to 1~GeV/$c$ and containing about half of the measured events. 
The high-\pmperp\,tail is predominantly due to resonance production that did not result in an additional pion or nucleon above the detection threshold. 
eGENIE using \susa{} reproduces the shape of the data moderately well, suggesting adequate reaction modeling, including the contribution of non-QE processes such as resonance production.
As expected, both data and \eGenie/\susa{} events with \pmperp\,\, $< 200$~MeV/$c$ almost all reconstruct to the correct incident energy. 
However, events with \pmperp $\ge 400$~MeV/$c$ do not reconstruct to the correct energy and are poorly reproduced by eGENIE.
This disagreement becomes even more pronounced at higher energies and heavier nuclei, and indicates that including high-\pmperp\,\,data in oscillation analyses could bias the extracted parameters. 
As high-\pmperp\,data accounts for $25 - 50\%$ of the measured events, care must be taken to improve the models implemented in GENIE, so that they can reproduce the high-\pmperp\,data.  
This will be especially true at the higher incident neutrino energies expected for DUNE.

\begin{figure} [htb!]
\begin{center}
\includegraphics[width=0.9\linewidth]{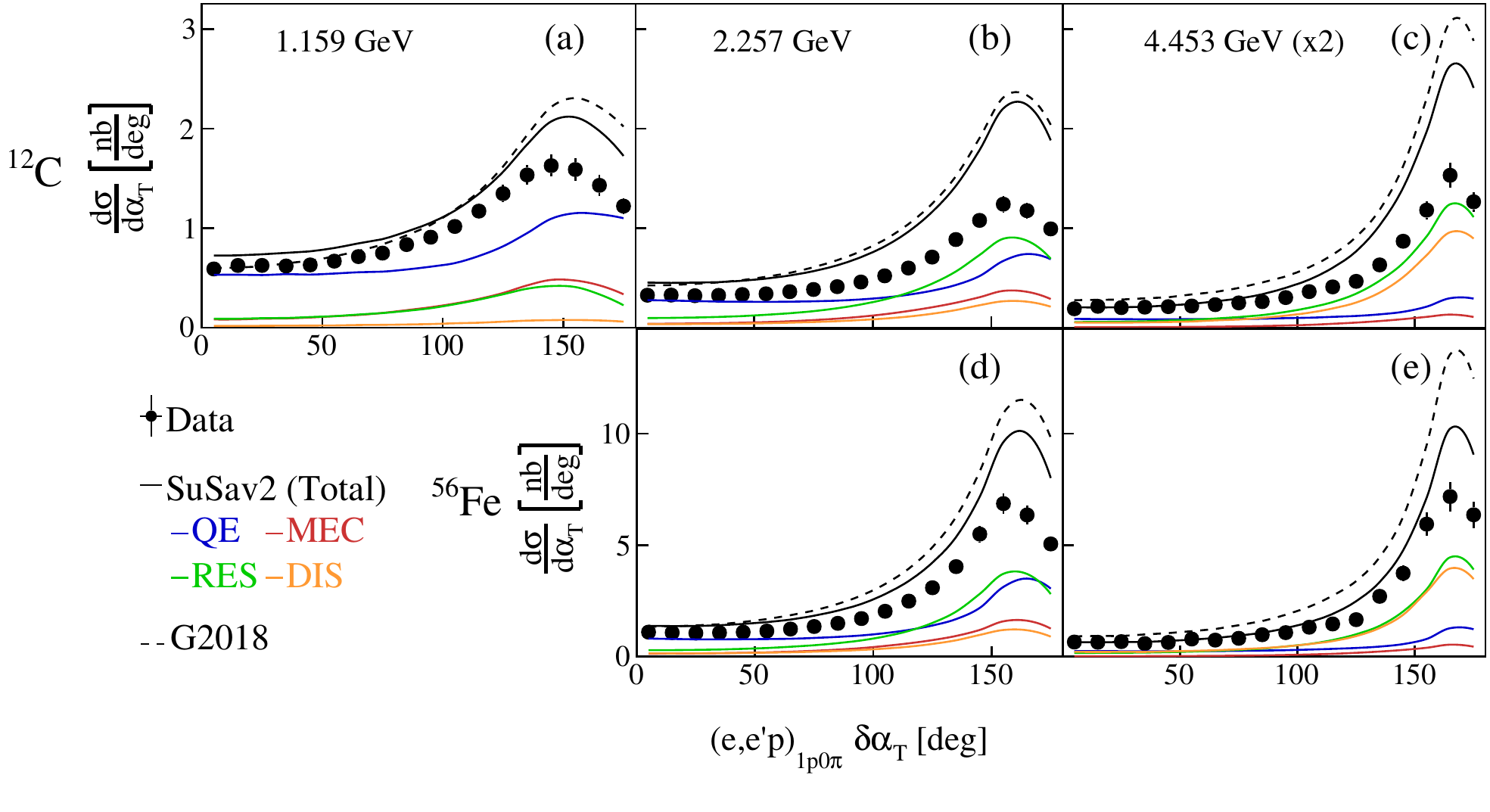}
\includegraphics[width=0.9\linewidth]{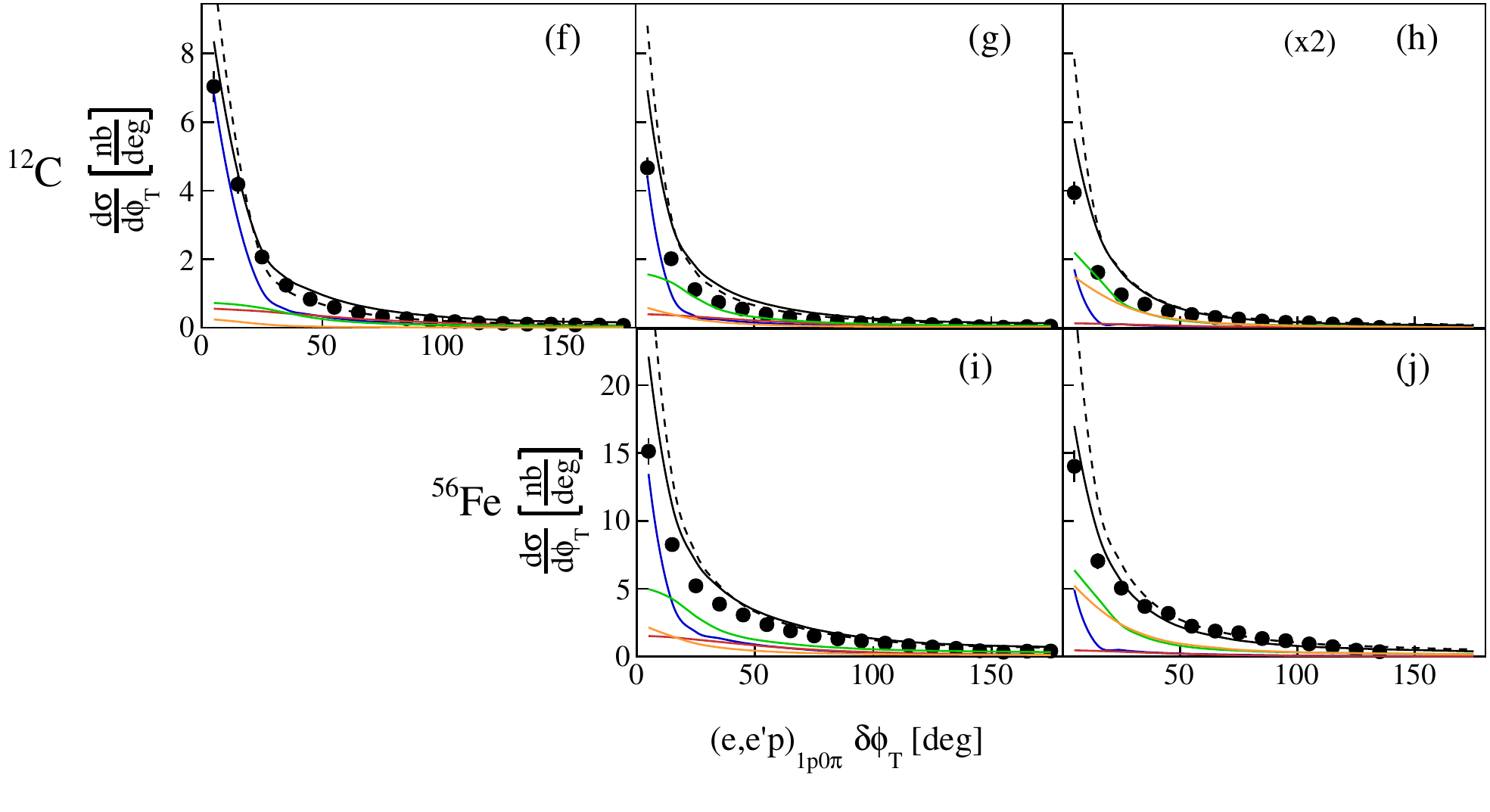}
\end{center}
\caption{\label{Efig:pperp2} The cross section plotted vs
  $\delta\alpha_{T}$ (a-e) and vs $\delta\phi_{T}$ (f-j) for data (black points), SuSAv2 (black solid
  curve) and G2018 (black dotted curve). Different panels show results
  for different beam energy and target nucleus combinations: (top row)
  Carbon target at (left to right) 1.159, 2.257 and 4.453 GeV, and
  (bottom) Iron target at (left) 2.257 and (right) 4.453 GeV. The
  4.453 GeV yields have been scaled by two to have the same vertical
  scale. Colored lines show the contributions of different processes
  to the SuSAv2 GENIE simulation: QE (blue), MEC (red), RES (green)
  and DIS (orange).}
\end{figure}

The opening angle $\delta\alpha_T$  measures the angle between \pmperp\,\,and the transverse momentum transfer ($\vec q_T = -\vec P_T^{\,e'}$) in the transverse plane  and is isotropic in the absence of final state interactions.  
$\delta\phi_T$ measures the opening angle between the detected proton momentum and the transverse momentum transfer and is forward peaked.  
The $\delta\alpha_T$ distributions become progressively less isotropic at higher energies and heavier targets, indicating the increasing importance of FSI and of non-QE reaction mechanisms.  
GENIE agrees best with data at the lowest beam energy.  
At the higher beam energies GENIE describes the relatively flat smaller angles much better than the back-angle peak.  
GENIE also describes the lowest energy $\delta\phi_T$ distribution.  
At higher energies, GENIE overestimates the height of the forward peak, as shown in figure \ref{Efig:pperp2}.

\begin{figure}[htb!]
\centering  
\includegraphics[width=0.49\linewidth]{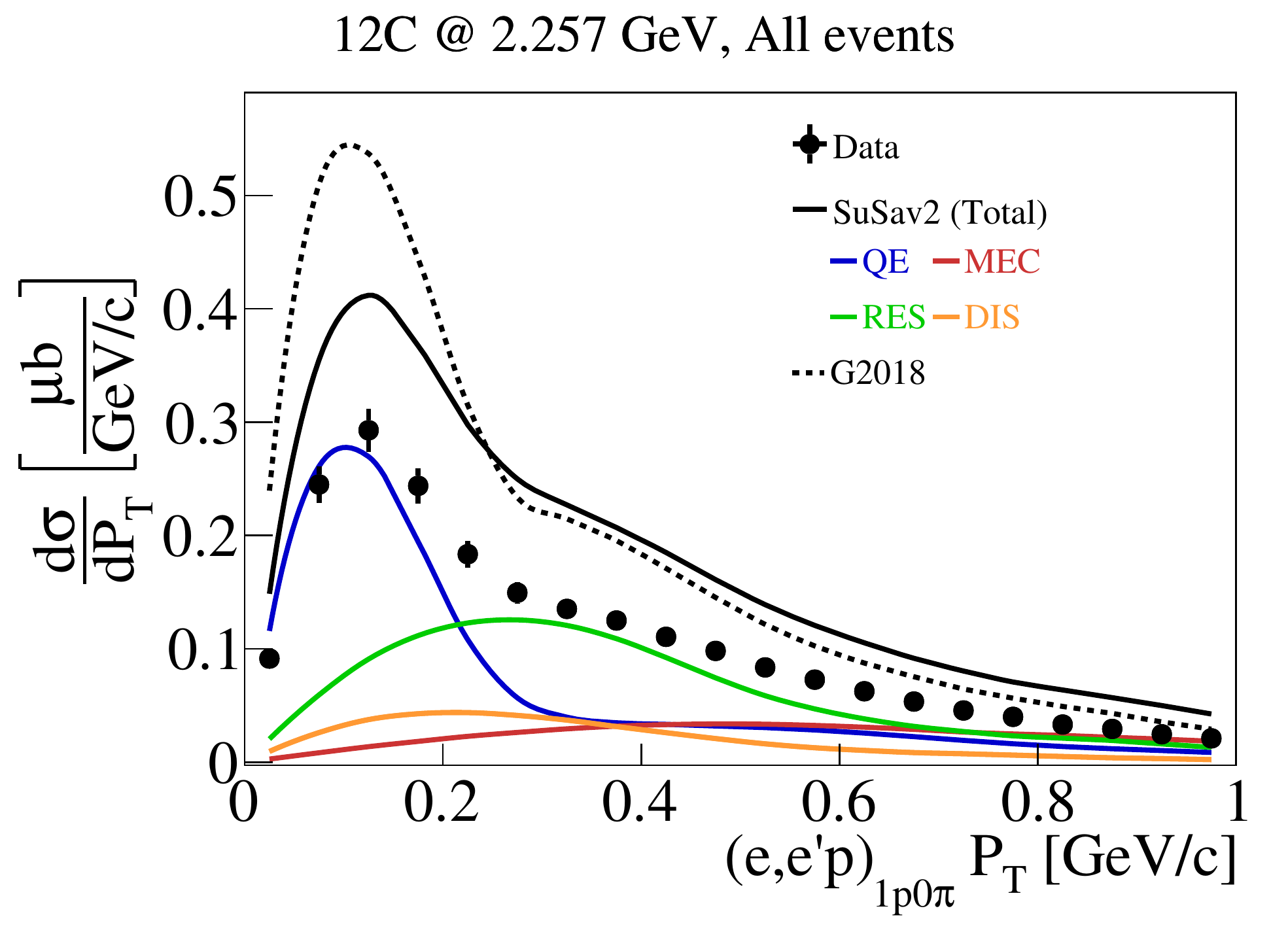}\\
\includegraphics[width=0.49\linewidth]{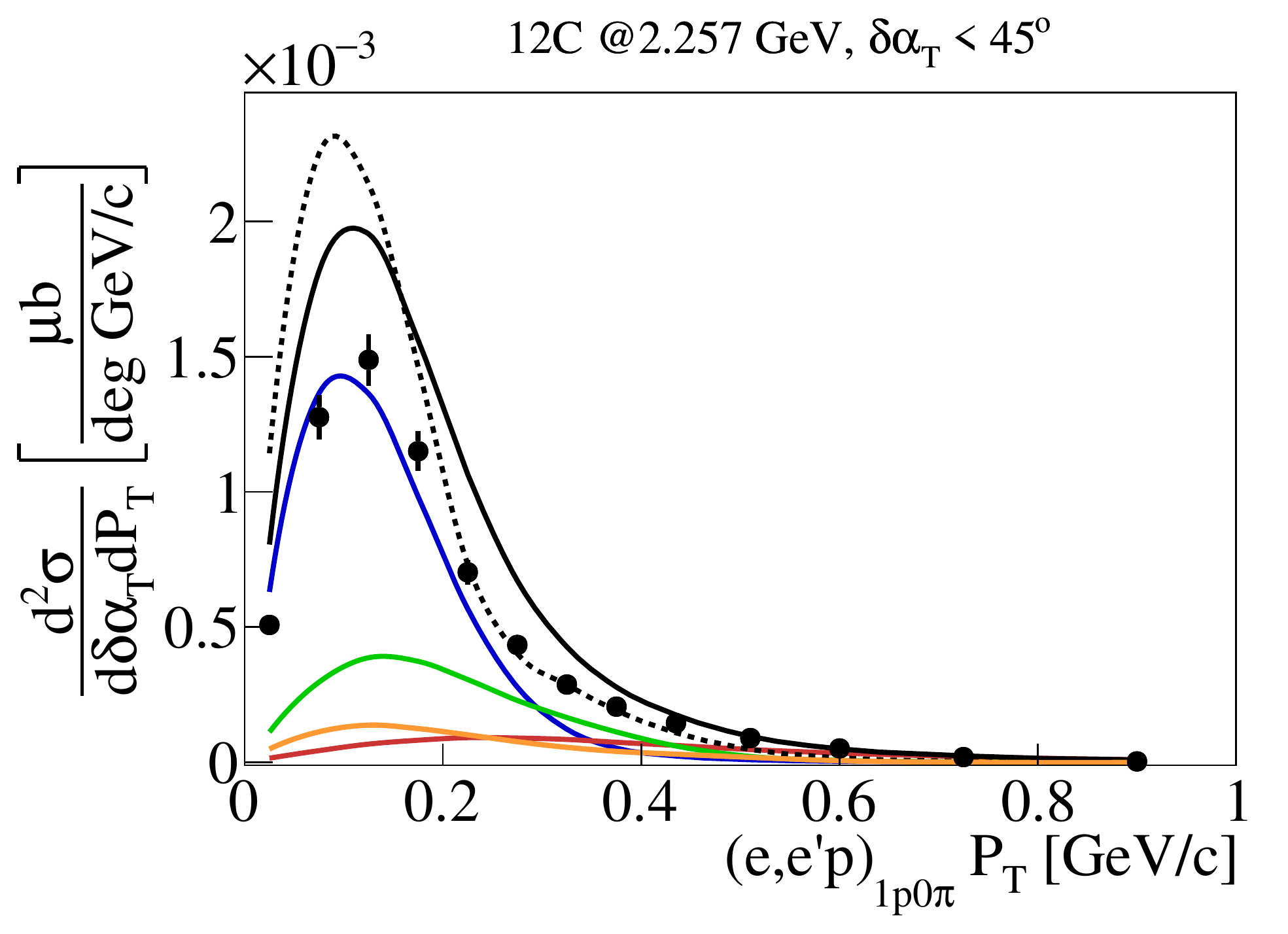}
\includegraphics[width=0.49\linewidth]{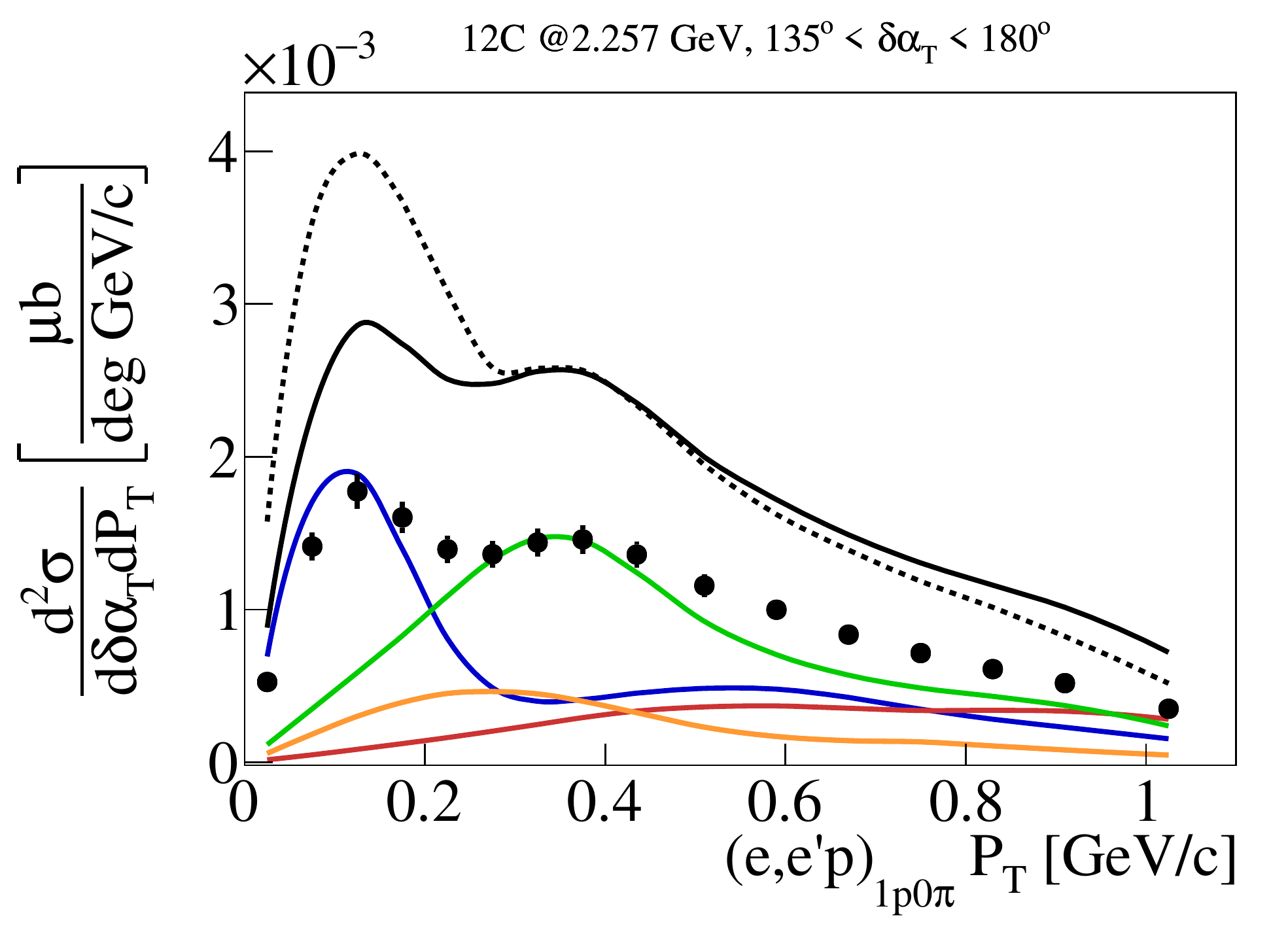}
\caption{\label{12C_DeltaPT_InDeltaAlphaT_Slices_2261}Data-simulation comparisons for $^{12}$C at 2.261 GeV showing the cross section results as a function of $P_{T}$ for (top) all the events, (bottom left) events with $\delta\alpha_{T} < 45^{o}$ dominated by QE interactions and no reinteractions, and (bottom right) events with $135^{o} < \delta\alpha_{T} < 180^{o}$ maximally affected by FSI and multi-hadron channels. 
Colored lines show the contributions of different processes
  to the SuSAv2 GENIE simulation: QE (blue), MEC (red), RES (green)
  and DIS (orange).}
\end{figure}

In line with the neutrino-based MicrobooNE analysis presented in section~\ref{CC1p1D}, these kinematic variables were further investigated in the form of multidimensional cross sections.
Figure~\ref{12C_DeltaPT_InDeltaAlphaT_Slices_2261} shows the data-simulation cross-section comparisons for $^{12}$C at 2.261 GeV as a function of $P_{T}$ for (top) all the events, (bottom left) events with $\delta\alpha_{T} < 45^{o}$ dominated by QE interactions and no reinteractions, and (bottom right) events with $135^{o} < \delta\alpha_{T} < 180^{o}$ maximally affected by FSI and multi-hadron channels.
Using all the events (top) that satisfy our selection yielded a QE-rich region up to $\approx$ 300\,MeV/c and a RES-dominated tail that extended to $\approx$ 1\,GeV/c.
Slicing the available $P_{T}$ phase-space in $\delta\alpha_{T}$ regions revealed regions with specific features.
More precisely, the region with $\delta\alpha_{T} < 45^{o}$ can be used to isolate primarily QE events and to test nuclear models in event generators.
Even in this region though, differences of 20-30\% in the QE strength are observed by both model configurations used for comparison in this analysis.
Meanwhile, the $135^{o} < \delta\alpha_{T} < 180^{o}$ region is dominated by RES interactions and could be used to tune FSI parameters and to improve the RES modeling. 

\begin{figure}[htb!]
\centering  
\includegraphics[width=0.49\linewidth]{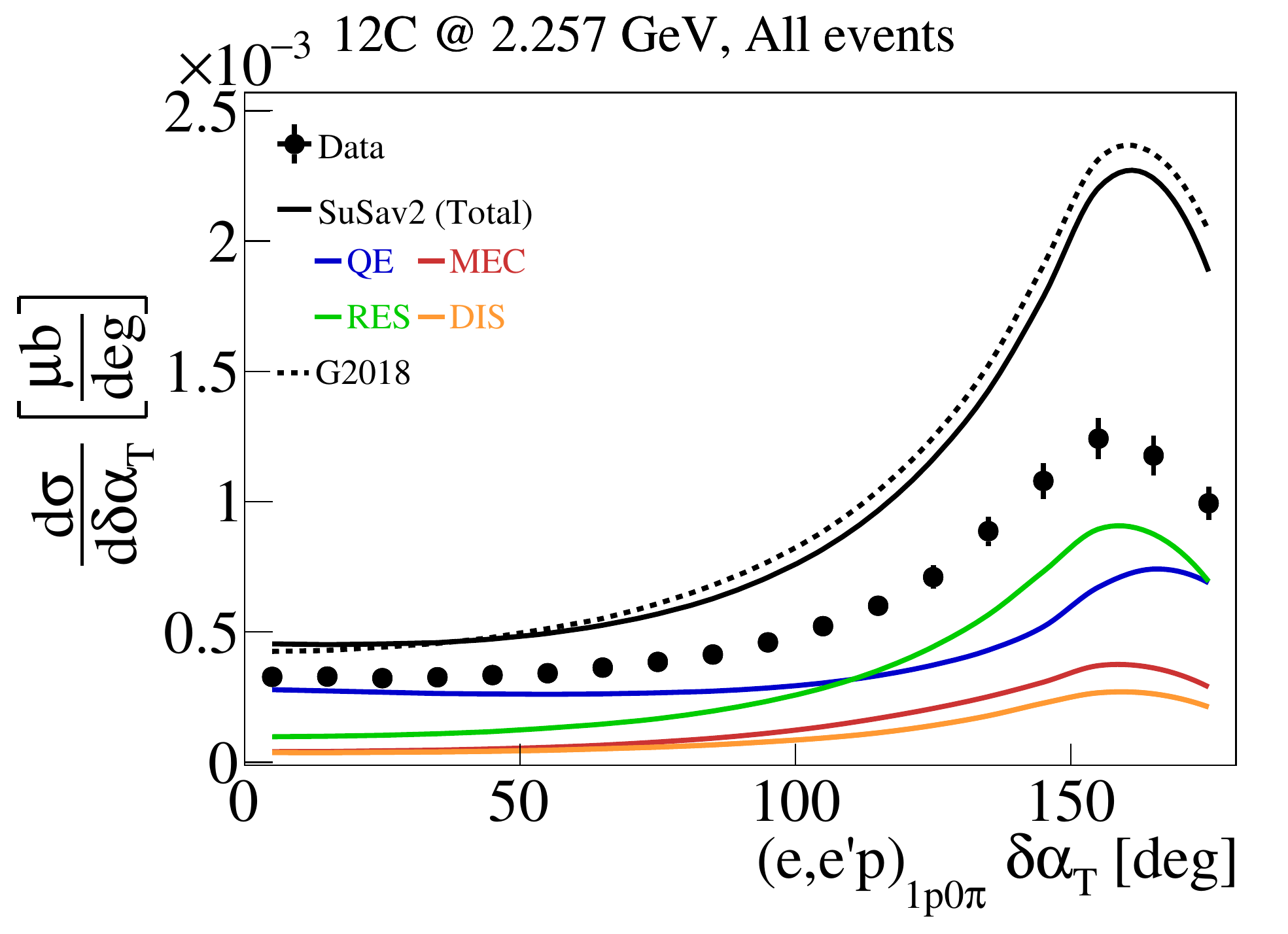}\\
\includegraphics[width=0.49\linewidth]{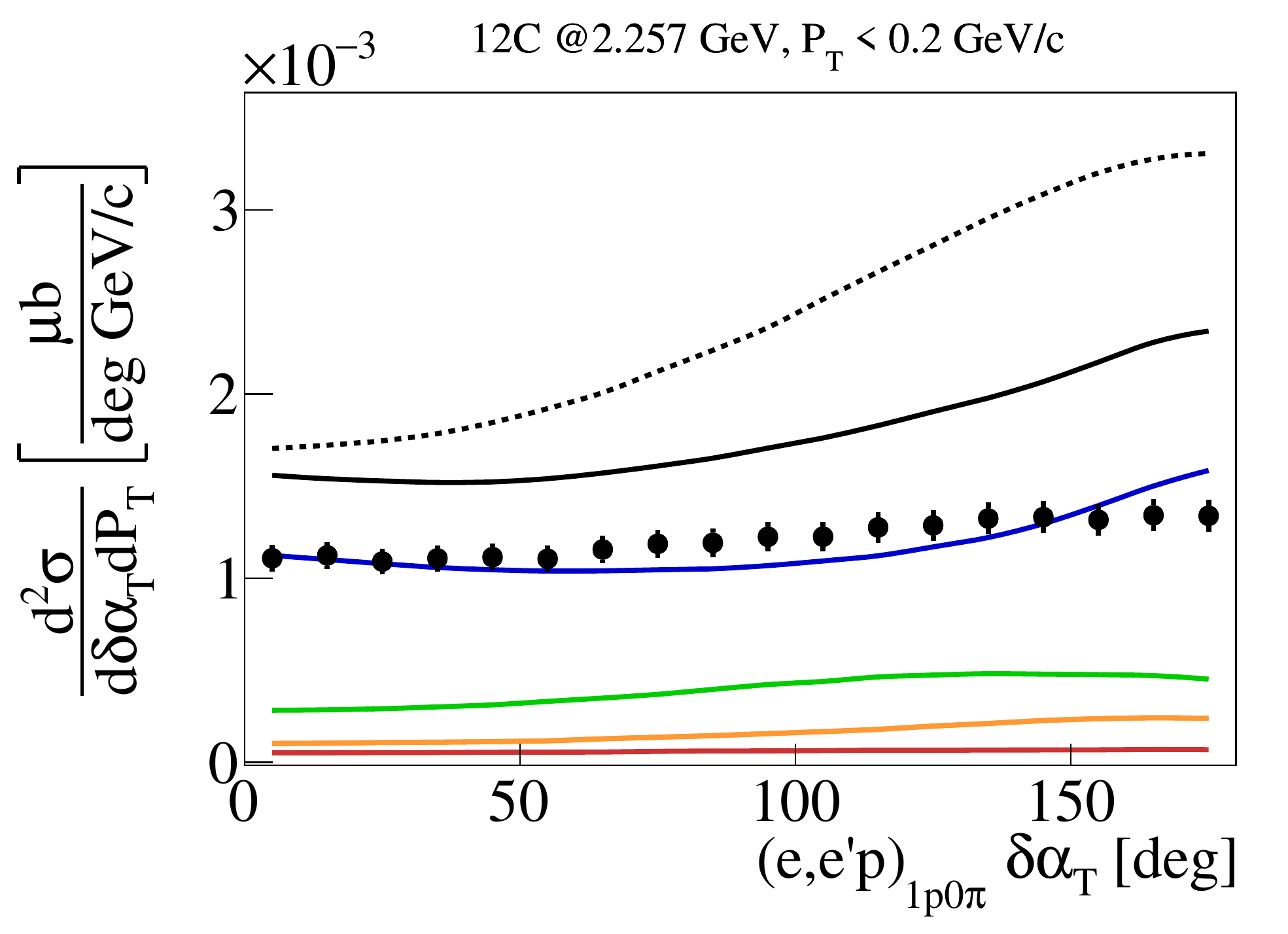}
\includegraphics[width=0.49\linewidth]{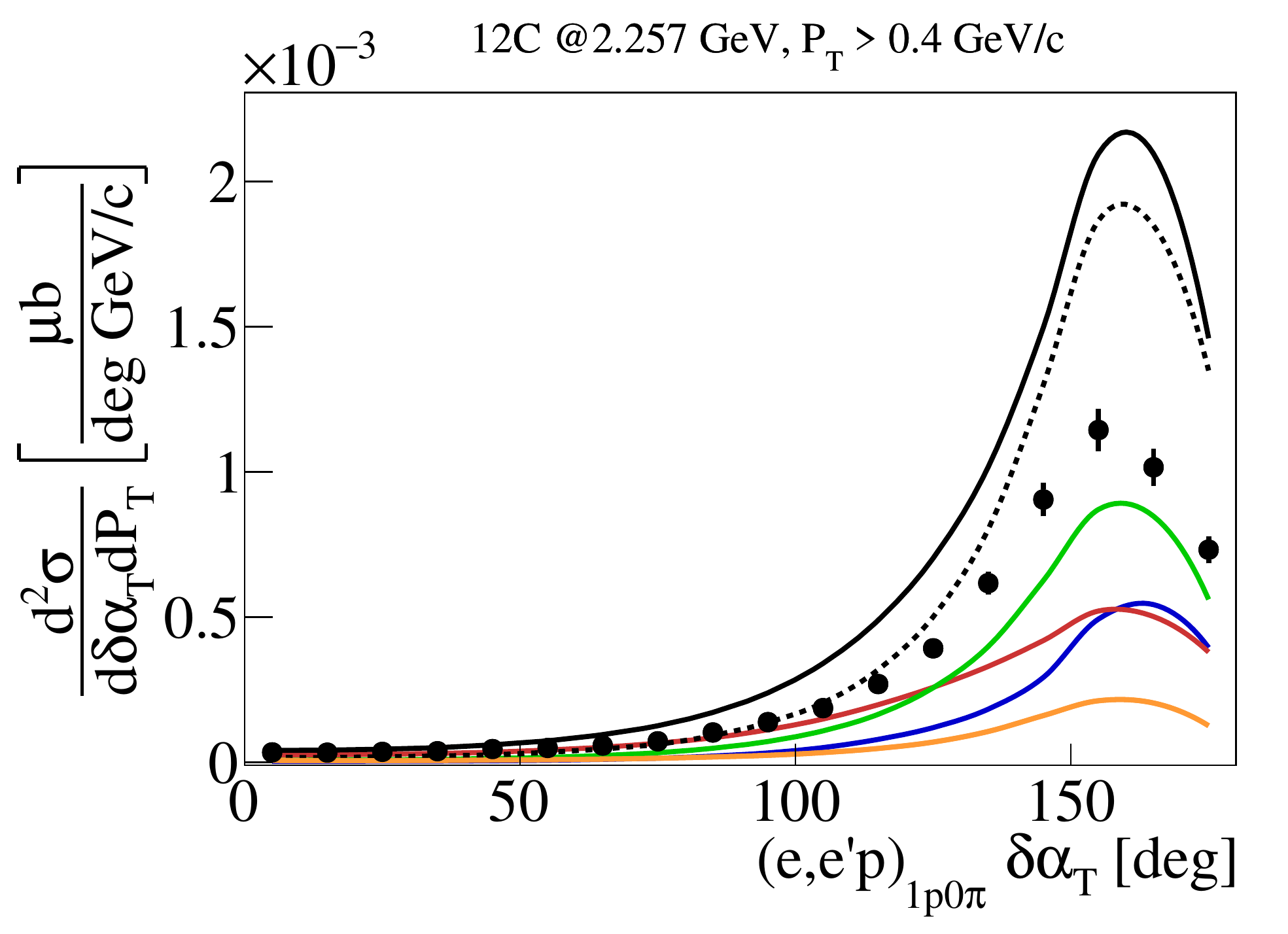}
\caption{\label{12C_DeltaAlphaT_InDeltaPT_Slices_2261}Data-simulation comparisons for $^{12}$C at 2.261 GeV showing the cross section results as a function of $\delta\alpha_{T}$ for (top) all the events, (bottom left) events with $P_{T} <$ 0.2\,GeV/c dominated by QE interactions and no reinteractions, and (bottom right) events with $P_{T} >$ 0.4\,GeV/c maximally affected by FSI and multi-hadron channels.
Colored lines show the contributions of different processes to the SuSAv2 GENIE simulation: QE (blue), MEC (red), RES (green) and DIS (orange).
}
\end{figure}

Figure~\ref{12C_DeltaAlphaT_InDeltaPT_Slices_2261} shows the data-simulation comparisons for $^{12}$C at 2.261 GeV as a function of $\delta\alpha_{T}$ for (left) all the events, (middle) events with $P_{T} <$ 0.2\,GeV/c, and (right) events with $P_{T} >$ 0.4\,GeV/c.
The $P_{T} <$ 0.2\,GeV/c slice is dominated by QE events with minimal FSI effects that result in a fairly uniform distribution with a slight enhancement in the forward direction due to the RES contamination.
The $P_{T} >$ 0.4\,GeV/c slice is dominated by multi-hadron and enhanced-FSI events that result in a sharp peak in the region close to $150^{o}$.
Therefore, this investigation of the $\delta\alpha_{T}$ phase-space in slices of $P_{T}$ is complimentary to the one illustrated in figure~\ref{12C_DeltaPT_InDeltaAlphaT_Slices_2261}.

Using the exact same $e4\nu$ data sets and cross-section extraction technique, the cross section as a function of the total struck nucleon momentum approximation, derived following the approach in section~\ref{CC1pDataAna}, was reported.
The only difference compared to the already outlined formalism on $^{40}$Ar is that B = 0.09216\,GeV is the binding energy for $^{12}$C, and $\epsilon^{N}$ = 0.024\,GeV is the corresponding removal energy~\cite{Bodek2019}.
This approximation, refered to as $P_{n,proxy}$, is compared to the true missing momentum $P_{Miss}$ calculated as shown in equation~\ref{pmiss},

\begin{equation}
P_{Miss} = |\vec{q} - \vec{p}|,
\label{pmiss}    
\end{equation}

where $\vec{q}$ is the 3-vector for the momentum transfer based on the difference between the kinematics of the incoming and the outgoing lepton, and $\vec{p}$ is the 3-vector of the outgoing proton.

\begin{figure}[htb!]
\centering  
\includegraphics[width=0.49\linewidth]{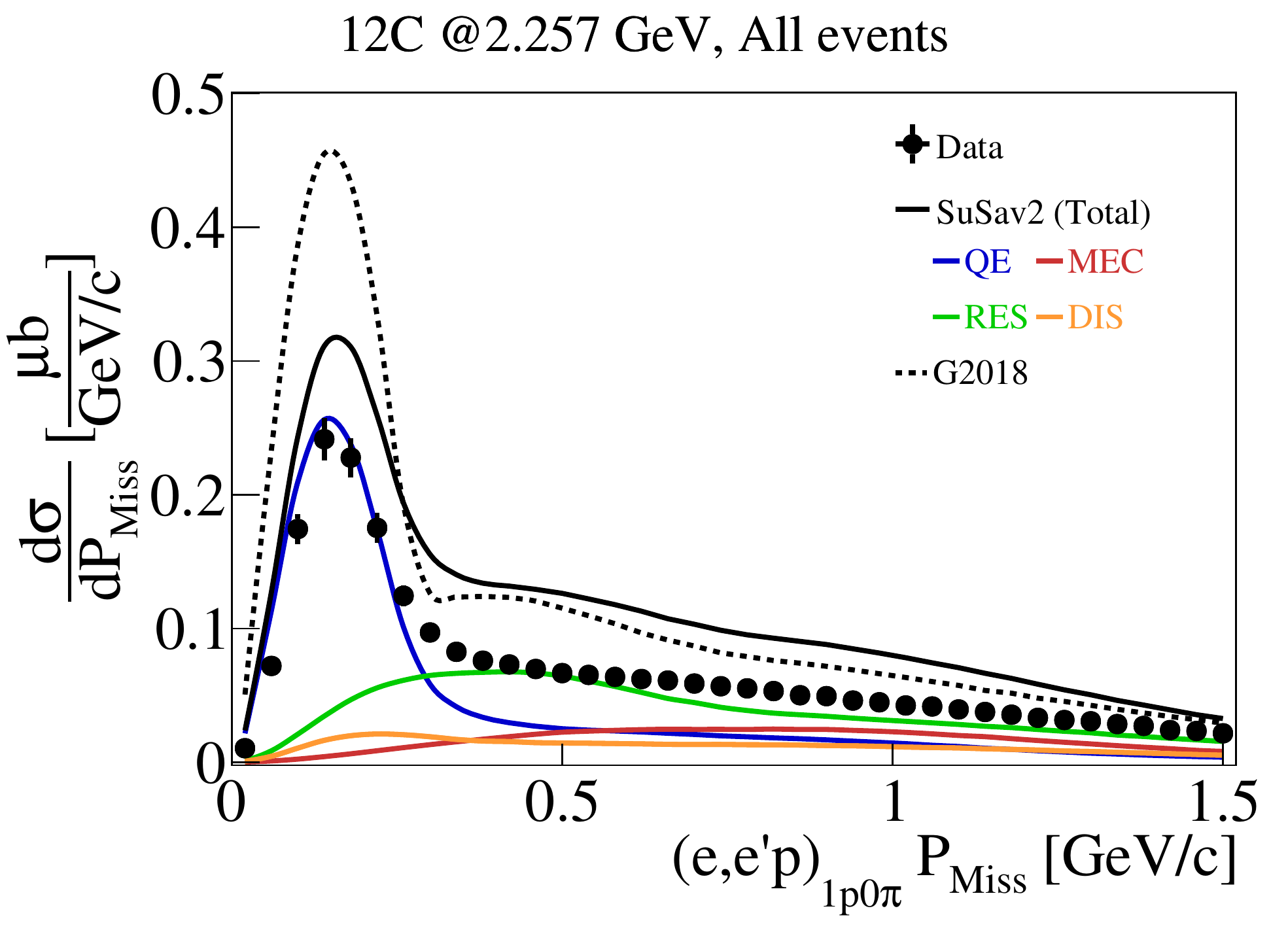}
\includegraphics[width=0.49\linewidth]{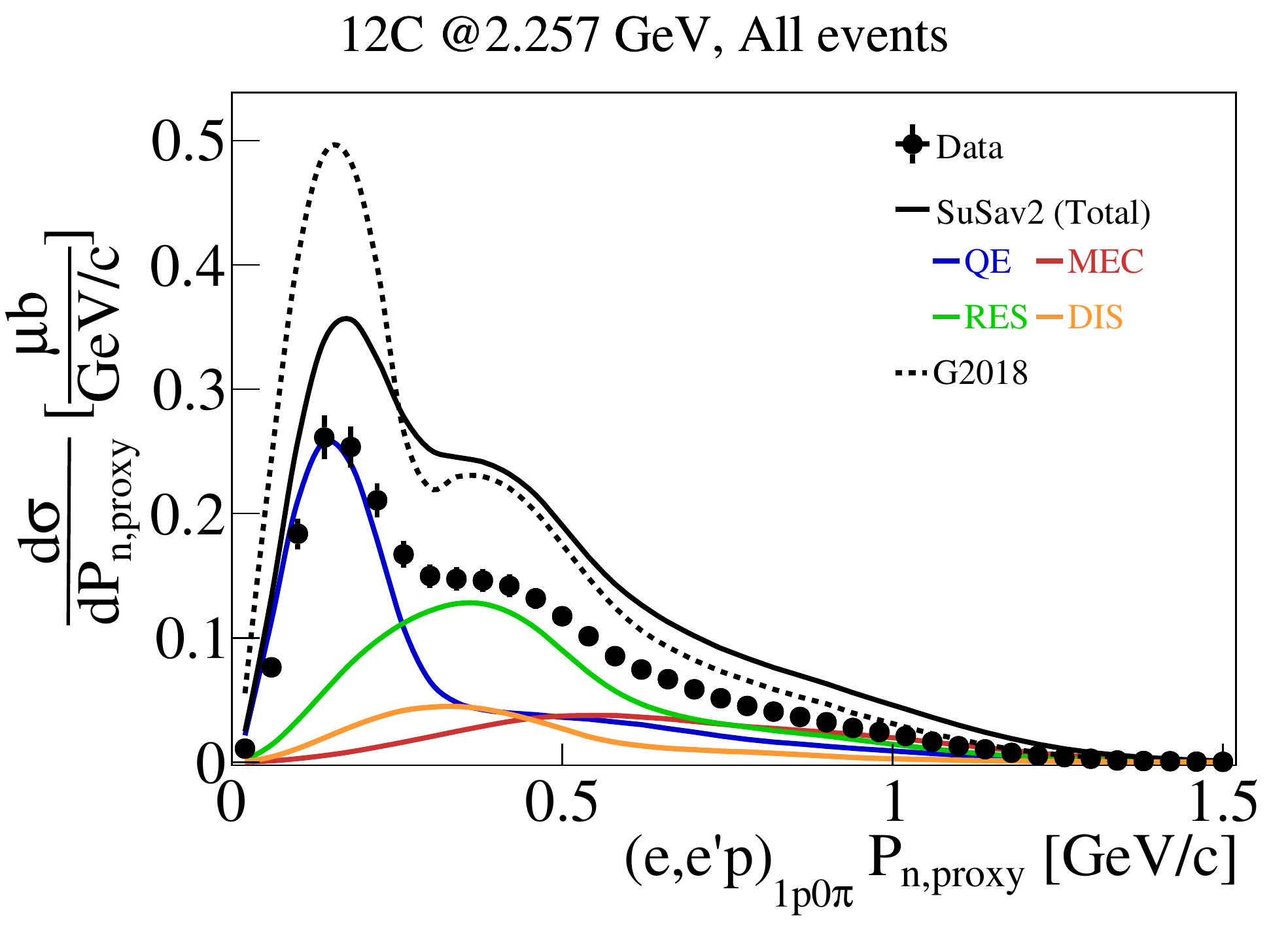}\\
\includegraphics[width=0.49\linewidth]{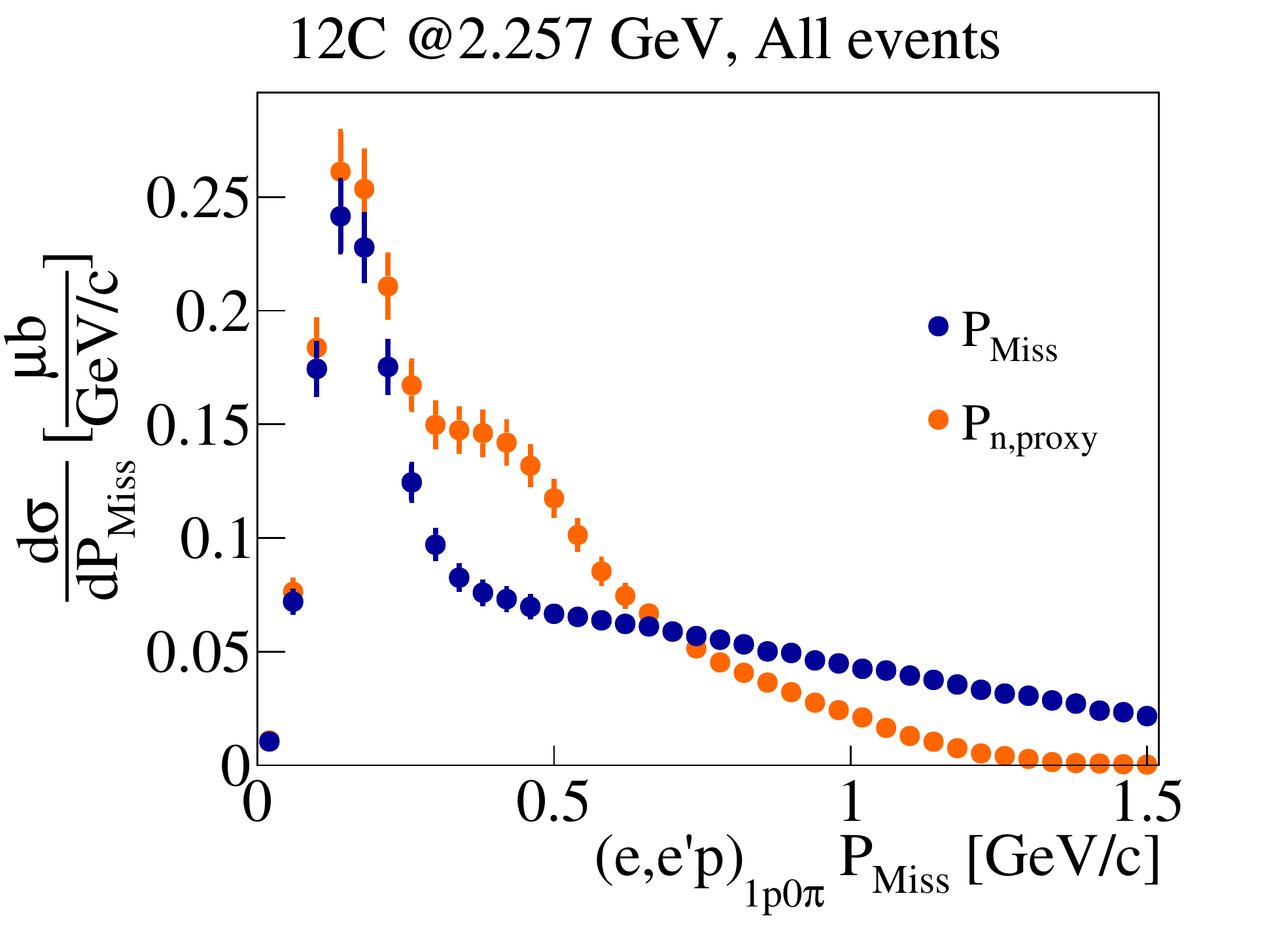}\\
\caption{\label{12C_XSec_2261}(Top) data-simulation comparisons on $^{12}$C at 2.261 GeV showing the cross section results as a function of (top left) the true missing momentum $P_{Miss}$ and (top right) the missing momentum approximation $P_{n,proxy}$ commonly used by neutrino experiments. (Bottom) overlay of the two extracted data cross sections illustrating the differences between $P_{Miss}$ and $P_{n,proxy}$.
%Colored lines show the contributions of different processes to the SuSAv2 GENIE simulation: QE (blue), MEC (red), RES (green) and DIS (orange).
}
\end{figure}

Figure~\ref{12C_XSec_2261} shows the data-simulation comparisons for $^{12}$C at 2.261 GeV as a function of (top left) the true missing momentum $P_{Miss}$ and (top right) the missing momentum approximation $P_{n,proxy}$ commonly used by neutrino experiments. 
The bottom panel shows the overlay of the two extracted data cross sections illustrating that $P_{n,proxy}$ fails to reproduce $P_{Miss}$.
As can be seen in the interaction breakdown plots, the differences are primarily driven by the RES interactions which are reconstructed at lower values when the QE-like assumption deployed for $P_{n,proxy}$ is used.

\begin{figure}[htb!]
\centering  
\includegraphics[width=0.49\linewidth]{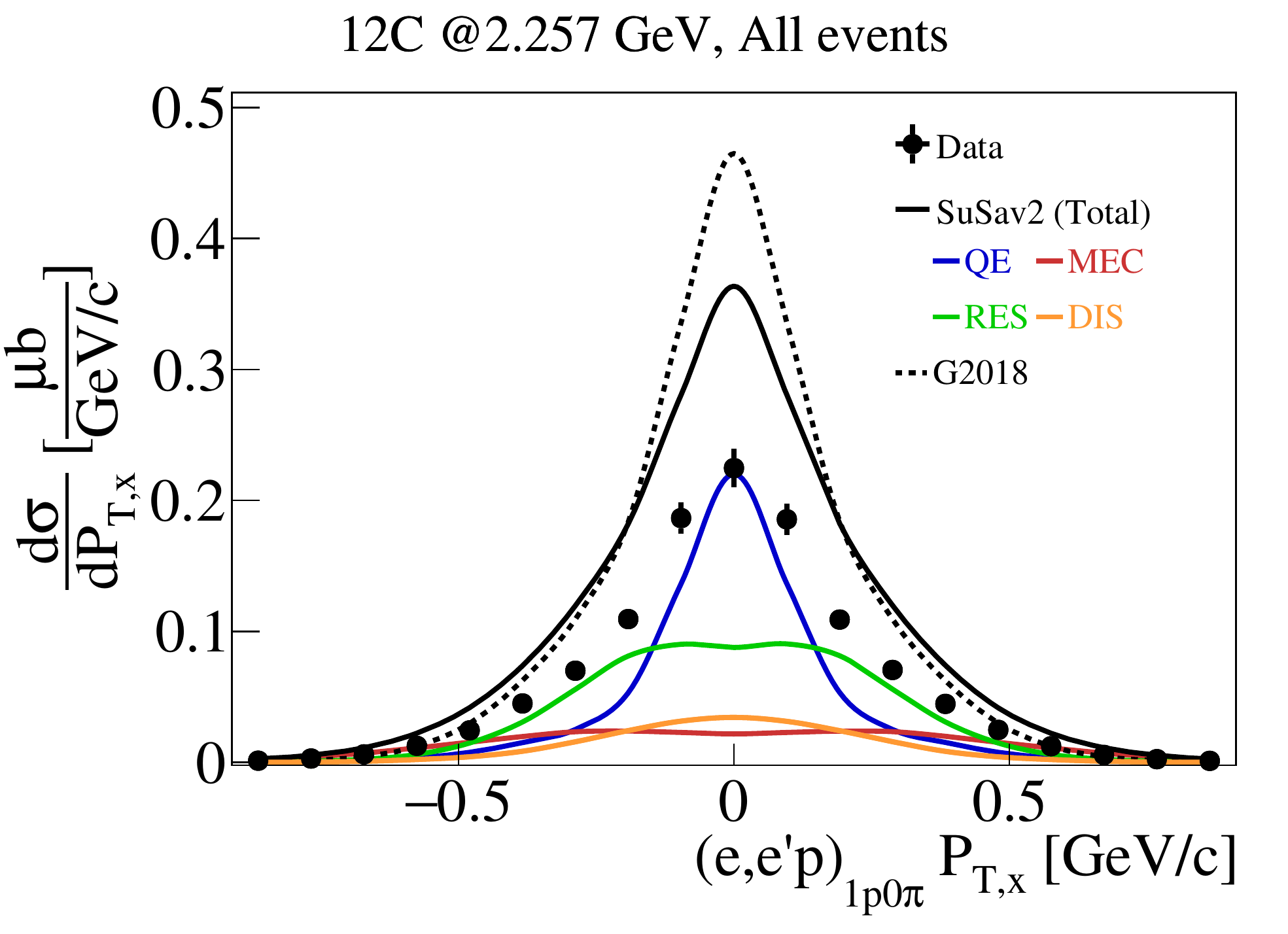}\\
\includegraphics[width=0.49\linewidth]{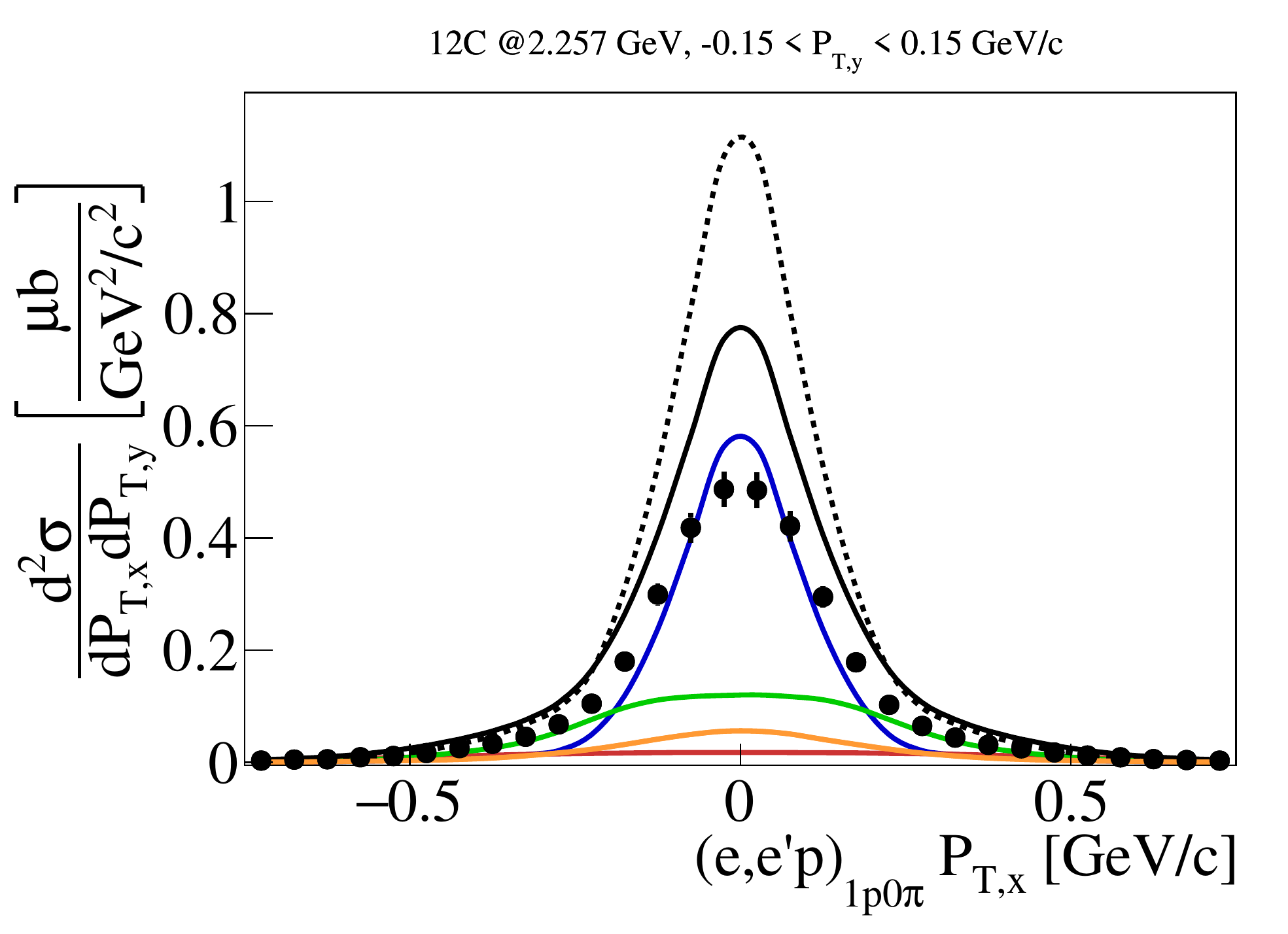}
\includegraphics[width=0.49\linewidth]{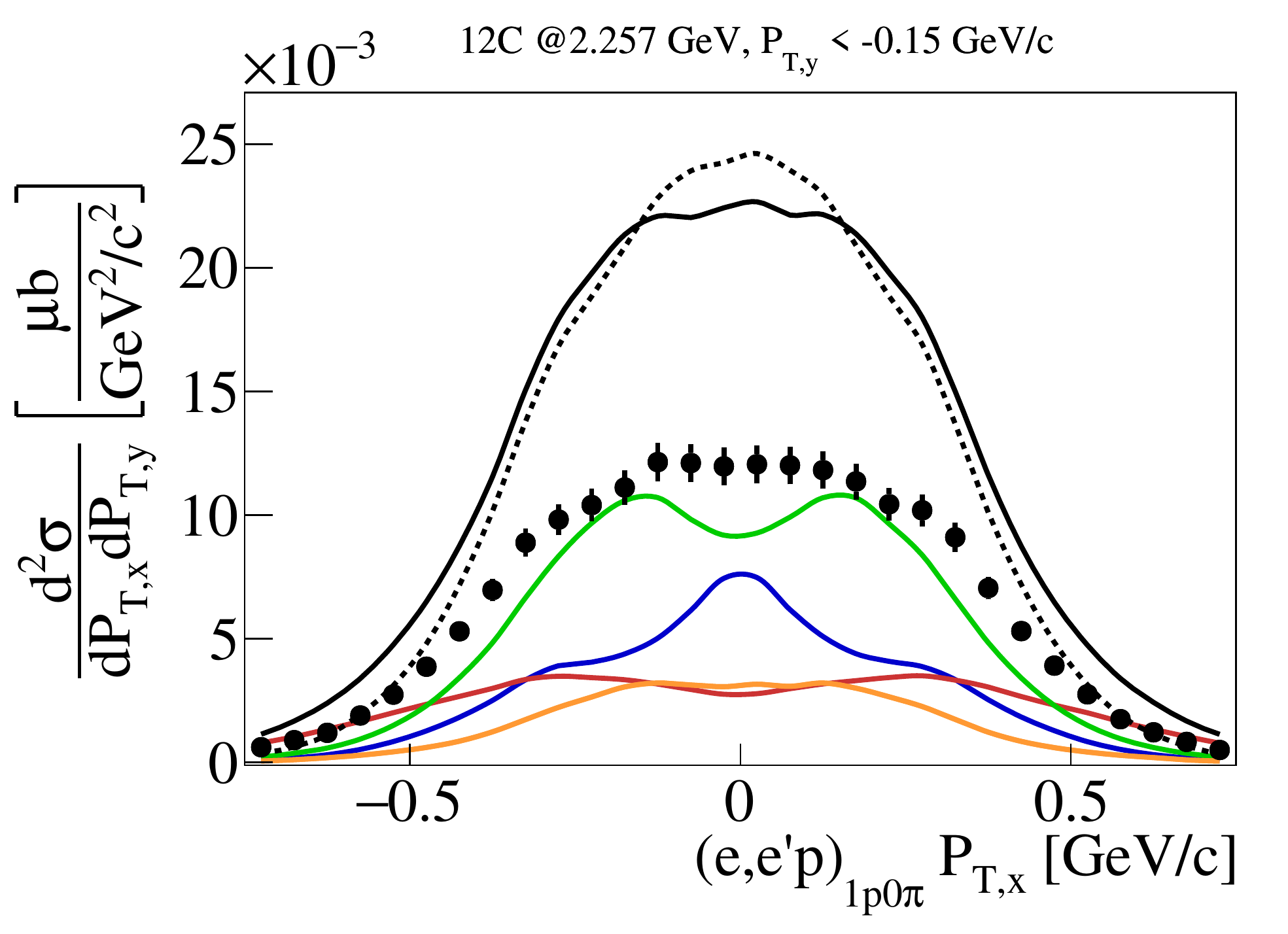}
\caption{\label{12C_DeltaPtxy_Slices_2261}Data-simulation comparisons for $^{12}$C at 2.261 GeV showing the cross section results as a function of $P_{T,x}$ for (top) all the events, (bottom left) events with -0.15 $< P_{T,y} <$ 0.15\,GeV/c dominated by QE interactions and no reinteractions, and (bottom right) events with $P_{T,y} <$ -0.15\,GeV/c maximally affected by FSI and multi-hadron channels.
%Colored lines show the contributions of different processes to the SuSAv2 GENIE simulation: QE (blue), MEC (red), RES (green) and DIS (orange).
}
\end{figure}

\begin{figure}[htb!]
\centering  
\includegraphics[width=0.49\linewidth]{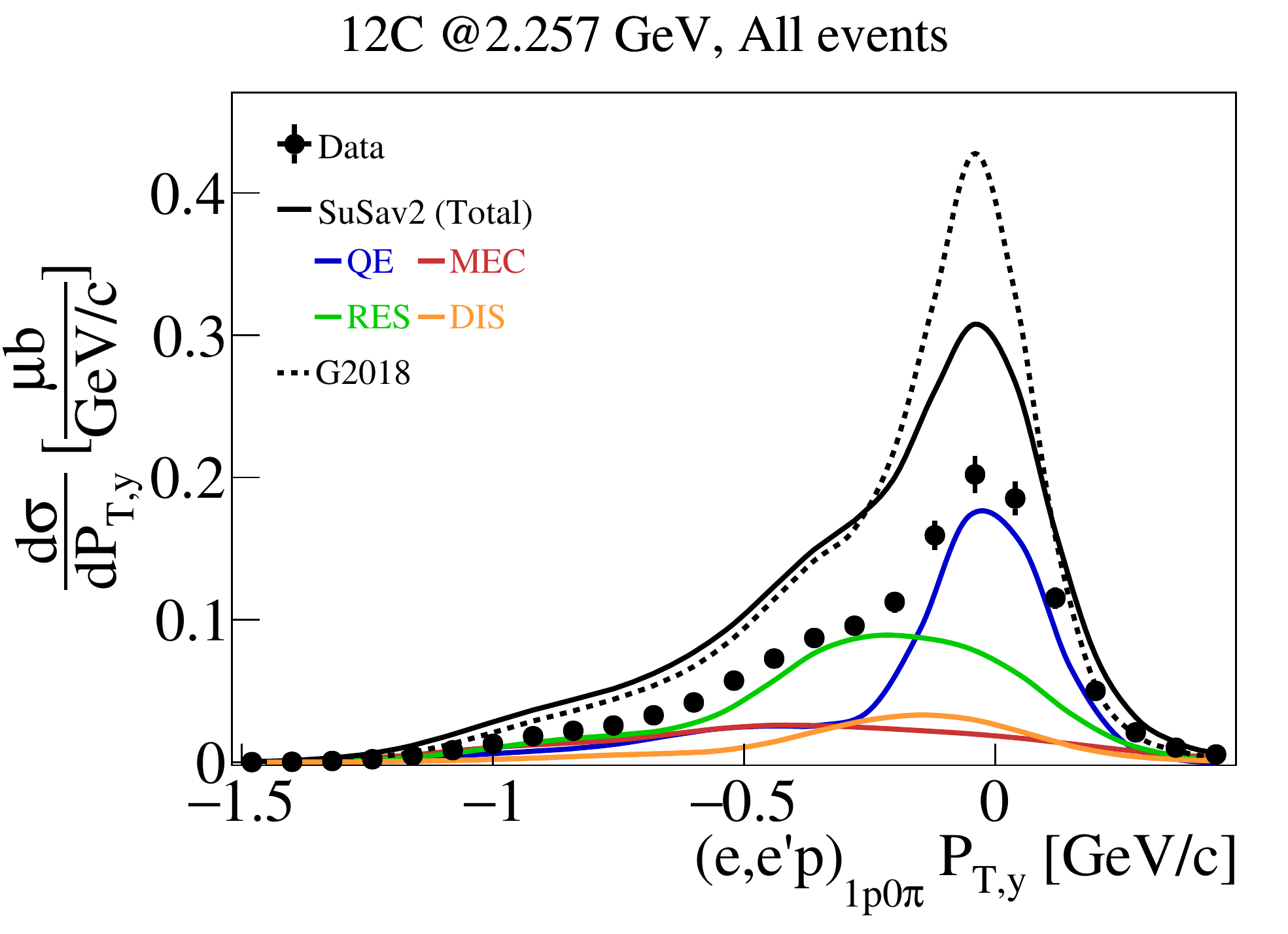}\\
\includegraphics[width=0.49\linewidth]{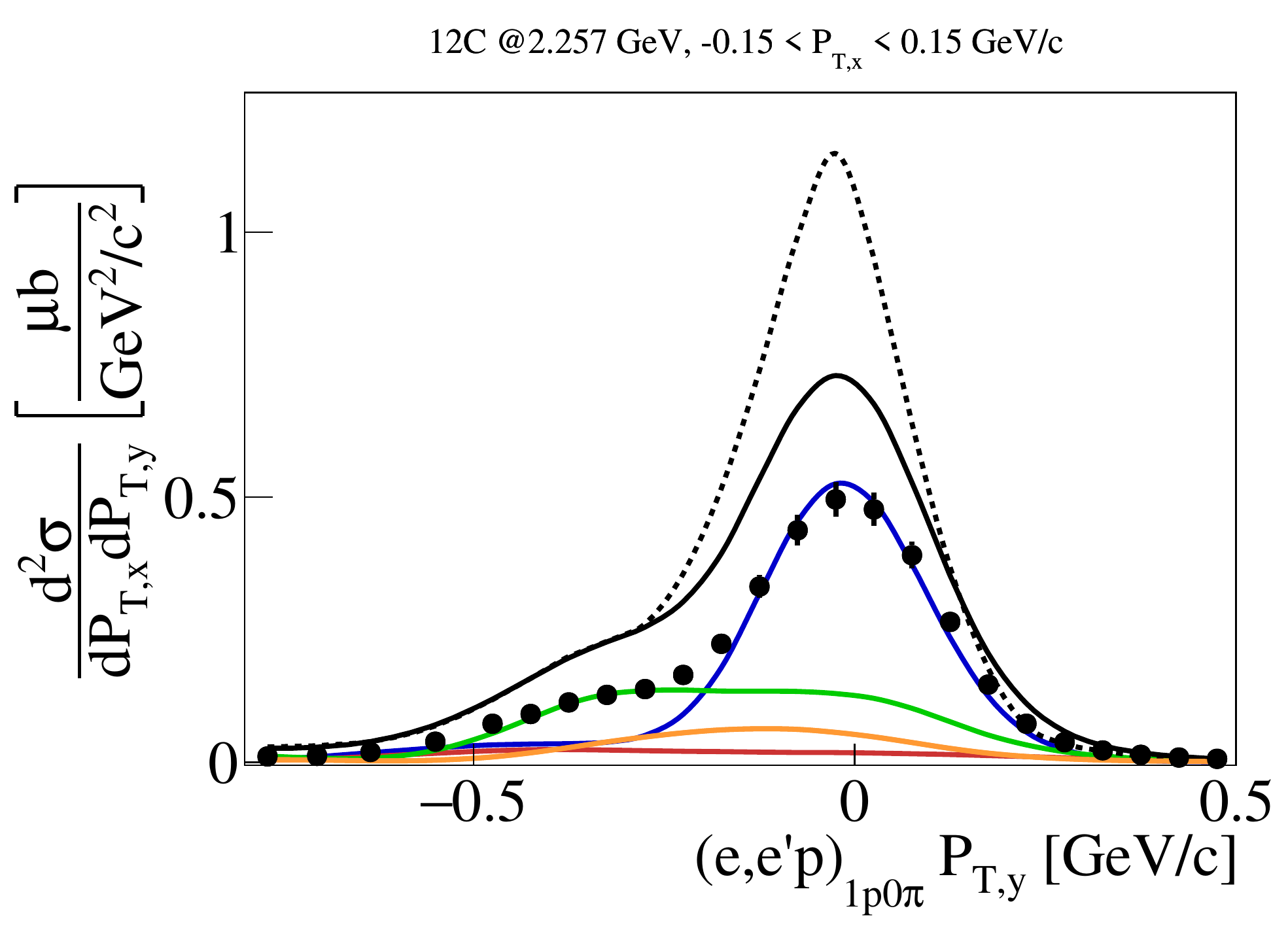}
\includegraphics[width=0.49\linewidth]{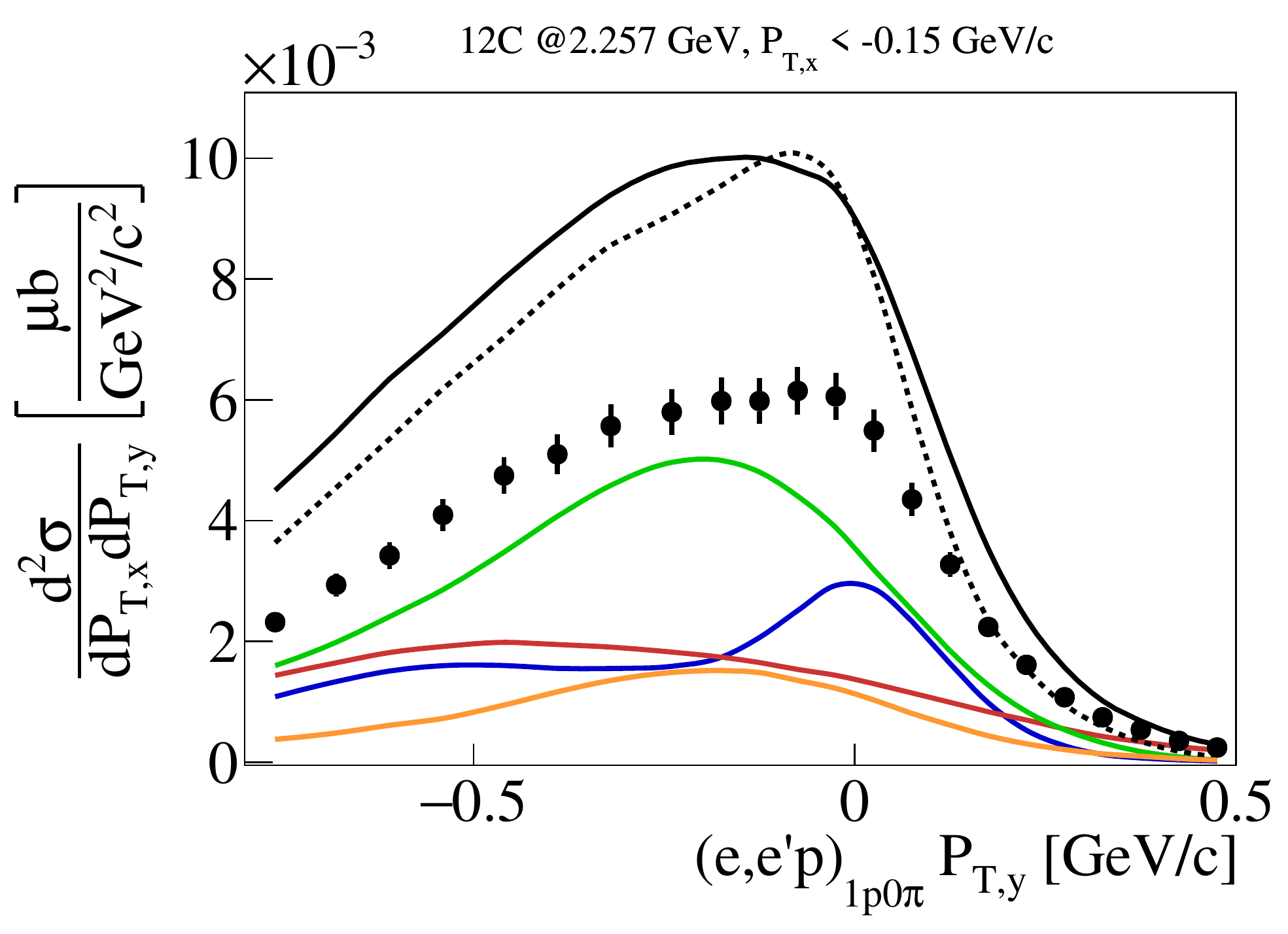}
\caption{\label{12C_DeltaPtyx_Slices_2261}Data-simulation comparisons for $^{12}$C at 2.261 GeV showing the cross section results as a function of $P_{T,y}$ for (top) all the events, (bottom left) events with -0.15 $< P_{T,x} <$ 0.15\,GeV/c dominated by QE interactions and no reinteractions, and (bottom right) events with $P_{T,x} <$ -0.15\,GeV/c maximally affected by FSI and multi-hadron channels.
%Colored lines show the contributions of different processes to the SuSAv2 GENIE simulation: QE (blue), MEC (red), RES (green) and DIS (orange).
}
\end{figure}

Figures~\ref{12C_DeltaPtxy_Slices_2261} and~\ref{12C_DeltaPtyx_Slices_2261} also illustrate the equivalent of the multidimensional MicroBooNE analysis using the $P_{T,x}$ cross sections in slices of $P_{T,y}$ (figure~\ref{12C_DeltaPtxy_Slices_2261}) and vice versa (figure~\ref{12C_DeltaPtyx_Slices_2261}).
As can be seen in the interaction breakdown plots, the regions close to 0 for both kinematic variables (bottom left panels) are the ideal place to isolate QE events with small contributions from more complex events.
Yet, even in this very QE-dominated region, significant data-simulation differences are observed, most likely due to the already-observed RES overestimation.
On the other hand, events occupying the phase-space outside that QE-rich band (bottom right) are dominated primarily by most complicated interactions (RES and DIS) and QE events with strong FSI effects, as indicated by the relevant pointy shapes.
These populations result in much more broader and smeared distributions.

Of particular interest is the asymmetric behavior observed in figure~\ref{12C_DeltaPtyx_Slices_2261}.
As discussed in section~\ref{CC1pDataAna}, this clearly-observed asymmetry is caused by multi-nucleon and FSI effects.
More precisely, the asymmetry in the QE-enhanced region (bottom left) is entirely driven by the overpredicted RES events.
In the non-QE-dominated slice (bottom right), data-simulation disagreements by a factor of $\sim$ 2 are further observed and are driven by RES/DIS events and FSI-enhanced QE interactions.

%%%%%%%%%%%%%%%%%%%%%%%%%%%%%%%%%%%%%%%%%%%%%%%%%%

%\clearpage
\section{Electrons-For-Neutrinos Conclusions}

In this electron-based analysis, the similarities between electron- and neutrino-nucleus interactions were exploited, and electron scattering data with known beam energies to test energy reconstruction methods and interaction models were used~\cite{e4vNature21}.
Even in simple interactions where no pions are detected, only a small fraction of events was found to reconstruct to the correct incident energy.  
More importantly, widely-used interaction models reproduced the reconstructed energy distribution only qualitatively and the quality of the reproduction varied strongly with beam energy.
This shows both the need and the pathway to improve current models to meet the requirements of next-generation, high-precision experiments such as DUNE~\cite{DUNE} and Hyper-Kamiokande (HK)~\cite{HK}.

%%%%%%%%%%%%%%%%%%%%%%%%%%%%%%%%%%%%%%%%%%%%%%%%%%

\section{Prospects With CLAS12}

\begin{figure} [htb!]
\begin{center}
\includegraphics[width=0.7\linewidth]{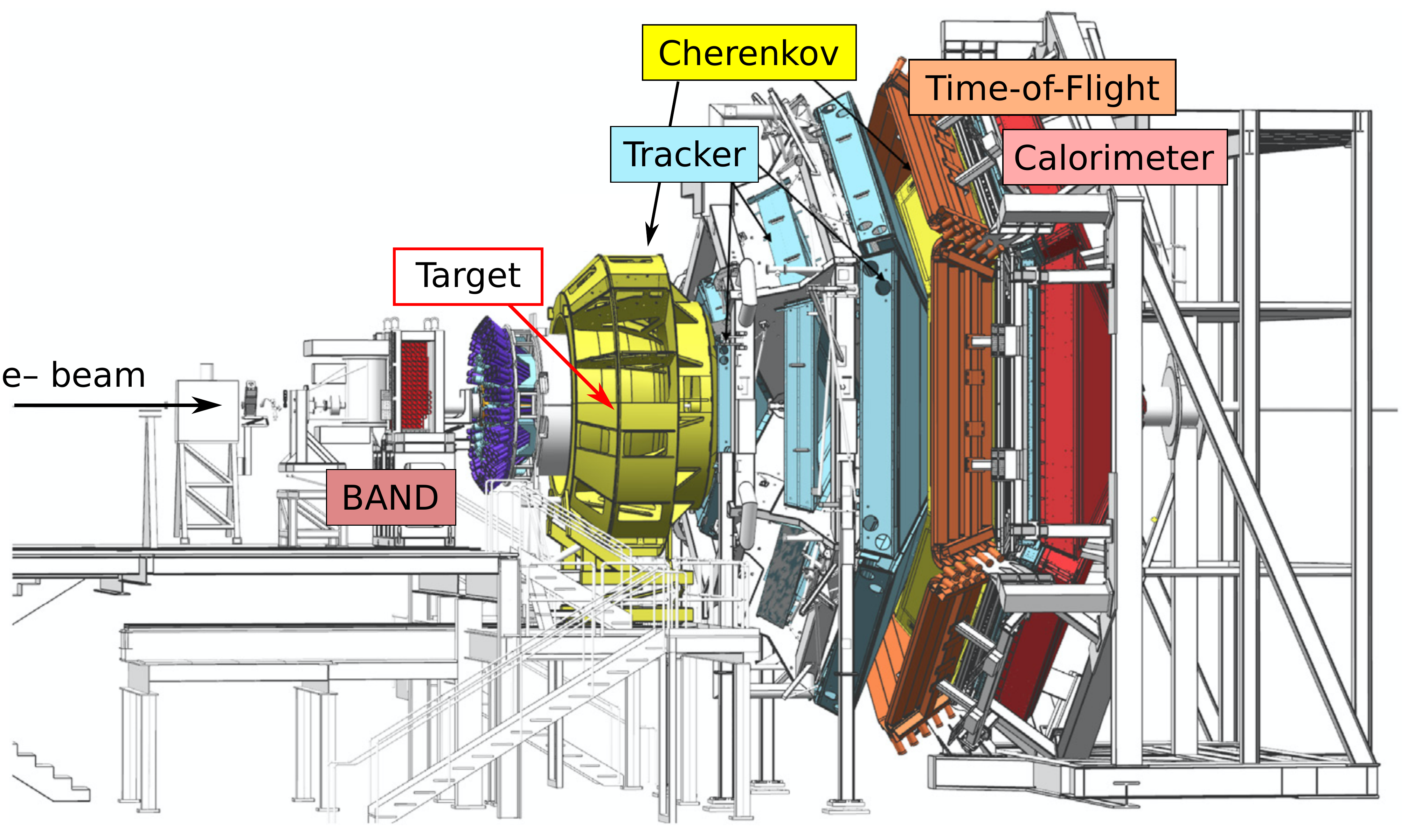}
\end{center}
\caption{\label{CLAS12Det}Schematic view of the upgraded CLAS12 detector components.}
\end{figure}

Jefferson Lab Experiment E12-17-006, ``Electrons for Neutrinos: Addressing Critical Neutrino–Nucleus Issues'' (scientific rating: A)~\cite{clas12proposal} has already taken further data on more targets with a greater kinematical range using the upgraded CLAS12 detector shown in figure~\ref{CLAS12Det}. 
The approved experiment includes measurements on $^{4}$He, C, Ar and Sn with 1-, 2-, 4-, and 6-GeV electron beams, as well as measurements on O with 1- and 2-GeV electron beams. 
The 1- and 2-GeV measurements will be performed with a minimum electron scattering angle of 5$^{o}$, compared to a minimum CLAS angle of about 15$^{o}$. 
This will extend the measurements down to the much lower momentum transfers, typical of some neutrino experiments, and to multi-hadron topologies. 
It will therefore allow comparisons with the lower beam-energy data of T2K and HK. 
The first part of the experiment ran in the second half of 2021 and the second one in the beginning of 2022.
The majority of the 2022 data sets have already been collected, with the exception of the 1\,GeV and $^{16}$O data sets. 

%\begin{figure} [htbp]
%\begin{center}
%%\includegraphics[width=\linewidth]{\figures clas12_3d_labels.pdf}
%\includegraphics[width=0.8\linewidth]{\figures clas12.pdf}
%\end{center}
%\caption{\label{CLAS12Det2D}2D schematic view of the CLAS12 detector components illustrating the wide angular acceptance in the polar angle along the beamline.}
%\end{figure}

%%%%%%%%%%%%%%%%%%%%%%%%%%%%%%%%%%%%%%%%%%%%%%%%%%%%%%%%%%%%%
\chapter{Summary}\label{summ}

The findings presented in this thesis report on both neutrino and electron cross-section modeling and analysis results in order to critically improve the understanding of lepton-nucleus interactions.
This knowledge will be used to significantly reduce the cross-section related uncertainties of forthcoming experiments that aim to extract the neutrino oscillation parameters with high-accuracy measurements.

The outlined work took a significant step towards this high-precision era with the use of neutrino data sets from the MicroBooNE liquid argon time projection chamber detector at Fermi National Laboratory.
The reported neutrino analyses isolated parts of the phase space where significant model improvements are required.
Furthermore, valuable kinematic variables were identified and established as tools to probe specific nuclear effects with multi-differential measurements.

This thesis further realized the connections between electron and neutrino interactions.
This realization resulted in significant improvements of the modeling used in neutrino oscillation analyses. 
Yet, these improved predictions failed to reproduce exclusive electron scattering results from the CLAS detector at Thomas Jefferson Laboratory using high-statistics data sets and monoenergetic beams.
However, such comparisons against electron scattering data sets can definitively constrain the vector part and the nuclear effects in lepton-nucleus interactions  in event generators that will be used for the forthcoming oscillation analyses.

%\appendix
\chapter{Appendices}\label{app}

\section{Total Struck Nucleon Momentum Derivation}\label{totstruckmom}

The total momentum of the struck nucleon can be obtained following the derivation outlined below for two-body interactions.
We followed the approach introduced by the Minerva collaboration to reconstruct the longitudinal and total nucleon momenta in~\cite{PhysRevLett.121.022504}. 
We focused on quasielastic-like (QE-like) \Signal processes,
 	
\begin{equation}
\begin{split}	
\nu_{\mu} + A \rightarrow \mu^{-} + p + (A-1),
\end{split}
\label{inter}
\end{equation}	
 	
while using the formalism detailed in~\cite{PhysRevC.95.065501} for the nuclear masses,
 	
\begin{equation}
\begin{split}
m_{A} = 22 \times M_{n} + 18 \times M_{p} - B\,[GeV]\\
m_{A-1} = m_{A} - M_{n} + \epsilon^{N}\,[GeV].
\end{split}
 \label{masses}
\end{equation}
 	
Here $M_{p}$ and  $M_{n}$ denote the proton and neutron masses, respectively, B = 0.34381 GeV is the argon binding energy (obtained from page 3 in~\cite{PhysRevC.95.065501} for 40 nucleons with an average binding energy of 9 MeV) and $\epsilon^{N}$ = 0.0309 GeV is the removal energy, with
 	
\begin{equation}
\begin{split}
\epsilon^{N} = S^{N} + E^{N}_{x} + \langle T_{A-1}\rangle
\end{split}
\label{epsilon}
\end{equation}	
 	
where $S^{N}$ = 9.9 MeV is the neutron separation energy obtained for argon from table 7 in~\cite{Bodek2019}, $E^{N}_{x}$ is the excitation energy, and $\langle T_{A-1}\rangle$ is the average kinetic energy of the remnant system, which is negligible.
 	
In equation~\ref{inter}, the incident neutrino energy $E_{\nu}$ is unknown, but the dependence of $\delta\vec{p}$ (struck nucleon momentum before the interaction) on $E_{\nu}$ can be removed under a QE-like approximation. This was achieved with the process detailed below.
    
First, we decomposed $\delta\vec{p}$ into longitudinal and transverse components with respect to the neutrino direction, used energy/momentum conservation equations and that $p_{\nu} = E_{\nu}$,
    
\begin{equation}
\begin{split}
\delta\vec{p} \equiv (\delta\vec{p}_{T}, \delta p_{L})
\end{split}
\label{Pdecomp}
\end{equation}   

\vspace{-2cm}
 	
\begin{equation}
\begin{split}
\vec{p}_{\nu} = \vec{p}_{\mu} + \vec{p}_{p} + \vec{p}_{A-1}%\\
%(total\,momentum)	
\end{split}
\label{Totaldecomp}
\end{equation}	

\vspace{-2cm}
 	
\begin{equation}
\begin{split}
E_{\nu} + \delta p_{L} = p_{L}^{\mu} + p_{L}^{p}%\\
%(longitudinal\,momentum)
\end{split}
\label{Ldecomp}
\end{equation}

\vspace{-2cm}
 	
\begin{equation}
\begin{split}
\delta\vec{p}_{T} = \vec{p}_{T}^{\mu} + \vec{p}_{L}^{p}%\\
%(transverse\,momentum)	
\end{split}
\label{Tdecomp}
\end{equation}	

\vspace{-2cm}
 	
\begin{equation}
\begin{split}
E_{\nu} + m_{A} = E^{\mu} + E^{p} + E_{A-1}%\\
%(energy\,conservation)	
\end{split}
\label{Edecomp}
\end{equation}	
 	
where $\vec{p}^{\mu}$ and $\vec{p}^{p}$ are the muon and proton momenta, respectively, $E_{\mu} = \sqrt{p_{\mu}^{2} + m_{\mu}^{2}}$ for the muon candidate and $E_{p} = \sqrt{p_{p}^{2} + m_{p}^{2}}$ for the proton candidate, where the corresponding momenta are obtained based on the particles' ranges~\cite{osti139791}, and $E_{A-1}$ is the energy of remnant nuclear system.
 	
Under the assumption that no final state interactions (FSI) take place,
 	
\begin{equation}
\begin{split}	
\vec{p}_{\nu} + \delta \vec{p} = \vec{p}_{\mu} + \vec{p}_{p}.
\end{split}
\label{nofsi}
\end{equation}	
 	
If we combine equations~\ref{Totaldecomp} and~\ref{nofsi}, $\delta\vec{p}$ gives the magnitude of its recoil momentum, $\delta\vec{p} = -\vec{p}_{A-1}$ and $\delta p = p_{A-1}$. Combining equations~\ref{Ldecomp} and~\ref{Edecomp} to eliminate $E_{\nu}$ yields
 	
\begin{equation}
\begin{split}
\delta p_{L} = m_{A} + p^{\mu}_{L} + p^{p}_{L} - E^{\mu} - E^{p} - E_{A-1}.
\end{split}
\label{DPLLong}
\end{equation}	
 	
Using the fact that $E_{A-1} = \sqrt{m_{A-1}^{2} + p_{A-1}^{2}}$, $\delta p = p_{A-1}$ and the decomposition of $\delta \vec{p}$ into longitudinal and transverse components in equation~\ref{Pdecomp},
 	
\begin{equation}
\begin{split}
\delta p_{L} = m_{A} + p^{\mu}_{L} + p^{p}_{L} - E^{\mu} - E^{p} - \sqrt{m_{A-1}^{2} + \delta p_{T}^{2} + \delta p_{L}^{2}}
\end{split}
\label{DPLReplace}
\end{equation}	
 	
For simplicity, we defined
 	
\begin{equation}
\begin{split}
R \equiv m_{A} + p^{\mu}_{L} + p^{p}_{L} - E^{\mu} - E^{p}
\end{split}
\label{Long}
\end{equation}
 	
Using R to simplify equation~\ref{DPLReplace},
 	
\begin{equation}
\begin{split}
\delta p_{L} = R - \sqrt{m_{A-1}^{2} + \delta p_{T}^{2} + \delta p_{L}^{2}}
\end{split}
\label{DPLSimple}
\end{equation}	
 	
Rearranging the terms, squaring each side and solving for $\delta p_{L}$ yields
 	
\begin{equation}
\begin{split}
\delta p_{L} = \frac{1}{2}R - \frac{m_{A-1}^{2} + \delta p_{T}^{2}}{2R}.
\end{split}
\label{DPLFinal}
 	\end{equation} 	
 	
Finally, combining the longitudinal and the transverse components, we obtain the total struck nucleon momentum. 
Given that this momentum is an approximation following the procedure and the assumptions mentioned above (QE-like scattering and no FSI), we will be referring to it as $p_{n,proxy}$ as opposed to $\delta p$, where
 	
\begin{equation}
\begin{split}
p_{n,proxy} = \sqrt{\delta p_{L}^{2} + \delta p_{T}^{2}}
\end{split}
\label{tot}
\end{equation}
\section{Wiener SVD Regularization Technique}\label{svdapp}

The procedure detailed below relies on~\cite{Tang_2017}.
We choose to work with a $\chi^{2}(s)$ metric.
    
\begin{equation}
\chi^{2}(s) = (\mathbf{m} - r \cdot s)^{T} Cov^{-1} (\mathbf{m} -r \cdot s) 
\label{metric}
\end{equation}
    
 \begin{center}
 \begin{table}[htb]
 \centering
 \begin{tabular}{ c c c }
 \hline
 \hline
 \makecell{Notation} & \makecell{Explanation} & \makecell{Dimension\\And Format} \\ 
 \hline
 \hline
 $\mathbf{m}$ & measured spectrum in data (signal and background events) & m $\times$ 1 vector\\
 s & free variable in $\chi^{2}(s)$ function & n $\times$ 1 vector\\
 $\hat{s}$ & estimator of true signal, obtained after minimizing $\chi^{2}$ & n $\times$ 1 vector\\
 $s_{true}$ & true signal & n $\times$ 1 vector\\
 $\overline{s}$ & nominal MC signal prediction & n $\times$ 1 vector\\
 r & response matrix & m $\times$ n matrix \\
 Cov & \makecell{symmetric covariance matrix\\with statistical and systematic uncertainties} & m $\times$ m matrix\\
 \hline
 \hline
 \end{tabular}
\label{MetricTable}
 \end{table} 
 \end{center}  
 
Note that we use $s_{true}$ to represent the true signal in order to differentiate from s which is a variable in the function $\chi^{2}(s)$. We use $\hat{s}$ to represent the estimator of the true signal $s_{true}$, which is obtained after minimizing the $\chi^{2}(s)$ function. We further restrict ourselves to the m $\geq$ n case.
 
Our objective is to minimize $\chi^{2}(s)$. Since the covariance matrix Cov is symmetric, the inverse of it $Cov^{-1}$ is also symmetric. Hence, $Cov^{-1}$ can be decomposed with Cholesky decomposition~\cite{cholesky} into
 
\begin{equation}
Cov^{-1} = Q^{T} \cdot Q 
\label{cholesky}
\end{equation}
    
where Q is a uniquely defined lower triangular matrix and $Q^{T}$ is its transpose. We further define
    
\begin{equation}
\begin{split}    
M \equiv Q \cdot \mathbf{m}\\
R \equiv Q \cdot r\\  
(pre-scaling)
\label{prescale}
\end{split}    
\end{equation}  
    
 \begin{center}
 \begin{table}[htb]
 \centering
 \begin{tabular}{ c c c }
 \hline
 \hline
 \makecell{Notation} & \makecell{Explanation} & \makecell{Dimension\\And Format} \\ 
 \hline
 \hline
 M & measured spectrum after pre-scaling & m $\times$ 1 vector\\
 R & response matrix after pre-scaling & m $\times$ n matrix \\
 Q & Lower triangular matrix from Cholesky decomposition of $Cov^{-1}$  & m $\times$ m matrix \\ 
 $\overline{M}$ & expectation spectrum after pre-scaling (R $\cdot$ $\overline{s}$) & m $\times$ 1 vector\\
 \hline
 \hline
 \end{tabular}
\label{MetricTablePreScaled}
 \end{table} 
 \end{center}    
    
Replacing the definitions in equation~\ref{prescale} into equation~\ref{metric} yields
    
\begin{equation}
\chi^{2}(s) = (M - R \cdot s)^{T} (M -R \cdot s) = \sum_{i} (M_{i} - \sum_{j} R_{ij} \cdot s_{j})^{2}
\label{PreScaledMetric}
\end{equation}   
    
Ideally, we want to achieve $\chi^{2}(s)$ = 0. That is obtained when
    
\begin{equation}
\begin{split}
M - R \cdot s = 0 \Rightarrow M = R \cdot s \Rightarrow R^{T} \cdot M = R^{T} \cdot R \cdot s
\label{ZeroOut}
\end{split}
\end{equation}    
    
We use the fact that $R^{T} \cdot R$ is a square n$\times$n invertable matrix. Thus, the exact solution $\hat{s}$ is uniquely defined as
    
\begin{equation}
\hat{s} = (R^{T} \cdot R)^{-1} \cdot R^{T} \cdot M 
\label{ExactSol}
\end{equation}
    
We decompose the measured spectrum into the signal (R $\cdot$ s) part and the noise / background (N).
    
\begin{equation}
M = R \cdot s_{true} + N 
\label{MDecomp}
\end{equation}  
    
Substituting equation~\ref{MDecomp} into equation~\ref{ExactSol} yields
    
\begin{equation}
\hat{s} = (R^{T} \cdot R)^{-1} \cdot R^{T} \cdot (R \cdot s_{true} + N)
\label{ExactSolExpand}
\end{equation}  
    
with N representing the ``noise'' coming from uncertainties (statistical and systematic uncertainties associated with both $\textbf{m}$ and r). 
Each term in the noise vector after pre-scaling follows a normal distribution with $\mu$ = 0 and $\sigma$ = 1, since the denominator of the $\chi^{2}$ function in equation~\ref{PreScaledMetric} (i.e. square of error) is unity. 
Given the fact that each term in the noise vector is independent (i.e. uncorrelated), we refer to the basis in this domain as orthogonal.
    
The response matrix after pre-scaling (R) can be decomposed, using the singular value decomposition (SVD) approach~\cite{decomp}, as
    
\begin{equation}
R = U \cdot D \cdot V^{T}
\label{RDecomp}
\end{equation}    
    
with both $U_{m\times m}$ and $V_{n\times n}$ being orthogonal matrices that satisfy $U^{T}\cdot U$ = $U\cdot U^{T}$ = $I_{m\times m}$ and $V^{T}\cdot V$ = $V\cdot V^{T}$ = $I_{n\times n}$, with U,V and D being uniquely defined. I is the identity matrix and the subscript represents the dimension. D is an m$\times$n diagonal matrix with positive definite diagonal elements (known as singular values) $D_{ii}$ = $d_{i}$ arranged in descending order as i increases.
    
  \begin{center}
 \begin{table}[htb]
 \centering
 \begin{tabular}{ c c c }
 \hline
 \hline
 \makecell{Notation} & \makecell{Explanation} & \makecell{Dimension\\And Format} \\ 
 \hline
 \hline
 $V^{T}$ & right orthogonal matrix from decomposition of R & n $\times$ n vector\\
 D & diagonal matrix from decomposition of R & m $\times$ n matrix \\
 U & left orthogonal matrix from decomposition of R & m $\times$ m matrix \\
 \hline
 \hline
 \end{tabular}
\label{Matrices}
 \end{table} 
 \end{center}   
 
Inserting equation~\ref{RDecomp} into~\ref{ExactSolExpand}, we have
 
\begin{equation}
\begin{split}    
\hat{s} = V \cdot D^{-1} \cdot U^{T} \cdot (R \cdot s_{true} + N)\\
= V \cdot D^{-1} \cdot (R_{U} \cdot s_{true} + N_{U})\\  
= V \cdot D^{-1} \cdot M_{U}
\label{TransSolution}
\end{split}    
\end{equation} 
    
where $R_{U}\equiv U^{T}\cdot R$, $N_{U}\equiv U^{T}\cdot N$, and $M_{U}\equiv U^{T}\cdot M$ are transformations of the smearing matrix R, the noise N, and the measured signal M, respectively. Note that, since U is an orthogonal matrix and the elements of the original noise vector N are uncorrelated, the elements of the new noise vector $N_{U}$ are still uncorrelated. Each element follows a normal distribution with $\mu$ = 0 and $\sigma$ = 1. Thus, the basis in this new domain is still orthogonal.
   
However, there are cases where the unbiased solution to an unfolding problem via a direct inversion can have catastrophic oscillations via the introduction of huge variances.
	
A proposed solution is the introduction of a trade-off term between the bias and the variance to suppress the oscillations. 
We refer to that term as ``regularization'' and is introduced in the form of an n$\times$n matrix $F$ that is applied on the exact solution $\hat{s}$.
    
\begin{equation}
\begin{split}    
\hat{s} = F \cdot V \cdot D^{-1} \cdot (R_{U} \cdot s_{true} + N_{U})\\
= F \cdot V \cdot D^{-1} \cdot U^{T} \cdot M\\
= F \cdot V \cdot D^{-1} \cdot M_{U}
\label{ExactSolTweaked}
\end{split}
\end{equation}    
    
Using equation~\ref{ExactSolTweaked} (focusing on the diagonal elements ii, using the fact that V is orthogonal $V \cdot V^{T}= I$, and that D is diagonal $D_{ii}^{-1} = 1/d_{i}$),
    
\begin{equation}
\begin{split}  
(V^{T} \hat{s})_{i} = F_{ii} \cdot \frac{M_{U,i}}{d_{i}}\\
\hat{S}(\omega) = F(\omega) \cdot \frac{M(\omega)}{R(\omega)}
\label{comparison2}
\end{split}
\end{equation}   
    
It is easy to see the similarities between the two lines in equation~\ref{comparison2}. Therefore, following the same terminology as that for the signal processing, we refer to $M_{U}$ after the SVD transformation as the measurement in the effective frequency domain in analogy to the frequency domain in the signal processing.
    
The additional matrix F can be decomposed as 
    
\begin{equation}
F = V \cdot W \cdot V^{T}.
\label{AddSmearMatrix}
\end{equation}    
    
At this point, the F and W matrices are still unknown.
    
Plugging equation~\ref{AddSmearMatrix} into equation~\ref{ExactSolTweaked} yields

\begin{equation}
\hat{s} = V \cdot W \cdot D^{-1} \cdot (R_{U} \cdot s_{true} + N_{U}).
\label{ExactSolReTweaked}
\end{equation}
    
We consider the expectation value of the signal in the effective frequency domain:
 
\begin{equation}
\overline{M_{U}} = U^{T} \cdot \overline{M} = U^{T} \cdot R \cdot \overline{s} 
\label{MU}
\end{equation} 
    
In general, the $s_{true}$ is unknown, so the expectation signal $\overline{s}$ using the nominal simulation prediction is used.
    
The construction of W is based on the Wiener filter $\frac{\overline{R^{2}(\omega) \cdot S^{2}(\omega)}}{\overline{R^{2}(\omega) \cdot S^{2}(\omega)} + \overline{N^{2}(\omega)}}$. Taking equation~\ref{MU} at bin i, we have
    
\begin{equation}
\overline{\langle R^{2} \cdot S^{2}\rangle} = \overline{M_{U,i}^{2}} =^{Eq.~\ref{comparison2}} d_{i}^{2} \cdot ( \sum_{j} V_{ij}^{T} \cdot \overline{s_{j}})^{2}
\label{MUWiener}
\end{equation}
    
\begin{equation}
\overline{\langle N^{2}\rangle} = 1
\label{NWiener}
\end{equation}  
    
resulting in a Wiener filter of 
    
\begin{equation}
W_{ik} = \frac{d_{i}^{2} \cdot (\sum_{j} V_{ij}^{T} \cdot \overline{s_{j}})^{2}}{d_{i}^{2} \cdot (\sum_{j} V_{ij}^{T} \cdot \overline{s_{j}})^{2} + 1} \cdot \delta_{ik}
\label{Wform}
\end{equation} 
    
At this point, W is uniquely defined. Here, equation~\ref{NWiener} is obtained, since each element of noise $N_{U}$ follows a normal distribution with $\mu$ = 0 and $\sigma$ = 1. We have
    
\begin{equation}
(W \cdot D^{-1})_{ij} = \frac{d_{i} \cdot (\sum_{j} V_{ij}^{T} \cdot \overline{s_{j}})^{2}}{d_{i}^{2} \cdot (\sum_{j} V_{ij}^{T} \cdot \overline{s_{j}})^{2} + 1} \cdot \delta_{ik}
\label{WDInvform}
\end{equation}   
    
The small value of $d_{i}$ is balanced by the finite value of the expectation value of $\overline{N^{2}} \equiv$ 1 and, thus, equation~\ref{ExactSolReTweaked} doesn't suffer from catastrophic oscillations. From equation~\ref{Wform}, the construction of the Wiener filter takes into account the strengths of both the signal and noise expectations and is independent of the regularization strength $\tau$ used in traditional regularization techniques~\cite{Hocker}.
    
As shown in~\cite{Hocker}, the regularization can be applied on the curvature of the spectrum instead of the strength of the spectrum, which involves an additional matrix C. 
This is also be achieved in the Wiener-SVD approach:
    
\begin{equation}
\overline{M} = R \cdot C^{-1} \cdot C \cdot \overline{s} 
\label{MC}
\end{equation}    
    
by including an additional matrix C that has the commonly used regularization forms, such as the first and second order derivatives. 
Since the effective frequency domain is determined by the smearing matrix R, the inclusion of C would alter the basis of the effective frequency domain. In this case, the SVD decomposition becomes (just like shown in equation~\ref{RDecomp})
    
\begin{equation}
R \cdot C^{-1} = U_{C} \cdot D_{C} \cdot V_{C}^{T}. 
\label{RC}
\end{equation}    
    
\begin{center}
 \begin{table}[htb]
 \centering
 \begin{tabular}{ c c c }
 \hline
 \hline
 \makecell{Notation} & \makecell{Explanation} & \makecell{Dimension\\And Format} \\ 
 \hline
 \hline
 $V^{T}_{C}$ & right orthogonal matrix from decomposition of R $\cdot C^{-1}$ & n $\times$ n vector\\
 $D_{C}$ & diagonal matrix from decomposition of R $\cdot C^{-1}$ & m $\times$ n matrix \\
 $U_{C}$ & left orthogonal matrix from decomposition of R $\cdot C^{-1}$ & m $\times$ m matrix \\
 \hline
 \hline
 \end{tabular}
\label{MatricesC}
 \end{table} 
 \end{center}      
    
The final solution of the regularisation becomes (just like shown in equation~\ref{TransSolution} by inserting it into equation~\ref{ExactSolExpand})
    
\begin{equation}
\hat{s} = C^{-1} \cdot V_{C} \cdot W_{C} \cdot V_{C}^{T} \cdot C \cdot (R^{T} R)^{-1} \cdot R^{T} \cdot M 
\label{Shat}
\end{equation}    
    
or, equivalently,
\begin{equation}
\hat{s} = A_{C} \cdot (R^{T} R)^{-1} \cdot R^{T} \cdot M 
\label{ShatAC}
\end{equation}   
    
where

\begin{equation}
A_{C} = C^{-1} \cdot V_{C} \cdot W_{C} \cdot V_{C}^{T} \cdot C
\label{ACDef}
\end{equation}  
    
The corresponding Wiener filter would be (once again, by realizing the similarities between the two expressions in equation~\ref{comparison2} and by using equation~\ref{ACDef})
    
\begin{equation}
W_{C,ii} = \frac{d_{Ci}^{2} \cdot (\sum_{j} V_{Cij}^{T} \cdot (\sum_{l} C_{jl}\overline{s_{l}}))^{2}}{d_{Ci}^{2} \cdot (\sum_{j} V_{Cij}^{T} \cdot (\sum_{l} C_{jl}\overline{s_{l}}))^{2} + 1}
\label{WCform}
\end{equation}    
    
where $C_{jl}$, $V^{T}_{Cij}$, and $d_{Ci}$ are matrix elements of matrices C, $V_{C}$, and $D_{C}$, respectively.
    
Since the unfolded results are a linear transformation of the measurement, we can easily evaluate the uncertainties associated with them. Equation~\ref{ShatAC} can be rewritten as 
    
\begin{equation}
\hat{s} = R_{tot} \cdot m
\label{Rot}
\end{equation}  
    
with
     
\begin{equation}
R_{tot} = A_{C} \cdot (R^{T} R)^{-1} \cdot R^{T} \cdot Q.
\label{RotDef}
\end{equation}   
    
Then, the covariance matrix of $\hat{s}$ can be deduced from the covariance matrix of $\mathbf{m}$ as
    
\begin{equation}
Cov_{\hat{s},\mathbf{m}} = R_{tot} \cdot Cov_{\mathbf{m}} \cdot R_{tot}^{T}.
\label{Covs}
\end{equation}
    
The variances of the unfolded data can also be easily calculated given that their origin N in equation~\ref{Shat} is uniquely defined. Defining N(i) as a vector with the i-th element being 1 and the rest of elements being 0, we can calculate the variance in s due to i-th element in N as:
    
\begin{equation}
T_{deviation}(i) = A_{C} \cdot (R^{T} R)^{-1} \cdot R^{T} \cdot N(i),
\label{Tdev}
\end{equation}    
    
with $T_{deviation}$(i) being a vector. The variance of the j-th element of $T_{deviation,j}$ can thus be written as:
    
\begin{equation}
T_{deviation,j} = \sqrt{\sum_{i}T^{2}_{deviation,j}(i)},
\label{TdevSqrt}
\end{equation} 
    
after summing the contribution from each independent noise source. 
The square of $T_{deviation,j}$ corresponds to the j-th diagonal element of the covariance matrix $Cov_{s}$ in equation~\ref{Covs}.
    
Given equation~\ref{Shat}, we can understand the entire process of unfolding as to ``remove'' the effect of R through multiplying $(R^{T}R)^{-1} \cdot R^{T}$ and then replace it with a new smearing matrix $A_{C}$. 
Therefore, it is straightforward to estimate the bias on the unfolded results:
    
\begin{equation}
\begin{split}
T_{bias} = (A_{C} - I) \cdot (R^{T}R)^{-1} \cdot R^{T} \cdot \overline{M}\\
= (A_{C} - I) \cdot \overline{s}
\end{split}    
\label{Tbias}
\end{equation}   
    
with I being identity matrix and $\overline{s}$ being the expectation of the nominal MC signal prediction.
    \section{Electrons-For-Neutrinos Fiducials}\label{fid}
    
    Fiducial cuts for e2a have been defined and used by several analyses \cite{Niyazov,Protop_elfidcut4,BZhang_elfid2} and follow the same general procedure for all charged particle species.
    A series of event selection cuts are applied to data to produce event samples suitable for defining the fiducial regions of CLAS.
    From these event samples, regions of good acceptance are found, defined as flat regions of $\phi$ space in bins of momentum and $\theta$, which are then parameterised into functions.
    The specific procedures for each particle species for which fiducial cuts have been defined are summarised in the following sections.    

    \begin{figure}[!htb]
    \centering
    \subfloat[]{\includegraphics[width=0.475\textwidth]{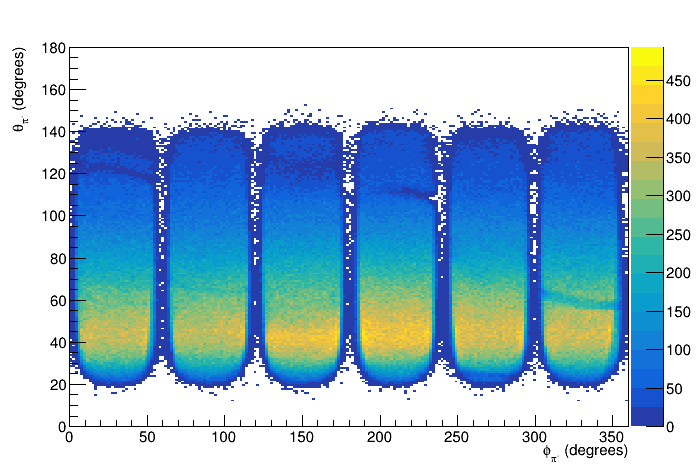}}
    \subfloat[]{\includegraphics[width=0.475\textwidth]{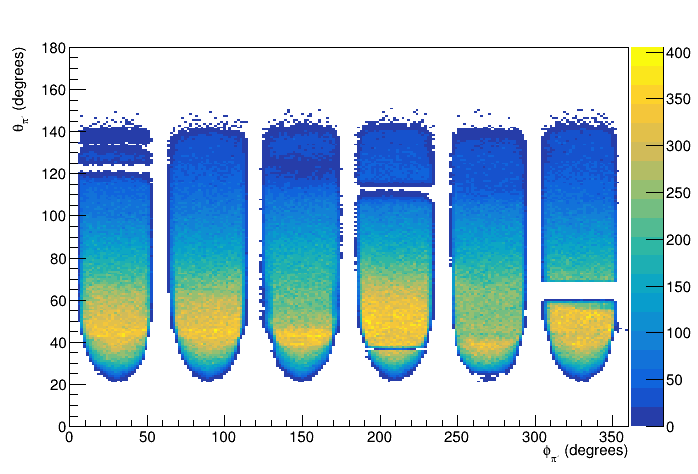}}
    \caption[The $\theta$ vs $\phi$ distributions for $\pi^{-}$ at $2.2\;\mathrm{GeV}$.]{The $\theta$ vs $\phi$ distributions for $\pi^{-}$, before (a) and after (b) the application of fiducial cuts as defined in \cite{e4nuQE} for $2.2\;\mathrm{GeV}$ analysis of $^{4}\mathrm{He}$.}
    \label{fig:pimi_fid}
    \end{figure}

	%%%%%%%%%%%%%%%%%%%%%%%%%%%%%%%%%%%%%%%%%%%%%%%%%%%%%%%%%%%%%%%%%%%%%%%

    %\section{Definition for electrons}
    
    Electron fiducial cuts were defined in~\cite{McLauchlan}, using electron candidates identified by the CLAS triggering requirements for e2a.
    These electron samples are then subjected to geometric cuts in the $u$, $v$, and $w$ ``views'' corresponding to the orientation of the scintillator layers of the electromagnetic calorimeter.
    This cut accounts for poorly understood electron acceptance at the edges of the calorimeter, and the difference in acceptance between the electromagnetic calorimeter and the Cerenkov Counter, which can be seen in data as a characteristic ``smile'' feature in the uncut $\theta$ vs $\phi$ distributions for electrons.
    Additionally, a cut is applied on the ratio of energy deposition by an electron in the calorimeter to electron momentum as measured by the drift chambers, $E_{tot}/p_{e'}$.
    This ratio is known as the `sampling fraction' and is fixed by the design of the detector to $0.3c$.
    At the detector edges, this value can decrease due to shower leakage, where the electron energy is not fully deposited in the calorimeter.

    The electrons distributions for each beam energy and torus field setting are then split into sectors and momentum bins $50~MeV/c$ wide, spanning the electron momentum range of the data.
    A plot of $\theta_{e}$ against $\phi_{e}$ is produced for each momentum bin, and for each degree in $\theta_{e}$, the plot was projected onto the $\phi_{e}$ axis and fitted with a trapezoid function.

    \begin{figure}[!htb]
    \centering
    \subfloat[]{\includegraphics[width=0.475\textwidth]{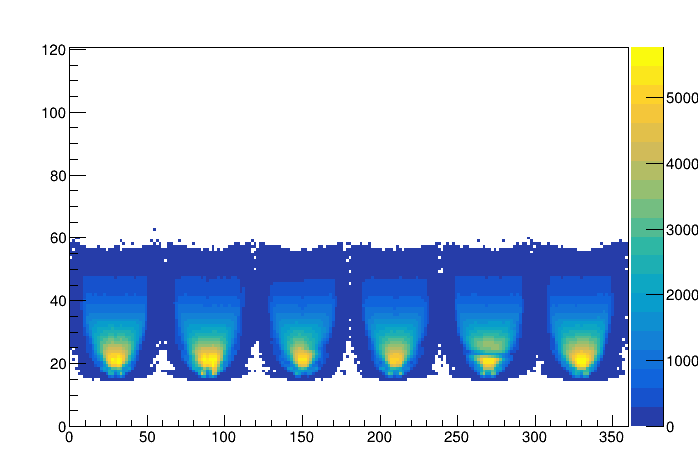}}
    \subfloat[]{\includegraphics[width=0.475\textwidth]{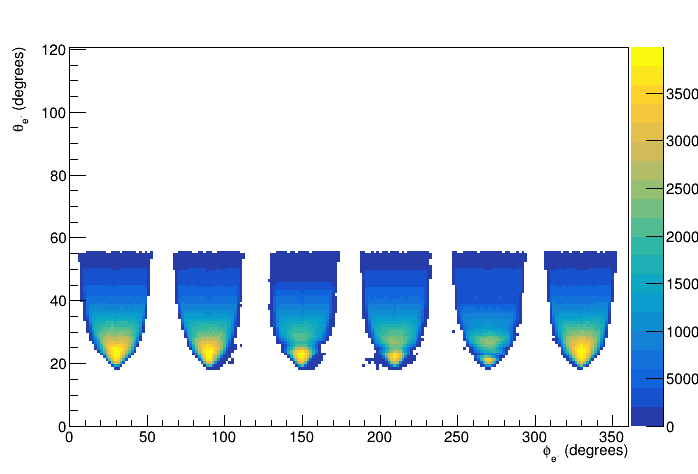}}
    \caption[The $\theta$ vs $\phi$ distributions for electrons at $2.2\;\mathrm{GeV}$.]{The $\theta$ vs $\phi$ distributions for electrons, before (left) and after (right) the application of fiducial cuts for $2.2\;\mathrm{GeV}$ analysis of $^{4}\mathrm{He}$.}
    \label{fig:electron_fid}
    \end{figure}
    
	%%%%%%%%%%%%%%%%%%%%%%%%%%%%%%%%%%%%%%%%%%%%%%%%%%%%%%%%%%%%%%%%%%%%%%%    

    %\section{Definition for positively charged hadrons (protons and positively charged pions)}

    Positively charged hadrons, i.e.~protons and $\pi^{+}$, have their fiducial cuts defined as a single species, under the assumption that their identical charge means their fiducial regions will be the same.
    The positive hadron fiducial cuts were defined in~\cite{Niyazov_posfidcut4}. 

    Event samples are identified as tracks with good drift chamber status, a good hit in the time-of-flight system, and particle identification via a $\chi^{2}$ cut on the DCPB bank

    In order to eliminate protons in quasi-free reactions, which are not uniformly distributed in $\phi$, an energy transfer cut is applied

    The positively charged hadron distributions for each beam energy and torus field setting are then split into sectors and momentum bins $50~MeV/c$ wide, spanning the momentum range of the data.
    A plot of $\theta$ against $\phi$ is produced for each momentum bin, and for each degree in $\theta$, the plot was projected onto the $\phi$ axis and fitted with a trapezoid function.

    \begin{figure}[!htb]
    \centering
    \subfloat[]{\includegraphics[width=0.475\textwidth]{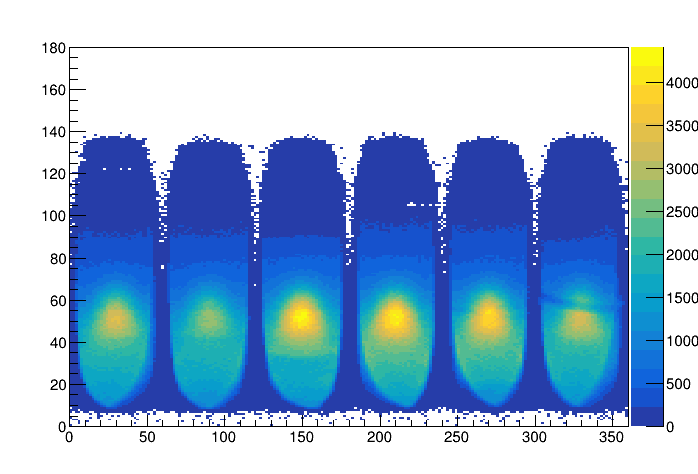}}
    \subfloat[]{\includegraphics[width=0.475\textwidth]{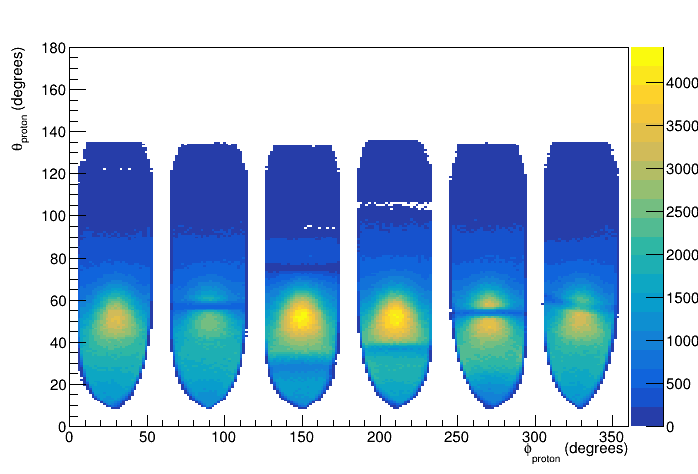}}
    \caption[The $\theta$ vs $\phi$ distributions for protons.]{The $\theta$ vs $\phi$ distributions for protons, before (left) and after (right) the application of fiducial cuts for $2.2\;\mathrm{GeV}$ analysis of $^{4}\mathrm{He}$.}
    \label{fig:proton_fid}
    \end{figure}
 
    \begin{figure}[!htb]
    \centering
    \subfloat[]{\includegraphics[width=0.475\textwidth]{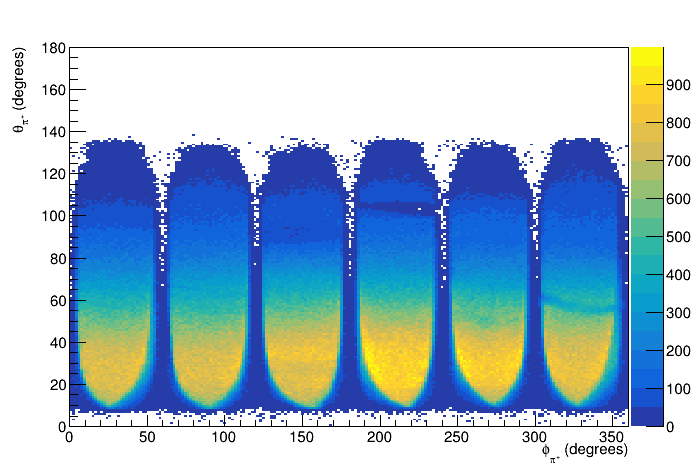}}
    \subfloat[]{\includegraphics[width=0.475\textwidth]{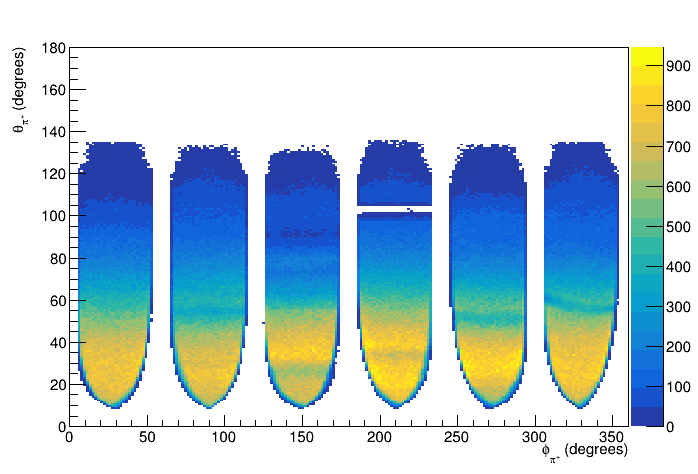}}
    \caption[The $\theta$ vs $\phi$ distributions for $\pi^{+}$.]{The $\theta$ vs $\phi$ distributions for $\pi^{+}$, before (left) and after (right) the application of fiducial cuts for $2.2\;\mathrm{GeV}$ analysis of $^{4}\mathrm{He}$.}
    \label{fig:pipl_fid}
    \end{figure}
    
	%%%%%%%%%%%%%%%%%%%%%%%%%%%%%%%%%%%%%%%%%%%%%%%%%%%%%%%%%%%%%%%%%%%%%%%    

    %\section{Definition for photons}
    
\begin{figure}[!htb]
    \centering
    \subfloat[$4.4\;\mathrm{GeV}$.]{\includegraphics[width=0.5\textwidth]{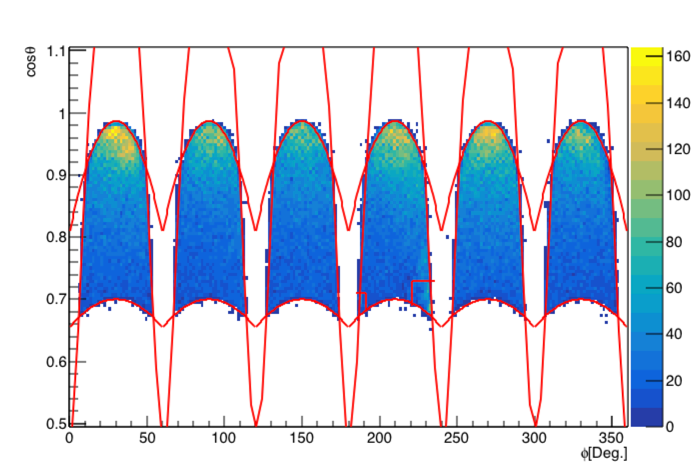}}
    \caption[The $\cos\theta$ vs $\phi$ distributions for photons.]{The $\cos\theta$ vs $\phi$ distributions for photons, with fiducial cut outline indicated by red for $4.4\;\mathrm{GeV}$ analysis for $^{3}\mathrm{He}$.}
    \label{fig:phot_fid_uvw}
    \end{figure}

    As neutral particles, the fiducial cuts for photons take the from of a cut on the detection area of the electromagnetic calorimeter, with no momentum dependence.
    Photon events are selected after applying the same $u$, $v$ and $w$ calorimeter cuts as used for electrons, and the fiducial regions for each energy and target defined on distributions of $cos\theta$ vs $\phi$.
    Two first order polynomials are used to describe the outline of the sides of the sector and two second order polynomials to describe the top and bottom edges, as shown in figure \ref{fig:phot_fid_uvw}.
    Additionally, two `hot spots' in the corners of sector four were removed.

	%%%%%%%%%%%%%%%%%%%%%%%%%%%%%%%%%%%%%%%%%%%%%%%%%%%%%%%%%%%%%%%%%%%%%%%

    %\section{Implementation in e2a/e4nu code}

    The originally defined fiducial cut functions for e2a were implemented in C++ and FORTRAN versions for the ROOT and PAW data analysis frameworks respectively.
    They were checked into the CLAS CVS repository for e2 analysis software, and have been applied as part of the various analyses performed on this data since it was collected in 1999.

    The e4nu analyses of e2a data have used the C++/ROOT implementation of the fiducial cuts, incorporating them into their analysis software, with minor modifications to ensure compatibility with contemporary software environments.

	%%%%%%%%%%%%%%%%%%%%%%%%%%%%%%%%%%%%%%%%%%%%%%%%%%%%%%%%%%%%%%%%%%%%%%%

    %\section{Reuse of electron cuts as a proxy for negatively charged pions}

    In the analysis of \cite{e4nuQE}, $\pi^{-}$ fiducial cuts were defined, initially by direct reuse of the electron fiducial cut parameters to cut fiducial regions for $\pi^{-}$.
    This was considered to offer a reasonable approximation for $\pi^{-}$, having the same charge and thus similar behaviour in the torus field of CLAS.
    However, because the electron fiducial cuts are only defined to a minimum momentum of 350 MeV/c, and the minimum momentum threshold used for pion detection in CLAS is 150 MeV/c, a new set of cut parameters had to be defined for the $\pi^{-}$ in the momentum range $150-350\;\mathrm{MeV/c}$.
    These used $50\;\mathrm{MeV/c}$ momentum bins on  $2.2\;\mathrm{GeV}$ $^{12}\mathrm{C}$ data, and follow the same procedure to obtain the $\theta$ vs $\phi$ outline cuts and the $\theta$ gaps corresponding to malfunctioning TOF paddles at this low momentum as described in\cite{BZhang_elfid2} for the $p>350\;\mathrm{MeV/c}$ region.

    At 1.1 GeV beam energy, this procedure was used to define $\pi^{-}$ fiducial cuts for the whole momentum range, as the electron cuts offer minimal coverage for $\pi^{-}$.

	%%%%%%%%%%%%%%%%%%%%%%%%%%%%%%%%%%%%%%%%%%%%%%%%%%%%%%%%%%%%%%%%%%%%%%%

    %\section{Refinement of negatively charged fiducial cuts}
    
    The theta gap functions for dead channels in $\pi^{-}$ were updated to account for previously missed gaps, and to skip electron gaps from the CC, a detector not used in determining the $\pi^{-}$.
    Several functions were refit to extend to the lower momentum range of the pions, removing the need to apply them only in specific momentum ranges, and allowing the clumsy box cuts previously used to be eliminated.

    \begin{figure}[!htb]
    \centering
    \includegraphics[width=0.5\textwidth]{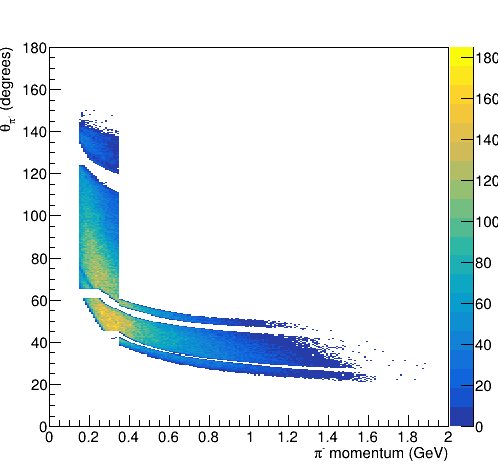}
    \caption[The $\theta$ vs momentum distributions for $\pi^{-}$.]{The $\theta$ vs momentum distributions for $\pi^{-}$ in sector 3, after the application of the original fiducial and theta gap cuts for $2.2\;\mathrm{GeV}$ analysis of $^{4}\mathrm{He}$. At low momentum, the theta gap cuts fail, and have been replaced box cuts that do not appropriately describe the gaps at low momenta.}
    \label{fig:pimi_fid_2gevOld}
    \end{figure}

\begin{figure}[!htb]
    \centering
    \includegraphics[width=0.9\textwidth]{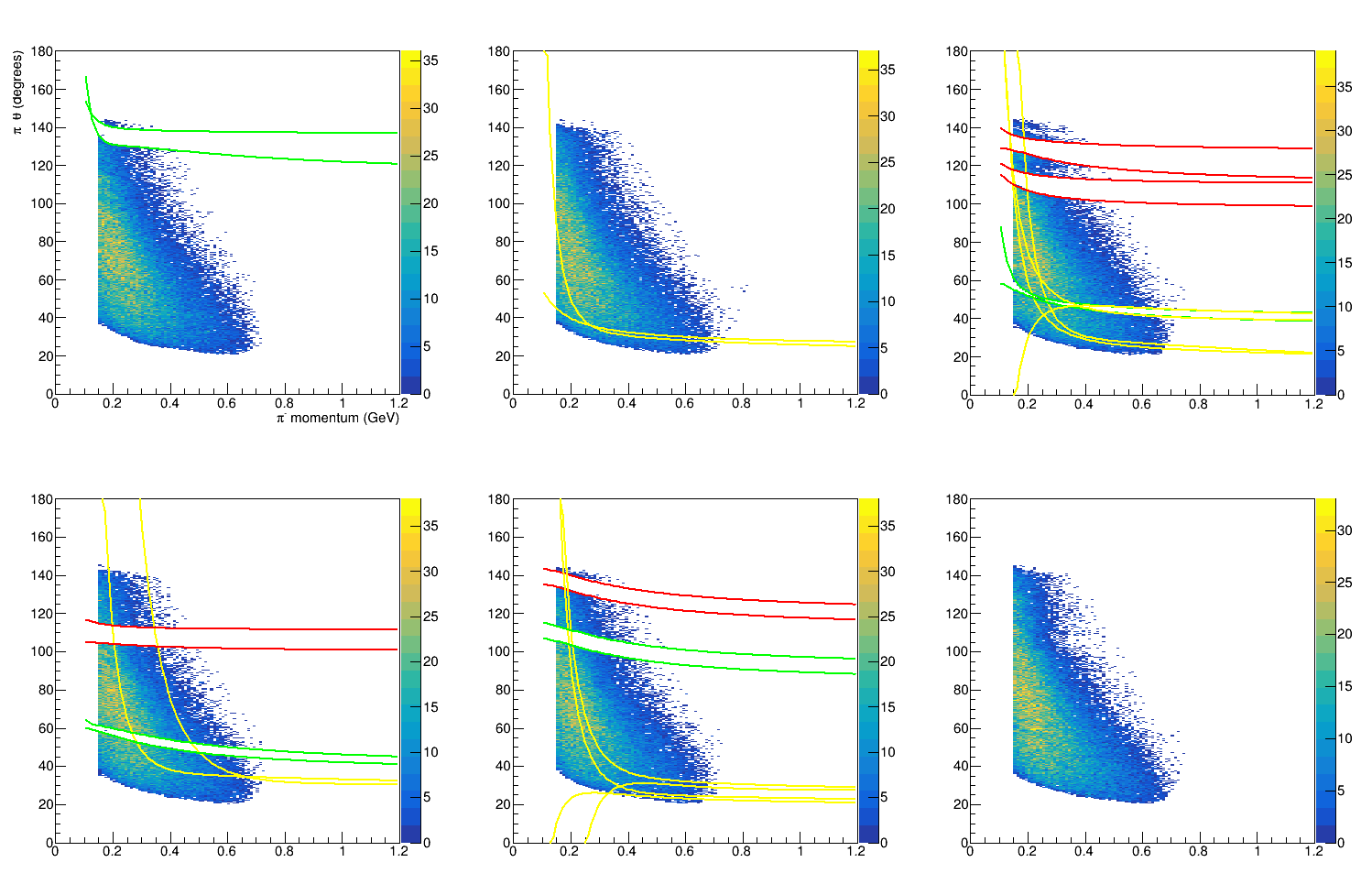}
    \caption[The $\theta$ vs momentum distributions for $\pi^{-}$.]{The $\theta$ vs momentum distributions for $\pi^{-}$, showing the $\pi^{-}$ updated theta gap cuts (green), retained (red) and not used (yellow) electron theta gap cuts for $1.1\;\mathrm{GeV}$ analysis of $^{3}\mathrm{He}$.}
    \label{fig:pimi_fid_gaps1}
    \end{figure}

    Figure \ref{fig:pimi_fid_gaps1} shows the polar angle versus momentum distributions for $\pi^{-}$ in each sector of CLAS, after the application of fiducial cuts.
    Of the theta gap functions used for the electrons, several are carried over unchanged, some are updated in order to appropriately apply to the full momentum range of $\pi^{-}$, while gaps defined for the CC are omitted, as this subsystem is not used to identify $\pi^{-}$.

    %\subsection{2 and 4 GeV runs}
    
\begin{figure}[!htb]
    \centering
    \includegraphics[width=0.9\textwidth]{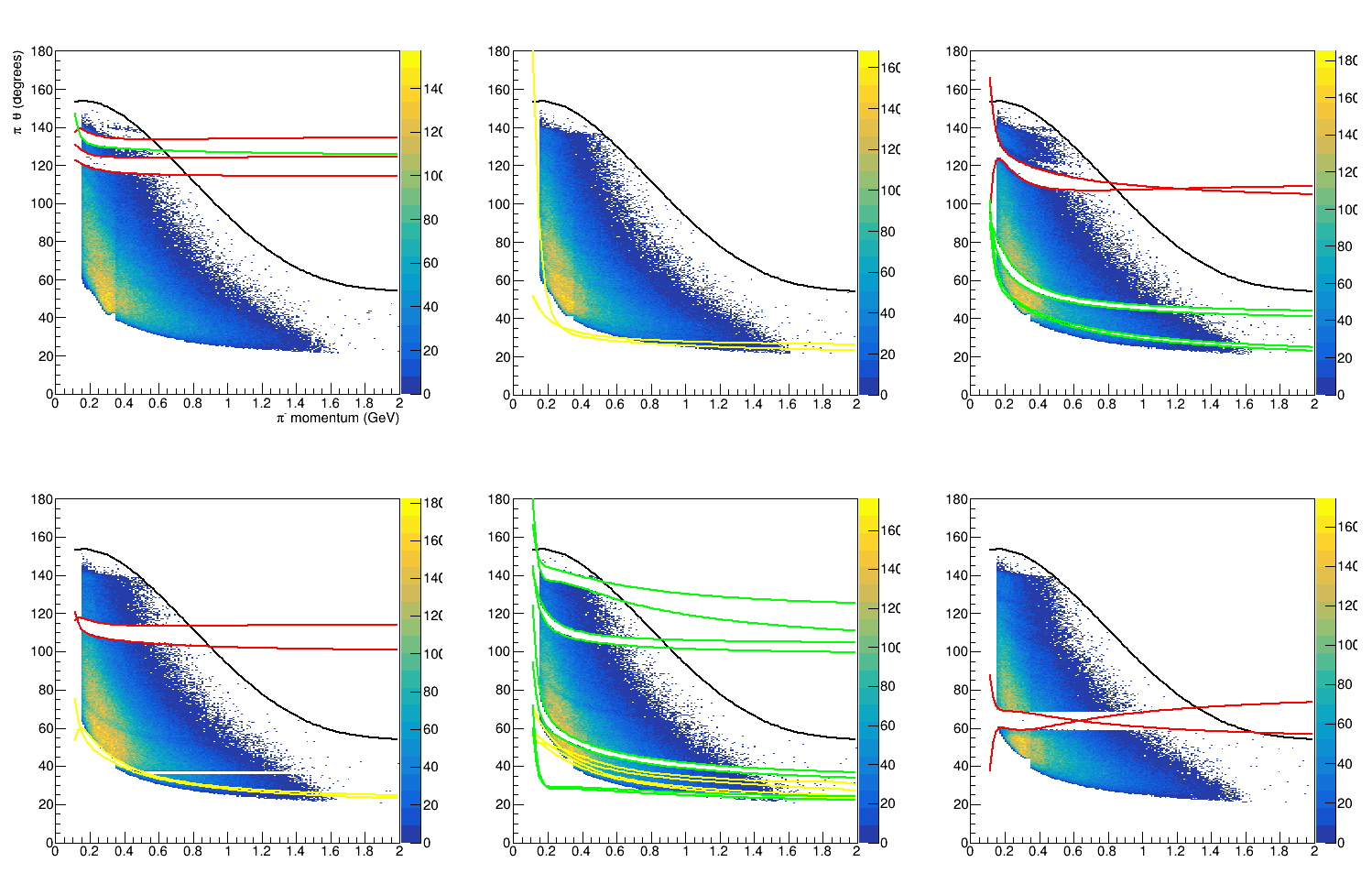}
    \caption[The $\theta$ vs momentum distributions for $\pi^{-}$.]{The $\theta$ vs momentum distributions for $\pi^{-}$, showing the $\pi^{-}$ updated theta gap cuts (green), retained (red) and not used (yellow) electron theta gap cuts for $2.2\;\mathrm{GeV}$ analysis of $^{3}\mathrm{He}$. A new parameterisation of the maximum polar angle cut is also shown (black line). The same gap functions are applied at $4.4\;\mathrm{GeV}$}
    \label{fig:pimi_fid_gaps2and4}
    \end{figure}

    Figure \ref{fig:pimi_fid_gaps2and4} shows the polar angle versus momentum distributions for $\pi^{-}$ in each sector of CLAS, after the application of fiducial cuts.
    As with the 1 GeV case, several theta gap functions are carried over unchanged from electrons, some are updated in order to appropriately apply to the full momentum range seen for $\pi^{-}$, and CC gaps are omitted.
    Additionally at $2.2\;\mathrm{and}\;4.4\;\mathrm{GeV}$, the maximum polar angle condition for $p>350\;\mathrm{MeV/c}$, imposed by the electron fiducial cut parameters, is removed.

	%%%%%%%%%%%%%%%%%%%%%%%%%%%%%%%%%%%%%%%%%%%%%%%%%%%%%%%%%%%%%%%%%%%%%%%    
    %\clearpage
    %\section{Production of new fiducial functions for negatively charged pions}

\begin{figure}[!htb]
    \centering
    \subfloat[]{\includegraphics[width=0.85\textwidth]{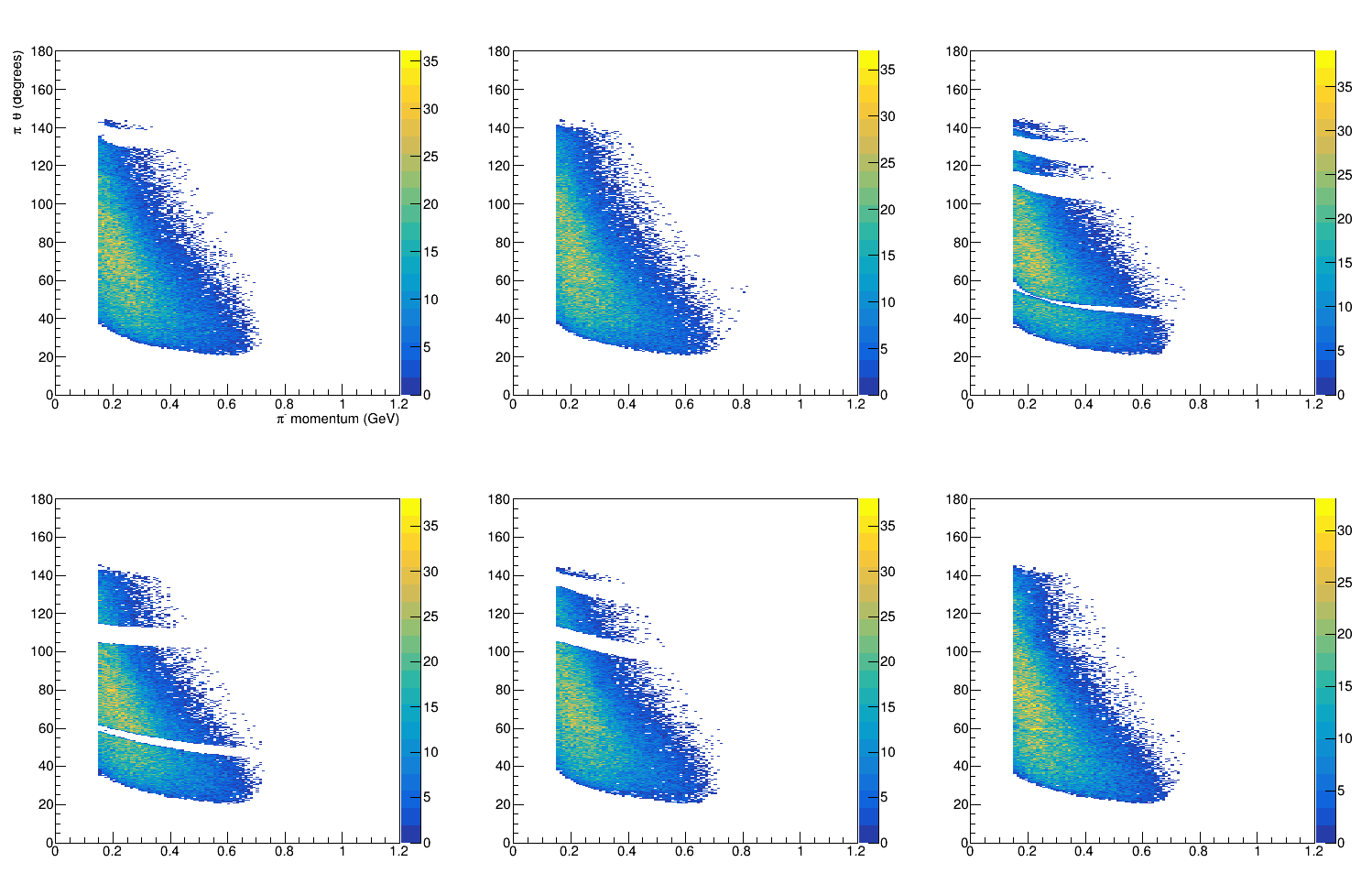}}
    \caption[The $\theta$ vs momentum distributions for $\pi^{-}$.]{The $\theta$ vs momentum distributions for $\pi^{-}$, after the application of the updated theta gap cuts for $1.1\;\mathrm{GeV}$ analysis of $^{3}\mathrm{He}$.}
    \label{fig:pimi_fid_1gevData}
    \end{figure}

    At 1 GeV beam energy, dedicated fiducial cut parameters for $\pi^{-}$ at 750 A torus current were defined as part of the analysis of~\cite{e4nuQE}.
    The 1500 A data was not used in the analysis of~\cite{e4nuQE}, therefore no bespoke $\pi^{-}$ fiducial cut parameters were produced for this field setting.
    Figure~\ref{fig:pimi_fid_1gevData} shows the $\theta$ vs momentum distributions for the $\pi^{-}$ at $1.1\;\mathrm{GeV}$, after the application of the updated fiducial cuts.

    %\clearpage
    
\begin{figure}[!htb]
    \centering
    \subfloat[]{\includegraphics[width=0.85\textwidth]{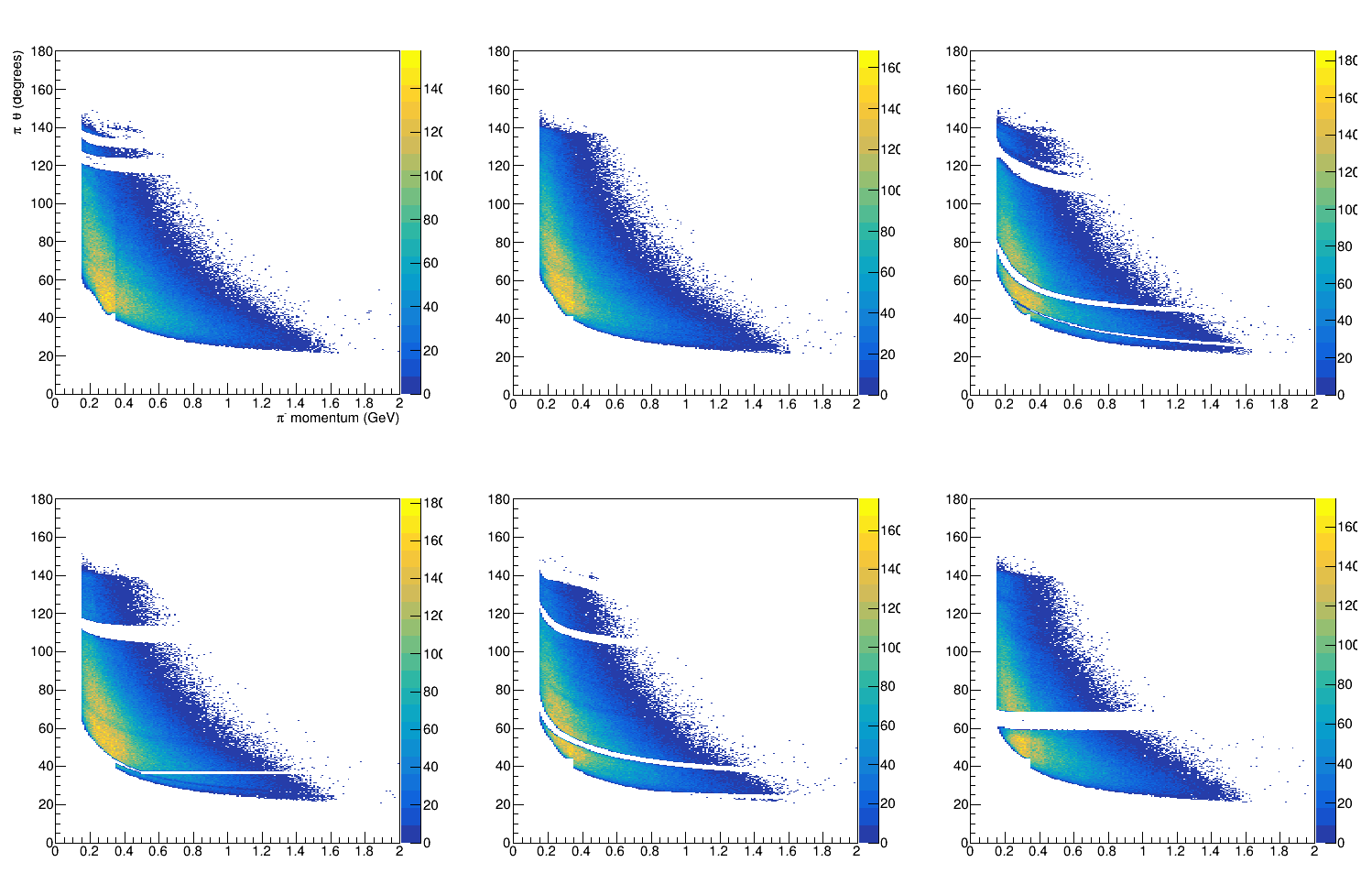}}
    \caption[The $\theta$ vs momentum distributions for $\pi^{-}$.]{The $\theta$ vs momentum distributions for $\pi^{-}$, after the application of the updated theta gap cuts for $2.2\;\mathrm{GeV}$ analysis of $^{3}\mathrm{He}$. The same cuts are applied in the 4.4 GeV analysis.}
    \label{fig:pimi_fid_2gevData}
    \end{figure}

    Figure \ref{fig:pimi_fid_2gevData} shows the $\theta$ vs momentum distributions for the $\pi^{-}$ at $2.2\;\mathrm{GeV}$, after the application of the updated fiducial cuts.  The same cuts are used in $4.4\;\mathrm{GeV}$ analyses.

	%%%%%%%%%%%%%%%%%%%%%%%%%%%%%%%%%%%%%%%%%%%%%%%%%%%%%%%%%%%%%%%%%%%%%%%    
    %\clearpage
    %\section{Refinement of electron and positively charged hadron fiducial cuts}
    
\begin{figure}[!htb]
    \centering
    \includegraphics[width=0.85\textwidth]{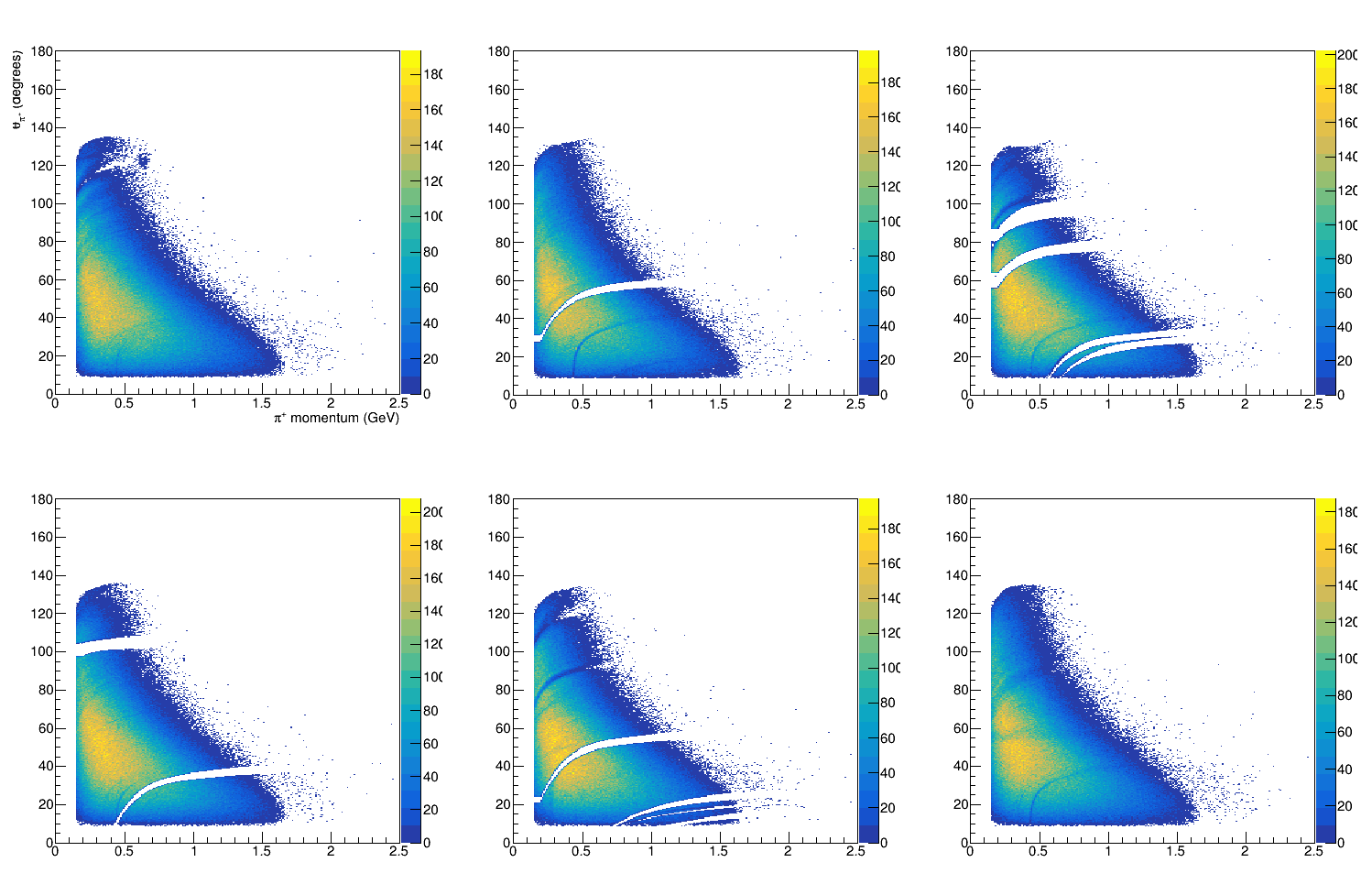}
    \caption[The $\theta$ vs momentum distributions for $\pi^{-}$.]{The $\theta$ vs momentum distributions for $\pi^{+}$, showing the original theta gap cuts for analysis of $2.2\;\mathrm{GeV}$ $^{4}\mathrm{He}$ data. The same gap functions are applied at $4.4\;\mathrm{GeV}$.}
    \label{fig:pip_fid_gaps2and4}
    \end{figure}

    As seen in the reuse of electron theta gap cuts on the $\pi^{-}$, where the gap functions were not defined to the lower momentum of the pions, a similar effect is seen in $\pi^{+}$, with the gap functions defined for protons directly reused.
    This has resulted in similar box cuts being applied at low momentum, as can be seen in figure~\ref{fig:pip_fid_gaps2and4}, removing good events rather than low and poorly understood acceptance regions.

    %\clearpage

\begin{figure}[!htb]
    \centering
    \subfloat[]{\includegraphics[width=0.45\textwidth]{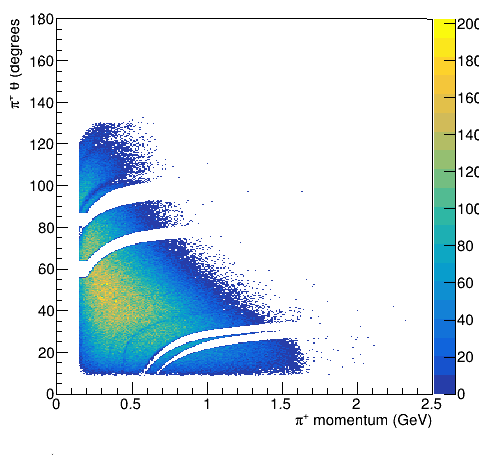}}
    \subfloat[]{\includegraphics[width=0.45\textwidth]{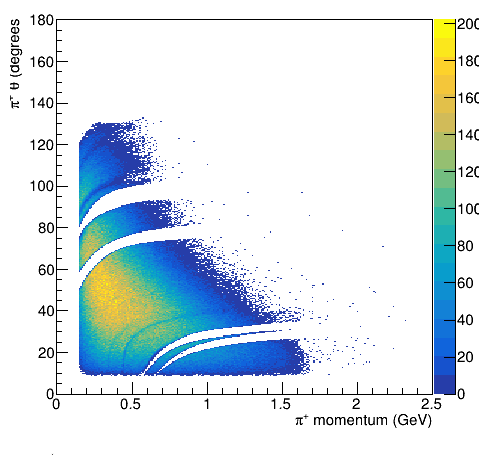}}
    \caption[The $\theta$ vs momentum distributions for $\pi^{+}$.]{The $\theta$ vs momentum distributions for $\pi^{+}$ in sector 3, showing the original (a) and new (b) theta gap cuts for $2.2\;\mathrm{GeV}$ analysis of $^{3}\mathrm{He}$. The same gap functions are applied at $4.4\;\mathrm{GeV}$ beam energy.}
    \label{fig:pip_fid_sector3}
    \end{figure}

    Unlike the $\pi^{-}$, the positive hadron fiducial cut functions, defined for protons, are for the most part valid at the lower pion momentum range.
    We have extended the application of several of these functions to define cuts at lower $\pi^{+}$ momenta, where the functions continue to appropriately describe the gap.  This is shown for sector 3 in figure \ref{fig:pip_fid_sector3}.

	%%%%%%%%%%%%%%%%%%%%%%%%%%%%%%%%%%%%%%%%%%%%%%%%%%%%%%%%%%%%%%%%%%%%%%%

\newpage
\addcontentsline{toc}{section}{\listfigurename}
\listoffigures
\newpage
\addcontentsline{toc}{section}{\listtablename}
\listoftables

\newpage
\addcontentsline{toc}{section}{References}

\begin{singlespace}
\bibliography{main}

\newcommand{\noopsort}[1]{} \newcommand{\printfirst}[2]{#1}
  \newcommand{\singleletter}[1]{#1} \newcommand{\switchargs}[2]{#2#1}
\begin{thebibliography}{100}

\bibitem{PhysRevD.98.030001}
M.~Tanabashi et~al.
\newblock Review of particle physics.
\newblock {\em Phys. Rev. D}, 98:030001, Aug 2018.

\bibitem{PhysRevLett.81.1562}
Y.~Fukuda et~al.
\newblock Evidence for oscillation of atmospheric neutrinos.
\newblock {\em Phys. Rev. Lett.}, 81:1562--1567, Aug 1998.

\bibitem{SNO:2002tuh}
Q.~R. Ahmad et~al.
\newblock {Direct evidence for neutrino flavor transformation from neutral
  current interactions in the Sudbury Neutrino Observatory}.
\newblock {\em Phys. Rev. Lett.}, 89:011301, 2002.

\bibitem{sciencealert}
{What Is The Standard Model of Particle Physics?}
\newblock \url{https://www.sciencealert.com/the-standard-model}.

\bibitem{PhysRev.109.1015}
M.~Goldhaber, L.~Grodzins, and A.~W. Sunyar.
\newblock Helicity of neutrinos.
\newblock {\em Phys. Rev.}, 109:1015--1017, Feb 1958.

\bibitem{VanDePontseele:2020tqz}
Wouter Van De~Pontseele.
\newblock {\em {Search for Electron Neutrino Anomalies with the MicroBooNE
  Detector}}.
\newblock PhD thesis, Oxford U., 2020.

\bibitem{PhysRevD.65.112002}
M.~Ahmed et~al.
\newblock Search for the lepton-family-number nonconserving decay
  ${\ensuremath{\mu}}^{+}\ensuremath{\rightarrow}{e}^{+}\ensuremath{\gamma}$.
\newblock {\em Phys. Rev. D}, 65:112002, Jun 2002.

\bibitem{NatureKATRIN}
M.~Aker et~al.
\newblock Direct neutrino-mass measurement with sub-electronvolt sensitivity.
\newblock {\em Nat. Phys.}, 18:160--166, 2022.

\bibitem{majorana}
E.~Majorana.
\newblock {Teoria simmetrica dell’elettrone e del positrone}.
\newblock {\em Nuovo Cim}, 14:171, 1937.

\bibitem{AKHMEDOV2000215}
E.Kh. Akhmedov, G.C. Branco, and M.N. Rebelo.
\newblock Seesaw mechanism and structure of neutrino mass matrix.
\newblock {\em Physics Letters B}, 478(1):215--223, 2000.

\bibitem{Freund:2001pn}
Martin Freund.
\newblock Analytic approximations for three neutrino oscillation parameters and
  probabilities in matter.
\newblock {\em Phys. Rev. D}, 64:053003, Jul 2001.

\bibitem{cervera2000}
A.~Cervera, A.~Donini, M.B. Gavela, J.J.~[Gomez Cádenas], P.~Hernández,
  O.~Mena, and S.~Rigolin.
\newblock Golden measurements at a neutrino factory.
\newblock {\em Nuclear Physics B}, 579(1):17 -- 55, 2000.

\bibitem{cervera2001}
A.~Cervera, A.~Donini, M.B. Gavela, J.J.~Gomez Cádenas, P.~Hernández,
  O.~Mena, and S.~Rigolin.
\newblock Erratum to “golden measurements at a neutrino factory”: [nucl.
  phys. b 579 (2000) 17].
\newblock {\em Nuclear Physics B}, 593(3):731 -- 732, 2001.

\bibitem{Fund}
C.~Giunti and C.~W. Kim.
\newblock Fundamentals of neutrino physics and astrophysics.
\newblock {\em University Press}, 2007.

\bibitem{2006257}
ALEPH, DELPHI, L3, OPAL, and SLD Collaborations.
\newblock Precision electroweak measurements on the z resonance.
\newblock {\em Physics Reports}, 427(5):257--454, 2006.

\bibitem{MSW}
A.~Yu. Smirnov.
\newblock The msw effect and matter effects in neutrino oscillations.
\newblock {\em Phys. Scr. 2005 57}, 2005.

\bibitem{10.3389/fspas.2018.00036}
Pablo~F. de~Salas, Stefano Gariazzo, Olga Mena, Christoph~A. Ternes, and Mariam
  Tórtola.
\newblock Neutrino mass ordering from oscillations and beyond: 2018 status and
  future prospects.
\newblock {\em Frontiers in Astronomy and Space Sciences}, 5:36, 2018.

\bibitem{PhysRevD.91.072004}
M.~G. Aartsen et~al.
\newblock Determining neutrino oscillation parameters from atmospheric muon
  neutrino disappearance with three years of icecube deepcore data.
\newblock {\em Phys. Rev. D}, 91:072004, Apr 2015.

\bibitem{AguilarArevalo:2008yp}
A.A. Aguilar-Arevalo et~al.
\newblock {The Neutrino Flux prediction at MiniBooNE}.
\newblock {\em Phys. Rev. D}, 79:072002, 2009.

\bibitem{Abi:2020qib}
B.~Abi et~al.
\newblock {Long-baseline neutrino oscillation physics potential of the DUNE
  experiment}.
\newblock {\em Eur. Phys. J. C}, 80(10):978, 2020.

\bibitem{fukugita86}
M.~Fukugita and T.~Yanagida.
\newblock Baryogenesis without grand unification.
\newblock {\em Phys. Lett. B}, 174:45, 1986.

\bibitem{T2KNature20}
K.~Abe et~al.
\newblock {Constraint on the matter–antimatter symmetry-violating phase in
  neutrino oscillations}.
\newblock {\em Nature}, 580:339, 2020.

\bibitem{T2K}
K.~Abe et~al.
\newblock {Search for CP Violation in Neutrino and Antineutrino Oscillations by
  the T2K Experiment with $2.2\times10^{21}$ Protons on Target}.
\newblock {\em Phys. Rev. Lett.}, 121(17):171802, 2018.

\bibitem{Alvarez-Ruso:2017oui}
L.~Alvarez-Ruso et~al.
\newblock {NuSTEC White Paper: Status and challenges of neutrino–nucleus
  scattering}.
\newblock {\em Prog. Part. Nucl. Phys.}, 100:1--68, 2018.

\bibitem{NOvA:2018gge}
M.A. Acero et~al.
\newblock {New constraints on oscillation parameters from $\nu_e$ appearance
  and $\nu_\mu$ disappearance in the NOvA experiment}.
\newblock {\em Phys. Rev. D}, 98:032012, 2018.

\bibitem{Ankowski:2015kya}
Artur~M. Ankowski, Pilar Coloma, Patrick Huber, Camillo Mariani, and Erica
  Vagnoni.
\newblock {Missing energy and the measurement of the CP-violating phase in
  neutrino oscillations}.
\newblock {\em Phys. Rev. D}, 92(9):091301, 2015.

\bibitem{Rocco2020}
Noemi Rocco.
\newblock Ab initio calculations of lepton-nucleus scattering.
\newblock {\em Frontiers in Physics}, 8:116, 2020.

\bibitem{PhysRevD.101.033003}
S.~Dolan, G.~D. Megias, and S.~Bolognesi.
\newblock Implementation of the susav2-meson exchange current 1p1h and 2p2h
  models in genie and analysis of nuclear effects in t2k measurements.
\newblock {\em Phys. Rev. D}, 101:033003, Feb 2020.

\bibitem{PhysRevLett.116.192501}
Noemi Rocco, Alessandro Lovato, and Omar Benhar.
\newblock Unified description of electron-nucleus scattering within the
  spectral function formalism.
\newblock {\em Phys. Rev. Lett.}, 116:192501, May 2016.

\bibitem{PhysRevD.94.092005}
L.~Aliaga et~al.
\newblock Neutrino flux predictions for the numi beam.
\newblock {\em Phys. Rev. D}, 94:092005, Nov 2016.

\bibitem{NovaNearDetector}
Kuldeep~K. Maan.
\newblock {Constraints on the Neutrino Flux in NOvA using the Near Detector
  Data}.
\newblock {\em PoS}, ICHEP2016:931, 2016.

\bibitem{T2KNearDetector}
L.~Haegel.
\newblock {T2K near detector constraints for oscillation results}.
\newblock In {\em {18th International Workshop on Neutrino Factories and Future
  Neutrino Facilities Search}}, 1 2017.

\bibitem{Genie2010}
C.~Andreopoulos, A.~Bell, D.~Bhattacharya, F.~Cavanna, J.~Dobson, S.~Dytman,
  H.~Gallagher, P.~Guzowski, R.~Hatcher, P.~Kehayias, A.~Meregaglia, D.~Naples,
  G.~Pearce, A.~Rubbia, M.~Whalley, and T.~Yang.
\newblock The genie neutrino monte carlo generator.
\newblock {\em Nucl. Instrum. Methods Phys. Res., Sect. A}, 614(1):87 -- 104,
  2010.

\bibitem{RevModPhys.84.1307}
J.~A. Formaggio and G.~P. Zeller.
\newblock From ev to eev: Neutrino cross sections across energy scales.
\newblock {\em Rev. Mod. Phys.}, 84:1307--1341, Sep 2012.

\bibitem{MarcoThesis}
M.~D. Tutto.
\newblock First measurements of inclusive muon neutrino charged current
  differential cross sections on argon at 0.8 gev average neutrino energy with
  the microboone detector.
\newblock {\em PhD thesis, University of Oxford}, 2019.

\bibitem{PhysRevD.23.1070}
A.~Bodek and J.~L. Ritchie.
\newblock Fermi-motion effects in deep-inelastic lepton scattering from nuclear
  targets.
\newblock {\em Phys. Rev. D}, 23:1070--1091, Mar 1981.

\bibitem{Carrasco:1989vq}
R.C. Carrasco and E.~Oset.
\newblock {Interaction of Real Photons With Nuclei From 100-{MeV} to
  500-{MeV}}.
\newblock {\em Nucl.\ Phys.\ A}, 536:445--508, 1992.

\bibitem{PhysRevC.74.054316}
Artur~M. Ankowski and Jan~T. Sobczyk.
\newblock Argon spectral function and neutrino interactions.
\newblock {\em Phys. Rev. C}, 74:054316, Nov 2006.

\bibitem{LarsThesis}
Lars Bathe-Peters.
\newblock Studies of single transverse kinematic variables for neutrino
  interactions on argon.
\newblock {\em Masters thesis, Harvard University}, 2020.

\bibitem{Katori_2017}
Teppei Katori and Marco Martini.
\newblock Neutrino{\textendash}nucleus cross sections for oscillation
  experiments.
\newblock {\em Journal of Physics G: Nuclear and Particle Physics},
  45(1):013001, dec 2017.

\bibitem{Amaro_2020}
J~E Amaro, M~B Barbaro, J~A Caballero, R~Gonz{\'{a}}lez-Jim{\'{e}}nez, G~D
  Megias, and I~Ruiz Simo.
\newblock Electron- versus neutrino-nucleus scattering.
\newblock {\em Journal of Physics G: Nuclear and Particle Physics},
  47(12):124001, nov 2020.

\bibitem{PhysRevD.103.113003}
A.~Papadopoulou, A.~Ashkenazi, S.~Gardiner, M.~Betancourt, S.~Dytman, L.~B.
  Weinstein, E.~Piasetzky, F.~Hauenstein, M.~Khachatryan, S.~Dolan, G.~D.
  Megias, and O.~Hen.
\newblock Inclusive electron scattering and the genie neutrino event generator.
\newblock {\em Phys. Rev. D}, 103:113003, Jun 2021.

\bibitem{KOPP2007101}
Sacha~E. Kopp.
\newblock Accelerator neutrino beams.
\newblock {\em Physics Reports}, 439(3):101--159, 2007.

\bibitem{OffAxis}
M.~Szleper A.~Para.
\newblock Neutrino oscillations experiments using off-axis numi beam.
\newblock {\em arXiv:0110032v1}, 2001.

\bibitem{fnalaccelerator}
Fermilab's accelerator complex.
\newblock
  https://www.fnal.gov/pub/science/particle-accelerators/accelerator-complex.html.

\bibitem{PhysRevD.79.072002}
Aguilar-Arevalo et~al.
\newblock Neutrino flux prediction at miniboone.
\newblock {\em Phys. Rev. D}, 79:072002, Apr 2009.

\bibitem{ADAMSON2016279}
P.~Adamson et~al.
\newblock The numi neutrino beam.
\newblock {\em Nuclear Instruments and Methods in Physics Research Section A:
  Accelerators, Spectrometers, Detectors and Associated Equipment},
  806:279--306, 2016.

\bibitem{Evans}
J.~Evans et~al.
\newblock The minos experiment: Results and prospects.
\newblock {\em Adv. High Energy Phys.}, page 182537, 2013.

\bibitem{doi:10.7566/JPSCP.12.010006}
Tomasz Golan, Leonidas Aliaga, and Mike Kordosky.
\newblock {\em MINERvA’s Flux Prediction}, chapter~12, page~8.
\newblock JPS, 2016.

\bibitem{PhysRevD.99.012002}
R.~Acciarri et~al.
\newblock Demonstration of mev-scale physics in liquid argon time projection
  chambers using argoneut.
\newblock {\em Phys. Rev. D}, 99:012002, Jan 2019.

\bibitem{PhysRevLett.123.151803}
M.~A. Acero et~al.
\newblock First measurement of neutrino oscillation parameters using neutrinos
  and antineutrinos by nova.
\newblock {\em Phys. Rev. Lett.}, 123:151803, Oct 2019.

\bibitem{PhysRevLett.122.091803}
P.~Adamson et~al.
\newblock Search for sterile neutrinos in minos and minos+ using a two-detector
  fit.
\newblock {\em Phys. Rev. Lett.}, 122:091803, Mar 2019.

\bibitem{PhysRevD.80.073001}
G.~Karagiorgi, Z.~Djurcic, J.~M. Conrad, M.~H. Shaevitz, and M.~Sorel.
\newblock Viability of $\ensuremath{\Delta}{m}^{2}\ensuremath{\sim}1\text{
  }\text{ }{\mathrm{ev}}^{2}$ sterile neutrino mixing models in light of
  miniboone electron neutrino and antineutrino data from the booster and numi
  beamlines.
\newblock {\em Phys. Rev. D}, 80:073001, Oct 2009.

\bibitem{PhysRevLett.121.221801}
A.~Aguilar-Arevalo et~al.
\newblock Significant excess of electronlike events in the miniboone
  short-baseline neutrino experiment.
\newblock {\em Phys. Rev. Lett.}, 121:221801, Nov 2018.

\bibitem{Adams:2018lzd}
C.~Adams et~al.
\newblock {Rejecting cosmic background for exclusive charged current quasi
  elastic neutrino interaction studies with Liquid Argon TPCs; a case study
  with the MicroBooNE detector}.
\newblock {\em Eur.\ Phys.\ J.\ C}, 79(8):673, 2019.

\bibitem{uBCRT}
C.~Adams et~al.
\newblock {Design and construction of the MicroBooNE Cosmic Ray Tagger system}.
\newblock {\em arXiv:1901.02862}, 2019.

\bibitem{corsika}
D.~Heck et~al.
\newblock Corsika: A monte carlo code to simulate extensive air showers.
\newblock {\em Forschungszentrum Karlsruhe Report FZKA}, 1998.

\bibitem{Geant4}
S.~Agostinelli et~al.
\newblock Geant4—a simulation toolkit.
\newblock {\em Nucl. Instrum. Meth.}, A 506, 2003.

\bibitem{TPC}
Jay~N. Marx and David~R. Nygren.
\newblock The time projection chamber, 1978.

\bibitem{LArTPCConcept}
C.~Rubbia.
\newblock The liquid argon time projection chamber: A new concept for neutrino
  detectors, 1977.

\bibitem{CAVANNA20181}
F.~Cavanna, A.~Ereditato, and B.T. Fleming.
\newblock Advances in liquid argon detectors.
\newblock {\em Nuclear Instruments and Methods in Physics Research Section A:
  Accelerators, Spectrometers, Detectors and Associated Equipment}, 907:1--8,
  2018.
\newblock Advances in Instrumentation and Experimental Methods (Special Issue
  in Honour of Kai Siegbahn).

\bibitem{grid}
O.~Bunemann, T.~E. Cranshaw, and J.~A. Harvey.
\newblock Design of grid ionization chambers.
\newblock {\em Canadian Journal of Research 27a.5}, page 191–206, 1949.

\bibitem{Acciarri_2017}
R.~Acciarri et~al.
\newblock Noise characterization and filtering in the microboone liquid argon
  tpc.
\newblock {\em Journal of Instrumentation}, 12(08):P08003--P08003, aug 2017.

\bibitem{read}
MicroBooNE Collaboration.
\newblock Design and construction of the microboone detector.
\newblock {\em JINST 12.P02017}, 2017.

\bibitem{Chen:2012ysa}
H.~Chen, G.~De~Geronimo, F.~Lanni, D.~Lissauer, D.~Makowiecki, V.~Radeka,
  S.~Rescia, C.~Thorn, and B.~Yu.
\newblock {Front End Readout Electronics of the MicroBooNE Experiment}.
\newblock {\em Phys. Procedia}, 37:1287--1294, 2012.

\bibitem{SCEPaper}
P.~Abratenko et~al.
\newblock Measurement of space charge effects in the {MicroBooNE} {LArTPC}
  using cosmic muons.
\newblock {\em Journal of Instrumentation}, 15(12):P12037--P12037, dec 2020.

\bibitem{ArMinEn}
Robert~S. Mulliken.
\newblock Potential curves of diatomic rare-gas molecules and their ions, with
  particular reference to xe2.
\newblock {\em The Journal of Chemical Physics}, 52(10):5170–5180, 1970.

\bibitem{gain}
MicroBooNE Collaboration.
\newblock Pmt gain calibration in microboone.
\newblock {\em MICROBOONE-NOTE-1064-TECH}, 2019.

\bibitem{ubpmtpic}
Microboone photomultiplier.
\newblock https://news.fnal.gov/2015/07/microboone-photomultiplier/.

\bibitem{DelTuttoThesis}
M.~Del Tutto.
\newblock {\em First measurements of inclusive muon neutrino charged current
  differential cross sections on argon at 0.8 GeV average neutrino energy with
  the MicroBooNE detector}.
\newblock PhD thesis, University of Oxford, 2019.

\bibitem{WireCellEventSelection}
P.~Abratenko et~al.
\newblock Neutrino event selection in the {MicroBooNE} liquid argon time
  projection chamber using wire-cell 3d imaging, clustering, and charge-light
  matching.
\newblock {\em Journal of Instrumentation}, 16(06):P06043, jun 2021.

\bibitem{WCcosmicpic}
Spotting accelerator-produced neutrinos in a cosmic haystack.
\newblock https://science.osti.gov/hep/Highlights/2022/HEP-2022-02-a.

\bibitem{Adams_2018}
C.~Adams et~al.
\newblock Ionization electron signal processing in single phase {LArTPCs}. part
  i. algorithm description and quantitative evaluation with {MicroBooNE}
  simulation.
\newblock {\em Journal of Instrumentation}, 13(07):P07006--P07006, jul 2018.

\bibitem{Adams:2018gbi}
C.~Adams et~al.
\newblock {Ionization electron signal processing in single phase LArTPCs. Part
  II. Data/simulation comparison and performance in MicroBooNE}.
\newblock {\em J. Instrum.}, 13(07):P07007, 2018.

\bibitem{Acciarri:2017hat}
R.~Acciarri et~al.
\newblock {The Pandora multi-algorithm approach to automated pattern
  recognition of cosmic-ray muon and neutrino events in the MicroBooNE
  detector}.
\newblock {\em Eur.\ Phys.\ J.\ C}, 78(1):82, 2018.

\bibitem{Acciarri_2013}
R~Acciarri et~al.
\newblock A study of electron recombination using highly ionizing particles in
  the {ArgoNeuT} liquid argon {TPC}.
\newblock {\em JINST}, 8(08):P08005--P08005, aug 2013.

\bibitem{caratelli_neutralpi}
C~Adams et~al.
\newblock Reconstruction and measurement of o(100) mev energy electromagnetic
  activity from neutral pion decays in the microboone lartpc.
\newblock {\em arXiv:1910.02166}, 2019.

\bibitem{kalman_filter}
MicroBooNE Collaboration.
\newblock Reconstruction performance studies with microboone data in support of
  summer 2018 analyses.
\newblock {\em MICROBOONE-NOTE-1049-PUB}, 2018.

\bibitem{geniev3highlights}
GENIE Collaboration.
\newblock Recent highlights from genie v3.
\newblock {\em Eur. Phys. J. Spec. Top.}, 2021.

\bibitem{Anderson:2012jds}
C.~Anderson et~al.
\newblock {First Measurements of Inclusive Muon Neutrino Charged Current
  Differential Cross Sections on Argon}.
\newblock {\em Phys.\ Rev.\ Lett.}, 108(10):161802, 2012.

\bibitem{Nakajima:2010fp}
Y.~Nakajima et~al.
\newblock {Measurement of Inclusive Charged Current Interactions on Carbon in a
  Few-GeV Neutrino Beam}.
\newblock {\em Phys.\ Rev.\ D}, 83:012005, 2011.

\bibitem{Aguilar-Arevalo:2013dva}
A.A. Aguilar-Arevalo et~al.
\newblock {First measurement of the muon antineutrino double-differential
  charged-current quasielastic cross section}.
\newblock {\em Phys.\ Rev.\ D}, 88(3):032001, 2013.

\bibitem{Abe:2014iza}
K.~Abe et~al.
\newblock {Measurement of the $\nu_{\mu}$ charged-current quasielastic cross
  section on carbon with the ND280 detector at T2K}.
\newblock {\em Phys.\ Rev.\ D}, 92(11):112003, 2015.

\bibitem{Carneiro:2019jds}
M.F. Carneiro et~al.
\newblock {High-Statistics Measurement of Neutrino Quasielastic-Like Scattering
  at $E_{\nu} \sim$ 6 GeV on a Hydrocarbon Target}.
\newblock {\em Phys.\ Rev.\ Lett.}, 124(12):121801, 2020.

\bibitem{Abratenko:2019jqo}
P.~Abratenko et~al.
\newblock {First Measurement of Inclusive Muon Neutrino Charged Current
  Differential Cross Sections on Argon at $E_\nu\sim$0.8 GeV with the
  MicroBooNE Detector}.
\newblock {\em Phys.\ Rev.\ Lett.}, 123(13):131801, 2019.

\bibitem{Fiorentini:2013ezn}
G.A. Fiorentini et~al.
\newblock {Measurement of Muon Neutrino Quasielastic Scattering on a
  Hydrocarbon Target at $E_\nu \sim 3.5$ GeV}.
\newblock {\em Phys.\ Rev.\ Lett.}, 111:022502, 2013.

\bibitem{Betancourt:2017uso}
M.~Betancourt et~al.
\newblock {Direct Measurement of Nuclear Dependence of Charged Current
  Quasielasticlike Neutrino Interactions Using MINER$\nu$A}.
\newblock {\em Phys.\ Rev.\ Lett.}, 119(8):082001, 2017.

\bibitem{Walton:2014esl}
T.~Walton et~al.
\newblock {Measurement of muon plus proton final states in $\nu_{\mu}$
  interactions on hydrocarbon at $\langle E_{\nu} \rangle = $ 4.2 GeV}.
\newblock {\em Phys.\ Rev.\ D}, 91(7):071301, 2015.

\bibitem{Abe:2018pwo}
K.~Abe et~al.
\newblock {Characterization of nuclear effects in muon-neutrino scattering on
  hydrocarbon with a measurement of final-state kinematics and correlations in
  charged-current pionless interactions at T2K}.
\newblock {\em Phys.\ Rev.\ D}, 98(3):032003, 2018.

\bibitem{Mosel:2013fxa}
U.~Mosel et~al.
\newblock {Energy reconstruction in the Long-Baseline Neutrino Experiment}.
\newblock {\em Phys. Rev. Lett.}, 112:151802, 2014.

\bibitem{Formaggio:2013kya}
J.A. Formaggio and G.P. Zeller.
\newblock {From eV to EeV: Neutrino Cross Sections Across Energy Scales}.
\newblock {\em Rev.\ Mod.\ Phys.}, 84:1307--1341, 2012.

\bibitem{PhysRevLett.125.201803}
P.~Abratenko et~al.
\newblock First measurement of differential charged current quasielasticlike
  ${\ensuremath{\nu}}_{\ensuremath{\mu}}$-argon scattering cross sections with
  the microboone detector.
\newblock {\em Phys. Rev. Lett.}, 125:201803, Nov 2020.

\bibitem{Acciarri:2016smi}
R.~Acciarri et~al.
\newblock {Design and Construction of the MicroBooNE Detector}.
\newblock {\em J. Instrum.}, 12(02):P02017, 2017.

\bibitem{Antonello:2015lea}
M.~Antonello et~al.
\newblock {A Proposal for a Three Detector Short-Baseline Neutrino Oscillation
  Program in the Fermilab Booster Neutrino Beam}.
\newblock {\em arXiv:1503.01520}, 2015.

\bibitem{Tortorici:2018yns}
F.~Tortorici, V.~Bellini, and C.M. Sutera.
\newblock {Upgrade of the ICARUS T600 Time Projection Chamber}.
\newblock {\em J. Phys. Conf. Ser.}, 1056(1):012057, 2018.

\bibitem{Abi:2020wmh}
Babak Abi et~al.
\newblock {Deep Underground Neutrino Experiment (DUNE), Far Detector Technical
  Design Report, Volume I Introduction to DUNE}.
\newblock {\em arXiv:2002.02967}, 2 2020.

\bibitem{Abi:2020evt}
Babak Abi et~al.
\newblock {Deep Underground Neutrino Experiment (DUNE), Far Detector Technical
  Design Report, Volume II DUNE Physics}.
\newblock {\em arXiv:2002.03005}, 2 2020.

\bibitem{Abi:2020oxb}
Babak Abi et~al.
\newblock {Deep Underground Neutrino Experiment (DUNE), Far Detector Technical
  Design Report, Volume III DUNE Far Detector Technical Coordination}.
\newblock {\em arXiv:2002.03008}, 2 2020.

\bibitem{Abi:2020loh}
Babak Abi et~al.
\newblock {Deep Underground Neutrino Experiment (DUNE), Far Detector Technical
  Design Report, Volume IV Far Detector Single-phase Technology}.
\newblock {\em arXiv:2002.03010}, 2 2020.

\bibitem{Abratenko:2017nki}
P.~Abratenko et~al.
\newblock {Determination of muon momentum in the MicroBooNE LArTPC using an
  improved model of multiple Coulomb scattering}.
\newblock {\em J. Instrum.}, 12(10):P10010, 2017.

\bibitem{Kaleko:2013eda}
D.~Kaleko et~al.
\newblock {PMT Triggering and Readout for the MicroBooNE Experiment}.
\newblock {\em J. Instrum.}, 8:C09009, 2013.

\bibitem{Adams:2016smi}
C.~Adams et~al.
\newblock {Calibration of the charge and energy loss per unit length of the
  MicroBooNE liquid argon time projection chamber using muons and protons}.
\newblock {\em J. Instrum.}, 15(03):P03022, 2020.

\bibitem{Andreopoulos:2009rq}
C.~Andreopoulos et~al.
\newblock The genie neutrino monte carlo generator.
\newblock {\em Nucl.\ Instrum.\ Meth.\ A}, 614:87--104, 2010.

\bibitem{Andreopoulos:2015wxa}
Costas Andreopoulos, Christopher Barry, Steve Dytman, Hugh Gallagher, Tomasz
  Golan, Robert Hatcher, Gabriel Perdue, and Julia Yarba.
\newblock {The GENIE Neutrino Monte Carlo Generator: Physics and User Manual}.
\newblock {\em arXiv:1510.05494}, 2015.

\bibitem{Roe:2007hw}
B.P. Roe.
\newblock {Statistical errors in Monte Carlo estimates of systematic errors}.
\newblock {\em Nucl.\ Instrum.\ Meth.\ A}, 570:159--164, 2007.

\bibitem{Pordes:2016ycs}
R.~Pordes and E.~Snider.
\newblock {The Liquid Argon Software Toolkit (LArSoft): Goals, Status and
  Plan}.
\newblock {\em PoS}, ICHEP2016:182, 2016.

\bibitem{Snider:2017wjd}
E.~Snider and G.~Petrillo.
\newblock {LArSoft: Toolkit for Simulation, Reconstruction and Analysis of
  Liquid Argon TPC Neutrino Detectors}.
\newblock {\em J.\ Phys.\ Conf.\ Ser.}, 898(4):042057, 2017.

\bibitem{LlewellynSmith:1971uhs}
C.H. Llewellyn~Smith.
\newblock {Neutrino Reactions at Accelerator Energies}.
\newblock {\em Phys. Rep.}, 3:261--379, 1972.

\bibitem{Katori:2013eoa}
Teppei Katori.
\newblock {Meson Exchange Current (MEC) Models in Neutrino Interaction
  Generators}.
\newblock {\em AIP Conf. Proc.}, 2015.

\bibitem{Rein:1980wg}
D.~Rein and L.~Sehgal.
\newblock {Neutrino Excitation of Baryon Resonances and Single Pion
  Production}.
\newblock {\em Ann. Phys. (N.Y.)}, 133:79--153, 1981.

\bibitem{Mashnik:2005ay}
S.G. Mashnik et~al.
\newblock {CEM03 and LAQGSM03: New modeling tools for nuclear applications}.
\newblock {\em J.\ Phys.\ Conf.\ Ser.}, 41:340--351, 2006.

\bibitem{Stowell_2017}
P.~Stowell, C.~Wret, C.~Wilkinson, L.~Pickering, S.~Cartwright, Y.~Hayato,
  K.~Mahn, K.S. McFarland, J.~Sobczyk, R.~Terri, L.~Thompson, M.O. Wascko, and
  Y.~Uchida.
\newblock {NUISANCE}: a neutrino cross-section generator tuning and comparison
  framework.
\newblock {\em Journal of Instrumentation}, 12(01):P01016--P01016, jan 2017.

\bibitem{GolanNuWro:2008yp}
T.~Golan et~al.
\newblock {NuWro: the Wroclaw Monte Carlo Generator of Neutrino Interactions}.
\newblock {\em Nucl.Phys.Proc.Suppl.}, 499:229--232, 2012.

\bibitem{Hayato:2008yp}
Y.~Hayato.
\newblock {A neutrino interaction simulation program library NEUT}.
\newblock {\em Acta Phys. Polon.}, B40:2477, 2009.

\bibitem{RPA}
J.~Nieves, J.~E. Amaro, and M.~Valverde.
\newblock Inclusive quasielastic charged-current neutrino-nucleus reactions.
\newblock {\em Phys. Rev. C}, 70:055503, Nov 2004.

\bibitem{Engel:1997fy}
Jonathan Engel.
\newblock {Approximate treatment of lepton distortion in charged current
  neutrino scattering from nuclei}.
\newblock {\em Phys. Rev. C}, 57:2004--2009, 1998.

\bibitem{Mosel:2008yp}
U.~Mosel.
\newblock {Neutrino event generators: foundation, status and future}.
\newblock {\em Phys. Rev. G}, 2019.

\bibitem{Bodek:2017hat}
A.~Bodek et~al.
\newblock {Neutrino Quasielastic Scattering on Nuclear Targets: Parametrizing
  Transverse Enhancement (Meson Exchange Currents)}.
\newblock {\em Eur.\ Phys.\ J.\ C}, 71:1726, 2011.

\bibitem{Graczyk:2008yp}
K.M. Graczyk et~al.
\newblock {C(5)**A axial form factor from bubble chamber experiments}.
\newblock {\em Phys. Rev. D}, 80:093001, 2009.

\bibitem{Berger:2008xs}
C.~Berger and L.~Sehgal.
\newblock {PCAC and coherent pion production by low energy neutrinos}.
\newblock {\em Phys. Rev. D}, 79:053003, 2009.

\bibitem{Nieves:2012yz}
J.~Nieves, F.~Sanchez, I.~Ruiz~Simo, and M.J. Vicente~Vacas.
\newblock {Neutrino Energy Reconstruction and the Shape of the CCQE-like Total
  Cross Section}.
\newblock {\em Phys.\ Rev.\ D}, 85:113008, 2012.

\bibitem{Schwehr:2016pvn}
J.~Schwehr, D.~Cherdack, and R.~Gran.
\newblock {GENIE implementation of IFIC Valencia model for QE-like 2p2h
  neutrino-nucleus cross section}.
\newblock {\em arXiv}, 1 2016.

\bibitem{Nowak:2009se}
J.~A. Nowak.
\newblock {Four Momentum Transfer Discrepancy in the Charged Current $\pi^+$
  Production in the MiniBooNE: Data vs. Theory}.
\newblock {\em AIP Conf.\ Proc.}, 1189(1):243--248, 2009.

\bibitem{Kuzmin:2003ji}
K.~Kuzmin et~al.
\newblock {Lepton polarization in neutrino nucleon interactions}.
\newblock {\em Phys.\ Part.\ Nucl.}, 35:S133--S138, 2004.

\bibitem{Berger:2007rq}
Ch. Berger and L.M. Sehgal.
\newblock {Lepton mass effects in single pion production by neutrinos}.
\newblock {\em Phys. Rev. D}, 76:113004, 2007.

\bibitem{Graczyk:2007bc}
K.~M. Graczyk and J.~T. Sobczyk.
\newblock {Form Factors in the Quark Resonance Model}.
\newblock {\em Phys.\ Rev.\ D}, 77:053001, 2008.
\newblock [Erratum: Phys.Rev.D 79, 079903 (2009)].

\bibitem{Leitner:2006ww}
Tina Leitner, L.~Alvarez-Ruso, and U.~Mosel.
\newblock {Charged current neutrino nucleus interactions at intermediate
  energies}.
\newblock {\em Phys. Rev. C}, 73:065502, 2006.

\bibitem{Mosel:2019vhx}
Ulrich Mosel.
\newblock {Neutrino event generators: foundation, status and future}.
\newblock {\em J. Phys. G}, 46(11):113001, 2019.

\bibitem{Sjostrand:2006za}
Torbjorn Sjostrand, Stephen Mrenna, and Peter~Z. Skands.
\newblock {PYTHIA 6.4 Physics and Manual}.
\newblock {\em JHEP}, 05:026, 2006.

\bibitem{WireMod}
MicroBooNE Collaboration.
\newblock Novel approach for evaluating detector-related uncertainties in a
  lartpc using microboone data.
\newblock {\em arXiv:2111.03556}, 2021.

\bibitem{GENIEKnobs}
P.~Abratenko et~al.
\newblock New $\mathrm{CC}0\ensuremath{\pi}$ genie model tune for microboone.
\newblock {\em Phys. Rev. D}, 105:072001, Apr 2022.

\bibitem{PhysRevLett.121.022504}
X.-G. Lu et~al.
\newblock Measurement of final-state correlations in neutrino muon-proton
  mesonless production on hydrocarbon at ev=3gev.
\newblock {\em Phys. Rev. Lett.}, 121:022504, Jul 2018.

\bibitem{PhysRevD.103.112009}
K.~et~al Abe.
\newblock First t2k measurement of transverse kinematic imbalance in the
  muon-neutrino charged-current single-${\ensuremath{\pi}}^{+}$ production
  channel containing at least one proton.
\newblock {\em Phys. Rev. D}, 103:112009, Jun 2021.

\bibitem{DolanTKITrento}
S.~Dolan.
\newblock Exploring nuclear effects with transverse imbalances.
\newblock
  https://indico.ectstar.eu/event/19/contributions/409/attachments/313/414/sdolanTalk.pdf.

\bibitem{PhysRevD.101.092001}
T.~et~al Cai.
\newblock Nucleon binding energy and transverse momentum imbalance in
  neutrino-nucleus reactions.
\newblock {\em Phys. Rev. D}, 101:092001, May 2020.

\bibitem{MuInAr}
{Table 289: muons in liquid argon}.
\newblock
  \url{http://pdg.lbl.gov/2012/AtomicNuclearProperties/MUON\_ELOSS\_TABLES/muonloss\_289.pdf}.

\bibitem{osti139791}
{Stopping powers and ranges for protons and alpha particles}.
\newblock
  \url{https://www.nist.gov/pml/stopping-power-range-tables-electrons-protons-and-helium-ions}.

\bibitem{log}
Classification of track-like particles in MicroBooNE - Paper Draft,
  https://arxiv.org/abs/2109.02460.

\bibitem{Tang_2017}
W.~Tang, X.~Li, X.~Qian, H.~Wei, and C.~Zhang.
\newblock Data unfolding with wiener-svd method.
\newblock {\em Journal of Instrumentation}, 12(10):P10002–P10002, Oct 2017.

\bibitem{H_cker_1996}
Andreas Höcker and Vakhtang Kartvelishvili.
\newblock Svd approach to data unfolding.
\newblock {\em Nuclear Instruments and Methods in Physics Research Section A:
  Accelerators, Spectrometers, Detectors and Associated Equipment},
  372(3):469–481, Apr 1996.

\bibitem{Schmitt_2012}
S~Schmitt.
\newblock Tunfold, an algorithm for correcting migration effects in high energy
  physics.
\newblock {\em Journal of Instrumentation}, 7(10):T10003–T10003, Oct 2012.

\bibitem{BaBarStat}
BaBar Statistics~Working Group.
\newblock Recommended statistical procedures for babar.
\newblock {\em BABAR Analysis Document 318}, 2002.

\bibitem{Ashery:1981tq}
D.~Ashery, I.~Navon, G.~Azuelos, H.K. Walter, H.J. Pfeiffer, and F.W.
  Schleputz.
\newblock {True Absorption and Scattering of Pions on Nuclei}.
\newblock {\em Phys.\ Rev.\ C}, 23:2173--2185, 1981.

\bibitem{hN2018}
L.~A. Harewood and R.~Gran.
\newblock Elastic hadron-nucleus scattering in neutrino-nucleus reactions and
  transverse kinematics measurements.
\newblock {\em arXiv:1906.10576}, 2019.

\bibitem{Wright:2015xia}
D.~H. Wright and M.~H. Kelsey.
\newblock {The Geant4 Bertini Cascade}.
\newblock {\em Nucl. Instrum. Meth. A}, 804:175--188, 2015.

\bibitem{DUNE}
K.~Abe et~al.
\newblock {The DUNE Far Detector Interim Design Report Volume 1: Physics,
  Technology and Strategies}.
\newblock {\em arXiv:1807.10334}, 2018.

\bibitem{nova19}
M.~A. Acero et~al.
\newblock First measurement of neutrino oscillation parameters using neutrinos
  and antineutrinos by nova.
\newblock {\em Phys. Rev. Lett.}, 123:151803, Oct 2019.

\bibitem{DUNEFlux}
Babak Abi et~al.
\newblock https://home.fnal.gov/~ljf26/DUNEFluxes/.

\bibitem{Ankowski:2020qbe}
Artur~M. Ankowski and Alexander Friedland.
\newblock {Assessing the accuracy of the GENIE event generator with
  electron-scattering data}.
\newblock {\em Phys. Rev. D}, 102(5):053001, 2020.

\bibitem{Amaro:2019zos}
J~E Amaro, M~B Barbaro, J~A Caballero, R~Gonz{\'{a}}lez-Jim{\'{e}}nez, G~D
  Megias, and I~Ruiz Simo.
\newblock Electron- versus neutrino-nucleus scattering.
\newblock {\em J. Phys. G}, 47(12):124001, nov 2020.

\bibitem{Barreau:1983ht}
P.~Barreau et~al.
\newblock {Deep Inelastic electron Scattering from Carbon}.
\newblock {\em Nucl. Phys. A402}, pages 515--540, 1983.

\bibitem{Bosted:2012qc}
P.E. Bosted and V.~Mamyan.
\newblock {Empirical Fit to electron-nucleus scattering}.
\newblock {\em arXiv:1203.2262}, 2012.

\bibitem{Dytman:2021ohr}
Steven Dytman, Yoshinari Hayato, Roland Raboanary, Jan Sobczyk, Julia
  Tena~Vidal, and Narisoa Vololoniaina.
\newblock {Comparison of Validation Methods of Simulations for Final State
  Interactions in Hadron Production Experiments}.
\newblock {\em arXiv2103.07535}, 2021.

\bibitem{hen15b}
O.~Hen, L.~B. Weinstein, E.~Piasetzky, G.~A. Miller, M.~M. Sargsian, and
  Y.~Sagi.
\newblock Correlated fermions in nuclei and ultracold atomic gases.
\newblock {\em Phys. Rev. C}, 92:045205, Oct 2015.

\bibitem{Hen:2016kwk}
O.~Hen, G.~A. Miller, E.~Piasetzky, and L.~B. Weinstein.
\newblock {Nucleon-Nucleon Correlations, Short-lived Excitations, and the
  Quarks Within}.
\newblock {\em Rev. Mod. Phys.}, 89(4):045002, 2017.

\bibitem{Bodek2003}
A~Bodek and U~K Yang.
\newblock Higher twist, $\xi_w$ scaling, and effective {LO} {PDFs} for lepton
  scattering in the few {GeV} region.
\newblock {\em J. Phys. G}, 29(8):1899--1905, jul 2003.

\bibitem{Bradford:2006yz}
R.~Bradford, A.~Bodek, Howard~Scott Budd, and J.~Arrington.
\newblock {A New parameterization of the nucleon elastic form-factors}.
\newblock {\em Nucl. Phys. B, Proc. Suppl.}, 159:127--132, 2006.

\bibitem{ValenciaModel}
J.~Nieves, I.~Ruiz Simo, and M.~J.~Vicente Vacas.
\newblock Inclusive charged-current neutrino-nucleus reactions.
\newblock {\em Phys. Rev. C}, 83:045501, Apr 2011.

\bibitem{Megias:2016lke}
G.~D. Megias, J.~E. Amaro, M.~B. Barbaro, J.~A. Caballero, and T.~W. Donnelly.
\newblock Inclusive electron scattering within the susav2 meson-exchange
  current approach.
\newblock {\em Phys. Rev. D}, 94:013012, Jul 2016.

\bibitem{Megias:2016fjk}
G.D. Megias, J.E. Amaro, M.B. Barbaro, J.A. Caballero, T.W. Donnelly, and
  I.~Ruiz~Simo.
\newblock {Charged-current neutrino-nucleus reactions within the superscaling
  meson-exchange current approach}.
\newblock {\em Phys. Rev. D}, 94(9):093004, 2016.

\bibitem{Caballero:2006wi}
J.A. Caballero.
\newblock {General study of superscaling in quasielastic (e,e') and (nu, mu)
  reactions using the relativistic impulse approximation}.
\newblock {\em Phys. Rev. C}, 74:015502, 2006.

\bibitem{Gonzalez-Jimenez:2019qhq}
R.~González-Jiménez, A.~Nikolakopoulos, N.~Jachowicz, and J.M. Udías.
\newblock {Nuclear effects in electron-nucleus and neutrino-nucleus scattering
  within a relativistic quantum mechanical framework}.
\newblock {\em Phys. Rev. C}, 100(4):045501, 2019.

\bibitem{Gonzalez-Jimenez:2019ejf}
R.~González-Jiménez, M.B. Barbaro, J.A. Caballero, T.W. Donnelly,
  N.~Jachowicz, G.D. Megias, K.~Niewczas, A.~Nikolakopoulos, and J.M. Udías.
\newblock {Constraints in modeling the quasielastic response in inclusive
  lepton-nucleus scattering}.
\newblock {\em Phys. Rev. C}, 101(1):015503, 2020.

\bibitem{GENIEValenciaMEC}
J.~Schwehr, D.~Cherdack, and R.~Gran.
\newblock {GENIE} implementation of {IFIC Valencia} model for {QE}-like 2p2h
  neutrino-nucleus cross section.
\newblock {\em arXiv:1601.02038}, 2016.

\bibitem{Simo_2017}
I~Ruiz Simo, J~E Amaro, M~B Barbaro, A~De Pace, J~A Caballero, and T~W
  Donnelly.
\newblock Relativistic model of 2p-2h meson exchange currents in (anti)neutrino
  scattering.
\newblock {\em Journal of Physics G: Nuclear and Particle Physics},
  44(6):065105, apr 2017.

\bibitem{RUIZSIMO2016124}
I.~{Ruiz Simo}, J.E. Amaro, M.B. Barbaro, A.~{De Pace}, J.A. Caballero, G.D.
  Megias, and T.W. Donnelly.
\newblock Emission of neutron–proton and proton–proton pairs in neutrino
  scattering.
\newblock {\em Physics Letters B}, 762:124--130, 2016.

\bibitem{PhysRevLett.82.3212}
T.~W. Donnelly and Ingo Sick.
\newblock Superscaling in inclusive electron-nucleus scattering.
\newblock {\em Phys. Rev. Lett.}, 82:3212--3215, Apr 1999.

\bibitem{PhysRevC.60.065502}
T.~W. Donnelly and Ingo Sick.
\newblock Superscaling of inclusive electron scattering from nuclei.
\newblock {\em Phys. Rev. C}, 60:065502, Nov 1999.

\bibitem{PhysRevC.99.042501}
M.~B. Barbaro, J.~A. Caballero, A.~De~Pace, T.~W. Donnelly,
  R.~Gonz\'alez-Jim\'enez, and G.~D. Megias.
\newblock Mean-field and two-body nuclear effects in inclusive electron
  scattering on argon, carbon, and titanium: The superscaling approach.
\newblock {\em Phys. Rev. C}, 99:042501, Apr 2019.

\bibitem{PhysRevD.94.013012}
G.~D. Megias, J.~E. Amaro, M.~B. Barbaro, J.~A. Caballero, and T.~W. Donnelly.
\newblock Inclusive electron scattering within the susav2 meson-exchange
  current approach.
\newblock {\em Phys. Rev. D}, 94:013012, Jul 2016.

\bibitem{PhysRevD.94.093004}
G.~D. Megias, J.~E. Amaro, M.~B. Barbaro, J.~A. Caballero, T.~W. Donnelly, and
  I.~Ruiz Simo.
\newblock Charged-current neutrino-nucleus reactions within the superscaling
  meson-exchange current approach.
\newblock {\em Phys. Rev. D}, 94:093004, Nov 2016.

\bibitem{Megias_2018}
G~D Megias, M~B Barbaro, J~A Caballero, J~E Amaro, T~W Donnelly, I~Ruiz Simo,
  and J~W~Van Orden.
\newblock Neutrino{\textendash}oxygen {CC}0$\uppi$ scattering in the
  {SuSAv}2-{MEC} model.
\newblock {\em Journal of Physics G: Nuclear and Particle Physics},
  46(1):015104, dec 2018.

\bibitem{PhysRevD.99.113002}
G.~D. Megias, M.~B. Barbaro, J.~A. Caballero, and S.~Dolan.
\newblock Analysis of the minerva antineutrino double-differential cross
  sections within the susav2 model including meson-exchange currents.
\newblock {\em Phys. Rev. D}, 99:113002, Jun 2019.

\bibitem{Megias:2014qva}
G.D. Megias et~al.
\newblock {Meson-exchange currents and quasielastic predictions for
  charged-current neutrino-$^{12}C$ scattering in the superscaling approach}.
\newblock {\em Phys. Rev. D}, 91(7):073004, 2015.

\bibitem{Amaro:2017eah}
J.~E. Amaro, M.~B. Barbaro, J.~A. Caballero, A.~De~Pace, T.~W. Donnelly, G.~D.
  Megias, and I.~Ruiz~Simo.
\newblock Density dependence of 2p-2h meson-exchange currents.
\newblock {\em Phys. Rev. C}, 95:065502, Jun 2017.

\bibitem{fkr1971}
R.~P. Feynman, M.~Kislinger, and F.~Ravndal.
\newblock Current matrix elements from a relativistic quark model.
\newblock {\em Phys. Rev. D}, 3:2706--2732, Jun 1971.

\bibitem{Yang2009}
T.~Yang, C.~Andreopoulos, H.~Gallagher, K.~Hofmann, and P.~Kehayias.
\newblock A hadronization model for few-{GeV} neutrino interactions.
\newblock {\em Eur. Phys. J. C}, 63(1):1--10, 2009.

\bibitem{PYTHIA6}
Torbj\"{o}rn Sj\"{o}strand, Stephen Mrenna, and Peter Skands.
\newblock {PYTHIA} 6.4 physics and manual.
\newblock {\em J. High Energy Phys. 06 (2006) 026}, 2006.

\bibitem{geniecollaboration2021neutrinonucleon}
J.Tena-Vidal et~al.
\newblock Neutrino-nucleon cross-section model tuning in genie v3.

\bibitem{dytman2011fsi}
S.A. Dytman and A.S. Meyer.
\newblock Final state interactions in genie.
\newblock {\em AIP Conf. Proc.}, 1405:213, 2011.

\bibitem{mashnik2006fsi}
S.~G. Mashnik, A.~J. Sierk, K.~K. Gudima, and M.~I. Baznat.
\newblock Cem03 and laqgsm03: New modelling tools for nuclear applications.
\newblock {\em J. Phys. Conf. Ser.}, 41:340, 2006.

\bibitem{PhysRevLett.62.1350}
R.~M. Sealock, K.~L. Giovanetti, S.~T. Thornton, Z.~E. Meziani, O.~A.
  Rondon-Aramayo, S.~Auffret, J.-P. Chen, D.~G. Christian, D.~B. Day, J.~S.
  McCarthy, R.~C. Minehart, L.~C. Dennis, K.~W. Kemper, B.~A. Mecking, and
  J.~Morgenstern.
\newblock Electroexcitation of the \ensuremath{\Delta}(1232) in nuclei.
\newblock {\em Phys. Rev. Lett.}, 62:1350--1353, Mar 1989.

\bibitem{Dai:2018xhi}
H.~Dai et~al.
\newblock {First Measurement of the Ti$(e,e^\prime){\rm X}$ Cross Section at
  Jefferson Lab}.
\newblock {\em Phys. Rev. C}, 98(1):014617, 2018.

\bibitem{Day:1993md}
D.B. Day et~al.
\newblock {Inclusive electron nucleus scattering at high momentum transfer}.
\newblock {\em Phys. Rev. C}, 48:1849--1863, 1993.

\bibitem{PhysRevC.99.054608}
H.~Dai et~al.
\newblock First measurement of the $\mathrm{Ar}(e,{e}^{\ensuremath{'}})x$ cross
  section at jefferson laboratory.
\newblock {\em Phys. Rev. C}, 99:054608, May 2019.

\bibitem{nicolescu2009}
M.~I. Niculescu.
\newblock Inclusive resonance electroproduction data from hydrogen and
  deuterium and studies of quark-hadron duality, Ph.D. thesis, Hampton
  University, 2009.

\bibitem{PhysRevLett.85.1186}
I.~Niculescu et~al.
\newblock Experimental verification of quark-hadron duality.
\newblock {\em Phys. Rev. Lett.}, 85:1186--1189, Aug 2000.

\bibitem{malace2006}
S.~P. Malace.
\newblock Measurements of inclusive resonance cross sections for quark-hadron
  duality studies, Ph.D. thesis, Hampton University, 2006.

\bibitem{PhysRevD.12.1884}
S.~Stein, W.~B. Atwood, E.~D. Bloom, R.~L.~A. Cottrell, H.~DeStaebler, C.~L.
  Jordan, H.~G. Piel, C.~Y. Prescott, R.~Siemann, and R.~E. Taylor.
\newblock Electron scattering at 4\ifmmode^\circ\else\textdegree\fi{} with
  energies of 4.5-20 gev.
\newblock {\em Phys. Rev. D}, 12:1884--1919, Oct 1975.

\bibitem{Christy:2007ve}
M.~E. Christy and Peter~E. Bosted.
\newblock {Empirical fit to precision inclusive electron-proton cross- sections
  in the resonance region}.
\newblock {\em Phys. Rev. C}, 81:055213, 2010.

\bibitem{Bosted:2007xd}
P.~E. Bosted and M.~E. Christy.
\newblock {Empirical fit to inelastic electron-deuteron and electron-neutron
  resonance region transverse cross-sections}.
\newblock {\em Phys. Rev. C}, 77:065206, 2008.

\bibitem{maid}
D.~Drechsel, O.~Hanstein, S.S. Kamalov, and L.~Tiator.
\newblock A unitary isobar model for pion photo- and electroproduction on the
  proton up to 1 gev.
\newblock {\em Nucl. Phys.}, A645:145, 1999.

\bibitem{Kamano:2016bgm}
H.~Kamano, S.~X. Nakamura, T.~S.~H. Lee, and T.~Sato.
\newblock {Isospin decomposition of $\gamma N \to N^*$ transitions within a
  dynamical coupled-channels model}.
\newblock {\em Phys. Rev. C}, 94(1):015201, 2016.

\bibitem{Nakamura:2013zaa}
S.~X. Nakamura, H.~Kamano, T.~S.~H. Lee, and T.~Sato.
\newblock {Neutrino-induced meson productions off nucleon at forward limit in
  nucleon resonance region}.
\newblock {\em AIP Conf. Proc.}, 1663(1):070005, 2015.

\bibitem{cebafref}
Accelerator Science.
\newblock https://www.jlab.org/accelerator.

\bibitem{cebafreview}
Christoph~W. Leemann, David~R. Douglas, and Geoffrey~A. Krafft.
\newblock The continuous electron beam accelerator facility: Cebaf at the
  jefferson laboratory.
\newblock
  https://www.jlab.org/div\_dept/physics\_division/talks/Background/Accelerator/CEBAF\_Ann\_Rev\_2001.pdf.

\bibitem{osti_282366}
A~J Street, J~S.H. Ross, and S~M Harrison.
\newblock Final site assembly and testing of the superconducting toroidal
  magnet for the cebaf large acceptance spectrometer (clas).
\newblock {\em IEEE Transactions on Magnetics}, 32, 7 1996.

\bibitem{DriftChambersCLAS}
M.D Mestayer, D.S Carman, Burin Asavapibhop, F.J. Barbosa, P~Bonneau, S.B
  Christo, G.E Dodge, T~Dooling, W.S Duncan, S.A Dytman, R~Feuerbach, Gerard
  Gilfoyle, Vardan Gyurjyan, K.H Hicks, R.S Hicks, C.E Hyde-Wright, G~Jacobs,
  Andi Klein, F.J Klein, and Jaycee Yun.
\newblock The clas drift chamber system.
\newblock {\em Nuclear Instruments and Methods in Physics Research Section A:
  Accelerators, Spectrometers, Detectors and Associated Equipment},
  449:81--111, 07 2000.

\bibitem{Amarian:2001kk}
M.~Amarian et~al.
\newblock The clas forward electromagnetic calorimeter.
\newblock {\em Nuclear Instruments and Methods in Physics Research Section A:
  Accelerators, Spectrometers, Detectors and Associated Equipment},
  460:239--265, 2001.

\bibitem{Adams:2001kk}
G.~Adams et~al.
\newblock {The CLAS Cherenkov detector}.
\newblock {\em Nucl. Instrum. Meth. A}, 465:414--427, 2001.

\bibitem{Smith:2001kk}
E.~Smith et~al.
\newblock {The time-of-flight system for CLAS}.
\newblock {\em Nucl. Instrum. Meth. A}, 432:265--298, 1999.

\bibitem{Osipenko:2010sb}
M.~Osipenko et~al.
\newblock {Measurement of the Nucleon Structure Function F2 in the Nuclear
  Medium and Evaluation of its Moments}.
\newblock {\em Nucl. Phys. A}, 845:1--32, 2010.

\bibitem{Egiyan:2005hs}
K.~S. Egiyan et~al.
\newblock {Measurement of 2- and 3-nucleon short range correlation
  probabilities in nuclei}.
\newblock {\em Phys. Rev. Lett.}, 96:082501, 2006.

\bibitem{Protopopescu:2004vx}
D.~Protopopescu et~al.
\newblock {Survey of A(LT-prime) asymmetries in semi-exclusive electron
  scattering on He-4 and C-12}.
\newblock {\em Nucl. Phys. A}, 748:357--373, 2005.

\bibitem{Stavinsky:2004ky}
A.~V. Stavinsky et~al.
\newblock {Proton source size measurements in the eA ---\ensuremath{>} e-prime
  ppX reaction}.
\newblock {\em Phys. Rev. Lett.}, 93:192301, 2004.

\bibitem{Niyazov:2003zr}
R.~A. Niyazov et~al.
\newblock {Two nucleon momentum distributions measured in He-3(e,e-prime pp)n}.
\newblock {\em Phys. Rev. Lett.}, 92:052303, 2004.
\newblock [Erratum: Phys.Rev.Lett. 92, 099903 (2004)].

\bibitem{Egiyan:2003vg}
K.~S. Egiyan et~al.
\newblock {Observation of nuclear scaling in the A(e, e-prime) reaction at x(B)
  greater than 1}.
\newblock {\em Phys. Rev. C}, 68:014313, 2003.

\bibitem{e4vNature21}
M.~Khachatryan, A.~Papadopoulou, et~al.
\newblock {Electron-beam energy reconstruction for neutrino oscillation
  measurements}.
\newblock {\em Nature}, 599:565, 2021.

\bibitem{Mecking:2003zu}
B.~A. Mecking et~al.
\newblock {The CEBAF Large Acceptance Spectrometer (CLAS)}.
\newblock {\em Nucl. Instrum. Meth.}, A503:513--553, 2003.

\bibitem{t2kcollaboration2021improved}
K.~Abe et~al.
\newblock Improved constraints on neutrino mixing from the t2k experiment with
  $\mathbf{3.13\times10^{21}}$ protons on target, 2021.

\bibitem{ALIAGA2014130}
L.~Aliaga et~al.
\newblock Design, calibration, and performance of the minerva detector.
\newblock {\em Nucl. Instrum. Methods}, A743:130 -- 159, 2014.

\bibitem{Abe:2018uyc}
K.~Abe et~al.
\newblock {Hyper-Kamiokande Design Report}.
\newblock {\em arXiv:1805.04163}, 2018.

\bibitem{Acciarri:2015uup}
R.~Acciarri et~al.
\newblock {Long-Baseline Neutrino Facility (LBNF) and Deep Underground Neutrino
  Experiment (DUNE)}.
\newblock {\em arXiv:1512.06148}, 2015.

\bibitem{Hen:2014nza}
O.~Hen et~al.
\newblock {Momentum sharing in imbalanced Fermi systems}.
\newblock {\em Science}, 346:614--617, 2014.

\bibitem{Adams:2019iqc}
P.~Abratenko et~al.
\newblock {First Measurement of Inclusive Muon Neutrino Charged Current
  Differential Cross Sections on Argon at $E_\nu\sim$0.8 GeV with the
  MicroBooNE Detector}.
\newblock {\em Phys. Rev. Lett.}, 123(13):131801, 2019.

\bibitem{SimcRadiation}
Simc~Monte Carlo.
\newblock Simc Monte Carlo, https://hallcweb.jlab.org/wiki/
  index.php/SIMC\_Monte\_Carlo, 2020.

\bibitem{Mo:1968cg}
Luke~W. Mo and Yung-Su Tsai.
\newblock {Radiative Corrections to Elastic and Inelastic e p and mu p
  Scattering}.
\newblock {\em Rev. Mod. Phys.}, 41:205--235, 1969.

\bibitem{PhysRevC.64.054610}
R.~Ent, B.~W. Filippone, N.~C.~R. Makins, R.~G. Milner, T.~G. O'Neill, and
  D.~A. Wasson.
\newblock Radiative corrections for ${(e,e}^{\ensuremath{'}}p)$ reactions at
  gev energies.
\newblock {\em Phys. Rev. C}, 64:054610, Oct 2001.

\bibitem{Cruz-Torres:2019bqw}
R.~Cruz-Torres et~al.
\newblock {Comparing proton momentum distributions in $A=2$ and 3 nuclei via
  $^2$H $^3$H and $^3$He $(e, e'p)$ measurements}.
\newblock {\em Phys. Lett. B}, 797:134890, 2019.

\bibitem{Katori:2016yel}
Teppei Katori and Marco Martini.
\newblock {Neutrino--nucleus cross sections for oscillation experiments}.
\newblock {\em J. Phys. G}, 45(1):013001, 2018.

\bibitem{Markov20}
N.~Markov et~al.
\newblock Exclusive ${\ensuremath{\pi}}^{0}p$ electroproduction off protons in
  the resonance region at photon virtualities
  $0.4\phantom{\rule{4pt}{0ex}}{\mathrm{gev}}^{2}\ensuremath{\le}{Q}^{2}\ensuremath{\le}1\phantom{\rule{4pt}{0ex}}{\mathrm{gev}}^{2}$.
\newblock {\em Phys. Rev. C}, 101:015208, Jan 2020.

\bibitem{osipenkoThesis}
M.~Osipenko.
\newblock {\em A Kinematically Complete Measurement of the Proton Structure
  Function $F_2$ in the Resonance Region and Evaluation of its Moments}.
\newblock PhD thesis, Moscow State University, 2002.

\bibitem{PhysRevC.94.015503}
X.-G. Lu, L.~Pickering, S.~Dolan, G.~Barr, D.~Coplowe, Y.~Uchida, D.~Wark,
  M.~O. Wascko, A.~Weber, and T.~Yuan.
\newblock Measurement of nuclear effects in neutrino interactions with minimal
  dependence on neutrino energy.
\newblock {\em Phys. Rev. C}, 94:015503, Jul 2016.

\bibitem{PhysRevD.98.032003}
K.~Abe et~al.
\newblock Characterization of nuclear effects in muon-neutrino scattering on
  hydrocarbon with a measurement of final-state kinematics and correlations in
  charged-current pionless interactions at t2k.
\newblock {\em Phys. Rev. D}, 98:032003, Aug 2018.

\bibitem{Bodek2019}
A.~Bodek and T.~Cai.
\newblock Removal energies and final state interaction in lepton nucleus
  scattering.
\newblock {\em Eur. Phys. J. C 79, 293}, 2019.

\bibitem{HK}
K.~Abe et~al.
\newblock {Hyper-Kamiokande Design Report}.
\newblock {\em arXiv:1805.04163}, 2018.

\bibitem{clas12proposal}
e4$\nu$ collaboration.
\newblock Electrons for neutrinos: Addressing critical neutrino-nucleus issues.
\newblock
  https://misportal.jlab.org/pacProposals/proposals/1377/attachments/104415/Proposal.pdf.

\bibitem{PhysRevC.95.065501}
Andrew~P. Furmanski and Jan~T. Sobczyk.
\newblock Neutrino energy reconstruction from one-muon and one-proton events.
\newblock {\em Phys. Rev. C}, 95:065501, Jun 2017.

\bibitem{cholesky}
Golub and Van Loan (1996, p. 143), Horn and Johnson (1985, p. 407), Trefethen
  and Bau (1997, p. 174).

\bibitem{decomp}
Banerjee, Sudipto; Roy, Anindya (2014), Linear Algebra and Matrix Analysis for
  Statistics, Texts in Statistical Science (1st ed.), Chapman and Hall/CRC,
  ISBN 978-1420095388.

\bibitem{Hocker}
A. Hocker and V. Kartvelishvili, SVD approach to data unfolding, Nucl. Instrum.
  Meth. A 372 (1996) 469 [hep-ph/9509307].

\bibitem{Niyazov}
Rustam Niyazov.
\newblock {\em Measurement of Correlated Pair Momentum Distributions on
  $^3He$(e,e'pp)n with CLAS}.
\newblock PhD thesis, Old Dominion University, 2003.

\bibitem{Protop_elfidcut4}
Dan Protopopescu.
\newblock Fiducial cuts for electrons in the {CLAS}/{E}2 data at
  $4.4\;\mathrm{GeV}$.
\newblock {\em CLAS-NOTE 2000-007 JLAB}, 2000.

\bibitem{BZhang_elfid2}
Bin Zhang.
\newblock Electron fiducial cuts.
\newblock \url{https://www.jlab.org/Hall-B/secure/e2/bzh/efiducialcut.html},
  2003.

\bibitem{e4nuQE}
M.~Khachatryan and L.~Weinstein.
\newblock Validation of neutrino energy estimation using electron scattering
  data.
\newblock Technical report, Old Dominion University, November 2019.
\newblock CLAS-Note.

\bibitem{McLauchlan}
Steven McLauchlan.
\newblock {\em $\Delta$ electroproduction in $^{12}C$}.
\newblock PhD thesis, University of Glasgow, January 2003.
\newblock Appendix C.

\bibitem{Niyazov_posfidcut4}
Lawrence~Weinstein Rustam~Niyazov.
\newblock Fiducial cut for positive hadrons in {CLAS}/{E}2 data at
  $4.4\;\mathrm{GeV}$.
\newblock {\em CLAS-NOTE 2001-013 JLAB}, 2001.

\end{thebibliography}
\bibliographystyle{unsrt}
\end{singlespace}

\end{document}